\documentclass[preprint,3p]{elsarticle}
\usepackage{caption,subcaption}
\usepackage[colorlinks=true,
linkcolor=blue,
filecolor=black,      
urlcolor=blue,
citecolor=magenta]{hyperref}
\usepackage{soul} 
\usepackage{graphics,graphicx}
\usepackage{amssymb,epsfig,amsmath,array,mathtools}
\usepackage{color,wasysym}
\usepackage{tensor}
\usepackage[dvipsnames]{xcolor}

\usepackage{tikz}
\usetikzlibrary{calc}
\usetikzlibrary{arrows,decorations.pathmorphing,patterns}

\numberwithin{equation}{section} 

\newcommand{\RT}{\mathrm{III}}
\newcommand{\RN}{\mathrm{IX}}
\newcommand{\Ht}{\mathrm{H}}

\newcommand{\bY}{Y}

\definecolor{myBlue}{rgb}{0,0,0.9}

\definecolor{myPurple}{rgb}{0.6,0.0,0.6}

\begin{document}

\hfill \hbox{\small CERN-TH-2023-108}
\vskip 0.1cm
\hfill \hbox{\small NORDITA 2023-026}
\vskip 0.1cm
\hfill \hbox{\small QMUL-PH-23-09}
\vskip 0.1cm
\hfill \hbox{\small UUITP-14/23}
\vskip 0.1cm

\begin{frontmatter}
	
	\title{The gravitational eikonal: from particle, string \\ and brane collisions to black-hole encounters}
	
	\author[1,2]{Paolo Di Vecchia}
	\address[1]{The Niels Bohr Institute, Blegdamsvej 17, DK-2100 Copenhagen, Denmark}
	\address[2]{NORDITA, KTH Royal Institute of Technology and Stockholm University,\\ Hannes Alfv{\'{e}}ns v{\"{a}}g 12, SE-11419 Stockholm, Sweden}
	\ead{divecchi@nbi.dk}
	
	\author[3,2]{Carlo Heissenberg}
	\address[3]{Department of Physics and Astronomy, Uppsala University, Box 516, SE-75120 Uppsala, Sweden}
	\ead{carlo.heissenberg@physics.uu.se}
	
	\author[4]{Rodolfo Russo}
	\address[4]{Centre for Theoretical Physics, Department of Physics and Astronomy, \\ Queen Mary University of London, Mile End Road, E1 4NS London, United Kingdom}
	\ead{r.russo@qmul.ac.uk}
	
	\author[5,6]{Gabriele Veneziano}
	\address[5]{Theory Department, CERN, CH-1211 Geneva 23, Switzerland}
	\address[6]{Coll\`ege de France, 11 place M. Berthelot, 75005 Paris, France}
	\ead{gabriele.veneziano@cern.ch}

	\begin{abstract}
		Motivated by conceptual problems in quantum theories of gravity, the gravitational eikonal approach, inspired by its electromagnetic predecessor, has been successfully applied to the transplanckian energy collisions of elementary particles and strings since the late eighties, and to string-brane collisions in the past decade. After the direct detection of gravitational waves from black-hole mergers, most of the attention has shifted towards adapting these methods to the physics of black-hole encounters. For such systems, the eikonal exponentiation provides an amplitude-based approach to calculate classical gravitational observables, thus complementing more traditional analytic methods such as the Post-Newtonian expansion, the worldline formalism, or the Effective-One-Body approach.
		In this review we summarize the main ideas and techniques behind the gravitational eikonal formalism. We discuss how it can be applied in various different physical setups involving particles, strings and branes and then
 we mainly concentrate on the most recent developments, focusing on massive scalars minimally coupled to gravity, for which we aim at being as self-contained and comprehensive as possible.
	\end{abstract}
	
	
\end{frontmatter}

\tableofcontents
\newpage
\section{Introduction}
\label{sec:intro}

\subsection{Aims and outline}
\label{ssec:aims}

Gravitational scattering, as seen from a Quantum-Field-Theory (QFT) perspective, was not a popular subject among particle theorists until the mid eighties. When in 1965 Steven Weinberg studied infrared gravitons in his now classic paper \cite{Weinberg:1965nx}, he felt obliged to provide some ``reasons for now attacking this question". One, he said, is ``because I can", the second was that ``something might go wrong" \dots but, he immediately added, ``does not".
For 20 more years or so particle theory was so much focused on the newly-formulated Standard Model of non-gravitational interactions---and of the elementary particles affected by them---that little attention was paid to the gravitational force in spite of the daring 1974  Scherk--Schwarz proposal  \cite{Scherk:1974ca,Scherk:1974mc} that the old (and unsuccessful) hadronic string theory should be recycled as a viable way to reconcile General Relativity (GR) with the quantum theory.

The situation changed dramatically in 1984 following the so-called first string revolution triggered by the breakthrough work by Green and Schwarz \cite{Green:1984sg} making it at least plausible that some fully consistent (i.e.~anomaly free) superstring theories could actually be candidate theories for all known forces and elementary particles. Although that idea still belongs to the dream category, the Green--Schwarz development gave a strong motivation for studying its implications as a full-fledged theory of quantum gravity.
Starting in 1987 several groups \cite{Soldate:1986mk,tHooft:1987vrq,Muzinich:1987in,Amati:1987wq,Gross:1987kza,Amati:1987uf,Gross:1987ar,Sundborg:1988tb,Amati:1988tn,Mende:1989wt,Amati:1990xe,Amati:1992zb,Amati:1993tb} started to analyze, in the spirit of the ``thought (gedanken) experiments" of the old quantum mechanics days, quantum gravitational scattering at transplanckian energies\footnote{For simplicity in this introduction we work in four spacetime dimension. We remind the reader that the very concept of a transplanckian energy is quantum-mechanical: classical GR has no intrinsic mass or length scale ($G^{-1}$ is a mass per unit length in units where $c=1$ which we use throughout unless explicitly stated).}
\begin{equation}\label{highenergylimit}
	\sqrt{s} \gg M_P \equiv \sqrt{\frac{\hbar}{G}}\,,\quad
	\text{i.e.}\quad
	\frac{Gs}{\hbar}\gg1\,.
\end{equation}
Among the initial aims of those investigations we would like to mention:
\begin{itemize}
	\item Understanding how unitarity bounds on partial waves, which are violated at tree level, get restored by loop corrections. This question can be asked both in QFT and in a string-theory context.
	\item Connecting the high-energy, fixed-angle behavior of string scattering amplitudes to modifications of gravity at short distance.
	\item Studying regimes in which the process is expected, classically, to lead to black-hole formation and, quantum mechanically, to subsequent black-hole evaporation. The construction of a unitary $S$-matrix in such a context would guarantee that information is preserved thus solving Hawking's famous paradox \cite{Hawking:1975vcx,Hawking:1976ra}. Alternatively, find a breakdown of unitarity.
\end{itemize}
For scattering of elementary particles or strings which are much lighter than the Planck mass, the transplanckian regime \eqref{highenergylimit} constitutes an ultra-relativistic limit, and is essential for dealing with the above questions. First, because gravity becomes the dominant force only at sufficiently high energies and this allows one to obtain (almost) theory-independent results. Second, because high energies are necessary (although, as we shall see, not always sufficient) in order to probe short distances. And, finally, because one would like to deal with black holes whose radius is (much) larger than the Planck length $\ell_P = \sqrt{ G \hbar}$  in order to apply to them the semiclassical approximations used in \cite{Hawking:1975vcx}.

A high-energy limit naturally leads to a semiclassical approximation to gravitational scattering, simplifying considerably all the calculations. In non-relativistic quantum mechanics the semiclassical limit goes under the name of WKB approximation (see e.g.~\cite{Sakurai:2011zz,WeinbergLectures2015}). In a more general relativistic framework, it is associated with the so-called eikonal approximation, the basic tool to be described in this review article.
The early applications of the eikonal approximation have met with considerable success in some regimes, including those in which point-particles and extended objects (strings, branes) behave differently \cite{Amati:1987wq,Amati:1987uf,Amati:1988tn,Amati:1990xe,Amati:1992zb,Amati:1993tb,DAppollonio:2010ae,Martinec:2020cml,Ceplak:2021kgl}. Parts of this review will be dedicated to these aspects of gravitational scattering. By comparison, less progress was made on the regime where gravitational collapse is classically expected to occur. In particular whether and how unitarity is preserved is still an unanswered question. We shall therefore make a less detailed account of those aspects of the problem.

In the last few years, in the wake of the first direct observation of gravitational waves  by ground-based interferometers \cite{Abbott:2016blz,Abbott:2016nmj},  the same semiclassical approximation turned out to be a useful tool for studying collisions of very heavy objects, provided they remain at sufficiently large distances with respect to their size. Typical candidates are, of course,  astrophysical black holes which, besides being extremely compact, are characterized by the same quantum numbers as those of an elementary particle, i.e.~mass and spin (plus possibly some conserved charges) as long as the process is only sensitive to the geometry outside their horizon.

Since semiclassical black holes, and a fortiori astrophysical ones, are much heavier than $M_P$, one does not have to consider, for them, a highly relativistic regime in order to justify the approximation \eqref{highenergylimit} and, in particular, the eikonal approach. And for current physical applications it is often rather the opposite, i.e.~a nonrelativistic approximation is viable.
For  black holes such as those featuring in the events observed by LIGO/Virgo \cite{Abbott:2016blz}, with masses of about $m_i\sim 30 M_{\odot}$ (with $M_\odot\simeq 2\times 10^{30}$kg the solar mass), the ratio between the Schwarzschild radii $\frac{2Gm}{c^2}$ and the reduced Compton wavelength $\frac{\hbar}{m c}$ obeys
\begin{equation}\label{scalesmass}
	\frac{2 Gm^2}{\hbar c} \simeq 10^{79} \gg 1\,,
\end{equation}
Therefore, for such systems, the classical limit is obviously an excellent approximation. What is less obvious is the use of a point-particle approximation. This is expected to be good for black hole collisions at distances much larger than their Schwarzschild radii while for neutron stars the detailed interior of the star matters and has to be incorporated through some non-minimal coupling to gravity of the effective point particle.

It thus comes as no surprise that the traditional techniques for studying the collision and merger of compact astrophysical objects have been based on classical GR. Among the most important ones are those using numerical relativity (as pioneered in \cite{Pretorius:2005gq} and reviewed e.g. in~\cite{Gourgoulhon:2007ue}), the self-force approach~\cite{Barack:2018yvs}, the Post-Newtonian expansion (see e.g.~the nice review in \cite{Blanchet:2013haa}), and the Effective-One-Body (EOB) framework \cite{Buonanno:1998gg,Buonanno:2000ef},  although the latter does make use of quantum mechanical concepts for the determination of effective potential's parameters.

More recently, however, the idea that scattering amplitudes---the bread and butter of quantum field theory calculations---could be recycled for use in the physics of compact astrophysical systems has made its way in the scientific community~\cite{Neill:2013wsa,Damour:2016gwp,Damour:2017zjx,Bjerrum-Bohr:2018xdl,Cheung:2018wkq,Bern:2019nnu,Bern:2019crd,Bern:2021dqo,Bern:2021yeh,Bjerrum-Bohr:2022blt}. Furthermore, as stressed by Damour in~\cite{Damour:2017zjx}, even the ultra-relativistic high-energy regime we have already mentioned could provide very valuable information on the classical GR problem. As is also immediately clear from \eqref{scalesmass}, in an amplitude-based approach, one is not dealing here with straight perturbation theory, which can be thought as an expansion for ``small-$G$''. One should therefore identify the appropriate way of resumming certain infinite sets of diagrams.  The essential, simplifying feature, which is shared by WKB and eikonal methods, is that in the semiclassical limit the amplitude is controlled by a large phase, typically representing a large, classical action in units of $\hbar$, in which the coupling constant sits in the exponent. To obtain this form, it is crucial that particular contributions of entire classes of loop diagrams ``exponentiate", i.e.~appear to the appropriate power and with the right combinatoric factor to all loop orders. Sometimes this can be explicitly checked, at least at the level under scrutiny, sometimes it can be justified by other methods (like in the world-line approaches), and sometimes will just be assumed. 
The validity of the eikonal approximation thus needs to be checked on a case-by-case basis, but, as we shall discuss, it has been successfully applied to semiclassical scattering for both  massive and massless scattering \cite{Amati:1990xe,Collado:2018isu,DiVecchia:2019myk,DiVecchia:2019kta,Bern:2020gjj,DiVecchia:2021bdo}, in various spacetime dimensions \cite{KoemansCollado:2019ggb,Cristofoli:2020uzm}. Amplitude techniques, including also the eikonal, have been applied the case in which the colliding objects can be subject to tidal deformations \cite{Cheung:2020sdj,Bern:2020uwk,Cheung:2020gbf,Aoude:2020ygw,AccettulliHuber:2020dal,Mougiakakos:2022sic,Jakobsen:2022psy,Heissenberg:2022tsn} or carry spin \cite{Arkani-Hamed:2017jhn,Vines:2017hyw,Guevara:2018wpp,Chung:2018kqs,Maybee:2019jus,Guevara:2019fsj,Arkani-Hamed:2019ymq,Johansson:2019dnu,Chung:2019duq,Damgaard:2019lfh,Bautista:2019evw,Aoude:2020onz,Chung:2020rrz,Bern:2020buy,Guevara:2020xjx,Kosmopoulos:2021zoq,Aoude:2021oqj,Bautista:2021wfy,Chiodaroli:2021eug,Haddad:2021znf,Chen:2021kxt,Aoude:2022trd,Bern:2022kto,Alessio:2022kwv,FebresCordero:2022jts,Bautista:2022wjf,Bjerrum-Bohr:2023jau}.
Tidal effects \cite{Damour:1992qi,Damour:1993zn,Goldberger:2005cd,Hinderer:2007mb,Flanagan:2007ix,Damour:2009vw,Binnington:2009bb,Hinderer:2009ca,Kol:2011vg,Damour:2012yf,Favata:2013rwa} may in particular provide clues on the equation of states of neutron stars \cite{Baiotti:2016qnr}, on the nature of black holes \cite{Barack:2018yly} and on possible exotic
astrophysical objects \cite{Buonanno:2014aza,Cardoso:2019rvt,Baumann:2019ztm}.

A very important feature of the semiclassical/eikonal approximation is thus that it allows one to take easily the classical limit itself, usually through a saddle-point approximation. On the one hand agreement with classical expectations provides a check of the quantum result. On the other hand, and perhaps more interestingly, the eikonal approximation offers a new tool for computing classical observables. 
The perturbative nature of such calculations is recovered by looking for an expansion of the exponent, after the resummation.
In this way, the eikonal exponentiation provides a simple way to recover the classical limit by matching to the Post-Minkowskian (PM) regime, in which the colliding objects remain sufficiently far apart and interact weakly, or, equivalently, undergo sufficiently small deflections,
\begin{equation}\label{}
	\frac{G\sqrt s}{b} \ll 1\,.
\end{equation}
For reference, to obtain a rough estimate of this parameter, we may consider again the situation of a merger event in the early inspiral phase, as measurable by LIGO/Virgo's detectors, where the typical relative separation $r_0$ between the two objects is such that
\begin{equation}\label{}
	\frac{Gm_i}{r_0} \simeq \frac16\,.
\end{equation} 
This small but non-negligible ratio motivates us to investigate higher orders in the PM expansion, which translate in higher-order approximations of the eikonal phase \cite{Kabat:1992tb,Akhoury:2013yua,KoemansCollado:2019ggb,Cheung:2020gyp,DiVecchia:2020ymx,AccettulliHuber:2020oou,DiVecchia:2021ndb,DiVecchia:2021bdo,Heissenberg:2021tzo,Bjerrum-Bohr:2021vuf,Bjerrum-Bohr:2021din,Damgaard:2021ipf,Brandhuber:2021eyq,DiVecchia:2022nna}.
In this way, the eikonal exponentiation provides an alternative approach for computing classical gravitational observables from scattering amplitudes, complementing various types of EFT setups \cite{Goldberger:2004jt,Porto:2016pyg,Levi:2018nxp,Cheung:2018wkq,Cristofoli:2020uzm}, the KMOC framework \cite{Kosower:2018adc,Herrmann:2021lqe,Herrmann:2021tct,Cristofoli:2021vyo,Cristofoli:2021jas,Adamo:2022rmp,Adamo:2022qci} as well as PM worldline EFT methods \cite{Goldberger:2016iau,Kalin:2020mvi,Kalin:2020fhe,Kalin:2020lmz,Mogull:2020sak,Jakobsen:2021smu,Mougiakakos:2021ckm,Liu:2021zxr,Dlapa:2021npj,Jakobsen:2021lvp,Jakobsen:2021zvh,Riva:2021vnj,Dlapa:2021vgp,Jakobsen:2022fcj,Mougiakakos:2022sic,Jakobsen:2022psy,Kalin:2022hph,Dlapa:2022lmu,Jakobsen:2022zsx,Dlapa:2023hsl} (see~\cite{Damgaard:2023vnx} for a comparison between these two approaches).

As already mentioned the gravitational eikonal approximation can be justified in a large variety of situations and this review will try to cover as many of them as possible indicating, in each case, both the achievements and the challenges lying ahead. 
One such challenge is represented by inelastic processes, such as gravitational radiation or internal excitations (as it happens in string theory), in which eikonal phase becomes an operator \cite{Ciafaloni:2018uwe,Addazi:2019mjh,Damgaard:2021ipf,DiVecchia:2022nna,DiVecchia:2022owy,Cristofoli:2021vyo,DiVecchia:2022piu,Brandhuber:2023hhy,Herderschee:2023fxh,Georgoudis:2023lgf}.
We organized the material according to the order in $G$ at which the eikonal phase (or operator) is computed. 
We believe that by starting from tree-level and then working our way up to two loops, we could organize the material in a pedagogical style and make the presentation as accessible as possible.  At each loop level we will consider the case of pure Einstein gravity and of some supersymmetric extensions of it. At tree level and one loop, we will discuss the case of string theory as well.
In the same spirit, we also describe first the two-to-two amplitude and only then turn to radiative processes and to radiation reaction. 

A very interesting development of the eikonal methods that we shall not cover in this report concerns their application to scattering processes taking place in asymptotically AdS spacetimes, and their connection to holographic CFTs. This analysis has been initiated in~\cite{Brower:2006ea,Cornalba:2006xk,Cornalba:2006xm,Cornalba:2007zb,Cornalba:2008qf} focusing on the high-energy scattering of light states as was done in the eighties in flat space. Through the AdS/CFT duality, the eikonal approach highlighted the existence of bounds on particular couplings in holographic CFTs, see for instance \cite{Camanho:2014apa,Kulaxizi:2017ixa,Li:2017lmh,Meltzer:2019pyl}. The eikonal regime has been also used in the analysis of the CFT correlators with heavy operators~\cite{Kulaxizi:2018dxo,Karlsson:2019qfi,Li:2019zba,Li:2020dqm,Giusto:2020mup,Ceplak:2021wak} and applied to the study of tidal excitations in AdS~\cite{Antunes:2020pof}, and of higher point correlators~\cite{Costa:2023wfz}. Another recent development which we will not discuss is the use of the eikonal approach in the context of the so-called celestial CFT~\cite{deGioia:2022fcn}.

With this general plan in mind, let us discuss how the report is organized. In the remainder of the present section, we shall spell out our kinematics conventions, review the definition of partial waves and discuss qualitative features of different regimes of gravitational scattering.
In Section~\ref{sec:leading}, we provide a self-contained, elementary treatment of the leading eikonal resummation for a gravitational $2\to2$ process involving massless scalar objects. We show how to calculate the leading eikonal phase $2\delta_0$, prove that it exponentiates and how it directly gives the deflection angle. This very simple example is mainly meant to whet the reader's appetite and lends itself to two quite different derivations that lead to the same result. One is based on the amplitude for a single graviton exchange and the other one is based on solving the geodesic equation in the Aichelburg--Sexl netric. We also illustrate how partial-wave unitarity, which is violated at finite loop order, is recovered via the resummation.

In Section~\ref{sec:treelevel}, we calculate the leading-order eikonal phase $2\delta_0$ in several different field theories, ranging from  minimally coupled massive scalars to graviton scattering off a massive scalar, dilaton gravity, $\mathcal N=8$ supergravity, higher-derivative corrections of GR and scattering of spinning objects. We then turn to tree-level string amplitudes, for which we mainly focus on string-brane scattering (the analog of a probe-limit calculation) and also introduce the eikonal operator, which accounts for transitions between different excited modes of the string.
Section~\ref{sec:oneloop} is instead devoted to one-loop calculations, which allow us on the one hand to check the first constraint arising from the exponentiation of the tree-level result and on the other hand to calculate the first sub-leading correction to the eikonal phase $2\delta_1$. Like for the previous one, we discuss both fields and strings.
In Section~\ref{sec:radiation}, we encounter for the first time $2\to3$ amplitudes, which enter the discussion of the unitarity cuts of the $2\to2$ amplitudes. We review in detail this point, focusing on the interplay between momentum-space convolutions and their impact-parameter Fourier transforms. This also serves as an occasion to anticipate the calculation of the imaginary part of the two-loop results, which is instead presented in Section~\ref{sec:twoloop} for $\mathcal N=8$ supergravity and for GR. There, we also comment more in detail on  the definition of the impact parameter and on the connection between eikonal phase, radial action and phase shifts.

The remainder of the report is devoted to the inclusion of radiation in the final state.
This is done first following soft theorems as in Section~\ref{sec:eikopsoft}, restricting one's attention to low-frequency radiation, and then, to leading order in the PM expansion, but capturing the full spectrum, by exponentiating the tree-level $2\to3$ amplitude as in Section~\ref{SemiclEik}. We summarize our conclusions and prospects for the future in Section~\ref{sec:outloop}.

Several appendices are included in order to make the material sufficiently self-contained. \ref{app:fieldtheory} collects conventions about Feynman rules and useful results concerning Fourier transforms from momentum to impact-parameter space.
In \ref{app:probelim} we present the calculation of the deflection angle in the probe limit for several theories.
A brief \ref{LeviCivita} collects useful identities involving the completely antisymmetric tensor. \ref{app:string} contains material that serves as background for the string-theory content of Sections~\ref{sec:treelevel}, \ref{sec:oneloop}. In \ref{KinRela1} we summarize a list of kinematic relations that apply to the $2\to3$ process involving graviton emissions and include expressions for the waveforms that complement those presented in Section~\ref{SemiclEik}. Finally, in  \ref{app:asymptoticlimit} we illustrate how the on-shell metric fluctuation is linked to the asymptotic waveform.
 
\subsection{Kinematics and conventions}
\label{sec:kinematics}

Before proceeding further, let us spell out here for later convenience the main conventions that will be employed in the rest of the review.

The basic object of study of this work is the collision of two energetic objects with masses $m_1$ and $m_2$. These define the initial states of a scattering process.
For convenience, all external momentum vectors will be regarded as outgoing, so that $-p_1$ and $-p_2$ represent the physical momenta for such incoming particles. 
We work with the mostly-plus signature for the metric,
\begin{equation}
	\label{eq:mostly+}
	\eta_{\mu \nu} = \operatorname{diag}(-1,1,\ldots,1)\,,\qquad \mu,\nu=0,1,\ldots,D-1\,,
\end{equation}
so that
\begin{equation}\label{}
	p_1^2=-m_1^2\,,\qquad
	p_2^2=-m_2^2\,.
\end{equation}
When the two incoming states have nonzero masses, we define their velocities according to\footnote{While $p_{1,2}^\mu$ are past-directed to comply with the all-outgoing convention for external momenta, $v_{1,2}^\mu$ are future-directed, hence the minus signs in \eqref{eq:velocities} and also in \eqref{eq:barpi} below.}
\begin{equation}\label{eq:velocities}
	p_1^\mu= -m_1^{\phantom{\mu}} v_1^\mu\,,\qquad
	p_2^\mu= -m_2^{\phantom{\mu}} v_2^\mu\,,\qquad
	v_1^2=-1\,,\qquad
	v_2^2=-1\,.
\end{equation}
Then their relative speed is sized by the invariant 
\begin{align}\label{eq:sigma}
	\sigma = -\frac{p_1\cdot p_2}{m_1 m_2} = - v_1\cdot v_2 = \frac{1}{\sqrt{1-v^2}}\,,
\end{align}
where $v$ is the velocity of either particle as seen from the rest frame of the other one. Of course, $\sigma\to1^+$ in the near-static limit, while $\sigma\to\infty$ in the ultra-relativistic limit.
It is also convenient to introduce the ``dual'' velocities
\begin{equation}\label{check12}
	\check{v}_1^\mu = \frac{\sigma v_2^\mu-v_1^\mu}{\sigma^2-1}\,,
	\qquad
	\check{v}_2^\mu = \frac{\sigma v_1^\mu-v_2^\mu}{\sigma^2-1}\,,
	\qquad
	v_1\cdot \check v_1=v_2\cdot \check v_2=-1\,,
	\qquad 
	v_1 \cdot \check v_2= v_2 \cdot \check v_1=0\,.
\end{equation}
In particular, these vectors allow one to conveniently decompose any given vector $\xi^\mu$ in terms of its longitudinal and transverse components according to
\begin{equation}\label{decomposition}
	\xi^\mu = \xi_{\parallel1} \check v_1^\mu +  \xi_{\parallel2} \check v_2^\mu + \xi_\perp^\mu\,,
\end{equation}
with
\begin{equation}\label{xiv}
	\xi_{\parallel1}  = -\xi \cdot v_1\,,\qquad
	\xi_{\parallel2}  = -\xi \cdot v_2\,,\qquad
	\xi_\perp \cdot v_{1}=0\,,\qquad
	\xi_\perp \cdot v_{2}=0\,.
\end{equation}
In a center-of-mass frame, the initial momenta take the form
\begin{equation}\label{CoM}
	-p_1= (E_1, \vec{p}\,)\,,\quad 
	-p_2 = (E_2, -\vec{p}\,)\,,
\end{equation}
where $\vec p$ is a $(D-1)$-dimensional spatial vector. 
Let us collect here a few useful relations that link the relativistic factor $\sigma$
and the masses $m_{1,2}$ to the total energy $E=E_1+E_2$, the single-particle energies $E_{1,2}$ and the absolute value $p = |\vec p\,|$ of the spatial momentum in such a frame,
\begin{align}
	\label{Ep}
	E p &= m_1 m_2 \sqrt{\sigma^2-1}\,,\\ 
	\label{EE1E2}
	E &= E_1+E_2= \sqrt{m_1^2+2m_1m_2\sigma+m_2^2}\,,\\
	\label{E1E}
	E_1 &= \frac{m_1}{E}(m_1+\sigma m_2)\,,
	\\
	\label{E2E}
	E_2 &= \frac{m_2}{E}(m_2+\sigma m_1)\,.
\end{align}

We define the scattering amplitude for the process $\alpha\to\beta$ in the standard way by decomposing the $S$ operator according to $S=1+iT$ and by letting
\begin{equation}\label{Adef}
	\langle \beta | T | \alpha \rangle = (2\pi)^D \delta^{(D)}(P_\alpha+P_\beta) \mathcal A_{\alpha\to\beta}\,,
\end{equation}
where the momentum conserving delta function takes into account that all momenta are formally outgoing. We  normalize single-particle momentum eigenstates with mass $m$ according to the Lorentz-invariant convention
\begin{equation}\label{norminv}
	2\pi\theta(p^0) \delta(p^2+m^2)\langle p|-p'\rangle = (2\pi)^D\delta^{(D)}(p+p')\,,
\end{equation}
which is equivalent to the more common normalization\footnote{One can multiply both sides of \eqref{norminv} by $2\pi\theta(-p'^0) \delta(p'^2+m^2)$ to make it manifestly symmetric.}
\begin{equation}
	\langle p|-p'\rangle  =2 \sqrt{|\vec{p}\,|^2+m^2} (2\pi)^{D-1} \delta^{(D-1)} (\vec p+\vec p\,')\,.
	\label{}
\end{equation}

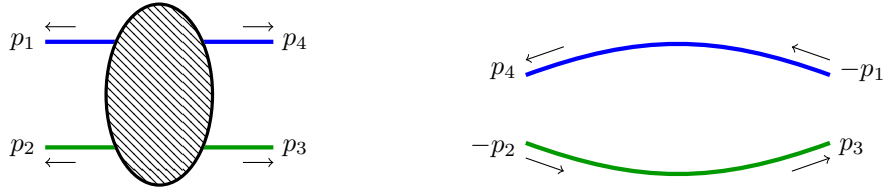
\begin{figure}
	\centering
	\begin{subfigure}{.45\textwidth}
		\centering
		\begin{tikzpicture}
			\path [draw, ultra thick, blue] (-4.5,2.2)--(-1.5,2.2);
			\path [draw, ultra thick, green!60!black] (-4.5,.8)--(-1.5,.8);
			\filldraw[white, very thick] (-3,1.5) ellipse (.7 and 1.2);
			\filldraw[pattern=north west lines, very thick] (-3,1.5) ellipse (.7 and 1.2);
			\draw [<-] (-4.5,2.4)--(-4.1,2.4);
			\draw [<-] (-1.5,2.4)--(-1.9,2.4);
			\draw [<-] (-4.5,.6)--(-4.1,.6);
			\draw [<-] (-1.5,.6)--(-1.9,.6);
			\node at (-1.5,2.2)[right]{$p_4$};
			\node at (-1.5,.8)[right]{$p_3$};
			\node at (-4.5,2.2)[left]{$p_1$};
			\node at (-4.5,.8)[left]{$p_2$};
		\end{tikzpicture}
	\end{subfigure}
	\begin{subfigure}{.45\textwidth}
		\centering
		\begin{tikzpicture}
			\draw[ultra thick, blue]  (-5,1.95).. controls (-3.5,2.5) and (-2.5,2.5) ..(-1,1.95);
			\draw[ultra thick, green!60!black] (-5,1.05).. controls (-3.5,.5) and (-2.5,.5) ..(-1,1.05);
			\draw [<-] (-5,2.15)--(-4.5,2.325);
			\draw [->] (-1,2.15)--(-1.5,2.325);
			\draw [->] (-5,.85)--(-4.5,.675);
			\draw [<-] (-1,.85)--(-1.5,.675);	
			\node at (-1,2)[right]{$-p_1$};
			\node at (-1,1)[right]{$p_3$};
			\node at (-5,2)[left]{$p_4$};
			\node at (-5,1)[left]{$-p_2$};
		\end{tikzpicture}
	\end{subfigure}
	\caption{\label{fig:kin4} To the left, a diagrammatic picture of the elastic $2\to2$ amplitude $\mathcal{A}$. To the right, a cartoon of two-body scattering in the center-of-mass frame.}
\end{figure}

A case that we will often consider in the following is that of an {\em elastic} $2\to2$ scattering depicted in Fig.~\ref{fig:kin4}.
We shall denote the corresponding amplitude simply by $\mathcal{A}$ and label the incoming particles by 1, 2 and the outgoing ones by 3, 4, in such a way that
\begin{equation}\label{}
	p_1^2=p_4^2=-m_1^2\,,\qquad
	p_2^2=p_3^2=-m_2^2\,.
\end{equation}
As already mentioned, all external momenta are regarded as outgoing so that they satisfy the conservation condition
\begin{equation}
	p_1+p_2+p_3+p_4=0
\end{equation}
and thus define the standard Mandelstam variables via
\begin{equation}
	\label{eq:mandvar}
	s=-(p_1+p_2)^2,\qquad t=-(p_1+p_4)^2,\qquad u=-(p_1+p_3)^2.
\end{equation}
As usual, $s$ is linked to the total energy in the center-of-mass frame $E$ by $E=\sqrt s$.
We also define the momentum transfer,
\begin{equation}
	\label{ppq}
	q=p_1+p_4=-p_2-p_3\,,
\end{equation} 
which is related to the Mandelstam invariant $t$ by
$t = -q^2$.
Of course, $u$ can be written in terms of $s$ and $t$ using momentum conservation and the mass-shell conditions,
\begin{equation}
	\label{eq:stum}
	s+t+u =2(m_1^2+m_2^2)\,.
\end{equation}
More explicitly, following the notation in \eqref{Adef},
\begin{equation}\label{2-to-2}
	\langle p_4,p_3 | T | -p_2,-p_1 \rangle = (2\pi)^D \delta^{(D)}(p_1+p_2+p_3+p_4) \mathcal A(s,t)\,.
\end{equation}
Let us note, for later convenience, that one can also factor the overall momentum-conserving delta function by recasting the $S$-matrix element as follows,
\begin{equation}\label{}
	2\pi\theta(p_4^0) \delta(p_4^2+m_1^2)
	2\pi\theta(p_3^0) \delta(p_3^2+m_2^2)
	\langle p_4,p_3| S | -p_2,-p_1 \rangle
	= 
	(2\pi)^D \delta^{(D)}(p_1+p_2+p_3+p_4) \mathcal S
\end{equation}
where, using $\delta^{(D)}(p_1+p_4)\delta^{(D)}(p_2+p_3)=\delta^{(D)}(p_1+p_2+p_3+p_4)\delta^{(D)}(q)$
for the disconnected piece, with $q$ as in \eqref{ppq},\footnote{For simplicity, we leave the $\theta$ functions implicit since they are irrelevant for sufficiently small $q$, given the fact that $p_4^0=-p_1^0+ q^0$ and $-p_1^0$ is positive. }
\begin{equation}\label{mathcalS}
	\mathcal S = \mathcal S(p_1,p_2;q) = (2\pi)^D \delta^{(D)}(q) + 2\pi \delta(2p_1\cdot q-q^2) 2\pi \delta(2p_2\cdot q+q^2) i \mathcal A(s,-q^2)\,.
\end{equation}
It can be convenient to introduce ``average'' momenta $\bar p_{i}^\mu$ and velocities $u_i^\mu$ according to \cite{Parra-Martinez:2020dzs}
\begin{equation}\label{eq:barpi}
	\begin{split}
		-p_1^\mu = \bar p_1^\mu - \frac12 q^\mu\,,\qquad
		p_4^\mu = \bar p_1^\mu + \frac12 q^\mu\,,\\
		-p_2^\mu = \bar p_2^\mu + \frac12 q^\mu\,,\qquad
		p_3^\mu = \bar p_2^\mu - \frac12 q^\mu\,,
	\end{split}
\end{equation}
and, for $i=1,2$,
\begin{equation}\label{averagevelocities}
	\bar p_i^\mu = \bar m_i u_i^\mu\,,\qquad u_i^2=-1\,.
\end{equation} 
In this way the mass-shell conditions turn into the following relations: by $p_1^2-p_4^2=0$ and $p_2^2-p_3^2=0$, 
\begin{equation}\label{}
	\bar p_i\cdot q=0\,,\qquad u_i\cdot q=0\,,
\end{equation}
for $i=1,2$,
while by $p_1^2+p_4^2=-2m_1^2$ and $p_2^2+p_3^2=-2m_2^2$,
\begin{equation}\label{}
	\bar m_1^2 = m_1^2 + \frac14 q^2\,,\qquad
	\bar m_2^2 = m_2^2 + \frac14 q^2\,.
\end{equation}
The Fourier transform of the $S$-matrix element $\mathcal S$ defined in \eqref{mathcalS} is then
\begin{equation}\label{FTexact}
	\int \frac{d^Dq}{(2\pi)^D}\,\mathcal S \,e^{ib\cdot q} = 1 + i  \int \frac{d^Dq}{(2\pi)^D}\, e^{ib\cdot q} 2\pi \delta(2\bar p_1\cdot q) 2\pi \delta(2\bar p_2\cdot q) \mathcal A(s,-q^2)
		=
	1+i\operatorname{FT}[\mathcal{A}](s,b)
	 \,.
\end{equation}

Going to a center-of-mass frame, one has 
\begin{equation}
	-p_1= (E_1, \vec{p}\,)\,,\quad 
	-p_2 = (E_2, -\vec{p}\,)\,,\quad
	p_3 = (E_2, - \vec{p}\,')\,,\quad
	p_4 = (E_1,  \vec{p}\,')\,,
	\label{comframe}
\end{equation}
where now $\vec{p}$ and $\vec{p}\,'$ are $(D-1)$-dimensional space vectors with $|\vec{p}\,|=|\vec{p}\,'|  \equiv p$. In this frame, $E_1$, $E_2$, $p$ obey \eqref{Ep}, \eqref{EE1E2}, \eqref{E1E}, \eqref{E2E},
and
\begin{equation}\label{}
	q^\mu= (0,\vec q\,)=(0,\vec p\,'-\vec p\,)\,,
	\qquad
	q^2 = |\vec q\,|^2 = 2p^2(1-\hat p\cdot \hat p\,')\,,
\end{equation}
where $\hat p = \vec p/p$ and similarly for $\hat p\,'$.
In fact, these properties correspond to further factorized forms of the $S$-matrix element \eqref{mathcalS},
\begin{align}\label{Scm}
	\mathcal S 
	&= 2\pi\delta(q^0) \left[
	(2\pi)^{D-1}\delta^{(D-1)}(\vec q\,)
	+
	i2\pi \delta(p-|\vec p+\vec q\,|) \frac{\mathcal A(s,-|\vec q\,|^2)}{4Ep}
	\right]\\
	\label{Scmangles}
	&=
	2\pi\delta(q^0) (2\pi)^{D-1} \frac{\delta(p-p')}{p^{D-2}}  \left[
	\delta^{(D-2)}(\hat p,\hat p'\,)
	+
	i\left(\frac{p}{2\pi}\right)^{D-2}\frac{\mathcal A(s,-2p^2(1-\hat p\cdot \hat p'))}{4Ep}
	\right]
\end{align}
where in the second line $\vec p\,' = \vec p + \vec q$ and $\delta^{(D-2)}(\hat p,\hat p')$ is the invariant delta function on the $(D-2)$-sphere (for instance $\delta^{(2)}(\hat p, \hat p'\,)=\delta(\theta-\theta')\delta(\phi-\phi')/\sin\theta$ in $D=4$ with standard spherical coordinates, which obviously  implies  $ \int_0^{2\pi} d\phi \int_0^\pi d \theta \, \sin \theta \, \delta^{(2)}(\hat p, \hat p'\,)=1$).

Aligning the ``average momenta'' so that $\bar p_1^\mu= (\bar E_1,0,\ldots,0,\bar p)$ and $\bar p_2^\mu= (\bar E_2,0,\ldots, 0,-\bar p)$,  one obtains
\begin{equation}\label{Breitframe}
	\begin{aligned}
		-p_1 &= \left(
		\bar E_1, -\frac{1}{2}\,\mathbf q, \bar p
		\right),\qquad p_4 =\left(
		\bar E_1, \frac{1}{2}\,\mathbf q, \bar p
		\right),\\
		-p_2 &= \left(
		\bar E_2, \frac{1}{2}\,\mathbf q, -\bar p
		\right),\qquad p_3 = \left(
		\bar E_2, -\frac{1}{2}\,\mathbf q, -\bar p
		\right),
	\end{aligned}
\end{equation}
and in this way
\begin{equation}
	q = (0,\mathbf q,0)\,,
\end{equation}
while the energies $\bar E_{1,2}$ can be obtained from the mass-shell conditions:
\begin{eqnarray}
	\bar E^2_{1,2} = {\bar{p}}^2 + \frac{q^2}{4} +m_{1,2}^2\,\,.
	\label{ener1}
\end{eqnarray}
As is clear from \eqref{Breitframe}, this parametrization corresponds to the so-called Breit (or brick-wall) frame, where each particle bounces back in the transverse directions and moves unperturbed along the longitudinal directions.

\begin{figure}
	\centering
	\begin{subfigure}{.45\textwidth}
		\centering
		\begin{tikzpicture}
			\path [draw, ultra thick, blue] (-4.5,2.2)--(-1.5,2.2);
			\path [draw, ultra thick, green!60!black] (-4.5,.8)--(-1.5,.8);
			\path [draw, thick, red] (-3,1.5)--(-1.5,1.5);
			\filldraw[white, very thick] (-3,1.5) ellipse (.7 and 1.2);
			\filldraw[pattern=north west lines, very thick] (-3,1.5) ellipse (.7 and 1.2);
			\draw [<-] (-4.5,2.4)--(-4.1,2.4);
			\draw [<-] (-1.5,2.4)--(-1.9,2.4);
			\draw [<-] (-4.5,.6)--(-4.1,.6);
			\draw [<-] (-1.5,1.7)--(-1.9,1.7);
			\draw [<-] (-1.5,.6)--(-1.9,.6);
			\node at (-1.5,2.2)[right]{$k_1$};
			\node at (-1.5,.8)[right]{$k_2$};
			\node at (-4.5,2.2)[left]{$p_1$};
			\node at (-4.5,.8)[left]{$p_2$};
			\node at (-1.5,1.5)[right]{$k$};
		\end{tikzpicture}
	\end{subfigure}
	\begin{subfigure}{.45\textwidth}
		\centering
		\begin{tikzpicture}
			\draw[ultra thick, blue]  (-5,1.95).. controls (-3.5,2.5) and (-2.5,2.5) ..(-1,1.95);
			\draw[ultra thick, green!60!black] (-5,1.05).. controls (-3.5,.5) and (-2.5,.5) ..(-1,1.05);
			\draw[thick,red] (.5,1.95) .. controls (.6,1.85) and (.6,1.15) .. (.5,1.05);
			\draw[thick,red] (.3,1.90) .. controls (.4,1.8) and (.4,1.2) .. (.3,1.10);
			\draw[thick,red] (.1,1.85) .. controls (.2,1.75) and (.2,1.25) .. (.1,1.15);
			\draw [<-] (-5,2.15)--(-4.5,2.325);
			\draw [->] (-1,2.15)--(-1.5,2.325);
			\draw [->] (-5,.85)--(-4.5,.675);
			\draw [<-] (-1,.85)--(-1.5,.675);	
			\draw [<-] (.7,1.5)--(.2,1.5);	
			\node at (-1,2)[right]{$-p_1$};
			\node at (-1,1)[right]{$k_2$};
			\node at (-5,2)[left]{$k_1$};
			\node at (-5,1)[left]{$-p_2$};
			\node at (.7,1.5)[right]{$k$};
		\end{tikzpicture}
	\end{subfigure}
	\caption{\label{fig:kin5} To the left, a diagrammatic picture of the inelastic $2\to3$ amplitude $\mathcal{A}^{(5)}$. To the right, a cartoon of two-body scattering with emission of gravitational radiation.}
\end{figure}
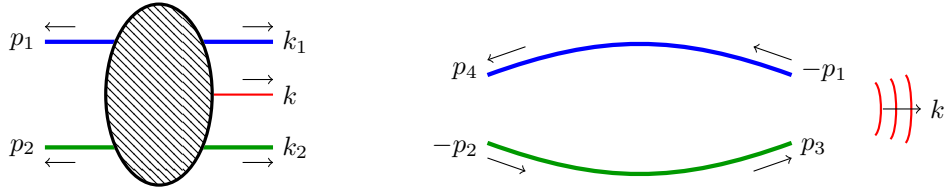

Another amplitude that we will employ in the following, especially when dealing with emitted radiation, is the $2\to3$ amplitude $\mathcal{A}^{(5)}$ describing the collision of two energetic particles with incoming momenta $-p_1$, $-p_2$ and outgoing momenta $k_1$, $k_2$ with the additional emission of a single graviton with momentum $k$, as depicted in Fig.~\ref{fig:kin5}
so that
\begin{equation}\label{averagesplitting}
	p_1^2=p_4^2=-m_1^2\,,\qquad 
	p_2^2=p_3^2=-m_2^2\,,\qquad
	k^2=0
\end{equation}
and
\begin{equation}
	p_1+p_2+k_1+k_2+k=0\,.
\end{equation} 
In this case, we find it convenient to define two ``momentum transfers''
\begin{equation}
	q_1 = p_1+k_1\,,\qquad q_2 = p_2+k_2\,,\qquad
	q_1+q_2+k=0\,.
\end{equation} 
Employing the velocities \eqref{eq:velocities}, the amplitude will depend on the invariants
\begin{equation}\label{invariants}
	\sigma = -v_1\!\cdot\! v_2 \ge 1\,,
	\qquad
	\omega_1 = -v_1\!\cdot\! k \ge 0\,,
	\qquad 
	\omega_2 = -v_2\!\cdot\! k \ge 0
\end{equation}
where $\omega_j$ is the graviton's frequency as seen in the rest frame of particle $j$ (for $j=1,2$).
Using \eqref{decomposition}, we can decompose $k^\mu$ as 
\begin{equation}\label{}
	k^\mu = \omega_1 \check v_1^\mu + \omega_2 \check v_2^\mu + k_\perp^\mu\,,
\end{equation}
and $k^2=0$ implies
\begin{equation}\label{calP}
	\frac{\mathcal P}{\sigma^2-1} = k_\perp^2\ge0 \,,\qquad \mathcal P = -\omega_1^2 + 2\omega_1\omega_2\sigma-\omega_2^2\,.
\end{equation}
As we shall discuss, the classical limit is obtained by considering $q_1$, $q_2$, $k$ simultaneously small compared with the incoming particles' momenta. Therefore the exact mass-shell conditions for the massive states
\begin{equation}\label{}
	2p_1\cdot q_1 = q_1^2\,,\qquad
	2p_2\cdot q_2 = q_2^2\,,
\end{equation}
read to leading order\footnote{We could introduce ``average'' momenta that are exactly orthogonal to $q_1$, $q_2$ respectively, analogous to $\bar p_{1,2}^\mu$ introduced in \eqref{averagevelocities}, but we will not need the relations \eqref{p1q1p2q2} beyond leading order.}
\begin{equation}\label{p1q1p2q2}
	p_1\cdot q_1\approx 0\,,\qquad
	p_2\cdot q_2\approx 0\,.
\end{equation}
In this way, we find
\begin{equation}\label{q1decompq2decomp}
	q_1^\mu \approx -\omega_2 \check v_2^\mu + q_{1\perp}^\mu\,,\qquad
	q_2^\mu \approx -\omega_1 \check v_1^\mu + q_{2\perp}^\mu\,,
\end{equation}
as follows from the last two equations in \eqref{check12}, and 
\begin{equation}\label{}
	q_1^2 \approx \frac{\omega_2^2}{\sigma^2-1} + q_{1\perp}^2\,,\qquad
	q_2^2 \approx \frac{\omega_1^2}{\sigma^2-1} + q_{2\perp}^2\,.
\end{equation}
This follows immediately from $\check{v}_1^2 = \check{v}_2^2= \frac{1}{\sigma^2-1}$ and from the last equation in \eqref{xiv} for $\xi= q_i$.

In a $2\to2+N$ process with emission of $N$ gravitons, the conservation condition would read
\begin{equation}\label{}
	p_1+p_2+k_1+k_2+P=0\,,
\end{equation}
with $P$ the sum of all graviton momenta. Squaring this relation one finds
\begin{equation}\label{sigmaprime}
	\sigma' \equiv -\frac{k_1\cdot k_2}{m_1m_2} = \sigma - \frac{2P\cdot (p_1+p_2)+P^2}{2m_1 m_2}\,,
\end{equation}
where $P\cdot (p_1+p_2)/\sqrt{s}=E_\text{rad}$ is the energy lost to graviton emissions in the (incoming) center-of-mass frame, while $-P^2>0$ represents the gravitons' invariant mass squared.
Clearly, by energy conservation $E_\text{rad}\le \sqrt s$, and thus 
\begin{equation}\label{}
	2P\cdot (p_1+p_2)+P^2=2\sqrt{s}\,E_\text{rad}-(E_\text{rad})^2+\vec P_\text{\,rad}^2\ge (E_\text{rad})^2+\vec P_\text{\,rad}^2\ge0\,.
\end{equation}
Therefore \eqref{sigmaprime} shows that the massive particles' relative velocity always decreases as a result of the emissions, $\sigma'\le \sigma$.
We shall return to this point in Section~\ref{SemiclEik} and in the outlook.

\subsection{Partial waves, unitarity, phase shifts and the eikonal phase}
\label{ssec:pwunitarity}

For later convenience, let us recall how the $ 2 \to 2$ scattering amplitude can be decomposed into partial waves by expressing it in terms of angular momentum eigenstates, and how unitarity holds for each partial wave. For simplicity we will discuss the procedure in $D=4$ although generalization to arbitrary $D$ is straightforward (see for instance~\cite{Muzinich:1987in,Giddings:2009gj}). We shall also focus on scalar massive particles, see e.g.~\cite{Bellazzini:2022wzv,Buric:2023ykg} for generalizations to the spinning case.

Let us start from the $S$-matrix element obtained in \eqref{Scmangles} by factoring out the motion of the center of mass, which we may regard as the definition of the ``reduced'' $S$ and $T$ operators in this frame,
\begin{equation}\label{Tdecomp}
	\langle \hat p\,'| \mathcal{S} |\hat p \rangle= \delta^{(2)}(\hat p,\hat p') + i\langle \hat p\,'| T |\hat p \rangle\,,\qquad\langle \hat p\,'| T |\hat p \rangle=\left(\frac{p}{2\pi}\right)^2\frac{\mathcal A(s,-q^2)}{4 E p}
	\,,
\end{equation}
where 
\begin{equation}\label{}
	\langle \hat p\,'|\hat p\rangle= \delta^{(2)}(\hat p,\hat p')\,,\qquad
	q = 2p\sin\frac{\theta}{2}\,,\qquad \hat p \cdot \hat p'\equiv\cos\theta\,.
\end{equation}
It is convenient to go to a basis with well-defined properties under rotations by means of the spherical harmonics 
$Y_{j m}(\hat p\,)$, 
\begin{equation}\label{jmtohatp}
	|j,m\rangle= \int d\Omega(\hat p)\,Y_{jm}(\hat p) |\hat p \rangle\,,
	\qquad
	\langle j',m'|j,m\rangle= \delta_{jj'} \delta_{mm'}\,.
\end{equation}
For concreteness, aligning $\hat p\,'$ along the $z$ axis, the three components of ${\hat{p}}$ can be expressed in terms of $\theta$ and $\phi$ as ${\hat{p}}= (\cos \phi \sin \theta, \sin \phi \sin \theta, \cos \theta)$ and $d \Omega ({\hat{p}})= \sin \theta\, d\theta\, d\phi$ with $0\leq \phi < 2\pi$ and $0\leq \theta \le \pi$.
The states $|j,m\rangle$ in \eqref{jmtohatp} are eigenstates of the total angular momentum $J^2$ and of its component along a given axis, say $J_z$,
\begin{equation}\label{}
	J^2	|j,m\rangle = \hbar^2 j(j+1)	|j,m\rangle\,,\qquad
	J_z	|j,m\rangle = \hbar m	|j,m\rangle\,.
\end{equation}
Since we are dealing with collisions of  scalar particles, the invariance of $S$ and $T$ under rotations implies via the Wigner--Eckart theorem that their matrix elements take the following diagonal form,
\begin{equation}\label{WE}
	\langle j',m'| \mathcal{S} | j,m \rangle = s_j(s) \delta_{jj'} \delta_{mm'}\,,\qquad
	\langle j',m'| T | j,m \rangle = 2f_j(s) \delta_{jj'} \delta_{mm'}\,,
\end{equation}
where $f_j(s)$ are suitable, $m$-independent functions, which we may identify as partial waves, and of course
\begin{equation}\label{}
	s_j(s) = 1+2if_j(s)\,.
\end{equation}
It is also common to express $f_j(s)$ in terms of the so-called phase shifts $\delta_j(s)$, according to
\begin{equation}\label{2deltaj}
	1+2if_j(s) = e^{2i \delta_j(s)}\,,\qquad
	f_j(s) = \frac{e^{2i\delta_j(s)}-1}{2i} = e^{i\delta_j(s)}\,\sin\delta_j(s)\,.
\end{equation}
Therefore, using \eqref{jmtohatp} and the definition \eqref{WE} of $f_j(s)$
we obtain
\begin{equation}\label{}
	\langle \hat p\,'| T | \hat p \rangle = \sum_{j=0}^\infty \frac{2j+1}{2\pi} P_j(\hat p\cdot \hat p\,')\,f_j(s)\,,
\end{equation}
after recalling the addition theorem that links the spherical harmonics to the Legendre polynomials $P_j$, 
(see for instance  \href{https://dlmf.nist.gov/14.30}{DLMF (14.30.09)}) 
\begin{equation}\label{}
	\sum_{m=-j}^j Y^\ast_{jm}(\hat p) Y_{jm}(\hat p') = \frac{2j+1}{4\pi} P_j(\hat p\cdot \hat p\,')\,.
\end{equation}
Comparing with \eqref{Tdecomp}, we see that the original amplitude can be decomposed as follows,
\begin{align}
	\mathcal{A}(s,-q^2) &= \frac{8\pi E}{p} \sum_{j=0}^\infty (2j+1) f_j(s) P_j (\cos\theta)\,,
	\label{ACV17bdire}
	\\
	f_j(s) &= \frac{p}{16\pi E} \int_{-1}^{+1} d(\cos\theta)\, P_j(\cos\theta) \mathcal{A}(s,-q^2)\big|_{q = 2p\sin\frac{\theta}{2}},
	\label{ACV17b}
\end{align}
where the latter relation can be obtained from the former using the orthogonality properties of the Legendre polynomials,
\begin{equation}
	\label{eq:Ortho}
	\int_{-1}^{+1} dx \, P_j (x) P_k (x) = \frac{2}{2j+1} \delta_{jk}\,,\qquad 
	P_j (1) =1\,.
\end{equation}

The virtue of the partial waves $f_j(s)$ is that the full non-linear unitarity condition $S^\dagger S=1$, which takes the following form for the $T$-matrix,
\begin{equation}
	\label{eq:Tunit}
	-i (T - T^{\dagger}) = T^{\dagger} T\,,
\end{equation}
becomes diagonal in $j$ and its elastic contribution is particularly simple.
Indeed, by taking the expectation value of \eqref{eq:Tunit} on $\langle j,m| \cdots | j,m \rangle$ and inserting a complete set of states on the right-hand side, by \eqref{WE} one finds that the sum over intermediate two-body states simplifies to a single term, and the result is
\begin{equation}
	\label{eq:funit}
	\operatorname{Im}  f_j(s,j) = |f_j(s)|^2 + {\rm inelastic}\,,
\end{equation}
where each ($j$-dependent) inelastic contribution  is non negative. 
This implies the bound
\begin{equation}\label{boundfj}
	\operatorname{Im}  f_j(s,j) \ge |f_j(s)|^2 
\end{equation}
or the (weaker) inequality 
\begin{equation}\label{}
	|f_j(s)| \le 1\,.
\end{equation}
Indeed, \eqref{eq:funit} can be obtained starting directly from $S^\dagger S=1$ and recalling the definition of the phase shifts \eqref{2deltaj},
\begin{equation}\label{}
	|s_j(s)|^2 + {\rm inelastic} = 1\,,\qquad
	e^{-2\operatorname{Im}2\delta_j(s)}+ {\rm inelastic} = 1
\end{equation}
and thus one obtains the following bounds equivalent to \eqref{boundfj},
\begin{equation}
	\label{eq:fphshift}
	|s_j(s)|\le1\,,\qquad
	\operatorname{Im} 2\delta_j(s) \ge 0
\end{equation}
with the equality signs holding true for perfectly elastic scattering.
Note that, unlike the Froissart bound, which depends on the absence of massless particles, the bound on partial waves should also hold in the presence of long range forces, in particular for gravity.

The eikonal phase, which constitutes the focus of this report, is closely related to the phase shifts discussed above. 
The main difference is that the eikonal is introduced by performing the Fourier transform of the $S$-matrix elements rather than their partial wave decomposition
  \begin{equation}
    \label{eq:SFT1}
    \tilde{\mathcal{S}}(s,b) = 1+i\tilde{\mathcal{A}}(s,b) = 1 + \int \frac{d^{D-2}q}{(2\pi)^{D-2}}\,e^{ib\cdot q} \frac{i {\cal A}(s,-q^2)}{4Ep}\;,
  \end{equation}
 where we are assuming the kinematics of the Breit frame~\eqref{Breitframe}. The technical advantage is that one does not have to deal with Legendre (or Gegenbauer for $D>4$) polynomials, while the drawback is that the exact diagonal form in~\eqref{WE} is lost as we shall see in Eq.~\eqref{almostfactorized} below. However in the classical limit there is an approximate diagonalization and so, in analogy with~\eqref{2deltaj}, it is convenient to introduce the eikonal phase $2\delta(s,b)$
  \begin{equation}
    \label{eq:ep1st}
    \tilde{\mathcal{S}}(s,b) = 1+i\tilde{\mathcal{A}}(s,b) = (1 + \cdots ) \, e^{2i \delta(s,b)}\;,
  \end{equation}
  where the eikonal $2\delta$ scales as $1/\hbar$ while the dots in the prefactor stand for quantum corrections. The impact parameter $b$, unlike $j$, is a continuous variable directly related to a classical quantity. The precise relation between $2\delta_j(s)$ and $2\delta(s,b)$ will be discussed in Subsection~\ref{ssec:exponapp} and more generally in Subsection~\ref{ssec:bbj}, where we will also elaborate on the relation between~\eqref{eq:SFT1} and the covariant definition of the Fourier Transform introduced in~\eqref{invFT}.

\subsection{Regimes of \texorpdfstring{$2\to2$ }{2->2} scattering}

As intuitively clear, the classical impulse $Q$, i.e.~the total momentum transfer during a classical $2\to2$ collision, emerges from the quantum description via the exchange of a very large number of gravitons, each carrying a momentum of order $q$. 
The resummation mechanism described in the next sections will make this intuitive picture precise, effectively capturing an infinite number of graviton exchanges and thus predicting the classical limit of the amplitude from its conventional expansion in perturbation theory.
Expressing the amplitude as a function of the collision's impact parameter, this resummation leads to the exponential of the original perturbative amplitude. Of course, due to the multiple interactions, the transferred momentum $Q$ exchanged in the classical process is much larger than the perturbative momentum transfer $q$,
\begin{equation}
	Q\gg q\,.
\end{equation} 
In particular, the ratio $Q/q$ goes to infinity in the classical limit, i.e.~in the formal $\hbar\to0$ limit.

As far as the classical process is concerned, we are particularly interested in the regime where $Q$ is small in comparison with the centre-of-mass energy
\begin{equation}
	\label{eq:smggt}
	s \gg Q^2\,,
\end{equation}
while the masses are kept arbitrary so as to explore various limits.
A first regime is when the total center of mass energy is of the order of the masses
\begin{equation}
	\label{eq:BH scatt}
	s,\,m_1^2,\,m_2^2 \gg Q^2\quad \quad \mbox{black-hole scattering,}
\end{equation}
which we dub as ``black hole scattering''. Indeed, this can be seen as a relativistic, open-trajectory analogue of a black-hole binary system tracing out a  non-relativistic, closed trajectory motion during the inspiral phase. From the conceptual point of view, it is interesting to take the extreme case where the kinetic energy is much larger than the (large) rest mass energy of the external states
\begin{equation}
	\label{eq:ultrarel}
	s\gg m_i^2 \gg Q^2\quad \quad \mbox{high-energy scattering.}
\end{equation}
This ultrarelativistic limit is in principle different from the case where the masses are set to zero and one describes the scattering of two shock waves
\begin{equation}
	\label{eq:shockscatt}
	s\gg Q^2\,,\quad m_i^2 = 0 \qquad \mbox{shock-wave scattering,}
\end{equation}
as will become clear especially in Subsection~\ref{sec:eikopsoftur}.
It will be  interesting to compare the two cases above. In particular we will see that for some quantities, such as the deflection angle up to 3PM, the transition from $ m_i^2\gg Q^2 $ to $m_i^2\to 0$ is smooth, while others quantities have rather different behaviors in these two regimes. Finally another interesting case is that of an elastic process where the mass of one state is much larger than the mass and kinetic energy of the other one 
\begin{equation}
	\label{eq:problim}
	s,m_1^2\gg s-m_1^2-m_2^2,\, m_2^2 \gg Q^2\quad \quad \mbox{probe limit.}
\end{equation}
In this limit the light external state can be seen as a probe particle traveling in the gravitational background produced by the heavy object. Thus, in this case, the results can be checked by studying the geodetic motion of a test mass in a fixed classical metric.

\section{Leading exponentiation: an appetizer}
\label{sec:leading}
In this section
we describe, in the technically simpler massless case,
how to extract the leading eikonal phase $e^{2i\delta_0}$ both from perturbative quantum field theory, following \cite{Levy:1969cr,Kabat:1992tb,Akhoury:2013yua,Bjerrum-Bohr:2018xdl}, and from purely classical considerations, following \cite{tHooft:1987vrq,Fabbrichesi:1993kz}. We also show how $2\delta_0$ is directly linked to the deflection angle for the classical trajectory via a stationary-phase approximation, and illustrate some related properties of the amplitude that become manifest in its partial-wave decomposition.

\subsection{Minimally coupled massless scalars: from amplitudes to geometry} 
\label{ssec:delta0ampl} 

In order to illustrate the leading-order exponentiation in the simplest possible gravitational setup, let us start from the case of two massless scalar fields minimally coupled to gravity,
\begin{equation}\label{Sminimally2}
	S
	=
	\int 
	\frac{R}{2\kappa_D^2} \sqrt{-g}\, d^Dx
	-
	\frac12
	\sum_{a=1,2}
	\int
	\partial_\mu \phi_a\, g^{\mu\nu}\,\partial_\nu \phi_a\,
	\sqrt{-g}\,d^Dx\,,
\end{equation}
with $\kappa_D = \sqrt{8\pi G}_D$. 
Considering fluctuations about a flat Minkowski background $g_{\mu\nu} = \eta_{\mu\nu} + 2\kappa_D\, h_{\mu\nu}$ and adopting De Donder gauge leads to the Feynman rules described in \ref{appFeynRules}.
In particular, the propagators for the graviton and for the scalar with ``flavor'' index $a=1,2$ are given by
($G$ and $G_{a}$ are the same function, but it is somewhat helpful to keep track of their subscript)\footnote{More precisely, since the propagator is diagonal in ``flavor'' space, $G_{ab}(k) = G_a(k)\,\delta_{ab}$}
\begin{equation}\label{GGagrav2}
	G_{\mu\nu,\rho\sigma}(k) = P_{\mu\nu,\rho\sigma} G(k)\,, 
	\qquad
	G(k)=\frac{-i}{k^2-i0}\,,
	\qquad
	G_a(k_a) = \frac{-i}{k_a^2-i0}\,,
\end{equation}
where
\begin{equation}\label{Pmunurhosigma2}
	P_{\mu\nu,\rho\sigma}
	=
	\frac12\left(
	\eta_{\mu\rho}\eta_{\nu\sigma}+\eta_{\mu\sigma}\eta_{\nu\rho}-\,\frac{2}{D-2}\eta_{\mu\nu}\eta_{\rho\sigma}
	\right).
\end{equation}
Moreover, the leading scalar-graviton-scalar vertex reads
\begin{equation}\label{taupp2}
	\tau_a^{\mu\nu}(p,p') =- i\kappa_D
	\Big[
	p^\mu p'^\nu + p^\nu p'^\mu
	- \eta^{\mu\nu} (pp')
	\Big],
\end{equation}
where the scalar lines are regarded as both outgoing.
One may consider replacing $P_{\mu\nu,\rho\sigma}$ with the projector over physical graviton states (see \eqref{TTprojector} below), but this replacement is immaterial for our present purposes because $\tau^{\mu\nu}_a$ is transverse, i.e.~$\tau^{\mu\nu}_a p_\mu=0=\tau^{\mu\nu}_ap'_\mu$ (and thus also $\tau^{\mu\nu}_a k_\mu=0$ with $k^\mu=-p^\mu-p'^\mu$) when $p^\mu$ and $p'^\mu$ are on-shell. 

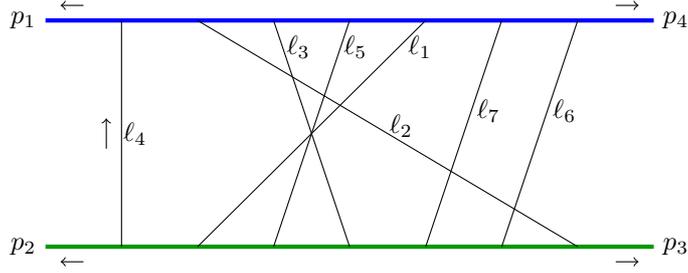
\begin{figure}
	\begin{center}
		\begin{tikzpicture}
			\draw[color=green!60!black, ultra thick] (-4,0) -- (4,0);
			\draw[color=blue, ultra thick] (-4,3) -- (4,3);
			\draw (-3,0) -- (-3,3);
			\draw (-2,0) -- (1,3);
			\draw (-1,0) -- (0,3);
			\draw (0,0) -- (-1,3);
			\draw (1,0) -- (2,3);
			\draw (2,0) -- (3,3);
			\draw (3,0) -- (-2,3);
			\node at (-4,3)[left]{$p_1$};
			\node at (-4,0)[left]{$p_2$};
			\node at (4,0)[right]{$p_3$};
			\node at (4,3)[right]{$p_4$};
			\node at (-3.1,1.5)[right]{$\ell_4$};
			\node at (-.95,2.65)[right]{$\ell_3$};
			\node at (-.2,2.65)[right]{$\ell_5$};
			\node at (.65,2.65)[right]{$\ell_1$};
			\node at (.4,1.6)[right]{$\ell_2$};
			\node at (1.55,1.8)[right]{$\ell_7$};
			\node at (2.55,1.8)[right]{$\ell_6$};
			\draw[<-] (-3.8,-.2)--(-3.5,-.2);
			\draw[<-] (3.8,-.2)--(3.5,-.2);
			\draw[<-] (-3.8,3.2)--(-3.5,3.2);
			\draw[<-] (3.8,3.2)--(3.5,3.2);
			\draw[<-] (-3.2,1.7)--(-3.2,1.3);
		\end{tikzpicture}
	\end{center}
	\caption{An example of (crossed) ladder topology. The thick line at the top (blue) refers to particle $1$ and the thick line at the bottom (green) refers to particle $2$. The exchanged graviton lines are attached and labeled by $\ell_1,\ldots,\ell_7$ in a random way. They flow from bottom to top, as indicated for the $\ell_4$ line.}
	\label{ladder}
\end{figure}

Let $\mathcal A(s,t)$ be the amplitude for the elastic scattering of $1$ and $2$. We shall consider its behavior in the Regge limit,
\begin{equation}\label{Regge-limit}
	s=-(p_1+p_2)^2\gg -t=-q^2\,.
\end{equation}
In this regime, the incoming particles are highly energetic and barely graze off each other, so that $p_1$, $p_2$ are formally very large compared to $q$, i.e.~the deflection is very small. 
Moreover, we shall focus on the contribution to the  amplitude at $L$ loops, $\mathcal A_L(s,t)$, that arises from the exchange of $n=L+1$
virtual gravitons between the two energetic lines as depicted in Fig.~\ref{ladder}. As we shall discuss shortly, these ``ladder'' diagrams indeed provide the dominant contribution to the $L$-loop amplitude in the Regge limit.
We further denote by $\ell_1,\ell_2,\ldots,\ell_n$ the momenta of such exchanges, flowing from $2$ to $1$ (see Fig.~\ref{ladder}).
Clearly this contribution takes the form
\begin{equation}\label{Angrav}
	i \mathcal A_{n-1}(s,t) = 
	\int 
	\left[\prod_{j=1}^{n}
	G(\ell_j)\,\frac{ d^D\ell_j}{(2\pi)^D} \right] (2\pi)^D\delta^{(D)}\left(q- \ell\right)\, J^{(n)},
\end{equation} 
where $\ell=\sum_j\ell_j$ and $J^{(n)}$ 
is given by a sum over all ladder topologies of suitable products of massless scalar propagators together with the appropriate contractions between the projectors $P_{\mu\nu,\rho\sigma}$ and the vertices $\tau^{\mu\nu}_a$. Before spelling out $J^{(n)}$ explicitly, let us examine one such contraction and retain only the leading order in the Regge or near-forward limit, $\ell_j \sim q\ll \sqrt s$. We find
\begin{equation}\label{tauapprox}
	\tau^{\mu\nu}_1(p_1-\ell_i,-p_1+\ell_j)  \simeq 2i\kappa_D p_1^\mu p_1^\nu
\end{equation}
and therefore
\begin{equation}\label{tauPtau}
	\tau^{\mu\nu}_1(p_1-\ell_i,-p_1+\ell_j) 
	P_{\mu\nu,\rho\sigma}
	\tau^{\rho\sigma}_2(p_2+\ell_k,-p_2-\ell_l) \simeq 
	-\kappa_D^2 s^2\,. 
\end{equation}
For instance, in the case of a single exchange (tree level),
\begin{equation}\label{explA1}
	\mathcal A_0(s,t)
	=
	\begin{gathered}
		\begin{tikzpicture}
			\draw[color=green!60!black, ultra thick] (-1.5,0) -- (-.5,0);
			\draw[color=blue, ultra thick] (-1.5,1) -- (-.5,1);
			\draw (-1,0) -- (-1,1);
		\end{tikzpicture}
	\end{gathered}
	\simeq
	\frac{\kappa_D^2 s^2}{q^2}
\end{equation}
where the subscript $0$ stands for tree level.
Consequently, to leading order, all dependence on vertices and index contractions factorizes according to
\begin{equation}\label{Jgrav}
	J^{(n)}
	\simeq
	\left(
	-\kappa_D^2s^2\right)^n
	I^{(n)},
\end{equation}
and we can rewrite \eqref{Angrav} as
\begin{equation}\label{Anapprox}
	i \mathcal A_{n-1}(s,t) 
	\simeq 
	\left(-\kappa_D^2 s^2\right)^{n}
	\int 
	\left[\prod_{j=1}^{n}
	G(\ell_j) \, \frac{ d^D\ell_j}{(2\pi)^D}\right] (2\pi)^D \delta^{(D)}\left(q- \ell\right)\, I^{(n)}
\end{equation}  
with $I^{(n)}$ a purely scalar object given by a sum over all ladder topologies of suitable products of $G_1$ and $G_2$ propagators, as we now describe.

\begin{figure}
	\centering
	\begin{subfigure}{\textwidth}
		\centering
		\begin{tikzpicture}
			\draw[color=green!60!black, ultra thick] (-1.5,0) -- (-.5,0);
			\draw[color=blue, ultra thick] (-1.5,1) -- (-.5,1);
			\draw (-1.4,0) -- (-1.4,1);
			\draw (-.6,0) -- (-.6,1);
			\draw[color=green!60!black, ultra thick] (1.5,0) -- (.5,0);
			\draw[color=blue, ultra thick] (1.5,1) -- (.5,1);
			\draw (.6,0) -- (1.4,1);
			\draw (1.4,0) -- (.6,1);
		\end{tikzpicture}
		\caption{Two-particle exchanges.}
		\label{twoexch}
	\end{subfigure}
	\par\bigskip
	\begin{subfigure}{\textwidth}
		\centering
		\begin{tikzpicture}
			\draw[color=green!60!black, ultra thick] (-1.5,0) -- (-.5,0);
			\draw[color=blue, ultra thick] (-1.5,1) -- (-.5,1);
			\draw (-1.4,0) -- (-.6,1);
			\draw (-1,0) -- (-1.4,1);
			\draw (-.6,0) -- (-1,1);
			\draw[color=green!60!black, ultra thick] (-3.5,0) -- (-2.5,0);
			\draw[color=blue, ultra thick] (-3.5,1) -- (-2.5,1);
			\draw (-3.4,0) -- (-3,1);
			\draw (-3,0) -- (-2.6,1);
			\draw (-2.6,0) -- (-3.4,1);
			\draw[color=green!60!black, ultra thick] (-5.5,0) -- (-4.5,0);
			\draw[color=blue, ultra thick] (-5.5,1) -- (-4.5,1);
			\draw (-5.4,0) -- (-5.4,1);
			\draw (-5,0) -- (-5,1);
			\draw (-4.6,0) -- (-4.6,1);
			\draw[color=green!60!black, ultra thick] (1.5,0) -- (.5,0);
			\draw[color=blue, ultra thick] (1.5,1) -- (.5,1);
			\draw (1.4,0) -- (1.4,1);
			\draw (1,0) -- (.6,1);
			\draw (.6,0) -- (1,1);
			\draw[color=green!60!black, ultra thick] (3.5,0) -- (2.5,0);
			\draw[color=blue, ultra thick] (3.5,1) -- (2.5,1);
			\draw (3.4,0) -- (3,1);
			\draw (3,0) -- (3.4,1);
			\draw (2.6,0) -- (2.6,1);
			\draw[color=green!60!black, ultra thick] (5.5,0) -- (4.5,0);
			\draw[color=blue, ultra thick] (5.5,1) -- (4.5,1);
			\draw (5.4,0) -- (4.6,1);
			\draw (5,0) -- (5,1);
			\draw (4.6,0) -- (5.4,1);
		\end{tikzpicture}
		\caption{Three-particle exchanges.}
		\label{threeexch}
	\end{subfigure}
	\caption{Ladder diagrams with two and three exchanged massless lines. The labels $\ell_1$, $\ell_2$ (and $\ell_3$) should be assigned in all possible ways.}
	\label{twoandthree}
\end{figure}
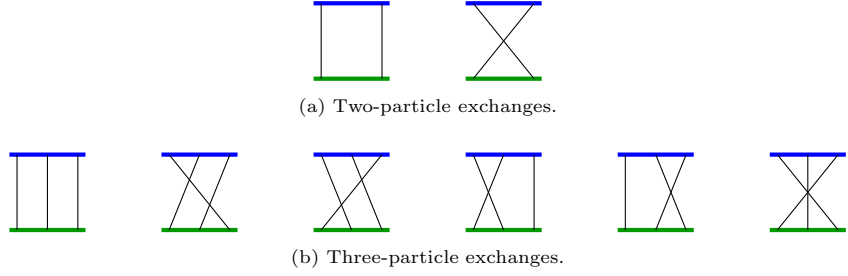

Summing all possible ways of attaching the exchanged lines and averaging over their labeling, we can write $I^{(1)}=1$ and
\begin{equation}\label{Inexch}
	\begin{split}
		I^{(n)} &= \frac{1}{n!}\sum_{\sigma\in\mathcal S_{n}}
		G_1(p_1-\ell_{\sigma_1})
		\cdots G_1(p_1-\ell_{\sigma_1}-\cdots-\ell_{\sigma_{n-1}})	\\
		&\times
		\sum_{\sigma'\in\mathcal S_{n}}
		G_2(p_2+\ell_{\sigma'_1})
		\cdots G_2(p_2+\ell_{\sigma'_1}+\cdots+\ell_{\sigma'_{n-1}})
	\end{split}
\end{equation}
for $n\ge2$,
where $\mathcal S_n$ is the permutation group for $n$ objects and $\frac{1}{n!}$ compensates for multiple counting. Note that, thanks to the explicit averages in eq.~\eqref{Inexch}, we solved for $\ell_{\sigma_n}=q-\ell_{\sigma_1}-\cdots -\ell_{\sigma_{n-1}}$ (and similarly for $\ell_{\sigma'_n}$) without spoiling the permutation symmetry of $I^{(n)}$. Two-particle and three-particle exchanges are illustrated in Fig.~\ref{twoandthree}.
Denoting by a subscript $\perp$ the components orthogonal to $p_1$ and $p_2$, to leading order in the Regge limit \eqref{Regge-limit} we can decompose the $D$-momentum conserving delta function as follows (the factor of $2s$ arises from the Jacobian of the appropriate change of variables, see Eq.~\eqref{detgmassless})
\begin{equation}\label{deltadecompose}
	\delta^{(D)}(q-\ell) \simeq  2s\ \delta(2p_1 \ell)\,
	\delta(2p_2 \ell)\, \delta^{(D-2)}(q_\perp-\ell_\perp)\,.
\end{equation}
Similarly, to leading order, we can make the approximation
\begin{equation}\label{G1approx}
	G_{1}(p_1-\ell_k) = \frac{-i}{-2p_1 \ell_k+\ell_k^2-i0}\simeq \frac{-i}{-2p_1 \ell_k-i0}\,,
\end{equation}
and similarly for $G_2(p_2+\ell_k)$.
In this way, we obtain
\begin{equation}\label{Insumsum}
	\begin{split}
		I^{(n)} &\simeq \frac{(-i)^{2(n-1)}}{n!}\,\sum_{\sigma\in\mathcal S_{n}}
		\frac{1}{-2 p_1 \ell_{\sigma_1}-i0}
		\cdots \frac{1}{-2 p_1 (\ell_{\sigma_1}+\cdots+\ell_{\sigma_{n-1}})-i0}	\\
		&\times
		\sum_{\sigma'\in\mathcal S_{n}}
		\frac{1}{2 p_2 \ell_{\sigma'_1}-i0}
		\cdots \frac{1}{2 p_2 (\ell_{\sigma'_1}+\cdots+\ell_{\sigma'_{n-1}})-i0}\,.
	\end{split}
\end{equation}
Letting
\begin{equation}\label{myf}
	f(a_1,\ldots,a_n)\equiv
	\sum_{\sigma\in\mathcal S_{n}} \frac{1}{a_{\sigma_1}}\cdots \frac{1}{a_{\sigma_1}+\cdots+a_{\sigma_{n-1}}}\,,
\end{equation}
we can then use the delta function cast in the form \eqref{deltadecompose} and the identity (see \ref{identities} and eq.~\eqref{indentitydeltas} in particular)
\begin{equation}\label{}
	\delta(\omega_1+\cdots + \omega_{n})\,
	f(\omega_1-i0,\ldots,\omega_n-i0) = (2i\pi)^{n-1} \delta(\omega_1)\cdots \delta(\omega_{n})
\end{equation}
to conclude that
\begin{equation}\label{iAnG}
	\begin{split}
		i \, \frac{\mathcal A_{n-1}(s,t)}{2s} &\simeq \frac{(-\kappa_D^2s^2)^{n}}{n!}
		\int 
		\left[\prod_{j=1}^{n}
		G(\ell_j)\delta(2 p_1 \ell_j)\delta(2 p_2 \ell_j)\,\frac{d^D\ell_j}{(2\pi)^{D-2}}\right] \\
		&\times
		(2\pi)^{D-2}\delta^{(D-2)}\left(q_\perp-\ell_\perp\right)
		.
	\end{split}
\end{equation}
This leads naturally to define the following Fourier transform with respect to the $(D-2)$-dimensional transverse momentum\footnote{See \ref{usefulFT}, in particular Eq.~\eqref{invFTLIN}, for more details concerning these Fourier transforms, including the relation between the two expressions appearing in \eqref{Aeibq}.} $q_\perp$,
\begin{equation}\label{Aeibq}
	\tilde{\mathcal A}(s,b)
	=
	\int
	 e^{ib q} 2\pi\delta(2 p_1 q)2\pi\delta(2 p_2 q)\,\mathcal A(s,-q^2)\frac{d^{D}q}{(2\pi)^{D}}
	=
	\int e^{ibq_\perp} \frac{\mathcal A(s,-q_\perp^2)}{2s}\,\frac{d^{D-2}q_\perp}{(2\pi)^{D-2}}\,,
\end{equation}
which yields
\begin{equation}\label{iAsb}
	i \tilde{\mathcal A}_{n-1}(s,b) \simeq \frac{(-\kappa_D^2s^2)^{n}}{n!} 
	\left[\int
	G(\ell) e^{ib \ell} 2\pi\delta(2 p_1 \ell)2\pi\delta(2 p_2 \ell)\,\frac{d^{D}\ell}{(2\pi)^{D}} \right]^n
	.
\end{equation}
The Fourier transform to the impact parameter $b$ simply had the effect of diagonalizing the convolution between single-particle exchanges occurring in momentum space.
The near-forward regime \eqref{Regge-limit} translates in impact-parameter space into the large-distance limit
\begin{equation}\label{largedistance}
	\frac{\hbar}{b}\ll \sqrt s\,,
\end{equation}
where $\hbar$ has been reinstated for clarity.
In conclusion, to leading order in this limit, adding up all contributions coming from $n$-graviton exchanges leads to
\begin{equation}\label{1+iAgrav}
	1+i\sum_{n=1}^\infty\tilde{\mathcal A}_{n-1}(s,b)
	\simeq
	e^{2i\delta_0}\,,
\end{equation}
with
\begin{equation}\label{eq:leik}
	2i\delta_0(s,b) = i \tilde {\mathcal{A}}_0(s,b) =  \frac{i \kappa_D^2 s}{2\hbar} \int \frac{e^{ib\ell_\perp}}{\ell_\perp^2}\,  \frac{d^{D-2}\ell_\perp}{(2\pi)^{D-2}}
	=
	\frac{iG_Ds}{\hbar}\,\frac{\Gamma\left(\frac{D-4}{2}\right)}{(\pi b^2)^{\frac{D-4}{2}}}\,.
\end{equation}
At this point, let us go back and explain what selects the $n$-graviton exchanges among all possible contributions to the elastic amplitude to all loop orders. To clarify this point, notice that when an extra graviton is attached to the energetic scalars the corresponding diagram is enhanced by a factor of $\kappa_D \sqrt s$, since the 3-point vertex itself brings a factor of $s$, see~\eqref{tauapprox} and~\eqref{tauPtau}, and the extra energetic propagator~\eqref{G1approx} scales as $\frac{1}{\sqrt s}$. If a graviton is added to the diagram in any other way, one obtains a contribution at the same order in the gravitational coupling, but without extra factors of $\sqrt s$. Notice that this is true also if higher-point vertices are used as for instance a contact interaction involving two gravitons and two scalars: since we are working with a two-derivative action~\eqref{Sminimally}, this vertex can scale at most as $\kappa_D^2 s$, while two 3-point vertices connected by a propagator~\eqref{G1approx} would yield $(\kappa_D s)^2 \frac{1}{\sqrt s}$. By dimensional analysis every time we lose a factor of $\sqrt s$, there must be an extra factor of $q\sim \frac{\hbar}{b}$ proving that non-ladder diagrams are subleading in the limit~\eqref{largedistance}.

In summary the classical, or \emph{eikonal}, limit requires $G_D s \,b^{4-D}\gg\hbar$ so the full amplitude oscillates infinitely rapidly as intuitively should be the case. By further taking the near-forward $q \ll \sqrt s$ or large-distance $b \sqrt s \gg \hbar $ limit, we can make the problem tractable, so  the regime of interest for our analysis is specified by the following hierarchy of length scales
\begin{equation}\label{hscales}
	\frac{\hbar}{\sqrt s} \ll G_D\sqrt s \,b^{4-D}\ll b \,,
\end{equation} 
where $\frac{\hbar}{\sqrt{s}}$ plays the role of a quantum wavelength and $G_D\sqrt s \, b^{4-D}$ is  the effective size of the curvature that characterizes the colliding objects. In the leading-order large $b$ approximation the full classical amplitude can be approximated by \eqref{1+iAgrav} and \eqref{eq:leik} which as we saw follows from the resummation of ladder diagrams.
The physically relevant $D=4$ of \eqref{hscales} is thus somewhat special, because the eikonal phase develops a $b$-independent infrared (IR) divergence: the combination $G_D \sqrt s$ coincides with (one half) the Schwarzschild radius of a black-hole of mass $\sqrt s$ and, from the finite part of the eikonal, one can read the effective size $G\sqrt{s}\,\log(b\mu)$ to be used in $D=4$ version of~\eqref{hscales}. It is natural to identify the IR cutoff $\mu$ with the inverse of the distance $r$ between the binary and a far-away observer.

In the regime~\eqref{hscales}, it is indeed natural to take the impact parameter $b$ and the energy $\sqrt{s}$ (or a  length scale $R$, defined in analogy with the Schwarzschild radius by $R^{D-3}=G_D\sqrt{s}$) as the classical quantities characterizing the  collision. In terms of these variables the leading eikonal is given by:
\begin{equation}
	2i \delta_0 = 	\frac{iG_Ds}{\hbar}\,\frac{\Gamma\left(\frac{D-4}{2}\right)}{(\pi b^2)^{\frac{D-4}{2}}} = i\frac{b\sqrt{s}}{\hbar}  \frac{G_D\sqrt{s}}{b^{D-3}} \frac{\Gamma\left(\frac{D-4}{2}\right)}{\pi^{\frac{D-4}{2}}}\,\,.
	\label{leadingeik}
\end{equation}
To reiterate, in the classical limit we must require that $R$ is much bigger than the Compton wavelength of a massless particle ($R \gg \frac{\hbar}{\sqrt{s}}$). On the other hand, in order to apply perturbation theory, we need that the interaction be weak and this corresponds to large values of $b$ ($b \gg R$).
In conclusion, Eq.~\eqref{leadingeik}
is valid for large values of $b$ and this is obtained by considering  the amplitude in the Regge limit ($|t| \ll s$).
The factor of $1/\hbar$ in~\eqref{leadingeik} signals that this quantity should indeed appear in an exponential $e^{2i \delta_0}$ so it can describe the value of the classical action, while all non-exponentiated terms are of order $\hbar^n$ with $n\geq 0$ and describe quantum corrections. 

\begin{figure}[ht]
	\centering
	\begin{tikzpicture}
		\draw[ultra thick, blue]  (-4,1.5).. controls (-1,2.1) and (1,2.1) ..(4,1.5);
		\draw[ultra thick, green!60!black] (-4,-1.5).. controls (-1,-2.1) and (1,-2.1) ..(4,-1.5);
		\draw [thick, <-] (-4,1.6)--(-3,1.8);
		\draw [dashed] (-3,1.8)--(0,2.4);
		\draw [thick, ->] (4,1.6)--(3,1.8);
		\draw [dashed] (3,1.8)--(0,2.4);
		\draw [thick, <-] (4,-1.6)--(3,-1.8);
		\draw [dashed] (3,-1.8)--(0,-2.4);
		\draw [thick, ->] (-4,-1.6)--(-3,-1.8);	
		\draw [dashed] (-3,-1.8)--(0,-2.4);
		\draw [thick, <-] (-4,-0.4)--(-3,-0.2);
		\draw [thick, <-] (-4,0)--(-3,-0.2);
		\draw [thick, ->] (-4.05,0)--(-4.05,-0.4);
		\draw[thick,red!80!black] (-3.5,-0.3) arc (210:150:0.2);
		\node at (-4,1.6)[left]{$\vec p\,'=\vec p_4$};
		\node at (-4,-1.6)[left]{$-\vec p=-\vec p_2$};
		\node at (4,-1.6)[right]{$- \vec p\,'=\vec p_3$};
		\node at (4,1.6)[right]{$\vec p=-\vec p_1$};
		\node at (-3.3,-0.2)[above]{$\Theta$};
		\node at (-4,-0.1)[left]{$\vec Q$};
		\node at (0,0)[left]{$\vec b$};
		\node at (0.4,0)[right]{$\vec b_J$};
		\node at (0,-1.4)[left]{$\dfrac{\Theta}{2}$};
		\draw [->] (0,-2.4)--(0,2.4);
		\draw [->] (0,-2.42)--(.9,2.2);
		\draw[red!80!black] (0,-1.4) arc (90:78.6901:1);
	\end{tikzpicture}
	\caption{Classical $2\to2$ scattering in the center-of-mass frame \eqref{comframe}. The difference between $b$ and $b_J$ is $\mathcal O(G^2)$ and thus negligible in this section.}
	\label{fig:theta}
\end{figure}
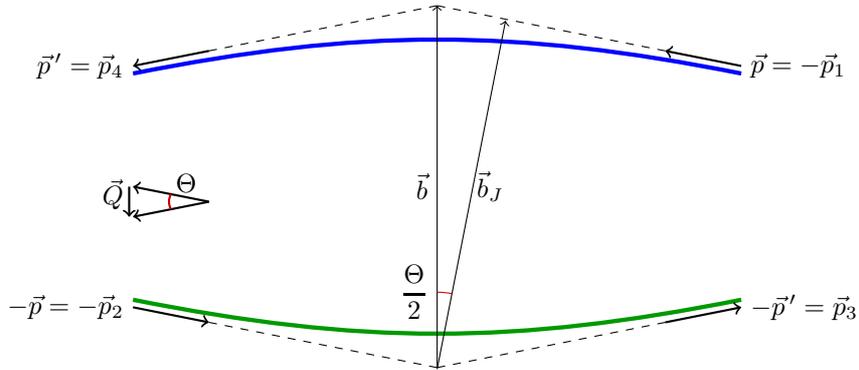

Let us now show how we can calculate from this semiclassical action two important classical observables: the impulse $Q$ and the deflection angle $\Theta$ depicted in Fig.~\ref{fig:theta}.
In the leading approximation, the exponentiation takes the form \eqref{1+iAgrav}, or briefly
\begin{equation}\label{1+iAtildereloaded}
	\tilde{\mathcal S}(s,b)=1+i \tilde {\mathcal{A}}(s,b) \simeq e^{2i\delta_0(s,b)}\,.
\end{equation}
After exponentiation in impact-parameter space, we can Fourier transform back to momentum space\footnote{$\mathcal S(s,-Q^2)$  should be distinguished from $\mathcal S(p_1,p_2;q)$ introduced in \eqref{mathcalS}, which also includes the mass-shell delta functions.}
\begin{equation}
	\label{eq:btq}
	\mathcal{S}(s,-Q^2) = 2 s \int d^{D-2} b\,  e^{-\frac{i}{\hbar} b Q+ 2i\delta_0(s,b)}\,.
\end{equation}
Here $Q$ describes the full exchanged momentum, or impulse, in the classical process, obtained after the eikonal resummation. Therefore it is important to distinguish it from $q$, the momentum exchanged in the perturbative amplitude in \eqref{explA1} via a single graviton. 
In the classical limit we can approximate \eqref{eq:btq} in terms of the stationary-phase contribution getting, at the saddle point,
\begin{equation}
	\label{eq:saddlebQ}
	Q^\mu  = \hbar \frac{\partial 2\delta_0}{\partial b_\mu}
	= - {2 G_Ds}\, \frac{\Gamma(\frac{D}{2}-1)}{\pi^{\frac{D}{2}-2} b^{D-3}}\, \frac{b^\mu}{b}\,,
\end{equation}
where $b$ denotes the magnitude of $b^\mu$.
Note that, unlike $2\delta_0$, the resulting leading-order impulse \eqref{eq:saddlebQ} is not singular as $D\to4$ thanks to the action of the derivative that eliminates the $b$-independent Coulombic pole at $D=4$.
In terms of the quantities introduced after Eq.~\eqref{hscales}, this becomes
\begin{equation}
	\label{eq:Rbsmall}
	\frac{Q}{\sqrt{s}} = \frac{2 \Gamma(\frac{D}{2}-1)}{\pi^{\frac{D}{2}-2}} \frac{G_D\sqrt{s}}{b^{D-3}} \ll 1 \;,
\end{equation}
where the last inequality holds in the limit of large distances where the impact parameter is much larger than $R$. 
On the other hand, from the Fourier transform~\eqref{Aeibq}, we have the usual relation $b \sim \hbar/q$ between the impact parameter and the momentum of a {\em single} graviton exchanged. Then the ratio between the classical impulse~\eqref{eq:saddlebQ}  and this single-graviton momentum transfer can be written as
\begin{equation}
	\label{eq:engr}
	\frac{Q}{q} \simeq \frac{2 G_Ds}{\hbar} \frac{\Gamma(\frac{D}{2}-1)}{\pi^{\frac{D}{2}-2} b^{D-4}} \simeq \frac{\Theta b}{\hbar} \gg 1
\end{equation}
and can be interpreted as the typical number of exchanged gravitons in the classical process. This makes it manifest that in the classical limit the number of gravitons exchanged is very large, as signaled by the factor of $\hbar^{-1}$, while classical observables, such as the impulse~\eqref{eq:Rbsmall} or the deflection angle (see \eqref{eq:Theta1PMmassless} below), remain small in the regime under consideration.

We can now use \eqref{eq:saddlebQ} and \eqref{eq:Rbsmall} to compute the classical deflection angle $\Theta$
defined as the angle between the space-like components of the momenta $-p_1$ and $p_4$ in the center-of-mass frame \eqref{comframe}, as shown in Fig.~\ref{fig:theta}.
In this frame $Q^2 =(p_1+p_4)^2= 2p^2 (1-\cos \Theta)$ where $\hat p\cdot \hat p\,'=\cos\Theta$, and thus
\begin{equation}
	\label{eq:thetap}
	Q = 2 p \sin\frac{\Theta}{2}  \quad \Rightarrow \quad  \sin\frac{\Theta}{2} = \frac{1}{2 p} \left(-\hbar \frac{\partial 2 \delta_0}{\partial b}\right),
\end{equation}
where the second expression is obtained by using~\eqref{eq:saddlebQ}. 
The minus sign comes from the fact that $|\partial_b 2\delta_0|=-\partial_b 2\delta_0$. Incidentally, this shows that the gravitational force is attractive because in this way the deflection of either particle points toward the other one (see Fig.~\ref{fig:theta}).
Since,
in the massless case,\footnote{The massive case will be discussed in Subsection~\ref{sec:grtree} below.}  $p=\frac{ \sqrt{s}}{2}$, and the condition \eqref{eq:Rbsmall}, which characterizes the eikonal regime, ensures that $Q/p$ is also small, we can expand $\sin\frac{\Theta}{2}\simeq \frac{\Theta}{2}$, obtaining the leading-order deflection angle
\begin{equation}
	\label{eq:Theta1PMmassless}
	\Theta \simeq \frac{4 \Gamma(\frac{D}{2}-1)}{\pi^{\frac{D}{2}-2}} \frac{G_D\sqrt{s}}{b^{D-3}} \ll 1\;.
\end{equation}

Similarly, one can go to the time domain by taking a Fourier transform over $E=\sqrt{s}$. Again a stationary phase approximation yields a relation between the eikonal and the Shapiro time delay $\Delta T$ which measures how much objects are slowed down by the gravitational force \cite{Maiani:1997pd,Ciafaloni:2014esa,Bellazzini:2021shn}
  \begin{equation}
    \label{eq:Std}
    \Delta T = \hbar\, \frac{\partial 2 \delta_0}{\partial E}\,,
  \end{equation}
as seen from the center-of-mass frame.
Contrary to what happens for the deflection angle~\eqref{eq:saddlebQ}, in this case the Coulomb divergence in $\delta_0$~\eqref{eq:leik} does not cancel, so a better IR-safe observable is for instance the variation of $\Delta T$ when $b$ changes.

As we shall further discuss in Section~\ref{ssec:ASmetric}, both deflection angle and time delay were found to agree with classical GR expectations in the approximation in which each particle produces a non-trivial curved metric that affects the geodetic motion of the other particle  \cite{tHooft:1987vrq,Ferrari:1988cc}. 
In conclusion, we have seen how the classical deflection and the time delay, which we may attribute to an effective metric, emerges from standard (quantum) scattering amplitude calculations on flat background spacetime. 
Notably, this avoids possible challenges, particularly in string theory, with quantization in curved space-time. See however \cite{Cornalba:2006xk,Cornalba:2006xm,Cornalba:2007zb,Cornalba:2008qf} for discussions of the eikonal exponentiation in AdS spacetime, as well as  \cite{Adamo:2021rfq,Adamo:2022rmp,Adamo:2022qci} for recent discussions in curved spacetimes. 

In the ensuing sections we will see that the eikonal exponentiation generalizes well beyond the simple case sketched here. While the basic idea is the same, it becomes more intricate when extended to the $R/b$ corrections (subleading eikonal) and  to theories beyond GR including more fields or higher derivative couplings or to string theory. This mechanism is expected to hold in all consistent gravitational theories, as we will discuss in some detail in this review.
However, first, we will comment on the connection between eikonal phase and partial waves, as well as on the interpretation of the eikonal phase as a classical action for shockwave scattering.

\subsection{Eikonal phase and phase shift}
\label{ssec:exponapp}

In this subsection we discuss the relation between the partial-wave decomposition introduced in Sect.~\ref{ssec:pwunitarity} and the eikonal phase. As we shall see, this will highlight how the eikonal exponentiation is also instrumental in solving an apparent tension between the explicit result for the tree-level amplitude \eqref{eq:leik} and the unitarity bound \eqref{eq:fphshift}.

The partial wave expansion of the resummed amplitude \eqref{eq:btq} is most easily performed for $D=4$, but in this case the leading eikonal  phase \eqref{leadingeik}  is divergent. On the other hand, after exponentiation, this divergence is an overall $b$-independent phase which can be ignored (this is the usual Coulomb divergence due to the long-range nature of the tree-level potential in three space dimensions). 
The finite $b$-dependent contribution is 
\begin{equation}\label{2idelta0finite}
	2 \delta_0(s,b) \to - \frac{G s}{\hbar} \log (\mu b)^2\,,
\end{equation}
with $\mu$ an inverse length scale introduced for dimensional reasons, so that the integral in~\eqref{eq:btq} can be performed exactly by using the general Fourier transform~\eqref{B1bis}\,,
\begin{equation}
	\mathcal{S}(s, -Q^2) \simeq i\, \frac{8 \pi G s^2}{Q^2} \left( \frac{4\mu^2 \hbar^2}{Q^2}\right)^{-i \alpha_G} \frac{\Gamma (1 -i \alpha_G)}{\Gamma(1+ i \alpha_G)}\,.
	\label{ACV2}
\end{equation}
Here   
\begin{equation}\label{}
	\alpha_G \equiv  \frac{Gs}{\hbar}
\end{equation}
is dimensionless and provides an analog of the fine structure constant of QED. We see that the expression \eqref{ACV2} for the resummed amplitude is equal to the one for single-graviton exchanges \eqref{explA1} with the substitution $q\rightarrow Q$ times two extra phase factors (the ratio of the two $\Gamma$-functions is also a pure phase).

In order to apply the partial wave projection \eqref{ACV17b}, we note that in the massless case $2p = \sqrt{s}$
and recall the Rodrigues formula,
\begin{equation}
	\label{eq:Rodigues}
	P_j(x) = \frac{1}{2^j j!} \frac{d^j}{dx^j} (x^2-1)^j\,.
\end{equation}
We then need to compute the following integral,
\begin{eqnarray}
	\frac{(-1)^j}{2^j j!}  \int_{-1}^{+1} dx \frac{d^j}{\partial x^j} (1-x^2)^j
	(1-x)^{-1 + i\alpha_G}\,.
	\label{ACV32}
\end{eqnarray}
Integrating by parts, one can easily bring all the derivatives with respect to $x$ on the term with $\alpha_G$,
then the integral can be performed by changing variable from $x$ to $z= \frac{1+x}{2}$ and recognizing that one obtains a quantity proportional to the Euler Beta-function. By including all the factors of the  amplitude~\eqref{ACV2} one obtains\footnote{At fixed $j$ that ratio has singularities along the imaginary $s$-axis, known as 't Hooft's poles. However, their presence depends on the large-$Q$ behavior of the resummed scattering amplitude which is beyond control of the leading eikonal approximation \cite{Amati:1992zb}.}
\begin{equation}
	\label{eq:iAj}
	{s}_j(s) = e^{2i\delta_j(s)} = \left( \frac{s}{4\mu^2\hbar^2} \right)^{i \alpha_G} \frac{\Gamma(1+j-i \alpha_G)}{\Gamma \left( 1 + j + i \alpha_G \right)}\,.
\end{equation}
We now consider the classical limit,
\begin{equation}\label{classicaljJ}
	j\sim \alpha_G \gg 1\,,\qquad J = \hbar j\,,
\end{equation}
where the quantum number $j$ takes large values, in order to lead to a sizable classical angular momenta $J$. 
Taking the limit \eqref{classicaljJ} in \eqref{eq:iAj}, and using the Stirling approximation, we find
\begin{equation}\label{radialactionLO}
	2\delta_j = -\left[
	\alpha_G \log(j^2+\alpha_G^2)
	+
	2j \arctan\frac{\alpha_G}{j}
	\right],
\end{equation}
where we used $\arctan\frac{\alpha_G}{j}=\frac{1}{2i} \log \frac{j+i\alpha_G}{j-i \alpha_G}$.
Moreover, to leading order in the PM limit $j\gg \alpha_G\gg1$, this simplifies to
\begin{equation}\label{largejnaive}
	2\delta_j(s) \simeq - \frac{G s}{\hbar} \log (j^2)\,,
\end{equation}
and thus, comparing with \eqref{2idelta0finite}, 
\begin{equation}\label{2delta0=2deltaJ}
	2\delta_0(s,b) \simeq 2\delta_j(s) 
	\,,\qquad
	\hbar j = J \simeq pb\,,
\end{equation}
up to subleading, $b$-independent corrections.   
At higher PM orders, as we shall see, it is instead important to maintain the distinction between $2\delta_j(s)$ and $2\delta(s,b)$. For this reason, we introduce a special symbol for $2\delta_j(s)$ in the classical limit,
\begin{equation}\label{2deltajchiJ}
	2\delta_{j}(s)  = \chi(s,J)\,,
\end{equation}
which is a function of the continuous variable $J=\hbar j$. We shall come back to this point in Subsection~\ref{ssec:bbj}.

This identification between $b$ and $j$-expansions to leading order, which is also natural by geometric considerations (see Fig.~\eqref{fig:theta}), seems to raise a problem with the partial wave unitarity bound \eqref{eq:fphshift}.
Indeed, focusing on the tree-level amplitude, $|\tilde{\mathcal{A}}_0(s,b)| \simeq |f_j(s)| \gg 1$ in the classical, high-energy regime, which seems to contradict \eqref{eq:fphshift}. 
However the resummed amplitude makes perfect sense since the large quantity $\tilde{\mathcal{A}}_0(s,b)=2\delta_0(s,b)\simeq 2\delta_j(s)$ appears only as a phase $e^{2i \delta_0}$. Thus, $\mathcal S(s,b)$ and $s_j(s)$ simply saturate the bound \eqref{eq:fphshift}, by neglecting inelastic contributions,
\begin{equation}\label{}
	|s_j(s)|^2=1\,,
\end{equation}
as one can also see directly from \eqref{eq:iAj}.
When taking inelastic processes into account, as we shall see in the following, $2\delta$ develops a {\em positive} imaginary part, so that
\begin{equation}\label{}
	|s_j(s)|^2 = e^{-2\operatorname{Im}2\delta_j(s)}<1
\end{equation}
and the bound is still respected, but no longer saturated.
In this way, the eikonal exponentiation is instrumental in resolving the apparent tension between the high-energy behavior of perturbative amplitudes in gravity and the bounds coming from the unitarity constraint. In fact, this was one of the main historical motivation for its study.

Let us now go back to the connection between partial waves and deflection angle.
Starting from \eqref{ACV17bdire}, we see that the $Q$-space representation of the amplitude is given by a sum of the type
\begin{equation}\label{sumj}
	\mathcal S(s,Q)\propto	\sum_j P_j(\cos\Theta) e^{2i\delta_j}\,,\qquad Q = 2p\sin\frac{\Theta}{2}\,.
\end{equation}
Using the large-$j$ limit of $P_j (\cos\Theta)$,
(see for instance  \href{https://dlmf.nist.gov/14.15}{DLMF (14.15.11)}
	with $\mu=0$ and $\nu=j$ and \eqref{J0large})
\begin{equation}
	P_j (\cos \Theta) \underset{j\to\infty}{\sim}  \sqrt{\frac{2 \Theta}{\pi \sin \Theta}} \cos \left( \left(j + \frac12\right) \Theta - \frac{\pi}{4} \right),
	\label{largej}
\end{equation}
the classical deflection angle can then be obtained by applying the saddle point condition to sum in \eqref{sumj}, obtaining\footnote{Here we use that, out of the two saddle-points $\Theta= \pm \partial_j(2\delta_j)$, since $\Theta\in[0,\pi]$, only $\Theta= - \partial_j(2\delta_j)$ can be realized in theories like gravity where the interaction is attractive, $- \partial_j(2\delta_j)>0$.}
\begin{equation}
	\Theta = - \frac{\partial 2 \delta_j(s)}{\partial j}=-\hbar \frac{\partial \chi(s,J)}{\partial J}\,,
	\label{thetafrompw}
\end{equation}
where we used the definition of $\chi(s,J)$ \eqref{2deltajchiJ} in the last step.
Naturally, substituting the approximate expression \eqref{largejnaive} for $j\gg \alpha_G\gg1$ we simply recover the result of \eqref{eq:Theta1PMmassless} for the leading-order deflection angle in $D=4$,
\begin{equation}\label{ASmetric}
	\Theta \simeq \frac{2Gs}{J} \simeq \frac{4G\sqrt s}{b}\,,\qquad pb\simeq J = \hbar j\,.
\end{equation}
Of course, in a physical quantity as the deflection angle, the dependence on $\mu$ drops out.

However, we can obtain a relation which formally does not rely on the PM expansion by substituting \eqref{eq:iAj} for $j \sim \alpha_G\gg1$, according to which
\begin{equation}
	\Theta =i \left(\psi (j+1 - i \alpha_G) -  \psi (j+1 + i \alpha_G)  \right)
	\label{thet}
\end{equation}
where $\psi$ is the logarithmic derivative of the $\Gamma$-function. 
Using the following equation for  $\psi$:
\begin{eqnarray}
	\psi (1+z) = - \gamma + \sum_{n=1}^\infty \left(\frac{1}{n} - \frac{1}{n+z} \right)
	\label{ACV45}
\end{eqnarray}
we can write $\Theta$ as follows:
\begin{equation}
	\Theta= \sum_{m=j+1}^\infty \frac{2 \alpha_G}{m^2 + \alpha^2_G} \sim  2\int_0^{\frac{\alpha_G}{j}}  \frac{dx}{1+x^2} =2\arctan\frac{\alpha_G}{j}
	\label{ACV46}
\end{equation}
where in the middle step we have approximated the sum with an integral by using $x=\alpha_G/m$ for large values of $j$. Equivalently, we could have substituted in \eqref{thet} directly the expression \eqref{radialactionLO}, obtaining the same conclusion. Then we get
\begin{equation}
	\tan \frac{\Theta}{2} =  \frac{\alpha_G}{j}\,.
	\label{ACV48}
\end{equation}
Taking into account that, in the classical limit, $j \hbar=  p b_J = J$ we can compare  this equation with \eqref{eq:thetap}. They have precisely the same form in $D=4$ except for the substitution of $\sin$ with $\tan$ and of $b$ with $b_J$. Taking into account the difference between $b$ and $b_J$ in Fig.~\ref{fig:theta} they are in agreement up to order $G^2$. Let us also mention that, as discussed in \ref{app:geon8}, the probe-limit scattering of massive objects in $\mathcal{N}=8$ obeys a relation formally very close to \eqref{ACV48}.

\subsection{Minimally coupled massless scalars: from geometry to amplitudes }
\label{ssec:ASmetric}

Interestingly enough, the result \eqref{leadingeik} was also obtained, and at exactly the same time, by 't Hooft \cite{tHooft:1987vrq} from a  calculation that follows exactly the opposite logic with respect to the one of the two previous subsections. The starting point of 't Hooft's approach is the \emph{classical} shock-wave metric produced by a massless particle of given energy $E^{(1)}$ moving, say, in the positive-$z$ direction at $u \equiv t -z = 0$.
That metric was obtained long ago by Aichelburg and Sexl (AS) in the classic paper \cite{Aichelburg:1970dh}. It can be easily generalized to arbitrary $D$ (see e.g.~\cite{Ferrari:1988cc}) and, in suitable coordinates, it reads
\begin{equation} 
	\label{eq:AS}
	ds^2 =  - du\, dv + f(x_\perp) \delta(u) du^2 + \sum_{i=1}^{D-2} dx^i \,dx^i\,,
	\qquad
	v \equiv t +z\,,
\end{equation}
where $x_\perp=(x^1,\ldots,x^{D-2})$ and, in order to satisfy Einstein's equations with a source,
\begin{equation} 
	T_{\mu \nu} = E^{(1)} \delta(u) \delta^{(D-2)} (x^i) \delta^u_{\mu} \delta^u_{\nu},
\end{equation}
$f(x_\perp)$ should satisfy the equation
\begin{equation}
	\label{eq:Eesw}
	\Box_{\perp} f (x_\perp) =-16 \pi G_DE^{(1)}  \delta^{(D-2)}(x_\perp)\,,\qquad
	\Box_{\perp}  \equiv \sum_{i=1}^{D-2} \partial_i^2 \, .
\end{equation}
Hence, $f$ is proportional to the well-known Green function in $D-2$ dimensions. More specifically, one finds\footnote{In \cite{Ferrari:1988cc} the AS metric was (almost trivially)  extended to a  ``beam" of null energy which is still concentrated at $u=0$ but has an arbitrary transverse profile. Such effective shock waves naturally appear  when the source is extended, like in the case of strings. See Section~\ref{ssec:string-brane-sup}.}
\begin{equation}
	\label{eq:finD}
	f(x_\perp)=-16 \pi G_DE^{(1)} \frac{r^{4-D}}{(4-D) \Omega_{D-2}}\,,
	\qquad 
	r^2 \equiv \sum_{i=1}^{D-2} (x^i)^2\,,
	\qquad  
	\Omega_{D-2} = \frac{2 \pi^{\frac{D-2}{2}}}{\Gamma ( \frac{D-2}{2})}
\end{equation}
(note that, in any $D$, $f$ has the correct dimensions of a length).

Consider now the null geodesics describing the worldline of massless (test) particles moving in the opposite direction, initially at $v=0$ and at transverse coordinates $x^i = b^i$. They describe straight lines at $t <0$, but  suffer both a $v$-delay and a deflection for $t \ge 0$.

These geodesics were considered in \cite{Dray:1984ha,tHooft:1987vrq} for the AS metric and generalized in \cite{Ferrari:1988cc} for arbitrary $D$ as well as for extended sources. It turns out that, in these convenient albeit rather singular coordinates, the $v$-coordinate of the test particle suddenly changes from zero to $f(b)$. This $v$-delay is instantaneous and simply given by (using $r \to  b$):
\begin{equation}
	\label{vdelay}
	v \sim \theta(u) f(b) \Rightarrow \Delta v = f(b)= 4G_DE^{(1)} b^{4-D} \frac{\Gamma ( \frac{D-4}{2})}{\pi^{\frac{D-4}{2}}}
\end{equation}
where $\theta(u)$ is the Heaviside step function. We refer to \cite{Ferrari:1988cc,Shore:2018kmt} for more details about the solution of the geodesic equation.

't Hooft next considered a Lorentz frame in which particle 1 is very energetic and produces the shock wave while particle 2 is very soft and behaves as a test particle suffering the $v$-shift \eqref{vdelay}. Quantum mechanically, particle 2 has a wave function $\Psi^{(2)} \sim \exp (i \frac{E^{(2)} v}{\hbar})$ for $t <0$ but, as it goes through $t=0$, its phase picks up a shift  given by
\begin{equation}
	\frac{E^{(2)}  \Delta v }{\hbar} = \frac{E^{(2)}  f(b) }{\hbar} =
	\frac{G_Ds }{\hbar \,\, b^{D-4}} 
	\frac{ \Gamma ( \frac{D-4}{2})}{ \pi^{\frac{D-4}{2}}}
	\label{phase}
\end{equation}
where we have used  that $4E^{(1)} E^{(2)} =s$. The above equation is precisely the phase shift in \eqref{leadingeik}.

As pointed out in~\cite{Fabbrichesi:1993kz} this raised a doubt: working in the boosted reference frame mentioned above, the shock wave produced by the soft particle ($2$) is extremely weak, but it cannot be neglected because, when it is used to calculate its contribution to the phase of the wave function of the energetic particle ($1$), it is multiplied by a large factor $E^{(1)}$ producing in the end exactly the same contribution~\eqref{phase}. So it seems that the total classical phase shift is twice the eikonal result \eqref{leadingeik}. This puzzle was solved in~\cite{Fabbrichesi:1993kz} by realizing that, in this geometric description, the classical action receives an extra contribution from the boundary term needed to define properly the Einstein-Hilbert variational principle. In the case at hand, this extra contribution cancels precisely half of the result obtained by simply summing the contributions of the two particles, thus restoring the agreement with the amplitude-based result. 

As we will see in \ref{app:probelim}, the same idea can be applied, even more straightforwardly, to the scattering of massive particles at the leading order in the probe limit~\eqref{eq:problim}, where particle $1$ can be described by a stationary geometry and the classical action is derived from the geodesic followed by the particle $2$. A similar approach also works, with the appropriate modifications, in the case of string-string collisions as discussed in detail in Sect.~\ref{ssec:string-brane-sup}. 

As discussed in Subsection~\ref{ssec:probeQ}, the probe limit actually determines not only the 1PM, i.e.~$\mathcal O(G)$ eikonal phase, but also the 2PM one, corresponding to a one-loop calculation on the amplitude side. Instead, new classical data that is not fixed by the probe limit appears at 3PM, two-loop order, both in the massless and in the massive setups.
Let us finally remark that this elegant method based on approximating the problem as the one of a particle scattering in the metric produced by the other particle represents the starting point of the so-called self-force expansion \cite{Mino:1996nk,Quinn:1996am,Poisson:2011nh,Barack:2018yvs}.

\subsection{Collapse criteria in GR}
\label{ssec:CTS}

Following up on the discussion of the previous subsection we would like to consider now the opposite situation in which not only the probe approximation breaks down, but even a perturbative expansion in $\frac{R}{b}$ is bound to fail (diverge?). This is when classical GR arguments lead to the conclusion that a finite fraction of the initial energy should collapse to form a black hole.

Of course, the quantitative study of such a regime is out of reach both by analytic and by numerical relativity methods.
Amusingly, however, some non trivial results can be obtained in the strictly massless case as first pointed out by Penrose \cite{PenroseUnpublished} for the axisymmetric situation ($b=0$). The basic simplifying feature of massless collisions is that, by causality, nothing  happens before the two front waves (at $u=0$ and $v=0$) meet at $t=0$. In other words, a simple linear superposition of two AS shock waves of the form \eqref{eq:AS} exactly satisfies Einstein's equations for $t < 0$.

The idea of Penrose, later extended to $b \ne 0$ by Eardley and Giddings \cite{Eardley:2002re}, is to check whether or not a marginally  trapped closed surface (MTCS)\footnote{A (marginally) closed trapped surface is, loosely speaking, a closed surface whose future directed light cone is (marginally) inside the surface itself. More precisely, it may be defined as a $(D-2)$-dimensional closed surface whose outer normals have zero ``convergence'' (or ``expansion'').}  can be constructed using the exact, known geometry at $t \rightarrow 0^{-}$. If the answer were positive, then we would conclude from general theorems that such a surface evolves later into the horizon of a black hole whose mass is bounded from below by the area of the initial surface. Unfortunately, the opposite statement cannot be claimed to be true: even if one can prove that at $t=0$ there is no MTCS, one cannot exclude that one such surface is formed at some later time $t >0$. A fortiori, only a lower limit on the mass of the future black hole (and therefore an upper limit on the total  energy reaching future null infinity) can be given.

In his work \cite{PenroseUnpublished}, Penrose  found that, at $b=0$, a MTCS can be easily constructed and, from the value of its area, he concluded that at least a fraction $\frac{1}{\sqrt{2}}$ of the total incoming energy goes into forming a black hole. Eardley and Giddings, instead, managed to construct a MTCS (in $D=4$) for $b < \gamma_\text{EG} G \sqrt{s}$ with $\gamma_{\text{EG}}\simeq 1.61$, meaning that the critical ratio for collapse $(\frac{G \sqrt{s}}{b})_\text{cr} \le1/\gamma_{\text{EG}} \simeq 0.62$. 

An interesting generalization \cite{Kohlprath:2002yh}  (particularly in view of a  collision of extended objects like strings) is the one in which one considers the collision on two null beams of massless particles (moving again in opposite directions along the $z$-axis). As physically expected, one finds that the transverse size of the beam ``adds up" to the impact parameter (defined as the transverse distance of the two beam's center of mass). One starts from a simple generalization of \eqref{eq:Eesw}:
\begin{equation}
	\label{eq:Eeswb}
	\Box_{\perp} f^{(i)} (x_\perp) =-16 \pi G \rho^{(i)}(x_\perp)\,,\qquad
	i = 1, 2 \, ,
\end{equation}
where $\rho^{(i)}(x_\perp)$ is the energy density per unit transverse area of each beam. Since  the solution \eqref{eq:finD} of  \eqref{eq:Eesw} is nothing but the Green function of the problem, the general solution of  \eqref{eq:Eeswb} is also known.

At this point one can check whether a MTCS is formed already at $t=0$. Different physical situations were discussed in 
\cite{Kohlprath:2002yh}. Let us  just mention two relatively simple cases:
\begin{itemize}
	\item {\bf Axisymmetric collision on two non-homogeneous axisymmetric beams}. This generalizes Penrose's construction to extended objects. In this case $\rho^{(i)}(x_\perp)$ depends just on the radial distance $r$ from the symmetry axis. Consider the energy in each beam below some given $r$:
	\begin{equation}
		\label{eq:Er}
		E^{(i)}(r) = \Omega_{D-2} \int_0^r  \rho^{(i)}(r') r'^{D-3} dr' ~;~ \Omega_d = \frac{2 \pi^{d/2} }{\Gamma(d/2)} \,.
	\end{equation}
	One finds that a MTCS can be constructed whenever one can find an $r_c$ such that:
	\begin{equation}
		\label{eq:axs}
		\Omega_{D-2} r_c^{D-3} \le 8 \pi G_D\sqrt{ E^{(1)}(r_c) E^{(2)}(r_c)}
	\end{equation}
	where $r_c$ plays the role of an effective impact parameter. In the special case of two identical homogeneous discs of radius $L$ \eqref{eq:axs} simply gives $\Omega_{D-2}L^{D-3} \le 4 \pi G_D\sqrt{s}$ ($L \le 2 G \sqrt{s}$ in $D=4$).
	
	\item {\bf Collision of two homogeneous  beams at $b \ne 0$}. Considering for simplicity the case of $D=4$ one finds that a MTCS can be constructed at $t=0$ provided:
	\begin{equation}
		\label{eq:bne0}
		G \sqrt{s} > 2 L  ~ {\rm and} ~ b \le G \sqrt{s} - 2 L \,.
	\end{equation}
	Comparing this to the point particle criterion of ref. \cite{Eardley:2002re}, we see that the role of $b$ is now played by 
	$b_\text{eff.} = b + 2 L$ suggesting that no black hole is formed if the total energy is such that $G \sqrt{s} < 2 L$, i.e.~that the associated Schwarzschild radius is smaller than the sizes of the two beams.
\end{itemize}

Many other interesting issues belong to this problematic. One is the nature of the ``phase transition" between the collapse and ``dispersive" regime, see e.g.~\cite{Gundlach:2002sx}. Is it first order, i.e.~with the mass of the black hole having a lower bound as soon as $\frac{R}{b}>(\frac{R}{b})_\text{cr}$, or second order, in which case we may talk about the critical exponent $\alpha$ appearing in $m_\text{BH} \sim (b_c-b)^{\alpha}$ for $b \to b_c^-$? This question is related to one raised recently by Don Page \cite{Page:2022bem} of whether at $b=b_c$ the fraction of initial energy being radiated in a massive point-particle collision approaches $1$ in the limit of infinite Lorentz boost.
These topics being somewhat outside the scope of this report, we refer the interested reader to the literature.

\section{Leading eikonal for generic gravitational 2--body scattering}
\label{sec:treelevel}

In the introductory section~\ref{sec:leading} we discussed how, in the massless case, the leading eikonal is obtained from the Fourier transform to impact parameter space of the tree diagram with one graviton exchange in the Regge limit where $s \gg |t|$.
In this section we want to extend this result to the massive case in several gravitational theories. We start from Einstein's GR coupled to massive matter fields, then consider extended theories of gravity with extra massless fields (such as the dilaton) or with higher derivative couplings and finally discuss the case of string theory. We focus on the tree-level approximation and on the limit where the momentum exchanged is much smaller than both the center of mass energy and the masses, since these contributions determine the leading eikonal. For minimally coupled massive scalars in GR the exponentiation of the leading eikonal works exactly as in the massless case discussed in Subsection~\ref{ssec:delta0ampl} (see \cite{Kabat:1992tb,Bjerrum-Bohr:2018xdl}). Here we assume that the leading eikonal exponentiates for all gravitational theories considered in this section and then discuss how to generalize the analysis of Subsection~\ref{ssec:exponapp} and extract the classical physics observables of interest. We will postpone to the next sections the study of the one and two loop amplitudes relevant to eikonal exponentiation: they will serve as a check of the assumption taken here and provide new information about the subleading terms in the eikonal expansion.

\subsection{Field theory at tree-level}
\label{sec:FTtree}

\subsubsection{Minimally coupled massive scalars}
\label{sec:grtree}

We consider here pure $D$-dimensional GR minimally coupled to two scalar fields with masses $m_1$ and $m_2$. This is a simple generalization of the equal-mass setup analyzed in~\cite{Kabat:1992tb}. The action reads
\begin{equation}
	\label{eq:mincphi}
	S = \int d^D x \sqrt{|g|} \left\{\frac{R}{2 \kappa_D^2} - \frac{1}{2}\sum_{i=1}^2 \left[\partial_\mu \phi_i \partial_\nu \phi_i g^{\mu\nu} + m_i^2 \phi_i^2\right] \right\}.
\end{equation}
As discussed in the introduction, the eikonal exponentiation also lends itself to describe the classical interactions between two Schwarzschild black holes at large distances, so we take the parameters $m_1$ and $m_2$ to be classically large, e.g.~of the order of ten solar masses. Because of the no-hair theorem, it is natural to use Eq.~\eqref{eq:mincphi} as a starting point, since the mass is the only feature that is relevant classically. In order to describe other compact objects, such as neutron stars, one needs to add non-minimal couplings parametrizing tidal deformations or other classical quantities besides the mass, for instance to describe spin.

To leading order in the conventional Born approximation, the interaction between the scalars $\phi_1$ and $\phi_2$ is captured by the diagram with a single graviton exchanged between the two massive lines. By using the Feynman propagator~\eqref{GGagrav} and the vertex~\eqref{tauppmassive}, which generalizes the one in~\eqref{taupp2},
\begin{equation}\label{taupp2m}
	\tau_a^{\mu\nu}(p,p') =- i\kappa_D \left[
	p^\mu p'^\nu + p^\nu p'^\mu
	- \eta^{\mu\nu} (p\cdot p' - m_a^2)
	\right],
\end{equation}
it is straightforward to calculate the contribution of this diagram, which in $D$ dimension reads
\begin{equation}
	\label{eq:Aphi12tl}
	\mathcal{A}_0(s,t) = \begin{gathered}
		\begin{tikzpicture}
			\draw[color=green!60!black, ultra thick] (-1.5,0) -- (-.5,0);
			\draw[color=blue, ultra thick] (-1.5,1) -- (-.5,1);
			\draw (-1,0) -- (-1,1);
		\end{tikzpicture}
	\end{gathered}
	=\frac{2 \kappa_D^2}{-t} \left[\frac{1}{2} (s-m_1^2-m_2^2)^2 - \frac{2\, m_1^2 m_2^2}{D-2} + \frac{t}{2} \left(s-m_1^2-m_2^2 \right) \right].
\end{equation}
We then focus on the non-analytic terms in the $t =- (p_1+p_4)^2 \sim 0$ limit which, as in the massless case, will capture the long range interaction in impact parameter space. In this approximation the amplitude~\eqref{eq:Aphi12tl} is dominated by the pole located at $t=-q^2=0$,
\begin{equation}
	\label{eq:Aphi12tlp}
	\mathcal{A}_0(s,-q^2) = 2 \kappa_D^2 \frac{\gamma(s,m_i)}{q^2} +\mathcal O((q^2)^0)\;,
\end{equation}
with
\begin{equation}
	\label{eq:gammasm}
	\gamma(s,m_i) \equiv \frac{1}{2} (s-m_1^2-m_2^2)^2 - \frac{2\, m_1^2 m_2^2}{D-2} 
	= 2m_1^2 m_2^2 \left(\sigma^2- \frac{1}{D-2}\right),
\end{equation}
where in the final step we give the result in terms of the variable $\sigma$ defined in~\eqref{eq:sigma}. The leading term in~\eqref{eq:Aphi12tlp} can be derived by replacing \eqref{taupp2m} with the effective vertex
  \begin{equation}
    \label{eq:taupp2meff}
    \tau_{\rm eff}^{\mu\nu}(p, p') = 2i\kappa_D \bar{p}^\mu \bar{p}^\nu\,, \qquad p^\mu=-\bar p^\mu + \frac12 q^\mu\,,\qquad p'^\mu=\bar p^\mu+\frac12 q^\mu\,,
  \end{equation}
where one introduces an ``average momentum'' $\bar p^\mu$ as in \eqref{eq:barpi}, so that $\bar p\cdot q = 0$, and \eqref{eq:taupp2meff} differs from~\eqref{taupp2m} by terms suppressed in the small $q$ limit (thanks to~\eqref{eq:barpi} and $(p\cdot p'-m_a^2) = \frac{q^2}{2}$). The result for the massless amplitude is recovered by considering the ultrarelativistic limit,
\begin{equation}\label{MasslessLimit}
	\sigma\to \infty\,,\qquad m_1 m_2 \sigma \sim \frac{s}{2}\,,
\end{equation}
and it is easy to check that, in this limit,~\eqref{eq:Aphi12tlp} reproduces~\eqref{explA1}.

Actually the Born approximation is not justified in the setup under analysis and, in the massive case, even the low velocity limit is outside the tree-level regime of validity. This can be seen by adapting the derivation in Section~\ref{ssec:delta0ampl} to this new kinematic setup. As before, we take the Fourier transform of~\eqref{eq:Aphi12tlp} by keeping the external states on-shell to rewrite the result in impact parameter space: as shown in~\ref{usefulFT}, we can decompose $q$ in the component $q_\perp$, which is perpendicular to the $p_{1,2}$ and the longitudinal components and the latter can be neglected since they are proportional to $q^2$, see~\eqref{massiveFT}, and kill the pole in~\eqref{eq:Aphi12tlp} yielding to irrelevant analytic terms. Thus we effectively deal with the $(D-2)$-dimensional Fourier transform appearing in the leading term of~\eqref{4EpFT}
\begin{equation}
	\label{eq:Atdefg}
	{\tilde{\cal A}_0}(s,b) = \!\int\! \frac{d^{D-2} q_\perp}{(2\pi)^{D-2}} \frac{{\cal A}_0(s,-{q^2_\perp})}{4Ep } \, {\rm e}^{i b q_\perp}\;,
\end{equation}
where $E$ and $p$ are respectively the total energy and the absolute value of the spatial momentum in the center-of-mass frame (see Eq.~\ref{comframe}). Because of the scaling \eqref{scalesmass}, ${\tilde{\cal A}_0}$ is large even at low velocities. Then, as discussed in detail in Section~\ref{sec:oneloop}, the 1-loop correction, arising from diagrams where two gravitons are exchanged between the scalar particles, includes a term involving the convolution of Eq.~\eqref{eq:Aphi12tl} with itself. This yields in impact parameter space a contribution proportional to ${\tilde{\cal A}^2_0}$ which signals a clear breakdown of standard perturbation theory since ${\tilde{\cal A}_0}$ is large. However we expect that the eikonal exponentiation resums this class of contributions as in~\eqref{1+iAgrav}, so we can identify the leading tree-level contribution~\eqref{eq:Aphi12tlp} in impact parameter space with the 1PM eikonal.
\begin{equation}
	\label{eq:tildeA0d0}
	{\tilde{\cal A}_0} = 2\delta_0\;.
\end{equation}
Then by using Eq.~\eqref{B1} to calculate~\eqref{eq:Atdefg}, we find
\begin{equation}
	\label{eq:d0m1m2}
	2\delta_0 = \frac{2 m_1m_2 G_D \left(\sigma^2- \frac{1}{D-2}\right) \Gamma \left(\frac{D-4}{2}\right) }{\sqrt{\sigma^2-1} (\pi b^2)^{\frac{D-4}{2}}}\;.
\end{equation}

In the ultrarelativistic limit \eqref{MasslessLimit}, Eq.~\eqref{eq:d0m1m2} reduces to~\eqref{eq:leik} and we recover the massless case once again. Notice that in this limit Eq.~\eqref{leadingeik} is the leading eikonal for any gravitational field theory in the two derivative approximation. This universality is a consequence of the Regge limit thanks to the following two properties. First, in the high energy regime amplitudes are dominated by the exchange of the states with the highest spin: indicating with $j$ the spin of such particles, their leading contribution to ${\cal A}_0$ scales as $s^j$. Then, when we are interested in long range effects, we can focus just on exchanges of massless particles. In gravitational theories the highest spin massless particle is the graviton which gives rise to a universal\footnote{As we will see in section~\ref{ssec:ethd} the universality of the graviton coupling can be spoiled in presence of higher derivative corrections.} contribution to ${\cal A}_0$ scaling as $s^2$. Thus, in the ultrarelativistic regime, the presence of other lower spin fields is irrelevant for the derivation of the leading eikonal.

Going back to general kinematics, we can proceed as in the massless case and use Eq.~\eqref{eq:thetap} to compute the contribution of the leading eikonal to the deflection angle
\begin{equation}
	\label{eq:Theta0}
	2p\sin \frac{\Theta_0}{2} = \frac{4 G_D m_1 m_2 \left(\sigma^2- \frac{1}{D-2}\right) \Gamma \left(\frac{D-2}{2}\right) }{\sqrt{\sigma^2-1} \pi^{\frac{D-4}{2}} b^{D-3}}\;.
\end{equation}
As discussed in detail in the later sections, the subleading eikonal terms will provide further contributions to the physical deflection angle $\Theta$  and so, in a PM approach, this result can be trusted up to ${\cal O}(G_D)$
\begin{equation}
	\Theta = \frac{4 G_D E \left(\sigma^2- \frac{1}{D-2}\right) \Gamma \left(\frac{D-2}{2}\right) }{(\sigma^2-1) \pi^{\frac{D-4}{2}} b^{D-3}} + {\cal O}(G_D^2)\;,
	\label{defle1}
\end{equation}
where we expanded the $\sin$ function to leading order and rewrote the momentum $p$ in terms of the center-of-mass energy $E$ thanks to~\eqref{Ep}. However, also the expression in~\eqref{eq:Theta0} is useful as it contains all the non-linear contribution to the angle arising from the leading eikonal. For instance it allows to recover the full leading answer in the standard $D=4$ Post-Newtonian (PN) limit $\frac{1}{j_{\rm PN}} \sim v_{\infty}\ll 1$ where\footnote{We follow the conventions of~\cite{Bini:2017wfr}.}
\begin{equation}
	\label{eq:PNdef}
	\sigma = \sqrt{1+ v^2_\infty}\,, \qquad \frac{1}{j_{\rm PN}} = \frac{G m_1 m_2}{J}\;. 
\end{equation}
One can easily recast the angle~\eqref{eq:Theta0} in terms of the angular momentum by using the relation\footnote{In section~\ref{ssec:bbj} we will present a derivation of this relation based on the standard partial wave decomposition as done in section~\ref{ssec:exponapp} for the leading eikonal.} $b_J=b\cos(\Theta/2)=b$, which is obvious from Fig.~\ref{fig:theta}, where by definition the angular momentum equals $J= p b_J$. Then we have
\begin{equation}
	\label{eq:tanTheta0}
	\tan \frac{\Theta_0}{2} = \frac{G m_1 m_2 \left(2\sigma^2- 1\right) }{J\,\sqrt{\sigma^2-1} }\,, \qquad \frac{\Theta_{\rm 0PN}}{2} = \arctan \left(\frac{1}{j_{\rm PN}\, v_\infty}\right)\;.
\end{equation}
where in the equation on the right we took the leading PN term finding the classic Newtonian result, see~\cite{Bini:2017wfr} and references therein. In this regime, even if the scattering angle is finite, the momentum transfer in~\eqref{eq:Theta0} is small because $p\to 0$. 

Instead, in the leading PM approximation, we can use $b_J=b + \mathcal O(G_D^2)$ obtaining
\begin{equation}
	\Theta = 
	\frac{4 G_D E \left(\sigma^2- \frac{1}{D-2}\right) \Gamma \left(\frac{D-2}{2}\right) }{(\sigma^2-1) \pi^{\frac{D-4}{2}}} \frac{p^{D-3}}{J^{D-3}} + {\cal O}(G_D^2) \;.\label{defle2}
\end{equation}
Let us also collect here for later convenience the $D=4$ expressions for the 1PM impulse and the deflection angle,
\begin{equation}\label{1PM}
	Q = \frac{4Gm_1m_2\left(\sigma^2-\tfrac12\right)}{b\sqrt{\sigma^2-1}}+\mathcal O(G^2)\,,\qquad
	\Theta = \frac{4GE\left(\sigma^2-\tfrac12\right)}{b(\sigma^2-1)}+\mathcal O(G^2)\,.
\end{equation}

Another interesting limit is to take one mass much larger than the energy of the other, for instance $m_2 \gg E_1$. In this regime, we can see particle $1$ as a probe propagating in the gravitational background of the heavy particle $2$. So it should be possible to obtain~\eqref{defle2} by a classical calculation studying the geodetic motion in the appropriate curved geometry. This probe limit provides a straightforward but important check on the diagrammatic calculations, in particular when testing the subleading terms in the eikonal expansion. In this limit the total energy coincides with the rest mass of the heavy particle ($M_h$) while $\sigma$ and $p$ can be written in terms of the energy and the mass of the light probe particle
\begin{equation}
	\label{eq:problim2}
	m_2\equiv M_h\;, \qquad \sigma \to \frac{E_1}{m_1} 
	\;,  \qquad E \to \sqrt{M_h^2+2 M_h E_1}\;, \qquad p\to \sqrt{E_1^2-m_1^2}\;.
\end{equation}
By using~\eqref{eq:problim2} in~\eqref{eq:Aphi12tlp} we have
\begin{equation}
	\label{eq:214pl}
	{\cal A}_0 \sim 4 \kappa_D^2 M^2_h \frac{(D-2) E_1^2 -m_1^2 }{(D-2) (-t)}\;.
\end{equation}
In this limit the factor of $4 Ep$ in~\eqref{eq:Atdefg} reduces to $4 M_h \sqrt{E_1^2- m_1^2}$, so in impact parameter space we have
\begin{equation}
	\label{eq:217pl}
	2 \delta_0 = \tilde{\cal A}_0 (s, b) = \frac{2 G_D M_h}{(D-2)} \frac{(D-2) E_1^2 -m_1^2 }{\sqrt{E_1^2-m_1^2}}\, \frac{\Gamma\left(\frac{D-4}{2}\right)}{(\pi b^2)^{\frac{D-4}{2}}} \;.
\end{equation}
Then, by using again~\eqref{eq:thetap}, we can derive the contribution from $\delta_0$ to the deflection angle
\begin{equation}
	\label{eq:pdeflampl1}
	\Theta =  \frac{\sqrt{\pi} \Gamma\left(\frac{D-2}{2}\right)}{2 \Gamma\left(\frac{D-1}{2}\right)} \frac{(D-2) E_1^2 -m_1^2 }{E_1^2-m_1^2} \left(\frac{R}{b}\right)^{D-3}+ {\cal O}(G_D^2)\;,
\end{equation}
where we used~\eqref{eq:RsD} to introduce the $D$-dimensional Schwarzschild radius. It is then easy to check that, at linear order in $G$, Eq.~\eqref{eq:pdeflampl1} agrees with the result of the classical geodesic calculation~\eqref{eq:chi1g} (since at leading order $J  = p b_J \simeq p b = \sqrt{E_1^2-m_1^2} \,b$).

\subsubsection{Graviton scattering off a massive scalar}
\label{sec:gravsctree}

\begin{figure}
	\centering
		\begin{tikzpicture}
			\draw[<-] (-1.9,-.15)--(-1.4,-.15);
			\draw[<-] (1.9,-.15)--(1.4,-.15);
			\draw[<-] (-1.1,1.4)--(-.7,1);
			\draw[<-] (1.1,1.4)--(.7,1);
			\draw[ultra thick, blue] (-2,0)--(2,0);
			\draw[thick, style=decorate, decoration=snake] (0,0) -- (1.3,1.3);
			\draw[thick, style=decorate, decoration=snake] (0,0) -- (-1.3,1.3);
			\filldraw[color=white, fill=white, very thick](0,0) circle (.8);
			\filldraw[pattern=north east lines, very thick](0,0) circle (.8);
			\node at (-2,0)[left]{$k_1$};
			\node at (2,0)[right]{$k_4$};
			\node at (-1.7,1.3){$\rho\sigma$};
			\node at (-.55,1.3){$q_2$};
			\node at (1.7,1.3){$\alpha\beta$};
			\node at (.55,1.3){$q_3$};
		\end{tikzpicture}
		\caption{The kinematics of a two scalar, two graviton amplitude.} \label{fig:2g2ms}
	\end{figure}
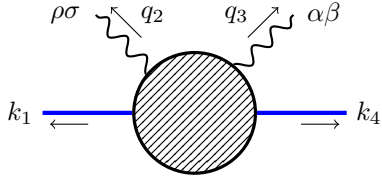

Another instructive tree-level result describes the amplitude between two scalar and two gravitons. A direct derivation from~\eqref{eq:mincphi} requires to evaluate four Feynman diagrams and involves the cumbersome three-graviton vertex. It is possible to obtain a more compact expression by taking the field theory limit of a string amplitude~\cite{KoemansCollado:2019ggb}
\begin{equation}
	\label{eq:2g2ms}
	i {\cal A}^{\alpha\beta;\rho\sigma} = -i \frac{2\kappa_D^2 (k_4 q_2) (k_1 q_2)}{(q_2q_3) }\left[ \frac{k_4^\rho k_1^\alpha}{k_4 q_2}  + \frac{k_4^\alpha k_1^\rho}{k_1 q_2} +\eta^{\rho \alpha} \right]\left[ \frac{k_4^\sigma k_1^\beta}{k_4 q_2}  + \frac{k_4^\beta k_1^\sigma}{k_1 q_2} +\eta^{\sigma \beta} \right]\,,
\end{equation}
where $k_i$ label the momenta of the massive scalars and $q_i$ those of the gravitons, see Fig.~\ref{fig:2g2ms}. The factorized form of the result is a by-product of using the KLT approach~\cite{Kawai:1985xq} at the string theory level.
	If we focus on the $D=4$ case, then it is possible to rewrite the result in terms of the compact spinor helicity formalism. As usual one can express the massless momenta in terms of a pair of commuting $SU(2)$  spinors\footnote{As usual $\sigma^\mu_{\alpha \dot{\alpha}}= (1,\sigma^i)_{\alpha \dot{\alpha}}$ and $\bar\sigma^{\mu\,\dot{\alpha} \alpha }= (1,-\sigma^i)^{ \dot{\alpha}\alpha}$ where $\sigma^i$ are the Pauli matrices. The $SU(2)$ indices are raised and lower as follows: $\lambda_\alpha = \epsilon_{\alpha\beta} \lambda^\beta$, $\lambda^\alpha =  \epsilon^{\alpha\beta} \lambda_\beta$ and similarly $\tilde\lambda^{\dot\alpha} =  \epsilon^{\dot\alpha\dot\beta} \tilde\lambda_{\dot\beta}$, $\tilde\lambda_{\dot\alpha} =  \epsilon_{\dot\alpha\dot\beta} \tilde\lambda^{\dot\beta}$ with $\epsilon^{12}=-\epsilon_{12}=1$.}
	\begin{equation}
		\label{eq:pltl}
		p_{\alpha \dot{\alpha}}=p_{\mu} \sigma^{\mu}_{\alpha \dot{\alpha}}=\lambda_\alpha \tilde\lambda_{\dot{\alpha}}\;,
	\end{equation}
	where $\tilde\lambda$ is the complex conjugate of $\lambda$ if one sticks to the standard Lorentzian signature. The two physical helicities of the gravitons can be written in terms of
	\begin{equation}
		\label{eq:polsh}
		\varepsilon_{\alpha \dot{\alpha}}^{+} = \sqrt{2}\, \frac{\mu_{\alpha} \tilde{\lambda}_{\dot{\alpha}}}{\langle \mu \lambda \rangle} \qquad \varepsilon_{\alpha \dot{\alpha}}^{-} = \sqrt{2}\, \frac{\lambda_{\alpha} \tilde{\mu}_{\dot{\alpha}}}{[\tilde{\lambda} \tilde{\mu}]}
	\end{equation}
	by using $\epsilon^{++}=(\varepsilon^{+})^2$ and $\epsilon^{--}=(\varepsilon^{-})^2$, where we defined 
	\begin{equation}
		\begin{split}
			\langle \lambda \mu\rangle &= \lambda^\alpha \mu_\alpha = - \lambda_\beta \mu^\beta = -\langle \mu \lambda \rangle\,, \\
			[\lambda \mu] &= \lambda_{\dot\alpha} \mu^{\dot\alpha} = -\mu^{\dot\beta} \lambda_{\dot\beta} = -[\mu \lambda] \,.
		\end{split}
	\end{equation}
	In~\eqref{eq:polsh} $\mu$ and $\tilde{\mu}$ are arbitrary spinors and the freedom in their choice reflects just the possibility of performing gauge transformation. Then one can contract the free indices in~\eqref{eq:2g2ms} with two graviton polarizations
	\begin{equation}
		\label{eq:2g2mse}
		\begin{aligned}
			i {\cal A}_0 = -i \frac{2\kappa_D^2 (k_4 q_2) (k_1 q_2)}{(q_2q_3) }
			& 
			\left[ \varepsilon_2 \varepsilon_3 + \frac{(\varepsilon_3 k_1) (\varepsilon_2 k_4)}{q_2 k_4} + \frac{(\varepsilon_2 k_1) (\varepsilon_3 k_4)}{q_2 k_1}\right] 
			\\  
			\times 
			& 
			\left[\tilde\varepsilon_2 \tilde\varepsilon_3 + \frac{(\tilde\varepsilon_3 k_1) (\tilde\varepsilon_2 k_4)}{q_2 k_4} + \frac{(\tilde\varepsilon_2 k_1) (\tilde\varepsilon_3 k_4)}{q_2 k_1}\right],
		\end{aligned}
	\end{equation}
	where the polarizations have been factorized according to $\epsilon_{\mu\nu}=\varepsilon_\mu \tilde\varepsilon_\nu$ and we rewrote the prefactor in~\eqref{eq:2g2ms} in terms of~\eqref{eq:gammasm}. In $D=4$, we can use~\eqref{eq:polsh} and we have $\varepsilon^{\pm}=\tilde\varepsilon^{\pm}$. 
	One can choose explicit expressions for the auxiliary spinors $\mu$, $\tilde\mu$ to simplify the various scalar product: in particular it is convenient to take $\mu_2=\lambda_3$, $\mu_3=\lambda_2$ and similarly $\tilde{\mu}_2=\tilde{\lambda}_3$, $\tilde{\mu}_3=\tilde{\lambda}_2$, where $q_{i\,\alpha\dot\alpha}=\lambda_{i\,\alpha}\tilde\lambda_{i\,\dot\alpha}$. By using\footnote{We have some unconventional signs since we are using the mostly plus metric and so $\operatorname{Tr}(\sigma_\mu \bar\sigma_\nu)=-2\eta_{\mu\nu}$.} 
	\begin{equation}\label{eq:angsqbra}
		s_{ij} = - 2 z_i z_j  = \langle i j \rangle [j i]\,,
	\end{equation}
	for null vectors $z_i$, $z_j$,
	and introducing the notation
	\begin{equation}\label{}
		\langle i\, w\, j] = \lambda_i^{\alpha} w_\mu \sigma^{\mu}_{\alpha \dot{\alpha}} \tilde\lambda_j^{\dot{\alpha}}\,,
	\end{equation}
	for any (not necessarily null) four-vector $w^\mu$, we obtain the following results
	\begin{equation}
		\label{eq:fortab}
		\begin{gathered}
			\varepsilon^+_2 k_1 = - \varepsilon^+_2 k_4= \frac{1}{\sqrt{2}} \frac{\langle 3 k_4 2]}{\langle 3 2\rangle}\;, \quad - \varepsilon^+_ 3k_1 = \varepsilon^+_3 k_4 = \frac{1}{\sqrt{2}} \frac{\langle 2 k_4 3]}{\langle 3 2\rangle}\;, \\
			\varepsilon^-_2 k_1 = - \varepsilon^-_2 k_4= - \frac{1}{\sqrt{2}} \frac{\langle 2 k_4 3]}{[3 2]}\;, \quad \varepsilon^-_ 3k_1 = -\varepsilon^-_3 k_4 = \frac{1}{\sqrt{2}} \frac{\langle 3 k_4 2]}{[3 2]}\;, \\
			\varepsilon^{\pm}_i q_j = 0 \;, \quad \varepsilon^+_2\varepsilon^+_3 = \frac{[23]}{\langle 23\rangle}\;, \quad \varepsilon^+_2\varepsilon^-_3 = 0\;,\quad \langle 3 k_4 2]\langle 2 k_4 3] = -(2 k_2 k_3) k_4^2 + (2 k_2 k_4)(2 k_3 k_4)\;. 
		\end{gathered}
	\end{equation}
	Then we can rewrite the tree-level amplitude separating the result for the various helicities of the gravitons we find\footnote{We take $s=-(k_1+q_2)^2$, $t=-(k_1+k_4)^2$, $u=-(k_1+q_3)^2$.}
	\begin{align}
		i\mathcal{A}^{++;++} & =  i \kappa_D^2 \frac{[23]^2}{\langle 23\rangle^2}\frac{m_1^4 t}{(s-m_1^2)(u-m_1^2)}\;, \\ 
		i\mathcal{A}^{++;--} & =  i \kappa_D^2 \frac{\langle 3 | k_4 | 2]^4}{t(s-m_1^2)(u-m_1^2)}\;,
	\end{align}
	in agreement with Eq.~(2.19) and~(2.20) of~\cite{Cachazo:2017jef}. Notice that the helicity violating amplitude $\mathcal{A}^{++;++}$ does not have a $\frac{1}{t}$ pole since the ratio $\frac{[23]}{\langle 23\rangle}$ is finite  as $t\to 0$. We are interested in terms dominated by the $t$-pole, which are automatically leading in the limit~\eqref{eq:smggt} and so only the helicity preserving structure $\mathcal{A}^{++;--}$ survives. We can extract the pole by taking the approximation $\langle 3 | k_4 | 2]^4 \simeq \langle 3 | k_4 | 3]^4 = (s-m_1^2)^4$, since $q_2 \simeq -q_3$ in the limit of small momentum transfer. Thus the leading contribution to the amplitude with two massive scalars and two gravitons is again captured by~\eqref{eq:Aphi12tlp} with of course $m_2=0$
	\begin{equation}
		\label{eq:2g2mhe}
		i {\cal A}^{++;--}_0 \simeq \frac{2 i \kappa_D^2}{-t} \,\gamma(s,m_1,m_2=0)\,\; .
	\end{equation}
	Moving then to the impact parameter space, we obtain the following eikonal phase describing the scattering of a graviton off a massive scalar
	\begin{equation}
		\label{eq:d0m1g}
		2\delta_0 = \frac{G_D \gamma(s,m_1,m_2=0)}{\sqrt{\sigma^2-1}}  \frac{\Gamma\left(\frac{D-4}{2}\right)}{(\pi b^2)^{\frac{D-4}{2}}}\;.
	\end{equation}
	
	\subsubsection{The effect of the dilaton on the eikonal scattering}
	\label{ssec:dilatonicth}
	
	In the two previous subsections we considered theories in which the only massless particle is the graviton. In this subsection and in the next one, we study how the tree-level result~\eqref{eq:Aphi12tlp} changes in theories with a richer spectrum of massless states. We start by considering the case where the massless spectrum, besides the graviton, includes also a massless scalar focusing in particular on the case of the string theory dilaton. In a QFT setup the dilaton is automatically included when considering a gravitational theory which is obtained via the double copy approach~\cite{Bern:2008qj}. This technique was used in~\cite{Goldberger:2016iau,Goldberger:2017frp,Goldberger:2017vcg,Goldberger:2017ogt,Li:2018qap,Shen:2018ebu,Goldberger:2019xef,Bastianelli:2021rbt,Shi:2021qsb} to derive classical quantities in various setups. Even if we will not make extensive use of this approach, we sketch briefly the idea in an explicit example,  since this has been a popular approach to derive (super)gravity amplitudes with the presence of extra massless fields like the dilaton.
	
	One starts from a {\em gauge} theory with no dynamical gravity -- in our example a minimally coupled massive scalar field in the adjoint representation of the gauge group. We will not need to specify the gauge group as only general properties, such as the Jacobi identities, are used. The three-point amplitude involving two scalars and a gluon is given by: 
	\begin{equation}
		i {\cal A}_{0\,\mu}^{(3)} = i g f^{a_1 a_3 b}  (k_1- k_3)_{\mu}= 2 i g f^{a_1 a_3 b}  
		(k_1 - \frac{q}{2})_{\mu} \;,
		\label{A3gluon}
	\end{equation}
	where $k_1$, $k_3$ are the momenta of the scalars and $q$ is the one of the gluon. We can use this ingredient to compute the four-point scalar amplitude with a gluon exchange
	\begin{align}
		i {\cal A}_{0} = & ~i g^2\left[ f^{a_1 a_3 b} f^{a_4 a_2 b}  
		\frac{(k_1 -k_3) (k_4 -k_2)}{(
			k_1+k_3)^2}+   f^{a_1 a_2 b} f^{a_3 a_4 b} \frac{(k_1 -k_2) (k_3 -k_4)}{(
			k_1+k_2)^2} \right. \nonumber \\
		&~ +  \left. f^{a_1 a_4 b} f^{a_2 a_3 b}  \frac{(k_1 -k_4) (k_2 -k_3)}{(
			k_1+k_4)^2} \right] \label{A4gluon} \\ \nonumber
		=&~ i g^2 \left[ f^{a_1 a_3 b} f^{a_4 a_2 b}  \frac{s-u}{(-t)} +
		f^{a_1 a_2 b} f^{a_3 a_4 b}  \frac{u-t}{(-s)} +  f^{a_1 a_4 b} f^{a_2 a_3 b} 
		\frac{t-s}{(-u)} \right]\,.
	\end{align}
	The basic idea of the double copy is that there exists a color-kinematics duality  based on the observation that, in each term of the expression above, the color factors and the momentum dependent numerators satisfy appropriate Jacobi identities.\footnote{For general amplitudes it is not straightforward to write the result that makes this feature manifest.} Indeed, the color factors obey the standard Jacobi identity:
	\begin{eqnarray}
		f^{a_1 a_3 b} f^{a_4 a_2 b} + f^{a_1 a_2 b} f^{a_3 a_4 b} + 
		f^{a_1 a_4 b} f^{a_2 a_3 b} =0
		\label{ffJacobi}
	\end{eqnarray}
	and corresponding kinematic numerators 
	\begin{gather}
		f^{a_1 a_3 b} f^{a_4 a_2 b}  ~\longleftrightarrow~  (k_1 -k_3) (k_4 -k_2) = s-u
		\nonumber \\
		f^{a_1 a_2 b} f^{a_3 a_4 b}  ~\longleftrightarrow~  (k_1 -k_2) (k_3 -k_4) =
		u-t \label{substi} \\  \nonumber 
		f^{a_1 a_4 b} f^{a_2 a_3 b}  ~\longleftrightarrow~  (k_1 -k_4) (k_2 -k_3) =
		t-s
	\end{gather}
	satisfy a ``kinematic'' Jacobi identity, i.e.~their sum vanishes. By following the double copy approach, one starts from~\eqref{A4gluon} and obtains a gravitational amplitude by substituting the color factors with the corresponding momentum-dependent factors~\eqref{substi}. This yields
	\begin{equation}
		i {\cal A}_0 = i \left(\frac{\kappa_D}{2}\right)^2 \left[\frac{(u-s)^2}{(-t)} +  
		\frac{(u - t)^2}{(-s)} +
		\frac{(s-t )^2}{(-u)}  \right]\;,
		\label{squabrac}
	\end{equation}
	where we mapped $g\to \kappa_D/2$. This result is appropriate to describe the interaction among identical external states and, for the case of two different scalars considered in~\eqref{eq:Aphi12tl}, we can focus just on the first term in~\eqref{squabrac}. By using $u=-s-t+2m_1^2+ 2m_2^2$, it is easy to see that even the $1/t$ part does not match the pure gravity result. The reason for this is that the gravitational theory obtained by this construction contains extra massless fields, in addition to the graviton:\footnote{Notice that the double copy amplitude contains also a contribution linear in $t$ which is absent in~\eqref{eq:Aphi12tl}. This signals that the action contains an extra contact term interaction between the four scalars.} a massless antisymmetric tensor and, crucially for us, a dilaton whose coupling to scalars is proportional to square of the mass. 
	
	In summary, the long-range interaction between two massive scalars in a theory containing both gravitons and dilatons takes the same form as~\eqref{eq:Aphi12tlp}, but with a different kinematic factor
	\begin{equation}
		\label{eq:gammasmdil}
		\gamma^{\text{(dil)}}(s,m_i) \equiv \frac{1}{2} (s-m_1^2-m_2^2)^2\;.
	\end{equation}
	Thus the tree amplitude to be used to construct the leading eikonal is
	\begin{equation}
		{\cal{A}}_0(s,-q^2) = \frac{32 \pi m_1^2 m_2^2 G_D \sigma^2}{q^2} +\mathcal O((q^2)^0)
		\label{A0dilaton}
	\end{equation}
	instead of the one in \eqref{eq:Aphi12tlp}. Then, by following the usual steps, we obtain the leading eikonal 
	\begin{equation}
		\label{eq:d0dil}
		2\delta_0 = \frac{2 m_1m_2 G_D \sigma^2 \Gamma \left(\frac{D-4}{2}\right) }{\sqrt{\sigma^2-1} (\pi b^2)^{\frac{D-4}{2}}}\;.
	\end{equation}

	It is easy to isolate the contribution due to the dilaton exchanges and 
	\begin{equation}
		{\cal{A}}_0(s,-q^2) = \frac{32 \pi m_1^2 m_2^2 G_D}{(D-2) q^2}
		\label{dss}
	\end{equation}
	from which we can also read off the dilaton-scalar-scalar coupling,
	\begin{equation}\label{}
		{\cal A}_0^{(3)} = - \frac{2 \kappa_D  m_{1,2}^2}{\sqrt{D-2}}\,.
	\end{equation}
	We can check the interpretation above simply by adding to the initial Lagrangian~\eqref{eq:mincphi} a dilaton $\varphi$ 
	\begin{equation}
		\label{eq:dilphi}
		S = \int d^D x \sqrt{|G|} \left\{\frac{R}{2 \kappa_D^2} - \frac{1}{2}\left[ \partial_\mu \varphi \partial_\nu \varphi G^{\mu\nu} + \sum_{i=1}^2 \left(\partial_\mu \phi_i \partial_\nu \phi_i G^{\mu\nu}  + m_i^2 {\rm e}^{\frac{2\kappa_D \varphi}{\sqrt{D-2}}} \phi_i^2\right)\right] \right\}.~~~
	\end{equation}
	Then it is straightforward that the exchange diagram involving the dilaton yields an extra contribution canceling the final term in~\eqref{eq:gammasm} and thus reproducing~\eqref{eq:gammasmdil}.

	\subsubsection{Maximally supersymmetric gravity}
	\label{sec:n8tree}
	
	We can further enrich the theory including extra fields. An interesting case is that of supergravities whose perturbative expansion has been studied in great details at high loop order~\cite{Bern:2006kd,Bern:2007hh,Bern:2009kd,Bern:2014sna,Bern:2018jmv}.  Focusing on the maximal supersymmetric case, the four point amplitude among massless states is determined by a single scalar function even in a string theory setup~\cite{green1988superstring}. In the field theory limit we have 
	\begin{equation}
		\label{eq:AN8tree}
		A_0 = 8 \pi G \frac{\cal K}{stu}\;,
	\end{equation}
	where the kinematic prefactor ${\cal K}$ can be written as a pair of 4-index tensors $K$ appearing also in the corresponding string amplitude, see for instance Eq.~(7.4.57) of~\cite{green1988superstring} and Section~7.4.2 of the same reference for the explicit expressions of $K$ (see also \cite{Bjerrum-Bohr:2021din}). It is convenient to organize the graviton multiplet in terms of on-shell realization of supersymmetry, see~\cite{Nair:1988bq} for a realization relevant to $D=4$ and~\cite{Boels:2012ie} for higher dimensional cases. The basic idea is that ${\cal K}$ encodes a super-momentum conserving delta function $\delta^{(16)}(Q)$ involving the sum of the supercharges of the external states. 
	
	In $D=4$, all $2^8=256$ states of the $\mathcal N=8$ supermultiplet can be packaged into the on-shell superfield 
	\begin{equation}\label{superPhi}
		\Phi(\eta) = h^{++} + \eta^{A}\psi_{A} + \cdots + \eta^A \eta^B \eta^C \eta^D \varphi_{ABCD} + \cdots + \eta^1 \eta^2 \eta^3 \eta^4 \eta^5 \eta^6 \eta^7 \eta^8  h^{--}\,,
	\end{equation}
	where $\eta^{A}$ denotes auxiliary Grassmann variables that saturate the $R$-symmetry indices $A,B,C,\ldots$ (taking values from $1$ to $8$) of the various fields. For instance, $h^{\pm\pm}$ denote the positive/negative-helicity gravitons $h^{++}$, while $\varphi_{ABCD}$ collect the $70$ scalars present in the theory.
	In this formalism, one can write a $2\to2$ ``super-amplitude'', which concisely encodes all $2\to2$ scattering amplitudes among any four given states of the supermultiplet, and takes the following very compact form \cite{Elvang:2015rqa,Parra-Martinez:2020dzs}
	\begin{equation}\label{superA0}
		A_0 = \frac{\kappa^2}{stu} \frac{[3 4]^4}{\langle  1 2\rangle^4} \,\delta^{(16)}(Q)\,,
	\end{equation}
	where the ``super-momentum'' conserving delta function is given by
	\begin{equation}\label{}
		\delta^{(16)}(Q)  = \frac{1}{2^4}\prod_{A=1}^{8} \sum_{i,j=1}^4 \langle ij \rangle\, \eta_i^A \eta_j^A\,.
	\end{equation}
	The super-amplitude \eqref{superA0} should be thought of as a function of four copies of the on-shell superfield \eqref{superPhi}, $\Phi(\eta_j)$, 
	with $i=1,2,3,4$ labeling the states as in Fig.~\ref{fig:N=8tree}, and of course
	\begin{equation}\label{mandelstamstu}
		s = -(p_1+p_2)^2\,,\qquad
		t = -(p_1+p_4)^2\,,\qquad
		u = -(p_1+p_3)^2\,.
	\end{equation}
	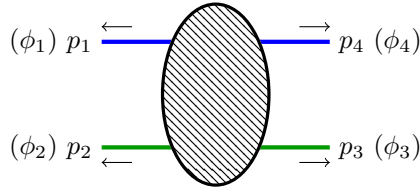
\begin{figure}
		\begin{center}
			\begin{tikzpicture}
				\path [draw, ultra thick, blue] (-4.5,2.2)--(-1.5,2.2);
				\path [draw, ultra thick, green!60!black] (-4.5,.8)--(-1.5,.8);
				\filldraw[white, very thick] (-3,1.5) ellipse (.7 and 1.2);
				\filldraw[pattern=north west lines, very thick] (-3,1.5) ellipse (.7 and 1.2);
				\draw [<-] (-4.5,2.4)--(-4.1,2.4);
				\draw [<-] (-1.5,2.4)--(-1.9,2.4);
				\draw [<-] (-4.5,.6)--(-4.1,.6);
				\draw [<-] (-1.5,.6)--(-1.9,.6);
				\node at (-1.5,2.2)[right]{$p_4$ $(\phi_4)$};
				\node at (-1.5,.8)[right]{$p_3$ $(\phi_3)$};
				\node at (-4.5,2.2)[left]{$(\phi_1)$ $p_1$};
				\node at (-4.5,.8)[left]{$(\phi_2)$ $p_2$};
			\end{tikzpicture}
		\end{center}
		\caption{Scattering of four states in $\mathcal N=8$ supergravity}
		\label{fig:N=8tree}
	\end{figure}
	We consider three possible situations. The first one is given by the choice
	\begin{equation}\label{}
		\phi_{1} = \phi_{2} = \varphi_{1234} \equiv \tau \,, \qquad
		\phi_{3} = \phi_{4} = \varphi_{5678} \equiv \bar\tau \,.
	\end{equation}
	The idea is to focus on a complex scalar $\tau$ (corresponding for instance to the axio-dilaton of type IIB via dimensional reduction) and its complex conjugate.
	In order to extract the contribution due to the corresponding states from the super-amplitude \eqref{superA0}, one should take the appropriate derivatives with respect to the auxiliary Grassmann variables as dictated by the way their fields feature in the superfield \eqref{superPhi}. For our case,
	since
	\begin{equation}\label{}
		\tau = \partial\Phi(\eta) \Big|_{\eta=0}\,,\qquad
		\bar\tau = \bar\partial \Phi(\eta) \Big|_{\eta=0}
	\end{equation}
	with
	\begin{equation}\label{}
		\partial \equiv
		\prod_{A=1}^{4}\frac{\partial}{\partial \eta^A} \,,\qquad
		\bar\partial \equiv \prod_{A=5}^{8}\frac{\partial}{\partial \eta^A} \,,
	\end{equation}
	we need to calculate
	\begin{equation}\label{}
		\mathcal A_0^{\tau\tau\to\bar\tau\bar\tau} = 
		\partial_1\partial_2\bar\partial_3\bar\partial_4
		A_0 \Big|_{\eta=0}\,.
	\end{equation}
	The differential operators, which could be also interpreted as Grassmann integrals, only act on $\delta^{(16)}(Q)$ and it is easy to see that the result is simply
	\begin{equation}\label{}
		\partial_1\partial_2\bar\partial_3\bar\partial_4
		\delta^{(16)}(Q)\Big|_{\eta=0}
		=
		\langle  1 2\rangle^4
		\langle3 4\rangle^4\,.
	\end{equation}
	In this way, we obtain
	\begin{equation}\label{eq:AN8treee1}
		\mathcal A_0^{\tau\tau\to\bar\tau\bar\tau} = \frac{\kappa^2 s^4}{stu}\,,
	\end{equation}
	which is of course symmetric in the exchange $1\leftrightarrow 2$.
	Similarly, for the choice 
	\begin{equation}\label{}
		\phi_{1} = \phi_{4} = \varphi_{1234} \equiv \tau \,, \qquad
		\phi_{2} = \phi_{3} = \varphi_{5678} \equiv \bar\tau\,,
	\end{equation}
	we have 
	\begin{equation}\label{t-supN8}
		\mathcal A_0^{\tau\bar\tau\to\tau\bar\tau} = \frac{\kappa^2 t^4}{stu}\,,
	\end{equation}
	and for the choice
	\begin{equation}\label{}
		\phi_{1} = \phi_{3} = \varphi_{1234} \equiv \tau \,, \qquad
		\phi_{2} = \phi_{4} = \varphi_{5678} \equiv \bar\tau\,,
	\end{equation}
	we have 
	\begin{equation}\label{u-supN8}
		\mathcal A_0^{\tau\bar\tau\to\bar\tau\tau} = \frac{\kappa^2 u^4}{stu}\,.
	\end{equation}
	Eqs.~\eqref{t-supN8}, \eqref{u-supN8} can be deduced directly from \eqref{eq:AN8treee1} by crossing symmetry in the $t$ and $u$-channel respectively, although it is instructive to derive them from the super-amplitude \eqref{superA0} by applying the appropriate differential operator,
	\begin{equation}\label{}
		\mathcal A_0^{\tau\bar\tau\to\tau\bar\tau} = 
		\partial_1\bar\partial_2\bar\partial_3\partial_4
		A_0 \Big|_{\eta=0}\,,\qquad
		\mathcal A_0^{\tau\bar\tau\to\bar\tau\tau} = 
		\partial_1\bar\partial_2\partial_3\bar\partial_4
		A_0 \Big|_{\eta=0}\,.
	\end{equation}
	
	Yet another choice is to consider linear combinations of the above states, for instance an axion and a dilaton,  as both in- and out-states, 
	\begin{equation}\label{}
		\phi_{1} = \phi_{4} = \frac{\tau+\bar\tau}{\sqrt2} \,, \qquad
		\phi_{2} = \phi_{3} = \frac{\tau-\bar\tau}{i\sqrt2}\,.
	\end{equation}
	The corresponding amplitude is given by
	\begin{equation}\label{}
		\mathcal A_0 = 
		-
		\frac14
		(\partial_1+\bar\partial_1)(\partial_2-\bar\partial_2)(\partial_3-\bar\partial_3)(\partial_4+\bar\partial_4)
		A_0 \Big|_{\eta=0}
	\end{equation}
	or, expanding the derivatives and recognizing the amplitudes calculated above,
	\begin{equation}\label{}
		\mathcal A_0 = \frac12 \left(
		\mathcal A_0^{\tau\tau\to\bar\tau\bar\tau}
		-
		\mathcal A_0^{\tau\bar\tau\to\tau\bar\tau}
		+
		\mathcal A_0^{\tau\bar\tau\to\bar\tau\tau}
		\right)
	\end{equation}
	and finally
	\begin{equation}
		\label{eq:AN8treee2}
		\mathcal A_0 = \kappa^2 \frac{s^4-t^4+u^4}{2stu}\,.
	\end{equation}
	In this way, by construction, the result is $s\leftrightarrow u$ symmetric. This will be useful in Section~\ref{sec:twoloop} since the $s\leftrightarrow u$ symmetry simplifies the analysis of the analytic properties of this amplitudes at higher loops. At tree level, of course \eqref{eq:AN8treee2} coincides with \eqref{eq:AN8treee1} up to irrelevant analytic terms as $t\to0$ that correspond to short-range contributions in $b$-space.
	
	One can introduce masses for the scalars by means of a Kaluza--Klein compactification. To this end, we can re-interpret \eqref{eq:AN8treee2} as (say) 10-dimensional, and perform a toroidal compactification of 6 spatial dimensions, letting
	\begin{equation}\label{eq:KKmasses}
		\begin{aligned}
			p_1^M &= (p_1^\mu,0,\ldots,0,0,m_1)\,,\qquad p_4^M = (p_4^\mu,0,\ldots,0,0,-m_1)\,,\\
			p_2^M &= (p_2^\mu,0,\ldots,0,m_2,0)\,,\qquad p_3^M = (p_3^\mu,0,\ldots,0,-m_2,0)\,,
		\end{aligned}
	\end{equation}
	in such a way that
	\begin{equation}\label{}
		0 = p_1^Mp_{1M} = p_1^\mu p_{1\mu} + m_1^2\,,\qquad 
		0 = p_2^Mp_{2M} = p_2^\mu p_{2\mu} + m_2^2
	\end{equation}
	and similarly for $p_3$, $p_4$. Likewise, from the 4D perspective,
	\begin{equation}\label{}
		-2p_1\cdot p_2 = s-m_1^2-m_2^2 \,,\qquad
		-(p_1+p_4)^2 = t\,,\qquad
		-2p_1\cdot p_3 = u-m_1^2 -m_2^2\,,
	\end{equation}
	and therefore
	\begin{equation}\label{atrees8afterKK}
		\mathcal A_0 = \kappa^2 \frac{(s-m_1^2-m_2^2)^4-t^4+(u-m_1^2-m_2^2)^4}{2(s-m_1^2-m_2^2)t(u-m_1^2-m_2^2)}\,.
	\end{equation}
	Calculating the leading eikonal by going to impact-parameter space is now straightforward, and neglecting contributions that lack the $1/t$ pole, one finds again the result 
	\begin{equation}
		\label{eq:2d0N=8}
		2\delta_0 = \frac{2 m_1m_2 G_D \sigma^2 \Gamma \left(\frac{D-4}{2}\right) }{\sqrt{\sigma^2-1} (\pi b^2)^{\frac{D-4}{2}}}\,,
	\end{equation}
	which we had obtained in \eqref{eq:d0dil} when discussing the $\mathcal N=0$ theory emerging from the double copy. Therefore, as far as the long-range effects captured by the leading eikonal exponentiation are concerned, switching on supergravity is equivalent to taking into account the additional dilaton exchanges between the two massive particles.\footnote{In the approach above, we assumed that the compactification scale is much smaller than the classical length scales relevant to the binary, such as the impact parameter and the effective size $GE$ of the colliding objects. Thus, long-range effects are captured just by the exchange of states with zero Kaluza--Klein numbers yielding for instance Eq.~\eqref{eq:2d0N=8}. See~\cite{KoemansCollado:2019lnh} for a treatment in which the dynamics of the Kaluza--Klein modes is included in an eikonal context.}
	
	Following the familiar steps, we can take $-p\frac{\partial}{\partial b}$ of \eqref{eq:2d0N=8} to obtain the leading deflection angle
	\begin{equation}\label{LOThetaN8}
		\Theta = \frac{4GE (\pi b^2)^{\epsilon}\sigma^2 \Gamma(1-\epsilon)}{(\sigma^2-1)b} + \mathcal O(G^2)
		\xrightarrow[D\to 4]{} \frac{4GE \sigma^2 }{(\sigma^2-1)b} + \mathcal O(G^2)
		\,.
	\end{equation}
	Let us conclude by looking again at the probe limit $m_1\ll m_2$, in which $E\simeq m_1 = M$ and $\sigma \simeq E_p/m_p$, so that from \eqref{LOThetaN8} we find
	\begin{equation}\label{}
		\Theta 
		 = \frac{4GM (\pi b^2)^{\epsilon}\sigma^2 \Gamma(1-\epsilon)}{(\sigma^2-1)b} + \mathcal O(G^2)
		 \xrightarrow[D\to 4]{} \frac{4GM \sigma^2 }{(\sigma^2-1)b} + \mathcal O(G^2)
	\end{equation}
in agreement with \eqref{eq:dangln8} in $D=4$ to leading order in $G$. 
In fact, show that the 1PM term determines completely the probe limit of the maximally supersymmetric case in four dimension \cite{Caron-Huot:2018ape}, and the result can be obtained by proceeding as we did in Subsection~\ref{ssec:exponapp}. Starting from the phase shift determined by \eqref{eq:2d0N=8},
\begin{equation}\label{}
	2\delta_j = \chi(J) = - \frac{2 m_1m_2 G \sigma^2}{\sqrt{\sigma^2-1}} \,\log J\,,
\end{equation}
one obtains an equation analogous to \eqref{ACV48},
\begin{equation}\label{}
	\tan\frac{\Theta}{2} =  \frac{2 m_1m_2 G \sigma^2}{\sqrt{\sigma^2-1}}\frac{1}{J} \simeq \frac{2MG\sigma^2}{b(\sigma^2-1)}\,,
\end{equation}
where we have used that $J = pb$ and, in the probe limit, $p\simeq m_p \sqrt{\sigma^2-1}$, thus recovering \eqref{eq:dangln8}.

	\subsubsection{Effective field theories beyond GR}
	\label{ssec:ethd}

	All cases discussed so far focus on gravitational theories whose action contains just two-derivative terms. At large distance this is certainly a reliable approximation, but it is interesting to see how higher derivative corrections can modify high energy scattering. By following~\cite{Camanho:2014apa} we start discussing the case where the standard gravitational action is modified by quadratic and cubic terms schematically as follows
	\begin{equation}
		\label{eqehr2r3}
		S_{gr} = \frac{1}{2 \kappa_D^2} \int d^D x \sqrt{|G|} \left[R +l_2^2 R^2 + l_4^4 R^3 \right]\;,
	\end{equation}
	where $l_i$ are the length scales determining when the new higher derivative terms start becoming relevant, which we take to be parametrically larger than $\ell_P$. From an effective field theory point of view, it may appear strange to assume that the $l_i$ are decoupled from $\ell_P$, but this is quite natural in setups where the modification in~\eqref{eqehr2r3} arise from a {\em classical} microscopic dynamics, as in string theory, or from integrating out non-gravitational degrees of freedom as in~\cite{Bellazzini:2021shn,Bellazzini:2022wzv}. In explicit string constructions the scales $l_i$ can be written in terms of the string length $\ell_s$, but in this section we will work within an effective field theory approach and keep them arbitrary just assuming $l_i \gg \ell_P$. It is interesting to focus on the corrections in~\eqref{eqehr2r3} because they are the only ones that can modify the on-shell 3-graviton amplitude. Strictly speaking this amplitude vanishes because the on-shell conditions force the momenta of the external states to be collinear. However one can either complexify the momenta or work in a signature with two times to define a non-trivial on-shell 3-graviton amplitude. So from~\eqref{eqehr2r3} we have
	\begin{equation}
		\label{eq:3grmod}
		A_3 = A_3^{(0)} + A_3^{(2)} + A_3^{(4)}\;,
	\end{equation}
	where $A_3^{(0)}$ is the contribution from the Einstein-Hilbert part of the action one and the other two terms are higher derivative corrections with two and four additional derivatives respectively. The Einstein-Hilbert contribution reads 
	\begin{equation}
		\label{eq:3am0}
		i A^{(0)}_3 = -2i \kappa_D \left[(\varepsilon_1 \varepsilon_2) (\varepsilon_3 q_1) + \mbox{cycl.}  \right] \left[(\tilde\varepsilon_1 \tilde\varepsilon_2) (\tilde\varepsilon_3 q_1) + \mbox{cycl.} \right]\;,
	\end{equation}
	where $\varepsilon_{i\,\mu} \tilde\varepsilon_{i\,\nu} = h_{i\,\mu\nu}$ is a symmetric tensor representing the polarization of one of the external gravitons and in each parenthesis one needs to sum over the cyclical permutations. The correction proportional to $l_2^2$ reads
	\begin{equation}
		\label{eq:3am2}
		i A^{(2)}_3 = 2 i\kappa_D l_2^2\; \Big\{[(\varepsilon_1 \varepsilon_2) (\varepsilon_3 q_1) + \mbox{cycl.} ] [(\tilde\varepsilon_1 q_2) (\tilde\varepsilon_2 q_3) (\tilde\varepsilon_3 q_1)] + \varepsilon \leftrightarrow \tilde\varepsilon\Big\}\;,
	\end{equation}
	while the one proportional to $l_4^4$ is
	\begin{equation}
		\label{eq:3am4}
		i A^{(4)}_3 = 2 i\kappa_D l_4^4 \left[(\varepsilon_1 q_2) (\varepsilon_2 q_3) (\varepsilon_3 q_1) \right] \left[(\tilde\varepsilon_1 q_2) (\tilde\varepsilon_2 q_3) (\tilde\varepsilon_3 q_1)\right]\;.
	\end{equation}
Since these on-shell couplings are written in the factorized form following from the KLT or the double copy relations, one can immediately see that they are obtained by taking the products of the space-time part of the standard 3-gluon coupling and its higher derivative modification related to a $\operatorname{Tr}(F^3)$ in the Lagrangian. In the supersymmetric theories the modification in~\eqref{eq:3am4} is not allowed and in the maximally supersymmetric case also~\eqref{eq:3am2} is set to zero and so, in this case, the 3-point vertex cannot be deformed. In the case of pure gravity in $D=4$, the modification~\eqref{eq:3am2} does not contribute to the eikonal since it can be derived by the Gauss--Bonnet term which is a total derivative in four dimension. In the following we will keep $D$ generic and we will not decouple the dilaton and the antisymmetric tensor arising from the double copy construction.	
	
	We can now discuss how the higher derivative corrections modify the tree-level amplitude~\eqref{eq:2g2ms} describing scattering of a graviton off a massive scalar. We focus in particular on the high-energy limit and look for terms that scale at least as the factor of $\gamma(s,m_1,m_2=0)$ present in the Einstein-Hilbert case~\eqref{eq:2g2mhe}. In order to do this we can simply combine, by using a de Donder propagator~\eqref{GGagrav}, an on-shell 3-graviton vertex obtained from~\eqref{eq:3grmod} and the vertex with the external scalars~\eqref{taupp}.
        As a warming up exercise one can derive~\eqref{eq:2g2mhe} from~\eqref{eq:3am0}: by combining the scalar vertex with the de Donder propagator, one obtains
	\begin{equation}
		\label{eq:ssgded}
		A_{\mu \nu}^{(1)} =- \frac{\kappa_D}{q^2} \left( k_{1\mu} k_{4\nu}+  k_{1 \nu} k_{4 \mu} - \eta_{\mu \nu } \frac{2m_1^2}{D-2} \right)\;,
	\end{equation}
	which can be saturated with the variation of~\eqref{eq:3am0} with respect to $\epsilon_{1\,\mu\nu}$ (identifying that graviton as the particle exchanged in the $t$-channel). Only the term proportional to $q_2 \epsilon_1 q_2$ can produce a leading contribution in the energy since, when saturated with~\eqref{eq:ssgded}, it yields the scalar product $2 (k_1 q_2) (k_4 q_2)\simeq - \gamma(s,m_1^2,m_2=0)$ where in the last step we neglected a term proportional to $t$. This reproduces exactly the result~\eqref{eq:2g2mhe}. One can follow the same approach by using~\eqref{eq:3am2} which provides the corrections proportional to $l_2^2$ to the scattering discussed above. Again in this effective vertex there is a term proportional to $q_2 \epsilon_1 q_2$ which for the same reasons above provides a contribution that scales with the energy exactly in the same way as the term coming from the standard Einstein-Hilbert 3-graviton vertex. The main qualitative difference that we find in this case is that the polarization of the external graviton appear dotted with the exchange momentum $q$
	\begin{equation}
		\label{eq:2g2mhel2}
		i {\cal A}_0 \sim \frac{2 i \kappa_D^2}{-t} \,\gamma(s,m_1,m_2=0)\,(2 l_2^2)\, [q \epsilon_1\epsilon_2 q]\; .
	\end{equation}
	By following similar steps, one can obtain the correction related to $l_4$
	\begin{equation}
		\label{eq:2g2mhel4}
		i {\cal A}_0 \sim -\frac{2 i \kappa_D^2}{-t} \,\gamma(s,m_1,m_2=0)\,(2 l_4^4)\, [(q \epsilon_1 q ) (q \epsilon_2 q)]\;.
	\end{equation}
	Seen as contributions to a tree-level amplitude the results in~\eqref{eq:2g2mhel2} and~\eqref{eq:2g2mhel4} do not have anything unusual, but since they behave as~\eqref{eq:2g2mhe} in the limit~\eqref{eq:smggt} we expect them to exponentiate and contribute to leading {\em classical} eikonal. As before we rewrite the result in impact parameter space:
	\begin{equation}
          \label{eq:d0m1gl2l4}
          \begin{aligned}
            2\delta_0 = \frac{G_D \gamma(s,m_1,m_2=0)}{\sqrt{\sigma^2-1}}
            \frac{\Gamma \left(\frac{D-4}{2}\right) }{ (\pi b^2)^{\frac{D-4}{2}}}
            & \left[\epsilon_{1\,ij} \epsilon_{2\,ij} + (D-4) \frac{2 l_2^2}{{b}^2} \epsilon_{1\,ij} \Pi_{jh} \epsilon_{2\,hi}\,-  \right. \\ & ~~ \left. (D-2)(D-4) \frac{l_4^4}{{b}^4} \epsilon_{1\,ih} \epsilon_{2\,jk} \Pi_{ijhk}  \right]\,,
            \end{aligned}
	\end{equation}
	where we labeled the physical polarizations with the indices $i,j,\ldots$ of the transverse space and 
	\begin{equation}
		\label{eq:Pi2i4i}
		\begin{aligned}
			\Pi_{ij} & =&&\delta_{ij} - (D-2) \frac{b_i b_j}{b^2}\;, \\
			\Pi_{ijhk} & =&&\delta_{hk} \delta_{ij} +
			\delta_{hj} \delta_{ik} + \delta_{jk} \delta_{ih} - \frac{D}{{b}^2}\Big({b}_h 
			{b}_k \delta_{ij} + {b}_h {b}_j \delta_{ik} 
			\\ & && \!+ {b}_i {b}_h \delta_{jk} + {b}_j {b}_k \delta_{ih} + 
			{b}_i {b}_k \delta_{jh} + {b}_i {b}_j \delta_{hk} -
			(D+2) \frac{{b}_i {b}_j {b}_h {b}_k}{{b}^2}\Big) \ .
		\end{aligned}
	\end{equation}
	Of course the first term in~\eqref{eq:d0m1gl2l4} is just the Einstein-Hilbert contribution of Eq.~\eqref{eq:d0m1g} which is identical for all the polarizations of the graviton. Notice, on the contrary, that the corrections related to $l_2$ and $l_4$ are not universal and depend on whether the polarizations involved have a trivial projection (or not) along the direction ${b}$ of the impact parameter. The main novelty of the eikonal~\eqref{eq:d0m1gl2l4} with respect to the cases discussed so far is that it acts non-trivially in the space of the polarizations of the incident massless particles. Thus, instead of being a phase it becomes an operator mixing different helicities of the graviton, the dilaton and the antisymmetric tensor. As discussed in Section~\ref{ssec:seikcc} this is the origin of causality violating contributions in the eikonal scattering for these modified theories when the impact parameter becomes of the order of $l_{2,4}$~\cite{Camanho:2014apa}.

	\subsubsection{Including classical spin}
\label{ssec:incspin}

In this subsection we generalize the calculation of the leading eikonal phase to the case in which the massive  particles also have a ``classical spin'', i.e.~an intrinsic angular momentum as opposed to the one associated to their orbital motion. 
At the level of solutions of Einstein's equations, introducing spin corresponds to going  from the Schwarzschild solution, characterized by its mass, to the Kerr solution describing a black hole with both a mass $m$ and spin $J_a$, provided the inequality $J_a \le G m^2$ is satisfied (with the equality sign corresponding to the ``extremal" case). Both the Schwarzschild and Kerr solutions involve the full non-linear structure of GR. 

One can, however, construct a linearized version of the Kerr black hole by keeping in the GR Lagrangian only the kinetic term of the gravitational field and a term that describes its interaction with the energy-momentum  tensor of the spinning matter. Then, from it, one can extract the three-point amplitude involving two massive particles with spin and a graviton~\cite{Vines:2017hyw,Guevara:2019fsj,Guevara:2018wpp} (the round parenthesis indicates symmetrization without additional factors, $A^{(\alpha}B^{\beta)} = A^\alpha B^\beta+A^\beta B^\alpha$):
\begin{equation}\label{spinvertex}
\tau^{\mu\nu}(p,p',k;a) = \kappa_D \bar{p}^{(\mu}\mathrm{exp}(i\epsilon^{\nu)}{}_{\rho\alpha\beta}a^{\alpha}k^{\beta})\bar{p}^{\rho}\,,
\end{equation}
where all three momenta entering the vertex are regarded as outgoing, and can
take complex values in order to obey both momentum conservation and the mass shell relations,
\begin{equation}\label{}
	p^\mu + p'^\mu + k^\mu= 0\,,
	\qquad
	\frac12 \left(
	p'^\mu-p^\mu
	\right)
	= \bar p^\mu\,,
	\qquad
	p^2 = -m^2 = p'^2\,,\qquad
	k^2 = 0=
	\bar p\cdot k
	\,.
\end{equation}
The effective vertex \eqref{spinvertex} 
involves 
the spin vector $a^\mu$ of the massive object. This is related to the spin tensor $S^{\mu \nu}$ through the following relations,\footnote{To leading order in $k^\mu$, we may disregard the difference between $\bar p^\mu$, $p'^\mu$ and $-p'^\mu$.}
\begin{equation}
	a^\rho =- \frac{1}{2m^2} \epsilon^{\rho \alpha \beta \gamma} \bar{p}_{\alpha} S_{\beta \gamma}\,,\qquad 
	S^{\mu \nu} =\epsilon^{\mu \nu \rho \sigma} \bar p_\rho a_{\sigma}
	\label{IS3}
\end{equation} 
and satisfies in particular
\begin{equation}\label{}
	a\cdot \bar p =  0\,,
	\qquad
	J_a = m a\,,
\end{equation}
where $J_a$ is the magnitude of the spin angular momentum, as can be seen by going to the rest frame. Of course,
in the $a_i\to 0$ limit, Eq.~\eqref{spinvertex} reduces to the non-spinning effective vertex~\eqref{eq:taupp2meff}. 

Using the above properties as well as the identities involving $\epsilon_{\mu\nu\rho\sigma}$ collected for convenience in \ref{LeviCivita}, one can show that
\begin{equation}\label{}
	\epsilon\indices{^\nu_{\mu\alpha\beta}} a^\alpha k^\beta \epsilon\indices{^\mu_{\rho\gamma\delta}}\, a^\gamma k^\delta\ {\bar{p}}^\rho
	= - (a\cdot k)^2 \delta\indices{^\nu_\rho} \,\bar p^\rho\,.
\end{equation}
As a result, the matrix exponential in \eqref{spinvertex} can be rewritten in the more explicit form
\begin{equation}\label{}
e^{i \epsilon\indices{^\nu_{\rho\alpha\beta}}\, a^\alpha k^\beta} {\bar{p}}^\rho
=
	\left[\cosh(a\cdot k) \delta\indices{^{\nu}_\rho} + i \frac{\sinh(a\cdot k)}{a\cdot k} \epsilon\indices{^\nu_{\rho \alpha\beta}} a^\alpha k^\beta 
	\right] \bar p^\rho \,,
\end{equation}
and the vertex reads 
\begin{equation}\label{threepointspin}
	\tau^{\mu\nu}(p,p',k;a) = 
	i\kappa
	\left[\cosh(a\cdot k) 2 \bar p^\mu \bar p^\nu + i \frac{\sinh(a\cdot k)}{a\cdot k} \left(  \bar p^\mu 
\epsilon\indices{^\nu_{\rho \alpha\beta}} a^\alpha k^\beta 	 \bar p^\rho + \bar p^\nu 
\epsilon\indices{^\mu_{\rho \alpha\beta}} a^\alpha k^\beta \bar p^\rho \right)\right] .
\end{equation}
We note that the trace of this vertex only comes from the  $\cosh$ part,
\begin{equation}\label{tracespin}
	\eta_{\mu\nu} \tau^{\mu\nu}(p,p',k;a) = 2i \kappa \bar p^2 \cosh(a\cdot k)\,.
\end{equation}

We can now sew together two copies of the on-shell vertex \eqref{threepointspin}, describing each the motion of a distinct particle labeled by 1 and 2, with the de Donder propagator~\eqref{GGagrav}, obtaining the following $2\to2$ amplitude for the scattering of two massive spinning objects,
\begin{equation}\label{a0sewn}
	i\mathcal A_0 = \tau^{\mu\nu}(p_1,p_4,-q;a_1) G_{\mu \nu,\rho \sigma} (q)
	\tau^{\rho\sigma}(p_2,p_3,q;a_2)\,.
\end{equation}
Here of course $q^2$ is not zero, unlike for $k^\mu$ in \eqref{spinvertex}. Therefore, the expression \eqref{a0sewn} is only accurate up to contact terms, i.e.~contributions whose residue at $q^2\to0$ vanishes, which only contribute to short-range effects.
Vice-versa, the relations
\begin{equation}\label{}
	\bar p_{1,2} \cdot q  =0\,,\qquad \bar p_{1}\cdot a_1 = 0\,,\qquad \bar p_2\cdot a_2 = 0
\end{equation}
still hold and we may now consider real kinematics. Writing explicitly \eqref{a0sewn} we get
\begin{align}
		\label{IS21}
		{\cal A}_0 &=  \frac{2\kappa^2}{q^2}   \left[\cosh(a_1 q) \bar{p}_1^{\mu} \bar{p}_1^{\nu} - \frac{i}{2}(\epsilon^{\mu}{}_{\gamma\alpha\beta}\bar{p}_1^{\nu}+\epsilon^{\nu}{}_{\gamma\alpha\beta}\bar{p}_1^{\mu})\bar{p}_1^{\gamma}a^{\alpha}_1q^{\beta}\frac{\sinh(a_1 q)}{q a_1} \right] \\
		&\times  (2\eta_{\mu\rho}\eta_{\nu\sigma}-\eta_{\mu\nu}\eta_{\rho\sigma}) 
		\left[\cosh(a_2 q) \bar{p}_2^{\rho} \bar{p}_2^{\sigma} +\frac{i}{2}(\epsilon^{\rho}{}_{\lambda\delta\eta}\bar{p}_2^{\sigma}+\epsilon^{\sigma}{}_{\lambda\delta\eta}\bar{p}_2^{\rho})\bar{p}_2^{\lambda}a^{\delta}_2q^{\eta}\frac{\sinh(a_2 q)}{q a_2}\right]\;. \nonumber
	\end{align}     
The term with  $\cosh(qa_1) \cosh(qa_2)$ can be easily computed and one gets 
	\begin{equation}
		\frac{2\kappa^2}{q^2} m_1^2 m_2^2 (2\sigma^2-1) \cosh(qa_1) \cosh(qa_2)\,,
		\label{IS22}
	\end{equation}
For the computation of the term with 	$\sinh(qa_1) \sinh(qa_2)$ we can take advantage of  the fact that only the first term of the de Donder propagator contributes (the second term vanishes according to the observation before  \eqref{tracespin}) and one gets:
\begin{align}
&\frac{2 \kappa^2}{q^2} \frac{\sinh(a_1 q)}{q a_1}\frac{\sinh(a_2 q)}{q a_2} \nonumber \\
&\times \Bigg( - (\epsilon\indices{_{ \gamma \sigma \alpha\beta}} {\bar{p}}_{1}^{\gamma} {\bar{p}}_2^\sigma  a_{1}^\alpha q^\beta)  	 (\epsilon\indices{_{ \rho \lambda \delta \eta}} {\bar{p}}_{1}^{\rho} {\bar{p}}_2^\lambda  a_{2}^\delta q^\eta)  +(p_1p_2) \epsilon\indices{^\rho_{\gamma \alpha\beta}} {\bar{p}}_1^\gamma a_1^\alpha q^\beta   \epsilon_{\rho \lambda \delta \eta} {\bar{p}}_2^\lambda a_2^\delta q^\eta 
\Bigg)	 
\label{IS24}
\end{align}
The second term in the big round parenthesis can be computed using the relation
\begin{equation}
	\label{firstone}
\epsilon\indices{^\rho_{\gamma \alpha\beta}}{\bar{p}}_1^\gamma a_1^\alpha q^\beta \epsilon_{\rho \lambda \delta \eta} {\bar{p}}_2^\lambda a_2^\delta q^\eta=
(\bar{p}_1 \cdot \bar{p}_2) (a_1 \cdot q) (a_2 \cdot q)+\left(a_1 \cdot \bar{p}_2\, a_2 \cdot \bar{p}_1-\bar{p}_1 \cdot \bar{p}_2\, a_1 \cdot a_2\right) q^2
\end{equation}
that can be obtained from the formulas in  \ref{LeviCivita}, while for the first term we need the following relation
\begin{equation}\label{|eps|}
	-
	(\epsilon_{\mu\nu\alpha\beta} \bar p_1^\mu \bar p_2^\nu a_1^\alpha q^\beta)
	(\epsilon_{\mu\nu\alpha\beta} \bar p_1^\mu \bar p_2^\nu a_2^\alpha q^\beta)
	=
	\left[\left(\bar{p}_1 \cdot \bar{p}_2\right)^2-\bar{p}_1^2 \bar{p}_2^2\right]\left(a_1 \cdot q a_2 \cdot q-a_1 \cdot a_2 q^2\right)+\bar{p}_1 \cdot \bar{p}_2 a_1 \cdot \bar{p}_2 a_2 \cdot \bar{p}_1 q^2
\end{equation}
Actually any term proportional to $q^2$ in \eqref{firstone} and \eqref{|eps|} can be neglected because it will give an analytic piece. 
This means that   the term with $\sinh(qa_1) \sinh(qa_2)$ is equal to
\begin{equation}
		\frac{2\kappa_D^2}{q^2} m_1^2 m_2^2 (2\sigma^2-1) \sinh(qa_1) \sinh(qa_2)\,.
		\label{IS23}
	\end{equation}
Finally, the two mixed terms can be computed similarly, and one gets the following final result
\begin{equation}\label{a0intermediate}
	\begin{aligned}
		\mathcal A_0 
		&= \frac{2\kappa^2m_1^2m_2^2 \sigma}{q^2}
		\Big[
		\sigma
		(1+v^2) \cosh(a\cdot q)
		\\
		&
		-2i \epsilon_{\mu\nu\alpha\beta} v_1^\mu v_2^\nu \left(
		a_2^\alpha q^\beta \frac{\sinh(a_2\cdot q)}{a_2\cdot q}\,\cosh(a_1\cdot q) 
		+
		a_1^\alpha q^\beta \frac{\sinh(a_1\cdot q)}{a_1\cdot q}\,\cosh(a_2\cdot q) 
		\right)
		\Big],
	\end{aligned}
\end{equation}
where the term in the first line comes from the sum of the terms in \eqref{IS22} and \eqref{IS23} and the identity $\sigma^2 (1+ v^2)= 2\sigma^2-1$.
In the previous equation we used
\begin{equation}\label{}
	\bar p_1^\mu \simeq m_1 v_1^\mu \,,\qquad
	\bar p_2^\mu \simeq m_2 v_2^\mu \,,\qquad
	\sigma = - v_1\cdot v_2
\end{equation}
and introduced 
\begin{equation}\label{}
	a^\mu = a_1^\mu + a_2^\mu\,,\qquad
	\sigma= \frac{1}{\sqrt{1-v^2}}\,.
\end{equation}

In the center of mass,
\begin{equation}\label{}
	\bar p_1^{\mu} \simeq (E_1,\vec p\,)\,,\qquad 
	\bar p_2^{\mu} \simeq (E_2,-\vec p\,)\,,
	\qquad
	q^\mu=(0,\vec q\,)\,,
\end{equation}
letting
\begin{equation}\label{}
	{\vec{p}} = (0,0,p)\,,\qquad
	\vec q = (\mathbf q,0) =  (q^x,q^y,0)\,,
\end{equation}
we have
\begin{equation}\label{eps}
	\epsilon_{\mu\nu\alpha\beta} \bar p_1^\mu \bar p_2^\nu a_1^\alpha q^\beta
	=-m_1m_2\sigma v (\hat{p}\times \vec a_1)\cdot \vec q\,,
\end{equation}
so that \eqref{a0intermediate} can be rewritten as follows
\begin{equation}\label{a0intermediate2}
	\begin{split}
		\mathcal A_0 
		&= \frac{2\kappa^2m_1^2m_2^2 \sigma^2}{q^2}
		\Big[
		(1+v^2) \cosh(a\cdot q)\\
		&+
		2iv
		\left( (\hat{p}\times \vec a_2)\cdot \vec q\, \frac{\sinh(a_2\cdot q)}{a_2\cdot q}\,\cosh(a_1\cdot q) 
		+
		(\hat{p}\times \vec a_1)\cdot \vec q \, \frac{\sinh(a_1\cdot q)}{a_1\cdot q}\,\cosh(a_2\cdot q) 
		\right)
		\Big].
	\end{split}
\end{equation}
Comparing \eqref{|eps|} and \eqref{eps}, we see that
\begin{equation}\label{residue}
	(a_1\cdot q)^2 = a_1^2 q^2-((\hat{p}\times \vec a_1)\cdot \vec q\,)^2\,.
\end{equation}
We note that taking a square root of \eqref{residue} and setting $q^2=0$, one can effectively replace 
\begin{equation}\label{substitution}
	a_1\cdot q
	\to 
	\pm i (\hat{p}\times \vec a_1)\cdot \vec q
\end{equation}
in \eqref{a0intermediate} at the price of short-range corrections (similarly for $a_2$). One can check this in the $m_2=0$ case~\cite{Bautista:2021wfy} by writing $q^2=(p_2+p_3)^2$ in terms of the spinor formalism as in~\eqref{eq:angsqbra}. Then one can extract the leading term for small $q$ by sending either $\langle 2 3\rangle$ or $[23]$ to zero. This choice is related to the potential sign ambiguity introduced by the square root~\eqref{substitution}, see~\cite{Bautista:2021wfy} for a discussion of this point. However this sign is immaterial in~\eqref{a0intermediate2} and the terms in the second line of that equation recombine via $\sinh\alpha\cosh\beta+\sinh\beta\cosh\alpha=\sinh(\alpha+\beta)$, leading to 
\begin{equation}\label{}
	\mathcal A_0 
	= \frac{2\kappa^2m_1^2m_2^2 \sigma^2}{q^2}
	\Big[
	(1+v^2) \cosh( i(\hat{p}\times \vec a)\cdot \vec q\,)
	+
	2v
	\sinh( i(\hat{p}\times \vec a)\cdot \vec q\,)
	\Big],
\end{equation}
which can be also recast in the compact form
\begin{equation}\label{a0compact}
	\mathcal  A_0 = \frac{\kappa^2m_1^2m_2^2\sigma^2}{q^2} \sum_{\eta= \pm1}(1+\eta v)^2 e^{i\eta \vec c \cdot \vec q}\,,\qquad \vec c = \hat{p}\times \vec a\,.
\end{equation}

Starting from \eqref{a0compact}, and going to impact-parameter space in the 
usual way, 
\begin{equation}\label{}
	2\delta_0= 
	\tilde{\mathcal{A}}_0 = \frac{\kappa^2m_1 m_2 \sigma}{4v} \sum_{\eta= \pm1}(1+\eta v)^2 
	\int\frac{d^{2-2\epsilon} q}{(2\pi)^{2-2\epsilon}}\frac{e^{i(\vec b+\eta \vec c\,)\cdot \vec q}}{q^2}\,,
\end{equation}
and, using \eqref{B1}, we obtain the leading eikonal phase including classical spin,
\begin{equation}\label{eikonalwithspin}
	2\delta_0
	= \frac{\kappa^2m_1 m_2 \sigma}{4v}
	\frac{1}{4\pi^{1-\epsilon}} \sum_{\eta= \pm1}(1+\eta v)^2 
	\frac{\Gamma(-\epsilon)}{\big(\big|\vec{b}+\eta \vec{c}\,\big|^2\big)^{-\epsilon}}\,.
\end{equation}
We remark that $\vec c$ is orthogonal to $\vec p$ by its definition \eqref{a0compact}.
The impulse is therefore given by
\begin{equation}\label{deflectionwithspin}
	-\vec Q =
	- \frac{\partial 2\delta_0}{\partial \vec{b}}
	=
	\frac{\kappa^2m_1 m_2 \sigma}{2v}
	\frac{1}{4\pi} \sum_{\eta= \pm1}(1+\eta v)^2 
	\frac{\vec {b}+\eta {\vec{c}}}{\big|\vec {b}+\eta {\vec{c}}\,\big|^2}\,.
\end{equation}
We see that the entire spin dependence is encoded in the shift $\vec{b}\rightarrow \vec{b}\pm {\vec{c}}$, which is reminiscent of the Newman--Janis shift \cite{Newman:1965tw}, relating Kerr to Schwarzschild black holes.

\begin{figure}
	\centering
	\begin{subfigure}{.45\textwidth}
		\centering
		\begin{tikzpicture}[scale=.6]
			\draw[help lines,->] (-5,0) -- (5,0) coordinate (zaxis);
			\draw[help lines,->] (-3,-2) -- (3,2) coordinate (xaxis);
			\draw[help lines,->] (0,-3) -- (0,3) coordinate (yaxis);
			\draw[ultra thick, blue, ->] (2.1,1.4) -- (-1.8,-1.2);
			\draw[ultra thick, ->] (-3.8,-1.2) -- (-1.8,-1.2); 
			\draw[dashed]  (-3.8,-1.2).. controls (-1,-1.1) and (1,-1.3) ..(4.5,-.1);
			\draw[ultra thick, ->] (4.1,1.4) -- (2.1,1.4);
			\draw[dashed]  (4.1,1.4).. controls (1.3,1.3) and (-.7,1.5) ..(-4.2,.3); 
			\draw[ultra thick, red, ->] (-3.8,-1.2) -- (-3.8,-.5);
			\filldraw[black] (-3.8,-1.2) circle (5pt);
			\draw[ultra thick, red, ->] (4.1,1.4) -- (4.1,2);
			\filldraw[black] (4.1,1.4) circle (5pt);
			\draw[ultra thick, green!60!black, ->] (0,0) -- (0,2); 
			\node at(5,0)[below]{$z$};
			\node at(0,3.1)[left]{$y$};
			\node at(3,2)[right]{$x$};
			\node at(-2.8,-1.2)[below]{$\vec p$};
			\node at(3.1,1.4)[below]{$-\vec p$};
			\node at(-3.8,-.5)[left]{$\vec a_1$};
			\node at(4.1,2)[right]{$\vec a_2$};
			\node at(0,-.7)[left]{$\vec{b}$};
			\node at(0,2)[left]{$\vec{L}$};
		\end{tikzpicture}
		\caption{Scattering with aligned spins.\label{fig:spinsaligned}}
	\end{subfigure}
	\begin{subfigure}{.45\textwidth}
		\centering
		\begin{tikzpicture}[scale=.6]
			\draw[help lines,->] (-5,0) -- (5,0) coordinate (zaxis);
			\draw[help lines,->] (0,-3) -- (0,3) coordinate (zaxis);
			\draw[help lines,->] (-3,-2) -- (3,2) coordinate (xaxis);
			\draw[ultra thick, blue, ->] (2.1,1.4) -- (-1.8,-1.2);
			\draw[ultra thick, ->] (-3.8,-1.2) -- (-1.8,-1.2); 
			\draw[ultra thick, ->] (4.1,1.4) -- (2.1,1.4); 
			\draw[ultra thick, red, ->] (-3.8,-1.2) -- (-3.6,-.5);
			\draw[dashed]  (-3.8,-1.2).. controls (-1,-1.1) and (1,-1.3) ..(4.5,-.1);
			\filldraw[black] (-3.8,-1.2) circle (5pt);
			\draw[ultra thick, red, ->] (4.1,1.4) -- (4.2,.8);
			\draw[dashed]  (4.1,1.4).. controls (1.3,1.3) and (-.7,1.5) ..(-4.2,.3); 
			\filldraw[black] (4.1,1.4) circle (5pt);
			\draw[ultra thick, green!60!black, ->] (0,0) -- (0,2); 
			\node at(5,0)[below]{$z$};
			\node at(0,3.1)[left]{$y$};
			\node at(3,2)[right]{$x$};
			\node at(-2.8,-1.2)[below]{$\vec p$};
			\node at(3.1,1.4)[below]{$-\vec p$};
			\node at(-3.6,-.5)[left]{$\vec a_1$};
			\node at(4.2,.8)[right]{$\vec a_2$};
			\node at(0,-.7)[left]{$\vec{b}$};
			\node at(0,2)[left]{$\vec{L}$};
		\end{tikzpicture}
		\caption{Scattering with generic spin orientations.\label{fig:spinsgeneric}}
	\end{subfigure}
	\caption{Scattering of spinning objects in the center-of-mass frame.\label{fig:spins}}
	\label{scatteringplanespin}
\end{figure}
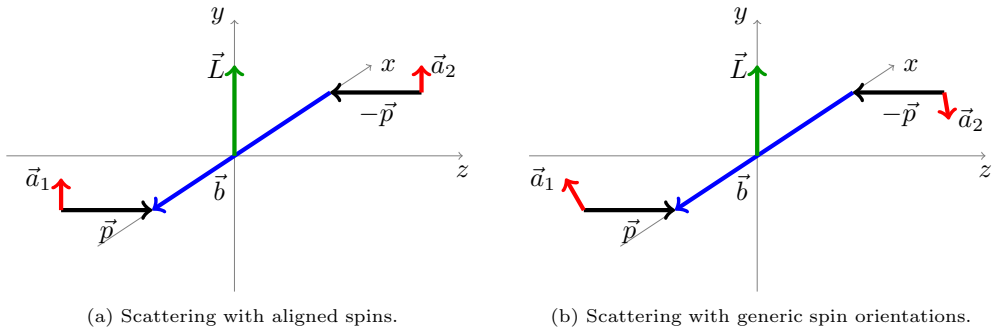

These results are valid for generic spin orientations (see Fig.~\ref{fig:spins}).
Let us now consider the case in which both spins are parallel to the angular momentum in the center of mass frame as in Fig.~\ref{fig:spinsaligned}. That is, we align the impact parameter by letting
\begin{equation}\label{bbbb}
	b^{\mu} = (0,-b,0,0)\,, 
\end{equation}
so that $\vec L = \vec b \times \vec p = (0,pb,0)$,
then
\begin{equation}\label{aaaa}
	\vec a_{1,2} = (0,a_{1,2},0)\,,\qquad
	a_{1,2}>0\,,\qquad
	\vec c = (-a,0,0)\,,
\end{equation}
as summarized in Fig.~\ref{scatteringplanespin}(a).
In this case, \eqref{eikonalwithspin} reduces to 
\begin{equation}\label{}
	2\delta_0 = \frac{\kappa^2m_1m_2\sigma}{2v}
	\left[
        -\frac{1+v^2}{4\pi^{1-\epsilon}}\left(\frac{1}{\epsilon} + \log (b^2-a^2) + \gamma \right)
	-
	\frac{2v}{4\pi}
	\log\frac{b+a}{b-a}
	\right]
	+\mathcal O(\epsilon)\,,
\end{equation}
and the impulse is given by 
\begin{equation}\label{impulsealigned}
	- Q^\mu = Q\, \frac{{b}^\mu}{b}\,,\qquad
	Q  = \frac{\kappa^2m_1m_2\sigma}{v} \frac{1}{4\pi b} \frac{(1+v^2)-\frac{2va}{b}}{1-\frac{a^2}{b^2}}\,,
\end{equation} 
which agrees with Eq.~(82) of Ref.~\cite{Vines:2017hyw}. The previous result corresponds to the case where the two spins are parallel to the orbital angular momentum (see~\eqref{aaaa} and the equation  for  $\vec{L}$ after \eqref{bbbb}). If we had instead taken them to be anti-parallel, we would have obtained a plus sign in the last term in the numerator of  \eqref{impulsealigned}. This implies that the sign in the last term in the numerator  of  \eqref{impulsealigned}   is such that for parallel (antiparallel) orbital angular momentum and spin the deflection angle decreases (increases) in agreement with expectations~\cite{Zhang:2022rnn}.

We remark that, in view of the bound on the spin, $a\le Gm$, and of the PM limit $Gm \ll b$, the parameter $a/b$ is always very small. For this reason, the final results \eqref{deflectionwithspin}, \eqref{impulsealigned} should be thought of only as a convenient way to package their power series around $a/b=0$.

	\subsection{String Theory}
	\label{sec:string-disk}
	
	Because of the  presence of a massless spin-two state in its spectrum, string theory naturally contains gravity and, at sufficiently low energies (large distances), it reduces to an effective gravitational field theory. It is therefore natural to study the gravitational 2-body scattering in a string theory framework\footnote{Although the large distance physics discussed in this report appears to be insensitive to the ultraviolet completion of the theory, it is also desirable to work in a framework in which such a completion is explicitly implemented.}. The focus of this section is to derive the leading eikonal in a truly stringy regime, {\rm i.e.} when an effective field theory description is not reliable and the results obtained in the previous subsections cannot be used. Particular emphasis is given to the new phenomena arising in a string context which are related to the existence of Regge trajectories and to the extended nature of the fundamental objects. This analysis was first performed in~\cite{Amati:1987wq,Amati:1987uf} focusing on the string theory version of the high-energy massless scattering\footnote{As we will see, this high-energy limit is dominated by the leading Regge trajectory of the graviton, while subleading Regge trajectories, such as the one of the dilaton, become irrelevant.} discussed in Section~\ref{sec:leading}. Here we start from the slightly simpler fixed-target setup where a light string scatters off a stack of D$p$-branes~\cite{DAppollonio:2010ae}. Then we will summarize the results obtained in the string-string collision case in Section~\ref{ssec:string-brane-sup}.
	
	Conceptually, we follow exactly the same approach discussed in the field theory context: we use scattering amplitudes to derive the dynamical information relevant for the leading eikonal and then, as a second step, show how to interpret the result as the motion of a probe in a curved background (in the spirit of Section~\ref{ssec:ASmetric}). In the main text we focus mostly on type II superstring theories, which are free of tachyonic instabilities, and we refer to the original references for the derivations of several string results that are the starting point of our analysis. In ~\ref{app:string} we provide some of these derivations in the context of bosonic string theory where it is possible to focus on the key features by using a simpler formalism. It turns out that the tachyonic instabilities do not play an important role in the definition of the string eikonal, since as we shall see they do not contribute to long-range effects, which instead arise from the expansion of the leading Regge trajectory around $t=0$. Thus bosonic string theory provides the perfect arena both for developing the technical analysis and for discussing some interesting features of non-supersymmetric setups (see for instance Section~\ref{ssec:causality-rest}). In order to make the comparison between supersymmetric and bosonic string theories easier we indicate the number of spacetime dimensions with $d$: of course we have $d=10$ for type II theories or $d=26$ for the bosonic theory, but several results at high energies take the same functional form if the spacetime dimension is formally indicated with $d$ (while $D$ will denote the number of noncompact dimensions).
	
	\subsubsection{String-brane scattering at tree level}
	\label{ssec:string-brane}
	
	We consider the scattering of a massless closed string off a stack of $N$ coincident D$p$-branes in type II theories where some space-like directions can be compactified on circles. While we will stick to this simple setup, it should be possible to extend this analysis to more complicated configurations with orbifold compactifications and more general boudary conditions~\cite{Angelantonj:2002ct}. The leading contribution is captured by a worldsheet with the topology of the disk and two punctures in its interior (see Fig.~\ref{fig:StringOpenClosed} and Fig.~\ref{fig:StringProbe}).
\begin{figure}
	\centering
	\begin{tikzpicture}
		\draw (-5,2) .. controls (-4,1.5) and (-3,0) .. (-3,-1);
		\draw (5,2) .. controls (4,1.5) and (3,0) .. (3,-1);
		\fill [green!10!white] (0,-1) ellipse (3 and .5);
		\draw [thick, black!30!green] (0,-1) ellipse (3 and .5);
		\draw[red] (-1,-1.46) .. controls (-1,-.8) and (-1,2.25) .. (0,2.25);
		\draw[red,dashed] (1,-.485) .. controls (1,-.8) and (1,2.25) .. (0,2.25);
		\draw (-4,3) .. controls (-3,2) and (3,2) .. (4,3);
		\draw [thick,dotted] (0,.8) ellipse (3.67 and .5);
		\draw [ultra thick] (-3.67,.8) .. controls (-3.5,.14) and (3.5,.14) .. (3.67,.8);
		\filldraw[blue!10!white] (-5,2) .. controls (-5.3,2.3) and (-4.3,3.3) .. (-4,3);
		\filldraw[blue!10!white] (-5,2) .. controls (-4.7,1.7) and (-3.7,2.7) .. (-4,3);
		\draw[thick,blue] (-5,2) .. controls (-5.3,2.3) and (-4.3,3.3) .. (-4,3);
		\draw[thick,blue] (-5,2) .. controls (-4.7,1.7) and (-3.7,2.7) .. (-4,3);
		\filldraw[blue!10!white] (5,2) .. controls (5.3,2.3) and (4.3,3.3) .. (4,3);
		\filldraw[blue!10!white] (5,2) .. controls (4.7,1.7) and (3.7,2.7) .. (4,3);
		\draw[thick,blue] (5,2) .. controls (5.3,2.3) and (4.3,3.3) .. (4,3);
		\draw[thick,blue] (5,2) .. controls (4.7,1.7) and (3.7,2.7) .. (4,3);
	\end{tikzpicture}
	\caption{\label{fig:StringOpenClosed} Scattering of a massless string off a D$p$-brane. The blue circles represent punctures associated to the incoming and outgoing closed string and the green disc rests on the D$p$-brane. The thin red line represents a heavy open string produced in the $s$-channel (Fig.~\ref{fig:StringReg2} below), while the thick black  line represents a closed string exchanged in the $t$-channel (Fig.~\ref{fig:StringReg1} below).}
\end{figure}
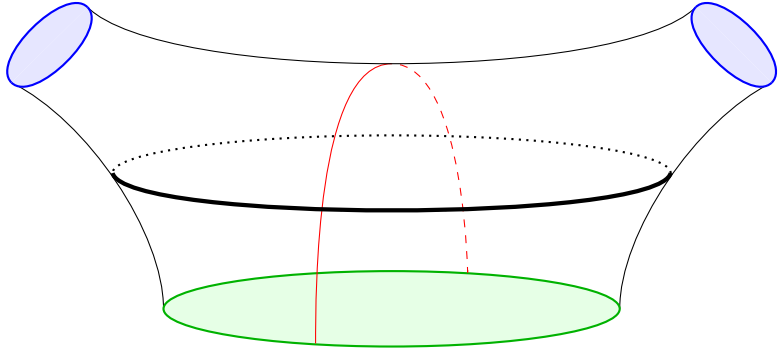
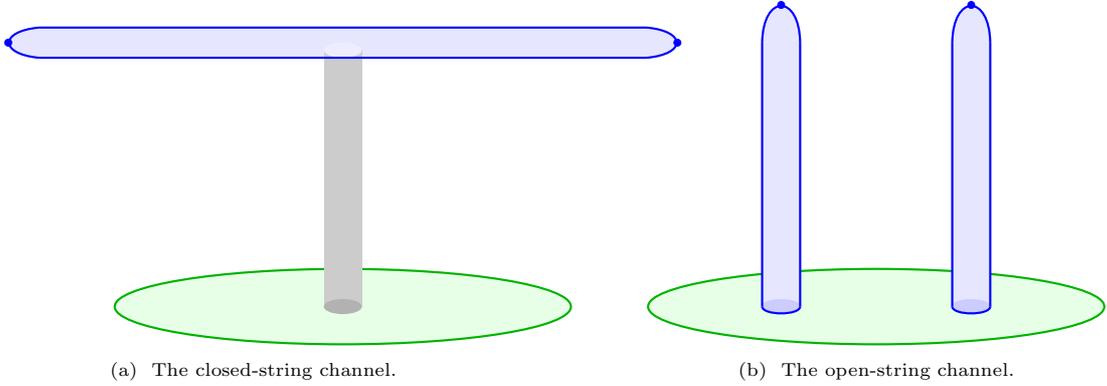
\begin{figure}
	\centering
	\begin{subfigure}{.45\textwidth}
		\centering
		\begin{tikzpicture}
		\fill [green!10!white] (0,-1) ellipse (3 and .5);
		\draw [thick, black!30!green] (0,-1) ellipse (3 and .5);
		\fill [blue!10!white] (-4,2.3) rectangle (4,2.7);
		\fill [blue!10!white] (-4,2.5) ellipse (.4 and .2);
		\fill [blue!10!white] (4,2.5) ellipse (.4 and .2);
		\fill [black!20!white] (-.25,-1) rectangle (.25,2.4);
		\fill [black!20!white] (-.25,2.29) rectangle (.25,2.4);
		\fill [blue!7!white] (0,2.4) ellipse (.25 and .1);
		\fill [black!30!white] (0,-1) ellipse (.25 and .1);	
		\draw[thick, blue] (-4,2.7)--(4,2.7);
		\draw[thick, blue] (-4,2.3)--(4,2.3);
		\draw[thick,blue] (-4,2.3) .. controls (-4.52,2.35) and (-4.52,2.65) .. (-4,2.7);
		\draw[thick,blue] (4,2.3) .. controls (4.52,2.35) and (4.52,2.65) .. (4,2.7);
		\filldraw[blue] (-4.4,2.5) circle (1.3pt); 	
		\filldraw[blue] (4.4,2.5) circle (1.3pt); 	
		\end{tikzpicture}
	\caption{\label{fig:StringReg1} The closed-string channel.}
	\end{subfigure}
\hfill
\begin{subfigure}{.45\textwidth}
	\centering
	\begin{tikzpicture}
		\fill [green!10!white] (0,-1) ellipse (3 and .5);
		\draw [thick, black!30!green] (0,-1) ellipse (3 and .5);
		\fill [blue!10!white] (-1.5,-1) rectangle (-1,2.5);
		\fill [blue!10!white] (1.5,-1) rectangle (1,2.5);
		\fill [blue!10!white] (-1.25,2.5) ellipse (.25 and .5);
		\fill [blue!10!white] (1.25,2.5) ellipse (.25 and .5);
		\fill [blue!20!white] (-1.25,-1) ellipse (.265 and .1);
		\fill [blue!20!white] (1.25,-1) ellipse (.265 and .1);
		\draw[thick, blue] (-1.5,-1)--(-1.5,2.5);
		\draw[thick, blue] (1.5,-1)--(1.5,2.5);
		\draw[thick, blue] (-1,-1)--(-1,2.5);
		\draw[thick, blue] (1,-1)--(1,2.5);
		\draw[thick,blue] (-1.5,2.5) .. controls (-1.49,3.15) and (-1.01,3.15) .. (-1,2.5);
		\draw[thick,blue] (1.5,2.5) .. controls (1.49,3.15) and (1.01,3.15) .. (1,2.5);
		\draw[thick,blue] (-1.5,-1) .. controls (-1.49,-1.12) and (-1.01,-1.12) .. (-1,-1);
		\draw[thick,blue] (1.5,-1) .. controls (1.49,-1.12) and (1.01,-1.12) .. (1,-1);
		\filldraw[blue] (-1.25,3) circle (1.3pt); 	
		\filldraw[blue] (1.25,3) circle (1.3pt); 	
	\end{tikzpicture}
	\caption{\label{fig:StringReg2} The open-string channel.}
\end{subfigure}
	\caption{\label{fig:StringProbe} Two degeneration limits of Fig.~\ref{fig:StringOpenClosed}.
    The blue dots represent punctures associated to the incoming and outgoing string states, while the boundary of the green disc lies on the D$p$-branes. The picture on the left depicts the corner of the worldsheet moduli space relevant in the large distance regime where the closed strings in gray are those of the leading Regge trajectory, see Section~\ref{ssec:reggeon}. The picture on the right depicts the region capturing the closed/open string transition, see Section~\ref{ssec:opcltrans}.}
\end{figure}
        The boundary of the disk lies on one of the D$p$-branes and the punctures describe the ingoing/outgoing closed string that, for sake of simplicity, we take to be a massless NS-NS state. We focus on the case $p\leq 6$ so that the gravitational backreaction of the D$p$-branes decays far away (see Section~\ref{ssec:PLstract} for a discussion of this viewpoint). The leading tree-level contribution to this $1\to 1$ scattering takes the usual form of a ratio of $\Gamma$-functions~\cite{Klebanov:1995ni,Garousi:1996ad,Hashimoto:1996bf}
	\begin{equation}
		\label{eq:sstree}
		{\cal A}_0(p_i,\epsilon_i) = -\frac{\kappa_{d} N T_p}{2}
		{\cal K}(p_i, \epsilon_i) \,
		\frac{\Gamma \left(- \alpha' E_s^2 \right) \Gamma \left( -
			\frac{\alpha'}{4}t \right)}{\Gamma \left(1- \alpha' E_s^2-\frac{\alpha'}{4}t \right)} \equiv \frac{{\cal K}(p_i, \epsilon_i)}{(\alpha' E_s^2)^2} \,{\cal A}_0~,
	\end{equation}
	where our string conventions on the gravitational coupling $\kappa_d$ and the D-brane tensions $T_p$ are summarized in~\ref{app:stringnormaliszations}. The closed string momentum is not conserved in the transverse directions $\mu=p+1,\ldots,d-1$ and $p_1+p_2=q$ is the momentum transferred to the D$p$-branes with $t=-q^2$. Momentum is conserved along the D$p$-branes, in the directions $\mu=0,1,\ldots,p$, so we introduce $E^2_s= - (p_i)^2_\parallel$ and we will be interested in the high-energy eikonal regime corresponding to the limit $E_s^2\gg |t|$. The result~\eqref{eq:sstree} is closely related to the tree-level amplitude for four open strings, as is the case for general mixed open/closed string amplitudes, see~\cite{Stieberger:2009hq} for a detailed discussion at tree-level. Then the kinematic factor ${\cal K}$ is the same one discussed for supergravity in Section~\ref{sec:n8tree}
	and the closed string polarizations $(\epsilon_i)_{\mu\nu}$ should be thought as pairs of open string polarizations $(\varepsilon_i)_\mu$, as done for scalar-graviton scattering in Section~\ref{sec:gravsctree}. The two $\Gamma$-functions in the numerator of \eqref{eq:sstree} encode different sets of poles: the Gamma that depends on $t$ contains the poles that lie on the closed string Regge trajectory $\alpha_\text{closed} (t) = 2+ \frac{\alpha't}{2}$, while the other contain the $s$-channel poles on the open string Regge trajectory $\alpha_\text{open}(E_s^2) = 1 + \alpha' E_s^2$. We prefer to indicate the total energy in the center of mass for the string brane scattering as $E_s$ rather than $\sqrt{s}$ as perhaps would be more natural by using the relations with the 4-point open string amplitude mentioned above. The reason is that one can see this process as the probe limit of a $2\to 2$ scattering where the D-brane stack is the heavy state and we reserve $s$ for the total center-of-mass energy of this scattering, including the rest energy of the D-branes.
	
	At leading order in the limit $E_s^2\gg |t|$ we have 
	\begin{equation}
		{\cal K}(p_1, \epsilon_1; p_2,\epsilon_2)  = (\alpha' E_s^2)^2 \epsilon_{1\,{\mu\nu}} \, \epsilon_2^{\mu\nu}+\ldots\;, \label{eq:Klim}
	\end{equation}
	where $\epsilon_{1,2\mu\nu}$ are the polarizations of bosons in the NS-NS,
	and then, by using
	\begin{equation}
		\label{eq:gstirl}
		\frac{\Gamma(a+b)}{\Gamma(a+c)} \sim a^{b-c}\,,\qquad \mbox{as}\quad a\gg b,c\;,
	\end{equation}
	we get
	\begin{equation}
		\label{eq:disregss}
		{\cal A}_0  \simeq \frac{\kappa_{d} N T_p}{2}
		\Gamma \left(-\frac{\alpha'}{4}t \right) e^{-i \pi \frac{\alpha' t}{4}} (\alpha' E_s^2)^{1+\frac{\alpha' t}{4}}\;.
	\end{equation}
	The overall normalization 
	is directly related to the classical length scale $R_p$ of the geometry describing the gravitational backreaction of the D$p$-branes, see Eq.~\eqref{eq:RpTp}. Notice that we wrote explicitly the phase factor of $(-\alpha'E_s^2)^{\frac{\alpha' t}{4}}$ arising from the Stirling approximation~\eqref{eq:gstirl}: a simple way to make the appropriate choice of branch cut in the factor of $e^{-i \pi \frac{\alpha' t}{4}}$ is to make sure that the imaginary part of the amplitude is positive (recalling that $t < 0$). As usual, this follows from the relation between this imaginary part and the production rate of new degrees of freedom, in this case open strings attached to the  D$p$-branes. We will discuss this interpretation of the phase in Section~\ref{ssec:opcltrans} showing that there is a quantitative link between the imaginary part of~\eqref{eq:disregss} and the amplitudes describing the transition from the external closed string and a particular class of open string states.
      
	When the momentum transferred is small in comparison to the string scale $\alpha' |t|\ll 1$, the string amplitude is dominated by the exchange of massless states and, since we already took the high energy limit, we are just isolating the graviton propagating between the D$p$-branes and the external closed strings
	\begin{equation}
		\label{eq:SeikLft}
		{\cal A}_0 \simeq \frac{4 \kappa_{d} N T_p}{2} \left(-\frac{E_s^2}{t}\right)\;.
	\end{equation}
	It is possible to make contact with the probe limit discussed at the end of Section~\ref{sec:grtree} by identifying the D$p$-branes with the heavy particle. We can wrap the extended space-like direction of the D$p$-branes on a $p$-dimensional torus of volume $V_c$ and then they would describe a target of mass $M_h$ in the noncompact $D=(d-p)$ dimensions (see also the comments around~\eqref{eq:kappaD})
		\begin{equation}
			\label{eq:T2Mh}
			M_h= N \tau_p V_c = \frac{N T_p V_c}{\kappa_{d}}\,,\qquad \frac{1}{\kappa^2_D} = \frac{V_c}{\kappa_d^2}\;.   
		\end{equation}
	The string calculation already includes the factor $1/(2M_h)$  which is part of the Fourier transform~\eqref{eq:Atdefg} used when both particles in the scattering are dynamical, as in the probe limit we have
		$4E p \to (2 M_h) \, (2 E_s)\,,$
	where we used~\eqref{eq:problim2} with $E_1\to E_s$ and $m_1\to 0$ since we are considering a massless probe. So we have to compare \eqref{eq:SeikLft} with~\eqref{eq:214pl} divided by $2 M_h$ and  for $m=0$. By using~\eqref{eq:T2Mh}, we find full agreement. This approximation is reliable when the impact parameter $b$ of the process is much larger that the string length $\ell_s=\sqrt{\alpha'\hbar}$, as it follows from the usual relation $|q|\sim \hbar/b$ connecting the impact parameter $b$ and the momentum $|q|\sim \sqrt{-t}$ of a single graviton exchanged between the string probe and the D$p$-branes.
	
	For general values of $\alpha'|t|$ there are two novelties of the string result with respect to the field theory one: the first is the appearance of a phase and the second is the fact that the simple $t$-dependence related to the field theory propagator for the graviton is substituted by a $\Gamma$-function. Both are string effects: the first related to the existence of an open string sector supported by the D$p$-branes and the second related to the existence of higher-spin massive closed string states. Of course these features are a consequence of the extended nature of the elementary objects used in this case and appear generically in all string theories\footnote{Details change depending on the particular setups: for instance in the case of the string graviton-graviton scattering of Section~\ref{ssec:string-brane-sup} the degrees of freedom responsible for the imaginary part of the tree-level amplitude are closed strings in the $s$-channel.}. Then it is natural to expect that the string eikonal is qualitatively different from the field theory one when the impact parameter is of the order of the string scale, however it turns out that there are deviations at distances that are {\em parametrically} larger. In order to provide a first example of such phenomenon let us introduce 
	\begin{equation}
		\label{eq:DbraneFT}
		2 \delta_0 = {\tilde{\cal A}}_0(E_s,b) = \!\int\! \frac{d^{d-p-2} q}{(2\pi)^{d-p-2}} \frac{{\cal A}_0(E_s,{q}^2)}{2E_s} \, {\rm e}^{i b q}\;,
	\end{equation}
	where, as mentioned above, we need to include just the factor of $1/(2E_s)$ because the factor of $(2 M_s)$ is already included in the amplitude.
	Following the treatment in \cite{Amati:1987uf}, we make the massless pole of the string amplitude explicit
	\begin{equation}
		\label{calAsuper}
		{\cal{A}}_0 =  \frac{\kappa_d N T_p}{2} \frac{4 E_s^2}{- t} \, \Gamma \left(1 - \frac{\alpha' t}{4}\right) {\rm e}^{  \frac{t}{4} \bY},
	\end{equation}
	where we introduced
	\begin{equation}
		\label{eq:lsE}
		\bY = l^2_s(E_s)- i \pi\alpha'\;,\quad \mbox{ with }\quad l_s(E_s) = \sqrt{\alpha' \ln(\alpha' E^2_s)}\;.
	\end{equation}
	The appearance of the effective string length $l_s(E_s)$, with the extra factor of $\ln(\sqrt{\alpha'} E_s)$, will play an important role and signals that elementary energetic strings are larger than expected from naive dimensional analysis. When~\eqref{calAsuper} is used in~\eqref{eq:DbraneFT}, we can rewrite the dependence on $q$ of all factors that are analytic as $q\to 0$ in terms of derivatives with respect to the impact parameter. For instance we have
	\begin{equation}
		\label{eq:ttopartialb}
		t = - {q}^2 = \frac{\partial^2}{\partial {b}^i \partial {b}^i} \equiv \nabla^2 \; \ .
	\end{equation}
	Then we find the following integral representation of the leading string eikonal
	\begin{align}
		2\delta_0 & = \Gamma\! \left( 1 + \frac{\alpha'}{4} \nabla^2 \right ) \int  \frac{d^{d-p-2} {q}}{(2\pi)^{d-p-2}} \frac{ \kappa_d N T_p E_s}{q^2} \, \exp \left(- \frac{\bY}{4} {q}^2  + i {q} {b}   \right) 
		\nonumber 
		\\
		& = \kappa_d N T_p E_s \,\Gamma\! \left ( 1 + \frac{\alpha'}{4} \nabla^2 \right )\int_0^{\infty} dT \int  \frac{d^{d-p-2} {q}}{(2\pi)^{d-p-2}}  {\rm e}^{-q^2 \left(T + \frac{\bY}{4}  \right) + i {q} {b} } \nonumber \\
		& = \kappa_d N T_p E_s \,\Gamma\! \left ( 1 + \frac{\alpha'}{4} \nabla^2 \right ) \left( \frac{b^2}{4} \right)^{-\frac{d-p-4}{2}}\int_{0}^{ \frac{b^2}{\bY}} \frac{d \hat{T}}{(4\pi)^{\frac{d-p-2}{2}}} \,\, {\hat T}^{\frac{d-6-p}{2}} {\rm e}^{-\hat{T}}\; \ ,
		\label{impactb}
	\end{align}
	where in the last step we introduced $\hat{T}=b^2/[4(T+\frac{\bY}{4})]$. For real values of $Y$, the integral yields the incomplete $\Gamma$-function
	$\gamma(z,a)$ 
	with
	\begin{equation}
		\label{eq:incomgamma}
		\gamma(z,a) = \int_0^a dt\, t^{z-1} e^{-t} = \sum_{k=0}^\infty \frac{a^{z+k} e^{-a}}{z (z+1) \ldots (z+k)}\;,
	\end{equation}
	so by analytical continuation we can write
	\begin{equation}
		\label{eq:incg2}
		2\delta_0 = \frac{\kappa_d N T_p E_s}{4\pi}\, \Gamma\!\left(1+\frac{\alpha' \nabla^2}{4}\right)  \left[(\pi b^2)^{-\frac{d-4-p}{2}} \gamma\left(\frac{d-4-p}{2},\frac{b^2}{\bY}\right)\right],
	\end{equation}
	also when $Y$ is given by~\eqref{eq:lsE}. As usual, we expect that this result exponentiates after including the leading contributions of worldsheet diagrams with $h+1$ boundaries on the D$p$-branes. This is based on the same counting used in the QFT case: by using~\eqref{eq:RpTp} we see that the first factor in~\eqref{eq:incg2} is proportional to the classical scale $R_p^{d-p-3}$, so by dimensional analysis $\delta_0$ scales as $(R_p/b)^{d-p-3} E_s b/\hbar$ and is a large quantity in the classical limit. In Section~\ref{ssec:1lstrng} we will provide an explicit check of this exponentiation at the string level, focusing just on a particular regime since an exhaustive analysis is not yet available in the literature. Here we assume the eikonal exponentiation and discuss the behavior of~\eqref{eq:incg2} at large and small distances with respect to the effective string scale introduced in~\eqref{eq:lsE}. In the large distance regime $b \gg l_s(E_s)$ we can use
	\begin{equation}
		\label{eq:gamlab}
		\gamma\left(z,\frac{b^2}{\bY}\right) \sim \Gamma(z) - e^{-\frac{b^2}{\bY}} \left(\frac{b^2}{\bY}\right)^{z-1}\;.
	\end{equation}
	To leading order, the above expression does not depend on the impact parameter and yields a real contribution to $\delta_0$ in Eq.~\eqref{eq:incg2}. In the large distance limit we can neglect the overall $\Gamma$-function in~\eqref{eq:incg2} as it yields corrections suppressed by $\alpha'/b^2$,
        \begin{equation}
          \label{eq:bbigls}
          2\delta_0 \sim \frac{\sqrt{\pi} E_s}{2} \frac{\Gamma \left(\frac{d-p-4}{2}\right)}{\Gamma\left( \frac{d-p-3}{2}\right)} \frac{R_p^{d-p-3}}{b^{d-p-4}} + \frac{i\pi E_s}{2\Gamma\left( \frac{d-p-3}{2}\right)}\sqrt{\frac{\pi\alpha'}{\ln(\alpha' E_s^2)}} \left(\frac{R_p}{l_s(E_s)}\right)^{d-p-3} e^{-\frac{b^2}{l_s^2(E_s)}}\,,
        \end{equation}
        where we wrote the leading real and imaginary terms and used~\eqref{eq:RpTp}. This is obtained by expanding the $\gamma$-function around the first term of the following equation
\begin{equation}
	\label{eq:bYexp}
	\frac{b^2}{\bY} \simeq \frac{b^2}{l_s^2(E_s)} + i\frac{\pi}{\ln(\alpha' E_s^2)} \frac{b^2}{l_s^2(E_s)}\;,
\end{equation}
 since the second one is suppressed by $1/\ln(\alpha'E_s^2)$ at high energy. Then the expansion of the $\gamma$-function is obtained from the integral expression in~\eqref{eq:incomgamma}
\begin{equation}
	\label{eq:ingamexp}
	\gamma\left(\frac{d-4-p}{2},\frac{b^2}{\bY}\right)\simeq \gamma\left(\frac{d-4-p}{2},\frac{b^2}{l_s^2(E_s)}\right)+i\pi \frac{e^{-\frac{b^2}{l_s^2(E_s)}} \left(\frac{b}{l_s(E_s)}\right)^{d-4-p}}{\ln(\alpha'E_s^2)}\;.
\end{equation}
The real part of~\eqref{eq:bbigls} can also be written as $2 G_D M_h E_s \Gamma\left(\frac{D-4}{2}\right) (\sqrt{\pi} b)^{-(D-4)}$ with the identifications $\kappa^2_D = 8 \pi G_D$, $D=d-p$ and~\eqref{eq:T2Mh}. This reproduces the QFT result in Eq.~\eqref{eq:217pl} for $m_1=0$, as expected since in this regime the field theory setup of Section~\ref{sec:grtree} is a good effective description. The exponentially suppressed contributions are relevant for the imaginary part and can be derived directly from the amplitudes~\eqref{eq:disregss} as done in~\eqref{eq:imdelta0Strbis}. Taking $-\frac{1}{E_s}\frac{\partial}{\partial b}$ of the real part of \eqref{eq:gamlab}, one finds the deflection angle \eqref{eq:3.10} obtained in~\ref{app:geoDp} by solving the geodesic equation. 
	
	In the opposite regime, where the impact parameter is small with respect to the effective string length $b < l_s(E_s)$, we can keep just the first term in the expansion~\eqref{eq:incomgamma} obtaining
	\begin{equation}
          2 \delta_0  \sim  \frac{\kappa_d N T_p E_s}{4\pi}  \, 
          \frac{(\pi\bY)^{-\frac{d-4-p}{2}} }{\frac{d-4-p}{2}}
		\left( 1+ O \left (  \frac{b^2}{\bY},\frac{1}{\ln(\alpha' E_s^2)} \right)\right) \ . 
		\label{lowb}
	\end{equation}
	It is interesting to consider the high-energy stringy regime where $b^2 \ll |\bY|$, which is possible because of the $\ln(\alpha' E_s^2)$ enhancement in~\eqref{eq:lsE}. Then, the $\Gamma$-function  in~\eqref{lowb} can be approximated to $1$ (since the shift proportional to $\alpha' \nabla^2$ yields corrections suppressed by $\alpha'/\bY$) and the leading contribution to $2\delta_0$ takes a simple $b$-independent form. Notice, however, that this result is not real: by using~\eqref{eq:lsE} and expanding the factor $Y^{}$ in the regime $\sqrt{\alpha'}\sim b< l_s(E_s)$, one can see that the leading imaginary contribution takes the same form as in~\eqref{eq:bbigls} without the exponential factor which is negligible in this regime
	\begin{equation}
		\label{eq:leadimSR}
		\operatorname{Im}(2\delta_0) \simeq \frac{\pi}{2 \Gamma\left(\frac{d-p-3}{2}\right)} \left(\frac{R_p}{l_s(E_s)}\right)^{d-p-3} \frac{\sqrt{\pi\alpha'} E_s}{\sqrt{\ln(\alpha' E_s^2)}}\;.
	\end{equation}
	Because of the presence of this imaginary part the elastic partial wave unitarity discussed in Section~\ref{ssec:pwunitarity} is violated. This is not surprising since, when the  probe gets close to the D$p$-branes, it can excite the open strings degrees of freedom, {\rm i.e.} the closed string can ``touch'' the target and make a transition to an open string whose endpoints are anchored to one of the D$p$-branes. The less obvious aspect of this phenomenon is that it starts when $b\sim l_s(E_s)$ rather than simply when the impact parameter is of the order of the string length. In order to restore unitarity, one would need to include explicitly the open string degrees of freedom and promote $2\delta_0$ to an operator that can describe a closed/open string transition as discussed in Section~\ref{ssec:opcltrans}. 
	
	\subsubsection{The closed/open string transition}
	\label{ssec:opcltrans}
	
	Let us discuss in some more detail the imaginary part of~\eqref{eq:disregss}
	\begin{equation}
		\label{eq:imA}
		{\rm Im}({\cal A}_0) \simeq \frac{\kappa_{d} N T_p}{2} (\alpha' E_p^2)^{1+\frac{\alpha' t}{4}} \sin\left(-{\pi \frac{\alpha' t}{4}}\right) \Gamma\left(-\frac{\alpha' t}{4}\right)
		=\pi \frac{\kappa_{d} N T_p}{2} \frac{ (\alpha' E_p^2)^{1+\frac{\alpha' t}{4}}}{\Gamma\left(1+\frac{\alpha' t}{4}\right)} \;,
	\end{equation}
	where in the final step we used
\begin{equation}
	\label{eq:gar1}
	\Gamma(z) \Gamma(1-z) = \frac{\pi}{\sin(\pi z)}\;.
\end{equation}        
	A similar formula holds also for the imaginary part of the bosonic string amplitude~\eqref{eq:TTeikL}
	\begin{equation}
		\label{eq:imATTbis}
		{\rm Im}({\cal A}_0) \simeq \pi \frac{\kappa_{d} N T_p}{2} \frac{ (\alpha' E_p^2)^{1+\frac{\alpha' t}{4}}}{\Gamma\left(\xi+\frac{\alpha' t}{4}\right)} \;,
	\end{equation}
	with $\xi=2$, instead of $\xi=1$ as in~\eqref{eq:imA}.
	
	This imaginary part is clearly a string effect since it vanishes in the $\alpha'\to 0$ limit and we will now show that it is due to the propagation of new degrees of freedom in the $s$-channel which in our setup are open strings representing excitations of the target D$p$-branes. Let us illustrate this by starting from the full amplitude~\eqref{eq:sstree} (or~\eqref{eq:TTtree} in the bosonic case). For simplicity in the superstring case~\eqref{eq:sstree} we choose the external states to be a Kalb--Ramond field with polarizations along the spacelike directions of the D-brane worldvolume: then the factor ${\mathcal K}(p_i,\epsilon_i)$ takes the form given in the square parenthesis of~\eqref{eq:sstreeschpolbis}~\cite{Garousi:1998bj}. However, as expected, only the leading term in $E_s$ will be relevant for the final result, so one can also focus directly in ${\cal A}_0$ in~\eqref{eq:sstree}. By using the relation
\begin{equation}
		\label{eq:gammarat}
		\frac{\Gamma(x)}{\Gamma(x+a)} = \sum_{n=0}^\infty \frac{(-1)^n}{n! \Gamma(a-n)} \frac{1}{x+n} \;,
	\end{equation}
it is possible to rewrite the string amplitudes mentioned above as a series of the $s$-channel poles
\begin{subequations}    \label{eq:treeschpolbis}
\begin{align}		\label{eq:TTtreeschpolbis}
	{\cal A}^{T}_0  & = \frac{\kappa_{d} N T_p}{2} \sum_{n=0}^\infty \frac{1}{n!} \frac{\Gamma\left(2+\frac{\alpha' t}{4} + n\right)}{\Gamma\left(2+\frac{\alpha' t}{4}\right)} \frac{1}{-\alpha' E_s^2 + n - 1}\;,
	\\ \label{eq:sstreeschpolbis} {\cal A}_0^{B} & = \frac{\kappa_{d} N T_p}{2}  \left[(\alpha' E^2_s)^2 + (\alpha' E_s^2) \left(\frac{\alpha' t}{4}\right)\right] \sum_{n=0}^\infty \frac{1}{n!} \frac{\Gamma\left(\frac{\alpha' t}{4} + n\right)}{\Gamma\left(1+\frac{\alpha' t}{4}\right)} \frac{1}{-\alpha' E_s^2 + n}\;,
\end{align}
\end{subequations}
where we used again~\eqref{eq:gar1}. The amplitude in the first line refers to bosonic string theory where the probe particle is a tachyon, while the second line refers to the superstring case with the Kalb--Ramond state mentioned above. 
The square parenthesis in superstring case follows from the prefactor in~\eqref{eq:sstree} as one can see with a calculation similar to the one discussed {for supergravity} in Section~\ref{sec:n8tree}. The poles in $\alpha' E_s^2$ correspond to open strings of mass $M_n^2 =n/\alpha'$ (or $M_n^2 =(n-1)/\alpha'$ in the bosonic case) propagating on the disk and the residue is a polynomial in $t$ of degree $n$ which is the maximum spin for the states of mass $M_n$. While the result in~\eqref{eq:treeschpolbis} seems to be real, there is actually an imaginary part which is hidden in the $i \epsilon$ prescription which is not written explicitly and should be reinstated. This can be done as follows
	\begin{equation}
		\label{eq:iereins}
		\frac{1}{-E_s^2 + \frac{n - (\xi-1)}{\alpha'}} \to \frac{1}{-E_s^2 + M_n^2 - i \epsilon} = {\rm PV}\left[\frac{1}{-E_s^2 + M_n^2}\right] + i \pi \delta(-E_s^2 + M_n^2)\;,
	\end{equation}
	where the first term in the last step indicates the Cauchy Principal Value and the second one is the imaginary contribution due to $i\epsilon$ Feynman's prescription. Notice that this fixes the sign of the imaginary part of~\eqref{eq:iereins} providing a first principle justification for the prescription discussed after~\eqref{eq:disregss}. Then we can write the imaginary part of the amplitude in a unified way for the bosonic and the superstring case 
	\begin{equation}
		\label{eq:imATTsumpol}
		{2}\, {\rm Im}({\cal A}_0) \simeq \pi {\kappa_{d} N T_p} \sum_{n=0}^\infty \frac{1}{\Gamma(n+\xi-1)} \frac{\Gamma\left(\xi+\frac{\alpha' t}{4} + n\right)}{\Gamma\left(\xi+\frac{\alpha' t}{4}\right)} \delta(-\alpha' E_s^2 + n-(\xi-1))\;,
	\end{equation}
	where we used the delta function to simplify the square parenthesis in the second equation of~\eqref{eq:sstreeschpolbis}. Because of perturbative unitarity, each term in the sum represents the modulus square of a 2-point amplitude describing a closed string transforming into a linear combination of open strings at level $n$. Since we are interested in the limit $\alpha' E_s^2 \gg 1$, we can approximate the sum as an integral over the continuous variable $x = \frac{n}{\alpha' E_s^2}$
	\begin{equation}
		\label{eq:sprl}
		\sum_{n=0}^\infty \delta(-\alpha'E_s^2+ n -(\xi-1)) \to \int_0^\infty\!\! dx\, \delta(-1+x)
	\end{equation}
	and rewrite the ratio of the $n$-dependent $\Gamma$-functions by using the limit~\eqref{eq:gstirl} obtaining 
	\begin{equation}
		\label{eq:imATTsumpol2}
		{\rm Im}({\cal A}_0) \simeq \pi \frac{\kappa_{d} N T_p}{2} \frac{\left(\alpha' E_s^2\right)^{1+\frac{\alpha' t}{4}}}{\Gamma\left(\xi+\frac{\alpha' t}{4}\right)} \;,
	\end{equation}
	in agreement with~\eqref{eq:imATTbis}. 
	
	Exactly as in the field theory case, one can extract the long-range description of this string scattering amplitude by taking its Fourier transform to impact parameter space~\eqref{eq:DbraneFT}. Starting from the imaginary part, we can extract the leading large distance behavior by approximating $\Gamma\left(\xi+\frac{\alpha' t}{4}\right)\simeq 1$ and then, since $t=-q^2$, the Fourier transform translates into a Gaussian integral
	\begin{equation}
		\label{eq:FTtacStrbis}
		{\rm Im}(2\delta_0) = \!\int\! \frac{d^{d-2-p} q}{(2\pi)^{d-2-p}} \frac{{\rm Im}\,{{\cal A}_0}(E_s,{q}^2)}{2E_s} \, {\rm e}^{i b q}\simeq \pi \frac{\kappa_{d} N T_p}{4} \frac{\alpha' E_s \, e^{-\frac{b^2}{\alpha' \ln(\alpha' E_s^2)}}}{(\pi\alpha' \ln(\alpha' E_s^2))^{\frac{d-2-p}{2}}}\;.
	\end{equation}
	Then the standard eikonal $e^{2i \delta_0}$ is not a phase as it includes a real factor $e^{-{\rm Im}2 \delta_0}$ with
	\begin{equation}
		\label{eq:imdelta0Strbis}
		{\rm Im}(2 \delta_0 ) \simeq \frac{\pi}{2 \Gamma\left(\frac{d-p-3}{2}\right)}  \frac{\sqrt{\pi\alpha'} E_s }{\sqrt{\ln(\alpha' E_s^2)}}\left(\frac{R_p}{l_s(E_s)} \right)^{{d-3-p}} e^{-\frac{b^2}{l^2_s(E_s)}}\;,
	\end{equation}
	where we used~\eqref{eq:RpTp}. This result is consistent with the one obtained in~\eqref{eq:leadimSR} from~\eqref{lowb}. 
	
	It is interesting to notice that the $\xi$ dependence drops out  in~\eqref{eq:imdelta0Strbis}, so the stringy imaginary part of the tree-level eikonal takes the same form both in the superstring and in the bosonic case. Clearly supersymmetry does not play a crucial role in the closed/open string transition at high energies. Thus for the rest of this subsection we can focus on the slightly easier setup of bosonic string theory  without losing any physically interesting qualitative feature. Our aim is to provide a characterization of the open strings produced when the probe is captured by the D$p$-branes. Since the incident closed string probe has zero space-like momentum along the D$p$-brane worldvolume, the open strings responsible for the imaginary part~\eqref{eq:imdelta0Strbis} will be ``static'' and have just a non-zero energy component $\alpha' E_s^2 = n_o-1$, where $n_o$ indicates the level of the open string produced. So one should be able to retrieve~\eqref{eq:imdelta0Strbis} from the ``square'' of the operator $\hat{V}$ describing the tree level transition between a generic closed string $V_c$ and a generic open string $V_o$.

To be more specific,  by following~\cite{DAppollonio:2015oag}, we introduce the open-closed string vertex $\hat{V}$ that, when saturated with a particular open string state $\langle V_o|$ on the left and a particular closed string state $|V_c \bar{V}_c\rangle$ on the right,  provides their coupling
\begin{equation}
		\label{eq:clopgv}
		\langle V_0|\hat{V}|(V_c \bar{V}_c) (p)\rangle = \sqrt{\frac{\kappa_d N T_p}{2\alpha'}}  \langle  V_c(p)  \bar{V}_c(p) V_o \rangle,
\end{equation}
where $p$ is the momentum of the closed state which is described as a ket-vector $|V_c \bar{V}_c\rangle$ on the left hand side to stress that it contains a holomorphic and an antiholomorphic part. In formulae we expect
	\begin{equation}
		\label{eq:Vpsqima}
		{\rm Im}({\cal A}_0) = \pi \alpha' \langle (V_c \bar{V}_c)(p_2) |\hat{V}^{\dagger} \, \hat{V}|(V_c \bar{V}_c)(p_1) \rangle\;.
	\end{equation}
	The explicit form of $\hat{V}$ was first discussed in~\cite{Cremmer:1973ig, Clavelli:1973uk} (in the case of all Neumann boundary conditions) and can also be derived from the overlap of generic Del Giudice, Di Vecchia, Fubini (DDF) states~\cite{DelGiudice:1971fp} by following~\cite{Ademollo:1974kz}. This is the same formalism that we {will} use in Section~\ref{ssec:seikop} to describe the {\em closed} string transitions that are relevant for the tidal excitations. We refer to the literature for the derivation (see~\cite{DAppollonio:2015oag} and references therein) and here it suffices to say that the operator describing the closed/open transition takes the form $\hat{V} \sim \exp[\hat{a}_k N_{kl} \hat{a}_l]$, where the operators $\hat{a}$ indicate the light-cone creation (for the open string) or annihilation (for the closed string) oscillators of level $k,l$ and $N_{kl}$ are explicit coefficients known as Neumann coefficients. The index for the level includes also zero where it represents the energy/momentum of the states. The key idea is to start from the exact expression of the Neumann coefficients and take the Regge limit to obtain simplified expressions in the gauge where the space-like part of the light-cone vector is along the direction of momentum of the closed string (this choice will also be used later in the closed string context, see~\eqref{eq:epmlc}). For instance we can write the transition between an incident closed string tachyon with vanishing transverse momentum and a generic open string state as follows
	\begin{equation}
		\label{eq:tachopg}
		\hat{V}_{p_\perp=0}|0\rangle = \sqrt{\frac{\kappa_d N T_p}{2\alpha'}} \, \delta\left(\alpha' E_s^2+1 - \sum \hat{a}^i_{-n} \hat{a}^i_n\right) \, e^{\frac{1}{2} \sum\limits_{k,l} N^{33}_{k l} \hat{a}^i_{-k} \hat{a}^i_{-l}}|0\rangle \,,
	\end{equation}
where here the $\hat{a}$'s represent the open string oscillators satisfying $[\hat{a}^i_{-n},\hat{a}^j_k]=n \delta_{k+n,0} \delta^{ij}$. The delta function enforces the energy conservation and, as mentioned, means that the level of the open string produced is large $n_o \sim \alpha' E_s^2$. In the high energy regime we have $n_o\gg 1$ and, as shown in~\cite{DAppollonio:2015oag}, the Neumann coefficient
\begin{equation}
		\label{eq:N33ex}
		N^{33}_{k l} = -\frac{\alpha_2}{\alpha_1 (k+l)}\binom{-k\frac{\alpha_1}{\alpha_3}}{k}\binom{-l\frac{\alpha_1}{\alpha_3}}{l}\;,
\end{equation}
where $\alpha_1$ ($\alpha_2$) are the dimensionless left (right) moving light-cone momenta for the incident closed string, while $\alpha_3$ is the momentum of the open string produced. It is convenient to align the light-cone along the direction of motion of the closed string which we take to be a tachyon. Then we have
\begin{align}
	\alpha_1 &  = \sqrt{\alpha'} \left( E_s + p\right) = \sqrt{n_o-1}  + \sqrt{n_o +3 } \sim 2\sqrt{n_o} \left(1+\frac{1}{2n_o}\right) , 
	\nonumber \\ 
	\alpha_2 &   = \sqrt{\alpha'} \left( E_s - p\right) = \sqrt{n_o-1}  - \sqrt{n_o+3  } \sim - \frac{2}{\sqrt{n_o}}  \left(1-\frac{1 }{2n_o}\right) , 
	\label{alpha1alpha2} \\ \nonumber 
	\alpha_3 &  = -2 \sqrt{\alpha'} E_s  =  -2 \sqrt{n_o-1} \sim -2\sqrt{n_o} \left(1-\frac{1}{2n_o}\right) , 
\end{align}
where we used $\alpha' E_s^2=n_o-1$ and $\alpha' p^2 = n_o+3$ for the tachyon and took the large energy limit. Then one can check that the Neumann coefficients~\eqref{eq:N33ex} are suppressed as $1/n_o$ in this limit if $k,\,l$ are kept fixed. Instead if both $k,\,l$ scale with $n_o$ the Neumann coefficients stay finite
\begin{equation}
	\label{eq:N33as}
	N^{33}_{k l}\sim  n^{x+y-2} \, \frac{x^{ x}y^{ y}}{\Gamma(1 + x)\Gamma(1 + y)}
	\frac{1}{x+y} \;,\quad  k = n_o x \;, \quad l = n_o y\;.
\end{equation}
This means that we can describe the final open string state in terms of few oscillators in the light-cone gauge~\eqref{alpha1alpha2}.
It is difficult to prove analytically~\eqref{eq:Vpsqima} even in the case $q=0$, but, in this case, it is interesting to notice that it is easy to obtain a convincing numerical check in the light-cone gauge mentioned above: one can expand the exponential in~\eqref{eq:tachopg} and focus on the contribution that is linear in $N^{33}$ for each vertex. Then in \eqref{eq:Vpsqima} we obtain a contribution with two sums over the open string levels which at high energy can be approximated with two integrals. By recalling that in the bosonic case the indices $i,j$ run from $1$ to $24$, we obtain the following numerical estimate for \eqref{eq:Vpsqima}
	\begin{equation}
		\label{eq:1stcont}
		12 \pi \alpha' \frac{\kappa_d N T_p}{2\alpha'} n_o \, \int_0^1 dx \int_0^1 dy  \frac{x^{2x+1}y^{2y+1}\delta(x+y-1)}{\Gamma^2(1+x)\Gamma^2(1+y)} \sim 0.929 \,\frac{\kappa_d N T_p}{2} \alpha' E_s^2 \;,
	\end{equation}
which is already very close to the full result~\eqref{eq:imATTsumpol2}.  By going to next order and including in~\eqref{eq:tachopg} the quadratic terms in the Neumann coefficients coming from the expansion of the exponential, one can check numerically that this new contribution is subleading and changes the numerical coefficient~\eqref{eq:1stcont} from $0.929$ to $0.998$~\cite{DAppollonio:2015oag}. It is clear that in this gauge the expansion is converging very quickly so one can describe the open string produced in the transition at $t=0$ as a linear combination of states with a pair of $\hat{a}^i_k$ oscillators plus small corrections with the insertion of few other pairs (with the levels summing up to $n_o$ globally).
	
We refer again to~\cite{DAppollonio:2015oag} for the generalization of~\eqref{eq:tachopg} to the case of a non-vanishing transferred momentum $q$. By indicating the resulting closed/open vertex as ${\hat V}_{q_\perp}$, one can introduce its impact parameter version via the usual Fourier transform obtaining
	\begin{equation}
		\label{eq:Vclopftb}
		\hat{V}_b|0\rangle =  \int \frac{d^{d-p-2}q_\perp}{(2\pi)^{d-p-2}}
		\, e^{i b \vec q_\perp} \,  \hat{V}_{q_\perp}|0\rangle \;,
	\end{equation}
	where $\hat{V}_{q_\perp}$ is similar to~\eqref{eq:tachopg}, but contains also linear terms in the exponential with the open string oscillators~\cite{DAppollonio:2015oag}. In the high energy limit, the final result takes the form of a squeezed coherent state 
	\begin{equation}
		\label{eq:tachopg2}
		\hat{V}_b|0\rangle  = \frac{1}{\left(\pi \alpha' \log \alpha' E^2_s \right)^{\frac{d-p-2}{2}}} \hat{V}\, e^{-\frac{1}{\alpha' \log \alpha' E^2_s}\left(b^i + i  \sqrt{\frac{\alpha'}{2}}\sum_k \frac{1}{k} \hat{a}^i_{-k}  \right)^2}  |0\rangle\;,
         \end{equation}
         with the expected exponential factor $e^{-\frac{b^2}{l_s^2(E_s)}}$~\cite{DAppollonio:2015oag} which explicitly shows that the closed/open string transition is suppressed unless the impact parameter is smaller than the effective string length $l_s(E_s)$. It would of course be interesting to generalise this analysis to the case where two open strings are produced by looking at the cuts of the annulus contribution to the closed string scattering. In this case there is a richer phase space for the final states and we expect that the total energy is shared preferably equally among the open strings. However, as far as we know, this analysis has not been carried out explicitly even in the bosonic case.
	
	\subsubsection{The Reggeon vertex formalism}
	\label{ssec:reggeon}

        As usual in tree-level string theories, Eq.~\eqref{eq:sstree} displays an infinite set of equally spaced poles. As discussed in the previous sections, in the high energy regime $E_s\gg |t|$ the poles in the $s$-channel merge to form a cut, while the poles in $t$-channel appear explicitly in the result~\eqref{eq:disregss} and in the eikonal~\eqref{lowb}. In this regime the dominant states exchanged between the D$p$-branes and the scattered string have maximal spin at each mass level, {\rm i.e.} they should lie along the leading Regge trajectory of the closed string spectrum. In~\ref{app:reggeonb} we show this quite explicitly by following the approach of~\cite{Ademollo:1989ag,Ademollo:1990sd} and even if our summary there focuses on the bosonic theory, the generalization to the superstring case is straightforward. Here we follow the approach taken in~\cite{Brower:2006ea} that was applied to the string-brane scattering in~\cite{Black:2011ep,Bianchi:2011se}.
	
	The basic idea is that, at high energy, the dominant region in the integral over the worldsheet moduli (see for instance~\eqref{eq:TTDp}) corresponds to diagrams where two closed string vertex operators are very close: the size of the relevant region scales as  $ {(\alpha 'E}^2_s)^{-1}$, as pointed out also in the discussion after \eqref{eq:TTder} for the bosonic string.  Thus instead of calculating the full string amplitude and then take the high energy limit, it is possible to first take a generalized OPE to define a new ingredient, the ``Reggeon vertex'', that then can be attached to the D$p$-branes (or another Reggeon vertex in the case of the string-string scattering) to obtain directly the high energy limit of the amplitude.  In this discussion we can keep the vertex operators ${\cal V}_i$ general as the only feature we need is the exponential factor $e^{i p_{1,2} X}$ that is always present for states with nonzero momentum. A disk amplitude with two closed strings takes a form similar to~\eqref{eq:TTDp} 
	\begin{equation}
		\label{eq:TTDpbis}
		{\cal A}_0 = C_{S_2} \frac{\alpha' \kappa_d}{8\pi} N \int \frac{d^2z_1 d^2z_2}{dV_{SL(2,R)}} \langle 0| {\cal V}^{(-1)}_1(z_1,\bar{z}_1) {\cal V}^{(0)}_2(z_2,\bar{z}_2) | B \rangle\;,
	\end{equation}
	but with generic vertex operators\footnote{Focusing on NS-NS states we need to take one vertex, say ${\cal V}_1$ in the superghost picture $(-1,-1)$ and the other in the picture $(0,0)$.} ${\cal V}_i$. In \eqref{eq:TTDpbis},  $C_{S_2}$ is fixed as in \eqref{eq:Cs2kap} and the vertex operators have a normalization factor $\frac{\kappa_d}{2\pi}$ as in~\eqref{eq:tachvo}.  By introducing $z=\frac{z_1+z_2}{2}$ and $w=z_1-z_2$ we can write the contribution from the exponential to the OPE as follows
	\begin{equation}
		\label{eq:opre}
		{\cal V}^{(-1)}_1(z_1,\bar{z}_1) {\cal V}^{(0)}_2(z_2,\bar{z}_2) 
			\sim |w|^{\alpha' p_1 p_2} e^{iq X(z,\bar{z})+i\frac{p_1-p_2}{2} (w\partial_z X + \bar{w} \partial_{\bar{z}} X)}  \frac{{\cal O}(z,\bar{z})}{|w|^{2 n_1} |w|^{2 n_2}} + \ldots\;, 
		\end{equation}
	where ${\cal O}(z,\bar{z})$ comes from the contribution of the non-exponential part of the vertices ${\cal V}_i$ and $\alpha' p_i^2= -4n_i$. We are focusing only on the most singular term as $w\to 0$ as the subleading contributions are suppressed at high energies. As mentioned, the motivation for taking the OPE approximation above is that the high energy result is dominated by the region of $|w|^2 \leq 1/(\alpha' E_s^2) \ll 1$. However this means that we need to treat exactly the terms where $w$ is enhanced by factor of $E_s$, as in the combination proportional to $(p_1-p_2)$ in the exponential of~\eqref{eq:opre}. By using the on-shell conditions we can rewrite the first contribution in this exponential as $\alpha' p_1 p_2 = -\frac{\alpha' t}{2}+2n_1 +2n_2 $. In order to obtain a non-trivial result in the high energy limit, the OPE between the polynomial part of the vertices should compensate the factor $|w|^{2n_1 +2n_2}$ coming from contraction of the exponential part leaving a pole and this is why we are focusing on the leading term in~\eqref{eq:opre}. One can check that such contribution can exist by considering states in the leading Regge trajectory ${\cal V}_j \sim {\cal P}^{\{M_i\}}_j \bar{\cal P}^{\{N_i\}}_j  e^{i p_j X}$ that contain a polynomial of degree $n_1$ in the picture $-1$ (or $n_2+1$ in the picture $0$) in $\partial X$ and similarly for the antiholomorphic sector
		\begin{equation}
			\label{eq:lrt}
			\begin{aligned}
			{\cal P}^{\{M_i\}}_1 & = \psi^{(M_0} \sqrt{\frac{2}{\alpha'}} i\partial X^{M_1}\ldots \sqrt{\frac{2}{\alpha'}} i\partial X^{M_{n_1})}\,  e^{-\varphi}\,, \\
			{\cal P}^{\{M_i\}}_2 & = \left(\sqrt{\frac{2}{\alpha'}} i \partial X^{(M_0} \sqrt{\frac{2}{\alpha'}} i \partial X^{M_1} + (p_2 \psi) \psi^{(M_0} i \partial X^{M_1} -n_2 \psi^{(M_0} \partial \psi^{M_1} \right) \\ &\qquad  \sqrt{\frac{2}{\alpha'}} i \partial X^{M_1}\ldots \sqrt{\frac{2}{\alpha'}} i \partial X^{M_{n_2})}\;.
		\end{aligned}
		\end{equation}
		The leading contribution is when all factors of $\partial X$ and  $\bar\partial X$ in both vertices are contracted either among themselves or with the exponential part of the other vertex, yielding a factor of $|w|^{-2n_1-2n_2}$, and finally there is an extra factor of $|w|^{-2}$ from the contraction of the terms with $p_2\psi$ (and $p_2\bar\psi$) in the second term of ${\cal P}^{\{M_i\}}_2$ (and $\bar{\cal P}^{\{M_i\}}_2$). When the leading term in the sum~\eqref{eq:opre} is trivial, then the transition between the states described by ${\cal V}_1$ and ${\cal V}_2$ is suppressed at high energies. Thus we can isolate the dependence on $w$ obtaining
		\begin{equation}
			\label{eq:prereve}
			\int \frac{d^2z_1 d^2z_2}{dV_{SL(2,R)}}  {\cal V}_1(z_1,\bar{z}_1) {\cal V}_2(z_2,\bar{z}_2) \sim  {\cal O}(z,\bar{z})\, e^{iq X}\! \int \! \frac{d^2 w}{2\pi} \,|w|^{\frac{-\alpha' t}{2}-2} e^{i\frac{p_1-p_2}{2} (w\partial_z X + \bar{w} \partial_{\bar{z}} X)}\;,
		\end{equation}
where we can set $z=1 $ and the factor of $2\pi$ in the measure comes from the residual conformal invariance. We can then perform the integral over $w$ and ${\bar{w}}$ by using\footnote{The factor $2$ in front comes from the normalization of $d^2w$ discussed after \eqref{eq:clospr}.}
		\begin{equation}
			\label{eq:intwwb}
			\int d^2u \,(|u|^2)^{-A-\frac{\alpha' t}{4}} e^{B (u +\bar u)} =2 \pi \frac{\Gamma\left(1-A-\frac{\alpha' t}{4}\right)}{\Gamma\left(A+\frac{\alpha' t}{4}\right)} (-B^2)^{A-1+\frac{\alpha' t}{4}}\;,
		\end{equation}
		which can be checked by introducing a Schwinger parameter $\tau$ to rewrite the first factor as an exponential and then by carrying out the Gaussian integration and the integration over $\tau$. 
		
		In order to extract the energy dependence it is convenient to introduce a set of two light-cone vectors $e^\pm$ such that
		\begin{equation}
			\label{eq:epmlc}
			(e^-)^\mu = \frac{1}{\sqrt{2}} \lim_{E_s\to \infty}\frac{p_2^\mu}{E_s} = -\frac{1}{\sqrt{2}} \lim_{E_s\to \infty}\frac{p_1^\mu}{E_s} \;, \qquad (e^+ e^-) = 1\;.
		\end{equation}
		Then in the case at hand, ${\cal O}$ contains just the superghost contributions and a factor of $(p_2 \psi) (p_2 \bar\psi)$  from the vertex in the zero picture, and we can use~\eqref{eq:intwwb} to capture the high energy behavior
		\begin{align} \label{eq:prereve2}
                \int \frac{d^2z_1 d^2z_2}{dV_{SL(2,R)}}{\cal V}^{(-1)}_1(z_1,\bar{z}_1) {\cal V}^{(0)}_2(z_2,\bar{z}_2) & \simeq e^{-i\pi\frac{\alpha' t}{4}}(\alpha' E_s^2)^{1+\frac{\alpha' t}{4}} \frac{\Gamma\left(-\frac{\alpha' t}{4}\right)}{\Gamma\left(1+\frac{\alpha' t}{4}\right)}\, {\cal O}'\, e^{iq X} \\ \nonumber                                                                                                &  {\psi^+ e^{-\varphi}} {\bar\psi^+ e^{-\bar\varphi}} \left(i\sqrt{\frac{2}{\alpha'}}\partial X^+\right)^{\frac{\alpha' t}{4}} \! \left(i\sqrt{\frac{2}{\alpha'}} \bar\partial X^+\right)^{\frac{\alpha' t}{4}} \;,
	\end{align}    
	where ${\cal O}'$ is the result of the contractions between the string coordinates in ${\cal V}_1$ and ${\cal V}_2$ as discussed after~\eqref{eq:lrt}. When inserted in~\eqref{eq:TTDpbis}, the operatorial part of~\eqref{eq:prereve2} is trivial as the holomorphic and antihomorphic parts are contracted among them, while ${\cal O}'$ is a $c$-number that by construction follows from the correlator $\langle {\cal V}^{(-1)}_1 e^{iq X} {\cal V}^{(0)}_2 \rangle$  Thus we can factorize the amplitude~\eqref{eq:TTDpbis}
	\begin{equation}
			\label{eq:factre1she}
			{\cal A}_0 \simeq \langle {\cal V}^{(-1)}_1 {\cal V}^{(0)}_2  {\cal V}_R^{(-1)}\rangle \, \Pi_R \, \langle {\cal V}_R^{(-1)}| B\rangle\;,
		\end{equation}
		where we introduced a Reggeon vertex ${\cal V}^{(-1)}_R$ describing the collective contributions of the states exchanged in the closed-string channel, the Reggeon propagator $\Pi_R$ and the couplings to the D$p$-branes described by the boundary state $|B\rangle$~\cite{DiVecchia:1999mal,DiVecchia:1999fje}. The Reggeon vertex is basically given by~\eqref{eq:prereve2}
		\begin{equation}
			\label{eq:Rem1s}
			{\cal V}^{(-1)}_R = \kappa_d\, \left[{\psi^+ {\rm e}^{-\varphi}} \left(\sqrt{ \frac{2}{\alpha'}} {i \partial X^+} \right)^{\frac{\alpha' t}{4}}\right]  \left[{\bar\psi^+{\rm e}^{-\bar\varphi}} \left(  \sqrt{ \frac{2}{\alpha'}}  {i \bar\partial X^+} \right)^{\frac{\alpha' t}{4}} \right] \,e^{iq X}\;.
		\end{equation}
		The factors of $\sqrt{\alpha'} E_s$ in~\eqref{eq:prereve2} are automatically produced when inserting the Reggeon vertex in the $3$-point correlator in~\eqref{eq:factre1she} by performing the contraction between $\partial X^+$ and the exponential factors $e^{ip_{1,2} X}$, see~\eqref{eq:dXpeix}. The split between the propagator and the boundary state contribution is defined as follows for later convenience, see the discussion after~\eqref{grsup},
		\begin{equation}\label{eq:rpst}
			\Pi_R = \frac{1}{2\pi}\,e^{-i\pi\frac{\alpha' t}{4}}\,\frac{\Gamma\!\left(-\frac{\alpha' t}{4}\right)}{\Gamma\!\left(1+\frac{\alpha' t}{4}\right)}\;, \quad
			\langle {\cal V}^{(-1)}_R|B\rangle =2\pi \frac{N T_p}{2} \Gamma\!\left(1+\frac{\alpha' t}{4}\right)\;.
		\end{equation}
We can use the OPE with the supercurrent to obtain the Reggeon vertex in the $(0,0)$ picture from the expression in the $(-1,-1)$ picture given in~\eqref{eq:Rem1s} obtaining 
		\begin{align}
			\label{rv0}  
			{\cal  V}^{(0)}_R & = \kappa_d\, \left[  - 
			\frac{2}{\alpha'} {\partial X^+ \partial X^+} - i q \psi 
			{\psi^+ \partial X^+}
			- \frac{\alpha' t}{4} {\psi^+  \partial \psi^+} \right]
			\left (\sqrt{ \frac{2}{\alpha'}} {i \partial X^+} 
			\right)^{\frac{\alpha' t}{4}-1} \\ \nonumber &  \left[  - 
			\frac{2}{\alpha'} {\bar\partial X^+ \bar\partial X^+} - i q \psi {\bar\psi^+ \bar\partial X^+}
			- \frac{\alpha' t}{4} {\bar\psi^+  \bar\partial \bar\psi^+} \right] \left (\sqrt{ \frac{2}{\alpha'}} {i \bar\partial X^+} \right)^{\frac{\alpha' t}{4}-1} \!\!\! {\rm e}^{i q X} \,.
		\end{align}
In conclusion we have constructed a Reggeon vertex operator ${\cal{V}}_R$  both in the picture $0$ and in the picture $-1$ that, when inserted in \eqref{eq:factre1she}, gives the correct high energy behavior of the amplitude  for any choice of the two vertex operators ${\cal{V}}_1$ and ${\cal{V}}_2$.

		\subsubsection{The string eikonal operator: the closed string sector}
		\label{ssec:seikop}
		
		The main advantage of the approach discussed in the previous section is that the external states in the string-brane scattering are arbitrary. So it is possible to go beyond the scattering of massless states discussed in Section~\ref{ssec:string-brane} and consider inelastic transitions of course focusing always on the leading contribution at high energy that has the same scaling as the elastic amplitudes~\eqref{eq:disregss}. As we will see, such transitions are unavoidable in string theory and promote the eikonal phase discussed in Section~\ref{ssec:string-brane} to an eikonal operator that acts on the Hilbert space of the closed string excitations~\cite{Amati:1987wq,Amati:1987uf}. This is an effect of the tidal forces acting on the scattered string that get enhanced at high energies~\cite{Giddings:2006vu}. As we will see explicitly in Section~\ref{ssec:tidalrev}, in the case at hand these tidal forces are related to the gravitational field produced by the D$p$-branes.
		
		There are several ways to derive the eikonal operator related to tidal excitations. In the original approach~\cite{Amati:1987wq,Amati:1987uf}, one obtains the result by factorizing elastic loop amplitudes, where arbitrary tidally excited states appear in the intermediate channel. Here we follow the opposite approach and study directly inelastic tree-level amplitudes. Various techniques can be used to connect the eikonal operator and the string amplitudes~\cite{Black:2011ep,Bianchi:2011se,DAppollonio:2013mgj}: here we will use the results obtained in the previous section and use the Reggeon vertex as the object that encodes all the contributions relevant at high energies. In practice we will derive an explicit expression for the correlator $\langle {\cal V}_2  {\cal V}_R {\cal V}_1\rangle$ appearing in~\eqref{eq:factre1she} (even though it will be simpler to use the $(0,0)$ picture version in~\eqref{rv0} so that both external states ${\cal V}_{1,2}$ can be written in the $(-1,-1)$ picture). Alternative approaches include the direct evaluation of the inelastic correlators by using vertex operators or the use of the Green--Schwarz 3-string vertex~\cite{Green:1982tc,Green:1983hw}, as discussed in detail in~\cite{DAppollonio:2013mgj}.
		
		The basic idea is to use the DDF formalism~\cite{DelGiudice:1971fp} (see also~\cite{Brower:1973iz,Schwarz:1974ix} and~\cite{Hornfeck:1987wt} for an explicit discussion of the NS sector of superstring theory) to deal with the vertex operators ${\cal V}_{1,2}$. For more recent work on the DDF states see~\cite{Skliros:2009cs,Hindmarsh:2010if,Bianchi:2019ywd,Aldi:2019osr,Firrotta:2022cku}. So we briefly review the key ideas of such construction.  The starting point is an auxiliary tachyonic state with momentum $p_T$ 
		\begin{equation}
			\label{eq:auxtac}
			|p_T;0\rangle \,, \qquad \frac{\alpha'}{2} p^2_T=1 \;.
		\end{equation}
		Then we need to introduce a null vector $k$ whose scalar product with $p_T$ is one, and $(d-2)$ space-like vectors $\epsilon_j$ that are perpendicular to $k$. In summary we have
		\begin{equation}
			\label{eq:kesum}
			{\frac{\alpha'}{2}} p_T k =1\,,\qquad \epsilon_j k =0\,, \qquad \epsilon_i \epsilon_j =\delta_{ij}\;.
		\end{equation}
		Focusing on the NS sector for simplicity, the physical states are constructed by acting on the ground state~\eqref{eq:auxtac} with the following DDF oscillators 
		\begin{gather}
			\label{ABigen}
			A_{-n,j} = -i \oint_{0} dw \,\, (\epsilon_j)_{\mu} \left(\sqrt{\frac{2}{\alpha'}}\partial X^{\mu} +i  
			 n (k \psi)  \psi^{\mu}   \right)
			{\rm e}^{-i n k X_L (w)}\;, \\
			B_{-r,j} = i \oint_{0} dw \,\, (\epsilon_j)_{\mu} \left(\sqrt{\frac{2}{\alpha'}} \partial X^{\mu} \, (k \psi) -  \psi^{\mu} 
			(k \partial X) + \frac{1}{2} 
			\psi^{\mu} (k \psi) \frac{(k \partial \psi)}{(k \partial X) } \right)
			\frac{ {\rm e}^{-i r k X_L (w)}  }{(i k \partial
				X)^{\frac{1}{2}}} \;,
			\nonumber
		\end{gather}
		where $n$ ($r$) is a positive integer (half-integer). Of course a similar definition holds for the anti-holomorphic part with the exchange $X_L \to X_R$, $\psi \to \bar\psi$, see~\ref{app:modeexp} for our string conventions. As usual, it is necessary to impose the GSO projection: in order to describe the matter part of the states in the $(-1,-1)$ picture, we have to select only the states containing an odd number  of $B_{-r,j}$, so the first non trivial physical state is obtained by applying the operator $B_{- \frac{1}{2},j}$.
		
		We are now ready to write explicitly the Reggeon vertex~\eqref{rv0} in the DDF basis for the incoming and outgoing states
		\begin{equation}
			\label{eq:VRDDF1}
			\langle {\cal V}^{(-1)}_2 {\cal V}_R^{(0)}  {\cal V}^{(-1)}_1\rangle \;,
		\end{equation}
		where $|{\cal V}^{(-1)}_i\rangle$ are the operators corresponding to the DDF states introduced above. The general overlap between three DDF states was discussed in~\cite{Ademollo:1974kz} in the bosonic case and in~\cite{Hornfeck:1987wt} for the NS string and evaluating the contour integrals in the definition of the DDF oscillators~\eqref{ABigen} yields the Neumann coefficients defining the generic couplings among three string states\footnote{The same result is obtained in the light-cone approach, see~\cite{Cremmer:1974ej} for the bosonic case and~\cite{Green:1982tc,Green:1983hw} for the superstring.}. Evaluating~\eqref{eq:VRDDF1} turns out to be much simpler and the key point is that we can choose $k$ to simplify the high-energy limit of~\eqref{eq:VRDDF1}, which can be done by choosing $k$ to be along $e^+$ introduced in \eqref{eq:epmlc}. 
		Then things become particularly simple in the fermionic sector: the Reggeon vertex does not contain any $\psi^-$ (or $\bar\psi^-$) insertions, so no contractions are possible for the terms proportional to $k\psi$ (or $k\bar\psi$) which then can be set to zero. Then for our propose we can approximate the DDF oscillators as follows
		\begin{gather}
			\label{ABigensim}
			A_{-n,j} \to -i \oint_{z} dw \,\, (\epsilon_j)_{\mu} \sqrt{\frac{2}{\alpha'}} \partial X^{\mu} {\rm e}^{-i n k X_L (w)}\;, \\
			B_{-r,j} \to -i \oint_{z} dw \,\, (\epsilon_j)_{\mu}   \psi^{\mu} 
			(k \partial X)^{\frac{1}{2}}  {\rm e}^{-i r k X_L (w)}   \;\,
			\nonumber
		\end{gather}
		where the contour integrals are around the tachyon exponential factor $e^{i p_T X(z)}$. We can also approximate the Reggeon vertex in \eqref{rv0} keeping only:
\begin{equation}
\label{simplyreggeon}
{\cal  V}^{(0)}_R \simeq \kappa_d \left (\sqrt{ \frac{2}{\alpha'}} {i \partial X^+} 
			\right)^{\frac{\alpha' t}{4}+1}  \left (\sqrt{ \frac{2}{\alpha'}} {i \bar\partial X^+} \right)^{\frac{\alpha' t}{4}+1} \!\!\! {\rm e}^{i q X} \,.
\end{equation}
The Reggeon vertex does not depend on the transverse fermionic coordinates $\epsilon_j \psi $, so the oscillators $B$ and $\bar{B}$ must pair up between the incoming and the outgoing state and the contour integrals in their definition reduces to the ones appearing in the 2-point function. In summary the Reggeon vertex in the fermionic sector acts simply as the identity operator. For the bosonic transverse oscillators~\eqref{ABigensim} there are two options: they can be paired between the two external states as in the fermionic case, which of course requires that they appear in identical pairs, or they can be contracted with the exponential factor of the Reggeon vertex~\eqref{simplyreggeon}. The latter option is the technical origin for the inelastic transitions we are interested in. Each one of such contractions yields a factor of $\pm\sqrt{\frac{\alpha'}{2}} \epsilon_j q$ times an integral over the insertion $w$ which turns out to be equal to one.  The same result holds also  for the anti-holomorphic modes. Finally the level matching condition implies that the difference of the total holomorphic and anti-holomorphic mode number vanishes for each external state. So we can equate this difference for the first and the second state and then take away the contribution of all modes that are paired since they contribute equally to both sides. Thus we see that the bosonic modes that are contracted with the exponential factor in the Reggeon vertex satisfy the following constraint
		\begin{equation}
			\label{eq:sumnA}
			\sum n_{1}-\sum \bar{n}_{1}=  \sum n_{2}-\sum \bar{n}_{2}\;.
		\end{equation}
		
		We can then summarize the action of the Reggeon vertex in~\eqref{eq:VRDDF1} as an exponential constructed with the DDF bosonic oscillators and the transferred momentum $q$
		\begin{subequations}
                  \label{eq:Vrexpf}
                  \begin{align}
			{\cal V}^{(0)}_R \to \kappa_d \left(\alpha' E_s^2\right)^{1+\frac{\alpha' t}{4}} \int_0^{2\pi}\! \frac{d\sigma}{2\pi} :e^{i q \hat{X}}:\;,\\  \hat{X}^j = i \sqrt{\frac{\alpha'}{2}} \sum_{n\not=0} \left(\frac{A_{n,j}}{n} e^{i n \sigma}+\frac{\bar{A}_{n,j}}{n} e^{-i n \sigma}\right),                  \end{align}
                      \end{subequations}
                      where the integral over $\sigma$ enforces the constraint~\eqref{eq:sumnA} and the energy dependent factors follows from the contraction of the fields $\partial X^+$ in~\eqref{simplyreggeon} with the exponential part of the external states. The exponential is normal ordered so the positive modes are contracted with the incoming state $|{\cal V}_1\rangle$ and the negative modes with the outgoing one $\langle {\cal V}_2|$. 
		
		An advantage of the approach based on the Reggeon vertex discussed here is that it is fully covariant and so it can be used to provide a full characterization (in terms of the little group $SO(d-1)$) of the (massive) excited states produced by tidal excitations. It is then possible to carry out explicit checks between the results obtained by using~\eqref{eq:Vrexpf} and those obtained by the direct evaluation of the corresponding covariant amplitudes, see~\cite{DAppollonio:2013mgj} for a detailed discussion of the transition between the ground state to the first and the second massive level. All results are consistent with the key properties of~\eqref{eq:Vrexpf}: the excitations added or damped by the tidal forces involve only the bosonic oscillators and are always in the spatial direction perpendicular to the (fast) motion of the scattering string 
	
		We can use~\eqref{eq:Vrexpf} to write the high-energy result for the string-brane scattering as an operator instead of a matrix element between two specified states as in~\eqref{eq:factre1she}
		\begin{equation}
			\label{eq:Ahat0}
			\hat{\cal A}_0 \simeq \frac{N T_p \kappa_d}{2} e^{-i \frac{\alpha' t}{4}}\, \Gamma\!\left(-\frac{\alpha' t}{4}\right) \, (\alpha' E^2_s )^{1+\frac{\alpha' t}{4}} \int_0^{2\pi}\! \frac{d\sigma}{2\pi} :e^{i q \hat{X}}:\;,
                \end{equation}
                where we used~\eqref{eq:rpst}. Notice that the non-operatorial overall factor matches that of Eq.~\eqref{eq:disregss}. As usual we can derive the eikonal in the impact parameter space by using~\eqref{eq:DbraneFT}. As we discussed in Section~\ref{ssec:string-brane}, when the impact parameter becomes of the order of $l_s(E) = \sqrt{\alpha' \ln(\alpha' E_s^2)}$, the dynamics is dominated by the open strings attached to the D$p$-branes and the picture of a closed string to closed string scattering is not reliable. So, let us focus on the case $b\gg l_s(E_s)$ where $\alpha' t$ is very small and the elastic factor in \eqref{eq:Ahat0} reduces to the field theory result~\eqref{eq:SeikLft}. Then it is straightforward to perform formally the Fourier transform~\eqref{eq:DbraneFT} at the operatorial level, since the last factor in~\eqref{eq:Ahat0} provides just a shift $b \to b+\hat{X}$ in the result
		\begin{equation}
			\label{eq:hatdelta0cl}
			\begin{aligned}
				2\hat{\delta}_0 & = \!\int\! \frac{d^{d-p-2} q}{(2\pi)^{d-p-2}} \frac{\hat{\cal A}_0(E_s,{q}^2)}{2E_s} \, {\rm e}^{i b q}\\ & \simeq \!\int\! \frac{d^{d-p-2} q}{(2\pi)^{d-p-2}} N T_p \kappa_d E_s  \int_0^{2\pi}\! \frac{d\sigma}{2\pi} \frac{:e^{i q (b+\hat{X})}:}{q^2} =  \int_0^{2\pi}\! \frac{d\sigma}{2\pi} :2\delta_0 (b+\hat{X}):   \,,
			\end{aligned}    
\end{equation}
where we followed the steps in~\eqref{impactb} for $b^2\gg l_s^2(E_s)$ and wrote explicitly the normal ordering prescription. Thus, at the level of the leading eikonal, the generalization to the full string level takes the appealing form of an average of the effective QFT result, but with the appropriate ``local'' impact parameter for each point along the string: the part of the closed string that are closer to the D$p$-branes feel a stronger gravitational force than those that are further apart. As we will see in Sections~\ref{ssec:tidalrev} and~\ref{ssec:PLstract} this is at the basis of the tidal effects that are enhanced in the high energy limit.

		\subsubsection{Graviton scattering in bosonic string theory}
		\label{ssec:causality-rest}
	{In this subsection we show that, in sharp contrast with what happens in field theory discussed in Section~\ref{ssec:ethd}, the terms  appearing in the graviton scattering in the bosonic string that come from     a quadratic $(\mathrm{Riemann})^2$ and a cubic term $(\mathrm{Riemann})^3$ do not contribute to leading order  at small impact parameter. Therefore in the bosonic string there is no risk of negative Shapiro time delay.}
		The disk amplitude with two closed string tachyons is discussed in~\ref{ssec:string-brane-bos-1}. In the Regge limit it reduces to Eq.~\eqref{eq:TTeikL} from which one can derive the eikonal~\eqref{eq:incg1} for the tachyon scattering  off a stack of $N$ coincident D$p$-branes. In the same regime the disk amplitude of a massless state is given instead by:
		\begin{eqnarray}
			\label{grbos}
			{\cal{A}}_{0}^{{\cal G}{\cal G}} & \sim & {\frac{\kappa_{d} T_p N}{2}} 
			{\rm e}^{-i \pi \frac{\alpha't}{4}} (\alpha' E^2_s)^{1 +\frac{\alpha't}{4}}  \Gamma \left(- 1- \frac{\alpha' t }{4} \right) \\
			& \times & \left( (\epsilon_1 \epsilon_2)  - \frac{\alpha'}{2}  ({{\epsilon}}_1 q) ({{\epsilon}}_2 q)   \right)\left( ({\bar{\epsilon}}_1 {\bar{\epsilon}}_2) -
			\frac{\alpha'}{2} ({\bar{\epsilon}}_1 q) ({\bar{\epsilon}}_2 q) \right) \nonumber ~,
		\end{eqnarray}
		where, as usual, we split the closed string polarization into its holomorphic and anti-holomorphic part ${\cal G}_{\mu\nu}=\epsilon_{\mu} \bar{\epsilon}_\nu$. Notice that the quantity in the first line in~(\ref{grbos}) is just the elastic scattering of a tachyon on the  D$p$-branes~\eqref{eq:TTeikL}. This result can be compared with that obtained for the supersymmetric case in Section~\ref{ssec:string-brane}, which we report below by combining~\eqref{eq:Klim} and~\eqref{eq:disregss}
		\begin{eqnarray}
			{\cal{A}}_{0} \sim (\epsilon_1 \epsilon_2) ({\bar{\epsilon}}_1 {\bar{\epsilon}}_2) \frac{\kappa_{d} T_p N}{2} \Gamma \left(- \frac{\alpha' t }{4} \right)  {\rm e}^{-i \pi \frac{\alpha't}{4}} (\alpha' E^2_s)^{1 +\frac{\alpha't}{4}}~.
			\label{grsup}
		\end{eqnarray}                
		A  first qualitative difference between~\eqref{grsup} and~\eqref{grbos} is that the leading Regge trajectory in the latter includes the tachyonic (ground) state of the bosonic theory.
		The second difference is that even 
		in the high energy Regge limit the bosonic amplitude has a non-trivial dependence on the polarization tensors, see the second line in~(\ref{grbos}).  As emphasized in~\cite{Camanho:2014apa}, this is a direct consequence of the modification of the three-graviton vertex in the bosonic theory which yields a quadratic $(\mathrm{Riemann})^2$ and a cubic term $(\mathrm{Riemann})^3$ in the effective action, while in the maximally supersymmetric case these corrections are forbidden by supersymmetry. Because of this, the Lorentz structure in~(\ref{grbos}) is the same as the one appearing in Section~\ref{ssec:ethd} except that in the tree-level graviton-brane scattering there is a single parameter ($\sqrt{\alpha'}$) in the three-point vertex, while in the effective QFT description~\eqref{eq:3grmod} there are in general two independent parameters ($l_2$ in~\eqref{eq:3am2} and $l_4$ in~\eqref{eq:3am4}).
		
		We can follow the same approach discussed in Section~\ref{ssec:string-brane} for the superstring and~\ref{ssec:string-brane-bos-1} for the bosonic case and derive the eikonal for the graviton scattering: it is sufficient to rewrite the momentum transfer $q$ appearing in the second line of~\eqref{grbos} in terms of a derivative with respect to the impact parameter and use the result~\eqref{eq:incg1} for the first line of~\eqref{grbos}. Thus we obtain
		\begin{equation}
			\label{eq:incgGR}
			\begin{aligned}
				2\delta_0  & = \left( (\epsilon_1 \epsilon_2) + \frac{\alpha'}{2}  ({{\epsilon}}_1 \partial_b) ({{\epsilon}}_2 \partial_b)   \right)\left( ({\bar{\epsilon}}_1 {\bar{\epsilon}}_2) +
				\frac{\alpha'}{2} ({\bar{\epsilon}}_1 \partial_b) ({\bar{\epsilon}}_2 \partial_b) \right) \\ & \quad \times \frac{\kappa_d N T_p E_s}{4\pi}\, \frac{\Gamma\left(1+\frac{\alpha' \nabla^2_b}{4}\right)}{1-\frac{\alpha'  \nabla^2_b}{4}} \left[(\pi b^2)^{-\frac{d-4-p}{2}} \gamma\left(\frac{d-4-p}{2},\frac{b^2}{\bY}\right)\right]\;.
			\end{aligned}
		\end{equation}
		For the current analysis we ignore the absorptive effects, related to the imaginary part of $Y$, and concentrate our attention on the leading real part by replacing ${Y}$ with $l^2_s(E_s)$, see Eq~\eqref{eq:lsE}. By focusing on the regime $\sqrt{\alpha'}\ll b \ll l_s(E_s)$ we can use~\eqref{eq:incomgamma}, keeping only the term with $k=0$, and approximate the square parenthesis in~\eqref{eq:incgGR} as follows
		\begin{equation}
			\label{eq:red0grb}
			\left[(\pi b^2)^{-\frac{d-4-p}{2}} \gamma\left(\frac{d-4-p}{2},\frac{b^2}{\bY}\right)\right] = \frac{(\pi l_s^2(E_s))^{-\frac{d-4-p}{2}} }{\frac{d-4-p}{2}} +{\cal O} \left (  \frac{b^2}{l_s^2(E_s)} \right)\;.
		\end{equation}
		This means that the differential operators in the $\Gamma$-functions and in the polarization dependent prefactor act on a function of $\frac{b^2}{l^2_s(E_s)}$ that starts with a constant, thus they do not contribute to leading order at small impact parameter. Notice that also the tachyonic pole of the bosonic string becomes harmless.\footnote{This is essentially due to the fact that tachyon exchange is suppressed by two powers of the energy with respect to graviton exchange and therefore it is negligible in the high-energy limit.}
		
		The pattern discussed above is in sharp contrast with the QFT case discussed in Section~\ref{ssec:ethd}, since in that case the eikonal~\eqref{eq:d0m1gl2l4} for the scattering of massless states contained terms that grow in the regime where the impact parameter is smaller than the length scales weighting the higher derivative corrections $b\ll l_{2,4}$ (as we mentioned $\ell_s$ in the string analysis plays the role of both $l_{2}$ and $l_4$ of the effective description). As we will see in Section~\ref{ssec:seikcc} this plays a crucial difference in the behavior of the deflection angle and the Shapiro time delay obtained in two cases.
		
	\subsubsection{String-string scattering at tree level}
	\label{ssec:string-brane-sup}
	
\begin{figure}
	\centering
	\begin{tikzpicture}
		\draw (-5,2) .. controls (-3.5,1.5) and (-3.5,-1.5) .. (-5,-2);
		\draw (5,2) .. controls (3.5,1.5) and (3.5,-1.5) .. (5,-2);
		\draw[red] (0,-1.875) .. controls (-1,-1.875) and (-1,1.875) .. (0,1.875);
		\draw[red,dashed] (0,-1.875) .. controls (1,-1.875) and (1,1.875) .. (0,1.875);
		\draw (-4,3) .. controls (-3,1.5) and (3,1.5) .. (4,3);
		\draw (-4,-3) .. controls (-3,-1.5) and (3,-1.5) .. (4,-3);
		\draw [thick,dotted] (0,0) ellipse (3.85 and .5);
		\draw [ultra thick] (-3.85,0) .. controls (-3.5,-.65) and (3.5,-.65) .. (3.85,0);
		\filldraw[blue!10!white] (-5,2) .. controls (-5.3,2.3) and (-4.3,3.3) .. (-4,3);
		\filldraw[blue!10!white] (-5,2) .. controls (-4.7,1.7) and (-3.7,2.7) .. (-4,3);
		\draw[thick,blue] (-5,2) .. controls (-5.3,2.3) and (-4.3,3.3) .. (-4,3);
		\draw[thick,blue] (-5,2) .. controls (-4.7,1.7) and (-3.7,2.7) .. (-4,3);
		\filldraw[blue!10!white] (5,2) .. controls (5.3,2.3) and (4.3,3.3) .. (4,3);
		\filldraw[blue!10!white] (5,2) .. controls (4.7,1.7) and (3.7,2.7) .. (4,3);
		\draw[thick,blue] (5,2) .. controls (5.3,2.3) and (4.3,3.3) .. (4,3);
		\draw[thick,blue] (5,2) .. controls (4.7,1.7) and (3.7,2.7) .. (4,3);
		\filldraw[green!10!white] (-5,-2) .. controls (-5.3,-2.3) and (-4.3,-3.3) .. (-4,-3);
		\filldraw[green!10!white] (-5,-2) .. controls (-4.7,-1.7) and (-3.7,-2.7) .. (-4,-3);
		\draw[thick,black!30!green] (-5,-2) .. controls (-5.3,-2.3) and (-4.3,-3.3) .. (-4,-3);
		\draw[thick,black!30!green] (-5,-2) .. controls (-4.7,-1.7) and (-3.7,-2.7) .. (-4,-3);
		\filldraw[green!10!white] (5,-2) .. controls (5.3,-2.3) and (4.3,-3.3) .. (4,-3);
		\filldraw[green!10!white] (5,-2) .. controls (4.7,-1.7) and (3.7,-2.7) .. (4,-3);
		\draw[thick,black!30!green] (5,-2) .. controls (5.3,-2.3) and (4.3,-3.3) .. (4,-3);
		\draw[thick,black!30!green] (5,-2) .. controls (4.7,-1.7) and (3.7,-2.7) .. (4,-3);
	\end{tikzpicture}
	\caption{\label{fig:StringString} Scattering of two massless closed  strings represented by blue and green circles. The thick black line corresponds to closed strings exchanged in the $t$-channel. The thin red line to very massive closed strings in the $s$-channel.}
\end{figure}
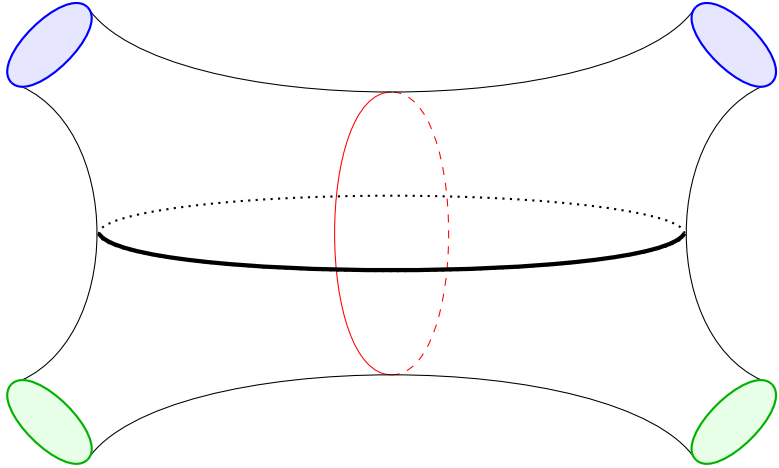

	Although for pedagogical reasons we have started our discussion of gravitational scattering in string theory from the case of string-brane collisions, historically the first case considered was the one of (massless) string-string collisions at transplanckian energy in  (Type II) critical (i.e.~$d=10$) superstring theory (see Fig.~\ref{fig:StringString}). 
Such a theory is ghost and tachyon free and looks like a fully consistent quantum theory of gravity at least in perturbation theory. In that same approximation it differs from GR, even at large distance, because of the presence of other gravitationally coupled massless modes, besides the graviton. However, by going to high-energy, graviton exchange (or better the Regge trajectory on which it lies) dominates because of its higher spin and therefore by studying string-string collisions at transplanckian energy one can draw interesting lessons for generic theories of quantum gravity involving extended objects. For the original motivations for studying such gedanken experiments we refer to the introductory Section~\ref{ssec:aims}.
	
	The original idea, in the late eighties, was to consider the collision of massless strings (e.g. gravitons or dilatons) and to study the process in a parameter space containing three relevant length scales: the fundamental length $\ell_s$ of string theory, see~\ref{app:string}, the impact parameter $b$ of the process, and the characteristic scale of the geometry associated with the total center-of-mass energy $R \sim (G E)^\frac{1}{D-3}$. Note that this latter scale depends on Newton's constant $G$, which, in string theory can be traded for the string coupling $g_s$. By considering the regime of weakly-coupled string theory, $g_s \ll1$, one can arguably dispose of a fourth length scale, the Planck length $\ell_P$, since it will be much smaller than $\ell_s$, see again~\ref{app:string}. Since the physics of the process can only depend on dimensionless ratios, the final parameter space is effectively two-dimensional.\footnote{One can also neglect the mass of the colliding strings, first by taking the initial states to be massless, and then by using the dynamical fact that only the excitation of relatively light massive strings is induced by tidal forces (see Section~\ref{ssec:tidalrev}).} The other free parameter is the number $D$ of non-compact spatial dimensions, having assumed the remaining $(d-D)$ to be very small and static.
	
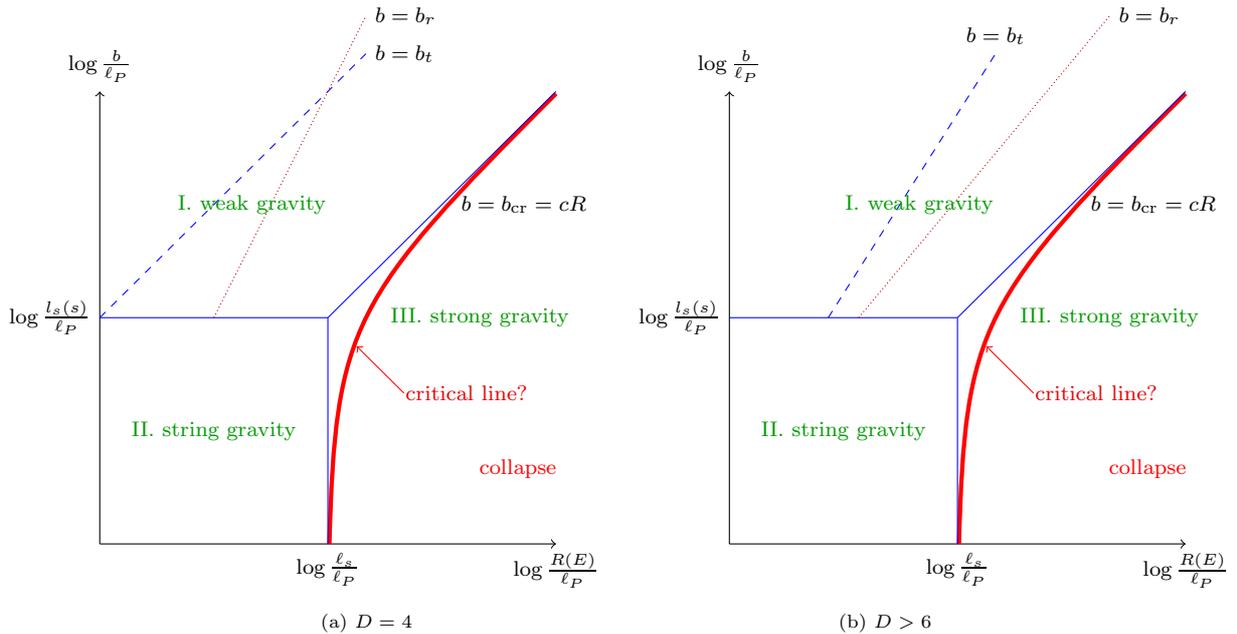
\begin{figure}
	\begin{center}
	\begin{subfigure}{.45\textwidth}
		\centering
		\hspace{-50pt}
		\begin{tikzpicture}
			\draw[->, thin] (-3,-3)--(3,-3) node[below]{\footnotesize  $\log\frac{R(E)}{\ell_P}$};
			\draw[->, thin] (-3,-3)--(-3,3) node[above]{\footnotesize  $\log\frac{b}{\ell_P}$};
			\draw[ultra thick, red]  (.02,-3).. controls (0.1,-0.1) and (0.3,0.25) ..(3,2.97);
			\draw[thin, blue] (-3,0)--(0,0);
			\draw[thin, blue] (0,-3)--(0,0);
			\draw[thin, blue] (0,0)--(3,3);
			\draw[thin, dashed, blue] (-3,0)--(.5,3.5);
			\node at (.5,3.5)[right]{\footnotesize $b=b_t$};
			\draw[densely dotted, red!70!black] (-1.5,0)--(.5,4);
			\node at (.5,4)[right]{\footnotesize $b=b_r$};
			\draw [->,red] (1,-1)--(0.38,-0.38);	
			\node at (.9,-1)[right]{\color{black!10!red}\footnotesize critical line?};
			\node at (2.5,-2)[]{\color{red}\footnotesize collapse};
			\node at (-1,1.5)[]{\color{green!60!black}\footnotesize I. weak gravity};
			\node at (-1.5,-1.5)[]{\color{green!60!black} \footnotesize II. string gravity};
			\node at (2,0)[]{\color{green!60!black} \footnotesize III. strong gravity};
			\node at (1.63,1.5)[right]{\footnotesize$b=b_\text{cr}=cR$};
			\node at (-3,0)[left]{\footnotesize $\log\frac{l_s(s)}{\ell_P}$};
			\node at (0,-3)[below]{\footnotesize $\log\frac{\ell_s}{\ell_P}$};
		\end{tikzpicture}
	\subcaption{$D=4$}
	\end{subfigure}
\begin{subfigure}{.45\textwidth}
	\centering
\begin{tikzpicture}
	\draw[->, thin] (-3,-3)--(3,-3) node[below]{\footnotesize  $\log\frac{R(E)}{\ell_P}$};
	\draw[->, thin] (-3,-3)--(-3,3) node[above]{\footnotesize  $\log\frac{b}{\ell_P}$};
	\draw[ultra thick, red]  (.02,-3).. controls (0.1,-0.1) and (0.3,0.25) ..(3,2.97);
	\draw[thin, blue] (-3,0)--(0,0);
	\draw[thin, blue] (0,-3)--(0,0);
	\draw[thin, blue] (0,0)--(3,3);
	\draw[thin, dashed, blue] (-1.7,0)--(.5,3.5);
	\node at (.5,3.5)[above]{\footnotesize $b=b_t$};
	\draw[densely dotted, red!70!black] (-1.3,0)--(2,4);
	\node at (2,4)[right]{\footnotesize $b=b_r$};
	\draw [->,red] (1,-1)--(0.38,-0.38);
	\node at (.9,-1)[right]{\color{black!10!red}\footnotesize critical line?};
	\node at (2.5,-2)[]{\color{red}\footnotesize collapse};
	\node at (-.5,1.5)[]{\color{green!60!black}\footnotesize I. weak gravity};
	\node at (-1.5,-1.5)[]{\color{green!60!black} \footnotesize II. string gravity};
	\node at (2,0)[]{\color{green!60!black} \footnotesize III. strong gravity};
	\node at (1.63,1.5)[right]{\footnotesize$b=b_\text{cr}=cR$};
	\node at (-3,0)[left]{\footnotesize $\log\frac{l_s(s)}{\ell_P}$};
	\node at (0,-3)[below]{\footnotesize $\log\frac{\ell_s}{\ell_P}$};
\end{tikzpicture}
\subcaption{$D>6$}
\end{subfigure}
\end{center}
	\caption{\label{fig:StringRegimes} A broad-brush phase diagram from collapse criteria, where the effective string scale $l_s(s)$ appearing on the vertical axis is defined in~\eqref{eq:lscsv}. We also show, qualitatively, the impact parameters $b_t$ and $b_r$ below which string-size (tidal) and radiative corrections, respectively, start to be relevant  in the weak gravity regime. We also illustrate in the two panels how the relative importance of the two kinds of corrections strongly depends on $D$. See Section~\ref{beyondtree} for further details.}
\end{figure}

	The parameter space is naturally divided in three regions, each one characterized by which length scale dominates over the other two (see Fig.~\ref{fig:StringRegimes}).
In region I, one has $b \gg l_s(s), R$ corresponding to a weak-gravity, point-like regime. In region II, the string scale $l_s(s)$ dominates over $b$ and $R$. We may call it the string-gravity regime because it is here that  string-size effects are strongly enhanced and deviations from GR are most evident. Finally, when $R > {\ell}_s, b$ we enter in the strong-gravity regime where, classical  gravitational collapse  is expected to occur. A sketch of a possible critical line separating the collapse from the ``dispersion" regime is shown in Fig.~\ref{fig:StringRegimes}. It uses the physical expectation --following from the collapse criteria discussed in Section~\ref{ssec:CTS} -- according to which the size of the colliding objects plays the role of an additional contribution to the effective impact parameter. This is why the critical line is expected to bend downwards and hit the real axis when $R \sim {\ell}_s$ (or $\sqrt{s} \sim g_s^{-2} M_s$).  
	Obviously, if one were able to solve completely the problem in the collapse region by constructing a unitary $S$-matrix, one would  solve the (in)famous information paradox raised by Hawking in the seventies \cite{Hawking:1976ra}.  
	
	We should finally mention  that dividing the parameter space in just three regions is a gross approximation. There are interesting sub-regions, as we will discuss in Section~\eqref{beyondtree}. 
	
	This being said, let us start by considering, as the simplest example, 
	the tree-level four-dilaton amplitude. It is fully symmetric in all three Mandelstam variables and given by
	\begin{eqnarray}
		\mathcal{A}_0(s,t) = 8 \pi G \left( \frac{tu}{s} + \frac{su}{t} + \frac{st}{u} \right)\frac{\Gamma (1 - \frac{\alpha' s}{4}) 
			\Gamma (1 - \frac{\alpha' u}{4}) \Gamma (1 - \frac{\alpha' t}{4})}{\Gamma (1 + \frac{\alpha' s}{4})
			\Gamma (1 + \frac{\alpha' u}{4}) \Gamma (1 + \frac{\alpha' t}{4})}     
		\label{Astsuper}
	\end{eqnarray}
	with $s + t + u =0$.  In  the Regge limit, $s \to \infty$ at fixed $t$, using \eqref{eq:gstirl} on the four  $\Gamma$-functions that depend on $s$,
	it becomes:
	\begin{eqnarray}
		\mathcal{A}_0 (s,t)  \sim  \frac{32 \pi G}{\alpha' }  \frac{\Gamma (- \frac{\alpha'}{4}t )}{\Gamma (1 + \frac{\alpha'}{4}t )}
		\left( \frac{\alpha'}{4} s \right)^{2 + \frac{\alpha'}{2}t} {\rm e}^{-i \pi \frac{\alpha'}{4}t }\, .
		\label{T1acv}
	\end{eqnarray}
        It is possible to obtain directly the high-energy formula above by using the approach of Section~\ref{ssec:reggeon} where the amplitude is constructed by gluing together two $3$-point correlators involving a Reggeon vertex
                \begin{equation}
                  \label{eq:Regq}
                   C_{S_2} \langle {\cal V}_{1} {\cal V}_{4} {\cal V}_R\rangle \,\left(\frac{\alpha'}{8\pi} \Pi_R\right) C_{S_2} \langle {\cal V}_R {\cal V}_{2} {\cal V}_{3} \rangle  = \frac{4}{\alpha'} \frac{\Gamma\!\left(-\frac{\alpha' t}{4}\right)}{\Gamma\!\left(1+\frac{\alpha' t}{4}\right)} {\rm e}^{-i \pi \frac{\alpha' t}{4} } \left[\kappa_d \left(\alpha' E^2_s\right)^{1 + \frac{\alpha' t}{4}} \right]^2.
                \end{equation}
The square parenthesis follows from the two correlators involving the Reggeon vertex, where each field along the plus light-cone direction takes the leading value $\sqrt{\alpha'} E_s$, where $E_s$ is the energy of each initial state so $s=(2E_s)^2$ (recalling that we are working in the center-of-mass frame). The overall prefactor combines the various normalizations and Reggeon propagator~\eqref{eq:rpst} and, by using~\eqref{eq:Cs2kap} and~\eqref{eq:kappad}, one can check that~\eqref{eq:Regq} agrees with~\eqref{T1acv}. In the field theory limit this result reduces to 
	\begin{equation}
		\mathcal{A}_0 (s,t) = 8 \pi G  \frac{s^2 }{(-t)}\,.
		\label{as}
	\end{equation}
	In the same limit, the leading eikonal phase is given by 
	\begin{equation}
		2 \delta_0 (s,b) = \int \frac{d^{D-2} q}{(2\pi)^{D-2}} {\rm e}^{i q b} \frac{\mathcal{A}_0(s,t)}{2s} = \pi G \frac{ s \Gamma (\frac{D-4}{2})}{  \pi^{\frac{D-2}{2}} b^{D-4}}\,.
		\label{delta}
	\end{equation}
	This can be compared with \eqref{eq:leik} showing that, indeed, GR and string theory share the same ultra-relativistic large impact parameter limit. 
	
	However, even at arbitrarily high energy, this agreement fails to persist  at somewhat lower values of $b$. Naively, one would expect this to happen when $b$ becomes ${\cal O}({\ell}_s)$, possibly modulo a logarithmic enhancement as in \eqref{eq:lsE}. This turns out not to be the case.
	As we will discuss in Sect.~\ref{beyondtree}, at a $b_t$ parametrically larger than $\ell_s$ (see Eq.~\eqref{bt}), the phenomenon of tidal excitation  kicks in\footnote{This phenomenon was first discussed in \cite{Amati:1987uf} where it was called diffractive excitation in analogy with a well-known phenomenon in hadronic physics. The true physical interpretation in terms of tidal forces was first given by Giddings \cite{Giddings:2006sj}.} as a result of the gravi-Reggeon exchange already discussed in Sections~\ref{ssec:reggeon}, \ref{ssec:seikop} in the case of string-brane collisions, with the only difference that now the Reggeon vertex affects both strings.

One can follow the same steps discussed in~\eqref{impactb} and rewrite~\eqref{T1acv} in impact parameter space as follows
\begin{equation}
2 \delta_0 (s, b) = G s  \frac{ \Gamma (1- \frac{\alpha'}{4} \nabla^2)}{\Gamma (1+ \frac{\alpha'}{4} \nabla^2)}
\left[ (b^2 \pi)^{-\frac{D-4}{2}} \gamma\left( \frac{D-4}{2}; \frac{b^2}{Y_c}\right) \right]\;,
\label{impactb1}
\end{equation}
where
\begin{equation}
  \label{eq:lscsv}
  Y_c = l^2_s(s)- i \pi \alpha'\;,\quad \quad l^2_s(s) = 2\alpha' \log \frac{\alpha' s}{4}\;.
\end{equation}
For $b\gg l_s(s)$ we can ignore the ratio of two $\Gamma$-functions and use \eqref{eq:gamlab} obtaining
\begin{align}
  2 \delta_0 (s, b)  \sim \frac{Gs\, \Gamma ( \frac{D-4}{2})}{(\sqrt{\pi}b)^{D-4}} -\; Gs \,(\pi b^2)^{-\frac{D-4}{2}}  e^{-\frac{b^2}{Y_c}} \left( \frac{b^2}{Y_c}\right)^{ \frac{D-6}{2}} + \cdots
\label{impb2}
\end{align}
From~\eqref{impb2} we have 
\begin{equation}\label{eq:redcs}
\operatorname{Re} 2\delta_0 (s,b) \sim \frac{Gs\, \Gamma ( \frac{D-4}{2})}{(\sqrt{\pi}b)^{D-4}} -\; Gs \,(\pi b^2 )^{-\frac{D-4}{2}} e^{-\frac{b^2}{l_s^2(s)}} \left( \frac{b^2}{l_s^2(s)}\right)^{ \frac{D-6}{2}} 
\end{equation}
for the real part and
\begin{equation}
\operatorname{Im} 2\delta_0 (s,b) \sim  \frac{\pi \alpha' }{l^2_s(s)} Gs (\pi l_s^2(s))^{-\frac{D-4}{2}}  e^{-\frac{b^2}{l_s^2(s)}} \Bigg( 1-  \frac{D-6}{2}\, \frac{l_s^2(s)}{b^2}\Bigg)
\label{eq:ima}
\end{equation}
for the imaginary part. As in the string-brane case, the exponentially suppressed term provides the leading imaginary part. 

The real part of \eqref{impactb1} has an amusing shock-wave interpretation in terms of the generalized  Aichelburg--Sexl metrics discussed in Section~\ref{ssec:CTS}. Neglecting again the correction coming from  the ratio of the two Gamma-functions,  it can be shown \cite{Veneziano:1988aj} to correspond to a shock-wave metric  where
 the function $f (x_\perp)$ is obtained from Eq. \eqref{eq:Eesw} by substituting, as in Eq. \eqref{eq:Eeswb},   $E^{(1)}$ times the transverse $\delta$-function with  a Gaussian profile:
\begin{equation}
\rho(x_\perp) = E^{(1)}~ e^{-\frac{x_{\perp}^2}{l_s^2(s)}} (\sqrt{\pi}\, l_s (s))^{2-D}\,,
\label{profile}
\end{equation}
i.e.~with an approximately uniform-density beam of size $\sim l_s(s)$ that reduces to the delta function of \eqref{eq:Eesw} for $l_s (s) \rightarrow 0$. In other words, while  the delta-function profile provides only the first term of \eqref{impb2} as shown in \eqref{phase}, the Gaussian profile provides instead the entire real part of \eqref{impactb1}.  The proof of the above statement is straightforward. Following 't Hooft's derivation of the leading eikonal phase sketched in Section~\ref{ssec:ASmetric}, we know that what determines  the phase is, up to an energy factor, the time delay $\Delta v$ suffered by one particle as it moves in the shock-wave produced by the other particle. According to Eq. \eqref{vdelay}, this time delay is determined by
 the quantity $f(x_\perp)$ appearing in the (generalized) AS metric~\eqref{eq:AS}. Therefore, we can either extract   $f(x_\perp)$ directly from \eqref{impactb1} and determine the transverse profile of the beam by using  \eqref{eq:Eesw} with the profile  \eqref{profile} or, given the profile \eqref{profile}, by using \eqref{eq:Eesw} to fix first $f(b)$ and then, from it,  the real part of \eqref{impactb1}.  

Without repeating here the discussion given above in the case of string-brane collisions we now simply give the analog of \eqref{eq:hatdelta0cl} for string-string scattering:
\begin{equation}
\label{tidalop}
\hat{\delta}_0  = \frac{1}{4 \pi^2}\int_0^{2 \pi} \int_0^{2 \pi} d\sigma_1 d \sigma_2 :\delta_0 (E, b + \hat{X}_1(\sigma_1) - \hat{X}_2(\sigma_2)): 
~ ,
\end{equation}
where $\hat{X}_{1,2}$  are Hermitian (and commuting) closed-string position operators for each incoming string, and $\delta_0$ is the leading eikonal~\eqref{delta}.

For $b \gg l_s(s)$, $\delta_0(E,b)$ is real and thus $\hat{\delta}_0$ is Hermitian up to exponentially small corrections. The physical interpretation of \eqref{tidalop} is that the graviton is exchanged between one point on one string and one point on the other string. The (transverse) coordinates of each string are however operators and \eqref{tidalop} picks up an expectation value of the transverse distance between those two points. The physical consequences of the shift in $b$, tidal excitations due to the extended nature of quantum strings, will be discussed in Section~\ref{ssec:tidalrev}. Here we just mention that,
unlike the singular point-like case, the string-corrected eikonal has a finite  $b \to 0$ limit and a smooth expansion in $b$ around  it. This is the string-string counterpart to what we have already discussed in the string-brane collision, see~\eqref{lowb}. One finds \cite{Amati:1987uf}:
\begin{equation}
\label{lowbstst}
 \operatorname{Re} 2 \delta_0  \sim \frac{2 G s}{\hbar} \frac{1}{(\pi l_s^2(s))^{\frac{D-4}{2}}} \left( \frac{1}{D-4} - \frac{b^2}{(D-2) l_s^2(s)} + \cdots \right)\,.
\end{equation}

Another, even more distinctive feature of string theory as opposed to traditional  QFT, is that, already at tree level, graviton exchange does not correspond to a real scattering amplitude. This can be seen immediately in \eqref{T1acv} through its last factor, typical of Regge-pole behavior. 
This imaginary part comes from the presence of $s$-channel poles in the tree-level Shapiro--Virasoro amplitude corresponding to the heavy (i.e.~of mass $\sqrt{s}$) closed-string intermediate states produced in that channel. It is the very essence of Dolen--Horn--Schmid duality \cite{Dolen:1967zz, Dolen:1967jr}, as it is incorporated in string theory, that this imaginary part matches, on average, the imaginary part corresponding to the Regge-pole behavior given in~\eqref{T1acv}.

Note, however, that the imaginary part of the amplitude, unlike the real part, is  order $\alpha'$ and lacks  the graviton pole at  $t=0$.
		This lack of a singularity corresponds, in $b$-space to an exponential cut-off at large impact parameter. More quantitatively one finds \cite{Amati:1987uf}:
\begin{equation}
\operatorname{Im} 2 \delta_0 (s,b) \sim {Gs \pi^2 \alpha'}\, {(\pi l^2_s(s))^{-\frac{D-2}{2}}} 
e^{- \frac{b^2}{l_s^2(s)}}\,.
\label{Imdeltastst}
\end{equation}
		This result fully justifies having neglected $\operatorname{Im}2 \delta_0$ in the previous, weak-gravity (large $b$) regime. 		
		At higher string-loop level we expect multiple string formation in the $s$-channel. In Section~\ref{beyondtree}  we will discuss how one can have a qualitative idea of the dominant process as a function of the total initial energy up to the already mentioned expected  threshold of gravitational collapse, $\sqrt{s} \sim M_s g_s^{-2}$. 		

From~\eqref{eq:redcs} and \eqref{lowbstst} we can compute the deflection angle for small and large impact parameter. We get
\begin{subequations} \label{eq:ascsf}
  \begin{align}
\Theta &=\frac{4G\sqrt{s}}{b^{D-3}} \frac{\Gamma (\frac{D-2}{2})}{\pi^{\frac{D-4}{2}}} \rightarrow  \frac{4G\sqrt{s}}{b}\,,\qquad  b \gg l_s(s)\,, \\
\Theta &= \frac{8G \sqrt{s} \,b}{(D-2) \pi^{\frac{D-4}{2}}\, l_s^{D-2}(s)} \rightarrow \frac{4G \sqrt{s}b}{l_s^2(s)}\,,\qquad b \ll l_s(s)\,,
  \end{align}
\end{subequations}  
where the arrow indicates the limit $D \rightarrow 4$.
The first, for large impact parameter, coincides with the one of a massless point-particle moving in the Aichelburg--Sexl metric (see \eqref{ASmetric} that can be derived from \eqref{phase}). The second instead is intrinsically stringy because of the presence of $\alpha'$. Since the deflection angle increases with $b$ for small $b$ and decreases with $b$ for large $b$, it is to be expected that it will have a maximum at some intermediate $b$. We will discuss this and what happens in the intermediate region in Section  \ref{beyondtree}.

	\subsubsection{Brane-brane scattering at leading order}
	\label{ssec:brane-brane-sup}

It is possible, and of course interesting, to study the scattering between non-perturbative objects in string theory. When dealing with D-branes, one can use an exact CFT description in terms of the open strings attached to them as we have already done in the string-brane analysis. The calculation by Polchinski~\cite{Polchinski:1995mt} for two static parallel D-branes was immediately generalised to the case of moving D-branes~\cite{Bachas:1995kx}, see~\cite{Billo:1997eg} for a derivation in the closed string channel by using the boundary state formalism briefly summarised in~\ref{app:boundarystate}. The case of D$0$-brane scattering is of particular interest because in this setup one can see the emergence of the 11-dimensional Planck length $\ell_{11}$ which is the fundamental scale of M-theory~\cite{Douglas:1996yp}. Here we will not provide a detailed analysis of the eikonal scattering for the brane-brane case, but will just highlight some similarities and differences with the more detailed analysis of the previous sections.

Technically the leading order eikonal for the D-brane scattering is captured by a world-sheet whose topology is an annulus. The boundary conditions are given by reflection matrices that generalise the one discussed in~\eqref{matS}, as they now have to encode the information about the velocities of the D-branes. At large distances, this diagram is best described as a long thin cylinder and is dominated by the exchange of massless closed strings. Focusing on the case of the scattering between two D$0$-brane bound states, the fields mediating such interaction are the dilaton, the Ramond-Ramond vector and of course the graviton. Since type IIA string theory can be viewed as M-theory compactified on a circle and D$0$-branes are just 11-dimensional supergravity fields with non trivial Kaluza--Klein numbers, it is not surprising that the scattering between D$0$-branes falls in the setup described in Section~\ref{sec:n8tree}. A difference with respect to the configuration summarised in \eqref{eq:KKmasses} (beside the fact that the starting point is now a 11-dimensional theory) is that the KK-modes lie in the same direction. As discussed in~\cite{Caron-Huot:2018ape,Parra-Martinez:2020dzs} this amounts to generalising the factor of $\sigma^2$ in the numerator of \eqref{eq:2d0N=8} to $(\sigma-\cos\phi)^2$ with $\phi=0$ for KK-modes in parallel directions (rather $\phi=\pi/2$ which is appropriate for the choice made in~\eqref{eq:KKmasses}). Thus for D$0$-brane scattering we have the following large distance eikonal phase
	\begin{equation}
		\label{eq:2d0gD}
		2\delta_0 = \frac{2 m_1m_2 G_d (\sigma-1)^2 \Gamma \left(\frac{d-4}{2}\right) }{\sqrt{\sigma^2-1} (\pi b^2)^{\frac{d-4}{2}}}\,.
	\end{equation}
In this case we need to set $d=10$ for type IIA theory and the mass of each D$0$-brane bound state is simply $m_i= n_i \tau_0$, where $n_i$ is the number of D$0$-branes constituents (identified with the KK-mode in the M-theory picture) and $\tau_0$ is the mass of a single D$0$-brane see~\eqref{eq:TpTen}. In the discussion below we focus for simplicity on the case $n_i \sim{\mathcal O}(1)$.

By lowering the impact parameter, we expect to see the inelastic channels typical of string theory. For instance at high energy ($s\!\gg \!m_i^2$), there should be a scale $b_t$ below which the internal degrees of freedom of each D$0$-brane bound state are excited and the elastic amplitudes is suppressed. This is the analogue of what was discussed in Sect.~\ref{ssec:seikop} for the case of elementary strings, but now the excitations are open strings starting and ending on the same D$0$-brane bound state. From the argument given in Eq.~\eqref{UB2} we expect that, in the case of D-brane scattering, the scale for these tidal excitations is $b^8_t\sim g_s G_{10} \alpha' s$. The extra factor of the string coupling $g_s$ in comparison to the string case is due to the tension of the D-branes which is larger by $1/g_s$ with respect to that of fundamental strings (equivalently the production of a D-brane excitation requires a closed-open string vertex and so it is suppressed by $g_s$). It would be interesting to check this with an explicit calculation and provide a precise description for these inelastic transitions in terms of an eikonal operator. To our knowledge, this has not been done yet in the literature. 

Instead, the string gravity regime depicted in Fig.\ref{fig:StringRegimes} was already discussed in the original paper~\cite{Bachas:1995kx} by looking at the imaginary part of the annulus partition function in the full string result. In this case the degrees of freedom responsible for this suppression of the elastic D-brane scattering are the open strings stretched between the two colliding D-brane bound states. Naively one would expect that this channel opens when the impact parameter is of the order of the string scale, but exactly as it happens in the string-brane (Sect.~\ref{ssec:string-brane}) and the string-string (Sect.~\ref{ssec:string-brane-sup}) cases, there is an enhancement factor scaling logarithmically with the center of mass energy and so again the relevant scale is $l_s(s)$ defined in~\eqref{eq:lscsv}. Thus in the ultrarelativistic regime string effects dominate short-distance brane-brane scattering.

 Finally, at low velocities the interaction between D$0$-branes is weaker as already suggested by~\eqref{eq:2d0gD} which vanishes as $(\sigma-1)^2\sim p_\infty^4$. In this regime the scattering between two D$0$-branes remains essentially elastic, although quantum mechanical, up to 11-dimensional Planck length~\cite{Douglas:1996yp}. It remains to be seen whether the process becomes (semi)classical down to $b = \ell_{11}$  if one considers the collision of two bound states with $n_i \gg 1$ D$0$-branes. 

\section{Exponentiation and the subleading 2--body eikonal}
\label{sec:oneloop}

In the previous two sections we have extracted the leading eikonal $\delta_0$ in various gravitational theories from the tree-level amplitude. Its exponentiation, as we have seen, is equivalent to the statement that, in the $(n-1)$-loop amplitude in impact-parameter space $\tilde{\mathcal A}_{n-1}$, the leading classical term equals $-\frac{i}{n!}(2i\delta_0)^n$. From the leading asymptotics of the resummed amplitude $1+i\tilde{\mathcal A} \simeq e^{2i\delta_0}$, one can then compute classical observables, like the deflection angle and the Shapiro time delay, for collisions with large impact parameter, $b\gg R$ with $R$ the typical size of the colliding objects. 

However, the result obtained in this way is only accurate to leading order in the small parameter $\frac{R}{b}$. How can we compute the corrections to this leading  behavior? Since this regime is the one characterized by weak gravitational interactions between the colliding objects, an equivalent question is: How do we retrieve the classical PM expansion of the deflection angle? The answer is that we have to calculate higher-order corrections in the eikonal phase $\delta=\delta_0+\delta_1+\cdots$ and, as we shall see, this can be done by looking at subleading terms in the classical expansion of loop diagrams. 
In this section, we perform this analysis for the one-loop amplitude $\tilde{\mathcal{A}}_1$. 

Indeed, by the exponentiation of $2i\delta_0$, the leading term in the classical limit of  $i\tilde{\mathcal A}_1$ must be given by $\frac{1}{2!}(2i\delta_0)^2$ and is thus proportional to $\frac{1}{\hbar^2}$, but the one-loop amplitude contains also an additional sub-leading term that is  again proportional to $\frac{1}{\hbar}$ as it happened for  the tree diagram itself.
The former term is sometimes referred to as ``super-classical'' or as an ``iteration'', because it is the most singular in the classical limit and it does not provide any new information compared to the tree-level amplitude. The latter identifies instead a new classical term and one can extract from it the sub-leading eikonal $2i\delta_1$.
It is  natural to conjecture that also this sub-leading eikonal $2i\delta_1$  exponentiates and this can be argued on general grounds \cite{Akhoury:2013yua,Bjerrum-Bohr:2018xdl,Fazio:2021aai}, although we will not discuss this proof in detail here.
As we will see, its contribution is indeed subdominant for large impact parameter with respect to the leading one, i.e.~it is  suppressed by one more power of $\frac{R}{b}$.

Of course, the one-loop amplitude also contains quantum terms that scale like $\hbar^n$ with $n \geq0$.  We will see that, although these extra terms do not contribute to the eikonal phase at one loop, they play an important role in the extraction of the sub-sub-leading parts of the  eikonal  at two loops. 

More concretely, since at each order of the perturbative expansion we find additional classical terms that ought to exponentiate contributing to higher sub-leading parts of the  eikonal and quantum terms that do not need to exponentiate,\footnote{One could exponentiate also the quantum part $1+2i\Delta = e^{2i\delta^\text{quant}}$ which amounts to a redefinition of the coefficients of the expansion~\eqref{eq:quant-term}. We prefer not to do so in order to define an object, $2\delta$, which directly yields the classical observables such as the deflection angle.} we are then led to the conclusion that, in the classical limit, the full amplitude in impact parameter space is encoded in the following expression:
\begin{equation}
	1+i {\tilde{\mathcal A}} (s, b) =\left[1+2i\Delta (s,b) \right]  {e}^{2i \delta}
	\label{fullampli}
\end{equation}
where, schematically, in $D=4$ 
\begin{equation}\label{}
	2\delta=  2\delta_0+2\delta_1+2\delta_2 +\cdots = \frac{GE^2}{\hbar} \left(
	\log b + \frac{G E }{b} + \left(\frac{G E}{b}\right)^2 + \cdots
	\right)
\end{equation}
is the classical eikonal and 
\begin{equation}\label{eq:quant-term}
	2\Delta = 2\Delta_1+2\Delta_2+\cdots = \left(\frac{GE}{b}\right)^2 \left[1+\frac{\hbar}{Eb}+\cdots \right] + \left(\frac{GE}{b}\right)^3 \left[1+\frac{\hbar}{Eb}+\cdots \right] 
\end{equation}
is the quantum remainder.
In the bottom-up approach we are going to adopt, we shall first calculate the perturbative amplitude order by order in the loop expansion $\mathcal A =\mathcal A_0 + \mathcal A_1 + \mathcal A_2 +\cdots$, concentrating on the limit of small momentum transfer, take the Fourier transform from momentum space to impact-parameter space, and then determine $\delta_n$ and $\Delta_n$ by solving
\begin{align}\label{eikexp0}
	i\tilde{\mathcal A}_0 &= 2i\delta_0\,,\\
	\label{eikexp1}
	i\tilde{\mathcal A}_1 &= \frac{1}{2!}(2i\delta_0)^2 + 2i\delta_1 + 2i\Delta_1\,,\\
	\label{eikexp2}
	i\tilde{\mathcal A}_2 &= \frac{1}{3!}(2i\delta_0)^3+ 2i\delta_0\,2i\delta_1 + \left[
	2i\delta_2+2i\delta_0\,2i\Delta_1
	\right]
	+2i\Delta_2\,.
\end{align}
These relations are predicted by the formal re-expansion of \eqref{fullampli} for small $G$ and define $2\delta_n$, $2\Delta_n$ at an operative level, while also dictating the structure of super-classical terms, which only arise, at each level, from iterations of lower-order terms.
In particular, the $\mathcal O(b^{-1})$ terms of $\tilde{\mathcal A}_1$ determine $2\delta_1$ and its  $\mathcal O(b^{-2})$ terms determine $2\Delta_1$.
Moreover, while $2\Delta_1$ is irrelevant for determining $2\delta_1$ at one-loop level, it is needed at two-loop level, together with $2\delta_0$, in order to solve \eqref{eikexp2} for the unknown $2\delta_2$, once the $\mathcal O(b^{-2})$ terms of $\tilde{\mathcal A}_2$ are computed.

Let us also mention that, while the leading eikonal is always real, sub-leading eikonals have, in general, also an imaginary part that is connected to the existence of inelastic channels. In this case, the classical deflection angle can be computed from the real part of the sub-leading eikonals, since the real part is the one contributing to classical phase oscillations, while the imaginary part only gives rise to an overall exponential suppression. This issue is going to be particularly acute in the two-loop eikonal that we will discuss in the next section, where the imaginary part actually contains long-range infrared divergences associated to soft graviton emissions.

In this section we restrict ourselves to the one-loop level, however, and analyze in detail the classical limit of the one-loop scattering amplitude $i\tilde{\mathcal A}_1$ in massive ${\cal{N}}=8$ supergravity and in GR. This gives us the opportunity to check that the leading super-classical term is indeed obtained from the quadratic term of the expansion of the leading eikonal ${\rm e}^{2i \delta_0}$, i.e.~that $i\tilde{\mathcal A}_1\sim \frac{1}{2!}\,(2i\delta_0)^2$ as in \eqref{eikexp1}. We will then extract, from the  next to the leading term, the sub-leading eikonal $2\delta_1$ and also, from the next to the next to the leading  term,  the quantum $2\Delta_1$ that is important in reproducing the two-loop amplitude according to Eq. \eqref{fullampli}. 

\subsection{1-loop (2PM) in QFT }
\label{oneloop}

This section is divided in two sub-sections. In the first one we study the case of massive ${\cal{N}}=8$ supergravity and in the second one the case of GR.
In the massive case under consideration it is convenient to use the relative Lorentz factor $\sigma$, already introduced in \eqref{eq:sigma}, and also to define a related variable $z$ which has the advantage of rationalizing some square roots, via
\begin{eqnarray}
	\sigma = - \frac{p_1p_2}{m_1m_2}= \frac{s-m_1^2-m_2^2}{2m_1m_2}= \frac{1}{2}\left(z+\frac{1}{z}\right)\,,
	\qquad
	z =\sigma - \sqrt{\sigma^2-1}\,,
	\label{sigmax}
\end{eqnarray}
where $s$ is the Mandelstam variable.
In particular $1\le \sigma <\infty$ and $0<z\le1$, with $\sigma=1$ or $z=1$ corresponding to the case of two particles mutually at rest, and $\sigma\to \infty$ or $z\to 0^+$ to the case of an ultrarelativistic collision.

\subsubsection{Massive \texorpdfstring{${\cal{N}}=8$}{N=8} supergravity }
\label{sec:n81loop}

\begin{figure}[h!]
	\centering
	\begin{tikzpicture}
		\draw[color=green!60!black,ultra thick] (-1.8,0)--(1.8,0);
		\draw[color=blue,ultra thick] (-1.8,2)--(1.8,2);
		\draw (-1,0)--(-1,2);
		\draw (1,0)--(1,2);
		\node at (-1.8,2)[left]{$p_1$};
		\node at (-1.8,0)[left]{$p_2$};
		\node at (1.8,0)[right]{$p_3$};
		\node at (1.8,2)[right]{$p_4$};
	\end{tikzpicture}
	\hspace{2cm}
	\begin{tikzpicture}
		\draw[color=green!60!black,ultra thick] (-1.8,0)--(1.8,0);
		\draw[color=blue,ultra thick] (-1.8,2)--(1.8,2);
		\draw (-1,0)--(1,2);
		\draw (-.05,1.05)--(-1,2);
		\draw (1,0)--(.05,.95);
		\node at (-1.8,2)[left]{$p_1$};
		\node at (-1.8,0)[left]{$p_2$};
		\node at (1.8,0)[right]{$p_3$};
		\node at (1.8,2)[right]{$p_4$};
	\end{tikzpicture}
	\caption{The box and crossed-box topologies.}
	\label{fig:BoxandCrossedBox}
\end{figure}
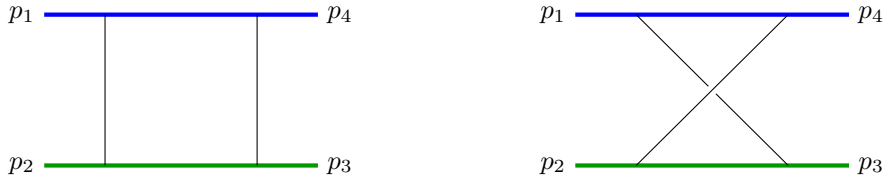

Let us start from the maximally supersymmetric case, introducing the masses via Kaluza--Klein compactification as discussed in Subsection~\ref{sec:n8tree}.
For the $s$-$u$ symmetric collision discussed there, corresponding to the elastic scattering of an axion and a dilaton, the one-loop amplitude in ${\cal{N}}=8$ supergravity with massive external states is given by \cite{Caron-Huot:2018ape,Parra-Martinez:2020dzs}
\begin{eqnarray}
	\label{1loopmass}
	{\mathcal{A}}_1= (8\pi G)^2   \frac{c(\epsilon)}{2} \left[ \left( s- m_1^2 -m_2^2\right)^4 +\left( u- m_1^2- m_2^2\right)^4 - t^4 \right] \left( I^{\phantom{}}_{\mathrm{II}} + I_{\overline{\mathrm{II}}} \right) 
\end{eqnarray}
where
\begin{equation}\label{cepsilon}
	c(\epsilon) = \frac{ e^{-\gamma_E \epsilon} }{(4\pi)^{2-\epsilon}}\,,
\end{equation}
while
$I^{\phantom{}}_{\mathrm{II}}$ stands for the box integral,
\begin{equation}\label{IsubII}
	I^{\phantom{}}_{\mathrm{II}}
	=
	\int_\ell
	\frac{1}{(-2p_1\ell+\ell^2-i0)(2p_2\ell+\ell^2-i0)(\ell^2-i0)((\ell-q)^2-i0)}\,,
\end{equation} 
with the shorthand notation
\begin{equation}\label{intsubell}
	\int_\ell= e^{\gamma_E\epsilon}\int \frac{d^{4-2\epsilon}\ell}{i\pi^{2-\epsilon}}\,,
\end{equation}
and $I_{\overline{\mathrm{II}}}$ stands for the crossed box integral, obtained from $I^{\phantom{}}_{\mathrm{II}}$ replacing $p_1$ with $p_4$. 
A schematic representation of these topologies is given in Fig.~\ref{fig:BoxandCrossedBox}.
Let us also recall that $q=p_1+p_4$. 

We then face the task of evaluating the integrals $I^{\phantom{}}_{\mathrm{II}}$,  $I_{\overline{\mathrm{II}}}$. A key simplification in this respect is that we do not need their exact expression, but, since the typical perturbative momentum exchange is $q\sim \frac{\hbar}{b}$ and we focus on the regime of large impact parameters, we can restrict our attention to the limit 
\begin{equation}\label{}
	s,\, m_1^2,\, m_2^2 \gg |t|=q^2
\end{equation}
characterizing the classical PM regime.  
More precisely, since we are interested in the long-range terms in $b$-space after Fourier transform, we can concentrate on non-analytic terms in the small-$q^2$ expansion.
A systematic strategy that is well suited to this type of calculations is the method of regions \cite{Beneke:1997zp,Smirnov:2002pj}. In this approach, the guiding principle is that the asymptotic expansion of the desired integrals as $q\to0$ is captured by Taylor-expanding the \emph{integrand} with respect to all possible scalings of the loop momentum $\ell$ that give rise to non-scaleless integrals in dimensional regularization. In our case, the nontrivial scaling choices are $\ell\sim \mathcal O(q^0)$, which defines the hard region, and $\ell \sim \mathcal O(q)$, which defines to the soft region.
The hard region is effectively a power series in $q^2$ and thus only gives rise to strictly localized contributions in $b$-space. Conversely, the soft region is the one responsible for the non-analytic terms we are after, so we can safely discard the former and concentrate on the latter.

Although we refrain here from exhibiting the full soft-region calculation of $I^{\phantom{}}_{\mathrm{II}}$, which was performed in detail for instance in \cite{KoemansCollado:2019ggb,Cristofoli:2020uzm,Parra-Martinez:2020dzs}, let us show how the leading soft term in $I^{\phantom{}}_{\mathrm{II}} + I_{\overline{\mathrm{II}}}$ can be retrieved, following \cite{Bjerrum-Bohr:2021din}.
Starting from \eqref{IsubII}, the leading soft term in $I^{\phantom{}}_{\mathrm{II}}$ reads
\begin{equation}\label{IsubIIleading}
	I^{\phantom{}}_{\mathrm{II}}
	\simeq 
	\int_\ell \frac{1}{(-2p_1\ell-i0)(2p_2\ell-i0)(\ell^2-i0)((\ell-q)^2-i0)}\,,
\end{equation}
where we used the scaling condition $\ell\sim q$, while $p_1$, $p_2$ are $q$-independent to leading order. Sending $\ell\to q-\ell$, recalling $q=p_1+p_4=-p_2-p_3$ and using $2p_1q = q^2$, $2p_2q= -q^2$, eq.~\eqref{IsubIIleading} can be rewritten to leading order in the form
\begin{equation}\label{}
	\begin{split}
		I^{\phantom{}}_{\mathrm{II}}
		&\simeq 
		\frac12
		\int_\ell \frac{1}{(\ell^2-i0)((\ell-q)^2-i0)}
		\\
		&
		\times
		\left[
		\frac{1}{(-2p_1\ell-i0)(2p_2\ell-i0)}
		+
		\frac{1}{(2p_1\ell-i0)(-2p_2\ell-i0)}	
		\right].
	\end{split}
\end{equation}
Adding the crossing symmetry partner $I_{\overline{\mathrm{II}}}$ and applying very similar manipulations, one then finds
\begin{equation}\label{}
	\begin{split}
		I^{\phantom{}}_{\mathrm{II}}
		+
		I_{\overline{\mathrm{II}}}
		&\simeq 
		\frac12
		\int_\ell \frac{1}{(\ell^2-i0)((\ell-q)^2-i0)}
		\\
		&
		\times
		\left[
		\frac{1}{2p_1\ell-i0}
		+
		\frac{1}{-2p_1\ell-i0}	
		\right] \left[
		\frac{1}{2p_2\ell-i0}
		+
		\frac{1}{-2p_2\ell-i0}	
		\right].
	\end{split}
\end{equation}
This localizes the integration on a $(D-2)$-dimensional subspace
\begin{equation}\label{}
	I^{\phantom{}}_{\mathrm{II}}
	+
	I_{\overline{\mathrm{II}}}
	\simeq 
	-2\pi^2
	\int_\ell \frac{\delta(2p_1\ell) \delta(2p_2\ell)}{(\ell^2-i0)((\ell-q)^2-i0)}\,.
\end{equation}
Choosing $p_1=(-E_1,0,\ldots,0,-p)$, $p_2 = (-E_2,0,\ldots,0,p)$, $q = (0,q_\perp,0)$ one can use the delta functions to perform the integrals with respect to $\ell^0$ and $\ell_\parallel$ in $\ell= (\ell^0,\ell_\perp,\ell_\parallel)$, obtaining
\begin{equation}\label{}
	I^{\phantom{}}_{\mathrm{II}}
	+
	I_{\overline{\mathrm{II}}}
	\simeq
	e^{\gamma_E \epsilon} 
	\,
	\frac{i\pi^{\epsilon}}{2pE} 
	\int \frac{d^{2-2\epsilon}\ell_\perp}{\ell_\perp^2(\ell_\perp-q_\perp)^2}\,,
\end{equation}
with $E = E_1+ E_2$. The last integral is straightforward to perform using Schwinger or Feynman parameters, and one gets
\begin{equation}\label{}
	I^{\phantom{}}_{\mathrm{II}}
	+
	I_{\overline{\mathrm{II}}}
	\simeq
	e^{\gamma_E \epsilon} 
	\,
	\frac{i\pi}{2pE} \frac{\Gamma(1+\epsilon)}{(q^2)^{1+\epsilon}}\frac{\Gamma(-\epsilon)^2}{\Gamma(-2\epsilon)}
	\,.
\end{equation}
When substituted into \eqref{1loopmass}, recalling that $p E = m_1m_2\sqrt{\sigma^2-1}$ as in \eqref{Ep},
we thus obtain the leading contribution
\begin{equation}\label{}
	\frac{ i \mathcal{A}_1 (s,  q^2)}{4pE} \simeq \frac{(8 \pi G )^2}{(4\pi)^2}
	\left( \frac{4\pi}{q^2}\right)^{\epsilon} \frac{ (-2   \pi m_1^2 m_2^2)}{q^2} \frac{\sigma^4}{ \sigma^2-1} \frac{\Gamma (1+\epsilon) \Gamma^2 (-\epsilon)}{\Gamma (-2\epsilon)}\,.
\end{equation}
Sub-leading and sub-sub-leading terms in $q$ are obtained by retaining higher-order contributions in the soft-region expansion. This can be done  via direct integration in the soft region as exemplified here \cite{KoemansCollado:2019ggb,Cristofoli:2020uzm,Bjerrum-Bohr:2021vuf,Bjerrum-Bohr:2021din} or, more systematically, using Integration-By-Parts identities, reduction to master integrals and soft differential equations adapted to the soft region \cite{Parra-Martinez:2020dzs,DiVecchia:2021bdo}.

The resulting expression, complete to sub-sub-leading order, reads
\begin{align}
	\label{A1sugra}
	& \frac{ i \mathcal{A}_1 (s,  q^2)}{4pE} = 4G^2
	\left( \frac{4\pi}{q^2}\right)^{\epsilon} \Bigg\{  \frac{ -2   \pi m_1^2 m_2^2}{q^2} \frac{\sigma^4}{ \sigma^2-1} \frac{\Gamma (1+\epsilon) \Gamma^2 (-\epsilon)}{\Gamma (-2\epsilon)} \\
	\nonumber
	+&\frac{2 i \sqrt{\pi} m_1 m_2 (m_1+m_2)}{ \sqrt{q^2}} \frac{\sigma^4}{(\sigma^2-1)^{\frac{3}{2}}} \frac{\Gamma ( \epsilon+ \frac{1}{2}) \Gamma^2 ( \frac{1}{2}-\epsilon)}{\Gamma (-2\epsilon)} -   \frac{i \sigma^3}{(\sigma^2-1)^{2}} \frac{\Gamma^2 (-\epsilon) \Gamma (1+\epsilon)}{\Gamma(-2\epsilon)} \\
	\nonumber
	\times \! & \Bigg[ m_1 m_2\! \left[\!(1+2\epsilon)\!\left(\sigma^2 \log z \! + \! \sigma \sqrt{\sigma^2-1} \right) \! + \!2 i \pi (\sigma^2-1) \right] \!  + \! \frac{i\pi \epsilon}{2}  s\,\sigma \Bigg] \Bigg\} .
\end{align}
Let us now introduce a notation to distinguish leading, sub-leading and sub-sub-leading terms in the $q$-expansion of $\mathcal A_1(s,q^2)$ according to \begin{equation}\label{}
	\mathcal A_1 
	= \mathcal A_1^{[2]} 
	+ \mathcal A_1^{[1]}
	+\mathcal A_1^{[0]}+\cdots\,,
\end{equation}
in such a way that $\mathcal A_1^{[k]}\sim \mathcal O(q^{-k-2\epsilon})$.
The $\mathcal{A}_1^{[2]}$ term, corresponding to the first line of Eq.~\eqref{A1sugra}, is the super-classical term. Its Fourier transform in impact parameter space (see Eq. \eqref{B1}) indeed reproduces the quadratic iteration term of the expansion of the leading-order eikonal,
\begin{equation}
	i{\tilde{\mathcal A}}_1^{[2]} (s, b)= \frac{1}{2} \left( \frac{2 i m_1m_2 G (\pi b^2)^{\epsilon}\sigma^2 \Gamma (-\epsilon) }{ \sqrt{\sigma^2-1}}\right)^2= \frac{1}{2} (2 i \delta_0)^2\,\,,
	\label{A11}
\end{equation}
in agreement with the leading-order term in \eqref{eikexp1}.
The Fourier transform of $\mathcal{A}_1^{[1]}$, which appears in the second line of Eq.~\eqref{A1sugra} and is sub-leading for small $q$, gives instead the sub-leading eikonal according to
\begin{equation}\label{}
	2 i \delta_1
	=
	i {\tilde{\mathcal A}}_1^{[1]}
\end{equation}
so that\footnote{Let us point out a typo in the expression for $2\delta_1$ in Ref.~\cite{DiVecchia:2021bdo}, which should be multiplied by $1/2$.}
\begin{equation}
	\begin{split}\label{A12}
		2 \delta_1
		&= \frac{4(\pi b^2)^{2\epsilon} G^2 m_1m_2 (m_1+m_2)}{\sqrt{\pi b^2}}
		\frac{\sigma^4}{(\sigma^2-1)^{\frac{3}{2}}} \frac{\Gamma ( \frac{1}{2} - 2\epsilon) \Gamma^2 ( \frac{1}{2} -\epsilon)}{ \Gamma (-2 \epsilon)}\,.
	\end{split}
\end{equation}
Note that this next-to-leading eikonal, arising from box integrals, represents the first correction to \eqref{eq:2d0N=8} and is proportional to the sum of the masses.
Taking the by now familiar derivative with respect to $b$, this translates into the following expression for the deflection angle, 
\begin{equation}\label{theta2PMN8}
	\Theta= \frac{4GE (\pi b^2)^{\epsilon}\sigma^2 \Gamma(1-\epsilon)}{(\sigma^2-1)b}
	+
	\frac{4G^2 E (\pi b^2)^{2\epsilon} \sigma^4 (m_1+m_2)}{(\sigma^2-1)^{2}\sqrt{\pi}\,b^2}
	\frac{\Gamma ( \frac{1}{2} - 2\epsilon) \Gamma^2 ( \frac{1}{2} -\epsilon)}{ \Gamma (-2 \epsilon)}
	+
	\mathcal O(G^3)\,.
\end{equation}
Consistently with  the above considerations, the $\mathcal O(G^2)$ correction vanishes identically for the collision of two (massless) shockwaves. Moreover, even for massive objects, it happens to be zero in $D=4$ \cite{Caron-Huot:2018ape}, while it does provide a nontrivial $\mathcal O(\frac{R^2}{b^2})$ correction to the deflection angle in higher-dimensional spacetimes.

Finally, going to impact parameter with sub-sub-leading term $\mathcal A_1^{[0]}$ one gets the  first contribution to the quantum   remainder,
\begin{equation}\label{}
	2i\Delta_1= i\tilde{\mathcal{A}}_1^{[0]}\,,
\end{equation}
finding a real part given by
\begin{equation}
	\begin{split}\label{1L7}
		\operatorname{Re} 2\Delta_1 
		& =\frac{8 G^2 m_1 m_2 (\pi b^2)^{2\epsilon}}{\pi b^2}
		\frac{\sigma^4  \left(\sigma \log z +  \sqrt{\sigma^2-1} \right) }{(\sigma^2-1)^2}  
		(1+2\epsilon) \Gamma^2 (1-\epsilon)\, ,
	\end{split}
\end{equation}
and an imaginary part given by
\begin{equation}
	\begin{split}
		\label{1L71}
		\operatorname{Im} 2\Delta_1 
		& = \frac{8 G^2\sigma^3 (\pi b^2)^{2\epsilon} \Gamma^2 (1-\epsilon)}{ b^2 (\sigma^2-1)^2 } 
		\left[ \frac{ \epsilon}{2} s \sigma +  2m_1 m_2 (\sigma^2-1) \right]. \end{split}
\end{equation}

In this section we focused on the elastic amplitude~\eqref{1loopmass}. Looking more in general at the $2\to 2$ scattering of massless states in ${\cal N}=8$ supergravity, we remark that there can be \emph{inelastic} contributions at the level of $\mathcal{A}_1^{[1]}$. However this does not imply that the 2PM eikonal becomes an operator, because there is an $\mathcal O(G)$ contribution to $\Delta$ that encodes the transition between different massless states (so this $\Delta_0$ is nontrivial, in this more general setup, and becomes an operator). By using the natural generalization of~\eqref{fullampli}, one can verify that the  amplitudes in the massless sector are consistent with the 2PM eikonal~\eqref{A12}, see~\cite{Collado:2018isu} for an explicit check of this point in the probe limit where massless states scatter off a stack of D$p$-branes.

\subsubsection{Real-analytic, crossing-symmetric reformulation }
\label{analyticN81loop}

In the previous subsection we have collected the contribution of the different one-loop diagrams noticing that they combine into much simpler expressions than those of individual diagrams. They are in the form of an expansion in powers of $q^2$ i.e.~of the quantum level at which they contribute.
Furthermore, as  computer outputs, they simply collect independently real and imaginary terms and express the result in terms of a choice of the two independent Mandelstam variables, $q^2 = -t$ and $s$ (or of quantities like $\sigma$ and $z$, themselves functions of $s$ and the masses).

On the other hand, we expect the full amplitude to satisfy two exact properties:
\begin{itemize}
	\item Real analyticity i.e.~$\mathcal{A} (s^*, q^2) = (\mathcal{A} (s, q^2))^*$;
	\item Crossing symmetry  i.e.~$\mathcal{A} (s, q^2) = \mathcal{A} (u \equiv -s +q^2 + 2m_1^2 + 2m_2^2, q^2)$.
\end{itemize}
Those two properties, which should hold order by order in $q^2$,  are not at all apparent in the formulae of the previous subsection, but they must be hidden somewhere, of course up to the order in $q^2$ at which we stop our expansion. Here we will explicitly show how to rewrite  the tree and one-loop  amplitudes, given in the previous section
in a real-analytic and crossing-symmetric form. Besides its usefulness for the interpretation of the result this also serves as a rather stringent test of the calculations themselves. A similar analysis will be carried out at the two-loop level in Section~\ref{sec:twoloop} where analyticity and crossing will turn out to be very useful for actually fixing the entire two-loop amplitude from its imaginary part. 
Let us mention in passing that analyticity and crossing also played a key role in recent developments \cite{Caron-Huot:2023vxl,Caron-Huot:2023ikn}.

To this purpose it is useful to introduce variables which are related to those introduced earlier in Section~\ref{sec:kinematics} by $s \leftrightarrow u$ exchange:
\begin{equation}
	\begin{split}
		& \bar{\sigma} =  \frac{u - m_1^2-m_2^2}{2m_1m_2}  = -\left(\sigma - \frac{q^2}{2 m_1m_2}\right),    \\
		& \bar{z} = \bar{\sigma} - \sqrt{\bar{\sigma}^2-1} = - \frac{1}{z}\left( 1 - \frac{q^2/(2m_1m_2)}{\sqrt{\sigma^2-1}}\right) + O(q^4)\,,  \\
		& 2 \bar{\sigma}  = \left(\bar{z} + \frac{1}{\bar{z}}\right),\qquad  2 \sqrt{\bar{\sigma}^2-1} = \left(\frac{1}{\bar{z}}- \bar{z}\right).
		\label{defs}
	\end{split}
\end{equation}
We also have to bear in mind that the factor $\frac{1}{4 p E}$ appearing in some formulae cannot be ignored when discussing analyticity and crossing  symmetry of the amplitude.

The tree-level amplitude \eqref{atrees8afterKK} is already manifestly crossing symmetric and real analytic,
\begin{equation}\label{}
	\mathcal A_0 = -\frac{\pi G}{q^2} \frac{16 m_1^4 m_2^4 (\sigma^4 + \bar\sigma^4)-q^4}{m_1^2 m_2^2 \sigma  \bar\sigma}\,.
\end{equation}
Up to terms that lack the pole at $q^2=0$, we may equivalently write
\begin{equation}
	\mathcal{A}_0 (s, q^2) = \frac{16\pi Gm_1^2m_2^2}{q^2} \left(\sigma^2 + \bar{\sigma}^2\right).
	\label{T1sym}
\end{equation}
At one-loop, equation \eqref{A1sugra} suggests trying the following analytic, crossing-symmetric ansatz for the super classical term:
\begin{equation}
	\begin{split}
	\mathcal{A}_1^{\text{scl.}} (s , q^2) 
	=  \frac{16 G^2 m_1^3 m_2^3}{q^2} \left( \frac{4\pi}{q^2}\right)^{\epsilon} 
	\frac{\Gamma^2 (-\epsilon) \Gamma (1+\epsilon)}{\Gamma (-2\epsilon)} \left[ (\sigma^4 + \bar{\sigma}^4 ) \left(  \frac{\log(-z)}{\sqrt{\sigma^2-1}} + \frac{\log(-\bar{z})}{\sqrt{\bar{\sigma}^2-1}}\right) \right]\,.
	\label{Ansatz1scl}
	\end{split}
\end{equation} 
It is easy to check that, when expanded in $q^2$ using \eqref{defs} and the branch choice $\log (-z) = \log z+i\pi$,  \eqref{Ansatz1scl} reproduces both the super-classical (first line in \eqref{A1sugra}) and the $O(\epsilon^0 m_1 m_2)$ terms in the square bracket of \eqref{A1sugra}.
	
	The remaining terms  are as follows:
	\begin{itemize}
		\item A classical contribution proportional to $(q^2)^{-1/2}$ (second line in \eqref{A1sugra}). It is real and can be trivially symmetrized in $s-u$ since the error in doing so is of order $(q^2)^{1/2}$;
		\item A quantum contribution of $O(\epsilon)$ that can be written in the crossing-symmetric form:
		\begin{align}
			\mathcal{A}_1^{\text{qu.}} (s , q^2) 
			&=  
			16 G^2 m_1^2 m_2^2 \left( \frac{4\pi}{q^2}\right)^{\epsilon} 
			\frac{\Gamma^2 (-\epsilon) \Gamma (1+\epsilon)}{\Gamma (-2\epsilon)} \hat{\mathcal{A}}_1^{\text{qu.}} (s , q^2) \,,
			\\
			\begin{split}
			\hat{\mathcal{A}}_1^{\text{qu.}} (s , q^2) 
			&= 
			- \epsilon \left[ \left(  \frac{\sigma^5 \log(-z)}{(\sigma^2-1)^{3/2}} + \frac{\bar{\sigma}^5 \log(-\bar{z})}{(\bar{\sigma}^2-1)^{3/2}}\right) +\left(\frac{\sigma^4 }{\sigma^2-1} + \frac{\bar{\sigma}^4}{\bar{\sigma}^2-1}   \right) \right]  
			\\
			&  
			-\epsilon\, \frac{(m_1^2 + m_2^2)}{2 m_1 m_2} \left( \frac{\sigma^4 \log(-z)}{(\sigma^2-1)^{3/2}} + \frac{\bar{\sigma}^4\log(-\bar{z})}{(\bar{\sigma}^2-1)^{3/2}}   \right).
			\label{A1q2}
			\end{split}
		\end{align}   
	\end{itemize}
	Note that in  \eqref{A1q2} we can again neglect the difference between $\bar{\sigma}$ and $-\sigma$ as well as the difference between $\bar{z}$ and $- 1/z$. As a result, the last line of  \eqref{A1q2}  gives a purely imaginary contribution in the physical region.
	
	In the above equations crossing symmetry is manifest. Concerning real analyticity, it can be checked by using, in particular, the analytic properties of the combination $\log(-z)(\sigma^2-1)^{-1/2}$. We will come back to the checks of real analyticity after we include the two-loop results in Section~\ref{sec:twoloop}.
	
	\subsubsection{General Relativity}

	Let us now turn to the collision of two massive scalars minimally coupled to GR. 
	The logic we will apply is the same as for the $\mathcal N=8$ example discussed above. That is, we want to calculate the one-loop amplitude for the collision of two massive scalars minimally coupled to gravity, focusing on the non-analytic terms in the limit of small momentum transfer $q^2$.
	
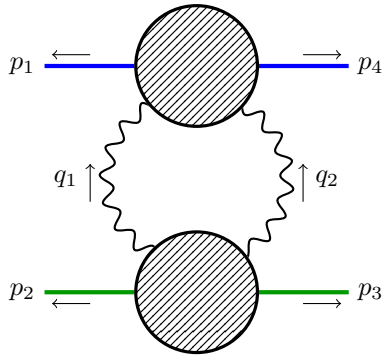
\begin{figure}[h!]
		\centering
		\begin{tikzpicture}
			\draw[<-] (-1.9,-.15)--(-1.4,-.15);
			\draw[<-] (1.9,-.15)--(1.4,-.15);
			\draw[<-] (-1.9,3.15)--(-1.4,3.15);
			\draw[<-] (1.9,3.15)--(1.4,3.15);
			\draw[<-] (-1.4,1.7)--(-1.4,1.2);
			\draw[<-] (1.4,1.7)--(1.4,1.2);
			\draw[color=green!60!black,ultra thick] (-2,0)--(2,0);
			\draw[color=blue,ultra thick] (-2,3)--(2,3);
			\draw[thick, style=decorate, decoration=snake] (0,.4) arc (-90:90:1.1);
			\draw[thick, style=decorate, decoration=snake] (0,.4) arc (270:90:1.1);
			\filldraw[color=white, fill=white, very thick](0,3) circle (.8);
			\filldraw[pattern=north east lines, very thick](0,3) circle (.8);
			\filldraw[color=white, fill=white, very thick](0,0) circle (.8);
			\filldraw[pattern=north east lines, very thick](0,0) circle (.8);
			\node at (-2,3)[left]{$p_1$};
			\node at (-2,0)[left]{$p_2$};
			\node at (2,0)[right]{$p_3$};
			\node at (2,3)[right]{$p_4$};
			\node at (-2,1.5)[right]{$q_1$};
			\node at (2,1.5)[left]{$q_2$};
		\end{tikzpicture}
		\caption{The two-graviton cut needed to obtain the one-loop integrand for the  amplitude in the classical limit. Each blob represents a tree-level amplitude with two scalars and two gravitons.}
		\label{fig:2gravitons}
	\end{figure}

	Of course, in the absence of supersymmetry, the amplitude integrand is not as simple as \eqref{1loopmass}, and deriving it from Feynman diagram techniques would pose a nontrivial challenge. Following  \cite{KoemansCollado:2019ggb,Bjerrum-Bohr:2021vuf}, a powerful integrand construction technique that we can apply to overcome this problem is the method of generalized unitarity. The one loop integrand we are after is a rational function, and its residues at the poles corresponding to the on-shell limits for certain internal lines are nothing but (sums of products of) tree-level amplitudes obtained by ``cutting'' such lines.
	A further simplification arises from the observation that, in order to capture all non-analytic terms associated to the long-range eikonal dynamics, we do not need to determine the full integrand, but rather only the contributions coming from the two-graviton cut depicted in Fig.~\ref{fig:2gravitons}.
	Indeed, we can neglect any topology involving contact interactions between the two massive lines, which would correspond to strictly localized effects. 
	
	We shall disregard topologies associated to vertex corrections and self-energy diagrams, which give rise to integrals that always vanish in the soft region.
	This may seem in principle not justified, because, although arising from the hard region only, these dressings of the one-graviton exchange naively appear in the amplitude with a kinematic prefactor $1/q^2$ (non-analytic).
	The key observation is that they also give rise to divergent terms that ought to be treated by appropriate coupling and wave-function renormalization.
	However one can check that the IR-divergences due to the two-graviton cut in Figure~\ref{fig:2gravitons} in fact exhaust the full IR-divergence of the one-loop amplitude predicted by the general exponential pattern \cite{Weinberg:1965nx,Heissenberg:2021tzo}.
	Therefore, all infrared divergences arising from vertex corrections and self-energy diagrams must cancel against one another, and we need not worry about them.
	The cancellation of the corresponding finite terms then ought to follow from the Ward identity linking the renormalization constants for charge and wavefunction, as discussed  in Ref.~\cite{Elkhidir:2023dco} for the case of electrodynamics, although we will not analyze it in detail here.

	Let us then turn to the evaluation of the two-graviton cut.
	The amplitude for each blob in Fig.~\ref{fig:2gravitons} is the one already discussed in Subsection~\ref{sec:gravsctree},
	\begin{equation}\label{}
		\mathcal A_{\rho\sigma,\alpha\beta}(k_1,k_2,q_1,q_2)
		=
		\begin{gathered}
			\begin{tikzpicture}
				\draw[<-] (-1.9,-.15)--(-1.4,-.15);
				\draw[<-] (1.9,-.15)--(1.4,-.15);
				\draw[<-] (-1.1,1.4)--(-.7,1);
				\draw[<-] (1.1,1.4)--(.7,1);
				\draw[ultra thick] (-2,0)--(2,0);
				\draw[thick, style=decorate, decoration=snake] (0,0) -- (1.3,1.3);
				\draw[thick, style=decorate, decoration=snake] (0,0) -- (-1.3,1.3);
				\filldraw[color=white, fill=white, very thick](0,0) circle (.8);
				\filldraw[pattern=north east lines, very thick](0,0) circle (.8);
				\node at (-2,0)[left]{$k_1$};
				\node at (2,0)[right]{$k_2$};
				\node at (-1.7,1.3){$\rho\sigma$};
				\node at (-.55,1.3){$q_1$};
				\node at (1.7,1.3){$\alpha\beta$};
				\node at (.55,1.3){$q_2$};
			\end{tikzpicture}
		\end{gathered}
	\end{equation} 
	can be taken as follows, using momentum conservation to eliminate $k_2$, \cite{KoemansCollado:2019ggb}
	\begin{equation}\label{Arsab}
		\begin{split}
&\mathcal A_{\rho\sigma,\alpha\beta}
= 
2\kappa^2\,
\frac{q_1\cdot(k_1+q_2)\, q_1\cdot k_1}{q_1\cdot q_2}\\
\times 
&\left[\frac{(k_1+q_2)^\rho k_1^\alpha}{q_1\cdot (k_1+q_2)}-\frac{(k_1+q_1)^\alpha k_1^\rho}{k_1\cdot q_1}+\eta^{\rho\alpha}\right]
\left[\frac{(k_1+q_2)^\sigma k_1^\beta}{q_1\cdot (k_1+q_2)}-\frac{(k_1+q_1)^\beta k_1^\sigma}{k_1\cdot q_1}+\eta^{\sigma\beta}\right].
		\end{split}
	\end{equation}
	This expression is of course highly non-unique, because one can always perform gauge transformations, i.e.~shifts proportional to $q_1^\rho\xi^\sigma+q_1^\sigma\xi^\rho$ or to $q_2^\alpha\zeta^\beta+q_2^\beta\zeta^\alpha$ obtaining an equivalent amplitude. The advantage of the specific form \eqref{Arsab} is that it is exactly transverse with respect to each graviton momentum~\cite{KoemansCollado:2019ggb,Kosmopoulos:2020pcd},
	\begin{equation}\label{}
		q_1^\rho \mathcal A_{\rho\sigma,\alpha\beta}=q_1^\sigma \mathcal A_{\rho\sigma,\alpha\beta}=0\,,\qquad
		q_2^\alpha \mathcal A_{\rho\sigma,\alpha\beta}=q_2^\beta \mathcal A_{\rho\sigma,\alpha\beta}=0\,,
	\end{equation}
	as can be checked using the mass-shell conditions. 
	This simplifies the sum over intermediate graviton polarizations involved in the cut in Fig.~\ref{fig:2gravitons}, and one can cast the amplitude of interest in the form
	\begin{equation}\label{2gc}
		\begin{split}
			&i\mathcal A_1(s,q) = \frac12\int\frac{d^Dq_1}{(2\pi)^D}\frac{d^Dq_2}{(2\pi)^D} (2\pi)^D\delta^{(D)}(q-q_1-q_2)\\
			&\times\mathcal A_{\rho\sigma,\alpha\beta}(p_1,p_4,-q_1,-q_2)
			G^{\rho\sigma,\rho'\sigma'}(q_1)
			G^{\alpha\beta,\alpha'\beta'}(q_2)
			\mathcal A_{\rho'\sigma',\alpha'\beta'}(p_2,p_3,q_1,q_2)\,,
		\end{split}
	\end{equation}
	where $G^{\mu\nu,\rho\sigma}$ denotes the De Donder propagator \eqref{GGagrav},
	\begin{equation}\label{}
		G^{\mu\nu,\rho\sigma} (\ell) = \frac{-i}{2(\ell^2-i0)}\left(
		\eta^{\mu\rho}\eta^{\nu\sigma}
		+
		\eta^{\mu\sigma}\eta^{\nu\rho}
		-
		\frac{2}{D-2}
		\eta^{\mu\nu}\eta^{\rho\sigma}
		\right).
	\end{equation}
	The overall factor of $\tfrac12$ in \eqref{2gc} is due to the reflection symmetry of Fig.~\ref{fig:2gravitons} about the vertical axis, or equivalently to the two possible choices of labeling for the loop momentum, either $q_{1} = \ell$, $q_{2}=\ell-q$ or $q_2=\ell$, $q_1=q-\ell$. 
	
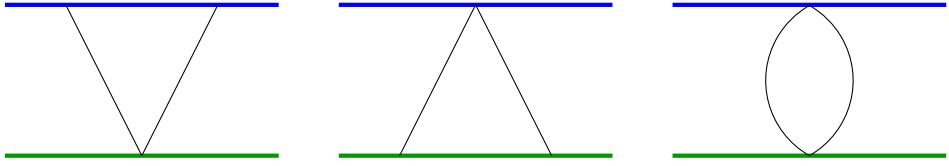
\begin{figure}[h!]
		\centering
		\begin{tikzpicture}
			\draw[color=green!60!black,ultra thick] (-1.8,0)--(1.8,0);
			\draw[color=blue,ultra thick] (-1.8,2)--(1.8,2);
			\draw (-1,2)--(0,0);
			\draw (1,2)--(0,0);
		\end{tikzpicture}
		\hspace{.5cm}
		\begin{tikzpicture}
			\draw[color=green!60!black,ultra thick] (-1.8,0)--(1.8,0);
			\draw[color=blue,ultra thick] (-1.8,2)--(1.8,2);
			\draw (-1,0)--(0,2);
			\draw (1,0)--(0,2);
		\end{tikzpicture}
		\hspace{.5cm}
		\begin{tikzpicture}
			\draw[color=green!60!black,ultra thick] (-1.8,0)--(1.8,0);
			\draw[color=blue,ultra thick] (-1.8,2)--(1.8,2);
			\draw (0,0) arc (-60:60:1.15);
			\draw (0,0) arc (240:120:1.15);
		\end{tikzpicture}
		\caption{Triangle and bubble topologies.}
		\label{fig:TrianglesBubble}
	\end{figure}

	The resulting integrand involves not only box and crossed box topologies (Fig.~\ref{fig:BoxandCrossedBox}), as its $\mathcal N=8$ counterpart, but also triangle and bubble topologies (Fig.~\ref{fig:TrianglesBubble}). Moreover, while the numerators of box and crossed box are constant in the loop momentum, the remaining topologies involve nontrivial numerators. As we already mentioned above, the task of performing such integrals is considerably simplified by employing integration-by-parts identities after expansion in the soft region.
	Using these techniques, one can reduce the calculation to a simple set of three master integrals \cite{Parra-Martinez:2020dzs}, the leading-order box and triangle, plus a trivial scalar bubble integral.
	
	For illustrative purposes, let us consider a scalar triangle integral
	(let us recall the shorthand notation $\int_\ell$ introduced in \eqref{intsubell})
	\begin{equation}\label{}
		I_{\Delta} 
		=
		\int_\ell \frac{1}{(-2p_1\ell+\ell^2-i0)(\ell^2-i0)((\ell-q)^2-i0)}\,.
	\end{equation}
	Again focusing on the soft region, where $\ell\sim\mathcal O(q)$, and on the leading-order contributions, we can massage this integral as we did for the leading box, obtaining
	\begin{equation}\label{}
		I_{\Delta} 
		\simeq
		\frac{1}{2}
		\int_\ell 
		\left[\frac{1}{2p_1\ell-i0}+\frac{1}{-2p_1\ell-i0}\right]
		\frac{1}{(\ell^2-i0)((\ell-q)^2-i0)}
	\end{equation}
	so that
	\begin{equation}\label{}
		I_\Delta \simeq  i\pi \int_\ell 
		\frac{\delta(2p_1\ell)}{(\ell^2-i0)((\ell-q)^2-i0)}\,.
	\end{equation}
	To evaluate this integral it is convenient to pick a reference frame where 
	\begin{equation}\label{}
		p_1 = \left(-\sqrt{m_1^2+\tfrac14|\vec{q}\,|^2},\tfrac12\,\vec q\,\right)\,,\qquad
		p_4 = \left(\sqrt{m_1^2+\tfrac14|\vec{q}\,|^2},\tfrac12\,\vec q\,\right)
	\end{equation}
	so that $q = (0,\vec q\,)$ and, to leading order for small $q$,
	\begin{equation}\label{}
		I_\Delta \simeq  e^{\gamma_E\epsilon}\,\frac{\pi^{-1+\epsilon}}{2m_1} \int 
		\frac{d^{3-2\epsilon}\vec \ell}{|\vec\ell\,|^2|\vec\ell-\vec q\,|^2}
		=
		e^{\gamma_E\epsilon}\,\frac{\sqrt{\pi}}{2m_1} 
		\frac{\Gamma\left(\tfrac12+\epsilon\right)}{(q^2)^{\tfrac12+\epsilon}}
		\frac{\Gamma\left(\tfrac12-\epsilon\right)^2}{\Gamma\left(1-2\epsilon\right)}\,.
	\end{equation}
	Combining this result with the box and bubble integrals, one can obtain an expression for the amplitude complete up to sub-sub-leading order in the small-$q$ expansion, which we now illustrate. For simplicity, let us again break down the result as follows
	\begin{equation}\label{}
		\mathcal A_1 = \mathcal A_1^{[2]} + \mathcal A_1^{[1]}
		+\mathcal A_1^{[0]}\,,
	\end{equation}
	where each term scales according to $\mathcal A_1^{(n)}\sim q^{-n-2\epsilon}$.
	The leading term, which is entirely determined by the box and crossed-box contributions, and is given by
	\begin{equation}\label{a12GR}
		\frac{i \mathcal{A}_1^{[2]}}{4p E} =
		-\frac{(8\pi G)^2}{(4\pi)^2}
		\left(\frac{4\pi}{q^2}\right)^{\epsilon}   \frac{2\pi   m_1^2 m_2^2 (\sigma^2 - \frac{1}{D-2})^2}{q^2 (\sigma^2-1)} 
		\frac{  \Gamma (1+\epsilon) \Gamma^2 (-\epsilon)}{  \Gamma (-2\epsilon)}\,.
	\end{equation}
	The subleading one reads instead
	\begin{equation}
		\label{a11GR}
		\begin{split}
			\frac{i \mathcal{A}^{[1]}_1}{4p E} 
			&=\frac{i(8\pi G)^2}{(4\pi)^2} \left(\frac{4\pi}{q^2}\right)^{\epsilon} 
			\frac{\Gamma (\epsilon+\frac{1}{2}) \Gamma (\frac{1}{2} - \epsilon)^2}{\Gamma (1-2\epsilon)}
			\Bigg[-\frac{4\epsilon\sqrt{\pi}  m_1 m_2  (m_1+m_2) (\sigma^2- \frac{1}{D-2})^2}{q(\sigma^2-1)^{\frac{3}{2}}}
			\\ 
			& + \frac{2\sqrt{\pi} m_1 m_2  (m_1+m_2)}{q (\sigma^2-1)^{\frac{1}{2} } } \left( \sigma^2 - \frac{4+ (1-2\epsilon)(\sigma^2-1)}{16 (1-\epsilon)^2}\right)
			\Bigg],
		\end{split}
	\end{equation}
	where the first line of this expression comes from the subleading expansion of box and crossed box, while the second line comes from the leading-order triangle contributions. Note that the first line vanishes in $D=4$, as $\epsilon\to0$.
	The subsubleading term instead combines all  three types of topologies, and we find it  convenient separate box/crossed box, triangle and bubble contributions according to
	\begin{equation}\label{a10GR}
		\mathcal{A}^{[0]}_1=\mathcal A_{1\Box}^{[0]} + \mathcal A_{1\triangle}^{[0]} + \mathcal A_{1\ocircle}^{[0]}\,.
	\end{equation}
	We find, from the box and crossed box topologies \cite{Collado:2018isu},
	\begin{equation}\label{bcb1GR}
		\begin{split}
			&\frac{i \mathcal A_{1\Box}^{[0]}}{4 p E}
			=
			\frac{(8\pi G)^2}{(4\pi)^2}\left(
			\frac{4\pi}{q^2}
			\right)^\epsilon\!\!
			\left(\sigma^2-\tfrac{1}{D-2}\right)
			\frac{\Gamma (1-\epsilon )^2 \Gamma (1+\epsilon )}{\Gamma (1-2 \epsilon )}
			\Bigg[
			\frac{-\pi  s\left(\sigma ^2-\frac{1}{D-2}\right) }{\left(\sigma ^2-1\right)^2 }
			\\
			&
			+
			\frac{2i m_1 m_2}{\epsilon\sqrt{\sigma ^2-1}}
			\Bigg(
			\frac{4 \sigma\operatorname{arccosh} \sigma }{\sqrt{\sigma ^2-1}}+\frac{ (1+2\epsilon) \left(\sigma ^2-\frac{1}{D-2}\right)}{\sigma ^2-1}
			\left(1-\frac{  \sigma\operatorname{arccosh}\sigma }{\sqrt{\sigma ^2-1}}\right)
			\Bigg)
			\Bigg],
		\end{split}
	\end{equation}
	from the triangle topologies,
	\begin{equation}\label{tr1GR}
		\begin{split}
			\frac{i \mathcal A_{1\triangle}^{[0]}}{4 p E}
			&=
			\frac{(8\pi G)^2}{(4\pi)^2}\left(
			\frac{4\pi}{q^2}
			\right)^\epsilon
			\frac{i m_1 m_2}{2 \sqrt{\sigma^2-1}}
			\frac{\Gamma(1-\epsilon)^2\Gamma(2+\epsilon)}{\epsilon(1-\epsilon)^2\Gamma(1-2\epsilon)}
			\\
			&\times
			\left[
			1-2\sigma^2(1-\epsilon)\frac{11-18\epsilon+8\epsilon^2}{(1+\epsilon)(1-2\epsilon)}
			\right]
			,
		\end{split}
	\end{equation}
	and, from the bubble,
	\begin{equation}\label{b1GR}
		\begin{split}
			&\frac{i \mathcal A_{1\ocircle}^{[0]}}{4 p E}
			=
			\frac{(8\pi G)^2}{(4\pi)^2}\left(
			\frac{4\pi}{q^2}
			\right)^\epsilon
			\frac{i m_1 m_2}{\sqrt{\sigma^2-1}}
			\frac{\Gamma(1-\epsilon)^2\Gamma(1+\epsilon)}{4\epsilon(5-2\epsilon)(3-2\epsilon)\Gamma(1-2\epsilon)}
			\\
			&
			\times
			\Bigg[
			\frac{\sigma ^2 \left(16 \epsilon ^3-210 \epsilon ^2+633 \epsilon -522\right)}{2 \epsilon -1}-\frac{17 \epsilon ^3-68 \epsilon ^2+65 \epsilon -2}{2 (\epsilon -1)^2}
			\Bigg].
		\end{split}
	\end{equation}

	We can now go to impact parameter space using Eq. \eqref{B1}. Let us start from the superclassical term $\mathcal A_1^{[2]}$ in  \eqref{a12GR}, which gives
	\begin{equation}
		i {\tilde{\mathcal{A}}}_1^{[2]} (s, b)= \frac{1}{2} \left( \frac{2 i m_1m_2 G (\pi b^2)^{\epsilon}(\sigma^2 - \frac{1}{D-2} )\Gamma (-\epsilon) }{ \sqrt{\sigma^2-1}}\right)^2= \frac{1}{2} (2 i \delta_0)^2\,\,.
		\label{exp1}
	\end{equation}
	Thus, as in the previous case, we again obtain a cross check of the exponentiation of $\delta_0$, according to which the leading superclassical term in the one-loop amplitude must be given by $\frac{1}{2!}(2i\delta_0)^2$ as dictated by \eqref{eikexp1}.
	
	The Fourier transform of the subleading, classical term $\mathcal A_1^{[1]}$, appearing in \eqref{a11GR}, 
	gives the next to the leading classical eikonal
	\begin{equation}
		2i \delta_1 = i{\tilde{\mathcal{A}}}_1^{[1]} (s, b)\,.
		\label{exp4}
	\end{equation}
	so that
	\begin{align}
		2i \delta_1
		&= \frac{iG^2 m_1 m_2 (m_1+m_2)}{(\pi b^2)^{\tfrac12-2\epsilon}}
		\Bigg[
		\frac{\Gamma^2(\frac{1}{2} -\epsilon) \Gamma ( \frac{1}{2} -2\epsilon)  \left(\sigma^2 - \frac{1}{D-2}\right)^2}{\Gamma (-2\epsilon) (\sigma^2-1)^{\frac{3}{2}}}
		\label{exp2}
		\\
		&+ \frac{4\Gamma(\frac{1}{2} -\epsilon)^2 \Gamma ( \frac{1}{2} -2\epsilon)
		}{\Gamma (1-2\epsilon)  \sqrt{\sigma^2-1} } 
		\left( \!\sigma^2\! - \frac{4+ (1-2\epsilon)(\sigma^2-1)}{16 (1-\epsilon)^2}\right) 
		\Bigg].
		\label{exp3}
	\end{align}
	Like the leading eikonal, $2\delta_1$ is also a purely real quantity.
	Taking a derivative with respect to $b$ according to
	\begin{equation}\label{}
		2p\sin\frac{\Theta}{2} = -\frac{\partial2\delta}{\partial b} \implies \Theta^\text{2PM} =- \frac{1}{p} \frac{\partial2\delta_1}{\partial b}\,,
	\end{equation}
	we can then obtain the 2PM correction to the deflection angle, in generic spacetime dimensions. Combining with the 1PM result \eqref{defle1}, we thus obtain the following expression for the deflection angle,
	\begin{equation}\label{Thexp3}
	\begin{aligned}
		\Theta
		&= 
		\frac{4 G E \left(\sigma^2- \frac{1}{D-2}\right) \Gamma \left(1-\epsilon\right) }{(\sigma^2-1) \pi^{-\epsilon} b^{1-2\epsilon}} \\
		&
		+
		\frac{(1-2\epsilon)G^2 E (m_1+m_2)}{\pi^{\tfrac12-2\epsilon}b^{2-2\epsilon}}
		\Bigg[
		\frac{\Gamma^2(\frac{1}{2} -\epsilon) \Gamma ( \frac{1}{2} -2\epsilon)  \left(\sigma^2 - \frac{1}{D-2}\right)^2}{\Gamma (-2\epsilon) (\sigma^2-1)^{2}}
		\\
		&+ \frac{4\Gamma(\frac{1}{2} -\epsilon)^2 \Gamma ( \frac{1}{2} -2\epsilon)
		}{\Gamma (1-2\epsilon) (\sigma^2-1)} 
		\left( \!\sigma^2\! - \frac{4+ (1-2\epsilon)(\sigma^2-1)}{16 (1-\epsilon)^2}\right) 
		\Bigg]
		+\mathcal O(G^3)\,.
	\end{aligned}
\end{equation}
	Note that the 2PM correction is present only if both masses are non vanishing.  In contrast with the $\mathcal N=8$ case, although the box and cross-box contributions \eqref{exp2} do not contribute for $D=4$, the triangle contribution \eqref{exp3} survives, and we get:
	\begin{equation}
		2  \delta_1 =  \frac{ 3  \pi G^2 m_1 m_2 (m_1+m_2)(5 \sigma^2-1)}{4 b \sqrt{\sigma^2-1}}
		\label{delta1a}
	\end{equation}
	and for the deflection angle~\cite{Westpfahl:1985tsl}
	\begin{eqnarray}
		\Theta = \frac{4G E ( \sigma^2- \frac{1}{2})}{b (\sigma^2-1)} + \frac{3\pi G^2E (m_1+m_2) (5 \sigma^2-1)}{4 (\sigma^2-1) b^2} + {\cal O}(G^3)\,.
		\label{theta2PMd4}
	\end{eqnarray}
	
	Finally the Fourier transform of the sub-sub-leading term $\mathcal A_1^{[0]}$ given by \eqref{a10GR} identifies the leading quantum remainder
	\begin{equation}
		2 i \Delta_1= i{\tilde{\mathcal{A}}}_1^{[0]}\,.
	\end{equation}
	This is a quantum term and indeed will not contribute to the classical eikonal. However, it will be needed at two-loop level in order to solve \eqref{eikexp2} for the unknown $2\delta_2$. Actually, since we will be able to obtain the two-loop amplitude only to the first few orders in the $\epsilon$ expansion around $\epsilon=0$, i.e.~$D=4$, let us evaluate this quantum remainder for small $\epsilon$, although performing the Fourier transform of the full expression \eqref{a10GR} is straightforward using \eqref{B1}. We thus obtain
	\begin{equation}\label{2D1GR}
		\begin{aligned}
			&2i\Delta_1=-\epsilon\,\frac{G^2 s  \left(2 \sigma^{2}-1\right)^{2}}{b^2\left(\sigma^{2}-1\right)^{2}}
			\\
			&
			+
			\frac{iG^{2}m_1m_2\left(\pi b^{2} e^{\gamma_{E}}\right)^{2 \epsilon}}{b^{2}\pi (\sigma^2-1)^{3/2}}\left(
			\frac{2 \sigma\left(2 \sigma^{2}-1\right)\left(6 \sigma^{2}-7\right) \operatorname{arccosh}\sigma}{\sqrt{\sigma^{2}-1}}
			-
			\frac{1-49 \sigma^{2}+18 \sigma^{4}}{15}
			\right)
			\\
			&
			+i \epsilon\, \frac{G^2 m_1 m_2 }{\pi b^2 \sqrt{\sigma^2-1}}
			\left(
			-\frac{8\sigma(2\sigma^2+1)\operatorname{arccosh}\sigma}{(\sigma^2-1)^{3/2}}
			+
			\frac{9234\sigma^2-1783}{450}
			\right)+\mathcal O(\epsilon^2)\,.
		\end{aligned}
	\end{equation}
	This expression for $2i\Delta_1$ is accurate through $\mathcal O(\epsilon)$ for both real and imaginary part. In particular its $\mathcal O(1)$ real part vanishes, while its imaginary part is nontrivial to $\mathcal O(1)$ and $\mathcal O(\epsilon)$. The terms involving $\operatorname{arccosh}\sigma$ of course only come from the box/crossed box contributions \eqref{bcb1GR} already obtained in \cite{KoemansCollado:2019ggb}. Moreover, the $\mathcal O(1)$ imaginary part and the $\mathcal O(\epsilon)$ real part agree with \cite{Bjerrum-Bohr:2021din}. As we will discuss in the next chapter, this quantum piece will play a role in our discussion of the two-loop eikonal. On the one hand, it is needed to perform the subtractions dictated by \eqref{eikexp2}. Moreover, it will be instrumental in determining the amplitude itself from its analyticity and crossing-symmetry properties. 
	
	\subsubsection{Real-analytic, crossing-symmetric reformulation }
	\label{analyticGR1loop}
	
	Following the procedure we used for the ${\cal N}=8$ case in Subsection~\ref{analyticN81loop}, let us recast the 3PM GR result in an explicitly real-analytic and crossing-symmetric form. Using the same notations as in \eqref{defs} the tree level amplitude \eqref{eq:Aphi12tlp} can be written, up to higher orders in $q^2$, as
	\begin{equation}
		\mathcal{A}_0 (s, q^2) = \frac{16\pi G}{q^2} m_1^2 m_2^2 \Big[  \left(\sigma^2 - \tfrac{1}{D-2}\right) + \left(\bar{\sigma}^2- \tfrac{1}{D-2}\right)  \Big]\,.
		\label{T1symGR}
	\end{equation}
	At one-loop, in analogy with equation \eqref{A1sugra}, we can deal separately and rather trivially with the purely real classical terms \eqref{a11GR} behaving like $\frac{1}{\sqrt{q^2}}$. For the remaining terms, inspired again by the ${\cal N}=8$ case of Subsection~\ref{analyticN81loop}, we try the ansatz:
	\begin{align}	\label{Ansatz1sclGR}
		& \mathcal{A}_1^{\text{scl.}} (s , q^2) =  \frac{(8\pi G)^2}{(4\pi)^2}  \left( \frac{4\pi}{q^2}\right)^{\epsilon} 
		\frac{\Gamma^2 (-\epsilon) \Gamma (1+\epsilon)}{\Gamma (-2\epsilon)} \frac{4 m_1^3 m_2^3}{q^2} \\
		&
		\left[  \left(\sigma^2- \frac{1}{2(1-\epsilon)}\right)^2 + \left(\bar{\sigma}^2- \frac{1}{2(1-\epsilon)}\right)^2 \right] \left[ \frac{\log(-z)}{\sqrt{\sigma^2-1}} + \frac{\log(-\bar{z})}{\sqrt{\bar{\sigma}^2-1}} \right]\,.  \nonumber
	\end{align} 
	This reproduces, in $b$-space, the iteration of the tree contribution \eqref{a12GR} and, furthermore, takes care of some quantum terms. Up to purely real pieces, the full structure of \eqref{a10GR} is reproduced by the following additional real-analytic crossing-symmetric terms:
	\begin{align}	\label{Ansatz1addGR}
		\mathcal{A}_1^{\text{add.}} (s , q^2) 
		&=  \frac{(8\pi G)^2}{(4\pi)^2}  \left( \frac{4\pi}{q^2}\right)^{\epsilon}  
		\frac{\Gamma^2 (-\epsilon) \Gamma (1+\epsilon)}{\Gamma (-2\epsilon)} 4 m_1^2 m_2^2 \hat {\mathcal{A}}_1^{\text{add.}} \,,\\ \nonumber
		\begin{split}
			\hat {\mathcal{A}}_1^{\text{add.}} 
			&=   \left[ \left(\sigma^2- \frac{1}{2(1-\epsilon)}\right) + \left(\bar{\sigma}^2- \frac{1}{2(1-\epsilon)}\right)\right]  \left[  \frac{\sigma \log(-z)}{\sqrt{\sigma^2-1}} + \frac{\bar{\sigma}\log(-\bar{z})}{\sqrt{\bar{\sigma}^2-1}} \right]   \\
			& - \frac{\epsilon}{2}  \left[  \left(\sigma^2- \frac{1}{2(1-\epsilon)}\right)^2 + \left(\bar{\sigma}^2- \frac{1}{2(1-\epsilon)}\right)^2 \right]  \left[  \frac{\sigma \log(-z)}{(\sigma^2-1)^{3/2}} + \frac{\bar{\sigma}\log(-\bar{z})}{(\bar{\sigma}^2-1)^{3/2}} \right]   \\
			& - \epsilon \frac{m_1^2 + m_2^2}{4 m_1 m_2}  \left[  \left(\sigma^2- \frac{1}{2(1-\epsilon)}\right)^2 + \left(\bar{\sigma}^2- \frac{1}{2(1-\epsilon)}\right)^2 \right]\left[  \frac{ \log(-z)}{(\sigma^2-1)^{3/2}} + \frac{\log(-\bar{z})}{(\bar{\sigma}^2-1)^{3/2}} \right]\,.   
		\end{split} 
	\end{align}
	We have checked that the sum of  \eqref{Ansatz1sclGR} and \eqref{Ansatz1addGR} reproduces the superclassical term \eqref{a12GR}, as well as the box and crossed box contributions \eqref{bcb1GR}, modulo a single real term given by:
	\begin{equation}
		\hat {\mathcal{A}}_1^{\text{res.}}= - 2 \epsilon \frac{ \left(\sigma^2 - \frac{1}{D-2}\right)^2}{\sigma^2-1}\,.
		\label{realleft}
	\end{equation}
	We also note that the triangle \eqref{tr1GR} and bubble \eqref{b1GR} contributions are real (indeed they do not exhibit any on shell intermediate states). Together with \eqref{realleft} they can be made crossing symmetric trivially.
	This exhausts the analytic study of the one-loop terms in GR.

\subsubsection{The probe limit}
\label{ssec:probeQ}

It is instructive to study the regime where one mass is much larger than all other energy scales as was done at leading PM order at the end of Section~\ref{sec:grtree} for GR and of Section~\ref{sec:n8tree} for $\mathcal N=8$ supergravity. Of course in this limit one should reproduce the results obtained from the classical motion of a probe in a fixed background describing the heavy object, see \ref{app:geoschw}. This is easily checked by taking the limit $m_2\gg E_1$ summarized in~\eqref{eq:problim2}. Focusing first on the case of GR, it is sufficient to take \eqref{Thexp3} in the probe limit, $m_1=m_p\ll M=m_2$,  to obtain
\begin{equation}
\begin{aligned}\label{eq:2PMpl}
 \Theta & 
    = 
    \frac{\sqrt{\pi} \,\Gamma\left(\frac{D-2}{2}\right)}{2 \Gamma\left(\frac{D-1}{2}\right)} \frac{(D-2) E^2 -m_p^2 }{E^2-m_p^2} \left(\frac{R_s}{b}\right)^{D-3}\\
    &+ \frac{\sqrt{\pi }\, \Gamma \left(D-\frac{5}{2}\right)}{8 \Gamma (D-2)} \left(\frac{R_s}{b}\right)^{2(D-3)} 
    \frac{(2 D-5)(2 D-3) E_p^4-6 (2 D-5) E_p^2 m_p^2+3 m_p^4}{(E_p^2-m_p^2)^2}
    \\
    &+ {\cal O}(G^3)
\end{aligned}
\end{equation}
where we used~\eqref{eq:RsD} for $R_s$. The leading term matches \eqref{eq:chi1g} while the subleading correction matches~\eqref{eq:chi2g}.
It is straightforward to repeat the check in the case of ${\cal N}=8$ supergravity by starting from~\eqref{theta2PMN8}, for which we get
\begin{equation} \label{eq:2PMpln8}
\begin{aligned} 
  \Theta 
  & = \frac{\sqrt{\pi} \, \Gamma\left(\frac{D-2}{2}\right)}{\Gamma\left(\frac{D-3}{2}\right)} \frac{E^2}{E^2-m_p^2} \left(\frac{R_s}{b}\right)^{D-3} \!\!\!\!
  + \frac{\sqrt{\pi} \, \Gamma\left(D-\frac{5}{2}\right)}{\Gamma\left(D-4\right)} \left(\frac{E^2}{E^2-m_p^2}\right)^2
  \!\! \left(\frac{R_s}{b}\right)^{2(D-3)} \!\!\!\!\!\!+ {\cal O}(G^3) \,.
\end{aligned}
\end{equation}
As already noticed, for $D=4$ the 2PM correction in~\eqref{eq:2PMpln8} vanishes and the 1PM result agrees with the probe limit \eqref{eq:dangln8}.

At 2PM order, it is actually possible to make the link in the opposite direction and reconstruct the deflection angle for generic masses from the result in the probe limit. The idea is to consider the following ansatz for the classical impulse $Q$ in an elastic scattering in the PM expansion~\cite{Damour:2019lcq},
\begin{equation}
  \label{eq:D215}
  Q= \frac{2 G m_1 m_2}{J/p} \sum_{k_1,k_2=0}^\infty Q_{k_1k_2}(\sigma) \left(\frac{2 G m_1}{b_J}\right)^{k_1} \left(\frac{2G m_2}{b_J}\right)^{k_2}
\end{equation}
where we restrict to $D=4$.
The key assumption here, sometimes referred to as ``good mass polynomiality'', is that the dependence on the masses $m_i$ always enters via the combination $2Gm_i/J$ and it is thus tied to the PM expansion itself. Each coefficient $Q_{k_1k_2}(\sigma)$ is only a function of $\sigma$, which contributes to the result at PM order $k_1+k_2+1$. Moreover, since the impulse $Q$ must be symmetric under particle-interchange symmetry, in the elastic case, we must have
\begin{equation}\label{}
	Q_{k_1k_2}(\sigma)=Q_{k_2k_1}(\sigma)\,.
\end{equation}
In the probe limit $m_1\ll m_2$, in Eq.~\eqref{eq:D215} only the terms with $k_1=0$ survive  to leading order, and thus one can deduce the functions $Q_{0k}(\sigma) = Q_{k0}(\sigma)$ from the deflection angle calculated in \ref{app:geoschw} by studying the motion of particle 1, with mass $m_1 = m_p$, in the background sourced by particle 2, with mass $m_2 = M$. In the case of GR, by \eqref{eq:chi1g4D}, \eqref{eq:chi2g4D} we have at the first two orders
\begin{align}\label{Q00Q11}
  Q_{00}(\sigma) & = \frac{2\sigma^2-1}{\sqrt{\sigma^2-1}}\,,\qquad
    Q_{01}\left(\sigma\right)  = \frac{3\pi}{16}  \frac{5\sigma^2-1}{\sqrt{\sigma^2-1}}\,,
\end{align}
where we used $E/m_p\simeq \sigma$ in the probe limit. Of course,  substituting back $Q_{00}$ and $Q_{01}$ into \eqref{eq:D215} one recovers precisely the 2PM accurate deflection angle \eqref{theta2PMd4}. On the contrary, at 3PM a new function $Q_{11}(\sigma)$ appears that cannot be fixed by looking at the probe limit, and needs to be determined by studying the problem where both particles are fully dynamical. This will be the subject of Section~\ref{sec:twoloop} for both $\mathcal{N}=8$ supergravity and GR.
Moreover, as we shall see in Section~\ref{SemiclEik}, in general the impulse $Q_1$ of particle 1 and that $Q_2$ of particle 2 can differ in the presence of radiative effects, thus introducing additional structures.
For recent developments concerning the assumption of good mass polynomiality to $\mathcal O(G^4)$ see \cite{Dlapa:2022lmu,Bini:2022enm,Dlapa:2023hsl}.

\subsubsection{Tidal effects in field theory}
Following Refs.~\cite{Cheung:2020sdj,Haddad:2020que,Bern:2020uwk,AccettulliHuber:2020dal}, one can conveniently include finite-size effects associated to tidal deformations in the scattering amplitude approach by introducing higher derivative operators that are quadratic in the scalar field and involve powers of the Weyl tensor. Focusing for simplicity on the leading, quadratic order, these can be decomposed into the so-called $E$ (``electric'') or mass-type and $B$ (``magnetic'') or current-type tidal operators. The resulting ``tidal vertex'' involves two massive lines and two graviton lines, and thus produces corrections to the sewing procedure of Fig.~\ref{fig:2gravitons} (while it doesn't affect the single-graviton, tree-level exchange). Without going into details, let us quote here for completeness the expression for the resulting one-loop correction to the impulse, which are of course equivalent to the tidal modifications of $2\delta_1$ up to a derivative with respect to $b$. For the case in which particle $1$ is subject to tidal deformations (the analogous case in which object $2$ can be deformed is obtained trivially by interchanging particle labels),
\begin{equation}\label{leadingtidalcorrections}
		Q_{E_1^2}=\frac{R_f b}{G} \frac{3 c_{E_1^2}}{m_1^2} \frac{35 \sigma^4-30 \sigma^2+11}{\sqrt{\sigma^2-1}} \,,\qquad
		Q_{B_1^2}=\frac{R_f b}{G} \frac{15 c_{B_1^2}}{m_1^2} \sqrt{\sigma^2-1}\left(7 \sigma^2+1\right),
\end{equation}
with ($i=1,2$)
\begin{equation}\label{}
	R_f=15 \pi G^3 m_1^2 m_2^2 /\left(64 b^7\right)\,,\qquad
	c_{E_i^2}=\frac{1}{6} k_i^{(2)} R_i^5 / G \,,\qquad c_{B_i^2}=\frac{1}{32} j_i^{(2)} R_i^5 / G
\end{equation}
and $k_i$, $j_i$ are the colliding objects' Love numbers, while $R_i$ is the radius of object $i$. For typical compact objects, $R_i = G m_i/K_i$ where $K_i$ is the body's ``compactness'' of order $0.1$, $0.2$ for neutron stars \cite{Blanchet:2013haa}.
Thus, we see that the leading tidal corrections  \eqref{leadingtidalcorrections} are weighted by the dimensionless power-counting parameter $(Gm/b)^5/K^5$ (for $m\simeq m_{1,2}$ and $K\simeq K_{1,2}$) relative to the leading-order impulse \eqref{1PM}. This shows how PM effects measured by $Gm/b$ can compete with finite-size effects measured by the compactness parameter $K$.

\subsection{Eikonal exponentiation in String Theory at one loop}
\label{ssec:1lstrng}

The first goal of this section is to analyze an explicit example of a one-loop amplitude in string theory. We will see that also in this case the leading term in the classical limit does not contain new information as it provides just a contribution towards the exponentiation of the tree-level result. The main difference with the QFT cases analyzed so far is that already the leading eikonal $\hat{\delta}_0$ \eqref{eq:hatdelta0cl} is an operator: as discussed in Section~\ref{ssec:seikop}, this operator acts on the space of the possible string states. The non-trivial check presented here is that the leading term of the one-loop result is the square of $\hat{\delta}_0$, extending Eq.~\eqref{eikexp2} to the more general setup where the colliding objects have a non-trivial structure. After gaining confidence that the full eikonal operator exponentiates, we revisit the physics related to tidal excitation. We show that, at the leading order, this classical effect is captured by a probe analysis focusing on the string dynamics in a non-trivial background. This generalizes to the string case the idea that the leading deflection angle for a point-like particle can be derived by solving the geodesic equation, see~\eqref{eq:pdeflampl1} and~\eqref{eq:chi1g}. We conclude this part on string theory by discussing more in detail how the exponentiated form of the leading eikonal captures the tidal effects on the string probe and how this can  exponentially suppress the elastic scattering even at very large distances if the probe energy is large enough.

\subsubsection{1-loop in String Theory}
\label{ssec:1loopstr}

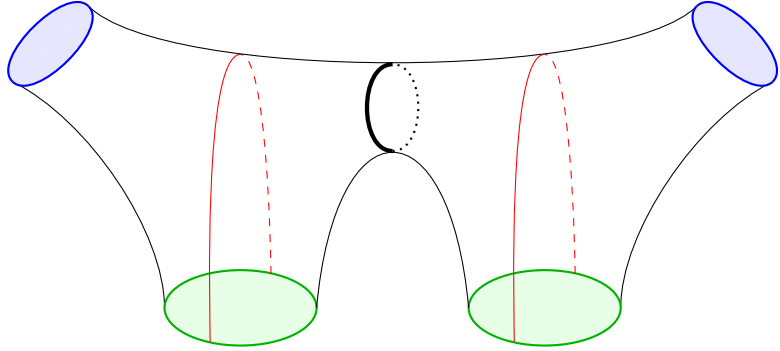
\begin{figure}
	\centering
	\begin{tikzpicture}
		\draw (-5,2) .. controls (-4,1.5) and (-3,0) .. (-3,-1);
		\draw (5,2) .. controls (4,1.5) and (3,0) .. (3,-1);
		\fill [green!10!white] (-2,-1) ellipse (1 and .5);
		\draw [thick, black!30!green] (-2,-1) ellipse (1 and .5);
		\fill [green!10!white] (2,-1) ellipse (1 and .5);
		\draw [thick, black!30!green] (2,-1) ellipse (1 and .5);
		\draw (-1,-1) .. controls (-.75,1.75) and (.75,1.75) .. (1,-1);
		\draw[red] (-2.4,-1.45) .. controls (-2.4,-1.1) and (-2.5,2.36) .. (-2,2.36);
		\draw[red,dashed] (-1.6,-.55) .. controls (-1.6,-.3) and (-1.65,2.36) .. (-2,2.36);
		\draw (-4,3) .. controls (-3,2) and (3,2) .. (4,3);
		\draw[red] (1.6,-1.45) .. controls (1.6,-1.1) and (1.5,2.36) .. (2,2.36);
		\draw[red,dashed] (2.4,-.55) .. controls (2.4,-.3) and (2.35,2.36) .. (2,2.36);
		\draw[thick,dotted] (0,1.08) .. controls (.45,1.08) and (.45,2.23) .. (0,2.23);
		\draw[ultra thick] (0,1.08) .. controls (-.45,1.08) and (-.45,2.23) .. (0,2.23);
		\filldraw[blue!10!white] (-5,2) .. controls (-5.3,2.3) and (-4.3,3.3) .. (-4,3);
		\filldraw[blue!10!white] (-5,2) .. controls (-4.7,1.7) and (-3.7,2.7) .. (-4,3);
		\draw[thick,blue] (-5,2) .. controls (-5.3,2.3) and (-4.3,3.3) .. (-4,3);
		\draw[thick,blue] (-5,2) .. controls (-4.7,1.7) and (-3.7,2.7) .. (-4,3);
		\filldraw[blue!10!white] (5,2) .. controls (5.3,2.3) and (4.3,3.3) .. (4,3);
		\filldraw[blue!10!white] (5,2) .. controls (4.7,1.7) and (3.7,2.7) .. (4,3);
		\draw[thick,blue] (5,2) .. controls (5.3,2.3) and (4.3,3.3) .. (4,3);
		\draw[thick,blue] (5,2) .. controls (4.7,1.7) and (3.7,2.7) .. (4,3);
	\end{tikzpicture}
	\caption{\label{fig:StringBrane1loop} Scattering of a massless string off a D$p$-brane at  one loop.}
\end{figure}

\begin{figure}
	\centering
	\begin{tikzpicture}
		\draw[help lines, color=gray!40,thick,->] (-4,0) -- (4,0);
		\node[right] at (4,0) {$\operatorname{Re}z$};
		\draw[help lines, color=gray!40,thick,->] (0,-3.5) -- (0,3.5);
		\node[right] at (0,3.5) {$\operatorname{Im}z$};
		\draw[thick, black!30!green] (0,0) ellipse (1 and 1);
		\draw[thick, black!30!green] (0,0) ellipse (3 and 3);
		\filldraw[thick, blue] (-2,.8) circle (.03);
		\node[below] at (-2,.8) {$z_2$};
		\filldraw[thick, blue] (1.8,1.6) circle (.03);
		\node[above] at (1.8,1.6) {$z_1$};
		\draw[->] (0,-.03) -- (3,-0.03);
		\node[below] at (3.2,0) {$1$};
		\draw[->] (0,.03) -- (1,0.03);
		\node[above] at (1.5,0) {$e^{-\pi\lambda}$};
	\end{tikzpicture}
	\caption{\label{fig:StringBrane1loopFlattened} Alternative representation of the one-loop scattering of a closed string off a $D$p-brane. The green circles are the boundaries of the worldsheet and rest on the $D$p-brane, while the blue dots represent punctures associated to the asymptotic closed-string states.}
\end{figure}
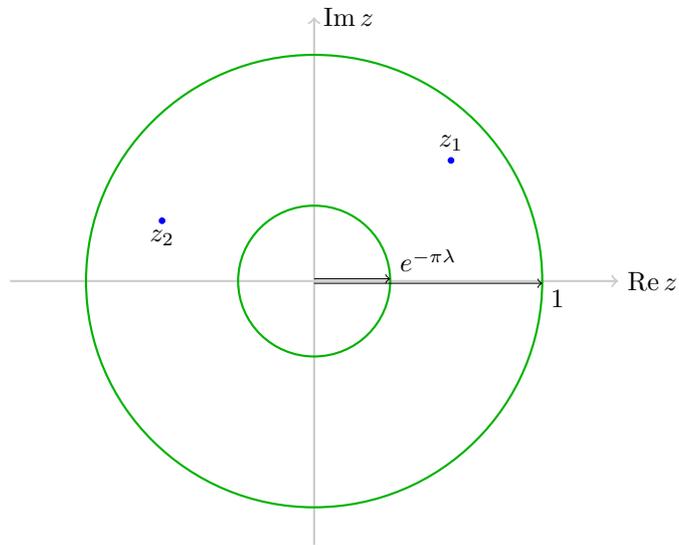

We work in the setup discussed in Section~\ref{ssec:string-brane} and consider the first correction to the $1\to 1$ scattering of a NS-NS massless state in presence of a stack of  $N$ coincident D$p$-branes in type II theories. From the world-sheet point of view, this amplitude is captured by a diagram with the topology of an annulus with 
two boundaries (which are supported by the D$p$-branes) and two punctures in the interior of the annulus representing the external closed string states, as depicted in Figure~\ref{fig:StringBrane1loop}. 
We will parameterize this surface as done in~\cite{DAppollonio:2010ae}: the ``thickness'' of the annulus is related to $e^{-\pi \lambda}$, with $0\leq \lambda<\infty$, while for the location of the punctures we use 
	\begin{equation}
		z_i=e^{2\pi (-\lambda \rho_i+i\omega_i)} \,,\qquad 
		0<\rho_i<\frac{1}{2}\,,\qquad 
		0\leq \omega_i <1\,.
		\label{changevari}
	\end{equation}
(see Figure~\ref{fig:StringBrane1loopFlattened}).
In type II theories, the behavior of this amplitude is constrained by supersymmetry and the kinematic dependence is the same as in the tree-level case \cite{green1988superstring}, so we can extract a scalar function ${\cal A}_1$
\begin{equation}
	\label{eq:strA1}
	{\cal A}_1(p_i, \epsilon_i) = {\cal K}(p_i, \epsilon_i) \,{\cal A}_1\;
\end{equation}
that captures all the dynamical information. The derivation of ${\cal A}_1$ is conceptually similar to the one summarized in \ref{ssec:string-brane-bos-2} for bosonic string theory, but technically requires to deal with some of the complications of superstring theory such as the sum over the spin structures. Here we start from the final  result below and refer to the literature~\cite{Pasquinucci:1997di,Lee:1997gwa} for a derivation and further references. In the superstring case we have
\begin{equation}
	\label{eq:BVVB2}
	{\cal A}_1 = \left(\frac{\kappa_{d} N T_p}{2}\right)^2 \frac{\alpha'}{16\pi} (2\pi^2 \alpha')^{-\frac{d-p-1}{2}} \int\limits_0^\infty d\lambda \int d^2 z_1 \, d^2 z_2\,\lambda^{-\frac{d-p-1}{2}}\, e^{- \alpha' E^2_s V_s - \frac{\alpha' t}{4} V_t }\;,
\end{equation}
with
\begin{subequations}\label{eq:Vsti}
	\begin{align}
		V_s  & = -  {2 \pi} \lambda  \rho^2 + \log\left[ \frac{\theta_1 ( i \lambda (\zeta+ \rho) | i \lambda) \theta_1 ( i \lambda(\zeta - \rho)| i \lambda)  }{ \theta_1 (i\lambda \zeta + \omega | i \lambda) \theta_1 (i \lambda \zeta - \omega| i \lambda) }\right] \ , \label{eq:Vsi}
		\\ V_t  & =  8\pi \lambda \rho_1 \rho_2 + \log\left[ \frac{\theta_1 (i \lambda \rho + \omega | i \lambda) \theta_1 (i\lambda \rho - \omega| i \lambda)  }{ \theta_1 (i \lambda \zeta +  \omega | i \lambda) \theta_1 (i \lambda  \zeta - \omega| i \lambda) }\right] \label{eq:Vti} \ ,
	\end{align}
\end{subequations}
where for the Jacobi theta-functions $\theta_i$ we follow the conventions of~\cite{Polchinski:1998rq} where
\begin{equation}
	\label{eq:th1P}
	\theta_1(\nu|\tau) = -2 e^{\frac{\pi i \tau}{4}} \sin(\pi\nu) \prod_{n=1}^\infty \left[ (1-e^{2\pi in \tau}) (1 - e^{2\pi i\nu} e^{2\pi i n \tau}) (1 - e^{-2\pi i\nu} e^{2\pi i n \tau}) \right]
\end{equation}
and we used the change of variables
\begin{equation}
	\label{eq:rhoomega}
	\omega=\omega_1-\omega_2\;,\quad \rho = \rho_1-\rho_2\;,\quad \zeta = \rho_1+\rho_2\;.
\end{equation}

Since we are interested in the long-range interaction between the stack of D$p$-branes and the external probe, we wrote the amplitude in the so-called closed string channel where the large-$\lambda$ limit is simple and describes a superposition of closed string states exchanged between the D$p$-branes and the scattered closed string. In this limit the world-sheet has the shape of a sphere connected to the D$p$-branes by two thin tubes and with two punctures representing the external states. The factor of $\lambda^{-\frac{d-p-1}{2}}$ in~\eqref{eq:BVVB2} follows from the Gaussian integration of the center-of-mass momentum of the virtual closed string propagating between the two boundaries, which can be non-trivial only along the $d-p-1$ directions where the D$p$-branes impose Dirichlet boundary conditions. After implementing the change of variables in \eqref{changevari} and \eqref{eq:rhoomega},   Eq.~\eqref{eq:BVVB2} reads
\begin{equation}
	\label{eq:BVVB2bis}
	{\cal A}_1 =  \left(\frac{\kappa_{d} N T_p}{2}\right)^2 \frac{\alpha'}{16\pi} (2\pi^2 \alpha')^{-\frac{d-p-1}{2}} 2(2\pi)^4 \int\limits_0^\infty d\lambda \int\limits_0^1 d\zeta \int\limits_{\mathcal R(\zeta)}\!\!\!\!\! d\rho \int\limits_0^1 d\omega \,\lambda^{-\frac{d-p-5}{2}}\, e^{- \alpha' E^2_s V_s - \frac{\alpha' t}{4} V_t }\;,
\end{equation}
where the region of integration for $\rho$ depends on $\zeta$: $\operatorname{max}\{-\zeta, -(1-\zeta)\}<\rho<\operatorname{min}\{\zeta, 1-\zeta\}$. The precise definition of ${\mathcal R}(\zeta)$ will not play any role in our analysis, since we focus on the leading contribution dominated by the $\rho\sim 0$ region, see however~\cite{Bianchi:2011se} for a more general setup where this point becomes relevant. As expected the integrand in~\eqref{eq:BVVB2} does not depend on $\omega_1+\omega_2$ and so, in \eqref{eq:BVVB2bis},  we have trivially performed the integration over this variable.
Our goal is to extract the leading contribution in the classical limit of the integral~\eqref{eq:BVVB2} and show that it provides the first term needed to exponentiate the 1PM stringy eikonal operator \eqref{eq:hatdelta0cl}. In this context the classical parameters are $R_p$ in~\eqref{eq:Rpscale}, describing the gravitational backreaction of the D$p$-branes, and $E_s$, describing the probe state. So, as in the tree-level analysis, we are interested in the large $E_s$ limit which from~\eqref{eq:BVVB2bis} is suppressed unless one takes also $\rho$ small. Thus, for our purposes, we can use in~\eqref{eq:BVVB2bis} the approximate expressions
\begin{align}
	\label{eq:VsVtap}
	V_s &\simeq -2\pi \lambda \rho^2 -4 \sin^2 \pi \omega \left( e^{- 2 \pi \lambda \zeta} + e^{- 2 \pi \lambda (1- \zeta)} \right )~ , \\
	V_t &\simeq - 2 \pi \lambda \zeta (1 - \zeta)  + \ln \left(4 \sin^2 \pi \omega\right) ~.  \nonumber  
\end{align}
In the expression for $V_t$ we kept only the terms that do not have any exponential factor involving $\lambda$, while for $V_s$ we kept also the first exponential terms that become relevant in the region $\lambda \sim \ln(\alpha' E^2_s)$ and considered only the $\rho$-dependent terms that are enhanced by a factor of $\lambda$. In order to extract the classical terms contributing to the 2PM eikonal operator $\hat{\delta}_1$, one would need a better approximation for $V_s$ and $V_t$ as partially discussed in~\cite{DAppollonio:2010ae}. The complete expression for $\hat\delta_1$ at string level is not known, so we will not pursue further this analysis here.

By following closely the derivation discussed in~\ref{ssec:string-brane-bos-2} for the bosonic string, we can rewrite~\eqref{eq:BVVB2bis} in terms of $d_\perp=(d-p-2)$ integral involving the full tree-level amplitude~\eqref{eq:disregss} and a kernel written in the square bracket below
\begin{equation}
	\label{eq:A1sfac}
	{\cal A}_1 \simeq \frac{i}{4 E_s} \int  \frac{ d^{d_\perp} \mathbf{k} }{ (2\pi)^{d_\perp}} {\cal A}_0(E_s,\mathbf{k}) {\cal A}_0(E_s,\mathbf{q}-\mathbf{k})\left[ \frac{2^{\alpha' \mathbf{k} (\mathbf{q}-\mathbf{k})}}{\pi}B \left( \frac{1}{2} + \frac{\alpha'}{2} \mathbf{k}(\mathbf{q}-\mathbf{k}) , \frac{1}{2} \right)\right]\;.
\end{equation}
The structure is familiar from the QFT case: the convolution over $\mathbf{k}$ becomes just a standard product in impact parameter space, the factor of $2E_s$ is the one needed to pass from the amplitude to  the eikonal (see Eq.~\eqref{eq:DbraneFT}) and the extra factor of $\frac{i}{2}$ comes from the expansion at the second order of the exponential with the leading eikonal (see the first term on the r.h.s. of~\eqref{eikexp1}). The kernel in the square parenthesis is a string effect as one can see by checking that in the $\alpha'\to 0$ limit it becomes the identity plus subleading corrections
\begin{equation}
	\label{eq:kerapp}
	\left[ \frac{2^{\alpha' \mathbf{k} (\mathbf{q}-\mathbf{k})}}{\pi}B \left( \frac{1}{2} + \frac{\alpha'}{2} \mathbf{k}(\mathbf{q}-\mathbf{k}) , \frac{1}{2} \right)\right] \simeq 1 + \frac{\pi^2}{6} \left(\frac{\alpha'}{2} {\bf k}\cdot ({\bf q-k})\right)^2 +\cdots ~.
\end{equation}
In order to completely factorize the result in~\eqref{eq:A1sfac}, it is sufficient to rewrite the kernel as an expectation value of operators written in terms of the string coordinates $\hat{X}$ introduced in Eq.~\eqref{eq:Vrexpf}
\begin{equation}
	\label{eq:evBX}
	\frac{1}{\pi}B \left( \frac{1}{2} + \frac{\alpha'}{2} \mathbf{k}_1 \mathbf{k}_2 ,  \frac{1}{2} \right) = 2^{- \alpha' \mathbf{k}_1 \mathbf{k}_2} \langle 0| \prod_{i=1}^{2} \int\limits_0^{2 \pi} \frac{ d \sigma_i}{2 \pi} :e^{i \mathbf{k}_1 {\hat{X}}(\sigma_1)} :\,:e^{i \mathbf{k}_2 {\hat{X}}(\sigma_2)} : | 0 \rangle~,
\end{equation}
where the fields $\hat{X}$ were introduced in~\eqref{eq:Vrexpf} and contain the physical excitations of the string involving the DDF oscillators, see Section~\ref{ssec:seikop} for a general discussion. Then the interpretation of the kernel appearing in this string theory case is clear: it is related to the possibility of having excited string states in the intermediate steps. Even when focusing on a particular elastic amplitude, where the initial and the final states are dilatons, the gravitational field of the D$p$-brane can stretch the string probe and excite it to a different state while it propagates. As already discussed, this is an entirely classical effect capturing the tidal forces on an extended probe~\cite{Giddings:2006sj}.
As a further check of this interpretation, this phenomenon can be quantitatively described by quantizing the string worldsheet action in the background produced by the D$p$-branes, as we will discuss in Section~\ref{ssec:PLstract}. However, as usual with the eikonal exponentiation, classical effects provide constraints on loop amplitudes and in particular in this case the leading contribution (as $\alpha' E^2_s$ becomes large) of the $h$-loop diagram (i.e.~a worldsheet with $h+1$ boundaries and no handles\footnote{As in Section~\ref{sec:string-disk}, in the string-brane case we are taking the probe approximation where the string coupling is small, but with $g_s N$ fixed, so we ignore the contributions of worldsheets with handles.}) should match the following convolution involving the tree-level result and the kernel discussed above
\begin{equation}
	\label{eq:Ahstr}
	\frac{{\cal A}_h^L}{2E_s}  = \frac{i^{h}}{(h+1)!} \langle 0| \prod_{i=1}^{h+1} \int \frac{d^{d_\perp}\mathbf{k}_i}{(2\pi)^{d_\perp}} \int\limits_0^{2 \pi} \frac{ d \sigma_i}{2 \pi}  \frac{{\cal A}_0(E_s,\mathbf{k}_i)}{2E_s}  :e^{i \mathbf{k}_i {\hat{X}}(\sigma_i)}: | 0 \rangle\, \delta^{d_\perp}\!\!\left(\sum_{i=1}^{h+1} \mathbf{k}_i-\mathbf{q}\right)\,.
\end{equation}
This indeed reduces to~\eqref{eq:A1sfac} for $h=1$ (one loop) thanks to~\eqref{eq:evBX}. Thanks to this factorized form of the leading contributions, it is straightforward to resum them in terms of the leading eikonal operator that was derived by studying the inelastic transition at tree level in Section~\ref{ssec:seikop}. As usual, after going to impact parameter space by taking the Fourier transform~\eqref{eq:DbraneFT} the convolution becomes a product and we obtain
\begin{equation}
	\label{eq:resLstr}
	\sum_{h=0}^\infty  \frac{i\tilde{\cal A}_h^L}{2E_s} = \langle 0| \left[e^{2i \hat{\delta}_0} - 1 \right] |0\rangle\quad \Rightarrow \quad S = e^{2i \hat{\delta}_0} \;,
\end{equation}
where the operator $\hat{\delta}_0$ was introduced in Eq.~\eqref{eq:hatdelta0cl} where the impact parameter $b$ is shifted by the corresponding string coordinate $\hat{X}$. Then, at leading PM order, the evolution of an initial (light) string probe interacting with the D$p$-branes is captured by the S-matrix $S$ above. Notice that while $\hat\delta_0$ is normal ordered, this is not the case for the evolution operator $S$, and this plays an important role in the analysis of tidal forces as discussed in the next section.

\subsubsection{Tidal forces in String Theory}
\label{ssec:tidalrev}

In this subsection we mainly discuss the suppression of the elastic channel as a consequence of the tidal excitations of the colliding string. This is a direct consequence of the eikonal operator derived in the previous section: as we will see, when this operator is used to estimate an elastic amplitude a dumping factor arises from the normal ordering of the oscillators. A brief summary on the nature of the inelastic channels is postponed to Section~\ref{beyondtree} where the same problem is considered for the case of string-string collisions. Thus, while leading eikonal for a point-like scalar object is a pure phase and describes an elastic scattering, in the string case a new dynamical scale $b_t$ appears: when the impact parameter is below this scale the elastic scattering is exponentially suppressed as the internal degrees of freedom of the string are excited during the scattering. In formulae we have
\begin{equation}
	\label{eq:bDscale}
	|\langle 0| S |0\rangle| \sim \exp{\left[-\left(\frac{b_t}{b}\right)^{d-p-2}\right]},
	\qquad b^{d-p-2}_t = \frac{\sqrt{\pi} E_s}{4 T} (d-p-3) \frac{\Gamma \left(\frac{d-p-2}{2} \right)}{\Gamma \left(\frac{d-p-3}{2} \right)} R_p^{d-p-3} \, ,
\end{equation}
where we approximated $S$ by the leading eikonal operator as in~\eqref{eq:resLstr} and wrote $b_t$ in terms of the string tension $T=\frac{1}{2\pi\alpha'}$ to stress that it is a classical quantity. See also the discussion around Eq.~\eqref{bt} below.

Notice that at high energies $b_t$ is parametrically larger than the string scale so tidal forces can become relevant also at large distances. In this regime we can expand the leading eikonal operator~\eqref{eq:hatdelta0cl} by taking $|\hat{X}|\ll b$: at the zeroth order, one of course recovers the point-like eikonal~\eqref{delta}, the linear term is absent since the integral over $\sigma$ of $\hat{X}$ in~\eqref{eq:Vrexpf} vanishes, and the leading effects are encoded by the quadratic term. We have
\begin{equation}
	\label{eq:3.11}
	\frac{1}{2} \frac{\partial^2 \delta_0(b)}{\partial b_i \partial b_j} =\frac{1}{2} \left\{\mu^2_{\hat{y}} \, \left [ \delta_{ij} - \frac{b_i b_j}{b^2} \right ] + \,
	\mu^2_0 \ \frac{b_ib_j}{b^2}\right\} ,
\end{equation}
where
\begin{equation}
	\label{eq:mu0muy}
	\begin{aligned}
		\mu_{\hat{y}}^2 & = -\frac{\sqrt{\pi} E_s}{2} \frac{\Gamma \left(\frac{d-p-2}{2} \right)}{\Gamma \left(\frac{d-p-3}{2} \right)} \frac{R_p^{d-p-3}}{b^{d-p-2}}\;, \\
		\mu_0^2 & = \frac{\sqrt{\pi} E_s}{2} (d-p-3) \frac{\Gamma \left(\frac{d-p-2}{2} \right)}{\Gamma \left(\frac{d-p-3}{2} \right)} \frac{R_p^{d-p-3}}{b^{d-p-2}}= -(d-p-3)\mu_{\hat{y}}^2\;.
	\end{aligned}
\end{equation}
Note that $\mu_{\hat y}^2<0$ while $\mu_0^2>0$, but this will not play an important role in the following.
Then at the second order in the $|\hat{X}|\ll b$ expansion we have
\begin{equation}
	\label{eq:dhat2or}
	e^{2i\hat\delta_0} \simeq e^{2i \delta_0} \, \exp\left\{i \int\limits_0^{2\pi}\! \frac{d\sigma}{2\pi}\,  \hat{X}^i\hat{X}^j\,\left[\mu^2_{\hat{y}} \left( \delta_{ij} - \frac{b_i b_j}{b^2} \right) + \mu^2_0 \frac{b_ib_j}{b^2}\right] \right\} .
\end{equation}
By using the mode expansion~\eqref{eq:Vrexpf} we have
\begin{equation}
	\label{eq:xixjav}
\int_0^{2\pi}\! \frac{d\sigma}{2\pi}\, \hat{X}^i \hat{X}^j = \alpha'\sum_{n=1}^\infty \frac{1}{n} \left[2 T^{ij}_{0,n} - T^{ij}_{+,n} - T^{ij}_{-,n}-\delta^{ij} \right]\;,
\end{equation}
where, by following~\cite{Amati:1987uf}, we introduced the combinations
\begin{equation}
\label{eq:TpmT0}
T^{ij}_{\pm, n} = \frac{1}{2n} A^{(i}_{\mp n} \bar{A}^{j)}_{\mp n} \ ,\quad
2T^{ij}_{0,n} = \delta^{ij} + \frac{1}{2n}\left (A^{(i}_{- n} A^{j)}_{n} +
\bar{A}^{(i}_{- n} \bar{A}^{j)}_{n} \right ) \ .
\end{equation}
In order to evaluate the first string corrections to the leading eikonal, it is convenient to choose a basis parametrized by $\hat{y}_i$, where  $i=1,\ldots, (d-p-3)$ indicate the coordinates orthogonal to the scattering plane and $i=0$ indicates the direction along the impact parameter. This diagonalizes the structure in~\eqref{eq:dhat2or} and we can consider $(d-p-2)$ decoupled families of generators $T^{\hat{y}_i}_n$ that satisfy the commutation relations of the $sl(2,R)$ algebra 
\begin{equation}
\label{eq:Tcomm}
[T^{\hat{y}_i}_{-, n} , T^{\hat{y}_j}_{+, m}] =  \delta^{ij} \, \delta_{n m} \, 2 T^{\hat{y}_i}_{0, n} \ , \quad [T^{\hat{y}_i}_{0, n} , T^{\hat{y}_j}_{\pm, m}] = \pm \delta^{ij}  \, \delta_{n m} \, T^{\hat{y}_i}_{\pm, n} \ .
\end{equation}
It is now possible to use the properties of the $sl(2,R)$ algebra to derive  the exponential suppression of the elastic scattering of an unexcited string mentioned in~\eqref{eq:bDscale}. Starting from
\begin{equation}
\label{eq:3.17}
e^{2i\hat{\delta}_0} \simeq e^{2i\delta_0}\, \exp\left\{ \sum_{n=1}^\infty \frac{i \alpha'}{n} \left[\mu_0^2 \left(2 T^{\hat{y}_0}_{0,n} - T^{\hat{y}_0}_{+,n} - T^{\hat{y}_0}_{-,n} -1\right) + \mu_{\hat{y}}^2 \sum_{i=1}^{d-p-3} \left(2 T^{\hat{y}_i}_{0,n} - T^{\hat{y}_i}_{+,n} - T^{\hat{y}_i}_{-,n} -1 \right) \right] \right\}
\end{equation}
we can use the identity
\begin{equation}
\label{eq:Tsplit}
e^{x \left(2 T^{\hat{y}_i}_{0,n} - T^{\hat{y}_i}_{+,n} - T^{\hat{y}_i}_{-,n} \right)} =  e^{-\frac{x}{1-x} T^{\hat{y}_i}_{+,n}} \ e^{- 2\ln(1 - x) T^{\hat{y}_i}_{0,n}}  \ e^{- \frac{x}{1-x} T^{\hat{y}_i}_{-,n}} 
\end{equation}
to calculate the elastic amplitude $\langle 0| S |0\rangle$. The terms proportional to $T^{\hat{y}_i}_{\pm,n}$ vanish when acting on the unexcited initial/final state and the same happens for the oscillator contribution to $T^{\hat{y}_i}_{0,n}$, see~\eqref{eq:TpmT0}. Then \eqref{eq:3.17} reads
\begin{equation}
\label{eq:3.20}
\langle 0 | e^{2i\hat{\delta}_0}| 0\rangle \simeq e^{2i\delta_0}\,\prod_{n=1}^\infty \left\{  \frac{e^{-\frac{i \alpha' \mu_0^2}{n}}}{1-\frac{i \alpha' \mu_0^2}{n}}  \left(\frac{e^{-\frac{i \alpha' \mu_{\hat{y}}^2}{n}}}{1-\frac{i \alpha' \mu_{\hat{y}}^2}{n}}\right)^{d-p-3} \right\}\;.
\end{equation}
By using Weierstrass's formula for the $\Gamma$-function and then taking the absolute value square which is directly related to the probability of the transition, we obtain
\begin{equation}
\label{eq:3.21}
\begin{aligned}
	|\langle 0 | e^{2i\hat{\delta}_0}| 0\rangle|^2 & \simeq \left| e^{2i\delta_0} \Gamma\left(1-i\alpha' \mu_{0}^2\right) \Gamma^{d-p-3}\left(1-i\alpha' \mu_{\hat{y}}^2\right) \right|^2 \\
	& \simeq e^{-2{\rm Im} 2\delta_0}\left( \frac{\pi \alpha' \mu_0^2}{\sinh\left(\pi\alpha' \mu_0^2\right)} \right)  \left( \frac{\pi \alpha' \mu^2_{\hat{y}}}{\sinh\left(\pi\alpha' \mu_{\hat{y}}^2\right)} \right)^{d-p-3} 
\end{aligned}
\end{equation}
where we used $\Gamma(1 + ix) \Gamma(1 - ix) = \frac{\pi x}{\sinh \pi x}$. As discussed in Section~\ref{ssec:string-brane}, the imaginary part of the leading eikonal is negligible when $b \gg l_s(E_s)$, so in this regime we can focus on the last two factors. In the leading eikonal approximation $b\gg R_p$, we have
\begin{equation}
\label{eq:3.25}
|\langle 0 | e^{2i\hat{\delta}_0}| 0\rangle|^2 \simeq   (d-p-3)\, (2\pi \alpha' |\mu_{\hat{y}}^2|)^{d-p-2}\, e^{-2\pi\alpha' |\mu_0^2|}
\end{equation}
where we used the last identity in the second equation \eqref{eq:mu0muy}.
Then, by using~\eqref{eq:mu0muy}, we obtain~\eqref{eq:bDscale} in the regime where $\mu_0^2$ (and $|\mu_{\hat{y}}^2|$) is large, i.e.~$b_t\gg b$.  

In summary, when the impact parameter is below the scale~\eqref{eq:bDscale}, internal bosonic excitations of the string along the directions $\hat{y}$ and $\hat{b}$ are excited according to parameters summarized in~\eqref{eq:mu0muy}, while no excitations are created at this order along the remaining direction. As we will see in the next section, the same result can be recovered by studying the propagation of a string in a curved background.

\subsubsection{A geometric description of the eikonal scattering}
\label{ssec:PLstract}

In the QFT setup  the probe limit of the  eikonal has a simple geometric interpretation as it is entirely captured by the classical motion in the background sourced by the heavy object, see Section~\ref{ssec:probeQ} and~\ref{app:probelim}. It is natural to expect that a similar pattern holds also for the string eikonal operator and the aim of this section is to show that this is indeed the case. Again we focus on the case of the string-brane scattering that has been the main example analyzed, but the same idea applies more generally (see Section~\ref{beyondtree} below). It is natural to see stack of D$p$-branes as the heavy object and the scattering string state as the probe. Thus the first ingredient we need is the supergravity solution describing the gravitational backreaction induced by the D$p$-branes. In the string frame it reads~\cite{Duff:1994an}
\begin{equation}
\label{eq:pbrsol}  
\begin{aligned}
	d s^2  & = \left[{H(r)}\right]^{-\frac{1}{2}}\, \eta_{\alpha\beta}dx^\alpha dx^\beta + \left[{H(r)}\right]^{\frac{1}{2}} \, \delta_{ij}d x^idx^j \ ,
	\\
	{\rm e}^{2{\phi}(x)} & = g_s^2 \left[ H(r) \right]^{\frac{3-p}{2}} \ , \quad
	C_{01\cdots p}(x) =  \frac{1}{H(r)} - 1 \ ,
\end{aligned}
\end{equation}
where the indices $\alpha,\beta=0,\ldots,p$ are along the D$p$-brane worldvolume while the indices $i,j=p+1,\ldots,d-1$ span the transverse directions. The fields in the second line are the dilaton ($\phi$) and the Ramond--Ramond (RR) $p+1$-form potential that couples minimally to the D$p$-branes and the harmonic function $H(r)$ is
\begin{equation}
\label{eq:Hhf}
H(r) = 1 + \left(\frac{R_p}{r}\right)^{d-p-3},\qquad r^2 = x_i^2\,,
\end{equation}
where the scale $R_p$ is given by~\eqref{eq:Rpscale} in terms of the string coupling $g_s$ and the number of D$p$-branes $N$. The classical motion of a point-like object in the geometry above is discussed in~\ref{app:geoDp}, but we now need to generalize this analysis to the full string case starting from a string action that includes the couplings to the non-trivial background fields. The presence of RR fields makes the exact description more complicated as one would need to resort to the Green--Schwarz~\cite{Green:1981yb,Green:1982tc,Green:1983hw,green1988superstring} or the pure-spinor (see e.g.~\cite{Berkovits:2002zk,Mafra:2022wml}  and references therein) formalisms. However the analysis of the Reggeon vertex in Section~\ref{ssec:reggeon} suggests that the fermionic degrees of freedom on the worldsheet do not play any role at the level of the leading eikonal as the string excitations induced by the tidal forces are fully captured by the bosonic field $X^\mu$. Thus we can focus just on the universal part of the (gauge fixed) string action
\begin{equation}
\label{eq:stract_u}
S = -\frac{1}{4\pi \alpha'} \int d\tau d\sigma \left[\partial_\tau X^\mu \partial_\tau X^\nu - \partial_\sigma X^\mu \partial_\sigma X^\nu\right] g_{\mu\nu}\;,
\end{equation}
where the metric is the one given in the first line of~\eqref{eq:pbrsol}. Even this simplified starting point contains much more information than needed for our purposes since we are interested in the classical dynamics of an energetic string probe. In this case we can follow the approach of~\cite{Blau:2002mw} and approximate the background metric by taking just its behavior around the null geodesic describing at leading order the trajectory of the string center of mass. We sketch the main steps of this approach below.

First it is convenient to choose coordinates adapted to the motion of the center of mass which takes place in a plane of the transverse space. In polar coordinates~\eqref{eq:Dpmetpol} this plane is parametrized by $r$ and $\phi$ and it is convenient to choose
\begin{equation}
\label{eq:adapcoor}
du = \frac{\sqrt{H}}{F} dr\;,\qquad dv = -dt + b_J\, d\phi + F dr\,,\qquad dz = d\phi + \frac{d\bar\phi}{du}\, du\,,
\end{equation}
where we introduced the shorthand notation $F=\sqrt{H(r) -\frac{b_J}{r^2}}$ and $\bar\phi \equiv \phi(u)$ is the value of the angle along the geodesic parametrized by $u$. So, from~\eqref{eq:adapcoor} and~\eqref{eq:4.2} we have
\begin{equation}
\label{eq:dphidu}
\frac{d\bar\phi(u)}{du} = \frac{d\bar\phi}{d\bar r} \frac{d\bar r}{du} = - \frac{b_J}{\bar{r}^2 \sqrt{H(\bar r)}}\;,
\end{equation}
where, following the convention above, $\bar r$ is the value of $r$ along the geodesic describing the motion of the string center of mass. This is at constant values of $v$, $z$ and the $(d-p-3)$ coordinates orthogonal to the scattering plane that we indicate with $y^i$. In these coordinates the metric in~\eqref{eq:pbrsol} takes the following form
\begin{equation}
\label{eq:4.12}
ds^2 = 2 du\, dv - \frac{dv^2}{\sqrt{H}} + \frac{2b\, dv\, dz}{\sqrt{H}} + \frac{(r F)^2}{\sqrt{H}} dz^2+ \frac{dx_a^2}{\sqrt{H}} + r^2 \sqrt{H} \sin^2(\phi-\bar\phi) d\Omega^2_{d-p-3}\;.
\end{equation}
A similar change of coordinates can be implemented on the RR gauge field and then it is possible to take the Penrose limit on the full solution by introducing a small parameter $\epsilon$ and rescaling the coordinates $(x^a,y^i,z)\to \epsilon (x^a,y^i,z)$ together with $v \to \epsilon^2 v$. It was shown in~\cite{Blau:2002mw} that in the limit $\epsilon \to 0$, the leading term is ${\cal O}(\epsilon^2)$ and yields a solution of the same supergravity field equations relevant for the original configuration~\eqref{eq:pbrsol}. In this limit the geometry~\eqref{eq:4.12} reduces to 
\begin{equation}
\label{eq:4.12bis}
ds^2 \simeq 2 du dv + \frac{(\bar{r} F(\bar{r}))^2}{\sqrt{H(\bar{r})}} dz^2+ \frac{dx_a^2}{\sqrt{H(\bar{r})}} + \bar{r}^2 \sqrt{H(\bar{r})} \sin^2(\bar\phi) dy^{i 2}\;.
\end{equation}
We can take this result in the canonical form by introducing the coordinates $\hat{y}_0$, $\hat{y}_i$ and $\hat{x}_a$ 
\begin{equation}
\label{eq:coorha}
\begin{aligned}  
	& z = \frac{H^{\frac{1}{4}}}{\bar{r} F} \hat{y}_0\;, \quad y^i = \frac{\hat{y}_i }{H^{\frac{1}{4}} \bar{r} \sin\bar\phi} \;, \quad x^a = H^{\frac{1}{4}} \hat{x}_a \\
	v = & \hat{v} + \frac{1}{2} \left[ \hat{y}_0^2 \, \partial_u \ln\left(\frac{\bar{r} F}{H^{\frac{1}{4}}}\right) + \hat{y}_i^2 \, \partial_u \ln\left(\bar{r} \sin\bar\phi H^{\frac{1}{4}}\right) + \hat{x}_a^2\,\partial_u \ln\left(H^{-\frac{1}{4}} \right) \right]
\end{aligned}
\end{equation}
obtaining
\begin{equation}
\label{eq:4.14}
ds^2 \simeq 2 du d\hat{v} + d\hat{y}_0^2+ d\hat{x}_a^2 + d\hat{y}_i^2 + \left({\cal G}_{\hat{x}} \hat{x}_a^2 + {\cal G}_{\hat{y}} \hat{y}_i^2 +  {\cal G}_{0} \hat{y}_0^2 \right) du^2\;,
\end{equation}
with
\begin{align} \label{eq:Gxy0}
{\cal G}_{\hat{x}} & =  \partial_u^2 \ln H^{-\frac{1}{4}} + \left(\partial_u \ln H^{\frac{1}{4}}\right)^2\;, \quad {\cal G}_{0} = \partial^2_u \ln\left(\frac{\bar{r} F}{H^{\frac{1}{4}}} \right) +  \left(\partial_u \ln \frac{\bar{r} F}{H^{\frac{1}{4}}} \right)^2\;, \\
& {\cal G}_{\hat{y}} = \partial^2_u \ln\left(\bar{r} \sin\bar\phi H^{\frac{1}{4}} \right) +  \left(\partial_u \ln \bar{r} \sin\bar\phi H^{\frac{1}{4}} \right)^2\;.
\end{align}
In order to capture the dynamics of energetic semiclassical string probes it is then sufficient to use the action
\begin{equation}
\label{eq:4.15}
S = S_\eta  - \frac{1}{4\pi\alpha'} \int d\tau \int_0^{2\pi}\!\! d\sigma\, \eta^{\alpha\beta} \partial_\alpha U \partial_\beta U \, \left({\cal G}_{\hat{x}} \hat{X}_a^2 + {\cal G}_{\hat{y}} \hat{Y}_i^2 +  {\cal G}_{0} \hat{Y}_0^2 \right)\;,
\end{equation}
where $S_\eta$ is the string action in flat space ({\rm i.e} Eq.~\eqref{eq:stract_u} with $g_{\mu\nu}=\eta_{\mu\nu}$). As standard, we are using uppercase symbols to indicate the string embedding fields corresponding to the various coordinates. The string analysis is greatly simplified by choosing a light cone gauge where the string oscillations along $U(\sigma,\tau)$ are gauged away so that we have
\begin{equation}
\label{eq:4.16}
U(\sigma,\tau) = \alpha' E_s \tau\;,
\end{equation}
so we can keep using a lowercase symbol to indicate this coordinate. Then, at high energies, the interaction between the string probe and the background becomes localized at $\tau \sim 0$ since $\tau = u/(\alpha' E_s)$, so by changing variables from $\tau$ to $u$, we can write
\begin{equation}
\label{eq:4.17}
S \simeq S_\eta  - \frac{E_s}{2} \int_{-\infty}^\infty du \int_0^{2\pi}\!\! d\sigma\, \left({\cal G}_{\hat{x}} \hat{X}_a^2(\sigma,0) + {\cal G}_{\hat{y}} \hat{Y}_i^2(\sigma,0) +  {\cal G}_{0} \hat{Y}_0^2(\sigma,0) \right).
\end{equation}
As usual for semiclassical strings captured by a Penrose limit, the worldsheet action is described by free massive fields where the effective mass square parameters are
\begin{subequations}
\label{eq:cxy0}
\begin{align}
	\frac{E_s}{2} \int_{-\infty}^\infty \!\! du \; {\cal G}_{\hat{x}} & = E_s \int_0^\infty \! du\, \left[\partial_u^2 \ln H^{-\frac{1}{4}} + \left(\partial_u \ln H^{\frac{1}{4}}\right)^2\right] \simeq 0 \;, \label{eq:cxy0a} \\
	\frac{E_s}{2} \int_{-\infty}^\infty \!\!du \; {\cal G}_{\hat{y}} & = E_s \int_0^\infty \! du\,\left[\partial^2_u \ln\left(\bar{r} \sin\bar\phi H^{\frac{1}{4}} \right) +  \left(\partial_u \ln \left(\bar{r} \sin\bar\phi H^{\frac{1}{4}} \right)\right)^2\right]\simeq
	\mu_{\hat{y}}^2\;, \label{eq:cxy0b} \\
	\frac{E_s}{2} \int_{-\infty}^\infty \!\!du\;  {\cal G}_{0} & = E_s \int_0^\infty \! du\,\left[\partial^2_u \ln\left(\frac{\bar{r} F}{H^{\frac{1}{4}}} \right) +  \left(\partial_u \ln \frac{\bar{r} F}{H^{\frac{1}{4}}} \right)^2\right] \simeq 
	\mu_{0}^2 \;, \label{eq:cxy0c}
\end{align}
\end{subequations}
where the final results are valid at 1PM (i.e.~at order $(R_p/b)^{d-p-3}$) and the parameters $\mu_0^2$, $\mu_{\hat{y}}^2$ have been introduced in~\eqref{eq:mu0muy}. In the equations above, we restricted the integration over $u$ from infinity to the inversion point $r_*$ (which corresponds to $u=0$) and included the overall factor of two to account for the contribution of the second part of the trajectory from $r_*$ back to infinity. The two  integrals in~\eqref{eq:cxy0a} vanish at 1PM order. 
The second  one  can be neglected since it scales as $(R_p/b)^{2(d-p-3)}$, while the first one can be written as a total derivative of a function that vanishes at the extremes, as done below
\begin{equation}
\label{eq:coev}
\int_0^\infty \! du\, \partial_u^2 \ln H^{-\frac{1}{4}} = \int_{r_*}^\infty \!\! dr\; \partial_r \left(\frac{F}{\sqrt{H}} \partial_r \ln H^{-\frac{1}{4}}\right) = 0 \,.
\end{equation}
Here we used~\eqref{eq:adapcoor} to rewrite $\partial_u$ as $(\partial_u r) \partial_r$ and the special values $F(r_*)=0$, while $H(r_*)$ is finite, $F(\infty)=1$ and $\partial_r H\to 0$ as $r\to\infty$. Evaluating the remaining two integrals requires a slightly more detailed discussion which is summarized at the end of~\ref{app:geoDp} and the result is summarized in~\eqref{eq:cxy0b} and~\eqref{eq:cxy0c}. Thus, as promised, we rederived from a geometric analysis the parameters \eqref{eq:mu0muy} determining the strength of the tidal excitations. It would be interesting to extend the comparison between the eikonal operator and geometric description of this section at subleading PM order to see whether the geometric interpretation of the string excitations holds beyond the leading eikonal.

\subsubsection{String eikonal and classical causality}
\label{ssec:seikcc}

In the previous sections we discussed how the leading string eikonal exponentiates as an operator and what novelties this implies with respect to the field theory case. This exponentiation holds in the case of bosonic string theory, see~\ref{app:string} for more details, which provides an interesting setup to study the higher derivative corrections to the on-shell 3-graviton vertex discussed in~\ref{ssec:ethd} from an EFT point of view. The main point we wish to convey is the following: the string-brane scattering in bosonic string theory provides {\em classical} observables that are affected by $\alpha'$-corrections that take the form of~\eqref{eq:3am2} and~\eqref{eq:3am4} with $\ell_2 \sim \ell_4\sim \ell_s$. A naive field-theory limit yields causality violations, in the form of negative time delays, when the impact parameter is of the order of $\ell_{2,4,s}$~\cite{Camanho:2014apa}. As discussed below, in the eikonal context, this can be seen by looking at the energy derivative of the leading eikonal. Equivalently, one can look at deflection angle and notice that the higher derivative corrections seem to make gravity repulsive when $b\sim \ell_{2,4,s}$. We will show how this pathological behavior is avoided in string theory.

A first observation is that, in bosonic string theory, the 3-graviton amplitude takes the form of~\eqref{eq:3grmod} with $\ell_4^2 = -2\ell_2^2 = \frac{\alpha'}{2}$. Then we saw that, at the level of the leading eikonal, string effects are captured by the shift $b\to b+\hat{X}$~\eqref{eq:hatdelta0cl}. By applying this recipe to the long-range eikonal~\eqref{eq:bbigls}
\begin{equation}
  \label{eq:ldisd0}
  2\delta_0 \sim \frac{\sqrt{\pi} E_s}{2} \frac{\Gamma \left(\frac{d-p-4}{2}\right)}{\Gamma\left( \frac{d-p-3}{2}\right)} \frac{R_p^{d-p-3}}{b^{d-p-4}}\;
\end{equation}
one obtains an eikonal operator that is directly related to the EFT eikonal~\eqref{eq:d0m1gl2l4}: the latter can be obtained from the former simply by taking the expectation value in the massless sector
\begin{equation}
  \label{eq:d0smles}
  \langle \varepsilon_1, \tilde{\varepsilon}_1| 2 \hat{\delta}_0(b+\hat{X}) |  \varepsilon_2, \tilde{\varepsilon}_2\rangle\;,
\end{equation}
where $| \varepsilon_a \tilde{\varepsilon}_a\rangle =  \varepsilon_{a\,i} \tilde{\varepsilon}_{a\,j} A_{-1}^i \bar{A}_{-1}^j |0\rangle$. 

However, this result does not describe the elastic amplitude in the classical regime since one needs instead to take the expectation value of the exponentiated eikonal $e^{2i\hat\delta_0}$ as discussed in Section~\ref{ssec:tidalrev}. The novelty here is that $e^{2i\hat\delta_0}$ is not a diagonal phase even in the massless sector. Thus in order to calculate the deflection angle of a massless state, one should first diagonalise $\hat\delta_0$ in the massless sector and then apply the usual relation~\eqref{eq:thetap} for the deflection angle or~\eqref{eq:Std} for the time delay.

Let us then start by analysing the EFT eikonal~\eqref{eq:d0m1gl2l4}. The key observation of~\cite{Camanho:2014apa} is that regardless of the signs of the higher derivative corrections ($\ell_2^2,\ell_4^4 \gtrless 0$) there always exists an eigenstate of the EFT eikonal with a {\em negative} eigenvalue. In order to show this let us focus on the eigenvetors that lie in the space of the two states with polarizations
\begin{equation}
  \label{eq:2deig}
  \varepsilon_i \tilde{\varepsilon}_j \to \epsilon^a_{ij} = \frac{1}{\sqrt{D-3}}\left(\delta_{ij} - \hat{b}_i \hat{b}_j\right)\;,\quad \varepsilon_i \tilde{\varepsilon}_j \to \epsilon^b_{ij} = \hat{b}_i \hat{b}_j
\end{equation}
(let us recall that $\hat b_i = b_i / b $).
This space is orthogonal to the other physical states and so it can be studied separately. From~\eqref{eq:d0m1gl2l4}, one obtains 
\begin{equation}
  \label{eq:M2t2}
\begin{aligned}
  \left(
    \begin{array}{cc}
      2\hat{\delta}_0^{aa} &  2\hat{\delta}_0^{ab} \\  2\hat{\delta}_0^{ba} &  2\hat{\delta}_0^{bb}
    \end{array}
\right)  = 2 \delta_0 & \left(
  \begin{array}{cc}
    1+ 2 c_2 + (D-1) c_4 & - (D-1)\sqrt{D-3} \, c_4 \\
    - (D-1)\sqrt{D-3} \,c_4 & 1 -2(D-3) c_2 + (D-1) (D-3) c_4 
    \end{array}
\right),
\\
& c_2 = (D-2) \frac{2\ell_2^2}{b^2} \;, \quad c_4 = D(D-2) \frac{\ell_4^4}{b^4} \;.
\end{aligned}
\end{equation}
Let us also emphasize that the matrix appearing in \eqref{eq:M2t2} is symmetric and real.
One can check that the determinant of~\eqref{eq:M2t2} always becomes negative when $|c_2|,|c_4| \gtrsim 1$, which means that the two eigenvalues 
\begin{equation}
  \label{eq:negeig}
  1 - (D- 4) c_2 + \frac{1}{2} (D - 2) (D - 1) c_4 \pm \frac{D-2}{2} \sqrt{4 c_2^2 -4 \frac{(D-4) (D-1)}{D-2} c_2 c_4 + (D-1)^2 c_4^2}\;
\end{equation}
have generically opposite sign and one of them must be negative. States corresponding to eigenvectors with a negative eigenvalue have a non-standard behavior: both the deflection angle and the Shapiro time delay due to the gravitational scattering would be negative as they are related to derivatives of the eikonal. This would happen when $b \gtrsim \ell_2, \ell_4$ which may seem to be within the range of validity of an EFT approach if $\ell_{2,4}\gg \ell_P$. However, the string theory analysis shows that, even when the higher derivative corrections in~\eqref{eqehr2r3} have a classical origin, this pathological behavior is avoided.

In order to see this, we must recall that the string eikonal~\eqref{eq:incg1} has a more complicated structure than the EFT eikonal used for the argument above. In particular when $b<l_s(E_s)\equiv\sqrt{\alpha'\ln(\alpha' E_s^2)}$ the string eikonal becomes a constant plus corrections proportional to $b^2$, see for instance~\eqref{lowb}. Then the corresponding eikonal operator, which is as before obtained with the shift $b\to b+\hat{X}$, differs from EFT eikonal at scales much larger than $b\sim \ell_s$: instead of growing at small distances, it becomes almost constant. The key ingredient for this is the Regge behavior which modifies the functional form  in the impact parameter as showed in~\eqref{impactb}. It is important that the effective string scale $l_s(E_s)$~\eqref{eq:lsE}, where the softer behavior kicks in, is enhanced by a factor of $\ln(\alpha' E_s^2)$ with respect to the scale of the higher derivative terms in the effective action. This is the effect of the exchange of the whole leading Regge trajectory and the same mechanism would not work in a theory with a finite number of extra higher spin states beyond the graviton \cite{Camanho:2014apa,DAppollonio:2015fly}.

\subsubsection{String-string scattering beyond tree level}
\label{beyondtree}

In Subsection~\ref{ssec:string-brane-sup} we have discussed the main features of high-energy string-string scattering at tree level and anticipated some results that follow from the exponentiation in $b$-space of the tree-level result discussed in this section.  Such an exponentiation indeed goes through in the case of string-string collisions as well: the only difference with respect to the case of string-brane collisions discussed in detail in Subsections~\ref{ssec:1loopstr} and \ref{ssec:tidalrev} is that now both incoming strings get excited through the exchange of the (gravi)reggeon: this is a consequence of factorization of Regge-pole residues, related, in turn, to $t$-channel unitarity. 
Without repeating here the discussion about tidal excitations given above in the case of string-brane collisions we simply give its analog for string-string scattering:
\begin{equation}
	\label{tidalS}
	S = \exp( 2 i \hat{\delta})
\end{equation}
with $\hat{\delta}$ now given in Eqns. (\ref{tidalop}), (\ref{delta}) (instead  of \eqref{eq:hatdelta0cl} ).
For  $b \gg l_s (s)$,  
$\hat{\delta}$ is Hermitian and thus  $S$ is unitary.
This result was first obtained in \cite{Amati:1987uf} by proving the exponentiation of the tree-level string scattering amplitude. An elegant rederivation \cite{deVega:1988ts, Veneziano:1988qq, Horowitz:1990sr, Giddings:2007bw} follows from considering the quantization of one string in the  Aichelburg--Sexl metric (see Section~\ref{ssec:ASmetric}) produced by the other string. The corresponding classical action takes the  form of \eqref{eq:stract_u} with $g_{\mu\nu}$ now given by (\ref{tidalop}).
	It is greatly simplified by choosing the gauge $ u = \sqrt{\alpha'} \tau$ which allows, thanks to the $\delta(u)$ in  \eqref{eq:AS}, to carry out the integration over $\tau$.
	This leaves just the $\sigma$ integral of the profile function $f(x_{\perp}(\sigma))$ of the transverse string coordinates as in (\ref{tidalop}).
	Actually,  the explicit form of the eikonal operator, including $\alpha'$ corrections, allows one to reconstruct \cite{Veneziano:1988qq} the  generalization of the AS metric produced by the profile \eqref{profile}.

At a qualitative level we can estimate the importance of the string corrections by noting that the normal ordering in $\hat{\delta}$ produces, at the level of $S$, corrections to the eikonal phase of order $\frac {\partial^2 2 \delta_0 }{(\partial b)^2}\ell_s^2$  which are, naively, of relative order $\frac{\ell_s^2}{b^2}$ with respect to $2\delta_0$ itself.
The crucial point, however, is that some of these corrections are imaginary and thus cannot be neglected as soon as 
$\frac {\partial^2  2 \delta_0 }{(\partial b)^2} \ell_s^2$ becomes ${\cal O}(1)$. A simple calculation, using $2\delta_0$ in \eqref{leadingeik},  shows that this happens at a critical value of $b$ --that we denote by $b_t$-- given by:
\begin{equation}
	\label{bt}
	b_t^{D-2} \sim  \frac{G s}{\hbar} \ell_s^2 \gg \ell_s^{D-2} ~~{\rm for}~~  \frac{G s}{\hbar} \gg \ell_s^{D-4}
\end{equation}
with $R^{D-3}\sim G \sqrt{s}$.
The above result can also be guessed from the two following qualitative physical arguments. One of them  goes as follows.
The Newton potential acting on two points of a string at a transverse distance $2\Delta b$ from each other is given by
\begin{equation}
	U (b \pm \Delta b) = \left( \frac{R}{b \pm \Delta b}\right)^{D-3} \Longrightarrow U (b+\Delta b) - U(b-\Delta b) = 2 \frac{R^{D-3}}{b^{D-2}} \Delta b\,.
	\label{Ub1}
\end{equation}
Taking the distance $2 \Delta b$ of the two points on the string to be of order $\ell_s$, and assuming that the two points  move at the speed of light, we can compute the difference  between the two forces acting on them. The tidal forces start to be relevant when such a difference is equal to the string tension times their distance $\ell_s$:
\begin{equation}
	2 \frac{R^{D-3}}{b_t^{D-2}}\sqrt{s} \ell_s \sim T \ell_s \Longrightarrow b_t^{D-2} \sim  2 \frac{ \sqrt{s} R^{D-3} }{T} \sim 4\pi  \alpha' G s \sim \frac{4\pi G s}{\hbar}\ell_s^2
	\label{UB2} 
\end{equation}
where we have used that $R^{D-3} \sim G \sqrt{s}$ and $T= \frac{1}{2\pi \alpha'}$.

Alternatively,  \eqref{bt}  can be obtained by considering how the different bits of one string (say the one of energy $E_1$) suffer, as a result of the metric produced by the other string (of energy $E_2$), slightly different deflection angles.  The  spread $\Delta \Theta_1$ of these deflection angles is roughly:  
\begin{equation}
	\Delta \Theta_1 \sim \frac{G E_2}{b^{D-3}} \frac{\ell_s}{b} = G E_2 \ell_s b^{2-D}\,.
\end{equation}
Such a spread corresponds to an  invariant excitation mass $\Delta M_1 \sim E_1 \Delta \Theta_1 \sim 
G E_1 E_2 \ell_s b^{2-D}$.  Requiring $\Delta M_1 \sim M_s$ (the mass of the first excited level) leads to the same estimate (\ref{bt}) for $b_t$ after using $s \sim E_1 E_2$ as in \eqref{phase}. 

Quantitatively, the phenomenon is described by the same gravi-reggeon vertex operator defined in \eqref{rv0}, acting now on both external strings.  After exponentiation of the tree-level amplitude in $b$-space, the resulting $S$-matrix is explicitly unitary within a Hilbert space containing just  two-(massive or massless) strings. We just stress here that, although single gravi-reggeon exchange can only excite a small number of string oscillators, the full $S$-matrix \eqref{tidalS} will populate a large number of excited states. Some details  of this $S$-matrix have already been discussed in Section~\ref{ssec:tidalrev}.  We just mention that the elastic amplitude gets suppressed as a result of the opening of the inelastic channels. One finds:
\begin{equation}
	\label{tidalabs}
	|A_\text{el}| \sim e^{ - \frac{G s}{\hbar}  \frac{\ell_s^2}{b^{D-2}}}\sim e^{- (\frac{b_t}{b})^{D-2}}  \ll 1 ~,~ ( b \ll  b_t)\,.
\end{equation}

The exponential suppression of the elastic amplitude for $b < b_t$ implies, by unitarity, that inelastic final states, consisting of two tidally excited strings, are copiously produced. The differential cross section $d\sigma / (dM_1 \, dM_2)$ for producing two massive strings of mass $M_1$ and $M_2$ (summed over the degeneracy of each mass level) was studied in  \cite{Amati:1987uf}. Interestingly enough, the cross section starts growing like the (exponential) degeneracy of each mass level as if all states of a given mass were democratically produced. However, such a growth stops when $M_i \sim M_s (b_t/b)^{D-2}$. The distribution has a maximum around that value of $M_{1,2}$ and then falls off as a Gaussian.

A much more detailed analysis of the tidal excitation spectrum was made in \cite{DAppollonio:2013mgj} for the case of string-brane scattering. This shows how the apparently incoherent sum of different contributions actually comes from a unitary evolution of the system. Indeed, each gravi-Reggeon exchange induces elementary transitions obeying very simple selection rules while the final spectrum takes its apparently statistical form  as a consequence of the eikonal exponentiation.

In the above discussion we have not considered graviton (and other massless particle) emission, a topic discussed in the following sections in the point particle limit. As one lowers $b$, radiation phenomena become important and things get increasingly complicated. The parameter controlling the importance of radiative corrections, relative to the leading order deflection, is formally $(R^2/b^2)^{D-3}$ but, once again, they become relevant as soon as their contribution to the imaginary part of the eikonal phase becomes ${\cal O}(1)$. A naive estimate of the critical impact parameter $b_r$ at which this happens is given by
	\begin{equation}
		\label{br}
		\frac{G s}{\hbar} b_r^{4-D}  \left(\frac{R}{b_r}\right)^{2(D-3)} \sim1 \,,\quad \text{i.e.}\quad b_r \gg R ~{\rm for} ~ \frac{G s}{\hbar} b_r^{4-D} \gg 1\,.
	\end{equation}
	This is a fair estimate of $b_r$ for quantities that do not suffer from infrared divergences which is generically the case for $D>4$. For $D=4$, instead, many quantities are infrared sensitive: for instance, the elastic amplitude goes to zero together with the IR regulator. The estimate (\ref{br}), however, should still apply to infrared safe quantities, such as the total radiated energy. Comparing now $b_t$ of (\ref{bt}) with $b_r$ of (\ref{br}) we see that which kind of corrections comes first as one lowers $b$, depends on $D$ and on the ratio $R/\ell_s$: as shown in Fig.~\ref{fig:StringRegimes}, radiative corrections take the upper (lower) hand at small (large) $D$. In any case the above discussion  would imply that, in going towards  $b \sim R$, we cannot keep just the string-size corrections and neglect the classical radiative corrections. Unfortunately, to this date, no calculation taking both corrections simultaneously into account is available.

Fortunately, a big simplification occurs if we lower $b$  as much as we want while keeping $R < \ell_s$ (thus entering region II in fig. \ref{fig:StringRegimes}). In this case, string-size effects should replace the expansion parameter $R^2/b^2$ by  $R^2/(b^2 + \ell_s^2) < R^2/ \ell_s^2 <1$ justifying a perturbative (yet stringy) approach.\footnote{This is still a physically motivated guess leading, as we shall see, to very sensible consequences. It would be interesting to have further explicit checks of its validity.} This amounts to saying that string-size corrections shield the bad short-distance regime of local quantum gravity.

Let us list now the most significative results in this string-gravity regime: 
\begin{itemize}
	\item The classical deflection angle reaches a maximal value $\Theta_M$ given by
	\begin{equation}
		\label{ThetaM}	
		\Theta_M \sim  (R /l_s(s))^{D-3} \sim \frac{G \sqrt{s}}{ l_s^{D-3}(s)}  \,  ,
	\end{equation}		
	which is reached for $b \sim l_s (s)$, i.e.~when the two strings graze each other. This result simply comes from the smooth behavior of the stringy eikonal at small $b$, Eq.~\eqref{lowbstst}. Indeed, taking the derivative of $\operatorname{Re} 2\delta$ with respect to $b$, one obtains (see \eqref{eq:ascsf}):
	\begin{equation}
		\label{smallbdefl}
		\Theta = \frac{8G \sqrt{s} \,b}{(D-2) \pi^{\frac{D-4}{2}}\, l_s^{D-2}(s)}  + \cdots \,,\qquad 
		b \ll l_s(s) \,  ,
	\end{equation}	
	which grows monotonically with $b$. It reaches the above-mentioned maximal value at  $ b \sim  l_s(s)$ before starting to decrease according to the second equation in \eqref{eq:ascsf}.

	\item For $\Theta > \Theta_M$ there is no real saddle point in $b$ and, as a result, the elastic amplitude is exponentially suppressed. In order to find the actual damping of the elastic amplitude one has to find the dominant complex saddle point. This was done in Ref. \cite{Amati:1988tn} of which we give below a more streamlined version. 
	Let us rewrite \eqref{impactb1} in the form:
	\begin{equation}
		2 \delta_0 ( s, b) =  \frac{Gs (b^2)^{\frac{4-D}{2}}}{\pi^{\frac{D-4}{2}}} \left[\Gamma \left(\frac{D-4}{2}\right) 
		- \int_{\frac{b^2}{Y_c}}^{\infty} dt \,e^{-t} t^{\frac{D-4}{2} -1}\right]
		\label{GMO1}~~;~~ Y_c = l_s^2 (s) - i \pi \alpha' \, .
	\end{equation}
	The first term is the field theory leading eikonal for the scattering of massless particles. It is easy to show that, after adding to it the tidal corrections, such term cannot provide a relevant complex saddle point. This can come, instead, from  the second term,  denoted by $2\delta_0^{(s)}$, which contains string-size (i.e.~finite-beam, see \eqref{profile})  effects. Indeed, at large complex values of $b$ (i.e.~at $|b|^2 > l_s^2 (s)$),  $2\delta_0^{(s)}$ can be exponentially enhanced. Its large $|b|$ behavior is easily obtained by integration by  parts and reads
	\begin{equation}
		2 \delta_0^{(s)} ( s, b) = - \frac{Gs (b^2)^{\frac{4-D}{2}}}{\pi^{\frac{D-4}{2}}}  e^{- \frac{b^2}{Y_c}}\left( \frac{b^2}{Y_c}\right)^{-3+ \frac{D}{2}}= - \frac{Gs}{\pi^{\frac{D-4}{2}} b^2 Y_c^{\frac{D}{2} -3}}  e^{- \frac{b^2}{Y_c}}
		\label{GMO2}
	\end{equation}
	We can now go back to the amplitude in momentum/angle space by a standard Fourier transform. Keeping $\Theta \ll1$ and fixed we get:
	\begin{equation}
		A (s, \Theta) = \int d^{D-2} b \,\, e^{-ib \frac{\sqrt{s} \Theta}{2}} e^{- i Gs \pi^{- \frac{D-4}{2}} b^{-2} 
			Y_c^{3- \frac{D}{2} }  e^{- \frac{b^2}{Y_c}}}\,.
		\label{GMO3}
	\end{equation}
	Looking for a (complex) saddle point gives the equation:
	\begin{equation}
		\frac{\sqrt{s}}{2} \Theta = \frac{2Gs}{\pi^{\frac{D-4}{2}}} 
		\frac{Y_c^{\frac{1}{2}} e^{-\frac{b^2_s}{Y_c}}}{Y_c^{\frac{D-3}{2}} b_s} = \frac{\sqrt{s}}{2} \Theta_M e^{-\frac{b^2_s}{Y_c}} \left(\frac{4Y_c^{\frac{1}{2}}}{ b_s\pi^{\frac{D-4}{2}}} \right) \,,\qquad 
		\Theta_M \equiv \frac{G \sqrt{s}}{ l_s^{D-3}(s)}\, .
		\label{GMO4}
	\end{equation}
	It is now clear that for $\Theta \gg \Theta_M$
	the saddle point has to be predominantly imaginary and with a negative imaginary part in order to produce damping.  Writing $b_s =-i \beta_s$ the previous equation becomes:
	\begin{equation}
		- i \frac{\Theta}{\Theta_M} = \frac{4}{\pi^{\frac{D-4}{2}}} \frac{Y_c^{\frac{1}{2}}}{\beta_s} e^{\frac{\beta_s^2}{Y_c}} \Longrightarrow \log \left( -i \frac{\Theta}{\Theta_M}\right) = \frac{\beta_s^2}{Y_c} + \log \left(\frac{4Y_c^{\frac{1}{2}}}{ \beta_s\pi^{\frac{D-4}{2}}} \right)\,.
		\label{GMO5}
	\end{equation}
	The last term is subleading for  large values of $\frac{\beta_s}{\sqrt{Y_c}}$ and thus we get:
	\begin{equation}
		\beta_s = Y_c^{\frac{1}{2}} \sqrt{ \log \frac{\Theta}{\Theta_M} - \frac{i\pi}{2}} \Longrightarrow b_s = - i Y_c^{\frac{1}{2}} \sqrt{ \log \frac{\Theta}{\Theta_M} - \frac{i\pi}{2}} \,.
		\label{GMO6}
	\end{equation}
	Inserting this saddle point in \eqref{GMO3} we find that the first exponential dominates over the second one by a $\log \frac{\Theta}{\Theta_M} \gg 1$ and arrive at the final expression \cite{Amati:1988tn}:
	\begin{equation}
		| \mathcal{A}(s, \Theta)| \equiv |\mathcal{A}_{ACV}| \sim e^{- \frac{\sqrt{s}}{2} \Theta Y_c^{\frac{1}{2}} \sqrt{ \log \frac{\Theta}{\Theta_M} - \frac{i\pi}{2}}} \sim e^{- \frac{\sqrt{s}}{2\hbar} \Theta  l_s (s)  \sqrt{  \log \frac{\Theta}{\Theta_M}}}\,.
		\label{ACV/GMO}
	\end{equation}
	Taking into account that $\frac{\Theta_M}{\Theta} \sim g_s^2 \ll 1$, we can also approximate \eqref{ACV/GMO} by
	\begin{equation}
		|\mathcal{A}_{ACV}|  \sim e^{- \frac{\sqrt{s}}{2 \hbar} \Theta  \sqrt{ 2\alpha'} \sqrt{ \log (\frac{\alpha' s}{4}) \log g_s^{-2}}}\,.
		\label{ACV/GMO1}
	\end{equation}
	The parametric form of the suppression is  consistent with the behavior found, using a completely different method,  by Gross--Mende--Ooguri (GMO) \cite{Gross:1987kza, Gross:1987ar, Mende:1989wt} in the overlapping kinematical regime in which $\Theta$ is small and kept fixed at small $g_s$. We recall that the Gross--Mende approach \cite{Gross:1987kza, Gross:1987ar}  is based on the behavior at high-energy, fixed-angle, of the elastic (closed) string-string scattering amplitude at each loop level, generalizing the original tree-level (open string) result in \cite{Veneziano:1968yb}. This regime exhibits an exponential damping of the amplitude:
	\begin{equation}
		\label{fixedangtree}
		\mathcal{A}(s,t)_\text{tree}  \sim g_s^2 e^{- \frac{\alpha'}{2} s f(\Theta)}  \,,\qquad 
		f(\Theta) = 
		-\sin^2 \frac{\Theta}{2} \log \sin^2 \frac{\Theta}{2} -	\cos^2 \frac{\Theta}{2} \log \cos^2 \frac{\Theta}{2}  \ge 0 \, .
	\end{equation}
	This tree-level behavior is in tension, at least at small $\Theta$, with expectations based on classical gravitational deflection which would suggest an unsuppressed amplitude at high energy.
	Thus, at fixed angle, one is actually in the opposite situation with respect to the fixed-$t$ large-$s$ behavior where, as already mentioned, the tree-level result is too large to be compatible with unitarity bounds. We have seen how the eikonal resummation of loops in that kinematical regime is capable of restoring unitarity bounds. Similarly, loops also come to the rescue at fixed angle. As shown in \cite{Gross:1987kza, Gross:1987ar}, loop amplitudes, in spite of being of higher order in the string coupling, are enhanced, relative to the tree:
	\begin{equation}
		\label{fixedangloop}
		\mathcal{A}(s,t)_{h-\text{loops}}  \sim g_s^{2(1+h)} e^{- \frac{\alpha' s f(\Theta)}{2(1+h)} }\, ,	
	\end{equation}
	where the milder exponential overcompensates the extra powers of $g_s$ at sufficiently large $s$. That means, of course, that the perturbative series diverges at high energy and that a non perturbative resummation is needed. 
	
	In \cite{Mende:1989wt} it was shown that, in a finite high energy window, such a divergent series can be  Borel resummed.  It is precisely in such an energy window that we can compare the result of \cite{Mende:1989wt} to the one based on the leading eikonal in the stringy regime at small deflection angle. One finds quite an amazing agreement between the two results:
	\begin{equation}
		\label{ACVvsGMO}
		|\mathcal{A}_{ACV}| \sim e^{ - \sqrt{\frac{ \alpha' s}{2 }}  \Theta  \sqrt{\log(\frac{\alpha' s}{4}) \log(g_s^{-2})}}\,,
		\qquad |\mathcal{A}_{GMO}| \sim e^{ - \sqrt{\frac{ \alpha' s}{2 }}  \Theta  \sqrt{\log (\frac{4}{\Theta^{2}}) \log(g_s^{-2})}}\, ,
	\end{equation}	
	where the second equation follows from Eq. (3.31) of~\cite{Mende:1989wt} (which uses units in which $\alpha' = \frac12$) for small $\Theta$: 
	\begin{equation}
		|\mathcal{A}_{GMO} | \sim e^{- \sqrt{ 2 \alpha'  sf (\Theta) \log (g_s^{-2})}} \, .
	\end{equation}
	
	The difference (a $ \log(\frac{\alpha' s}{4})$ replacing a  $\log( \frac{4}{\Theta^{2}})$)  is already apparent in the transition between the fixed-$t$ ($(\frac{\alpha' s}{4})^{\frac{\alpha' t}{2}}$ with $t= -s \sin^2 \frac{\theta}{2}$) and  small fixed-$\Theta$ regime $(e^{- \frac{\alpha' s}{8} \Theta^2 \log \frac{4}{\Theta^2}})$ at tree level.
	Let us stress that the above damping in $|\mathcal{A}_{ACV}| $ is only reliable in the stringy regime and has to be modified at higher energies when gravitational radiation further damps the elastic channel. Such effects are also (presumably) neglected in the GMO approach. Furthermore, the Borel resummation leading to $|\mathcal{A}_{GMO}| $ could only be justified in a finite energy range. In Eq.~(3.31) of Ref.~\cite{Mende:1989wt}  
	is claimed to be valid in the window $\log g_s^{-2}  <  s/M_s^2 <g_s^{-\frac43}$.

	\item The impact parameters that can be actually probed at  high energy at fixed $\Theta$ never go below the string-length scale. Indeed, for $\Theta < \Theta_M$ one finds two real saddle points in $b$, one above and one below $\ell_s$, but the former is found to be the dominant one \cite{Amati:1988tn}. On the other hand, as we have just seen, for $\Theta >\Theta_M$ the imaginary saddle point \eqref{GMO6}  too has an absolute value  larger than $\ell_s$. All this  suggests an effective generalized uncertainty principle (GUP) holding true in string theory and reading \cite{Veneziano:1986zf}, \cite{Amati:1988tn},  \cite{Gross:1988jb}, \cite{Veneziano:1989fc}:
	\begin{equation}
		\label{GUP}
		\Delta X \ge \frac{\hbar}{\Delta P} + \alpha' \Delta P \ge 2 \sqrt{\alpha' \hbar} = 2  \ell_s \,.
	\end{equation}
	
	\item
	Let us now turn to the consequences of exponentiating the full tree-level string-string scattering amplitude including its imaginary part given in \eqref{Imdeltastst}  at  $b \le l_s (s)$.\footnote{Interestingly, as $b \to l_s (s)$, the imaginary part due to tidal excitations ``saturates" and becomes of the same order as the one due to formation of $s$-channel resonances. } This problem was already addressed in \cite{Amati:1987uf}, was further analyzed in \cite{Veneziano:2004er} (see also \cite{Veneziano:2005du}), and was also used more recently in \cite{Addazi:2016ksu}.
	
	The basic point is how to interpret the imaginary part of the exponentiated scattering amplitude at a fixed loop order in terms of its $s$-channel cuts  through the exchanged Reggeized gravitons. Fortunately, old work by Abramovsky, Gribov and Kancheli (AGK) \cite{Abramovsky:1973fm} (see also \cite{Koplik:1975iu}) provides  very simple ``cutting rules'' for the relative weights when cutting $m$ out of the total number $n$ of exchanged gravitons at $(n-1)$ loop order\footnote{Given that the topology of the diagram is that of a sphere with $(n-1)$-handles, any value of $0 \le m \le n$ is allowed.}.
	These rules (when directly applied in impact parameter space) tell us that the full imaginary part of the $n$-graviton exchange graph is the result of a contribution
	\begin{equation}
		\label{AGK1}
		\sigma_m^n = (-1)^{n-m}   \frac{(4{\rm Im} \delta )^n}{m! (n-m)!}\,,\qquad n=1, 2, \ldots\,,\qquad m = 1, 2, \ldots , n-1, n
	\end{equation}
	due to cutting $m$ out of $n$ gravi-Reggeons, and a contribution
	\begin{equation}
		\label{AGK2}
		\sigma_0^n = (-1)^{n}    \frac{(4{\rm Im} \delta)^n}{n! } - 2 {\rm Re} S^{(n)} \,,\qquad n=1, 2, \ldots \,,
	\end{equation}
	when no gravi-Reggeon is cut, where $S^{(n)} $ is the full $n$-gravi-Reggeon exchange contribution to the S-matrix.
	As the symbol $\sigma_m$ indicates, these are also to be interpreted as 
	cross sections into inelastic channels with $m$ cut gravi-Reggeons (i.e.~$m$ closed strings).
	It is trivial to check that, for any given $n$, the sum of all contributions from $m=0$ to $m=n$ gives back twice the full imaginary part of $T^{(n)} \equiv i(1-S^{(n)}) $, as it should.
	
	In analogy with what was done for the description of tidal excitations, we shall again promote the eikonal phase to an eikonal operator \cite{Veneziano:2004er} acting  on both the tidally-excited pairs of strings and, at a more ``coarse-grained" level, on a Hilbert space labeled by the number of cut gravi-Reggeons.
	A new unitary $S$-matrix reproducing the AGK cutting rules takes the form:
	\begin{equation}
		\label{ShatI}
		S = {\rm exp}(i \hat{I}) ~,
	\end{equation}
	where the Hermitian operator $\hat{I}$ is given by:
	\begin{equation}
		\label{hatI}
		\hat{I} = (\hat{\delta} + \hat{\delta}^{\dagger}) + \sqrt{-2 i (\hat{\delta} - \hat{\delta}^{\dagger})}   (C + C^{\dagger}) =  \hat{I}^{\dagger} ~ ,
	\end{equation}
	and the operators $\hat{\delta}$, $\hat{\delta}^{\dagger}$, $C$ and $C^{\dagger}$ satisfy the commutation relations
	\begin{equation}
		\label{CR}
		[C,  C^{\dagger} ]  = 1 \; , \;  [\hat{\delta} , \hat{\delta}^{\dagger}] = [C, \hat{\delta}] =  [C, \hat{\delta}^{\dagger}] = 0 ~.
	\end{equation}
	Using well-known harmonic-oscillator formulae, Eqs. (\ref{ShatI}), (\ref{hatI}) lead to the more convenient form:
	\begin{equation}
		\label{ConvS}
		S  =  {\rm e}^{2 i \hat{\delta}}   {\rm e}^{ i \sqrt{-2 i (\hat{\delta} - \hat{\delta}^{\dagger})} ~ C^{\dagger}}  
		{\rm e}^{i \sqrt{-2 i (\hat{\delta} - \hat{\delta}^{\dagger})} ~  C}  ~.
	\end{equation}
	Equations (\ref{hatI}), (\ref{ConvS}) imply that, as long as $\delta$ is real (i.e. for $b \gg l_s (s)$),  $\hat{\delta}$ is essentially Hermitian, the oscillators $C$, $C^{\dagger}$ are turned off, and one recovers the unitary $S$-matrix (\ref{tidalS}) with $\hat{\delta}$  given by the usual recipe (\ref{tidalop}).  On the other hand, as one goes to values of $b$ of order $ l_s (s)$ or lower, $\delta$ in \eqref{eq:bbigls} picks up an imaginary part and consequently (\ref{tidalS}) is no longer unitary, signaling the opening up of new channels. These correspond to unitarity cuts going through the gravi-reggeons themselves. In order to recover unitarity we should further extend the Hilbert space by including, in the final state, whatever goes on shell after cutting the gravi-reggeons. A detailed account of this ``whatever" being unavailable, we limit ourselves to counting the number of such gravi-reggeons through the number-counting operator $N_C \equiv  C^{\dagger} C$.
	
	Indeed, from (\ref{ConvS}) one can calculate the average number of heavy strings produced. For $\operatorname{Im} 2 \delta$ of equation \eqref{Imdeltastst}  of ${\cal O}(1)$ or smaller, i.e.~for $\sqrt{s} < M_s g_s^{-1}$, a single closed string carrying the whole center of mass energy  is formed.  Instead, in the energy interval $ M_s g_s^{-1} < \sqrt{s} < M_s g_s^{-2}$ the average number of produced strings grows like $\operatorname{Im} 2 \delta$, i.e.~like $s$. Consequently, by energy conservation, the average energy of each produced string decreases like $1/\sqrt{s}$. More precisely:
	\begin{equation}
		\label{<E>}
		\langle E \rangle \sim \frac{M_s^2  g_s^{-2}}{\sqrt{s}}\,.
	\end{equation}
	
	As one approaches the supposed threshold of black hole production $s \sim  M_s g_s^{-2}$ the most probable final state will consist of $g_s^{-2}$ strings of energy $M_s = \frac{\hbar}{\ell_s} $ which appears to be consistent with a smooth transition to a Hawking-like behavior in which the Hawking temperature matches the Hagedorn temperature of string theory and the string entropy matches the Bekenstein-Hawking entropy of a black hole of radius $\ell_s$. Unfortunately, the approximations made cease to be valid when the energy going across each cut Reggeon goes below the string  scale.		
	
	\item Comparatively less work has been devoted to the study of heavy string scattering. An exception can be found in  \cite{Veneziano:2012yj} where one considers the high-energy collision of a light/massless string on a heavy one (with the energy of the light string smaller than the mass of the heavy one). The idea is to take the heavy string mass to lie as close as possible to the so-called correspondence point , $M = M_s g_s^{-2}$,  where fundamental strings and black holes share many properties, including their entropy $\sim g_s^{-2} \gg 1$ (see \cite {Horowitz:1997jc, Damour:1999aw} and references therein).
	
	One then checks whether the probe string can be sensitive to properties of the heavy one other than its mass or spin. One finds that, at least below the corresponding point (where the heavy string's Schwarzschild radius is smaller than $\ell_s$), the light strings is also sensitive to the quadrupole moment of the heavy string, which therefore acts as some kind of ``quantum hair". If that feature would persist all the way till the correspondence point, and possibly beyond, this would imply that some quantum hair can be detected for stringy black holes. Unfortunately, this conclusion cannot be reached on firm grounds since the approximations used become unjustified precisely as one approaches the correspondence point.
	
\end{itemize}

\section{Unitarity and Radiation-Reaction }
\label{sec:radiation}

Radiation reaction effects are contributions are due to the effective force acting on the system as a result of the emission of gravitational waves (and of additional massless modes, in the supersymmetric case). Intuitively, as the two colliding objects deflect due to the mutual ``potential'' attraction, they emit energy and angular momentum due to Bremsstrahlung. This in turn further bends their trajectories, giving rise to dissipative terms in the eikonal. This reflects in additional contributions to the deflection angle, which appear to lowest order at two loops (3PM).  
This chapter and the next are devoted to the calculation of these interesting new effects.
While in the next chapter we will derive them from a complete two-loop calculation of the elastic scattering amplitude, in this chapter we follow very closely the approach of Refs.~\cite{DiVecchia:2021ndb,DiVecchia:2021bdo} and obtain them by combining the contribution to the unitarity relation of the three-particle cut, involving two massive particles and a graviton, with the properties of real analyticity and crossing symmetry of the elastic scattering amplitude.  We apply this method to both GR and massive ${\cal{N}}=8$ supergravity.  In particular, to obtain the radiation-reaction we do not need the complete three-particle cut, but only the part that is divergent as $\frac{1}{\epsilon}$  that is obtained from the five-point amplitude involving four massive particles and a graviton keeping only the leading  term in the Weinberg limit of small graviton momenta $k^\mu\to0$, which behaves as $\frac{1}{\omega}$  with $\omega=k^0$. 

The chapter is organized as follows. In the first section, starting from the unitarity relation, we discuss the two-particle and three-particle cuts in momentum and impact-parameter space. In the second section we explicitly compute  $\operatorname{Im} 2 \delta_2$ for both GR and massive ${\cal{N}}=8$ supergravity.
Finally, in the third section, we discuss the constraints of real analyticity that relate this imaginary part to the radiation-reaction effects.

\subsection{Unitarity in momentum space and in impact-parameter space} 

The unitarity of $S=1+iT$, i.e.~$S^\dagger S=1$ translates to
\begin{equation}\label{unitarityOP}
	-iT+iT^\dagger = T^\dagger T\,,
\end{equation}
whose matrix elements between states $|a\rangle$, $|b\rangle$ yield
\begin{equation}\label{}
	-i (2\pi)^D \delta(P_a-P_b) \left(\mathcal M_{a\to b}- \overline{\mathcal M}_{b\to a}\right)= \langle b|
	T^\dagger T 
	|a\rangle\,.
\end{equation}
Inserting a complete set of free states on the right-hand side and restricting to $P_a=P_b$, then gives
\begin{equation}\label{}
	-i\left(\mathcal M_{a\to b}- \overline{\mathcal M}_{b\to a}\right)= \sum_I  (2\pi)^D\delta(P_a+P_I) \mathcal M_{a\to I}\overline{\mathcal M}_{b\to I}\,,
\end{equation}
where the sum includes Lorentz-invariant phase-space integrals over all the intermediate states $j\in I$ and, in general, sums over their spins/helicities.
Since we will focus on the case in which $a\to b$ is an elastic $2\to2$ process for (real) massive scalars, for which the amplitude is a function $\mathcal A(s,t)$ of the two Mandelstam invariants \eqref{eq:mandvar}, we have
\begin{equation}\label{unitarity}
	2\operatorname{Im} \mathcal M_{a\to b} = \sum_I (2\pi)^D \delta(P_a+P_I) \mathcal M_{a\to I}\overline{\mathcal M}_{b\to I}\,.
\end{equation}
Let us write,
\begin{equation}\label{cuts}
	\operatorname{Im} \mathcal A(s,t)=
	\operatorname{Im}_{2pc} \mathcal A(s,t)
	+
	\operatorname{Im}_{3pc} \mathcal A(s,t)
	+\cdots
\end{equation}
and discuss each cut separately (see below).
Taking the Fourier transform of \eqref{cuts} according to \eqref{FTexact}, and noting that $\operatorname{FT}[\operatorname{Im} \mathcal{A}](s,b)=\operatorname{Im} \operatorname{FT}[{\mathcal A}](s,b)$ because $\mathcal A (s,-q^2)$ is a symmetric function of $q$, we get
\begin{equation}\label{cutsb}
	\operatorname{Im} \operatorname{FT}[ \mathcal A](s,b)=
	\operatorname{Im}_{2pc} \operatorname{FT}[ \mathcal A](s,b)
	+
	\operatorname{Im}_{3pc} \operatorname{FT}[ \mathcal A](s,b)
	+
	\cdots\ .
\end{equation}

\subsubsection{Elastic unitarity}
The elastic contributions come from the two-particle cut
\begin{equation}\label{2pcbasic}
	2\operatorname{Im}_{2pc} \mathcal A = \int d(\text{LIPS})_2 \,\mathcal A_{\text{L.}} \mathcal A_\text{R.}^\ast
\end{equation}
(where L. and R. stand for ``left'' and ``right'') or in the pictorial form
\begin{equation}\label{2pcqspacesimpl}
	2\operatorname{Im}_{2pc}
	\begin{gathered}
		\begin{tikzpicture}[scale=.5]
			\draw[<-] (-4.8,5.17)--(-4.2,5.17);
			\draw[<-] (-1,5.15)--(-1.6,5.15);
			\draw[<-] (-1,.85)--(-1.6,.85);
			\draw[<-] (-4.8,.83)--(-4.2,.83);
			\draw[<-] (-2.85,3.4)--(-2.85,2.6);
			\path [draw, thick, blue] (-5,5)--(-3,5)--(-1,5);
			\path [draw, thick, color=green!60!black] (-5,1)--(-3,1)--(-1,1);
			\path [draw] (-3,1)--(-3,5);
			\draw[dashed] (-3,3) ellipse (1.3 and 2.3);
			\node at (-1,5)[right]{$p_4$};
			\node at (-1,1)[right]{$p_3$};
			\node at (-5,5)[left]{$p_1$};
			\node at (-5,1)[left]{$p_2$};
			\node at (-2.8,3)[left]{$q$};
		\end{tikzpicture}
	\end{gathered}
	=
	\int
	d(\text{LIPS})_2\,
	\begin{gathered}
		\begin{tikzpicture}[scale=.5]
			\draw[<-] (-4.8,5.17)--(-4.2,5.17);
			\draw[<-] (-1,5.15)--(-1.6,5.15);
			\draw[<-] (-1,.85)--(-1.6,.85);
			\draw[<-] (-4.8,.83)--(-4.2,.83);
			\draw[<-] (-2.85,3.4)--(-2.85,2.6);
			\path [draw, thick, blue] (-5,5)--(-3,5)--(-1,5);
			\path [draw, thick, color=green!60!black] (-5,1)--(-3,1)--(-1,1);
			\path [draw] (-3,1)--(-3,5);
			\draw[dashed] (-3,3) ellipse (1.3 and 2.3);
			\node at (-5,5)[left]{$p_1$};
			\node at (-5,1)[left]{$p_2$};
			\node at (-2.8,3)[left]{$q_1$};
			\draw[<-] (4.8,5.17)--(4.2,5.17);
			\draw[<-] (1,5.15)--(1.6,5.15);
			\draw[<-] (1,.85)--(1.6,.85);
			\draw[<-] (4.8,.83)--(4.2,.83);
			\draw[<-] (2.85,3.4)--(2.85,2.6);
			\path [draw, thick, red] (0,0)--(0,6);
			\path [draw, thick, blue] (5,5)--(3,5)--(1,5);
			\path [draw, thick, color=green!60!black] (5,1)--(3,1)--(1,1);
			\path [draw] (3,1)--(3,5);
			\draw[dashed] (3,3) ellipse (1.3 and 2.3);
			\node at (5,5)[right]{$p_4$};
			\node at (5,1)[right]{$p_3$};
			\node at (2.8,3)[right]{$q-q_1$};
		\end{tikzpicture}
	\end{gathered}
\end{equation}
with $d(\text{LIPS})_2$ the Lorentz-invariant phase space measure for the intermediate two-particle states.
For simplicity, we do not consider inelastic $2\to2$ processes for the time being.

Let us now check that the impact-parameter Fourier transform diagonalizes this two-particle phase space convolution.
We can start from \eqref{2pcqspacesimpl} and go to impact-parameter space via the Fourier transform for the $2\to2$ process \eqref{FTexact}.
When applying this Fourier transform to both sides of \eqref{2pcqspacesimpl}, we obtain
\begin{equation}\label{}
	\begin{split}
		&2\operatorname{Im}_{2pc}\operatorname{FT}[ \mathcal A]
		=
		\int \frac{d^Dq_1}{(2\pi)^D}2\pi\delta(-2p_1\cdot q_1+q_1^2)2\pi\delta(2p_2\cdot q_1+q_1^2)
		\!\!\!\!\!\!\!
		\begin{gathered}
			\begin{tikzpicture}[scale=.5]
				\draw[<-] (-4.8,5.17)--(-4.2,5.17);
				\draw[<-] (-1,5.15)--(-1.6,5.15);
				\draw[<-] (-1,.85)--(-1.6,.85);
				\draw[<-] (-4.8,.83)--(-4.2,.83);
				\draw[<-] (-2.85,3.4)--(-2.85,2.6);
				\path [draw, thick, blue] (-5,5)--(-3,5)--(-1,5);
				\path [draw, thick, color=green!60!black] (-5,1)--(-3,1)--(-1,1);
				\path [draw] (-3,1)--(-3,5);
				\draw[dashed] (-3,3) ellipse (1.3 and 2.3);
				\node at (-5,5)[left]{$p_1$};
				\node at (-5,1)[left]{$p_2$};
				\node at (-2.8,3)[left]{$q_1$};
			\end{tikzpicture}
		\end{gathered} \\
		&\times\int \frac{d^Dq}{(2\pi)^D}2\pi\delta(-2p_1\cdot q+q^2)2\pi\delta(2p_2\cdot q+q^2)
		e^{ib\cdot q}
		\left[
		\begin{gathered}
			\begin{tikzpicture}[scale=.5]
				\draw[<-] (-4.8,5.17)--(-4.2,5.17);
				\draw[<-] (-1,5.15)--(-1.6,5.15);
				\draw[<-] (-1,.85)--(-1.6,.85);
				\draw[<-] (-4.8,.83)--(-4.2,.83);
				\draw[<-] (-2.85,3.4)--(-2.85,2.6);
				\path [draw, thick, blue] (-5,5)--(-3,5)--(-1,5);
				\path [draw, thick, color=green!60!black] (-5,1)--(-3,1)--(-1,1);
				\path [draw] (-3,1)--(-3,5);
				\draw[dashed] (-3,3) ellipse (1.3 and 2.3);
				\node at (-1,5)[right]{$p_4$};
				\node at (-1,1)[right]{$p_3$};
				\node at (-2.8,3)[right]{$q-q_1$};
			\end{tikzpicture}
		\end{gathered}
		\right]^\ast
	\end{split}
\end{equation}
Then we change integration variables according to $q=q_1+q_1'$ and use
\begin{align}\label{}
	\delta(-2p_1\cdot q_1+q_1^2)\delta(-2p_1\cdot q+q^2)=
	\delta(-2p_1\cdot q_1+q_1^2)\delta(-2(p_1-q_1)\cdot q_1'+q_1'^2)
\end{align}
and similarly
\begin{equation}\label{}
	\delta(2p_2\cdot q_1+q_1^2)\delta(2p_2\cdot q+q^2)=
	\delta(2p_2\cdot q_1+q_1^2)\delta(2(p_2+q_1)\cdot q_1'+q_1'^2)
\end{equation}
which hold thanks to the properties of the delta function.
Then we get, 
\begin{equation}\label{unit2pc}
	\begin{split}
		&2\operatorname{Im}_{2pc}\operatorname{FT}[ \mathcal A]
		=
		\int \frac{d^Dq_1}{(2\pi)^D}2\pi\delta(-2p_1\cdot q_1+q_1^2)2\pi\delta(2p_2\cdot q_1+q_1^2)
		e^{ib\cdot q_1}\!\!\!\!\!\!\!
		\begin{gathered}
			\begin{tikzpicture}[scale=.5]
				\draw[<-] (-4.8,5.17)--(-4.2,5.17);
				\draw[<-] (-1,5.15)--(-1.6,5.15);
				\draw[<-] (-1,.85)--(-1.6,.85);
				\draw[<-] (-4.8,.83)--(-4.2,.83);
				\draw[<-] (-2.85,3.4)--(-2.85,2.6);
				\path [draw, thick, blue] (-5,5)--(-3,5)--(-1,5);
				\path [draw, thick, color=green!60!black] (-5,1)--(-3,1)--(-1,1);
				\path [draw] (-3,1)--(-3,5);
				\draw[dashed] (-3,3) ellipse (1.3 and 2.3);
				\node at (-1,5)[right]{$q_1-p_1$};
				\node at (-1,1)[right]{$-q_1-p_2$};
				\node at (-5,5)[left]{$p_1$};
				\node at (-5,1)[left]{$p_2$};
				\node at (-2.8,3)[left]{$q_1$};
			\end{tikzpicture}
		\end{gathered} \\
		&\times\int \frac{d^Dq_1'}{(2\pi)^D}2\pi\delta(-2(p_1-q_1)\cdot q_1'+q_1'^2)2\pi\delta(2(p_2+q_1)\cdot q_1'+q_1'^2)
		e^{ib\cdot q_1'}
		\left[
		\begin{gathered}
			\begin{tikzpicture}[scale=.5]
				\draw[<-] (-4.8,5.17)--(-4.2,5.17);
				\draw[<-] (-1,5.15)--(-1.6,5.15);
				\draw[<-] (-1,.85)--(-1.6,.85);
				\draw[<-] (-4.8,.83)--(-4.2,.83);
				\draw[<-] (-2.85,3.4)--(-2.85,2.6);
				\path [draw, thick, blue] (-5,5)--(-3,5)--(-1,5);
				\path [draw, thick, color=green!60!black] (-5,1)--(-3,1)--(-1,1);
				\path [draw] (-3,1)--(-3,5);
				\draw[dashed] (-3,3) ellipse (1.3 and 2.3);
				\node at (-1,5)[right]{$p_4$};
				\node at (-1,1)[right]{$p_3$};
				\node at (-2.8,3)[right]{$q_1'$};
			\end{tikzpicture}
		\end{gathered}
		\right]^\ast
	\end{split}
\end{equation}
As shown in \ref{usefulFT}, the Fourier transform FT[$\mathcal A$] differs from the linearized one $\tilde{\mathcal{A}}$ employed in the definition of the eikonal exponentiation \eqref{fullampli} by $G$-independent corrections suppressed by $1/b^2$ (more precisely, by powers of the dimensionless quantity $1/(pb)^2$), and thus we arrive at
\begin{equation}\label{almostexact}
		2\operatorname{Im}_{2pc}\tilde{\mathcal{A}}
		=
		|\tilde{\mathcal{A}}(s,b)|^2 \left(
		1+\mathcal O\left(\frac{1}{b^2}\right)
		\right)\,.
\end{equation}
Using the eikonal exponentiation \eqref{fullampli} on the right-hand side we find
\begin{equation}\label{}
	2\operatorname{Im}_{2pc}\tilde{\mathcal{A}}
	=
	\left|[1+2i\Delta(s,b)] e^{2i\delta(s,b)}-1 \right|^2
	\left(
	1+\mathcal O\left(\frac{1}{b^2}\right)
	\right)\,.
\end{equation}
We can expand this in $G$ to obtain various constraints. 
To leading order in $G$, of course we get that the tree-level amplitude is real
\begin{equation}\label{}
	2\operatorname{Im}_{2pc}\tilde{\mathcal{A}}_0=0
\end{equation} 
and hence the leading eikonal is real,
\begin{equation}\label{}
	\operatorname{Im} 2\delta_0=0\,.
\end{equation}

To first subleading order in $G$, i.e.~$\mathcal O(G^2)$, we get
\begin{equation}\label{ima1}
	2\operatorname{Im}_{2pc}\tilde{\mathcal{A}}_1
	=
	\left(2\delta_0\right)^2\left(
	1+\mathcal O\left(\frac{1}{b^2}\right)
	\right)\,.
\end{equation}
This is the first interesting constraint, since it reveals that the imaginary part of the one-loop amplitude up to $\mathcal O(1/(b^2)^{1-2\epsilon})$ is exhausted by the eikonal exponentiation. In this way, this constraint ensures that $2\delta_1$ is real
\begin{equation}\label{}
	\operatorname{Im} 2\delta_1=0\,,
\end{equation}
while in general $2\Delta_1$, being of the same order as the corrections in the right-hand side of \eqref{ima1} can itself develop an imaginary part.  
Additionally,  $\operatorname{Im}2\Delta_1$ receives contributions from inelastic $2\to2$ processes, e.g.~from intermediate two-gravitino states in supergravity.

To $\mathcal O(G^3)$, we find instead
\begin{equation}\label{}
	\operatorname{Im}_{2pc}\tilde{\mathcal{A}}_2
	=
	\left(2\delta_0\,2\delta_1+2\delta_0\,\operatorname{Re}2\Delta_1\right)
	\left(
	1+\mathcal O\left(\frac{1}{b^2}\right)
	\right)\,.
\end{equation}
Therefore, the imaginary part of the two-loop amplitude due to the two-particle cut up to $\mathcal O(1/(b^2)^{3/2-2\epsilon})$ is exhausted by the exponentiation.
On the other hand, the imaginary part of $2\delta_2$ is not captured by the two-particle cut. By power counting in $G$, it must be due to the three-particle cut,
\begin{equation}\label{imdelta23pc} 
		\operatorname{Im}2\delta_2
		=
		\operatorname{Im}_{3pc}\tilde{\mathcal{A}}_2\,,
\end{equation}
which we shall illustrate in the next subsection. 
	
Before proceeding, for illustrative purposes, let us explicitly calculate the $b$-space contribution of the two-particle cuts to order $G^2$. We start from Eq.~\eqref{unit2pc} and consider the situation in which the amplitudes appearing on its right-hand side are the tree-level ones, i.e.~such that
\begin{equation}\label{}
	\mathcal A_0(s,-q^2) = \frac{a_0}{q^2}\,,\qquad
	\tilde{\mathcal{A}}_0(s,b) = 2\delta_0(s,b) = \frac{1}{4Ep}\frac{a_0}{4 \pi}\frac{\Gamma(-\epsilon)}{(\pi b^2)^{-\epsilon}}\,.
\end{equation} 
We retain here a generic $q$-independent prefactor $a_0$, for the sake of generality; for instance
\begin{equation}\label{}
	a_0= 32\pi Gm_1^2m_2^2\left(\sigma^2-\tfrac{1}{2-2\epsilon} \right)
\end{equation}
for collisions of minimally coupled massive scalars in GR.
We start by evaluating the Fourier transform in the second line of \eqref{unit2pc}. We can write rewrite it as follows
\begin{equation}\label{}
\begin{split}
	&\int \frac{d^Dq_1'}{(2\pi)^D}2\pi\delta(2p_1'\cdot q_1'-q_1'^2)2\pi\delta(2 p_2'\cdot q_1'+q_1'^2)
e^{ib\cdot q_1'}
\mathcal A_0(s,-q_1'^2)\\
&=
\int \frac{d^Dq_1'}{(2\pi)^D}2\pi\delta(2p_1'\cdot q_1')2\pi\delta(2 p_2'\cdot q_1')
e^{ib\cdot q_1'}
\frac{a_0}{q_1'^2}
\end{split}
\end{equation}
where we have introduced the notation $p_1'=p_1-q_1$, $p_2'=p_2+q_1$ and we have used the fact that any correction involving positive integer powers of $q_1'^2$, such as those due to the application of Eq.~\eqref{4EpFT}, can be neglected because they give rise to contact terms in $b$.
At this point we may use \eqref{invFTLIN} and \eqref{B1} in the familiar way, up to an important point: the result only depends on the projection of $b^\mu$ orthogonal to $p_1'$ and $p_2'$ (and not to $p_1$, $p_2$), so that
\begin{equation}\label{intFTlinbprime}
	\int \frac{d^Dq_1'}{(2\pi)^D}2\pi\delta(2p_1'\cdot q_1')2\pi\delta(2 p_2'\cdot q_1')
	e^{ib\cdot q_1'}
	\frac{a_0}{q_1'^2}
	=
	\frac{1}{4Ep}\frac{a_0}{4\pi} \frac{\Gamma(-\epsilon)}{(\pi b'^2(q_1))^{-\epsilon}}\,,
\end{equation}
where
$b'(q_1)^2$ is the square of such projection,
\begin{equation}\label{bPRIME}
	b'^2(q_1) = b^2 - \frac{1}{p^2}(b\cdot q_1)^2\,.
\end{equation}
At this point we can substitute back into \eqref{unit2pc}, taking again into account that any positive integer power of $q_1^2$ can be safely dropped, so that
\begin{equation}\label{almostfactorized}
	2\operatorname{Im}\operatorname{FT}[ \mathcal A_1]
	=
	\int \frac{d^Dq_1}{(2\pi)^D}2\pi\delta(2p_1\cdot q_1)2\pi\delta(2p_2\cdot q_1)
	e^{ib\cdot q_1}\left[\frac{a_0}{q_1^2}\right]\left[\frac{1}{4Ep}\frac{a_0}{4\pi} \frac{\Gamma(-\epsilon)}{(\pi b'^2(q_1))^{-\epsilon}}\right].
\end{equation}
In this step, we need to expand \eqref{bPRIME} appearing in the second factor to subleading order for large $b$. This term also depends on $q_1$, which manifests the lack of complete factorization, if such subleading terms are taken into account. In the end, using \eqref{B1} and its derivatives with respect to $b^\mu$, one arrives at
\begin{equation}\label{2ImFTA1}
	2\operatorname{Im}\operatorname{FT}[ \mathcal A_1]
	 =  [2\delta_0(s,b)]^2
	 -
	  \frac{2(1-2\epsilon)}{p^2(b^2)^{1-2\epsilon}} \left[\frac{1}{4Ep}\frac{a_0}{4 \pi}\frac{\Gamma(1-\epsilon)}{\pi^{-\epsilon}}\right]^2 + \mathcal O(b^{-4+4\epsilon})\,.
\end{equation}
To summarize, the FT of the two-body convolution of the tree-level amplitude is the product of the FT only to leading order in $1/b^2$, while this factorization receives corrections to first subleading order. On the one hand, here we calculated it using the properties of the Fourier transform under considerations to achieve the ``almost factorized'' form \eqref{almostfactorized}.
On the other hand, in Eq.~\eqref{Re2d2} below we provide an expression for the two-particle convolution accurate to leading and subleading order in $q^2$. One can verify that, using \eqref{4EpFT}, taking FT of \eqref{Re2d2} one indeed recovers \eqref{2ImFTA1}.

\subsubsection{Inelastic unitarity} 

The first inelastic contributions come from the three-particle cut of the $2\to2$ amplitude $\mathcal A$. This is obtained by ``gluing together'' two copies of a $2\to3$ amplitude
\begin{equation}\label{ft223diag}
	\mathcal{A}^{(5)}(p_1,p_2,q_1,q_2,k)
		= 
		\begin{gathered}
			\begin{tikzpicture}[scale=.5]
				\draw[<-] (-4.8,5.17)--(-4.2,5.17);
				\draw[<-] (-1,5.15)--(-1.6,5.15);
				\draw[<-] (-1,3.15)--(-1.6,3.15);
				\draw[<-] (-1,.85)--(-1.6,.85);
				\draw[<-] (-4.8,.83)--(-4.2,.83);
				\draw[<-] (-2.85,1.7)--(-2.85,2.4);
				\draw[<-] (-2.85,4.3)--(-2.85,3.6);
				\path [draw, thick, blue] (-5,5)--(-3,5)--(-1,5);
				\path [draw, thick, color=green!60!black] (-5,1)--(-3,1)--(-1,1);
				\path [draw] (-3,3)--(-1,3);
				\path [draw] (-3,1)--(-3,5);
				\draw[dashed] (-3,3) ellipse (1.3 and 2.3);
				\node at (-1,5)[right]{$k_1$};
				\node at (-1,3)[right]{$k$};
				\node at (-1,1)[right]{$k_2$};
				\node at (-5,5)[left]{$p_1$};
				\node at (-5,1)[left]{$p_2$};
				\node at (-2.8,4)[left]{$q_1$};
				\node at (-2.8,2)[left]{$q_2$};
			\end{tikzpicture}
		\end{gathered}
\end{equation}
where blue and green lines represent classical states with masses $m_1$ and $m_2$, while massless lines are drawn in black.
The drawing inside the dashed bubbles does not represent a specific topology, but just provides a visual help to recall the definition of the $q_i$ variables, such that
\begin{equation}\label{}
	q_1+q_2+k=0\,.
\end{equation}
We will discuss momentarily on the precise form of the product of five point amplitudes appearing in this gluing, which of course involve index contractions with appropriate projector insertions. 
For the moment, let us write it schematically as
\begin{equation}\label{3pcbasic}
	2\operatorname{Im}_{3pc}\mathcal A = \int d(\text{LIPS}) \mathcal A_{\text{L.}}^{(5)} \mathcal A_\text{R.}^{(5)\ast}
\end{equation}
(where L. and R. stand for ``left'' and ``right'') or in the pictorial form
\begin{equation}\label{3pcqspace}
	2\operatorname{Im}_{3pc}
	\begin{gathered}
		\begin{tikzpicture}[scale=.5]
			\draw[<-] (-4.8,5.17)--(-4.2,5.17);
			\draw[<-] (-1,5.15)--(-1.6,5.15);
			\draw[<-] (-1,.85)--(-1.6,.85);
			\draw[<-] (-4.8,.83)--(-4.2,.83);
			\draw[<-] (-2.85,3.4)--(-2.85,2.6);
			\path [draw, thick, blue] (-5,5)--(-3,5)--(-1,5);
			\path [draw, thick, color=green!60!black] (-5,1)--(-3,1)--(-1,1);
			\path [draw] (-3,1)--(-3,5);
			\draw[dashed] (-3,3) ellipse (1.3 and 2.3);
			\node at (-1,5)[right]{$p_4$};
			\node at (-1,1)[right]{$p_3$};
			\node at (-5,5)[left]{$p_1$};
			\node at (-5,1)[left]{$p_2$};
			\node at (-2.8,3)[left]{$q$};
		\end{tikzpicture}
	\end{gathered}
	=
	\int
	d(\text{LIPS})
	\begin{gathered}
		\begin{tikzpicture}[scale=.5]
			\draw[<-] (-4.8,5.17)--(-4.2,5.17);
			\draw[<-] (-1,5.15)--(-1.6,5.15);
			\draw[<-] (-1,3.15)--(-1.6,3.15);
			\draw[<-] (-1,.85)--(-1.6,.85);
			\draw[<-] (-4.8,.83)--(-4.2,.83);
			\draw[<-] (-2.85,1.7)--(-2.85,2.4);
			\draw[<-] (-2.85,4.3)--(-2.85,3.6);
			\path [draw, thick, blue] (-5,5)--(-3,5)--(-1,5);
			\path [draw, thick, color=green!60!black] (-5,1)--(-3,1)--(-1,1);
			\path [draw] (-3,3)--(-1,3);
			\path [draw] (-3,1)--(-3,5);
			\draw[dashed] (-3,3) ellipse (1.3 and 2.3);
			\node at (-1,5)[right]{$k_1$};
			\node at (-1,3)[right]{$k$};
			\node at (-1,1)[right]{$k_2$};
			\node at (-5,5)[left]{$p_1$};
			\node at (-5,1)[left]{$p_2$};
			\node at (-2.8,4)[left]{$q_1$};
			\node at (-2.8,2)[left]{$q_2$};
			\draw[<-] (4.8,5.17)--(4.2,5.17);
			\draw[<-] (1,5.15)--(1.6,5.15);
			\draw[<-] (1,3.15)--(1.6,3.15);
			\draw[<-] (1,.85)--(1.6,.85);
			\draw[<-] (4.8,.83)--(4.2,.83);
			\draw[<-] (2.85,1.7)--(2.85,2.4);
			\draw[<-] (2.85,4.3)--(2.85,3.6);
			\path [draw, thick, red] (.5,0)--(.5,6);
			\path [draw, thick, blue] (5,5)--(3,5)--(1,5);
			\path [draw, thick, color=green!60!black] (5,1)--(3,1)--(1,1);
			\path [draw] (3,3)--(1,3);
			\path [draw] (3,1)--(3,5);
			\draw[dashed] (3,3) ellipse (1.3 and 2.3);
			\node at (5,5)[right]{$p_4$};
			\node at (5,1)[right]{$p_3$};
			\node at (2.8,4)[right]{$q_4$};
			\node at (2.8,2)[right]{$q_3$};
		\end{tikzpicture}
	\end{gathered}
\end{equation}
where
\begin{equation}\label{}
	d(\text{LIPS})
	=
	\frac{d^Dk}{(2\pi)^D}\,2\pi\theta(k^0)\delta(k^2)
	\frac{d^Dk_1}{(2\pi)^D}\,2\pi\theta(k_1^0)\delta(k_1^2+m_1^2) 2\pi\theta(k_2^0)\delta(k_2^2+m_2^2)
\end{equation}
is the Lorentz-invariant phase-space measure for the intermediate states. 
Using momentum conservation we can retain only, say, $k$ and $k_1$ as independent integration variables in the measure.

We enforce the classical limit as follows. As by now familiar, we assume that the momentum transfer $q=p_1+p_4$ for the full $2\to2$ process be small compared to the momenta of particle 1 and 2, depicted by thick colored lines in \eqref{3pcqspace}, so that to leading order
\begin{equation}\label{leadingp1p2k1k2}
	p_{1} \sim -m_1 u_1 + \mathcal O(q) \sim -k_1\,,\qquad 
	p_{2} \sim -m_2 u_2 + \mathcal O(q)\sim -k_2\,.
\end{equation}
By momentum conservation, this is tantamount to assuming the scaling
\begin{equation}\label{q1q2simq}
	q_1 \sim q_2 \sim k \sim \mathcal O(q)\,.
\end{equation}
We thus take the momenta $q_{1,2}$ and $k$ to be simultaneously small, of the order of the elastic momentum transfer $q$.
Equivalently, reinstating momentarily $\hbar$, this can be regarded as a formal $\hbar\to0$ limit in which the wavenumbers $q_{1,2}/\hbar$ and $k/\hbar$ are held fixed. 
Eq.~\eqref{q1q2simq} expresses the soft-region scaling that we discussed when evaluating loop integrals and is justified by the fact that it captures all non-analytic contributions in $q^2$. It also allows us to simplify the integration in a similar way as before. Using $q_1=p_1+k_1$ and $k$ as independent integration variables simplifies the LIPS measure as follows to leading order in the scaling \eqref{q1q2simq},
\begin{equation}\label{LIPSsoft}
	d(\text{LIPS})
	\simeq
	\frac{d^Dk}{(2\pi)^D}2\pi\theta(k^0)\delta(k^2)
	\frac{d^Dq_1}{(2\pi)^D}2\pi\delta(2p_1\cdot q_1)
	2\pi\delta(2p_2\cdot (q_1 + k))
	\,,
\end{equation}
and leads us to adopt the following  routing
\begin{equation}\label{3pcqspacesimpl}
	2\operatorname{Im}_{3pc}
	\begin{gathered}
		\begin{tikzpicture}[scale=.5]
			\draw[<-] (-4.8,5.17)--(-4.2,5.17);
			\draw[<-] (-1,5.15)--(-1.6,5.15);
			\draw[<-] (-1,.85)--(-1.6,.85);
			\draw[<-] (-4.8,.83)--(-4.2,.83);
			\draw[<-] (-2.85,3.4)--(-2.85,2.6);
			\path [draw, thick, blue] (-5,5)--(-3,5)--(-1,5);
			\path [draw, thick, color=green!60!black] (-5,1)--(-3,1)--(-1,1);
			\path [draw] (-3,1)--(-3,5);
			\draw[dashed] (-3,3) ellipse (1.3 and 2.3);
			\node at (-1,5)[right]{$p_4$};
			\node at (-1,1)[right]{$p_3$};
			\node at (-5,5)[left]{$p_1$};
			\node at (-5,1)[left]{$p_2$};
			\node at (-2.8,3)[left]{$q$};
		\end{tikzpicture}
	\end{gathered}
	=
	\int
	d(\text{LIPS})
	\begin{gathered}
		\begin{tikzpicture}[scale=.5]
			\draw[<-] (-4.8,5.17)--(-4.2,5.17);
			\draw[<-] (-1,5.15)--(-1.6,5.15);
			\draw[<-] (-1,3.15)--(-1.6,3.15);
			\draw[<-] (-1,.85)--(-1.6,.85);
			\draw[<-] (-4.8,.83)--(-4.2,.83);
			\draw[<-] (-2.85,1.7)--(-2.85,2.4);
			\draw[<-] (-2.85,4.3)--(-2.85,3.6);
			\path [draw, thick, blue] (-5,5)--(-3,5)--(-1,5);
			\path [draw, thick, color=green!60!black] (-5,1)--(-3,1)--(-1,1);
			\path [draw] (-3,3)--(-1,3);
			\path [draw] (-3,1)--(-3,5);
			\draw[dashed] (-3,3) ellipse (1.3 and 2.3);
			\node at (-1,3)[right]{$k$};
			\node at (-5,5)[left]{$p_1$};
			\node at (-5,1)[left]{$p_2$};
			\node at (-2.8,4)[left]{$q_1$};
			\node at (-2.8,2)[left]{$-q_1-k$};
			\draw[<-] (4.8,5.17)--(4.2,5.17);
			\draw[<-] (1,5.15)--(1.6,5.15);
			\draw[<-] (1,3.15)--(1.6,3.15);
			\draw[<-] (1,.85)--(1.6,.85);
			\draw[<-] (4.8,.83)--(4.2,.83);
			\draw[<-] (2.85,1.7)--(2.85,2.4);
			\draw[<-] (2.85,4.3)--(2.85,3.6);
			\path [draw, thick, red] (.3,0)--(.3,6);
			\path [draw, thick, blue] (5,5)--(3,5)--(1,5);
			\path [draw, thick, color=green!60!black] (5,1)--(3,1)--(1,1);
			\path [draw] (3,3)--(1,3);
			\path [draw] (3,1)--(3,5);
			\draw[dashed] (3,3) ellipse (1.3 and 2.3);
			\node at (5,5)[right]{$p_4$};
			\node at (5,1)[right]{$p_3$};
			\node at (2.8,4)[right]{$q-q_1$};
			\node at (2.8,2)[right]{$k-q+q_1$};
		\end{tikzpicture}
	\end{gathered}
\end{equation}
for the integrated momenta.
Since the massless momentum $k$ will mostly remain as a spectator in the following manipulations let us also introduce a shorthand notation for its on-shell phase space integration,
\begin{equation}\label{kphasespace}
	\int_{k} \equiv \int\frac{d^Dk}{(2\pi)^D}2\pi\theta(k^0)\delta(k^2)\,.
\end{equation}

Our basic ingredient \eqref{3pcbasic} will be the  $2\to 3$ tree-level amplitude in Eq.~(3.1) of~\cite{DiVecchia:2020ymx}, which had been obtained from the low energy limit of a string amplitude in a toroidal compactification. In this way it naturally includes the contributions of the dilaton, of vectors and scalars arising in the Kaluza--Klein compactification, and of the graviton.
To include all such contributions simultaneously, we find it convenient to formally promote all spacetime indices $\mu,\nu,\ldots$ to $10$-dimensional ones $M,N,\ldots$ according to conventions specified below.
The tree-level contribution to $\mathcal A^{(5)}$ in the classical limit 
reproduces the result of~\cite{Goldberger:2016iau,Luna:2017dtq,Mogull:2020sak} and can be then written in the following convenient form,\footnote{The full theory also allows for other $2\to3$ processes involving fermionic external states. However, these would yield contributions that are subleading in the limit of small momentum transfer.}
\begin{equation}
	\begin{split}
	& \mathcal A_0^{(5)MN}= (8\pi G)^{\frac{3}{2}} \Bigg\{ \beta
	\left[- \frac{p_1^M p_1^N (k\cdot q_1)}{(p_1\cdot k)^2 q_2^2} -  \frac{p_2^M p_2^N (k\cdot q_2)}{(p_2\cdot k)^2 q_1^2} \right. \\
	&
	+\frac{p_1^M (q_1-q_2)^N+p_1^N (q_1 -q_2)^M}{2(p_1\cdot k) q_2^2}  - \frac{p_2^M (q_1-q_2)^N+p_2^N (q_1- q_2)^M}{2(p_2\cdot k) q_1^2}   \\
	& + \left. \frac{(q_1 -q_2)^M (q_1 -q_2)^N}{2q_1^2 q_2^2} \right] +8 \frac{ \left( (p_1\cdot k)   p_2^M -(p_2 \cdot k) p_1^M \right) \left((p_1\cdot k)   p_2^N -(p_2\cdot k) p_1^N  \right)}{q_1^2 q_2^2} \\
	&+ (2p_1\cdot p_2) \left(\frac{4p_1^M p_1^N \frac{k\cdot p_2}{k\cdot p_1}- 2(p_1^M p_2^N+p_1^N p_2^M)}{q_2^2} +
	\frac{4p_2^M p_2^N \frac{k\cdot p_1}{k\cdot p_2}- 2(p_1^M p_2^N+p_1^N p_2^M)}{q_1^2} \right.  \\
	& \left. + \frac{(q_1-q_2)^M \left(-2(p_1 \cdot k) p_2^N +2(p_2\cdot k) p_1^N  \right) + (q_1-q_2)^N \left(-2 (p_1\cdot k) p_2^M +2(p_2\cdot k) p_1^M  \right) }{q_1^2 q_2^2} \right)\Bigg\}
	\label{GGG2}
	\end{split}
\end{equation} 
the quantity $\beta$ is defined in~\eqref{eq:betan8gr} below depending on the theory under consideration. The main feature of~\eqref{GGG2} 
is that it satisfies $k_M \mathcal{A}_{0}^{(5)MN} = k_N \mathcal A_{0}^{(5)MN}=0$ for arbitrary values of the free index, which makes the calculations in general dimensions easier~\cite{KoemansCollado:2019ggb,Kosmopoulos:2020pcd}. Notice that the terms proportional to $\beta$ and the remaining terms are independently gauge invariant.  It is also symmetric in its two spacetime indices, $\mathcal A_0^{(5)MN}=\mathcal A_{0}^{(5)N M}$.

When focusing on ${\cal N}=8$ supergravity, we choose the following 10D kinematics~\cite{DiVecchia:2021ndb}\footnote{For sake of simplicity here we focus directly on the case $\phi=\frac{\pi}{2}$ in the notations of that reference.} 
\begin{equation}
	\label{eq:pn8}
		p_1^M = (p_1^\mu;0,\ldots,0,0,m_1)\,,\qquad 
		p_2^M = (p_2^\mu;0,\ldots,0,m_2,0)\,,
\end{equation}
where $p_{1,2}^\mu$ are $(4-2\epsilon)$-dimensional momenta and the dots stand for $2+2\epsilon$ zero entries. In contrast, in the case of GR,
\begin{equation}
	\label{eq:pgr}
		p_1^M  = (p_1^\mu;0,\ldots,0,0,0)\,,\qquad
		p_2^M  = (p_2^\mu;0,\ldots,0,0,0)\,.
\end{equation}
The momenta $k^M$ and $q_{1,2}^M$ are always non-trivial only in the $4-2\epsilon$ non-compact directions so that they effectively coincide with $k^\mu$ and $q_{1,2}^\mu$, e.g.~$q_1^M=(q_1^\mu;0,\ldots,0,0,0)$. 
Another difference between ${\cal N}=8$ and GR is related to the contribution of the dilaton as an internal state exchanged between the massive objects. This exchange can only occur in ${\cal N}=8$ supergravity, while for GR the contribution of the dilaton has to be subtracted. This can be taken into account by specifying the parameter $\beta$ as follows in the two theories,
\begin{equation}
	\label{eq:betan8gr}
	\beta^{{\cal N}=8} = 4 m^2_1 m^2_2 \sigma^2\,,\qquad \beta^{\text{GR}} =  4 m^2_1 m^2_2 \left(\sigma^2 -\tfrac{1}{D-2}\right).
\end{equation}

Let us now describe more precisely how the ${\cal N}=8$ supergravity and GR amplitudes enter the integrand \eqref{3pcbasic}.
In both cases, the kinematics is dictated by the momentum flow depicted in Eq.~\eqref{3pcqspace}, so that
 \begin{equation}\label{}
 	\mathcal{A}^{(5)}_{\text{L.}}=\mathcal{A}^{(5)}(p_1,p_2,k_1,k_2,k)\,,\qquad
 	\mathcal{A}^{(5)\ast}_{\text{R.}}=\mathcal{A}^{(5)\ast}(p_4,p_3,-k_1,-k_2,-k)\,.
 \end{equation}
Index contractions are instead theory dependent.
In the maximally supersymmetric case, we simply work in $M,N,\ldots$ indices and saturate the Lorentz indices with the Minkowski metric, contracting the two amplutudes according to
\begin{equation}
	\label{eq:n8int}
	\mathcal{A}^{(5)}_{\text{L.}}
	\mathcal{A}^{(5)\ast}_{\text{R.}}
	=
		\mathcal{A}^{(5)MN}_{\text{L.}} \eta_{MR}\,\eta_{NS}\,	\mathcal{A}^{(5)\ast RS}_{\text{R.}}  \,,
\end{equation}
because in this case the amplitude $\mathcal A^{(5)MN}$ is transverse and traceless. As already stressed, the advantage of this trick is that it effectively encompasses graviton, dilaton and Kaluza--Klein vectors/scalars at the same time. 
For GR we work with $\mu,\nu, \ldots$ indices and on the top of this we need to subtract the contribution of the dilaton, and thus we employ the contraction
\begin{equation}
	\label{eq:grint} 
	\mathcal{A}^{(5)}_{\text{L.}}
	\mathcal{A}^{(5)\ast}_{\text{R.}}
	=
	\mathcal{A}^{(5)\mu\nu}_{\text{L.}} \left[\eta_{\mu\rho}\eta_{\nu\sigma}-\tfrac{1}{D-2} \eta_{\mu\nu} \eta_{\rho\sigma}\right]  \mathcal{A}^{(5)\ast\rho\sigma}_{\text{R.}}\,.
\end{equation}
The structure within square brackets arises from  the transverse-traceless projector over physical degrees of freedom  $\Pi_{\mu\nu,\rho\sigma}(k)$. Explicitly, for any reference null vector $r^\mu$, letting $\lambda^\mu=-r^\mu/(r\cdot k)$ and $\Pi^{\mu\nu}=\eta^{\mu\nu}+\lambda^\mu k^\nu+\lambda^\nu k^\mu$, one can construct this projector according to
\begin{equation}\label{TTprojector}
	\Pi^{\mu\nu,\rho\sigma}(k)
	=
	\frac12\left(
	\Pi^{\mu\rho}\Pi^{\nu\sigma}
	+
	\Pi^{\mu\sigma}\Pi^{\nu\rho}
	-
	\tfrac{2}{D-2}
	\Pi^{\mu\nu}\Pi^{\rho\sigma}
	\right).
\end{equation} 
The transversality condition $k_\mu \mathcal{A}^{(5)\mu\nu}=0$ then turns \eqref{TTprojector} into the conbination within square brackets in \eqref{eq:grint},
which is the one appearing also in the de Donder propagator, simplified by using $\mathcal A^{(5)\mu\nu}=\mathcal A^{(5)\nu\mu}$.
We will use \eqref{eq:n8int} and \eqref{eq:grint} as a convenient shorthand notation to suppress explicit index contractions between $2\to3$ amplitude also in Section~\ref{SemiclEik}.

We can now start from \eqref{3pcqspacesimpl} go to impact-parameter space via the usual Fourier transform for the $2\to2$ process (see \ref{usefulFT})
\begin{equation}\label{ft222diag}
	\tilde{\mathcal A} = \int \frac{d^Dq}{(2\pi)^D}\,2\pi\delta(2p_1\cdot q)2\pi\delta(2p_2\cdot q)\,e^{ib\cdot q}
	\begin{gathered}
		\begin{tikzpicture}[scale=.5]
			\draw[<-] (-4.8,5.17)--(-4.2,5.17);
			\draw[<-] (-1,5.15)--(-1.6,5.15);
			\draw[<-] (-1,.85)--(-1.6,.85);
			\draw[<-] (-4.8,.83)--(-4.2,.83);
			\draw[<-] (-2.85,3.4)--(-2.85,2.6);
			\path [draw, thick, blue] (-5,5)--(-3,5)--(-1,5);
			\path [draw, thick, color=green!60!black] (-5,1)--(-3,1)--(-1,1);
			\path [draw] (-3,1)--(-3,5);
			\draw[dashed] (-3,3) ellipse (1.3 and 2.3);
			\node at (-1,5)[right]{$p_4$};
			\node at (-1,1)[right]{$p_3$};
			\node at (-5,5)[left]{$p_1$};
			\node at (-5,1)[left]{$p_2$};
			\node at (-2.8,3)[left]{$q$};
		\end{tikzpicture}
	\end{gathered}.
\end{equation}
In the remainder of this section, we need not worry about the difference between $\operatorname{FT}[\mathcal{A}]$ \eqref{invFT} and $\tilde{\mathcal{A}}$ \eqref{invFTLIN}, since we will be working to leading order in $q\sim 1/b$. 
When applying this Fourier transform to both sides of \eqref{3pcqspacesimpl}, we change integration variable by letting $q=q_1+q_4$. After doing this, an important step is to use
\begin{align}\label{}
	\delta(p_1\cdot q_1)\delta(2p_1\cdot q)=
	\delta(p_1\cdot q_1)\delta(2p_1\cdot (q_1+q_4))
	=
	\delta(p_1\cdot q_1)\delta(2p_1\cdot q_4)
\end{align}
and similarly
\begin{equation}\label{}
	\delta(p_2\cdot (q_1+k))\delta(2p_2\cdot q)=
	\delta(p_2\cdot (q_1+k))\delta(2p_2\cdot (q_4-k))\,,
\end{equation}
which hold thanks to the properties of the delta function.
Then we get, 
\begin{equation}\label{}
	\begin{split}
		2\operatorname{Im}_{3pc}\tilde{\mathcal{A}}
		&=
		\int_{{k}} 
		\int \frac{d^Dq_1}{(2\pi)^D}2\pi\delta(2p_1\cdot q_1)2\pi\delta(2p_2\cdot (q_1+k))
		e^{ib\cdot q_1}\!\!\!\!\!\!\!
		\begin{gathered}
			\begin{tikzpicture}[scale=.5]
				\draw[<-] (-4.8,5.17)--(-4.2,5.17);
				\draw[<-] (-1,5.15)--(-1.6,5.15);
				\draw[<-] (-1,3.15)--(-1.6,3.15);
				\draw[<-] (-1,.85)--(-1.6,.85);
				\draw[<-] (-4.8,.83)--(-4.2,.83);
				\draw[<-] (-2.85,1.7)--(-2.85,2.4);
				\draw[<-] (-2.85,4.3)--(-2.85,3.6);
				\path [draw, thick, blue] (-5,5)--(-3,5)--(-1,5);
				\path [draw, thick, color=green!60!black] (-5,1)--(-3,1)--(-1,1);
				\path [draw] (-3,3)--(-1,3);
				\path [draw] (-3,1)--(-3,5);
				\draw[dashed] (-3,3) ellipse (1.3 and 2.3);
				\node at (-1,3)[below]{$k$};
				\node at (-5,5)[left]{$p_1$};
				\node at (-5,1)[left]{$p_2$};
				\node at (-2.8,4)[left]{$q_1$};
				\node at (-2.8,2)[left]{$-q_1-k$};
			\end{tikzpicture}
		\end{gathered} \\
		&\times\int \frac{d^Dq_4}{(2\pi)^D}2\pi\delta(2p_1\cdot q_4)2\pi\delta(2p_2\cdot (q_4-k))
		e^{ib\cdot q_4}
		\left[
		\begin{gathered}
			\begin{tikzpicture}[scale=.5]
				\draw[<-] (4.8,5.17)--(4.2,5.17);
				\draw[<-] (1,5.15)--(1.6,5.15);
				\draw[<-] (1,3.15)--(1.6,3.15);
				\draw[<-] (1,.85)--(1.6,.85);
				\draw[<-] (4.8,.83)--(4.2,.83);
				\draw[<-] (2.85,1.7)--(2.85,2.4);
				\draw[<-] (2.85,4.3)--(2.85,3.6);
				\path [draw, thick, blue] (5,5)--(3,5)--(1,5);
				\path [draw, thick, color=green!60!black] (5,1)--(3,1)--(1,1);
				\path [draw] (3,3)--(1,3);
				\path [draw] (3,1)--(3,5);
				\draw[dashed] (3,3) ellipse (1.3 and 2.3);
				\node at (5,5)[right]{$p_4$};
				\node at (5,1)[right]{$p_3$};
				\node at (2.8,4)[right]{$q_4$};
				\node at (2.8,2)[right]{$k-q_4$};
				\node at (1,3)[below]{$-k$};
			\end{tikzpicture}
		\end{gathered}
		\right]^\ast
	\end{split}
\end{equation}
We may reinterpret the $3\to2$ amplitude in the second line as a $2\to3$ one by flipping all external momenta and relabel $q_4=- q_1'$ in the integration. Since moreover to leading order $p_4\simeq -p_1$, $p_3\simeq -p_2$ we can write
\begin{equation}\label{3pcbspaceprel}
	\begin{split}
		2\operatorname{Im}_{3pc}\tilde{\mathcal{A}}
		=
		\int_{{k}} 
		&\left[
		\int \frac{d^Dq_1}{(2\pi)^D}2\pi\delta(2p_1\cdot q_1)2\pi\delta(2p_2\cdot (q_1+k))
		e^{ib\cdot q_1}\!\!\!\!\!\!\!
		\begin{gathered}
			\begin{tikzpicture}[scale=.5]
				\draw[<-] (-4.8,5.17)--(-4.2,5.17);
				\draw[<-] (-1,5.15)--(-1.6,5.15);
				\draw[<-] (-1,3.15)--(-1.6,3.15);
				\draw[<-] (-1,.85)--(-1.6,.85);
				\draw[<-] (-4.8,.83)--(-4.2,.83);
				\draw[<-] (-2.85,1.7)--(-2.85,2.4);
				\draw[<-] (-2.85,4.3)--(-2.85,3.6);
				\path [draw, thick, blue] (-5,5)--(-3,5)--(-1,5);
				\path [draw, thick, color=green!60!black] (-5,1)--(-3,1)--(-1,1);
				\path [draw] (-3,3)--(-1,3);
				\path [draw] (-3,1)--(-3,5);
				\draw[dashed] (-3,3) ellipse (1.3 and 2.3);
				\node at (-1,3)[below]{$k$};
				\node at (-5,5)[left]{$p_1$};
				\node at (-5,1)[left]{$p_2$};
				\node at (-2.8,4)[left]{$q_1$};
				\node at (-2.8,2)[left]{$-k-q_1$};
			\end{tikzpicture}
		\end{gathered}\right] \\
		\times
		&\left[\int \frac{d^Dq'_1}{(2\pi)^D}2\pi\delta(2p_1\cdot q'_1)2\pi\delta(2p_2\cdot (q_1'+k))
		e^{ib\cdot q'_1}
		\begin{gathered}
			\begin{tikzpicture}[scale=.5]
				\draw[<-] (-4.8,5.17)--(-4.2,5.17);
				\draw[<-] (-1,5.15)--(-1.6,5.15);
				\draw[<-] (-1,3.15)--(-1.6,3.15);
				\draw[<-] (-1,.85)--(-1.6,.85);
				\draw[<-] (-4.8,.83)--(-4.2,.83);
				\draw[<-] (-2.85,1.7)--(-2.85,2.4);
				\draw[<-] (-2.85,4.3)--(-2.85,3.6);
				\path [draw, thick, blue] (-5,5)--(-3,5)--(-1,5);
				\path [draw, thick, color=green!60!black] (-5,1)--(-3,1)--(-1,1);
				\path [draw] (-3,3)--(-1,3);
				\path [draw] (-3,1)--(-3,5);
				\draw[dashed] (-3,3) ellipse (1.3 and 2.3);
				\node at (-1,3)[below]{$k$};
				\node at (-5,5)[left]{$p_1$};
				\node at (-5,1)[left]{$p_2$};
				\node at (-2.8,4)[left]{$q'_1$};
				\node at (-2.8,2)[left]{$-k-q'_1$};
			\end{tikzpicture}
		\end{gathered}
		\right]^\ast
	\end{split}
\end{equation}
In this equation, the two quantities appearing within square brackets are equal to each other and can be identified as the Fourier transform of the five-point amplitude, $\tilde{\mathcal{A}}^{(5)}$.
Of course, \eqref{3pcbspaceprel} only fixes the definition of $\tilde{\mathcal{A}}^{(5)}$ up to an overall phase, which would of course cancel in the product $\tilde{\mathcal{A}}^{(5)} \tilde{\mathcal{A}}^{(5)\ast}$. In order to treat the two momenta $q_1$ and $q_2$ on the same footing, it is  natural to define
\begin{equation}\label{ft223DEF}
	\begin{split}
		\tilde {\mathcal{A}}^{(5)}
		= \int \frac{d^Dq_1}{(2\pi)^D}\, 2\pi\delta(2p_1\cdot q_1)2\pi\delta(2p_2\cdot q_2)
		e^{ib_1\cdot q_1+ib_2\cdot q_2}
		\begin{gathered}
			\begin{tikzpicture}[scale=.5]
				\draw[<-] (-4.8,5.17)--(-4.2,5.17);
				\draw[<-] (-1,5.15)--(-1.6,5.15);
				\draw[<-] (-1,3.15)--(-1.6,3.15);
				\draw[<-] (-1,.85)--(-1.6,.85);
				\draw[<-] (-4.8,.83)--(-4.2,.83);
				\draw[<-] (-2.85,1.7)--(-2.85,2.4);
				\draw[<-] (-2.85,4.3)--(-2.85,3.6);
				\path [draw, thick, blue] (-5,5)--(-3,5)--(-1,5);
				\path [draw, thick, color=green!60!black] (-5,1)--(-3,1)--(-1,1);
				\path [draw] (-3,3)--(-1,3);
				\path [draw] (-3,1)--(-3,5);
				\draw[dashed] (-3,3) ellipse (1.3 and 2.3);
				\node at (-1,5)[right]{$k_1$};
				\node at (-1,3)[right]{$k$};
				\node at (-1,1)[right]{$k_2$};
				\node at (-5,5)[left]{$p_1$};
				\node at (-5,1)[left]{$p_2$};
				\node at (-2.8,4)[left]{$q_1$};
				\node at (-2.8,2)[left]{$q_2$};
			\end{tikzpicture}
		\end{gathered}
	\end{split}
\end{equation}
with $b^\alpha=b_1^\alpha-b_2^\alpha$ and of course $q_1+q_2+k=0$ as indicated in the figure. We shall come back to this Fourier transform in Section~\ref{SemiclEik}. 
As we shall see, this is indeed the appropriate Fourier transform to leading order in the PM expansion.
We have thus related the imaginary part of the $b$-space $2\to2$ amplitude to the integral over the massless particle's phase space of the product of two $b$-space $2\to3$ amplitudes,
\begin{equation}\label{3pcbspace}
2\operatorname{Im}2\delta_2=	2\operatorname{Im}_{3pc}\tilde{\mathcal{A}}  = \int_{ k} \tilde{\mathcal{A}}^{(5)}\,\tilde{\mathcal{A}}^{(5)\ast}\,.
\end{equation} 
Let us recall that index contractions are left implicit and can be restored by using \eqref{eq:n8int} for $\mathcal N=8$ supergravity and \eqref{eq:grint} for GR.

To conclude, we have discussed how the imaginary part of the amplitude in $b$-space arising from the three-particle cut can be calculated in two different ways.
The first is to calculate the three-particle cut \eqref{3pcbasic} in momentum space and then take its Fourier transform at the very end. The second is to first take Fourier transforms of the five-point amplitudes via \eqref{ft223DEF}, contract them as in \eqref{eq:n8int}, \eqref{eq:grint} and integrate over the graviton phase space, according to \eqref{3pcbspace}.
The former method is more powerful as far as loop integration techniques are concerned. The reason is that, in momentum space, one can interpret the phase-space delta functions as ``cut propagators'' and apply the full machinery of master integrals and differential equations for two-loop topologies.
The latter makes the interpretation of this imaginary part more transparent, since \eqref{ft223DEF} is actually the waveform, as we will discuss in Section~\ref{SemiclEik},  and as we shall see it is particularly useful when studying the low-frequency limit.

\subsection{Calculation of \texorpdfstring{$\operatorname{Im}  2 \delta_2$}{Im2delta2} }

By the relation \eqref{3pcbspace}, it is easy to see that the divergent part of  
\begin{equation}\label{}
	\operatorname{Im} 2 \delta_2 = \operatorname{Im}_{3pc}\tilde{\mathcal{A}}_2
\end{equation}
can be obtained by only considering the leading term  in the soft limit $k^\mu\to0$ of the five-point amplitude in \eqref{GGG2} that diverges as $\frac{1}{\omega}$, with $\omega=k^0$, and that arises from the first two lines of \eqref{GGG2}. Since in this limit $q_1\simeq -q_2$, the Fourier transform \eqref{ft223DEF} can be easily performed and one obtains the results summarized in Section~4 of~\cite{DiVecchia:2021ndb}. 
In this way, one obtains 
\begin{eqnarray}
  \left({\rm Im} 2\delta_2\right)_{{\cal{N}}=8}(b) = -\frac{G^3 ({\beta}^{ {\cal{N}}=8 })^2}{\pi b^2 \epsilon (\sigma^2-1)^2} 
  \left[ \sigma^2 + \frac{\sigma (\sigma^2-2)}{(\sigma^2-1)^{\frac{1}{2}}} \operatorname{arccosh}\sigma  \right] + \mathcal O(\epsilon^0) 
\label{3.6}
\end{eqnarray} 
 for massive ${\cal{N}}=8$ supergravity, where $\beta^{{\cal{N}}=8}$ is given in \eqref{eq:betan8gr},
 and \cite{Damour:2020tta}
  \begin{equation}
 	({\rm Im} \,2 \delta_2)_{gr} (b) \simeq -\frac{1}{2 \epsilon } \frac{G^3 {(\beta^{\text{GR}})}^2}{\pi b^2 (\sigma^2-1)^2}
 	\left[ \frac{ 8 -5 \sigma^2}{3}  - \frac{\sigma(3-2\sigma^2)}{(\sigma^2-1)^{\frac{1}{2}}} \operatorname{arccosh}\sigma \right] 
 	+ \mathcal O(\epsilon^0)
 	\label{3.2}
 \end{equation}
 for the graviton contribution in GR, where $\beta^{\text{GR}}$ is given in \eqref{eq:betan8gr}.
 We refer to Section~5.1    
 	of~\cite{DiVecchia:2021ndb} for  more details on these steps, while in this report we shall obtain Eqs.~\eqref{3.2} and \eqref{3.6} as special cases of a more general setup from Eqs.~\eqref{eq:standpm} and \eqref{eq:standpmn8} below.

The complete expression for  $({\rm Im} \,2 \delta_2)$, including not only the divergent contribution but also the finite one, can be obtained from reverse unitarity. In GR one gets Eq.~(6.28) of~\cite{DiVecchia:2021bdo}
 \begin{equation}
	\label{GR15}
	\begin{split}
		\operatorname{Im}2\delta_2 &= \frac{2 m_1^2 m_2^2 
			(2 \sigma ^2-1)^2 G^3}{\pi b^2 \left(\sigma ^2-1\right)^2}\,(\pi b^2 e^{\gamma_E})^{3\epsilon} \\		
		&
		\times
		\Big[
		\left(
		-\frac{1}{\epsilon}
		+
		\log(4(\sigma^2-1))
		\right)
		\left(\frac{\sigma  \left(2 \sigma ^2-3\right)\operatorname{arccosh}\sigma}{\sqrt{\sigma ^2-1}}+\frac{8-5 \sigma ^2}{3}\right)\\
		&-
		(\operatorname{arccosh}\sigma)^2 \left(\frac{\sigma  \left(2 \sigma ^2-3\right)}{\sqrt{\sigma ^2-1}}+\frac{2 (\sigma^2-1)(4\sigma^4-12\sigma^2-3)}{\left(1-2 \sigma ^2\right)^2}\right)\\
		&+(\operatorname{arccosh}\sigma)
		\ \frac{\sigma  \left(88 \sigma ^6-240 \sigma ^4+240 \sigma ^2-97\right)}{3 \left(1-2 \sigma ^2\right)^2 \sqrt{\sigma ^2-1}}\\
		&+\operatorname{Li}_2(1-z^2)\frac{\sigma  \left(3-2 \sigma ^2\right)}{\sqrt{\sigma ^2-1}}\,
		+\frac{-140 \sigma ^6+220 \sigma ^4-127 \sigma ^2+56}{9 \left(1-2 \sigma ^2\right)^2}\Big].
	\end{split}
\end{equation}
while in massive ${\cal{N}}=8$ supergravity one gets Eq. (4.21) of~\cite{DiVecchia:2021bdo} 
\begin{align}
	\begin{split}
\operatorname{Im} 2 \delta_2^{{\cal{N}}=8}   =& - \frac{16 m_1^2m_2^2 G^3}{\pi b^2}
\frac{\sigma^4(\pi b^2  e^{\gamma_E})^{3\epsilon}}{(\sigma^2-1)^2}  \Bigg\{\frac{1}{\epsilon}\left( \sigma^2  + \frac{\sigma(\sigma^2-2)}{(\sigma^2-1)^{\frac{1}{2}}} \operatorname{arccosh}\sigma\right)
\\
& - \log (4(\sigma^2-1)) \left[\sigma^2 + \frac{\sigma (\sigma^2-2)}{(\sigma^2-1)^{\frac{1}{2}}} \operatorname{arccosh}\sigma\right]  \\ & + (\sigma^2-1)\left[ 1  + \frac{\sigma (\sigma^2 -2) }{(\sigma^2-1)^{\frac{3}{2}}}  \right](\operatorname{arccosh}\sigma)^2  \\
& + \frac{\sigma (\sigma^2-2)}{(\sigma^2-1)^{\frac{1}{2}}}  \operatorname{Li}_2 (1-z^2) + 2 \sigma^2  \Bigg\},
\label{CR15a}
\end{split}
\end{align}
Notice that the previous equations have both terms that, at high energy, behave as $(\log\sigma)^2$. However these terms cancel each others and  one is left with an  high energy behavior as $\log \sigma$ in agreement with the analysis of~\cite{DiVecchia:2020ymx} based on analyticity and crossing symmetry.  
 
\subsection{Radiation reaction from \texorpdfstring{$\operatorname{Im}2\delta_2$}{Im2delta2} by real analyticity }

In this section we follow the approach of Ref.~\cite{DiVecchia:2021ndb} and using unitarity and real analyticity, we obtain  a relation that connects two terms of $\operatorname{Im} 2 \delta_2$ to  the radiation reaction contribution, $\operatorname{Re} 2 \delta_2^{(\text{RR})}$,   of  the two-loop eikonal.  Such a relation is
\begin{equation}
	- \frac{i}{\pi\epsilon}
	{\rm Re}\,2\delta_2^{\text{RR}}
	\mapsto
 \left[
	1 + \frac{i}{\pi} 
	\left(
	- \frac{1}{\epsilon} + \log(\sigma^2-1)
	\right) 
	\right]  
	{\rm Re}\,2\delta_2^{\text{RR}}.
	\label{comb2}
\end{equation} 
The part in the round bracket comes from the integral over the frequency of the graviton given by\footnote{We need to keep $D=4-2\epsilon$ only for the integral over $|\vec{k}|$ while the integration over the angular variables can be done for $\epsilon=0$, so that effectively $d^{D-1} \vec k = |\vec{k}|^{2-2\epsilon} d|\vec{k}|\, \sin \theta \,d \theta\, d \varphi$.}
\begin{equation}
(b^2)^{-1+3\epsilon} \int_0^{\overline{\omega b}} \frac{d\omega}{\omega}  (\omega b)^{-2 \epsilon}
\label{combx}
\end{equation}
where the factor $ (b^2)^{-1+ 3 \epsilon}$ is precisely the one expected (also on dimensional ground) to appear in $2\delta_2$ and   $\overline{\omega b}$ is an appropriate upper limit on the classical dimensionless quantity $\omega b$.
 On the other hand, the integral over $\omega$ produces a $\frac{1}{\epsilon}$ divergence in the particular combination:
 \begin{equation}
 \int_0^{\overline{\omega b}} \frac{d \omega}{\omega} (\omega b)^{- 2 \epsilon} = - \frac{1}{2 \epsilon} (\,\overline{\omega b}\,)^{-2 \epsilon} =  -\frac{1}{2 \epsilon}  + \log \overline{\omega b}+ {\cal O}(\epsilon)
\label{comb}
\end{equation}
To obtain an estimate of the cutoff $\overline{\omega b}$, one can use \eqref{p1q1p2q2} for single-graviton exchanges, according to which
\begin{equation}\label{}
	|p_1^0\,q_1^0| \approx |\vec p_1\cdot \vec q_1|\le |\vec p_1|\,|\vec q_1|\,,\qquad
	|p_2^0\,q_2^0| \approx |\vec p_2\cdot \vec q_2|\le |\vec p_2|\,|\vec q_2|
\end{equation}
and therefore
\begin{equation}\label{}
	\omega = k^0 = -q_1^0 - q_2^0 \le |q_1^0| + |q_2^0|  \lesssim \frac{|\vec p_1|}{|p_1^0|}\,|\vec q_1| + \frac{|\vec p_2|}{|p_2^0|}\,|\vec q_2|\,.
\end{equation}
In the classical limit, $|\vec q_i|\sim 1 / b$ and thus $\omega b$ is bounded by
\begin{equation}\label{}
	\overline{\omega b} = \frac{|\vec p_1|}{|p_1^0|} + \frac{|\vec p_2|}{|p_2^0|}\,.
\end{equation}
Using the explicit expressions \eqref{Ep}, \eqref{E1E}, \eqref{E2E} in the CM frame,  one finds
\begin{equation}
\overline{\omega b} = \sqrt{\sigma^2-1}\,\left( 1 + {\cal O}(\sigma -1)\right)
\label{combz}
\end{equation}
In this way we have obtained the term in the round bracket in \eqref{comb2}. In order to obtain the last term outside of the round bracket one must observe that, because of real analyticity, the amplitude between the two branch points $\sigma = \pm 1$ must be real and therefore in this region the term in the round bracket should be $\log (1- \sigma^2)$. 
When we go from this region to the physical region $\sigma \geq1$ then we get
\begin{equation}
\log (1- \sigma^2-i0) = \log (\sigma^2-1) - i \pi
\label{combt}
\end{equation}
obtaining Eq.   \eqref{comb2} that allows us to determine the radiation reaction from the infrared divergent term of $\operatorname{Im} 2 \delta_2$.   For GR we obtain:
\begin{equation}
  \operatorname{Re} \,2 \delta^{\text{RR,GR}}(b) = \frac{8G^3 m_1^2 m_2^2 (\sigma^2- \frac{1}{2})^2}{\hbar b^2 (\sigma^2-1)^2} 
  \left[ \frac{ 8 -5 \sigma^2}{3}  - \frac{\sigma(3-2\sigma^2)}{(\sigma^2-1)^{\frac{1}{2}}} \operatorname{arccosh}\sigma \right],
\label{3.2b}
\end{equation}
while in the case of massive ${\cal{N}}=8$ supergravity one gets:
\begin{eqnarray}
  \operatorname{Re} 2\delta_2^{\text{RR},\mathcal{N}=8}(b) = \frac{ 16 G^3 m_1^2 m_2^2\sigma^4 }{\hbar b^2  (\sigma^2-1)^2} 
  \left[ \sigma^2 + \frac{\sigma (\sigma^2-2)}{(\sigma^2-1)^{\frac{1}{2}}} \operatorname{arccosh}\sigma\right]   \; .
\label{3.6bis}
\end{eqnarray}
In conclusion, we have derived the radiation reaction from Eq.~\eqref{3pcbspace} and from the properties of unitarity and real analyticity, without needing to construct the full two-loop amplitude. The arguments that brought us to the structure in the round bracket of \eqref{comb2} are confirmed by the first two terms of the explicit results for $\operatorname{Im} 2 \delta_2$ given in \eqref{GR15} and \eqref{CR15a}.

\section{The 2--body eikonal at 3PM }
\label{sec:twoloop}

The exponentiation pattern of Eq.~\eqref{eikexp0}, which we explicitly illustrated in Section~\ref{sec:oneloop} up sub-leading order, suffices to eliminate all super-classical redundancies arising from tree-level and one-loop order. Thus, it allows one to extract from the amplitude genuine classical information up to two loops, i.e.~$\mathcal O(G^3)$ or 3PM order. The purpose of this chapter is to perform this step explicitly. We will start by sketching the calculation of the two-loop amplitude in the classical limit, to then check that super-classical terms cancel out consistently with Eq.~\eqref{eikexp0} and finally obtain the 3PM correction to the eikonal, $2\delta_2$.

A novelty compared to lower loop-orders is the emergence of an imaginary part in the amplitude that is due, via the unitarity relation, to the presence of nontrivial three-particle cuts, as we discussed in Section~\ref{sec:radiation}. At one loop, the imaginary part of the amplitude is due to two-particle cuts only and is effectively subtracted in the exponentiation \eqref{eikexp0} at classical level, so that $2\delta_1$ is real and only the quantum remainder $2\Delta_1$ can develop an imaginary part. By contrast, at two loops, the three-particle inelastic channels leave behind a nonzero (in fact, infrared divergent!) imaginary part of $2\delta_2$, as in \eqref{imdelta23pc}. In ensuing chapters we will discuss a more systematic way to deal with such additional channels, restoring the expected Hermiticity of the eikonal $2\delta$, which requires to promote it to an operator.

The focus of this chapter is instead on illustrating an interesting physical phenomenon that is captured by $2\delta_2$: radiation-reaction effects, which first occur at 3PM order and can be singled out by their time-reversal odd expansion in the small-velocity limit. On the other hand, such radiation-reaction contributions are crucial in order to ensure consistency, finiteness and universality of the results at high energies (or high velocities). In particular they are instrumental in ensuring agreement between the high-energy behavior of massive scattering and the universal result for the deflection angle of massless objects. 

As for the one-loop level, we shall first focus on the technically simpler $\mathcal N=8$ case, considering the $s$-$u$ symmetric scattering of massive states obtained via Kaluza--Klein compactification, and then move to collisions of massive scalars minimally coupled to GR. We also provide more details on the relation between the eikonal phase, the phase shifts and the radial action, which as we shall see is also instrumental in performing the analytic continuation from the case of unbound orbits to bound systems.

\subsection{Partial wave decomposition: from \texorpdfstring{$b$ to $b_J$}{b to bJ}}
\label{ssec:bbj}

In Section~\ref{ssec:exponapp} we discussed how classical observables emerge in the leading eikonal approximation for the massless scattering. Starting from the resummed amplitude in the impact parameter representation, in the classical regime we can perform the Fourier transform back to momentum space~\eqref{eq:btq} by using a saddle-point approximation~\eqref{eq:saddlebQ}. The resummed amplitude can then be decomposed in partial waves, as in~\eqref{ACV17b}, and in the classical   limit \eqref{classicaljJ}, as expected, the orbital angular momentum of the system becomes large $J/\hbar = j \gg 1$. 
The partial wave analysis of that section is limited to the case of the leading eikonal in $D=4$, while the derivation based on the impact parameter representation seems more general and was already used in~\eqref{theta2PMd4} to derive the 2PM deflection angle in GR. The purpose of this section is to show that the same results can be obtained also by performing the partial wave decomposition in the classical regime starting from the eikonal amplitude at an arbitrary PM order~\cite{KoemansCollado:2020sxs} (see also \cite{Bellazzini:2022wzv}).

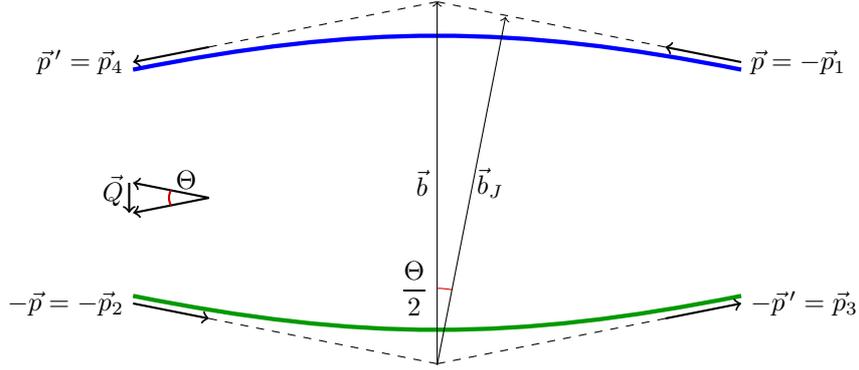
\begin{figure}[ht]
	\centering
	\begin{tikzpicture}
		\draw[ultra thick, blue]  (-4,1.5).. controls (-1,2.1) and (1,2.1) ..(4,1.5);
		\draw[ultra thick, green!60!black] (-4,-1.5).. controls (-1,-2.1) and (1,-2.1) ..(4,-1.5);
		\draw [thick, <-] (-4,1.6)--(-3,1.8);
		\draw [dashed] (-3,1.8)--(0,2.4);
		\draw [thick, ->] (4,1.6)--(3,1.8);
		\draw [dashed] (3,1.8)--(0,2.4);
		\draw [thick, <-] (4,-1.6)--(3,-1.8);
		\draw [dashed] (3,-1.8)--(0,-2.4);
		\draw [thick, ->] (-4,-1.6)--(-3,-1.8);	
		\draw [dashed] (-3,-1.8)--(0,-2.4);
		\draw [thick, <-] (-4,-0.4)--(-3,-0.2);
		\draw [thick, <-] (-4,0)--(-3,-0.2);
		\draw [thick, ->] (-4.05,0)--(-4.05,-0.4);
		\draw[thick,red!80!black] (-3.5,-0.3) arc (210:150:0.2);
		\node at (-4,1.6)[left]{$\vec p\,'=\vec p_4$};
		\node at (-4,-1.6)[left]{$-\vec p=-\vec p_2$};
		\node at (4,-1.6)[right]{$- \vec p\,'=\vec p_3$};
		\node at (4,1.6)[right]{$\vec p=-\vec p_1$};
		\node at (-3.3,-0.2)[above]{$\Theta$};
		\node at (-4,-0.1)[left]{$\vec Q$};
		\node at (0,0)[left]{$\vec b$};
		\node at (0.4,0)[right]{$\vec b_J$};
		\node at (0,-1.4)[left]{$\dfrac{\Theta}{2}$};
		\draw [->] (0,-2.4)--(0,2.4);
		\draw [->] (0,-2.42)--(.9,2.2);
		\draw[red!80!black] (0,-1.4) arc (90:78.6901:1);
	\end{tikzpicture}
	\caption{Classical $2\to2$ scattering in the center-of-mass frame \eqref{comframe}. 
		$\vec Q$ denotes the impulse and
		$\Theta$ the deflection angle.	
		The impact parameter $\vec b_J$ is perpendicular to the spatial momentum of either particle in the far past, and thus defines the angular momentum according to $J = p\, b_J$, with $p = |\vec p\,| = |\vec p\,'|$. Its magnitude differs by that of the eikonal impact parameter $\vec b$ by terms of order $\Theta^2$ as in \eqref{bJbThetadelta}.}
	\label{fig:scattering}
\end{figure}

We start by generalising~\eqref{eq:btq} to the massive scattering in $D=4$ by first defining
\begin{equation}\label{Sbspace}
	\tilde S(s,b) = 1+i\tilde{\mathcal{A}}(s,b) = e^{i\operatorname{Re}2\delta(s,b)}\,.
\end{equation}
Here we remove the $1/\epsilon$ pole in $2\delta_0$, associated to the Coulombic falloff of the field, and introduce an appropriate cutoff scale in the leftover logarithm, as we did in \eqref{2idelta0finite}, and we temporarily drop the imaginary part of $2\delta(s,b)$ that enters at 3PM, which will not enter the ensuing stationary-phase arguments.
In $Q$-space, Eq.~\eqref{Sbspace} translates to
\begin{equation}\label{}
	S(s,Q) = 4 E p \int d^2b\, e^{-iQ\cdot b}e^{i \operatorname{Re}2\delta(s,b)} = 8 \pi E p  \int db \,  b\, J_0\left(\frac{Qb}{\hbar}\right) e^{i \operatorname{Re}2\delta(s,b)}\,,
\end{equation}
where we used the fact that  $2\delta$ depends only on $b^2$ and the identity
\begin{equation}\label{J0}
	J_0(x) = \int_{-\pi}^{+\pi}\frac{d\phi}{2\pi} \,e^{ix\sin\phi} \,.
\end{equation}
(together with the fact that the integrand is a periodic function of $\phi$ with period $2\pi$).
For large arguments $x = Qb/\hbar \gg1$,
one also has
\begin{equation}\label{J0large}
	J_0(x) 
	\sim \frac{1}{\sqrt{2\pi}} \left(
	e^{i(x-\frac{\pi}{4})}
	+
	e^{-i(x-\frac{\pi}{4})}
	\right).
\end{equation}
Using this relation and picking the first term, which corresponds to $Q^\mu$ being anti-aligned to $b^\mu$, as in our conventions for gravitational scattering, 
we obtain
\begin{equation}\label{}
	S(s,Q) \sim \left(\,\cdots\right) \int db\, b\, e^{
		\frac{i}{\hbar} Qb+i\operatorname{Re}2\delta(s,b)}\,.
\end{equation}
Here and in the next few equations we need only worry about rapidly oscillating phase factors, and accordingly omit irrelevant slowly-varying prefactors. 
We then go to angular momentum-space by projecting on the $j$th Legendre polynomial as dictated by \eqref{ACV17b},
\begin{equation}
	\label{}
	S_j(s) \sim  \left(\cdots\right)  \int db\, b  \int_{-1}^{+1}d(\cos\Theta)\, 
	P_j(\cos\Theta) e^{\frac{i}{\hbar}\,2pb\sin\frac{\Theta}{2}+i\operatorname{Re}2\delta(s,b)}\,,
\end{equation}
where we employed the relation between $Q$ and $\Theta$, which is again the same as the one encountered in~\eqref{eq:thetap}, 
\begin{equation}\label{Qsaddle}
	Q = 2p \sin\frac{\Theta}{2}\,.
\end{equation}
Using finally the large-$j$ limit of the Legendre polynomial \eqref{largej}
we arrive at
\begin{equation}\label{}
	S_j(s) \sim \int_{0}^{\pi}d \Theta\   \int db \left(\cdots\right) e^{-ij\Theta+\frac{i}{\hbar}\,2pb\sin\frac{\Theta}{2}+i\operatorname{Re}2\delta(s,b)}\,.
\end{equation}
Collecting the terms appearing in the exponent, we recover the phase shift in the classical limit from \eqref{2deltajchiJ} and
\begin{equation}\label{PhaseShift}
	\chi(s,J) =  \frac{1}{\hbar}\left(
	-J \Theta + 2pb \sin\frac{\Theta}{2}
	\right)
	+
	\operatorname{Re}2\delta(s,b)\,,
\end{equation} 
which must be extremized by a double saddle point in $\Theta$ and $b$ at fixed  
\begin{equation}\label{bJdef}
	b_J = \frac{J}{p}\,.
\end{equation}
Taking derivatives of \eqref{PhaseShift}, the two saddle point conditions give
\begin{equation}\label{bJbThetadelta}
	b_J =  b \cos\frac{\Theta}{2}\,,\qquad
	2 p \sin \frac{\Theta}{2} =  -\hbar \,\frac{\partial \operatorname{Re}2\delta (s,b)}{\partial b}\,.
\end{equation}
Eqs.~\eqref{Qsaddle}, \eqref{bJdef} and \eqref{bJbThetadelta} identify the sought after relations among the properties of the trajectory depicted in Fig.~\ref{fig:scattering}, linking them to the eikonal phase.
In the classical limit, $S_j(s)$ is itself a rapidly oscillating phase factor, $S_j(s) \propto e^{2i\delta_j(s)}$ with  $2\delta_j(s) = \chi(s,J)$ given by \eqref{PhaseShift}.
Using the saddle point conditions, one can also recast it as follows in terms of $b$ and $b_J$,
\begin{equation}\label{PhaseShiftb}
	\chi(s,J)
	=
	\frac{2J}{\hbar}
	\left( 
	- \arccos \frac{b_J}{b} 
	+
	\sqrt{  \frac{b^2}{b_J^2} -1} \, \right) + \operatorname{Re}2\delta(s,b)\,.
\end{equation}
The derivative of $\chi(s,J)$ in \eqref{PhaseShift} with respect to $j=J/\hbar$ then takes the following form
\begin{equation}\label{}
	-\hbar\,\frac{\partial \chi(s,J)}{\partial J}
	=
	\Theta + 
	p
	\left[
	b_J-b\cos\frac{\Theta}{2}
	\right]\frac{\partial\Theta}{\partial J}
	-
	\left[
	2p\sin\frac{\Theta}{2} + \hbar\frac{\partial \operatorname{Re}2\delta(s,b)}{\partial b}
	\right]  \frac{\partial b}{\partial J}\,,
\end{equation}
so that, owing to the saddle point conditions \eqref{bJbThetadelta}, it 
gives back the saddle-point value of $\Theta$, 
\begin{equation}\label{Thetachi}
	\Theta = -\hbar\,\frac{\partial \chi(s,J)}{\partial J}\,.
\end{equation}
For this reason $\chi(s,J)$ is closely connected to the radial action $I(s,J)$, which obeys \cite{Kalin:2019rwq,Kalin:2019inp} 
\begin{equation}\label{ThetaI}
	\Theta + \pi = -\frac{\partial I(s,J)}{\partial J}
\end{equation}
so that
\begin{equation}\label{radialaction}
	I(s,J) = -\pi J + \hbar \chi(s,J)\,.
\end{equation}
The expression \eqref{PhaseShift} then provides the connection with the eikonal phase.

Eqs.~\eqref{PhaseShift} and \eqref{PhaseShiftb} provide this connection only implicitly, since one ought to solve the saddle point conditions in order to express the right-hand side as a function $J$. In particular, in \eqref{PhaseShiftb}, $b$ should be eliminated using the first condition \eqref{bJbThetadelta}, which in turn involves the deflection angle calculated using the second condition in \eqref{bJbThetadelta}. 
In order to illustrate how this works in practice let us consider the case of the perturbative PM expansion:\footnote{Although the relation \eqref{PhaseShiftb} is non-perturbative in $G$ it may actually break down at sufficiently small impact parameter if the relation between $b$ and $b_J$ cannot be straightforwardly inverted.}
\begin{equation}
	\label{eq:dexpcl}
	\operatorname{Re}(2\delta) = - d_0 \ln(b) + \sum_{n=1}^\infty \frac{d_n}{n b^n}\;,
\end{equation}
where $d_n$ captures the $(n+1)$PM correction.
It is straightforward to derive explicitly the first few terms of the classical deflection angle expressed as a function of $J$ perturbatively to any given order. For instance up to order $O(G^5)$ we have
\begin{equation}
	\begin{aligned}\label{phaseshift5}
		\chi(s,J)  = 
		&-d_0 \log{J} 
		+ \frac{d_1 p}{J} 
		+ \frac{d_2 p^2-\frac{d^3_0}{12}}{2J^2}
		\\
		&+ \frac{d_3 p^3-\frac{3}{8} d_0^2 d_1 p}{3J^3}
		+ \frac{d_4 p^4 - \frac{d_0 p^2}{2}\left(d_2 d_0 - d_1^2\right) +\frac{1}{80} d_0^5}{4J^4}
		+ \mathcal{O}(G^6)
	\end{aligned}
\end{equation} 
so that via \eqref{Thetachi},
\begin{equation}
	\begin{aligned}\label{dafpg5}
		\Theta & = \frac{d_0}{J} + \frac{d_1 p}{J^2} + \frac{d_2 p^2-\frac{d^3_0}{12}}{J^3} \\
		&+  \frac{d_3 p^3-\frac{3}{8} d_0^2 d_1 p}{J^4}+
		\frac{d_4 p^4 - \frac{d_0 p^2}{2}\left(d_2 d_0 - d_1^2\right) +\frac{1}{80} d_0^5}{J^5}+ {\cal O}(G^6)\,.
	\end{aligned}
\end{equation}
Likewise, via \eqref{radialaction}, Eq.~\eqref{phaseshift5} also gives the corresponding expansion of the radial action $I(s,J)$.

The derivation presented above linking $\chi(s,J)$ to the eikonal phase via saddle point conditions generalizes straightforwardly to generic $D$, as one can check thanks to known properties of Gegenbauer polynomials: [DLMF \S 18.15(i) using (18.7.1)] and [DLMF \S 10.17(i)].

\subsection{Massive \texorpdfstring{$\mathcal N=8$}{N=8} Supergravity }
\label{ssec:tree1L}

We go back to the elastic $2\to 2$ scattering in ${\cal{N}}=8$ supergravity where the external states are massive thanks to non-trivial Kaluza--Klein momenta in the compact directions~\cite{Caron-Huot:2018ape,Parra-Martinez:2020dzs}. We focus on the $s$-$u$ symmetric process already defined in Subsection~\ref{sec:n8tree}, where the states labeled by $p_1$ and $p_4$ are dilatons with KK momentum in one compact direction and those labeled by $p_2$ and $p_3$ are axions with KK momentum in another orthogonal compact direction \cite{Parra-Martinez:2020dzs,DiVecchia:2021ndb,DiVecchia:2021bdo}

As before, the starting ingredient for our analysis is the momentum-space amplitude in the small-$q$ expansion, where, as already emphasized, we can restrict to the terms that are non-analytic in $q^2$.
We will also neglect terms of order $\mathcal O(q^2)^{\frac12-2\epsilon}$ or smaller, which are irrelevant for determining $2\delta_2$ and would only enter the calculation of the quantum remainder $2\Delta_2$.
In view of this, the relevant part of the $s-u$ symmetric two-loop scattering amplitude~\cite{Herrmann:2021lqe} is given by the following sum of scalar Feynman integrals $I_T$ where $T$ indicates the integral's topology:
\begin{eqnarray}\label{CR1}
	&&\mathcal{A}_2 (s, q^2) = (8\pi G)^3 \frac{1}{2(4\pi)^{4-2\epsilon}} \left[ (s-m_1^2-m_2^2)^4 + (u-m_1^2-m_2^2)^4 \right]
	\\
	&& \times \Bigg[(s-m_1^2-m_2^2)^2 \left( I_{\RT} +2 I_{\RN}\right) + (u-m_1^2-m_2^2)^2 \left( I_{\overline{\RT}} +2I_{\overline{\RN}} \right) +t^2 (I_{\Ht}+ I_{{\overline{\Ht}}} )\Bigg].
	\nonumber
\end{eqnarray}
Let us write down the integrals explicitly and sketch their evaluation as $q\to0$. 
We follow the kinematic conventions of Section~\ref{sec:kinematics}, in particular the variables introduced in \eqref{averagesplitting}, \eqref{averagevelocities}.
In terms of these variables, the planar double box integral reads
\begin{equation}\label{}
	\begin{split}
		&I_{\RT} = \int \frac{d^D\ell_1}{i\pi^{D/2}} \frac{d^D\ell_2}{i\pi^{D/2}}
		\frac{1}{\left[2 \bar m_1 \ell_{1} \cdot u_{1} + (\ell_1^2-\ell_1\cdot q)\right]
			\left[-2\bar m_2 \ell_{1} \cdot u_{2} + (\ell_1^2-\ell_1\cdot q)\right]}\\
		&\frac{1}{\left[-2\bar m_1 \ell_{2} \cdot u_{1} + (\ell_2^2-\ell_2\cdot q)\right]\left[2\bar m_2 \ell_{2} \cdot u_{2} + (\ell_2^2-\ell_2\cdot q)\right]\ell_{1}^{2}\ell_{2}^{2} \left(\ell_{1}+\ell_{2}-q\right)^{2}}\,,
	\end{split}
\end{equation}
the non-planar double box integral reads
\begin{equation}\label{}
	\begin{split}
		& I_{\RN} = \int  \frac{d^D\ell_1}{i\pi^{D/2}} \frac{d^D\ell_2}{i\pi^{D/2}}
		\frac{1}{\left[-2\bar m_1\ell_{1} \cdot u_{1} + (\ell_1^2+\ell_1\cdot q)\right]
			\left[2\bar m_2 \ell_{1} \cdot u_{2} + (\ell_1^2+\ell_1\cdot q)\right]}
		\\
		&\frac{1}{\left[2\bar m_1 \ell_{2} \cdot u_{1}+ (\ell_2^2+\ell_2\cdot q)\right]\left[2\bar m_2 (\ell_1+\ell_{2}) \cdot u_{2} + ((\ell_1+\ell_2)^2+(\ell_1+\ell_2)\cdot q)\right]}\\
		& \frac{1}{\ell_{1}^{2}\ell_{2}^{2}\left(\ell_{1}+\ell_{2}+q\right)^{2}}\,,
	\end{split}
\end{equation}
and the $\Ht$ diagram reads
\begin{equation}\label{}
	\begin{aligned}\label{Hf9}
		I_{\Ht}
		=
		&
		\int  \frac{d^D\ell_1}{i\pi^{D/2}} \frac{d^D\ell_2}{i\pi^{D/2}}
		\frac{1}{[-\bar m_2 \ell_{1} \cdot u_{2}+(\ell_1^2-\ell_1\cdot q)][-2\bar m_1 \ell_{2} \cdot u_{1}+(\ell_2^2-\ell_2\cdot q)]}
		\\
		& \frac{1}{\ell_1^2 \ell_2^2 (\ell_1+\ell_2-q)^2(\ell_1-q)^2 (\ell_2-q)^2}\,.
	\end{aligned}
\end{equation}
The Feynman $-i0$ prescription is left implicit throughout.
The remaining  ``barred'' topologies are obtained from the previous three by crossing symmetry, i.e.~replacing $u_1$ by $-u_1$ (or equivalently $u_2$ by $-u_2$).

The expansion of these integrals for small $q$ can be simplified by resorting to the method of regions. The relevant, non-analytic contributions emerge from the expansion in the soft region, where $\ell_1\sim\ell_2\sim\mathcal O(q)$, while $\bar m_1u_1$, $\bar m_2u_2$ are formally held fixed.\footnote{The asymptotic expansion of the $\Ht$ topology also has non-analytic contributions that emerge from the mixed soft-hard region, where $\ell_1\sim\mathcal O(q)$, $\ell_2\sim\mathcal O(m)$. However, these only contribute to order $\mathcal O(q^{3-2\epsilon})$ to the amplitude.}
Conveniently, all dependence on the masses factorizes in this step: for instance, to leading order,
\begin{equation}\label{}
	I_{\Ht}
	\simeq \frac{1}{\bar m_1 \bar m_2}
	\int  \frac{d^D\ell_1}{i\pi^{D/2}} \frac{d^D\ell_2}{i\pi^{D/2}}
	\frac{1}{(-\ell_{1} \cdot u_{2})(-2 \ell_{2} \cdot u_{1})
		\ell_1^2 \ell_2^2 (\ell_1+\ell_2- q)^2(\ell_1- q)^2 (\ell_2- q)^2}\,.
\end{equation}
The overall dependence on $q^2$ can be also easily factorized by rescaling the loop momenta.
The resulting integrals are thus functions of the single invariant $y=-u_1\cdot u_2$. Still, such integrals are not elementary for generic $\epsilon$, in contrast with the situation at one loop. Convenient strategies to evaluate them are centered on the method of ordinary differential equations in canonical form in the variable $x=y-\sqrt{y^2-1}$. These equations afford simple iterative solutions order by order in $\epsilon$ in terms of polylogarithms in the variable $x$. The boundary conditions for these equations can be determined by performing yet another expansion by regions in the limit $y\to1^+$, which luckily does yield elementary integrals for generic $\epsilon$. \cite{DiVecchia:2021bdo} 

Performing the calculation according to this strategy, we get the following result:
\begin{eqnarray}
	&&\mathcal{A}_2 (s, q^2) = \frac{(8\pi G)^3}{(4\pi)^4} \left( \frac{4\pi e^{-\gamma_E}}{q^2} \right)^{2\epsilon} 
	\Bigg\{ - \frac{2\pi^2 m_1^4 m_2^4}{\epsilon^2 q^2}\left[ \left(\frac{1+z^2}{1-z^2}\right)^2\left( \frac{1+z^2}{z}\right)^4 \right] \nonumber \\
	&& + \frac{4\pi m_1^3 m_2^3 }{\epsilon^2} \frac{(1+z^2)^5}{(1-z^2)^4 z^3}\left[ \pi (1-z^2)^2+i \left( -1-4z^2 \log z +z^4\right)\right] \nonumber \\
	&& - \frac{i \pi m_1^2 m_2^2}{3\epsilon} \left[ \left( \frac{1+z^2}{1-z^2}\right)^4 \frac{1}{z^3}\right] \Bigg[ 6i\pi (m_1^2+m_2^2) z(1+z^2)^2 \nonumber \\
	&& + 6m_1m_2 \Big[ 4(1+z^2)^2 (1-z^2)+ 2(i \pi +2 \log z)  (1+z^2)^2  \nonumber \\
	&& + \left(\frac{\pi^2}{6} - \operatorname{Li}_2 (z^2) \right) (z^2+1)(z^4 -6z^2 +1)
	- 2 (1+z^2)^2 (1-z^2) \log (1-z^2) \nonumber \\
	&& +2 ( i\pi \log z +\log^2 z) \left(z^2 +4z^4 - z^6\right) \Big] \Bigg] \Bigg\}  ,
	\label{CR6}
\end{eqnarray}
where as in \eqref{sigmax}
\begin{equation}\label{}
	z = \sigma-\sqrt{\sigma^2-1}\,,\qquad \sigma = \frac12 \left(
	z+\frac1{z}
	\right).
\end{equation}
The above result can be organized in the following convenient form (we recall that $c(\epsilon)$ was defined in \eqref{cepsilon}): 
\begin{equation}\label{A222010}
	\mathcal{A}_2 = (8\pi G)^3 \frac{c(\epsilon)^2}{(q^2)^{2\epsilon}}
	\left[
	\frac{\mathcal A_2^{(2,2)}}{\epsilon^2 q^2}
	+
	\frac{\mathcal{A}_2^{(0,1)}}{q}
	+
	\frac{\mathcal A_2^{(2,0)}}{\epsilon^2}
	+
	\frac{\mathcal A_2^{(1,0)}}{\epsilon}
	+\cdots
	\right]
\end{equation}
up to terms that are further suppressed in $q^2$, i.e.~$\mathcal O((q^2)^{\frac12-2\epsilon})$ or smaller, or that are subleading in $\epsilon$. Note in particular that the coefficient of $\left((q^2)^{1+2\epsilon}\epsilon\right)^{-1}$ vanishes identically. 
A term of $\mathcal{O}(\epsilon^0 (q^2)^{-1-2\epsilon})$ is also needed in order to fully reproduce the ${\cal O}(\delta_0^3)$ exponentiation and will not be discussed further. Similarly, $\mathcal O((q^2)^{-\frac{1}{2}-2\epsilon})$ terms start contributing to order $\mathcal O(\epsilon^0)$ via $\mathcal{A}_2^{(0,1)}$. Using the method described in \cite{DiVecchia:2021bdo}, one can check that $\mathcal{A}_2^{(0,1)}$ is consistent with the iteration term $2\delta_0\, 2i\delta_1$, to leading order in $\epsilon$.
The fact that the $\mathcal O((q^2)^{-\frac{1}{2}-2\epsilon})$ contributions indeed reproduce such iteration term for generic $\epsilon$ has been checked in Ref.~\cite{Bjerrum-Bohr:2021vuf}.

Explicitly, in terms of the variable $z$ and $\sigma$,
the above coefficients read
\begin{align}
	\label{A22}
	\mathcal A_2^{(2,2)}
	&=-\frac{2 \pi ^2 m_1^4 m_2^4 \left(z^2+1\right){}^6}{z^4 \left(z^2-1\right){}^2}  = -\frac{32 \pi ^2 m_1^4 m_2^4 \sigma^6}{(\sigma^2-1)}\,,\\
	\mathcal A_2^{(2,0)}
	&=
	\frac{4 \pi ^2 m_1^3 m_2^3 \left(z^2+1\right){}^5}{z^3
		\left(z^2-1\right){}^2}	
	+
	i
	\,
	\frac{4 \pi  m_1^3 m_2^3 \left(z^2+1\right){}^5 \left(z^4-4 z^2 \log \left(z\right)-1\right)}{z^3
		\left(z^2-1\right){}^4} \, \nonumber \\
	& = \frac{32 \pi^2 m_1^3 m_2^3 \sigma^5}{(\sigma^2-1)}	
	-
	i
	\,
	32 \pi  m_1^3 m_2^3 \left( \frac{ \sigma^6 }{
		(\sigma^2-1)^{3/2}} +  \frac{ \sigma^5 }{
		(\sigma^2-1)^2} \log(z) \right) \,,	
	\label{A22bis}
\end{align}
and
\begin{equation}
	\label{RA10}
	\begin{aligned}
		\operatorname{Re}\mathcal A_2^{(1,0)}
		&=
		\frac{2 \pi ^2 m_1^2 m_2^2 \left(z^2+1\right){}^6
			\left(\left(m_1^2+m_2^2\right) z+2 m_1 m_2\right)}{z^3
			\left(z^2-1\right){}^4}\\
		&-\frac{4 \pi ^2 m_1^3 m_2^3
			\left(z^2+1\right){}^4 \left(z^4-4 z^2-1\right) \log \left(z\right)}{z \left(z^2-1\right){}^4}\,, \\
		&=
		8 \pi ^2 m_1^2 m_2^2 
		\left[(m_1^2+m_2^2)  \frac{\sigma^6}{(\sigma^2-1)^2} + 2 m_1 m_2 \left( \frac{\sigma^7}{(\sigma^2-1)^2}+ \frac{\sigma^6}{(\sigma^2-1)^{3/2}} \right)  \right]\\
		&+16 \pi ^2 m_1^3 m_2^3
		\sigma^4 \left(\frac{1}{(\sigma^2-1)^{1/2}} - \frac{\sigma (\sigma^2-2)}{(\sigma^2-1)^2} \right) \log (z) \,,
	\end{aligned}
\end{equation}
together with
\begin{equation} \label{IA10}
	\begin{aligned}
		\operatorname{Im}\mathcal A_2^{(1,0)}
		&=
		\frac{2 \pi  m_1^3 m_2^3 \left(z^4-6 z^2+1\right)
			\left(z^2+1\right){}^5 \operatorname{\operatorname{Li}}_2\left(z^2\right)}{z^3 \left(z^2-1\right){}^4}
		\\
		&-\frac{\pi 
			m_1^3 m_2^3 \left(-24(z^4-1) +\pi^2 (z^4 -6z^2 +1)
			\right) 
			\left(z^2+1\right){}^5}{3 z^3
			\left(z^2-1\right){}^4}\\
		&+\frac{4 \pi  m_1^3 m_2^3 \left(z^4-4 z^2-1\right) \left(z^2+1\right){}^4 \log
			^2\left(z\right)}{z \left(z^2-1\right){}^4}
		-\frac{4 \pi  m_1^3 m_2^3 \left(z^2+1\right){}^6
			\log \left(1-z^2\right)}{z^3 \left(z^2-1\right){}^3}\\
		&-\frac{8 \pi  m_1^3 m_2^3 \left(z^2+1\right){}^6
			\log \left(z \right)}{z^3 \left(z^2-1\right){}^4}\,, \\
		&=
		\frac{16 \pi  m_1^3 m_2^3 \sigma^5(\sigma^2 -2)
			\text{Li}_2\left(z^2\right)}{(\sigma^2-1){}^2}
		-64 \pi 
		m_1^3 m_2^3 \frac{\sigma^6}{(\sigma^2-1)^{3/2}} - \frac83 \pi^3 m_1^3 m_2^3 \frac{\sigma^5(\sigma^2-2)}{\left(\sigma^2-1\right){}^{2}}\\
		&- 16 \pi  m_1^3 m_2^3 \sigma^4 \left(\frac{1}{(\sigma^2-1)^{1/2}} - \frac{\sigma (\sigma^2-2)}{(\sigma^2-1)^2} \right) \log^2\left(z\right)
		+\frac{32 \pi  m_1^3 m_2^3 \sigma^6
			\log \left(1-z^2\right)}{\left(\sigma^2-1\right){}^{3/2}}\\
		&-32 \pi  m_1^3 m_2^3 \sigma^6
		\left(
		\frac{\sigma}{\left(\sigma^2-1\right)^2}
		+
		\frac{1}{(\sigma^2-1)^{3/2}}
		\right)\log \left(z \right)\,.
	\end{aligned}
\end{equation}

\subsubsection{Eikonal and deflection angle to 3PM order }
\label{ssec:deltachi3PM}

We have already computed $2\delta_0$ in Subsection~\ref{sec:n8tree} and $2\Delta_1$ in Subsection~\ref{sec:n81loop}.
This information can now be used to verify that the eikonal exponentiation takes place as expected up to two loops and to extract $2\delta_2$ from the previous amplitude through the following relations: 
\begin{eqnarray}
	&&\operatorname{Re} (2 \delta_2) = \operatorname{Re} {\tilde{\mathcal{A}}}_2 + \frac{1}{6} (2 \delta_0)^3 + 2\delta_0 		\operatorname{Im} 2\Delta_1 \,, \nonumber \\
	&& \operatorname{Im} (2 \delta_2) = \operatorname{Im} {\tilde{\mathcal{A}}}_2 - 2\delta_0 \operatorname{Re} 2 \Delta_1\,.
	\label{CR6a}
\end{eqnarray}
We obtain the following result, where we highlight in blue the terms associated to radiation-reaction,
\begin{equation}
	\begin{split}
		\operatorname{Re} (2\delta_2) &=\frac{4 m_1^2 m_2^2 G^3 (1+z^2)^4}{b^2 z^2 (1-z^2)^5} 
		\Bigg[ (1+z^2)^2 (1-z^2) + 2 z^2 (1 + 4z^2 -z^4 )\log z \Bigg]  \\
		&= \frac{16 G^3 m_1^2 m_2^2}{b^2} 
		\left\{ {\color{blue}\frac{\sigma^6}{(\sigma^2 -1)^2}} - 
		\operatorname{arccosh}\sigma 
		\Bigg[ \frac{\sigma^4}{\sigma^2-1} - 
		{\color{blue}\frac{\sigma^5 (\sigma^2-2)}{(\sigma^2-1)^{\frac{5}{2}}}  }  \Bigg]
		\right\}
		\label{CR8}
	\end{split}
\end{equation}
and
\begin{align}
	\begin{split}
		\operatorname{Im} 2 \delta_2 =&~  \frac{4 m_1^2 m_2^2 G^3 }{\pi b^2}\frac{(1+z^2)^4(\pi b^2  e^{\gamma_E})^{3\epsilon} }{ (1-z^2)^5 z^2} \Bigg\{ \frac{1}{\epsilon}(1+z^2)\left[  \left(1-6z^2 +z^4\right) \log z-1 +z^4 \right]  \\
		& - \Bigg[ 2 z^2 
		\left( 1+4z^2 -z^4\right) \log^2 z
		+ (1+z^2)^2 (1-z^2) 
		\left[ 2 - 2 \log (1-z^2) +2 \log z \right]    \\
		& + \left(\frac{\pi^2}{6} -  \operatorname{Li}_2 (z^2) \right) (z^2+1)(z^4 -6z^2 +1)
		\Bigg]\Bigg\}
	\end{split}
\end{align}
or equivalently
\begin{align}
	\begin{split}
		\operatorname{Im} 2 \delta_2 =&~ - \frac{16 m_1^2m_2^2 G^3}{\pi b^2}
		\frac{\sigma^4(\pi b^2  e^{\gamma_E})^{3\epsilon}}{(\sigma^2-1)^2}  \Bigg\{\frac{1}{\epsilon}\left( \sigma^2  + \frac{\sigma(\sigma^2-2)}{(\sigma^2-1)^{\frac{1}{2}}} \operatorname{arccosh}\sigma\right)
		\\
		& - \log (4(\sigma^2-1)) \left[\sigma^2 + \frac{\sigma (\sigma^2-2)}{(\sigma^2-1)^{\frac{1}{2}}} \operatorname{arccosh}\sigma\right]  \\ & + (\sigma^2-1)\left[ 1  + \frac{\sigma (\sigma^2 -2) }{(\sigma^2-1)^{\frac{3}{2}}}  \right](\operatorname{arccosh}\sigma)^2  \\
		& + \frac{\sigma (\sigma^2-2)}{(\sigma^2-1)^{\frac{1}{2}}}  \operatorname{Li}_2 (1-z^2) + 2 \sigma^2  \Bigg\},
		\label{CR15}
	\end{split}
\end{align}
where we made the branch-cut singularity for $z>1$ more explicit via
\begin{eqnarray}
	\operatorname{Li}_2 (z^2)
	= -\operatorname{Li}_2 (1-z^2) +\frac{\pi^2}{6} - \log(z^2) \log (1-z^2)\,.
	\label{enha3}
\end{eqnarray}
and used
\begin{equation}
	- \log z =\operatorname{arccosh}\sigma = \log \left( \sigma + \sqrt{\sigma^2-1} \right) = 2 \operatorname{arcsinh} \sqrt{\frac{\sigma-1}{2}}\,,
	\label{CR16}
\end{equation}
\begin{equation}
	\label{eq:l1mzsi}
	\log(1-z^2) = \frac{1}{2} \log(4(\sigma^2-1)) -\operatorname{arccosh}\sigma\,.
\end{equation}

An important observation is that, in the ultrarelativistic limit $\sigma \gg 1$ the terms proportional to $(\sigma \, \log \sigma)^2$ present in the second and third last lines of Eq.~\eqref{CR15} cancel out against each other, yielding
\begin{equation}\label{eq:ACV90Im}
	\operatorname{Im}2 \delta_2 
	\sim  \frac{\log(s)}{s}\,\frac{(8Gs)^3\Gamma(1-\epsilon)^3}{32 (\pi b^2)^{1-3\epsilon}}
	\left[
	-\frac{1}{4\epsilon}
	+
	\frac{1}{2} 
	+\mathcal O(\epsilon)
	\right].
\end{equation}
This is crucial in order to ensure agreement between the present calculation and the universal massless result \cite{Amati:1990xe} according to the general pattern discussed in~\cite{DiVecchia:2020ymx}.

For later convenience, let us also remark that the  function defined by
	\begin{equation}\label{fzeta}
		f(\zeta) = \frac{\operatorname{arccosh}\zeta}{\sqrt{\zeta^2-1}}
	\end{equation}
	for $\zeta>1$ can be actually analytically continued to all complex $\zeta$ except for a branch cut at $\zeta<-1$. In particular,  
		$f(\zeta)$ is analytic for $-1<\zeta<1$ where it takes the form
	\begin{equation}\label{fzetaANCONT}
		f(\zeta) = \frac{\operatorname{arccos}\zeta}{\sqrt{1-\zeta^2}}\,.
	\end{equation}
More explicitly, the Taylor series for $f(1+z)$ around $z=0$,
\begin{equation}\label{fzetaSERIES}
	f(1+z) = \sum_{n=0}^\infty \frac{n!\, (-z)^n}{(2n+1)!!}\,,
\end{equation}
is absolutely convergent for $|z|<2$, and convergent for $z=2$, where it yields $f(3)=\frac{1}{\sqrt 2}\log(1+\sqrt 2)$, while it diverges for $|z|>2$ and for $z=-2$.
These properties lie at the heart of the analytic continuation from unbound to bound trajectories \cite{Kalin:2019rwq,Kalin:2019inp,Cho:2021arx}. Indeed, $\zeta=\sigma>1$ for scattering kinematics, such that the center-of-mass energy reads $E=\sqrt{m_1^2+2m_1m_2\sigma+m_2^2} > m_1+m_2$, while $-1<\zeta<1$ for bound states, such that $E=\sqrt{m_1^2+2m_1m_2\zeta+m_2^2} < m_1+m_2$. The fact that $f(\zeta)$ has a branch point at $\zeta=-1$ will also provide useful information for the analysis of soft spectra carried out in Section~\ref{sec:eikopsoft}.

Let us  finally turn to the deflection angle $\Theta$, which follows straightforwardly from $2\delta$. Keeping in mind the link between $b$ and $b_J$ derived in Section~\ref{ssec:bbj}, we obtain
\begin{equation}
	\Theta  =\Theta_{1\mathrm{PM}} + \Theta_{2\mathrm{PM}}+ \Theta_{3\mathrm{PM}}
\end{equation}
with, to leading non-vanishing order in $\epsilon$ at each PM order,
\begin{align}
	\label{1PMN=8}
	\Theta_{1\mathrm{PM}} &= \frac{4 G m_1 m_2 \sigma^2}{J (\sigma^2-1)^{1/2}}\,,  \\
	\Theta_{2\mathrm{PM}} &= -\frac{8 \pi  m_1^2 m_2^2 (m_1+m_2)  G^2\epsilon \sigma^4 }{J^2\sqrt{m_1^2+m_2^2+2m_1m_2 \sigma} \,\,(\sigma^2-1)} \,,  \\
	\begin{split}
		\Theta_{3\mathrm{PM}} &= - \frac{16 m_1^3 m_2^3 \sigma^6 G^3}{3 J^3(\sigma^2-1)^{3/2}}\\ 
		&+\frac{ 32 m_1^4 m_2^4 G^3}{J^3(m_1^2 +m_2^2 +2m_1 m_2 \sigma)}
		\left\{
		{\color{blue}
		\frac{\sigma^6}{\sigma^2-1} 
		}  
		- 
		\Bigg [ \sigma^4 - 
		{\color{blue}
			\frac{\sigma^5 (\sigma^2-2)}{(\sigma^2-1)^{\frac{3}{2}}} 
		}  \Bigg]  \operatorname{arccosh}\sigma
		\right\}
		\,.
		\label{DA2}
	\end{split}
\end{align}
The 1PM contribution corresponds to a tree diagram where both the graviton and the dilaton are exchanged, the 2PM contribution is absent for $\epsilon=0$ in agreement with the results of Ref.~\cite{Caron-Huot:2018ape} at one loop. The first term in $\Theta_{3\mathrm{PM}}$ comes from the expansion of $\tan(\frac{\Theta}{2})$ at small $\Theta$ while the remaining terms are genuine new contributions from  $\operatorname{Re} \delta_2$. 

In the ultra-relativistic limit $\sigma \gg 1$ the leading term ${\cal O}(\sigma^4\log\sigma)$ in the second line of \eqref{DA2} cancels and only the first line survives reproducing the universal and finite ultra-relativistic result of \cite{Amati:1990xe},
\begin{equation}\label{eq:ACV90Re}
	\operatorname{Re}2\delta_2 \sim \frac{4 G^3 s^2}{b^2}\,.
\end{equation}
Thanks to analyticity and crossing arguments, as discussed in \cite{DiVecchia:2020ymx}, this cancellation can be seen as a consequence of the one occurring in the imaginary part as highlighted below Eq.~\eqref{eq:l1mzsi}. 

It is also instructive to look at the opposite limit $\sigma \to 1$ which is relevant to the PN  regime. 
In this respect, one should note that, as can be better appreciated looking at the second form of equation \eqref{CR8}, the first and last term (highlighted in blue color) contain only even powers of $p \sim  \sqrt{\sigma^2-1}$, while the second term has only odd powers. This means that the first and last term represent half-integer-PN corrections to the deflection angle, which are a consequence of dissipative processes, i.e.~the signature of radiation reaction. Instead,
the second term in \eqref{CR8} corresponds to the more conventional integer-PN expansion due to the conservative dynamics. 
More precisely, in the deflection angle \eqref{DA2}, such dissipative effects appear starting at $1.5$PN. This seems in tension with the $2.5$PN scaling of radiation reaction effects that is expected from GR calculations. As we shall see, this apparent mismatch is due to the presence of additional massless states propagating in the supersymmetric theory, while the same calculation in GR will lead to $2.5$PN radiation-reaction effects, as it should.

\subsubsection{Real-analytic, crossing-symmetric reformulation at two loops}
\label{analyticN=8}

We will now extend to two loops the procedure followed in Subsection~\ref{analyticN81loop} to recast the explicit results of the previous subsection  in  real-analytic and crossing-symmetric form.
We first combine the different contributions in the known form of Eq.~\eqref{CR6},
which suggests the ansatz: 
\begin{eqnarray}
	&&\mathcal{A}_2 (s, q^2) =  \frac{(8\pi G)^3 }{(4\pi)^4}  \left( \frac{4\pi {\rm e}^{-\gamma_E}}{q^2}\right)^{2\epsilon}  (\sigma^4 + \bar{\sigma}^4 ) \hat{A}_2 (s, q^2)  \,,  \nonumber \\
	&& \hat{A}_2 (s, q^2) =  \frac{\hat{A}_2^{[2]} (s, q^2)}{\epsilon^2} +  \frac{\hat{A}_2^{[1]} (s, q^2)}{\epsilon}\,,
	\label{Ansatz}
\end{eqnarray}
where ${\bar{\sigma}}$ has been introduced in \eqref{defs}.
In order to reproduce Eqs.~\eqref{A22} and \eqref{A22bis} we find:
\begin{eqnarray}
	&&  \hat{A}_2^{[2]} (s, q^2) = \Bigg[- \frac{8 \pi^2 m_1^4 m_2^4}{q^2} \left( \frac{\sigma^2}{\sigma^2-1} +  \frac{\bar{\sigma}^2}{\bar{\sigma}^2-1}\right) \nonumber \\
	&& -  8  m_1^3 m_2^3  \Bigg(  \frac{\sigma}{(\sigma^2-1)^2}  \log^2(-z) +   \frac{\bar{\sigma}}{(\bar{\sigma}^2-1)^2}  \log^2(-\bar{z})   \nonumber \\
	&& + 2\left[ \frac{\sigma^2}{(\sigma^2-1)^{3/2}}  \log(-z) +   \frac{\bar{\sigma}^2}{(\bar{\sigma}^2-1)^{3/2}}  \log(-\bar{z})  \right] \Bigg) \Bigg] 
	\label{Ansatz(2)}
\end{eqnarray}
and for Eqs.~\eqref{RA10} and   \eqref{IA10}:
\begin{subequations}
	\label{Ansatz(1bis)}
	\begin{eqnarray}
		&&  \frac{\hat{A}_2^{[1]} (s, q^2)}{m_1^3 m_2^3 } = 4 \pi^2  \frac{(m_1^2+ m_2^2)}{m_1 m_2} \frac{\sigma^2}{(\sigma^2-1)^2} \label{ImD1d0}\\
		&& - 32   \frac{\sigma^2}{(\sigma^2-1)^{3/2}} \left( \log(-z) - \log(z)  \right) \label{ReD1d0}\\
		&& + 4  \frac{\sigma (\sigma^2-2)}{(\sigma^2-1)^2}\left( \log(-z) - \log(z)  \right) \left(\operatorname{Li}_2(z^2) - \operatorname{Li}_2\left(\frac{1}{z^2}\right)\right) \label{newwlog} \\
		&& - \frac{8}{3}   \frac{1}{(\sigma^2-1)^{1/2}} \left[\left( \log^3(-z) - \log^3(z) \right) + \pi^2 \left( \log(-z) - \log(z) \right) \right] \label{P-MRZ} \\
		&& - 8  \frac{\sigma^3}{(\sigma^2-1)^2}\left( \log(-z) - \log(z)  \right)\left( \log(-z) + \log(z)  \right) \label{D1d0}\\
		&&+ 8    \frac{\sigma^2}{(\sigma^2-1)^{3/2}} \left( \log(-z) - \log(z)  \right) \left[ \log(1-z^2) + \log\left(1-\frac{1}{z^2}\right) \label{ACV} \right] 
	\end{eqnarray}
\end{subequations}
The two previous equations have been derived using    $\log (-z) = i\pi + \log z$ and the following identities and branch choices:
\begin{eqnarray}
	&& - \operatorname{Li}_2\left(\frac{1}{z^2}\right) =  \operatorname{Li}_2(z^2)+\frac{\pi^2}{6} + \frac12 \log^2(-z^2) = \operatorname{Li}_2(z^2) - \frac{\pi^2}{3} + 2 \log^2 z+2 i \pi \log z ; \nonumber\\
	&& \log\left(1-\frac{1}{z^2}\right) =  \log(1-z^2)  - 2 \log z - i \pi.
	\label{ident}
\end{eqnarray}
Combining these with \eqref{enha3} we can also write the combination appearing in  \eqref{newwlog} in a  form:
\begin{equation}
	\operatorname{Li}_2(z^2) - \operatorname{Li}_2\left(\frac{1}{z^2}\right) = - 2 \left( \operatorname{Li}_2(1- z^2) +  \log z^2 \log (1-z^2)  - \log^2 z -  i \pi \log z\right)\; ,
	\label{ident1}
\end{equation}
that is useful to discuss the non relativistic ($\sigma , z \to 1$) limit.

We can check that the expressions \eqref{Ansatz(2)} and \eqref{Ansatz(1bis)}  satisfy   crossing symmetry and real-analyticity.
The first property follows simply from the fact that, at this order in $q^2$, we can use the identification ${\bar{z}} = - \frac{1}{z}$. 

Checking real analyticity is a bit more subtle.   For \eqref{Ansatz(2)} one needs to take into account the subleading (in $q^2$) term origination from the first line. This (purely real) term exactly cancels a similar term coming from the second line. Then one is left with a purely imaginary term from the second line:
\begin{eqnarray}
	- \frac{16 m_1^3 m_2^3 \pi \sigma}{(\sigma^2-1)^2} i \log z \,\,,
	\label{replace1}
\end{eqnarray}
that can be written in real-analytic crossing-symmetric form as
\begin{equation}
	\label{replace}
	-  8  m_1^3 m_2^3  \frac{\sigma}{(\sigma^2-1)^2} \frac12 ( \log(-z) +   \log(-\bar{z}))( \log (z^2) -   \log(\bar{z}^2)) \,.
\end{equation}

At this point checking that the amplitude is real in the unphysical region $\sigma^2 < 1$ is straightforward once one realizes that, in that region, $|z|^2 = 1$, in other words $z$ is a pure phase and thus $\log z$ is purely imaginary.  This implies that the term in Eq. \eqref{replace1} is real
and the same is true for the last term in \eqref{Ansatz(2)} because of an extra factor $i$ coming from $\sqrt{\sigma^2-1}$.

Coming now to \eqref{Ansatz(1bis)}, real analyticity is easily checked along the same lines for \eqref{ImD1d0}, \eqref{ReD1d0}, \eqref{P-MRZ} and \eqref{D1d0}.
Concerning instead \eqref{newwlog} and \eqref{ACV} one has to remember
that, in the unphysical region, $z^{-1} = {\bar{z}}$. Since both $\operatorname{Li}_2(z^2)$ and $\log(1-z^2)$ are real-analytic functions, the combinations involving them, appearing in \eqref{newwlog} and \eqref{ACV} give a purely imaginary and purely real factor, respectively. Because of the different powers of $(\sigma^2-1)$ appearing in the two contributions this is exactly as needed for real analyticity. In conclusion, we have shown that Eq.~\eqref{Ansatz(2)} and
Eqs.~\eqref{Ansatz(1bis)} are real analytic functions. Note that this would not have been the case for the eikonal itself.

\subsection{General Relativity }

In this section we illustrate the calculation of the 3PM eikonal from the $2\to2$ amplitude for the collision of two massive scalars in GR. 
This calculation was carried out for the first time in full detail in \cite{Bjerrum-Bohr:2021din}. The underlying principles of this calculation are the same as those presented in the maximally supersymmetric case. One aims to reconstruct the amplitude $\mathcal A_2$ to its first few orders in the small-$q$ expansion, perform the Fourier transform to $b$-space, obtaining $\tilde{\mathcal A}_2$, and retrieve the classical information from $\operatorname{Re}2\delta_2$ by matching with the exponentiation. More explicitly, armed with the $q$-expansion of $\mathcal A_2$ or equivalently with the $1/b$-expansion of $\tilde{\mathcal A}_2$ and with the lower-loop order data $2\delta_0$, $2\delta_1$, $2\Delta_1$, one checks that in the equation
\begin{equation}\label{exp2AD}
	i\tilde{\mathcal A}_2 = \frac{(2i\delta_0)^3}{3!} + 2i\delta_0 \, 2i\delta_1  + 2i\delta_2 + i 2\delta_0 \,2i\Delta_1
\end{equation}
the first two terms on the right hand side cancel out against analogous superclassical terms of the right-hand side, and obtains $2\delta_2$ by subtracting $i 2\delta_0 \,2i\Delta_1$ from the rest.
The main novelty is the fact that the gravity integrand takes a much more involved form compared to the neat expression \eqref{CR1}, where only scalar integrals appeared. Still, one can reconstruct this integrand in a convenient way using generalized unitarity techniques \cite{Bern:2019nnu,Bern:2019crd,Herrmann:2021tct,Bjerrum-Bohr:2021din,Brandhuber:2021eyq} and then perform the integration using the same families of master integrals that were needed in the $\mathcal N=8$ case \cite{Bjerrum-Bohr:2021din}.

Instead of following this route, here we shall adopt a different strategy, which combines unitarity, analyticity and  crossing symmetry with the eikonal exponentiation itself. The idea is to start from \eqref{exp2AD}, which we may rewrite as
\begin{equation}\label{exp2ADReIm}
	i\tilde{\mathcal A}_2 = \frac{(2i\delta_0)^3}{3!} + 2i\delta_0 \, 2i\delta_1  + i [\operatorname{Re}2\delta_2 + i \operatorname{Im}2\delta_2 + i 2\delta_0 \,2\Delta_1]
\end{equation}
and regard $\tilde{\mathcal A}_2$ and $\operatorname{Re}2\delta_2$ as the unknown quantity. Indeed the tree-level and one-loop information determines $2\delta_0$, $2\delta_1$, $2\Delta_1$, and $\operatorname{Im}2\delta_2$ can be obtained by performing a relatively simpler integral over three-particle phase space, as we have discussed in the Section~\ref{sec:radiation}. This data determines all imaginary parts arising from the three-particle and two-particle cuts of the amplitude (plus a portion of the real part arising from $2\delta_0\,\operatorname{Im}2\Delta_1$). We can then make a real-analytic and cross-symmetric ansatz for $\mathcal A_2$, up to the $(q^2)^{-2\epsilon}$ order relevant for the classical limit, complete of its real and imaginary parts. As we shall see, the available data will be enough to fix this ansatz up to a single unknown function $f_a(\sigma)$, which however appears with an extra power of $s$ in front and is thus dominant in the limit of small mass ratio. Therefore, using the results of \ref{app:geoschw}, this last term can be fixed too by matching with the probe-limit calculation.
This will lead us to the complete result of $\mathcal A_2$ appropriate to then retrieve $\operatorname{Re}2\delta_2$ and the classical deflection angle.

Let us start by adopting a notation for the small-$q$ and small-$\epsilon$ expansion of the amplitude:
\begin{equation}\label{m2GR}
	\begin{split}
		\mathcal{A}_2 
		&= \mathcal{A}_2^{[2]} + \mathcal{A}_2^{[1]} + \mathcal{A}_2^{[0]} + \mathcal O(q^{1-4\epsilon})\\
		&= 
		\left(
		\frac{4\pi e^{-\gamma_E}}{q^2}
		\right)^{2\epsilon}
		\left[
		\frac{\mathcal{A}_2^{(2,2)}}{\epsilon^2 q^2}
		+
		\frac{\mathcal{A}_2^{(1,1)}}{\epsilon\, q}
		+
		\frac{\mathcal{A}_2^{(2,0)}}{\epsilon^2}
		+
		\frac{\mathcal{A}_2^{(1,0)}}{\epsilon}
		+\mathcal O(q)
		\right],
	\end{split}
\end{equation}
where $\mathcal{A}_2^{[k]}\sim q^{-k-4\epsilon}$ and in the second line we introduced a notation similar to the one employed for the $\mathcal N=8$ in \eqref{A222010} for convenience.
As is by now familiar, $\mathcal{A}_2^{[2]}$ and $\mathcal{A}_{2}^{[1]}$ are simply determined by the iteration of classical terms, i.e.~the first two terms on the right-hand side of \eqref{exp2ADReIm}. Therefore they must be given by the inverse Fourier transform of $- \frac{1}{3!}(2\delta_0)^3$ (see \eqref{eq:d0m1m2}),
\begin{equation}\label{m222GR}
	\mathcal{A}_{2}^{(2,2)}
	=
	-\frac{64 \pi G^{3} m_{1}^{4} m_{2}^{4}\left(\sigma^{2}-\tfrac{1}{2(1-\epsilon)}\right)^{3} \Gamma(1-\epsilon)^{3} \Gamma(1+2 \epsilon)}{ \left(\sigma^{2}-1\right) \Gamma(1-3 \epsilon)}\,
	e^{2\epsilon\gamma_E},
\end{equation}
and by the inverse Fourier transform of $2\delta_0 \,2i\delta_1$ (see \eqref{delta1a}), 
\begin{equation}\label{m211GR}
	\mathcal{A}_{2}^{(1,1)}=\frac{6 i \pi^{2} G^{3}\left(m_{1}+m_{2}\right) m_{1}^{3} m_{2}^{3}\left(2 \sigma^{2}-1\right)\left(1-5 \sigma^{2}\right)}{\sqrt{\sigma^{2}-1}}+\mathcal{O}\left(\epsilon\right),
\end{equation}
where we expanded for simplicity to leading order in $\epsilon$, although both $2\delta_0$ and $2\delta_1$ are of course known in any dimension.

The new classical information is contained in $\mathcal{A}_2^{[0]}$, which, comparing with \eqref{exp2ADReIm}, must take the following form in $b$-space
\begin{equation}\label{}
	\widetilde{\mathcal{A}}_2^{[0]} =  \operatorname{Re}2\delta_2+2\delta_0\,2i\Delta_1 + i \operatorname{Im}2\delta_2\,.
\end{equation} 
We may rewrite  this in $q$-space as
\begin{equation}\label{new2pc3pc}
	\mathcal{A}_{2}^{[0]}
	=
	\mathcal{A}_2^{\text{new}}
	+
	\mathcal{A}_{2pc}
	+
	\mathcal{A}_{3pc}\,.
\end{equation}
Our next task, following the above strategy, is to reconstruct the second and third term from two-particle and tree-particle cuts.

\subsubsection{Two-particle and three-particle cuts}

For the two-particle cut we can use
\begin{equation}\label{}
	i \mathcal{A}_{2pc} = 4m_1m_2\sqrt{\sigma^2-1} \int d^{2-2\epsilon}b\, e^{-ib\cdot q} 2i\delta_0\,2i\Delta_1\,.
\end{equation}
The leading eikonal is \eqref{eq:d0m1m2}
\begin{equation}\label{}
	2i\delta_0
	=
	\frac{2 i m_{1} m_{2} G\left(\sigma^{2}-\frac{1}{2-2\epsilon}\right) \Gamma(-\epsilon)}{\sqrt{\sigma^{2}-1} \left(\pi b^{2}\right)^{-\epsilon}}
\end{equation}
and the quantum remainder at one-loop order, $2\Delta_1$, is given by \eqref{2D1GR}.
At this stage, we can use the inverse Fourier transform \eqref{B1bis}
\begin{equation}\label{inverseFT2pc}
	4m_1 m_2 \sqrt{\sigma^2-1}\int d^{2-2\epsilon}b\ \frac{e^{-ibq}}{(b^2)^{1-3\epsilon}} = \frac{	4m_1 m_2 \sqrt{\sigma^2-1}}{(q^2)^{2\epsilon}}
	\frac{\pi^{1-\epsilon}\Gamma(2\epsilon)}{2^{-4\epsilon}\Gamma(1-3\epsilon)}
\end{equation}
to obtain $i\mathcal{A}_{2pc}$ after multiplying $2i\delta_0$ and $2i\Delta_1$. 
Since $2\delta_0$ is $\mathcal O(\epsilon^{-1})$ and in addition \eqref{inverseFT2pc} involves a singular factor $\Gamma(\epsilon)$, the first two nontrivial orders are captured by retaining up to $\mathcal O(\epsilon^{-1})$, for which we find
\begin{equation}\label{2ptc}
	\begin{split}
		&\mathcal{A}_{2pc} 
		=
		\frac{2   G^3\pi s m_1^2 m_2^2 \left(2\sigma ^2-1\right)^3}{\epsilon(q^2)^{2\epsilon}\left(\sigma^{2}-1\right)^{2}} 
		\\
		&
		-
		\frac{i2 G^3 m_1^3 m_2^3 \left(2\sigma ^2-1\right)}{\epsilon^2\left(\frac{q^2e^{\gamma_E}}{4\pi}\right)^{2\epsilon}\left(\sigma^{2}-1\right)^{\frac{3}{2}}} 
		\!
		\left(
		\frac{2 \sigma(2 \sigma^{2}-1)(6 \sigma^{2}-7) \operatorname{arccosh}\sigma}{\sqrt{\sigma^{2}-1}}
		-
		\frac{1-49 \sigma^{2}+18 \sigma^{4}}{15}
		\right)
		\\
		&
		+
		\frac{iG^3m_1^3m_2^3}{\epsilon(q^2)^{2\epsilon}(\sigma^2-1)^{\frac{3}{2}}}
		\Big(
		\frac{4 \sigma  \left(2 \sigma ^2-1\right) \left(8 \sigma ^4+2 \sigma ^2-11\right)\operatorname{arccosh}\sigma}{\sqrt{\sigma^2-1}}\\
		&
		\hspace{80pt}-
		\frac{18468 \sigma ^6-30728 \sigma ^4+13113 \sigma ^2-1753}{225}
		\Big).
	\end{split}
\end{equation}

Then, we need to add the imaginary part coming from the three-particle cut. As we know from Section~\ref{sec:radiation}, this is related to $\operatorname{Im}2\delta_2$ by Fourier transform. In $b$-space, we have \eqref{GR15},
\begin{equation}\label{Im2delta2toberef}
	\begin{split}
		\operatorname{Im}2\delta_2 &= \frac{2 m_1^2 m_2^2 
			(2 \sigma ^2-1)^2 G^3}{\pi b^2 \left(\sigma ^2-1\right)^2}\,(\pi b^2 e^{\gamma_E})^{3\epsilon} \\		
		&
		\times
		\Big[
		\left(
		-\frac{1}{\epsilon}
		+
		\log(4(\sigma^2-1))
		\right)
		\left(\frac{\sigma  \left(2 \sigma ^2-3\right)\operatorname{arccosh}\sigma}{\sqrt{\sigma ^2-1}}+\frac{8-5 \sigma ^2}{3}\right)\\
		&-
		(\operatorname{arccosh}\sigma)^2 \left(\frac{\sigma  \left(2 \sigma ^2-3\right)}{\sqrt{\sigma ^2-1}}+\frac{2 (\sigma^2-1)(4\sigma^4-12\sigma^2-3)}{\left(1-2 \sigma ^2\right)^2}\right)\\
		&+(\operatorname{arccosh}\sigma)
		\ \frac{\sigma  \left(88 \sigma ^6-240 \sigma ^4+240 \sigma ^2-97\right)}{3 \left(1-2 \sigma ^2\right)^2 \sqrt{\sigma ^2-1}}\\
		&+\operatorname{Li}_2(1-z^2)\frac{\sigma  \left(3-2 \sigma ^2\right)}{\sqrt{\sigma ^2-1}}\,
		+\frac{-140 \sigma ^6+220 \sigma ^4-127 \sigma ^2+56}{9 \left(1-2 \sigma ^2\right)^2}\Big].
	\end{split}
\end{equation}
Since $\widetilde{\mathcal{A}}_{3pc}=i\operatorname{Im}2\delta_2$, we can again go to $q$-space using \eqref{inverseFT2pc}, finding
\begin{equation}\label{3ptc}
	\begin{split}
		\mathcal{A}_{3pc} &= 
		\frac{4i m_1^3 m_2^3 (2 \sigma ^2-1)^2 G^3}{\epsilon\left(\sigma ^2-1\right)^{\frac{3}{2}}}\,\left(
		\frac{4\pi e^{-\gamma_E}}{q^2}
		\right)^{2\epsilon} \\		
		&
		\times
		\Bigg[
		\left(
		-\frac{1}{\epsilon}
		+
		\log(4(\sigma^2-1))
		\right)
		\left(\frac{\sigma  \left(2 \sigma ^2-3\right)\operatorname{arccosh}\sigma}{\sqrt{\sigma ^2-1}}+\frac{8-5 \sigma ^2}{3}\right)\\
		&-
		(\operatorname{arccosh}\sigma)^2 \left(\frac{\sigma  \left(2 \sigma ^2-3\right)}{\sqrt{\sigma ^2-1}}+\frac{2 (\sigma^2-1)(4\sigma^4-12\sigma^2-3)}{\left(1-2 \sigma ^2\right)^2}\right)\\
		&+(\operatorname{arccosh}\sigma)
		\ \frac{\sigma  \left(88 \sigma ^6-240 \sigma ^4+240 \sigma ^2-97\right)}{3 \left(1-2 \sigma ^2\right)^2 \sqrt{\sigma ^2-1}}\\
		&+\operatorname{Li}_2(1-z^2)\frac{\sigma  \left(3-2 \sigma ^2\right)}{\sqrt{\sigma ^2-1}}\,
		+\frac{-140 \sigma ^6+220 \sigma ^4-127 \sigma ^2+56}{9 \left(1-2 \sigma ^2\right)^2}\Bigg].
	\end{split}
\end{equation}

Comparing the decomposition \eqref{m2GR}, \eqref{new2pc3pc} with the above explicit formulas for $\mathcal{A}_{2pc}$ and $\mathcal{A}_{3pc}$, we discover that $\mathcal{A}_{2}^{(2,0)}$ must be  purely imaginary in order for $\mathcal A_2^\text{new}$ not to have $\mathcal O(\epsilon^{-2})$ real contributions that would result in a divergent eikonal \textit{phase}, and is given by
\begin{equation}\label{m220GR}
	\begin{split}
		i\mathcal{A}_2^{(2,0)}
		&=
		\frac{2 G^3 m_1^3 m_2^3 \left(2\sigma ^2-1\right)}{\left(\sigma^{2}-1\right)^{\frac{3}{2}}} 
		\!
		\left(
		\frac{2 \sigma(2 \sigma^{2}-1)(6 \sigma^{2}-7) \operatorname{arccosh}\sigma}{\sqrt{\sigma^{2}-1}}
		-
		\frac{1-49 \sigma^{2}+18 \sigma^{4}}{15}
		\right)
		\\
		&+
		\frac{4 G^3 m_1^3 m_2^3 (2 \sigma ^2-1)^2}{\left(\sigma ^2-1\right)^{\frac{3}{2}}}\,
		\left(\frac{\sigma  \left(2 \sigma ^2-3\right)\operatorname{arccosh}\sigma}{\sqrt{\sigma ^2-1}}+\frac{8-5 \sigma ^2}{3}\right).
	\end{split}
\end{equation}
Of course the imaginary part of $\mathcal{A}_2^{(1,0)}$ is also determined by the cuts, i.e.~by the $\mathcal O(\epsilon^{-1}(q^2)^{-2\epsilon})$ terms of \eqref{2ptc}, \eqref{3ptc}, 
\begin{equation}\label{Imm210}
	\begin{split}
		\operatorname{Im}\mathcal{A}_{2}^{(1,0)}
		&=
		\frac{G^3m_1^3m_2^3}{(\sigma^2-1)^{\frac{3}{2}}}
		\Big(
		\frac{4 \sigma  \left(2 \sigma ^2-1\right) \left(8 \sigma ^4+2 \sigma ^2-11\right)\operatorname{arccosh}\sigma}{\sqrt{\sigma^2-1}}\\
		&
		\hspace{80pt}-
		\frac{18468 \sigma ^6-30728 \sigma ^4+13113 \sigma ^2-1753}{225}
		\Big)
		\\
		&
		+\frac{4m_1^3 m_2^3 (2 \sigma ^2-1)^2 G^3}{\left(\sigma ^2-1\right)^{\frac{3}{2}}} \\		
		&
		\times
		\Bigg[
		\log(4(\sigma^2-1))
		\left(\frac{\sigma  \left(2 \sigma ^2-3\right)\operatorname{arccosh}\sigma}{\sqrt{\sigma ^2-1}}+\frac{8-5 \sigma ^2}{3}\right)\\
		&-
		(\operatorname{arccosh}\sigma)^2 \left(\frac{\sigma  \left(2 \sigma ^2-3\right)}{\sqrt{\sigma ^2-1}}+\frac{2 (\sigma^2-1)(4\sigma^4-12\sigma^2-3)}{\left(1-2 \sigma ^2\right)^2}\right)\\
		&+(\operatorname{arccosh}\sigma)
		\ \frac{\sigma  \left(88 \sigma ^6-240 \sigma ^4+240 \sigma ^2-97\right)}{3 \left(1-2 \sigma ^2\right)^2 \sqrt{\sigma ^2-1}}\\
		&+\operatorname{Li}_2(1-z^2)\frac{\sigma  \left(3-2 \sigma ^2\right)}{\sqrt{\sigma ^2-1}}\,
		+\frac{-140 \sigma ^6+220 \sigma ^4-127 \sigma ^2+56}{9 \left(1-2 \sigma ^2\right)^2}\Bigg].
	\end{split}
\end{equation}
For convenience, we may trade the unknown function $\mathcal{A}_2^{\text{new}}$ in \eqref{new2pc3pc} for an equivalent one $\mathcal{A}_2^{(1,\text{new})}$, obtained by removing the contribution of the 2-particle cut \eqref{2ptc} to the real part of $\mathcal{A}_2^{(1,0)}$:
\begin{equation}\label{Rem210}
	\begin{split}
		\operatorname{Re}\mathcal{A}_2^{(1,0)}
		&=
		\frac{2   G^3\pi s m_1^2 m_2^2 \left(2\sigma ^2-1\right)^3}{\left(\sigma^{2}-1\right)^{2}}  + \mathcal{A}_2^{(1,\text{new})}\,.
	\end{split}
\end{equation}

\subsubsection{Real-analytic, crossing-symmetric ansatz in GR }
\label{analyticGR2loops}

In this subsection we will follow a shortcut, based on analyticity and crossing symmetry, in order to reconstruct the full two-loop scattering amplitude (at the relevant classical level) from its imaginary part, the idea being that the latter is easier to compute from lower-loop-order on shell amplitudes. As it turns out, the only missing information is contained in purely real terms that can be easily computed from the probe limit.

In order to pursue this program we start with an ansatz using the analytic structures we have already seen in the $\mathcal{N}=8$ case discussed in subsection \ref{analyticN=8}.
Let us first notice that \eqref{m211GR} can be written in the analytic crossing symmetric form:
\begin{equation}\label{ra11}
	\mathcal{A}_{2}^{(1,1)}=\frac{6  \pi G^{3}\left(m_{1}+m_{2}\right) m_{1}^{3} m_{2}^{3}\left(2 \sigma^{2}-1\right)\left(1-5 \sigma^{2}\right)}{\sqrt{\sigma^{2}-1}}  \left( \log(-z) + \log(-\bar{z})  \right)+\mathcal{O}\left(\epsilon\right)\,.
\end{equation}
For the other terms we try, in analogy with \eqref{Ansatz1sclGR}, the following ansatz:
\begin{equation}	\label{AnsatzGR}
	\begin{split}
\mathcal{A}_2 (s, q^2) &=  \frac{(8\pi G)^3 }{(4\pi)^4}  \left( \frac{4\pi {\rm e}^{-\gamma_E}}{q^2}\right)^{2\epsilon}   \hat{A}_2 (s, q^2)\,, \\	
	 \hat{A}_2 (s, q^2) &=  \frac{\hat{A}_2^{[2]} (s, q^2, \epsilon)}{\epsilon^2} +  \frac{\hat{A}_2^{[1]} (s, q^2)}{\epsilon}\,.
	 \end{split}
\end{equation}
In order to reproduce Eqs.~\eqref{m222GR}  and  \eqref{m220GR} we try the same analytic structures appearing in \eqref{Ansatz(2)}. This uniquely determines:
\begin{align}\label{Ansatz(1bisGR)}
	\begin{split}
	& \frac{ \hat{A}_2^{[2]} (s, q^2)}{m_1^3 m_2^3} = - \frac{8 \pi^2 m_1 m_2}{q^2} \left( \frac{ \sigma^2 - \frac{1}{2(1-\epsilon)}}{\sigma^2-1} +  \frac{\bar{\sigma}^2- \frac{1}{2(1-\epsilon)}}{\bar{\sigma}^2-1}\right) \left[\left(\sigma^2-\tfrac{1}{2-2\epsilon}\right)^2+\left(\bar\sigma^2-\tfrac{1}{2-2\epsilon}\right)^2\right]
	c_\text{exp}(\epsilon)
	\\ 	
	& + 4 \left[\left(\sigma^2-\tfrac{1}{2}\right)^2+\left(\bar\sigma^2-\tfrac{1}{2}\right)^2\right]  \Bigg(  \left(\frac{\sigma (4 \sigma^2-5)}{(\sigma^2-1)^2}  \log^2(-z) +   \frac{\bar{\sigma}(4 \bar{\sigma}^2-5)}{(\bar{\sigma }^2-1)^2}  \log^2(-\bar{z}) \right)  \\
	& + \left[ \frac{118 \sigma^4 - 259 \sigma^2 +81}{60(\sigma^2-\frac12)(\sigma^2-1)^{3/2}}  \log(-z) +   \frac{118 \bar{\sigma}^4 - 259 \bar{\sigma}^2 +81}{60 (\bar{\sigma}^2-\frac12)(\bar{\sigma}^2-1)^{3/2}}  \log(-\bar{z})  \right] \Bigg)  \,, 
	\end{split}
\end{align}
with
\begin{equation}\label{}
	c_\text{exp}(\epsilon) = \frac{\Gamma(1-\epsilon)^{3} \Gamma(1+2 \epsilon)}{  \Gamma(1-3 \epsilon)}\,
	e^{2\epsilon\gamma_E} \,.
\end{equation}
Note that we have kept a more complete $\epsilon$ dependence in the first line of \eqref{Ansatz(1bisGR)}. In this way, when we expand that line to lowest order in $q^2$ (by setting $\bar{\sigma} = - \sigma$), it reproduces, at all $\epsilon$, the  superclassical (iteration) term $-  \frac{i}{6} (2 i \delta_0)^3$. At first order in $\bar{\sigma} + \sigma$ we get corrections which, at lowest order in $\epsilon$,  combine with the terms in the second and third line of  \eqref{Ansatz(1bisGR)} while at ${\cal O}(\epsilon)$ they contribute to $\hat{A}_2^{[1]}$.

Concentrating on the former, note that the last line of \eqref{Ansatz(1bisGR)} gives a purely imaginary contribution (up to negligible corrections of higher order in $q^2$). The second line, instead, is proportional to $ \log^2(-z) -  \log^2(-\bar{z})$ and therefore carries both a real and an imaginary part. Amusingly, the former is exactly canceled by the subleading (and obviously real) contribution from the first line and the final result reproduces exactly the purely imaginary result \eqref{m220GR}.

Turning now to   $\hat{A}_2^{[1]}$, we have to combine the above-mentioned leftover piece from the first line of \eqref{Ansatz(1bisGR)} with Eqs.~\eqref{RA10} and   \eqref{IA10} with an appropriate ansatz for the real-analytic, crossing symmetric $\hat{A}_2^{[1]}$ itself. Based on the analogy with the $\mathcal N=8$ case we try
\begin{subequations}	\label{Ansatz(unknowns)}
	\begin{align}  \label{Ansatz(unknowns)a}
		&\frac{ \hat{A}_2^{[1]} (s, q^2)}{m_1^3 m_2^3} = \pi^2 \frac{ (m_1^2+ m_2^2)}{m_1 m_2}\frac{\left[{\color{green!60!black}f_a}+(2\sigma^2-1)^3 \right]}{(\sigma^2-1)^2} \\
		& + \frac{f_b}{(\sigma^2-1)^{3/2}} \left( \log(-z) - \log(z)  \right) \\
		& +   \frac{f_c}{(\sigma^2-1)^2}\left( \log(-z) - \log(z)  \right) \left(\operatorname{Li}_2(z^2) - \operatorname{Li}_2\left(\frac{1}{z^2}\right)\right) \\
		& + \frac{f_d}{(\sigma^2-1)^{1/2}} \left[\left( \log^3(-z) - \log^3(z) \right) + \pi^2 \left( \log(-z) - \log(z) \right) \right] \\
		& + \frac{f_e}{(\sigma^2-1)^2}\left( \log^2(-z) - \log^2(z)  \right) \\
		&+ \frac{f_f}{(\sigma^2-1)^{3/2}} \left( \log(-z) - \log(z)  \right) \left[ \log(1-z^2) + \log\left(1-\frac{1}{z^2}\right)  \right] 
	\end{align}
\end{subequations}
where the unknown functions $f_a,\ldots,f_f$ are polynomials in $\sigma$. The first line of the ansatz~\eqref{Ansatz(unknowns)a} is motivated by the analogue equation for the ${\cal N}=8$ case, see~\eqref{ImD1d0}, but here we allow for an extra contribution, ${\color{green!60!black}f_a}$, besides the one from the 2-particle cut in the first line in~\eqref{2ptc}.

Matching onto the known terms in \eqref{Imm210}   fixes uniquely all the coefficients $f_i$ except for ${\color{green!60!black}f_a}$, which is purely real.
We thus obtain the following result consisting of six distinct real-analytic crossing-symmetric structures paralleling exactly the ones we found in ${\cal N} = 8$:

	\begin{subequations}\label{Ansatz(1bis)GR}
		\begin{align}  
			& \frac{ \hat{A}_2^{[1]} (s, q^2)}{m_1^3 m_2^3} = \pi^2 \frac{ (m_1^2+ m_2^2)}{m_1 m_2} \frac{\left[{\color{green!60!black}f_a}+(2\sigma^2-1)^3 \right]}{(\sigma^2-1)^2} \label{ImD1d0GR}\\
			& -   \frac{32468 \sigma^6 - 52728 \sigma^4 + 25813 \sigma^2 -7353 }{450 (\sigma^2-1)^{3/2}} \left( \log(-z) - \log(z)  \right) \label{ReD1d0GR} \\
			& +   \frac{\sigma (2 \sigma^2-1)^2(2 \sigma^2 -3)}{(\sigma^2-1)^2}\left( \log(-z) - \log(z)  \right) \left(\operatorname{Li}_2(z^2) - \operatorname{Li}_2\left(\frac{1}{z^2}\right)\right) \label{newwlogGR}  \\
			& - \frac{4}{3}   \frac{4\sigma^4 - 12 \sigma^2 -3}{(\sigma^2-1)^{1/2}} \left[\left( \log^3(-z) - \log^3(z) \right) + \pi^2 \left( \log(-z) - \log(z) \right) \right] \label{P-MRZGR}  \\
			& - \frac{4}{3}   \frac{\sigma(34 \sigma^6 -63 \sigma^4 + 42 \sigma^2 -16)}{(\sigma^2-1)^2}\left( \log^2(-z) - \log^2(z)  \right) \label{D1d0GR} \\
			&+ \frac{2}{3}      \frac{(2 \sigma^2-1)^2 (8- 5 \sigma^2)}{(\sigma^2-1)^{3/2}} \left( \log(-z) - \log(z)  \right) \left[ \log(1-z^2) + \log\left(1-\frac{1}{z^2}\right) \label{ACVGR} \right].
		\end{align}
\end{subequations}

\subsubsection{The 3PM eikonal \texorpdfstring{$2\delta_2$}{2delta2} }

Substituting \eqref{Ansatz(1bis)GR}, \eqref{Ansatz(1bisGR)} into \eqref{AnsatzGR}, and expanding for small $q^2$ as instructed by \eqref{m2GR}, we finally obtain
\begin{equation}\label{Rem210final}
	\begin{split}
		\operatorname{Re}\mathcal{A}_2^{(1,0)}
		&=
		\frac{2   G^3\pi s m_1^2 m_2^2 \left(2\sigma ^2-1\right)^3}{\left(\sigma^{2}-1\right)^{2}} \\
		&+2 \pi G^{3} m_{1}^{2} m_{2}^{2}
		\Bigg[{\color{green!60!black} \frac{s f_a(\sigma)}{(\sigma ^2-1)^2} }-\frac{4}{3} m_1 m_2 \sigma  \left(14 \sigma ^2+25\right)\\
		&+\frac{4 m_{1} m_{2}\left(3+12 \sigma^{2}-4 \sigma^{4}\right) \operatorname{arccosh}(\sigma)}{\sqrt{\sigma^{2}-1}} \\
		&+\frac{2 m_{1} m_{2}\left(2 \sigma^{2}-1\right)^{2}}{\sqrt{\sigma^{2}-1}}\left(
		\frac{8-5 \sigma ^2}{3 \left(\sigma ^2-1\right)}+\frac{\sigma  \left(2 \sigma ^2-3\right) \operatorname{arccosh}(\sigma )}{\left(\sigma ^2-1\right)^{3/2}}
		\right)
		\Bigg],
	\end{split}
\end{equation}
where the last three lines provide the sought after contribution $\mathcal A_2^{(1,\text{new})}$ in \eqref{Rem210}.

Following the same steps as for $\mathcal N=8$, we can start from the complete expression \eqref{m2GR}, go to $b$-space, subtract all iteration terms as dictated by \eqref{exp2AD}, in particular those due to $2\delta_0\,2\Delta_1$, and drop the imaginary part $\operatorname{Im}2\delta_2$.
However, since \eqref{Rem210final} is already the real part of the $\mathcal O(q^{-4\epsilon})$ term, this procedure is equivalent to starting from \eqref{Rem210final} and dropping the first term on the right-hand side.  This term is just the real part of \eqref{2ptc}, i.e.~the one arising from $2\delta_0\,2\Delta_1$. The final step is to restore the appropriate overall factors appearing in \eqref{m2GR} and go to $b$-space. 

In other words Eq. \eqref{Rem210final}, without the term in the first line, is the subtracted two-loop amplitude that, when translated in impact parameter space, directly gives the 3PM eikonal. In Subsection \ref{ssec:connectionradialaction} we will instead show which subtractions are appropriate for obtaining  the radial action and we will see that they are different from the ones for obtaining the eikonal (though of course closely related).

The result is the 3PM eikonal phase
\begin{equation}\label{re2d2}
	\begin{aligned}
		&\operatorname{Re} 2 \delta_{2}^{(g r)}=
		\frac{4 G^{3} m_{1}^{2} m_{2}^{2}}{b^{2}}\left\{
		{\color{blue}
			\frac{\left(2 \sigma^{2}-1\right)^{2}\left(8-5 \sigma^{2}\right)}{6\left(\sigma^{2}-1\right)^{2}}
		}
		-\frac{\sigma\left(14 \sigma^{2}+25\right)}{3 \sqrt{\sigma^{2}-1}}\right.\\
		&\left.
		{\color{green!60!black}	
			+\frac{s f_a(\sigma)}{4 m_{1} m_{2}\left(\sigma^{2}-1\right)^{\frac{5}{2}}}
		}	
		+\operatorname{arccosh}\sigma \left[
		{	\color{blue}
			\frac{\sigma\left(2 \sigma^{2}-1\right)^{2}\left(2 \sigma^{2}-3\right)}{2\left(\sigma^{2}-1\right)^{\frac{5}{2}}}
		}
		+\frac{-4 \sigma^{4}+12 \sigma^{2}+3}{\sigma^{2}-1}\right]\right\},
	\end{aligned}
\end{equation}
with ${\color{green!60!black}f_a(\sigma)}$ as in \eqref{fixingfa} below.
Eq.~\eqref{re2d2}, together with the imaginary part \eqref{Im2delta2toberef}, completes our discussion of the eikonal exponentiation up to 3PM.
While in the approach presented here they all appear on the same footing, the various terms in  \eqref{re2d2} have different physical interpretations that we can now illustrate. 

As already alluded to, the term in green in the second line will fixed by probe limit calculation momentarily.
The terms in black contain instead genuine 3PM dynamical information, associated to the so-called ``potential'' interaction between the two objects. They are due to the fact that each body perceives  gravitational attraction towards the other one, but this occurs in a fully relativistic manner as dictated by GR. Historically, they were the first 3PM effects to be calculated analytically \cite{Bern:2019nnu,Bern:2019crd}, a result that was achieved by using amplitude techniques.

The terms highlighted in blue instead have a different meaning. They are due to the fact that a system of two objects undergoing nontrivial deflections is not a conservative one. The two objects can in general lose energy and angular momentum that can be stored in the gravitational field. The effect of this phenomenon is that the two bodies feel an additional, non-conservative force called radiation-reaction force \cite{Thorne:1969rba,Burke:1970wx}. To 3PM order, this reaction force is captured by the terms in blue and is actually due to a 2PM order loss of angular momentum that the two-body system transmits to the gravitational field \cite{Bini:2012ji,Damour:2020tta}.

The interplay between potential and radiation-reaction terms is crucial in order to ensure that the high-energy limit of \eqref{re2d2} matches onto the corresponding result for the massless 3PM eikonal \cite{Amati:1990xe}. Taking $\sigma\to\infty$ with
\begin{equation}\label{}
	s = E^2 = m_1^2+2m_1m_2\sigma+m_2^2 \sim 2m_1m_2\sigma\,,
\end{equation} 
one finds again \eqref{eq:ACV90Re}
in precise agreement with Ref.~\cite{Amati:1990xe}. 
In particular, in this limit, a cancellation occurring in the square brackets of \eqref{re2d2} between potential and radiation-reaction terms is crucial in order to ensure that terms of order $\frac{G^3s^2}{b^2}\log\frac{s}{m_1m_2}$ drop out.
The universality of the result \eqref{eq:ACV90Re} for $\operatorname{Re}2\delta_2$ (and of the associated $\log(s)$ in the $\operatorname{Im}2\delta_2$ \eqref{eq:ACV90Im}) is due to the fact that, in the ultrarelativistic limit, graviton exchanges dominate the interactions because they couple with the highest power of the energy, both in $\mathcal N=8$ and in GR.

We can now go from the eikonal to the deflection angle using the familiar saddle-point equation
\begin{equation}\label{2psinfromdelta}
	2p\sin\frac{\Theta}{2} = -\frac{\partial\operatorname{Re}2\delta}{\partial b}\,.
\end{equation}
To this order, it is important to include in the left hand side the full deflection angle up to 3PM order, complete of its 1PM and 2PM terms, since the next-to-leading order term in the Taylor expansion gives rise to a contribution of order $(\Theta^\text{1PM})^3$.
Let us collect here the result for the deflection angle complete up to 3PM:
\begin{equation}\label{th123}
	\begin{split}
		\Theta
		&=
		\frac{4 G E\left(\sigma^{2}-\frac{1}{2}\right)}{b\left(\sigma^{2}-1\right)}+\frac{3 \pi G^{2} E\left(m_{1}+m_{2}\right)\left(5 \sigma^{2}-1\right)}{4\left(\sigma^{2}-1\right) b^{2}}
		\\
		&+
		\frac{8 G^{3}E m_{1} m_{2}}{b^3}\left\{
		{\color{blue}
			\frac{\left(2 \sigma^{2}-1\right)^{2}\left(8-5 \sigma^{2}\right)}{6\left(\sigma^{2}-1\right)^{\frac{5}{2}}}
		}
		-\frac{\sigma\left(14 \sigma^{2}+25\right)}{3 (\sigma^{2}-1)}
		{\color{green!60!black}	
			+\frac{sf_a(\sigma)}{4 m_{1} m_{2}\left(\sigma^{2}-1\right)^{3}}
		}	
		\right.\\
		&\left.
		+\operatorname{arccosh}\sigma \left[
		{	\color{blue}
			\frac{\sigma\left(2 \sigma^{2}-1\right)^{2}\left(2 \sigma^{2}-3\right)}{2\left(\sigma^{2}-1\right)^{3}}
		}
		+\frac{-4 \sigma^{4}+12 \sigma^{2}+3}{(\sigma^{2}-1)^{\frac{3}{2}}}\right]\right\}+
		\frac{G^3 s^{\frac{3}{2}} \left(2 \sigma ^2-1\right)^3}{3 b^3 \left(\sigma ^2-1\right)^3}\,,
	\end{split}
\end{equation}
with ${\color{green!60!black}f_a(\sigma)}$ as in \eqref{fixingfa} below.
To confirm the physical interpretation of the various contributions, it is instructive to take the small-velocity limit, i.e.~
\begin{equation}\label{}
	\sigma= \frac{1}{\sqrt{1-v^2}}\,,\qquad v\to0\,.
\end{equation}
Then, the green term in \eqref{th123}, which is associated to the probe limit, scales like $G^3v^{-4}$. In the PN counting where $G\sim v^2$, this corresponds to a 1PN effect. 
The potential interaction terms, appearing in black in the last two lines of \eqref{th123}, are proportional to $G^3v^{-2}$, so they are 2PN effects. Instead, the radiation-reaction terms in blue scale as $G^3v^{-1}$ indicating a 2.5PN effect. This half-odd PN order and the odd power of the velocity is indeed the hallmark of a dissipative effect. Comparing with the situation in $\mathcal N=8$, we note that in that case radiation-reaction effects started showing up already to 1.5PN order. This is due to the presence of additional states in the spectrum of  classical fields, in particular the  ones associated to Kaluza--Klein vectors that couple to the dipole of the system, unlike the graviton, which couples to its quadrupole.

In order to express the final result in terms of the system's angular momentum, it is important to recall that $b$ is not exactly orthogonal to the incoming particle velocities. It is instead related to the orthogonal impact parameter $b_J$ such that $J =p \, b_J$ by the additional saddle-point equation $b_J=b\cos(\Theta/2)$ as in \eqref{bJbThetadelta}. 
This difference of order $G^2$ is important in the factor of $1/b$ appearing in 1PM term of \eqref{th123}, and leads to an additional contribution to the 3PM result when expressed in these variables. The deflection angle complete up to 3PM order can be then cast in the form 
\begin{equation}\label{th123J}
	\begin{split}
		\Theta
		&=
		\frac{4 G m_1 m_2 \left(\sigma^{2}-\frac{1}{2}\right)}{J\sqrt{\sigma^2-1}}
		+\frac{3 \pi G^{2} m^2_1 m^2_2\left(m_{1}+m_{2}\right)\left(5 \sigma^{2}-1\right)}{4E J^2}
		\\
		&+
		\frac{8 G^{3} m^4_{1} m^4_{2}}{s J^3}\left\{
		{\color{blue}
			\frac{\left(2 \sigma^{2}-1\right)^{2}\left(8-5 \sigma^{2}\right)}{6\left(\sigma^{2}-1\right)}
		}
		-\frac{\sigma\left(14 \sigma^{2}+25\right)}{3 (\sigma^{2}-1)^{-\frac12}}
		{\color{green!60!black}	
			+\frac{sf_a(\sigma)}{4 m_{1} m_{2}(\sigma^{2}-1)^{3/2}}
		}	
		\right.\\
		&\left.
		+\frac{\operatorname{arccosh}\sigma}{\sqrt{\sigma^2-1}} \left[
		{	\color{blue}
			\frac{\sigma\left(2 \sigma^{2}-1\right)^{2}\left(2 \sigma^{2}-3\right)}{2\left(\sigma^{2}-1\right)}
		}
		-4 \sigma^{4}+12 \sigma^{2}+3\right]\right\}
		-\frac{2 G^3 m_1^3 m_2^3 \left(2 \sigma ^2-1\right)^3}{3 J^3 \left(\sigma ^2-1\right)^{3/2}}\,.
	\end{split}
\end{equation}
Finally, in the probe limit $m_1\ll m_2$, we 
find the following 3PM contribution\footnote{The other terms in \eqref{th123J}  are negligible because they are down by a factor $\frac{m_1 m_2}{s}\sim \frac{m_1}{m_2}$.}
\begin{equation}\label{th3PROBEf}
	\Theta_\text{3PM,probe}=
	\frac{2 G^{3} m^3_{1} m^3_{2} f_a(\sigma)}{J^3(\sigma^{2}-1)^{3/2}}
	-
	\frac{2 G^3 m_1^3 m_2^3 \left(2 \sigma ^2-1\right)^3}{3 J^3 \left(\sigma ^2-1\right)^{3/2}}
\end{equation}
and we can compare this with \eqref{eq:chi3g4D} which requires (taking into account the prefactor in \eqref{eq:chiperd} with $R_s=2Gm_2$ as in \eqref{eq:RsD})
\begin{equation}\label{th3PROBE}
	\Theta_\text{3PM,probe}
	=
	\left(\frac{2Gm_2\sqrt{E_p^2-m_p^2}}{J}\right)^3
	\frac{-120 E_p^4 m_p^2+60 E_p^2 m_p^4+64 E_p^6-5 m_p^6}{12 \left(E_p^2-m_p^2\right){}^3}
\end{equation}
with $E_p=m_1\sigma$ and $m_p=m_1$. As promised, equating \eqref{th3PROBEf} with \eqref{th3PROBE} fixes the unknown polynomial to be
\begin{equation}\label{fixingfa}
	{\color{green!60!black}	f_a(\sigma) = 2 (12 \sigma^4 - 10 \sigma^2 +1) (\sigma^2-1) }\,.
\end{equation}

\subsubsection{Extracting the radial action from the amplitude}
\label{ssec:connectionradialaction}

In this subsection we shall go back to the connection \eqref{PhaseShift} between the eikonal phase, which is the main focus of the present report, and the phase shift, which is more directly related to the radial action as highlighted by \eqref{radialaction}.
As reviewed in Subsection~\eqref{ssec:pwunitarity}, the main property of the states with well-defined angular momentum is that they diagonalize the $2\to2$ elastic $S$-matrix, and the corresponding diagonal elements define the phase shift (see in particular \eqref{WE} and \eqref{2deltaj}). Correspondingly, such a basis also diagonalizes the sum over intermediate two-particle states \eqref{eq:funit}.
On the other hand, the $b$-space Fourier transform at the basis of the eikonal formalism does not achieve this diagonalization exactly, as displayed in particular by Eq.~\eqref{almostexact}, but only up to terms that are further suppressed by powers of $1/b^2$ or, equivalently, of $q^2$.

In order to recover the phase shift from the amplitude one should subtract all super-classical terms by matching to \eqref{2deltaj}
and not the $b$-space expression \eqref{fullampli}. Following the notation introduced in \eqref{2deltajchiJ} for the phase shift in the classical limit, let us similarly define the symbol $\mathcal F(s,J)$
\begin{equation}\label{}
	2f_j (s) = \mathcal F(s,J)
\end{equation}
for the partial-wave amplitude in the same limit, when $J = \hbar j$ becomes classically sizable.
In this way, suppressing again $\hbar$ from now on, \eqref{2deltaj} becomes
\begin{equation}\label{}
	1+i \mathcal F(s,J) = e^{i\chi(s,J)}(1+i\rho(s,J))\,,
\end{equation}
where $\rho(s,J)$ denotes the quantum remainder in the $J$-basis.
The loop expansion for $\mathcal F(s,J)$,
\begin{equation}\label{}
	\mathcal F(s,J) = \mathcal F_0(s,J) + \mathcal F_1(s,J) + \mathcal F_2(s,J) +\cdots \,, \qquad \mathcal F_L(s,J) \sim \mathcal O(G^{L+1})\,,
\end{equation} 
translates into a PM expansion for $\chi(s,J)$,
\begin{equation}\label{}
	\chi(s,J) = \mathcal \chi_0(s,J) + \mathcal \chi_1(s,J) + \mathcal \chi_2(s,J) +\cdots \,, \qquad \mathcal \chi_L(s,J) \sim \mathcal O(G^{L+1})
\end{equation}
and similarly for the quantum remainder $\mathcal \rho(s,J)$.
This in turn dictates the following relations,
\begin{align}\label{}
	i\mathcal F_0 &= i\mathcal \chi_0\,,\\
	i\mathcal F_1 &= \frac{(i \chi_0)^2}{2!} + i\chi_0 + i \rho_1\,,\\
	\label{F3}
	i\mathcal F_2 &= \frac{(i \chi_0)^3}{3!} + (i\chi_1)(i\chi_0) + \left[ i \chi_2 + (i\chi_0)(i\rho_1)\right]\,.
\end{align}
Despite the resemblance with \eqref{eikexp0}, \eqref{eikexp1}, \eqref{eikexp2}, at loop level these equations dictate slightly different subtractions, which do remove the superclassical terms, but in general can leave behind different classical corrections, as we now turn to illustrate. 

As we checked in \eqref{2delta0=2deltaJ} in the massless setup, the leading-order expressions for the eikonal phase and the phase shift match,
\begin{equation}\label{chi0}
	\chi_0(s,J) =2\delta_0(b)\big|_{b \to  b_J}\,.
\end{equation} 
This is not surprising because corrections in the relation \eqref{bJbThetadelta} between $b$ and $b_J = J/p$ are suppressed by powers of $G^2$ in the PM expansion. Similarly we can conclude that
\begin{equation}\label{chi1}
	\chi_1(s,J) =2\delta_1(b)\big|_{b \to  b_J}\,.
\end{equation}
We had already seen this in \eqref{phaseshift5}, whose first two terms on the right-hand side are equivalent to \eqref{chi0}, \eqref{chi1}, as follows from the definition \eqref{eq:dexpcl} of $d_1$, $d_2$.
In fact, \eqref{chi0} and \eqref{chi1} can be easily checked by noting that the characterizing property \eqref{Thetachi} is trivially equivalent to \eqref{bJbThetadelta} to this order:
\begin{equation}\label{}
	2\sin\frac{\Theta^\text{1PM}+\Theta^\text{2PM}}{2} = \Theta^\text{1PM}+\Theta^\text{2PM} + \mathcal O(G^3)  
\end{equation}
and correspondingly
\begin{equation}\label{}
	- \frac{1}{p} \frac{\partial (2\delta_1(s,b) + 2\delta(s,b))}{\partial b} = -\frac{\partial (2\delta_1(s,b_J) + 2\delta(s,b_J))}{\partial J} + \mathcal O(G^3)\,.
\end{equation}

To subleading order in the amplitude, however, we encounter the first novelty. 
Indeed, repeating the steps in Section~\ref{sec:radiation} that lead to \eqref{almostexact}, but exploiting the exact diagonalization granted by the $J$-projection, we obtain the following consequence of the unitarity relation:
\begin{equation}\label{}
	2 \operatorname{Im}_{2pc}\mathcal F(s,J) = \left| \mathcal F(s,J) \right|^2
\end{equation} 
without corrections (in contrast with \eqref{almostexact}).
In turn, this ensures that not only that
\begin{equation}\label{}
	\operatorname{Im}\chi_0(s,J)=\operatorname{Im}\chi_1(s,J)=0
\end{equation}
(consistently with \eqref{chi0}, \eqref{chi1} and with the fact that $2\delta_0$, $2\delta_1$ are real), but also that the quantum remainder is real to this order
\begin{equation}\label{}
	\operatorname{Im}\rho_1(s,J)=0\,.
\end{equation}
This reflects at two loop level \eqref{F3} into the following link:
\begin{equation}\label{subtractchi03}
	\operatorname{Re}\chi_2	=\operatorname{Re}\mathcal F_2 + \frac1{3!}(\chi_0)^3 \,,
\end{equation}
where, being purely real, $\rho_1$ has dropped out.
In other words, there is no need to know the one-loop remainder in order to calculate $\operatorname{Re}\chi_2$: one only needs to take into account the cubic contribution involving the leading-order phase shift, $\chi_0$, in the above equation.

Rather than working in $J$-space, it is more convenient to calculate the subtraction appearing in \eqref{subtractchi03} in momentum space, where it translates to the (exact) triple convolution of the tree-level amplitude. Letting the subscript $M$ stand for ``momentum space'', we have
\begin{equation}\label{}
	(\chi_0)^3_M = 
	\int\int
	\begin{gathered}
		\begin{tikzpicture}[scale=.5]
			\draw[<-] (-4.8,5.17)--(-4.2,5.17);
			\draw[<-] (-1,5.15)--(-1.6,5.15);
			\draw[<-] (-1,.85)--(-1.6,.85);
			\draw[<-] (-4.8,.83)--(-4.2,.83);
			\draw[<-] (-2.85,3.4)--(-2.85,2.6);
			\path [draw, thick, blue] (-5,5)--(-3,5)--(-1,5);
			\path [draw, thick, color=green!60!black] (-5,1)--(-3,1)--(-1,1);
			\path [draw] (-3,1)--(-3,5);
			\draw[dashed] (-3,3) ellipse (1.3 and 2.3);
			\node at (-5,5)[left]{$p_1$};
			\node at (-5,1)[left]{$p_2$};
			\node at (-2.8,3)[right]{$q_1$};
		\end{tikzpicture}
	\end{gathered}
	d(\text{LIPS})_2
	\begin{gathered}
		\begin{tikzpicture}[scale=.5]
			\draw[<-] (-4.8,5.17)--(-4.2,5.17);
			\draw[<-] (-1,5.15)--(-1.6,5.15);
			\draw[<-] (-1,.85)--(-1.6,.85);
			\draw[<-] (-4.8,.83)--(-4.2,.83);
			\draw[<-] (-2.85,3.4)--(-2.85,2.6);
			\path [draw, thick, blue] (-5,5)--(-3,5)--(-1,5);
			\path [draw, thick, color=green!60!black] (-5,1)--(-3,1)--(-1,1);
			\path [draw] (-3,1)--(-3,5);
			\draw[dashed] (-3,3) ellipse (1.3 and 2.3);
			\node at (-2.8,3)[right]{$q_2$};
		\end{tikzpicture}
	\end{gathered}
	d(\text{LIPS})_2
	\begin{gathered}
		\begin{tikzpicture}[scale=.5]
			\draw[<-] (-4.8,5.17)--(-4.2,5.17);
			\draw[<-] (-1,5.15)--(-1.6,5.15);
			\draw[<-] (-1,.85)--(-1.6,.85);
			\draw[<-] (-4.8,.83)--(-4.2,.83);
			\draw[<-] (-2.85,3.4)--(-2.85,2.6);
			\path [draw, thick, blue] (-5,5)--(-3,5)--(-1,5);
			\path [draw, thick, color=green!60!black] (-5,1)--(-3,1)--(-1,1);
			\path [draw] (-3,1)--(-3,5);
			\draw[dashed] (-3,3) ellipse (1.3 and 2.3);
			\node at (-1,5)[right]{$p_4$};
			\node at (-1,1)[right]{$p_3$};
			\node at (-2.8,3)[right]{$q-q_1-q_2$};
		\end{tikzpicture}
	\end{gathered}
\end{equation}
This expression is easy to calculate to any desired accuracy in $q^2$, using the reverse unitarity strategy, since clearly it corresponds to a double cut of the double-box integral $I_\mathrm{III}$ appearing in the amplitude itself \eqref{CR1}. Its leading-order contribution must cancel the super-classical terms in the real part of the partial wave, while its subleading contribution will contribute to classical order. More concretely, since all internal massive lines must be cut, one can obtain the result by looking at the $q$-expansion in Eq.~(3.22) of \cite{DiVecchia:2021bdo} and the sought-for subleading contribution is given by (3.24g) of that reference.

Let us consider, for the sake of generality, a tree-level amplitude of the form 
\begin{equation}\label{generictreec0}
	\mathcal A_0(s,-q^2) = \frac{a_0}{q^2}\,,
\end{equation}
so that
\begin{equation}\label{}
	\tilde{\mathcal{A}}_0 = 2\delta_0= \frac{1}{4Ep}\frac{a_0}{4\pi}\frac{\Gamma(-\epsilon)}{(\pi b^2)^{-\epsilon}}\,,
	\qquad
	\Theta^\text{1PM}
	= -\frac{1}{p} \frac{\partial2\delta_0}{\partial b} =
	\frac{1}{4 Ep^2} \frac{a_0}{2\pi^{1-\epsilon}} \frac{\Gamma(1-\epsilon)}{b^{1-2\epsilon}}\,.
\end{equation}
In particular, for the case of minimally interacting massive scalars,
\begin{equation}\label{}
	a_0 = 32\pi G m_1^2 m_2^2 \left(\sigma^2-\tfrac{1}{2-2\epsilon}\right).
\end{equation}
Then using reverse unitarity as explained above (see Eq.~(3.11) of \cite{Damgaard:2021ipf}), one obtains 
\begin{equation}\label{}
	(\chi_0)^3_M 
	=
	\left[
	1- \frac{\epsilon q^2}{3p^2}+\mathcal O(q^4)
	\right]
	\frac{1}{16 E^2p^2}\frac{a_0^3\Gamma(-\epsilon)^3}{(4\pi)^{2-2\epsilon}\Gamma(-3\epsilon)}\frac{\Gamma(1+2\epsilon)}{(q^2)^{1+2\epsilon}}\,.
\end{equation}
Taking into account the small-$q$ expansion of the two-loop amplitude 
\begin{equation}\label{}
	\mathcal{A}_2 = \mathcal{A}_2^{[2]} + \mathcal{A}_2^{[1]} + \mathcal{A}_2^{[0]} + \mathcal O(q^{1-4\epsilon})
\end{equation}
where $\mathcal{A}_2^{[k]} \sim q^{-k-4\epsilon}$ as in \eqref{m2GR}, 
and going to $b$-space, we find
\begin{equation}\label{}
	\begin{split}
		\operatorname{Re}\tilde{\mathcal{A}}_2(s,b) + \frac{1}{3!} \operatorname{FT}[(\chi_0)^3_M] 
		&= 
		\tilde{\mathcal{A}}_2^{[2]}(s,b)
		+
		\operatorname{Re}\tilde{\mathcal{A}}_2^{[0]}(s,b)
		\\
		&+
		\frac{(2\delta_0)^3}{3!}-\frac{(\Theta^\text{1PM})^3}{3!}\,bp
		+\mathcal O(b^{-3-3\epsilon})
		.
	\end{split}
\end{equation}
In this way, the superclassical $\mathcal O(b^{3\epsilon})$ terms cancel out between the first term of each line. 
Note that the subleading corrections in the Fourier transform \eqref{4EpFT} do not play any role, precisely thanks to this cancellation.

The remaining classical terms that are left behind determine the 3PM phase shift according to
\begin{equation}\label{Rechi2}
	\operatorname{Re}\chi_2(s,J) = \left[\operatorname{Re}\tilde{\mathcal{A}}_2^{[0]}(s,b) - \frac{bp}{6}(\Theta^\text{1PM})^3 \right]_{b\to b_J}\,.
\end{equation}
Here, in the last step, we have used the fact that the Fourier transform to $b$-space and the $J$-projection agree to leading order, and that \emph{after the subtraction} the quantity we are after does not receive any subleading contribution (neither in $1/J$ nor in $G$).

This subtraction is not identical to the analogous one dictated by the eikonal exponentiation \eqref{eikexp1}, \eqref{eikexp2}. 
Indeed, as discussed in Section~\ref{sec:radiation}, provided that there are no available inelastic 2-particle channels, the imaginary part of $2\Delta_1$ is also dictated by the tree-level amplitude \eqref{generictreec0} via
\begin{equation}\label{}
	2\operatorname{Im}\mathcal A_1(s,-q^2)= 
	\int
	\begin{gathered}
		\begin{tikzpicture}[scale=.5]
			\draw[<-] (-4.8,5.17)--(-4.2,5.17);
			\draw[<-] (-1,5.15)--(-1.6,5.15);
			\draw[<-] (-1,.85)--(-1.6,.85);
			\draw[<-] (-4.8,.83)--(-4.2,.83);
			\draw[<-] (-2.85,3.4)--(-2.85,2.6);
			\path [draw, thick, blue] (-5,5)--(-3,5)--(-1,5);
			\path [draw, thick, color=green!60!black] (-5,1)--(-3,1)--(-1,1);
			\path [draw] (-3,1)--(-3,5);
			\draw[dashed] (-3,3) ellipse (1.3 and 2.3);
			\node at (-5,5)[left]{$p_1$};
			\node at (-5,1)[left]{$p_2$};
			\node at (-2.8,3)[right]{$q_1$};
		\end{tikzpicture}
	\end{gathered}
	d(\text{LIPS})_2
	\begin{gathered}
		\begin{tikzpicture}[scale=.5]
			\draw[<-] (-4.8,5.17)--(-4.2,5.17);
			\draw[<-] (-1,5.15)--(-1.6,5.15);
			\draw[<-] (-1,.85)--(-1.6,.85);
			\draw[<-] (-4.8,.83)--(-4.2,.83);
			\draw[<-] (-2.85,3.4)--(-2.85,2.6);
			\path [draw, thick, blue] (-5,5)--(-3,5)--(-1,5);
			\path [draw, thick, color=green!60!black] (-5,1)--(-3,1)--(-1,1);
			\path [draw] (-3,1)--(-3,5);
			\draw[dashed] (-3,3) ellipse (1.3 and 2.3);
			\node at (-1,5)[right]{$p_4$};
			\node at (-1,1)[right]{$p_3$};
			\node at (-2.8,3)[right]{$q-q_1$};
		\end{tikzpicture}
	\end{gathered}
\end{equation}
i.e using again reverse unitarity (see Eq.~(A.19) of Ref.~\cite{Damgaard:2021ipf}), 
\begin{equation}\label{Re2d2}
	2\operatorname{Im}2\mathcal A_1(s,-q^2)
	=
	\left[
	1- \frac{\epsilon q^2}{4p^2}+\mathcal O(q^4)
	\right]
	\frac{1}{4 Ep}\frac{a_0^2\Gamma(-\epsilon)^2}{(4\pi)^{1-\epsilon}\Gamma(-2\epsilon)}\frac{\Gamma(1+\epsilon)}{(q^2)^{1+\epsilon}}\,.
\end{equation}
As ensured by \eqref{almostexact}, going to $b$-space the leading term simply reproduces the contribution $(2\delta_0)^2$ dictated by the eikonal exponentiation, while the subleading term determines
\begin{equation}\label{}
	2\operatorname{Im}2\Delta_1(s,b) = \frac{2\epsilon}{(4Ep^2)^2} \frac{a_0^2\Gamma(1-\epsilon)^2}{16\pi^{2-2\epsilon}(b^2)^{1-2\epsilon}}\,.
\end{equation}
Therefore, using
\begin{equation}\label{}
	\operatorname{Re}2\delta_2(s,b) = \operatorname{Re}\tilde{\mathcal{A}}_2^{[0]}(s,b) + 2\delta_0(s,b)\operatorname{Im}2\Delta_1(s,b)\,,
\end{equation}
we obtain
\begin{equation}\label{Re2d2comparison}
	\operatorname{Re}2\delta_2(s,b) = \operatorname{Re}\tilde{\mathcal{A}}_2^{[0]}(s,b) - \frac{bp}{8} (\Theta^\text{1PM})^3\,.
\end{equation}
We are now in a position to make a precise comparison between the 3PM eikonal phase and phase shift. Subtracting \eqref{Rechi2} and \eqref{Re2d2comparison},
\begin{equation}\label{difference2d2chi2}
	\operatorname{Re}
	2\delta_2(s,b)\big|_{b\to J/p}-\operatorname{Re}\chi_2(s,J) = \frac{J}{24} (\Theta^\text{1PM})^3\,.
\end{equation}
	We remark that $\operatorname{Re}2\delta_2(s,b)$, which appears in \eqref{difference2d2chi2}, is the $\mathcal O(G^3)$ of $\operatorname{Re}2\delta_2(s,b)$ \emph{regarded as a function of} $b$. This should not be confused with the $\mathcal O(G^3)$ of $\operatorname{Re}2\delta_2(s,b_J/\cos\tfrac{\Theta}{2})$, regarded as a function of $b_J$. As discussed for instance above \eqref{th123J}, the substitution $b=b_J/\cos\tfrac{\Theta}{2}$ \eqref{bJbThetadelta} would induce additional $\mathcal O(G^3)$ terms, hence the need to pay attention to this distinction. These terms, however, do not cancel out against the right-side of \eqref{difference2d2chi2} so that the phase shift is not simply obtained from the eikonal phase by taking  $b=b_J/\cos\tfrac{\Theta}{2}$ into account,
	\begin{equation}\label{nottrue}
		\operatorname{Re}2\delta\left(s,b_J/\cos\tfrac{\Theta}{2}\right) \neq \operatorname{Re}\chi(s,J)\,.
	\end{equation}

The difference \eqref{difference2d2chi2} is precisely the one needed to reproduce the last term in the first line of \eqref{phaseshift5}, which we had derived formally from the saddle-point conditions. In turn, this is crucial to ensure that, while the eikonal phase is linked to the angle by \eqref{bJbThetadelta}, the phase shift obeys \eqref{Thetachi}. Indeed, in the notations of \eqref{eq:dexpcl}, \eqref{phaseshift5}, \eqref{dafpg5}, 
\begin{equation}\label{}
	\Theta^\text{1PM} = \frac{d_0}{J}\,p\,,
	\qquad
	\operatorname{Re}2\delta_2 = \frac{d_2\,p^2}{2 J^2}\,,
	\qquad
	\operatorname{Re}\chi_2 = \frac{d_2\,p^2-\frac{d_0^3}{12}}{2 J^2}\,.
\end{equation}
Hence the difference between eikonal phase and phase shift exactly matches the one derived in \eqref{difference2d2chi2}.

To complete the spectrum of possible definitions for the subtractions, let us consider employing the Fourier transform with the complete mass-shell delta functions. We begin by recalling that the standard form of the exponentiation is dictated by \eqref{fullampli}, were the $\tilde{\mathcal{A}}(s,b)$ is the Fourier transform with \emph{linearized} delta functions as in \eqref{eq:SFT1} and \eqref{invFTLIN}. In contrast, suppose we were to define
\begin{equation}\label{}
	1+i\operatorname{FT}[\mathcal A](s,b_J) = [1+2i\check\Delta(s,b_J)] e^{2i\check\delta(s,b_J)}\,,
\end{equation}
with $\operatorname{FT}[\mathcal A]$ as in \eqref{FTexact} and \eqref{invFT}.
Then, while of course
\begin{equation}\label{}
	2\check\delta_0(s,b_J) = 2\delta_0(s,b)\big|_{b\to b_J}\,,\qquad
	2\check\delta_1(s,b_J) = 2\delta_2(s,b)\big|_{b\to b_J}\,,
\end{equation}
one finds a nontrivial difference for $2\check\Delta_1$ and $2\check\delta_2$ to the effect that
\begin{equation}\label{Re2d2comparisontilde}
	\operatorname{Re}2\check\delta_2(s,b_J) =
	\left[
	\operatorname{Re}2\delta_2(s,b) 
	+  \frac{bp}{8} (\Theta^\text{1PM})^3
	\right]_{b\to b_J}
	=
	\left[
	\operatorname{Re}\tilde{\mathcal{A}}_2^{[0]}(s,b)
	\right]_{b\to b_J}
\end{equation}
where we used \eqref{Re2d2comparison} in the last step, and therefore
	\begin{equation}\label{true}
	\operatorname{Re}2\check\delta (s,b_J ) =
	\operatorname{Re}2\check \delta (s,b \cos\tfrac\Theta2 ) = \operatorname{Re}2\delta(s,b)
\end{equation}
\emph{up to} 3PM order.

Let us conclude this section by presenting the radial action up to 3PM in GR for the scattering of minimally coupled massive scalars in terms of the coefficients defined in \eqref{eq:dexpcl},
	\begin{equation}
		I(s,J) = -\pi J -d_0 \log J + \frac{d_1p}{J} + \frac{ d_2 p^2 - \frac{d_0^3}{12}}{2J^2}\,,
		\label{IR27}
	\end{equation}
	where
	\begin{equation}
		d_0 = \frac{2 G m_1 m_2 (2\sigma^2-1)}{\sqrt{\sigma^2-1}}
		\,,
		\quad
		d_1 = \frac{3\pi G^2 m_1^2 m_2^2 (m_1+m_2)(5\sigma^2-1)}{4Ep}\,,
		\label{IR28}
	\end{equation}
	and 
	\begin{equation}
	\begin{aligned}
		d_2 &= 8 G^{3} m_{1}^{2} m_{2}^{2}\left\{
		{\color{blue}
			\frac{\left(2 \sigma^{2}-1\right)^{2}\left(8-5 \sigma^{2}\right)}{6\left(\sigma^{2}-1\right)^{2}}
		}
		-\frac{\sigma\left(14 \sigma^{2}+25\right)}{3 \sqrt{\sigma^{2}-1}}\right.\\
		&\left.
		{\color{green!60!black}	
			+\frac{2s  (12 \sigma^4 - 10 \sigma^2 +1) }{ m_{1} m_{2}\left(\sigma^{2}-1\right)^{\frac{3}{2}}}
		}	
		+\operatorname{arccosh}\sigma \left[
		{	\color{blue}
			\frac{\sigma\left(2 \sigma^{2}-1\right)^{2}\left(2 \sigma^{2}-3\right)}{2\left(\sigma^{2}-1\right)^{\frac{5}{2}}}
		}
		+\frac{-4 \sigma^{4}+12 \sigma^{2}+3}{\sigma^{2}-1}\right]\right\}.
	\end{aligned}
\end{equation}

\subsubsection{Radial action, effective potential, PN limit and bound orbits}
\label{alternative}
In this subsection, we determine the deflection angle and the radial action in terms of a potential present in the Hamiltonian describing the relative motion  of two black holes. This serves as a tool to re-sum the PM contributions in a controlled way and access the PN limit that also applies to the case of bound orbits. We conclude by presenting the 2PN-accurate expressions for the periastron advance, revolution period for generic bound orbits and for the binding energy of  circular orbits. The idea of matching to an effective potential has its roots in the EOB approach, and here we will follow the particularly simple incarnation given in \cite{Damour:2022ybd}.

We start from the Hamiltonian describing the relative motion in a plane of two black holes:
\begin{equation}
	H  = p_r^2 + \frac{J^2}{r^2} + V(r)\,,\qquad V(r) =- \sum_{n=1}^\infty \frac{G^n}{r^n} f_n  \, ,
	\label{HH1}
\end{equation}
where  $V(r)$ is the potential given as an expansion in the Newton constant.
From the three Hamilton equations
\begin{equation}
	\frac{\partial H}{\partial \theta} =- \dot{p}_\theta=0\,,
	\qquad 
	{\dot{r}}= \frac{\partial H}{\partial p_r}= 2p_r\,,
	\qquad
	{\dot{\theta}}= \frac{\partial H}{\partial p_\theta} = \frac{2p_\theta}{r^2} 
	\label{HH2}
\end{equation}
we get  that the angular momentum $p_\theta=J$ is a constant of motion and the  relation
\begin{equation}
	\frac{\partial \theta}{\partial r} = \frac{J}{r^2 p_r}\;, \quad \mbox{with} \quad  p_r=\sqrt{  p^2- \frac{J^2}{r^2}- V(r)}\,.
	\label{HH3}
\end{equation}
Thus the deflection angle is 
\begin{equation}
	\Theta =-\pi  +2 J \int_{r_*}^\infty \frac{dr}{r^2 \sqrt{p^2-\frac{J^2}{r^2} - V(r)}}= -\pi - 2 \int_{r_*}^\infty dr \frac{\partial}{\partial J} \sqrt{  p^2- \frac{J^2}{r^2}- V(r)} \, ,
	\label{HH4}
\end{equation}
where $p^2$ is the constant value of the Hamiltonian and, in our case, $p$ is equal to the momentum in the center of mass frame.  The factor $-\pi$ is there to ensure that $\Theta=0$ if the potential vanishes and the factor $2$ takes care of the motion from infinity to the point of minimal distance $r_*$ and from $r_*$ back to infinity where $r_*$ is the largest positive root of the condition of turning point, $p_r(r_*)=0$. As shown in~\cite{Bjerrum-Bohr:2019kec} one can avoid to determine  it by computing the deflection angle by using instead the following equivalent expression (in \ref{app:geoschw} we review a similar approach  to calculate the deflection angle in the probe limit)
\begin{equation}
	\Theta = \sum_{k=1}^\infty \Theta_k (b_J) ~~;~~\Theta_k (b_J) = \frac{2b}{k!} \int_0^\infty du \left( \frac{\partial}{\partial b_J^2}\right)^k \frac{\left( V( \sqrt{u^2+b_J^2})\right)^k (u^2+b_J^2)^{k-1}}{p^{2k}}\, ,
	\label{HH5}
\end{equation}
where as usual $J=pb_J$. From the deflection angle we can reconstruct the radial action $I$ by using~\eqref{ThetaI}
	\begin{equation}
		I = -\int (\Theta+\pi) \, dJ
		\label{HH6}
	\end{equation}
	and ``the $+\pi$'' is  so that $I \to -\pi J$ when $\Theta\to0$; moreover the $J$-independent part of $I$ is fixed so as to agree with the 1PM expression for $2\delta_0$ via \eqref{radialaction} and \eqref{chi0}. Then, the relation between the radial action and the potential is given by
\begin{equation}
	I = -\pi J + 2 \int_{r_*}^{\infty} \sqrt{p^2 - \frac{J^2}{r^2} - V(r)}\,.
	\label{HH7}
\end{equation}
It can also be written in a more convenient way as
\begin{equation}
	I = -\pi J -\sum_{k=1}^\infty \frac{1}{k!} \int_0^\infty du \left( \frac{\partial}{\partial b_J^2}\right)^{k-1} \frac{\left( V( \sqrt{u^2+b_J^2})\right)^k (u^2+b_J^2)^{k-1}}{p^{2k-1}}
	\label{HH8}
\end{equation}
The integral over $u$ can be easily computed by changing variable to $u=b_J \sinh w$ and using 
\begin{equation}
	\int_0^\infty \frac{dw}{\cosh^{d-2} w}= \frac{\sqrt{\pi} \Gamma (\frac{d-2}{2})}{2 \Gamma (\frac{d-1}{2})}\, .
	\label{HH9}
\end{equation}
In this way we get
\begin{equation}
	I  (J, \sigma)= -\pi J + \frac{f_1 G_D b_J^{2\epsilon} \Gamma (-\epsilon)}{2p} + \frac{f_2 G^2 \pi}{2J} +\frac{G^3}{J^2} \left( p f_3  + \frac{f_1 f_2}{2 p} - \frac{f_1^3}{24 p^3} \right)  + {\cal{O}} (G^4)\;,
	\label{HH10}
\end{equation}
where for the $k=1$ contribution we reinstated the dimensional regularization parameter by using $- \frac{G_D f_1}{r^{D-3}}$ in place of $- \frac{G f_1}{r}$ in the potential. Then, by using~\eqref{ThetaI}, we can derive the deflection angle (in $D=4$)
\begin{equation}
	\Theta =\frac{f_1 G}{pJ} + \frac{f_2 G^2 \pi}{2J^2}+ \frac{2G^3}{J^3} \left( p f_3  + \frac{f_1 f_2}{2 p} - \frac{f_1^3}{24 p^3} \right) + {\cal{O}} (G^4)\,.
	\label{HH11}
\end{equation}
From the knowledge of the PM expanded deflection angle one can of course determine the coefficients $f_n$. Comparing the equation above with~\eqref{th123J}, we obtain up to 3PM
\begin{equation}
	f_1 = \frac{2m_1^2 m_2^2 (2\sigma^2-1)}{E}\;,\quad
	f_2=  \frac{ 3m_1^2 m_2^2 (m_1+m_2)(5\sigma^2-1)}{2E}
	\label{HH15}
\end{equation}
and
\begin{align}
	f_3 & = m_1^2 m_2^2  \Bigg\{\frac{2E (12\sigma^4 -10 \sigma^2 +1)}{ (\sigma^2-1)}- \frac{3(2\sigma^2-1) (5\sigma^2-1)(m_1+m_2)}{2(\sigma^2-1)} \nonumber \\
	& +  \frac{2m_1 m_2}{E} \Bigg[- \frac{2\sigma(14 \sigma^2 +25)}{3} - 2 \frac{4\sigma^4 -12 \sigma^2 -3}{\sqrt{\sigma^2-1}}
	\cosh^{-1} (\sigma) \nonumber \\
	& + \frac{(2\sigma^2-1)^2}{\sqrt{\sigma^2-1}} \left(\frac{8-5\sigma^2 }{3(\sigma^2-1)} 
	+ \frac{ (2\sigma^2-3)\sigma}{(\sigma^2-1)^{\frac{3}{2}} } \cosh^{-1}(\sigma) \right) \Bigg]\Bigg\} \, .
	\label{HH17}
\end{align}
It is possible also to derive the parameters $f_n$ directly by matching the gravity scattering amplitudes against those of an effective theory encoding the potential~\eqref{HH1} \cite{Cheung:2018wkq}. In this way one can derive at each PM order an object, $ A_{nPM} (\sigma, q)$, that satisfies the following impetus formula~\cite{Kalin:2019rwq,Bjerrum-Bohr:2019kec,Kalin:2019inp,Damour:2019lcq}:
\begin{equation}
	\frac{G^n}{r^n}f_n = \frac{1}{2E} \int \frac{d^3 q}{(2\pi)^3} e^{iqr} A_{nPM} (\sigma, q)\, .
	\label{HH12}
\end{equation}
At the tree level $A_{1PM}$ is the one in \eqref{eq:Aphi12tlp}.
	At one loop $A_{2PM}$ is given  in  \eqref{a11GR} for $\epsilon=0$.
Finally at two loops,
	\begin{align}
		& A_{3PM} (\sigma, q^2)= 4 G^3 m_1^2 m_2^2  \pi \log q^2 E\nonumber \\
		& \times  \Bigg[
		\frac{3(2\sigma^2-1)(5\sigma^2-1) (m_1+m_2)}{2(\sigma^2-1)}  - \frac{2E(12\sigma^4 -10\sigma^2 +1)}{ (\sigma^2-1)} \nonumber \\
		& +  \frac{2m_1 m_2}{ E} \Bigg( \frac{2(-3-12 \sigma^2+4\sigma^4) \cosh^{-1} (\sigma)}{\sqrt{\sigma^2-1}} + \frac{2\sigma (14\sigma^2 +25)}{3} \nonumber \\
		& - \frac{(2\sigma^2-1)^2}{\sqrt{\sigma^2-1}} \left(\frac{8-5\sigma^2 }{3(\sigma^2-1)} 
		+ \frac{ (2\sigma^2-3)\sigma}{(\sigma^2-1)^{\frac{3}{2}} } \cosh^{-1}(\sigma) \right)\Bigg) \Bigg]
		\label{HH16}
	\end{align}
	which  coincides with the part proportional to $\log q^2$ of Eq. (8) of~\cite{Bern:2019nnu}, with the addition of the two terms coming from radiation reaction. Once the parameters $f_n$, and so the potential $V(r)$~\eqref{HH1}, are determined, one can use this information in~\eqref{HH4} to find an ``improved'' deflection angle that resums a class of higher order corrections. A detailed analysis discussing also this approach is found in~\cite{Gold:2012tk,Damour:2014afa,Hopper:2022rwo,Khalil:2022ylj,Damour:2022ybd} showing excellent agreement with data from numerical relativity. The idea of resumming is at the basis of the EOB approach~\cite{Buonanno:1998gg,Buonanno:2000ef} that can be used to find accurate waveforms for bound systems by using an analytic approach.

A particularly simple case is the one in which the potential contains only  $f_1$ and $f_2$, which effectively reduces the calculations to the textbook Coulomb/Newton case. In this case the deflection angle can be computed exactly as a function of $G, J$ and $\sigma$. To show this it is convenient to introduce the two dimensionless quantities:
\begin{equation}
	{\hat{f}}_1 =  \frac{f_1}{p^2 M}~~;~~ {\hat{f}}_2 = \frac{f_2}{p^2 M^2}~~;~~M=m_1+m_2 \, . 
	\label{HH18}
\end{equation}
Then Eq. \eqref{HH4} becomes
\begin{eqnarray}
	\Theta (J,E) +\pi = \frac{2J}{p} \int_{r_\ast}^{\infty} \frac{dr}{r^2 \sqrt{ 1 + \frac{{\hat{f}}_1 GM}{r}  - \frac{{\hat{J}}^2}{p^2r^2}}}
	\label{DF3}
\end{eqnarray}
where
\begin{eqnarray}
	{\hat{J}}^2 = J^2 - {\hat{f}}_2 (GMp)^2
	\label{DF4}
\end{eqnarray}
and $r_\ast$ is the positive  root of the equation:
\begin{eqnarray}
	1+ \frac{{\hat{f}}_1 GM}{r}  - \frac{ {\hat{J}}^2}{p^2r^2} =0 
	\label{DF5}
\end{eqnarray}
that has the two roots:
\begin{equation}
	r_{\pm} = \frac{1}{2} \left( \pm \sqrt{ (f_1GM)^2 + \frac{4 {\hat{J}}^2}{p^2}} - f_1 GM \right)
	\label{DF6}
\end{equation}
and $r_\ast=r_+$. Rewriting the quantity in the square root in \eqref{DF3} in terms of the two roots and changing  the integration variable  as 
\begin{equation}
	\frac{r_+}{r}= 1-a z^2~~;~~ a = \frac{ r_{-}-r_{+}}{r_-} 
	\label{DF7}
\end{equation}
we get
\begin{equation}
	\frac{\Theta (J,E) +\pi}{2} = \frac{2J}{p \sqrt{r_+ (-r_-)} }\int_0^{\frac{1}{\sqrt{a}}} \frac{dz}{\sqrt{1-z^2}} =  \frac{2J}{p \sqrt{r_+ (-r_-)} } \arcsin \left( \frac{1}{\sqrt{a}}\right)\, .
	\label{DF8}
\end{equation}
By using~\eqref{DF6} we get
\begin{equation}
	2\sin^2 \Bigg[\frac{\Theta (J,E) +\pi}{4} \sqrt{ 1 - \frac{ {\hat{f}}_2 (GMp)^2}{J^2} } \Bigg]=   1 + 
	\frac{1}{\sqrt{1 + \frac{4{\hat{J}}^2}{(p {\hat{f}}_1 GM)^2} }}   
	\label{intf1f2}
\end{equation}
which, by using trigonometric identities and recalling that $0\leq \Theta\leq {\pi}$, can be shown to agree with Eq.~(7.6) of~\cite{Kalin:2019rwq} written below
\begin{equation}
	\frac{\Theta (J,E) +\pi}{2} =  \frac{1}{\sqrt{1- \frac{{\hat{f}}_2 y^2}{{\hat{f}}_1^2}}} \left( \frac{\pi}{2}+ \tan^{-1} 
	\frac{ \frac{y}{2}}{ \sqrt{ 1 - \frac{{\hat{f}}_2 y^2}{{\hat{f}}_1^2}} } \right)\;,
	\label{DF13}
\end{equation}
where we introduced $y=\frac{GM {\hat{f}}_1p}{J}$.

An interesting feature of~\eqref{DF13} is that its PN expansion captures the 1PN result to all order in the Newton constant $G$. We can see this by first using the results~\eqref{HH15} for GR to rewrite~\eqref{DF13} in terms of the parameters $j_{\rm PN}$ and $v_\infty$ introduced in~\eqref{eq:PNdef} 
and then by taking the limit $\frac{1}{j_\text{PN}}\sim v_\infty \ll 1$ up to the first subleading order with $\alpha^{-1} = j_{\rm PN} v_\infty$ fixed. Using, in this limit, the approximation   $\frac{y}{2} \simeq  \alpha ( 1+ \frac{2}{\alpha^2 j_\text{PN}^2})$ and  $\frac{{\hat{f}}_2 y^2}{{\hat{f}}_1^2} \simeq \frac{6}{j_\text{PN}^{2}}$ we obtain
\begin{equation}
	\label{eq:tanTheta1}
	\frac{\Theta}{2} = \arctan \alpha +       \frac{1}{j_{\rm PN}^2} \left(
	3 \left(\arctan\alpha +\frac{\pi}{2} \right)
	+
	\frac{3 \alpha ^2+2}{\alpha(\alpha^2+1) }\right) + {\cal O}\left(j_{\rm PN}^{-4}\right)\;.
\end{equation}
where we have used $ \tan^{-1} \alpha (1+x)= \tan^{-1} \alpha + \frac{\alpha x}{1+\alpha^2}+ \cdots$ with $x= \frac{3\alpha^2 +2}{\alpha^2 j_{PN}^2}$.
The first term above is the 0PN scattering angle obtained in~\eqref{eq:tanTheta0}, while the second term is the 1PN correction at all orders in the PM expansion, see for instance~\cite{Bini:2017wfr}. One can also obtain the result \eqref{eq:tanTheta1} from the probe limit calculation, as reviewed in \ref{app:geoschw}.
Integrating \eqref{eq:tanTheta1} as in \eqref{HH6}, we have
\begin{equation}\label{radialaction1PNugly}
	\begin{aligned}
		I
&
\simeq \frac{G m_1 m_2}{\alpha\,  v_{\infty }} 
\left[ 2 \alpha -2 \arctan\alpha -\alpha  \left(\log \left(\alpha ^2+1\right)+2 \log \left(\frac{G m_1 m_2}{\alpha }\right)
	-4 \log  v_{\infty }\right) \right.\\
	&
	\left.+\alpha\,  v_{\infty }^2 \left(-2 \log \left(\alpha ^2+1\right)+3 \pi  \alpha +6 \alpha  \arctan\alpha -4 \log \left(\frac{G m_1 m_2}{\alpha }\right)+8 \log v_{\infty }\right)-\pi \right].
	\end{aligned}
\end{equation}

As discussed in~\cite{Kalin:2019rwq,Kalin:2019inp,Cho:2021arx}, it is possible to extract information about bound binary systems starting from the resummed PN results for the scattering process. The basic idea is to analytically continue the unbound observables to the region where the angular momentum is kept fixed but the center-of-mass energy is below $m_1+m_2$, due to the presence of a nontrivial binding energy. In practice, we need to extend the Lorentz factor from the region $\sigma>1$ to $\tilde\sigma<1$, which can be done in two ways because of the branch cut $p \sim \sqrt{\sigma^2-1}$. The recipe for defining the radial action $\tilde{I}$ for the bound case is to take the linear combination of these two possibilities
\begin{equation}
	\label{eq:Irbound}
	\tilde{I} = I(\sqrt{\sigma^2-1}\to i \sqrt{1-\tilde{\sigma}^2},J) + I(\sqrt{\sigma^2-1}\to -i \sqrt{1-\tilde{\sigma}^2},J)\;.
\end{equation}
In the PN limit this analytic continuation is equivalent to sending $v_\infty \to \pm i v_\infty$. 
Applying this recipe to \eqref{radialaction1PNugly},  at 1PN order we find
\begin{equation}
	\label{eq:Irbound1PNa}
	\tilde{I} = -2 \pi J + 2\pi G m_1 m_2 \frac{2\tilde{\sigma}^2 -1}{\sqrt{1-\tilde{\sigma}^2}} + 6\pi \frac{G^2 m_1^2 m_2^2}{J}+ {\cal O}\big({\rm 2PN}\big) \;.
\end{equation}
This can be directly obtained from \eqref{eq:tanTheta1} by noting that all its terms are odd under $v_\infty \leftrightarrow -v_\infty$ (or equivalently $\alpha\to -\alpha$) except for $3\pi/(2j_\text{PN}^2)$ and therefore cancel out in \eqref{eq:Irbound}.
The $J$-independent term on the right hand side of \eqref{eq:Irbound1PNa}, which we can take from the PM expansion \eqref{HH10}, originates from the analytic continuation of $b_J^{2\epsilon}\Gamma(-\epsilon)$ in the second term of \eqref{HH10}: since $J$ is kept fixed in~\eqref{eq:Irbound}, we have to send $b_J \to \mp i b_J$ as $\sqrt{\sigma^2-1}\to \pm i \sqrt{1-\tilde{\sigma}^2}$.
The term ${\cal O}(G^2)$ in \eqref{eq:Irbound1PNa} follows from the only contribution in~\eqref{eq:tanTheta1} that is even under $v_\infty \leftrightarrow -v_\infty$. We can rewrite~\eqref{eq:Irbound1PNa} in the variables used by~\cite{Blanchet:2013haa}, introducing also the symmetric mass-ratio $\nu = m_1 m_2/(m_1+m_2)^2$,
\begin{equation}
	\label{eq:Blvar}
	\tilde{\sigma}= 1 - \varepsilon_B \left(\frac{1}{2}-\frac{\nu}{8} \varepsilon_B\right), \qquad J= \frac{G m_1 m_2 }{\sqrt{\varepsilon_B}} \sqrt{j_B},
\end{equation}
thus trading $J$, $\tilde\sigma$ for $j_B$, $\varepsilon_B$,
and obtain, in the small $\varepsilon_B$ limit (at fixed $j_B$),
\begin{equation}
	\label{eq:Irbound1PNb}
	\tilde{I} = -\pi G m_1 m_2 \left[2 \frac{\sqrt{j_B}-1}{\sqrt{\varepsilon_B}} - \sqrt{\varepsilon_B} \left(6 \frac{1}{\sqrt{j_B}}+\frac{\nu-15}{4} \right) \right] + {\cal O}\big({\rm 2PN}\big)\;.
\end{equation}
It was pointed out in~\cite{Kalin:2019rwq} that the radial action for the bound case vanishes in the case of circular trajectories which is the most relevant one for describing the typical black hole binaries~\cite{Blanchet:2013haa}. By using~\eqref{eq:Irbound1PNb} we obtain the following relation 
\begin{equation}
	\label{eq:jcric}
	\tilde{I} = 0 \;,\quad \Rightarrow\quad j_{ B,\mathrm{circ}}= 1+ \varepsilon_B \frac{\nu +9}{4} + {\cal O}\big({\rm 2PN}\big)\;,
\end{equation}
in agreement with~\cite{Blanchet:2013haa} and with the closely related derivation given in \cite{Adamo:2022ooq}.

Of course, it is also possible to introduce the bound radial action in the PM expansion. Substituting Eq.~\eqref{HH10} into Eq.~\eqref{eq:Irbound} we obtain at the 2PM approximation 
\begin{equation}
	\label{eq:Irbound2PM}
	\tilde{I} = - 2 \pi J + 2\pi G m_1 m_2 \frac{2\tilde{\sigma}^2 -1}{\sqrt{1-\tilde{\sigma}^2}} + \pi \frac{G^2 m^2_1 m^2_2}{J} (m_1+m_2) \frac{3 (5\tilde{\sigma}^2-1)}{2E}
	+ {\cal O}\big({\rm 3PM}\big) \;.
\end{equation}
By using the result above in the boundary-to-bound dictionary~\cite{Kalin:2019rwq,Kalin:2019inp}, one can derive the periastron advance $K=\frac{\Delta\Theta}{2\pi}$ and the period $T_b$ of the bound motion
\begin{equation}
	\label{eq:b2bdict}
	\Delta\Theta = -\partial_J \tilde{I} \,, \qquad T_b = \partial_E \tilde{I} = \frac{\partial_{\tilde\sigma}\tilde{I}}{\partial_{\tilde\sigma}\sqrt{m_1^2+2 m_1 m_2 \tilde\sigma+m_2^2}}\,.
\end{equation}
It is straightforward to check that, using~\eqref{eq:Irbound2PM} in~\eqref{eq:b2bdict} and rewriting the result in terms of the variables \eqref{eq:Blvar}, one obtains Eq.~(347b) of~\cite{Blanchet:2013haa} (at order $1/j_B$, but at all orders in $\varepsilon_B$) for the periastron advance $K$ and Eq.~(347a) of the same reference (at order $1/\sqrt{j_B}$, but at all orders in $\varepsilon_B$) for the angular frequency $n=2\pi/T_b$:\footnote{Let us recall that $\varepsilon_B \sim \frac{1}{c^2}$ and $j_B \sim c^0$ as in (344) of \cite{Blanchet:2013haa}.}
\begin{equation}
	\label{eq:347}
	\begin{aligned}
		n= & ~\frac{\varepsilon_B^{3 / 2} c^3}{G(m_1+m_2)}\left\{1+\frac{\varepsilon_B}{8}(-15+\nu)+\frac{\varepsilon_B^2}{128}\left[555+30 \nu+11 \nu^2+\frac{192}{j_B^{1 / 2}}(-5+2 \nu)\right]\right. \\
		& +\frac{\varepsilon_B^3}{3072}  \left[-29385-4995 \nu-315 \nu^2+135 \nu^3+\frac{5760}{j_B^{1 / 2}}\left(17-9 \nu+2 \nu^2\right)\right.  \\
		& \left.\left.+\frac{16}{j_B^{3 / 2}}\left(-10080+\left(13952-123 \pi^2\right) \nu-1440 \nu^2\right)\right]+\mathcal{O}\left(\frac{1}{c^8}\right)\right\},\\
		K = &~1+\frac{3 \varepsilon_B}{j_B}+\frac{\varepsilon_B^2}{4}\left[\frac{3}{j_B}(-5+2 \nu)+\frac{15}{j_B^2}(7-2 \nu)\right] \\
		+ & \frac{\varepsilon_B^3}{128} {\left[\frac{24}{j_B}\left(5-5 \nu+4 \nu^2\right)+\frac{1}{j_B^2}\left(-10080+\left(13952-123 \pi^2\right) \nu-1440 \nu^2\right)\right.} \\
		+ &\left.\frac{5}{j_B^3}\left(7392+\left(-8000+123 \pi^2\right) \nu+336 \nu^2\right)\right]+\mathcal{O}\left(\frac{1}{c^8}\right) .
	\end{aligned}
\end{equation}

\section{The eikonal operator in the soft limit }
\label{sec:eikopsoft}

The 3PM eikonal for the elastic $2\to2$ scattering presents a few interrelated puzzling features. The most evident one is perhaps the appearance of an infrared-divergent (positive) imaginary part $e^{2i\delta}=e^{-\operatorname{Im}2\delta}e^{i\operatorname{Re}2\delta}$ indicating that the probability for this process to take place is in fact infinitely suppressed in the limit $D\to4$. The appearance of an imaginary part also makes the eikonal manifestly non-unitary and this suggests a way out. It indicates that we neglected channels that are actually important also in the classical limit. Indeed, by considering a strictly $2\to2$ process, we neglected the fact that the two-body system emits radiation. The aim of this section is to take the first step in order to ameliorate this treatment by including the presence of soft radiation in the final state.
To this end, we will promote the eikonal phase to a Hermitian operator \cite{DiVecchia:2022nna} depending on the graviton creation/annihilation operators and on the Weinberg soft factor \cite{Weinberg:1964ew,Weinberg:1965nx}, taking inspiration from earlier approaches based on the Block-Nordsieck method  and more recently on dressed states \cite{Addazi:2019mjh,Mirbabayi:2016axw,Choi:2017ylo}.

As we will see, this will highlight how soft gravitons are responsible for the infrared divergent imaginary part of $2\delta$.
Moreover, another shortcoming of the elastic eikonal framework is that, by its very nature, it does not provide formulas to calculate observables associated to the gravitational field. Introducing soft radiation as well, via the soft eikonal operator, we will also gain access to the properties of gravitational waves at low frequencies: the Zero-Frequency Limit (ZFL) of the energy emission spectrum and the memory effect. In fact, thanks to the exact, nonperturbative nature of the soft theorem, the resulting formulas will even allow us to take a peek \emph{beyond} the conventional PM approximation considered so far. As we shall discuss, this will be crucial in resolving yet another apparent puzzle concerning the high-energy limit. In the na\"ive large-$\sigma$ limit the 3PM expression for the ZFL of the spectrum seems ill-behaved. However, this is not the signal of an actual pathological behavior, bur rather of the breakdown of the PM approximation. The exact expression derived thanks to the soft theorem instead provides the correct answer, even in the true ultrarelativistic limit, allowing us to make contact with the massless setup.

A subtle point is related to the effects of static gravitational fields, whose Fourier transform is localized exactly at zero frequency. 
Clearly these fields do not carry energy-momentum, so their inclusion is not relevant for evaluating the corresponding spectra. Instead, they can in general carry angular momentum and are therefore important for the angular momentum balance of the particles+field system. As shown in \cite{Damour:2020tta}, and as we will obtain below, their contribution to the angular momentum of the full gravitational field is in fact the leading one in the PM expansion and starts at $\mathcal O(G^2)$.
In order to accommodate for such effects, we will consider a different dressing of the elastic process, which essentially encodes static gravitational modes via the $-i0$ prescription.

Being based on the leading graviton theorem, the formulas presented in this section hold for generic deflections and are independent of the PM expansion.
Moreover they apply not only to the two-body scattering of spinless, point-like objects, but also to collisions involving spinning or tidally-deformable bodies as well as to multi-body scatterings. This is due to the universality of the soft factor, whose expression only depends on the momenta of the hard particles. Of course, spin and tidal effects for instance do enter the explicit expression of the results, but only insofar as they influence the relation between final and initial momenta, i.e.~the impulses.

In the present section, since we are restricting to leading order in the soft approximation, it is sufficient to work by using the momenta of energetic external particles in the elastic process as given, as in~\cite{Weinberg:1964ew,Weinberg:1965nx} and more recently in~\cite{Laddha:2018vbn,Sahoo:2018lxl,Saha:2019tub,Addazi:2019mjh,Sahoo:2021ctw}. Taking into account radiation back-reaction on the trajectories of the massive objects will be instead crucial when considering the full gravitational-wave spectrum in the ensuing section.

\subsection{Soft eikonal operator without static modes}
\label{sec:eikopwithout}

We can include soft radiation by following the method of Bloch-Nordsieck~\cite{Bloch:1937pw,Thirring:1951cz} and Weinberg~\cite{Weinberg:1964ew,Weinberg:1965nx} (see also \cite{Mirbabayi:2016axw,Choi:2017ylo,Arkani-Hamed:2020gyp}). Let us consider the $S$-matrix element $S^{(M)}_{\text{\textit{s.r.}},N}$ for the emission of $N$ soft gravitons on top of a background hard process involving particles with momenta $p_n$, where $n$ runs over the hard states.  
The subscript ``\textit{s.r.}'' emphasizes that we are restricting our attention to  soft radiation. The total $S$-matrix element
factorizes as the matrix element $S^{(M)}$ (the superscript $(M)$ stands for ``momentum space'') for the hard process times $N$ universal factors $f_j(k)$ expressed in terms of the polarization $j$ of the graviton and of  its momentum $k$, \cite{Bloch:1937pw,Thirring:1951cz,Weinberg:1964ew,Weinberg:1965nx} 
\begin{equation}\label{eq:SMNSM}
	S^{(M)}_{\text{\textit{s.r.}},N}
	=
	\prod_{r=1}^N w_{j_r}(k_r)\,  S^{(M)}
	\,,\quad
	w_j(k) = \varepsilon^{\ast\mu\nu}_j(k)w_{\mu\nu}(k)\,,\quad
	w^{\mu\nu}(k) = \sum_{n}\frac{\kappa\, p_n^\mu p_n^\nu}{p_n\cdot k}\,,
\end{equation}
where $\kappa= \sqrt{8\pi G}$. 
Of course an analogous formula holds for soft absorption processes, with $w_j(k)$ replaced by $-w_j^\ast(k)$. We keep graviton momenta always future-directed. For simplicity, we omit the $p_n$ from the arguments of $w_j(k)$.
As emphasized by Weinberg (see e.g. \cite{Weinberg:1995mt}), the formula \eqref{eq:SMNSM} applies to the case in which the ``bare" amplitude $S^{(M)}$ is connected, hence in our case to the $i T$ part of $S = 1 + i T$. Extending \eqref{eq:SMNSM} to the disconnected part of the $S$-matrix will be crucial for the inclusion of static effects and for the calculation of the angular momentum loss given in the next section. 

We introduce creation/annihilation operators for the gravitons, obeying canonical commutation relations 
\begin{equation}\label{key1}
	2\pi\theta(k^0)\delta(k^2)
	[ a^{\phantom{\dagger}}_i (k) , a_j^\dagger (k')] = (2\pi)^D\delta^{(D)}(k-k') \,\delta_{ij}\,,
\end{equation}
with $i,j$ labeling physical polarizations,
and we define
\begin{equation}
	\label{eq:aadcom}
	\int_k^{\ast}
	=
	\int \frac{d^Dk}{(2\pi)^D}\,2\pi\theta(k^0) \delta(k^2) \theta(\omega^\ast-k^0)\,.
\end{equation}
Following Weinberg \cite{Weinberg:1965nx}, we have introduced a frequency scale $\omega^\ast$ below which the approximation \eqref{eq:SMNSM} is valid, and in practice this can be taken $\omega^\ast \sim v/b$ for eikonal scattering. On the other hand we do not need an infrared frequency cutoff thanks to dimensional regularization.
We can then capture the factorization \eqref{eq:SMNSM} for soft emissions by defining an exponential operator depending on the oscillators, (the sum over repeated polarization indices is left implicit)
\begin{equation}\label{}
	e^{2i\hat\delta_{\text{\textit{s.r.}}}}
	=
	e^{
		\int_k^{\ast}  \left[ 
		w^{\phantom{\dagger}}_j(k)\, a^\dagger_j(k)
		-w^\ast_j(k)\, a^{\phantom{\ast}}_j(k)  \right]
	},
\end{equation}
and dressing the matrix element for the underlying process according to (here $|0\rangle$ denotes the oscillators' vacuum)
\begin{equation}
	\label{eq:Asr}
	S^{(M)}_{\text{\textit{s.r.}}}  = 
	e^{2i\hat\delta_{\textit{\text{\textit{s.r.}}}}}\,
	\frac{S^{(M)}}{\langle 0|e^{2i\hat\delta_{\textit{s.r.}}}|0\rangle} \,.
\end{equation}
Then, the matrix elements \eqref{eq:SMNSM} are recovered using the commutation relations \eqref{key1} in
\begin{equation}\label{SMajS}
	S_{\textit{s.r.},N}^{(M)}
	=
	\langle0| a_{j_1}(k_1)\cdots a_{j_N}(k_N)S^{(M)}_{\text{\textit{s.r.}}}|0\rangle\,,
\end{equation}
and similarly, for absorption processes $\langle 0 | S_{\textit{s.r.}}^{(M)} a_{j_1}^\dagger \cdots a_{j_N}^\dagger|0\rangle$.

Let us now focus on the case in which the background process is the elastic $2\to2$ scattering which has been the main object of study in the previous sections of this Report. 
We consider the Fourier transform to $b$-space of the two factors in~\eqref{eq:Asr} separately. By construction the second factor, which describes the elastic process, gives the eikonal. However, thanks to the division by $\langle 0|e^{2i\hat\delta_{\textit{s.r.}}}|0\rangle$ in \eqref{eq:Asr}, one needs only the real part of $2\delta$, as the infrared divergent contribution to the imaginary part is automatically encoded in the new operator part, as we shall see momentarily.\footnote{In this section, we focus on the infrared divergent contribution to this imaginary part. The finite contributions also involve non-soft modes and can be reproduced via a similar mechanism which will be discussed in Section~\ref{SemiclEik}.} The first factor in~\eqref{eq:Asr} is instead regular as $Q\to 0$, so we can write it as a differential operator acting on a delta-function $\delta^{D-2}(b)$ trading each $Q$ with a derivative
\begin{equation}
	\label{eq:qtodn}
	Q^\mu \to  -i \frac{\partial}{\partial b_\mu}
\end{equation}
in the Fourier transform.
Of course the product of these two factors in~\eqref{eq:Asr} becomes a convolution in $b$ space. However, since one factor is just a delta function, the integral of the convolution can be performed straightforwardly, and one obtains 
\begin{equation}
	\label{eq:eiksr}
	\begin{aligned}
		S_{\textit{s.r.}}  = e^{
			\int_k^{\ast}  \left[ 
			w^{\phantom{\dagger}}_j(k)\, a^\dagger_j(k)
			-w^\ast_j(k)\, a^{\phantom{\ast}}_j(k)  \right]
		}\left[1+2i\Delta(b)\right] e^{i \operatorname{Re}2\delta(b)} \,,
	\end{aligned}
\end{equation}
where the external momenta $p_n$ in the first line contain derivatives acting on the $b$-dependent functions in the second line.

Let us make a few general comments before using~\eqref{eq:eiksr} in some concrete calculations. First, the classical $S$-matrix obtained by neglecting the quantum remainder $\Delta$ is explicitly unitary since only the real part of $2\delta$ enters this equation and the inelastic prefactor is the exponential of an anti-Hermitian operator.
Second, we obtain the dominant contribution in the classical limit when the derivatives hidden in the external momenta $p_n$ due to \eqref{eq:qtodn} act on the rapidly oscillating eikonal phase, so that effectively
\begin{equation}
	\label{qtoQ2}
	Q^\mu \to \frac{\partial \operatorname{Re}2\delta (b)}{\partial b_\mu}\,.
\end{equation}
Since the soft factor $w_j(k)$ becomes proportional to $Q$ in the small-$Q$ limit (see for instance Eq. (2.11) of~\cite{DiVecchia:2021ndb}), we see explicitly how the disconnected term of the elastic $S$-matrix element drops out under the action of the derivative \eqref{qtoQ2} in \eqref{eq:eiksr}.  
Lastly, let us discuss how the factors involving graviton oscillators in \eqref{eq:eiksr} can be regarded as soft dressing of initial and final states. To this end, it is sufficient to define
\begin{equation}\label{outin}
	w_j^\text{out/in}(k) 
	= 
	\varepsilon^{\ast}_{j\mu\nu}(k)
	\sum_{n\in\text{out/in}}\eta_n \frac{\kappa\, p_n^\mu p_n^\nu}{p_n\cdot k}
\end{equation}
with $\eta_n=+1$ ($\eta_n=-1$) if $n$ is a final (initial) state of the background process,
and to introduce the dressed states
\begin{equation}\label{}
	|\text{out/in}\rangle  =  e^{\int_k^{\ast} \left(w_j^\text{out/in}(k) a_j^\dagger(k)-w_j^{\text{out/in}\ast}(k) a_j(k)\right)}|\Psi_{\text{out/in}}\rangle\,,
\end{equation}
where $|\Psi_{\text{out/in}}\rangle$ only involve massive (hard) states. In this way, if $|\Psi_\text{out}\rangle= e^{i\operatorname{Re}2\delta(b)}|\Psi_\text{in}\rangle$, then
we can rewrite this relation in terms of dressed states as follows,
\begin{align}
	|\text{out} \rangle &= e^{\int_k^{\ast} \left(w_j^\text{out}(k) a_j^\dagger(k)-w_j^{\text{out}\ast}(k) a_j(k)\right)}
	e^{\int_k^{\ast} \left(-w_j^\text{in}(k) a_j^\dagger(k)+w_j^{\text{in}\ast}(k) a_j(k)\right)}e^{i\operatorname{Re}2\delta (b)}
	|\text{in} \rangle  \nonumber \\
	&= e^{\int_k^{\ast} \left( (w_j^\text{out}(k) -w_j^\text{in}(k))   a_j^\dagger(k)-
		(w_j^{\text{out}}(k)- w_j^{\text{in}} (k))^\ast  a_j(k)\right)}e^{i\operatorname{Re}2\delta (b)}
	|\text{in} \rangle \,,
	\label{totdresssta}
\end{align}
since one can check that the two dressings for initial and final states commute as operators, owing to the reality of the combinations $w_j^{\text{out/in}}(k)$ themselves, and one obtains a total dressed state with $w_j(k) = w_j^\text{out}(k)-w_j^{\text{in}}(k)$. 
Therefore, $|\text{out} \rangle =S_{\textit{s.r.}} |\text{in}\rangle$ with $S_{\textit{s.r.}}$ precisely taking the overall dressing factor into account.

We can now apply the eikonal operator to discuss the contribution of low-energy gravitons to observables, including the waveforms, memory, and the particle-energy emission spectrum. The general procedure, given any quantum observable $\mathcal O$, is to take its expectation value according to
\begin{equation}\label{SOS}
	\langle \mathcal O \rangle = \langle \Psi_\text{in}| S_{\textit{s.r.}}^\dagger \mathcal O \, S_{\textit{s.r.}} |\Psi_\text{in}\rangle\,.  
      \end{equation}
Physically, this means to evaluate the mean value of $\mathcal O$ in the final state of the scattering event, obtained by the action of  $S_{\textit{s.r.}}$. Thus we follow the same strategy as in the KMOC approach~\cite{Kosower:2018adc}, but here we take the classical limit as a first step approximating the full $S$-matrix with the eikonal (operator)	.
Let us start from the classical field, which is obtained by inserting in the expectation value \eqref{SOS} the free gravitational field~\cite{Cristofoli:2021vyo}
\begin{equation}\label{eq:theGravitonField}
	H_{\mu\nu}(x)=
	\int_k  \left[ \varepsilon_{j \mu\nu}(k) a_j(k)\, e^{ik\cdot x} + \varepsilon^\ast_{j\mu\nu}(k) a^\dagger_j(k)\,\,e^{-ik\cdot x}\right].
\end{equation}
This yields
\begin{equation}
	\label{eq:hmnzfz}
	h^{\mu\nu}(x)
	= 
	\langle H^{\mu\nu}(x) \rangle
	=
	\int_k^{\ast}      \left[ w_\text{TT}^{\mu\nu}(k)\, e^{ikx} + w_\text{TT}^{\ast\mu\nu}(k)\,e^{-ikx}\right] \,,
\end{equation}
where
\begin{equation}\label{fTT}
	w_\text{TT}^{\mu\nu}(k) = \Pi^{\mu\nu}_{\rho\sigma}(k)\, w^{\rho\sigma}(k)\,,
	\qquad
	w^{\mu\nu}(k)
	=
	\sum_n \frac{\kappa\,  p_n^\mu p_n^\nu}{p_n\cdot k}\,,
\end{equation}
and $\Pi^{\mu\nu}_{\rho \sigma}(k)$ is the standard transverse-traceless projector over physical degrees of freedom \eqref{TTprojector}. 
Of course the prediction
\eqref{eq:hmnzfz} is only accurate for a detector placed at a large distance $r$ from the sources. Taking this limit (see \ref{app:asymptoticlimit}) at a fixed retarded time $u$, so that $r\gg |u|,b$, and moving along the null vector $n^\mu$, which characterizes the angular direction, \eqref{eq:hmnzfz} yields\footnote{We can send $\omega^\ast\to\infty$ as long as we focus on the value of the resulting integral for large $|u|\gtrsim b$.}
\begin{equation}\label{hmunufmunu}
	h^{\mu\nu}(x) \sim 
	\frac{1}{4\pi r}
	\int_{-\infty}^{+\infty} \frac{d\omega}{2i\pi}\, 
	w_\text{TT}^{\mu\nu}(\omega \,n)\, e^{-i\omega u} \,,
\end{equation}
where we used that $w_{\mu\nu}(k)=-w_{\mu\nu}^\ast(-k)$.
Adjusting the normalization by comparing
\begin{equation}\label{normWavesoft}
	g_{\mu\nu} = \eta_{\mu\nu} + 2 W_{\mu\nu}  = \eta_{\mu\nu} + 2\kappa h_{\mu\nu}\,,
\end{equation}
we define the waveform according to 
\begin{equation}\label{}
	W_{\mu\nu} =  \kappa\, h_{\mu\nu} \,.
\end{equation}
Performing the Fourier transform in \eqref{hmunufmunu}  requires in principle to specify how the $1/\omega$ singularity at $\omega=0$ is circumvented (see \eqref{thetatransform} below) \cite{Laddha:2018vbn,Sahoo:2018lxl,Saha:2019tub,Sahoo:2021ctw,DiVecchia:2022owy}. However, as stressed in \cite{Strominger:2014pwa,Strominger:2017zoo}, the key point is that the behavior of the waveform at large $|u|$ is completely determined by this pole at $\omega=0$, and possible ambiguities are in fact $u$-independent. 
Considering the invariant combination
\begin{equation}\label{}
	\Delta W_{\mu\nu}(n) = W_{\mu\nu}(u>0,n)-W_{\mu\nu}(u<0,n)\,,
\end{equation}
and recalling that $\Pi^{\mu\nu}_{\rho\sigma}(\omega n)=\Pi^{\mu\nu}_{\rho\sigma}(n)$,
we thus obtain the memory effect \cite{Zeldovich:1974gvh} 
\begin{equation}\label{memory}
	\Delta W_{\mu\nu}(n)
	=\frac{2G}{r}\,\Pi^{\mu\nu}_{\rho\sigma}(n)\,\sum_{a} \frac{p_a^\rho p_a^\sigma}{(-p_a \cdot  n)}\,,
\end{equation}
i.e.~the leading result of~\cite{Laddha:2018vbn,Sahoo:2018lxl,Saha:2019tub,Sahoo:2021ctw} or the term indicated as $A_{\mu\nu}$ in~\cite{Sahoo:2021ctw}.
However, this approach does not capture non-linear memory effects \cite{Christodoulou:1991cr,Wiseman:1991ss,Thorne:1992sdb,Damour:2020tta}.

Let us instead consider the projection of $S_{\textit{s.r.}}|\Psi_\text{in}\rangle$ back onto the initial state $|\Psi_\text{in}\rangle$ with no gravitons. Then one needs to normal order the inelastic exponential through the usual Baker--Campbell-Hausdorff formula
$e^{A + B}= e^{A} e^{B} e^{-\frac12 [A,B]}$,
so the amplitude for the elastic process is given by
\begin{equation}\label{FPIF}
	\langle \Psi_\text{in}| S_{\textit{s.r.}}|\Psi_\text{in}\rangle = \exp\left[-\frac{1}{2} \int_k^{\ast}  w^\ast_{\mu\nu}(k)\Pi^{\mu\nu,\rho\sigma}(k)w_{\rho\sigma}(k)\right] e^{i\operatorname{Re}2\delta(b)} \,,
\end{equation}
where focused on the classical contribution (ignoring $\Delta$).
The transversality condition $k^\mu w_{\mu\nu}=0$, which holds for gravity by momentum conservation, grants
\begin{equation}
	\label{cancellation}
	w^{\ast}_{\mu\nu}(k)\Pi^{\mu\nu,\rho\sigma}(k)w_{\rho\sigma}(k) = w^{\ast}_{\mu\nu}(k) \left(
	\eta^{\mu\rho}\eta^{\nu\sigma}-\tfrac{1}{D-2}\,\eta^{\mu\nu}\eta^{\rho\sigma}
	\right) w_{\rho\sigma}(k)
	\equiv
	w^{\ast}(k)\,w(k)
	\,,
\end{equation}
where we introduced a useful condensed notation according to which explicit index contractions are suppressed, as in \eqref{eq:grint}. We will use this repeatedly in the following.
Recasting \eqref{FPIF} as 
\begin{equation}\label{}
	\langle \Psi_\text{in}| S_{\textit{s.r.}}|\Psi_\text{in}\rangle 
	=
	e^{	
		i
		\left[
		\operatorname{Re}2\delta(b)
		+
		\frac{i}{2}\int_k^\ast w^\ast(k)w(k)
		\right]
	}
\end{equation}
we see that the damping factor that emerged from the reordering of the exponential factors can be interpreted as an imaginary contribution to the classical eikonal: the infrared-divergent one as $\epsilon = (4-D)/2\to0$,
\begin{equation}
	\label{eq:Imdeltadef}
	\operatorname{Im}2\delta(b) = \frac{1}{2} \int_k^{\ast}  w^\ast(k) w(k) + \mathcal O(\epsilon^0)
	=
	\frac{1}{2} \int_k  \theta(\omega^\ast-k^0) \,w^\ast(k) w(k) + \mathcal O(\epsilon^0)
	\,.
\end{equation}
Note that this integral is scale-invariant in the limit $\epsilon\to0$, which means that its $1/\epsilon$ term is in fact independent of the cutoff $\omega^\ast$.
We thus see the origin of the infrared-divergent imaginary part we had already encountered in the previous section. It emerges because the elastic amplitude neglects the contributions of soft-graviton emissions \cite{Weinberg:1965nx}.
The integral entering \eqref{eq:Imdeltadef} can be evaluated as follows to leading order in $\epsilon$, retaining also the logarithmic dependence on the cutoff. Introducing velocities analogous to \eqref{eq:velocities} for each state according to 
\begin{equation}\label{nvelocities}
	p_n^\mu=\eta_n m_n v_n^\mu
\end{equation}
so that $v_n^2=-1$ and $v_n^\mu$ is future-directed, we can use the following identity, 
\begin{equation}\label{p_nkp_mk}
	\int^\ast_k \frac{m_nm_m}{(p_n\cdot k)(p_m\cdot k)} = \int^\ast_k \frac{\eta_n \eta_m}{(v_n\cdot k)(v_m\cdot k)} = \left[
	\frac{(\omega^\ast)^{-2\epsilon}}{-2\epsilon} 
	\right]
	\frac{F_{nm}}{(2\pi)^2}
	+
	\mathcal O(\epsilon^0)
\end{equation}
where
\begin{equation}\label{keyf}
	F_{nm}=\frac{\eta_n \eta_m \operatorname{arccosh}{\sigma_{nm}}}{ \sqrt{\sigma_{nm}^2-1}}
	\,,\qquad
	\sigma_{nm}=-v_n\cdot v_m = - \eta_n \eta_m \frac{p_n\cdot p_m}{m_n m_m}\,.
\end{equation}
To show \eqref{p_nkp_mk}, it is convenient to perform a decomposition of the integrated momentum $k^\mu$ analogous to \eqref{decomposition} for each pair $n$, $m$, letting
\begin{equation}\label{komeganomegam}
	k^\mu = \omega_n\,\check v_n^\mu + \omega_m\,\check v_m^\mu + k_\perp^\mu\,,
	\qquad
	\check v_{n,m}^\mu = \frac{\sigma_{nm}v^\mu_{m,n}-v^\mu_{n,m}}{\sigma_{nm}^2-1}\,.
\end{equation}
Taking the Jacobian determinant $1/\sqrt{\sigma^2_{nm}-1}$ into account (see also \ref{usefulFT})
and focusing on the rest frame of particle $n$ where $k^0=\omega_n$, we obtain, 
\begin{equation}\label{}
	\int_k^\ast \frac{1}{(v_n\cdot k)(v_m\cdot k)} = \frac{1}{\sqrt{\sigma_{nm}^2-1}}\! 
	\int_0^{\omega^\ast}\! \frac{d\omega_n}{\omega_n}\!
	\int \!\frac{d \omega_m}{\omega_m}\! 
	\int \!\frac{ d^{2-2\epsilon} k_\perp}{ (2\pi)^{3-2\epsilon}}\,
	\delta\left(k_\perp^2-\frac{-\omega_n^2+2\omega_n\omega_m\sigma_{nm}-\omega_m^2}{\sigma_{nm}^2-1}\right)\!,
\end{equation}
where we have used $k\cdot v_{n, m} = - \omega_{n,m}$.
Changing integration variables by letting $\omega_n=\omega$, $\omega_m=\omega x$ and $k^\mu_\perp= \omega x^\mu_\perp$, performing the integral over $x_\perp^\mu$ and focusing on the leading term as $\epsilon\to0$, we find
\begin{equation}\label{}
	\int_k^\ast \frac{1}{(v_n\cdot k)(v_m\cdot k)} = \frac{1}{(2\pi)^2\sqrt{\sigma_{nm}^2-1}} 
	\int_0^{\omega^\ast}\!\!\frac{d\omega}{\omega^{1+2\epsilon}}
	\int\frac{dx}{2x} \,\theta\left(-x^2+2\sigma_{nm} x-1\right) + \mathcal O(\epsilon^0)\,.
\end{equation}
Noting that the Heaviside $\theta$ function restricts the integration over $x$ to lie between the two (positive) roots $\sigma_{nm}\pm\sqrt{\sigma_{nm}^2-1}$  finally leads to
\begin{equation}\label{}
	\int_k^\ast \frac{1}{(v_n\cdot k)(v_m\cdot k)} = \left[
	\frac{(\omega^\ast)^{-2\epsilon}}{-2\epsilon} 
	\right] \frac{\operatorname{arccosh}\sigma_{nm}}{(2\pi)^2\sqrt{\sigma_{nm}^2-1}}	+ \mathcal O(\epsilon^0) 
\end{equation}
where we used that $\operatorname{arccosh}\sigma_{nm}= \pm \log(\sigma_{nm}\pm\sqrt{\sigma_{nm}^2-1})$.
This shows \eqref{p_nkp_mk},
and using this equation in \eqref{eq:Imdeltadef}, one finds
\begin{equation}\label{Im2dDIV}
	\operatorname{Im}2\delta(b)
	= 
	\left[
	\frac{(\omega^\ast)^{-2\epsilon}}{-2\epsilon} 
	\right]
	\frac{G}{\pi} \sum_{n,m} m_n m_m  \left(\sigma_{nm}^2 - \tfrac{1}{2} \right) F_{nm}  + \mathcal O(\epsilon^0)\,.
\end{equation}

A similar calculation concerns the insertion of the energy-momentum operator
\begin{equation}\label{}
	P^\alpha 
	= 
	\int_k^\ast  k^\alpha\, a^\dagger_{j}(k) a_{j}(k)\,,
	\qquad
	\boldsymbol{P}^\alpha = \langle P^\alpha \rangle
\end{equation}
as in Eq.~\eqref{SOS}, which leads to 
\begin{equation}
	\label{eq:Erad}
	\boldsymbol{P}^\alpha = \int_k^{\ast}   k^\alpha w^\ast(k) w(k) \,,
\end{equation}
thanks to \eqref{cancellation}.
Since we employ the leading soft approximation, we can only resolve the ZFL of the energy emission spectrum,
\begin{equation}\label{spectrum}
	\lim_{\omega \to0} \frac{dE}{d\omega}\equiv \frac{\partial\boldsymbol P^0}{\partial \omega^\ast}\,,
\end{equation}
or, using \eqref{eq:Erad},
\begin{equation}\label{}
	\lim_{\omega \to0} \frac{dE}{d\omega}
	= 
	\int_k \delta(\omega^\ast-k^0) k^0 w^\ast(k) w(k) = \omega^\ast 	\int_k \delta(\omega^\ast-k^0) w^\ast(k) w(k)
\end{equation}
(with $\int_k$ as in \eqref{kphasespace}).
Comparing with \eqref{eq:Imdeltadef}, we then see that
\begin{equation}\label{spectrumint}
	\lim_{\omega \to0} \frac{dE}{d\omega}
	=
	\lim_{\epsilon\to0}
	2\omega^\ast \frac{\partial}{\partial \omega^\ast} \operatorname{Im}2\delta(b)
	+ \mathcal O(\epsilon^0)\,.
\end{equation}
Using \eqref{spectrumint} and the explicit result \eqref{Im2dDIV}, we see that this derivative cancels the divergence and extracts the coefficient of $\log\omega^\ast$ in the $\epsilon\to0$ expansion, leading to 
\begin{equation}
	\label{eq:ZFLspectrum}
	\lim_{\omega \to0}\frac{dE}{d\omega} = \lim_{\epsilon\to 0} \left[-4\epsilon \operatorname{Im}2\delta(b) \right]
\end{equation}
or, more explicitly,
\begin{equation}\label{Imdelta}
	\lim_{\omega \to0}\frac{dE}{d\omega} = \frac{2G }{\pi} \sum_{n,m} m_n m_m  \left(\sigma_{nm}^2 - \tfrac{1}{2} \right) F_{nm}\,.
\end{equation}
As already mentioned, the scope of validity of the formulas \eqref{Im2dDIV}, \eqref{Imdelta} is rather wide, since they retain an exact dependence on the background hard kinematics, they generalize straightforwardly to scatterings involving an arbitrary number of initial and finial states, and are also valid if the colliding objects have spin or an internal structure, e.g.~if they are subject to tidal deformations.
This highlights a general mechanism: the infrared divergences of the elastic amplitude determine the ZFL of the energy emission spectrum via massless quanta \cite{Weinberg:1965nx,DiVecchia:2021ndb}.

Focusing  on the $2\to2$ case, in order to obtain a more explicit formula it is sufficient to use $\sigma_{nn}=1$ and $F_{nn} = 1$, while for $n\not= m$ we have $\sigma_{nm}=\sigma_{mn}$ and
\begin{equation}\label{eq:sigmamn4p}
	\sigma_{12} = \sigma_{34} = \sigma\,,
	\quad
	\sigma_{13} = \sigma_{24} = \sigma_Q\,,\quad
	\sigma_{14} = 1+\frac{Q^2}{2m_1^2}\,,\quad
	\sigma_{23} = 1+\frac{Q^2}{2m_2^2}\,,
\end{equation}
where we introduced the shorthand notation
\begin{equation}\label{eq:sigmaq}
	\sigma_Q = \sigma -\frac{Q^2}{2m_1m_2} = - \frac{u - m_1^2 - m_2^2}{2 m_1m_2}\,.
\end{equation}
Then, Eq.~\eqref{Imdelta} becomes 
\begin{align}
	\nonumber
	\lim_{\omega \to0}\frac{dE}{d\omega} &=  \frac{4 G }{\pi}
	\Bigg\{
	2m_1m_2\left(\sigma^2-\tfrac12\right)\frac{\operatorname{arccosh}\sigma}{\sqrt{\sigma^2-1}}
	-
	2m_1m_2\left(\sigma_Q^2-\tfrac12\right)\frac{\operatorname{arccosh}\sigma_Q}{\sqrt{\sigma_Q^2-1}}
	\\  \label{eq:d2gr}
	&+
	\sum_{i=1,2}
	\Bigg[
	\frac{m_i^2}{2}
	-
	m_j^2 \Big(\left(
	1+\tfrac{Q^2}{2m_i^2}
	\right)^2-\tfrac12\Big)
	\frac{\operatorname{arccosh}\left(
		1+\tfrac{Q^2}{2m_i^2}
		\right)}{\sqrt{\left(
			1+\tfrac{Q^2}{2m_i^2}
			\right)^2-1}}
	\Bigg]\Bigg\},
\end{align} 
where, as discussed, the transferred momentum $Q$ should be interpreted by using~\eqref{qtoQ2}, which corresponds to the substitution $Q^\mu \rightarrow - {\hat{b}}^{\mu} 2p \sin \frac{\Theta}{2}$. As already emphasized, while later on we will focus on certain interesting kinematic limits, the dependence of this formula on the dynamics of the background elastic process, and in particular on $Q/m_i$, is exact.

\subsection{Eikonal operator including static modes }
\label{sec:eikopwith}

Let us now turn to the contribution of low-frequency gravitons to the waveform in position space, in particular its value at early retarded times, and to the angular momentum  \cite{Damour:2020tta,Veneziano:2022zwh}. As we shall discuss, both quantities are sensitive to static-field effects, and therefore to how one approach the $\sim 1/\omega$ singularity at $\omega=0$ in the radiation spectrum. 

As we emphasized, the soft eikonal operator \eqref{eq:eiksr} is based on the standard Weinberg soft theorem, which includes soft gravitons with low but nonzero frequency. 
As such, it does not include effects that arise due to exactly static fields, whose Fourier transform is localized at zero frequency.
To include them it is sufficient to replace the standard soft factor $w_j(k)$ in \eqref{eq:eiksr}  by\footnote{Let us note that the symmetric Lorentz tensor $F^{\mu\nu}$ in \eqref{softfactorF} should not be confused with the symmetric coefficients $F_{mn}$ introduced in \eqref{keyf} earlier on.}
\begin{equation}\label{softfactorF}
	f_j(k) = \varepsilon_{j\mu\nu}(k)^\ast F^{\mu\nu}(k)\,,\qquad
	F^{\mu\nu}(k) = \sum_{n} \frac{\sqrt{8\pi G}\,p_n^\mu p_n^\nu}{p_n\cdot k-i0}\,.
\end{equation}
and to consider the following operator (neglecting the quantum remainder for brevity) 
\begin{equation}
	\label{eq:eiksrstatic}
	\begin{aligned}
		\mathcal{S}_{s.r.}  = e^{\int_k^{\ast}  \left[ 
			f^{\phantom{\dagger}}_j(k)\, a^\dagger_j(k)
			-f^\ast_j(k)\, a^{\phantom{\ast}}_j(k)  \right]} e^{2i\tilde\delta(b)} \,,
	\end{aligned}
\end{equation}
where the definition of $2\tilde\delta$ will be specified momentarily.
By including the $-i0$ prescription in \eqref{softfactorF} even for real emissions of gravitons, we are now dressing the full $S$-matrix, including the identity term, and thus include possible ``emissions'' localized at $\omega=0$ from disconnected pieces of the hard matrix element.
This construction is not a standard application of Weinberg's theorem, which only holds for {\em connected} amplitudes \cite{Weinberg:1995mt,Weinberg:1972kfs}, and in this way it also captures static effects. In fact, the same static term in the asymptotic field also follows from the worldline approach in which one solves for the particle trajectories and the field using retarded propagators \cite{Mougiakakos:2021ckm,Riva:2021vnj}.

To see how this modification reflects the definition of dressed states, compared to the one discussed in the previous subsection, let us now consider 
\begin{equation}\label{}
	f_j^\text{out/in}(k) 
	= 
	\varepsilon^{\ast}_{j\mu\nu}(k)
	\sum_{n\in\text{out/in}} \eta_n\, \frac{\sqrt{8\pi G} p_n^\mu p_n^\nu}{p_n\cdot k-i0}
\end{equation}
and 
\begin{equation}\label{}
	|\text{\scriptsize OUT/IN}\rangle  =  e^{\int_k^{\ast} \left(f_j^\text{out/in}(k) a_j^\dagger(k)-f_j^{\text{out/in}\ast}(k) a_j(k)\right)}|\Psi_{\text{out/in}}\rangle
\end{equation}
(the notation $|\text{\scriptsize{OUT/IN}}\rangle$ is meant to distinguish these states from $|\text{out/in}\rangle$ in \eqref{outin}).
In this way, if we start again from 
$|\Psi_\text{out}\rangle= e^{i\operatorname{Re}2\delta (b)}|\Psi_\text{in}\rangle$ and we rewrite it in terms of dressed states, we get
\begin{equation}\label{}
	|\text{\scriptsize OUT}\rangle 
	= 
	e^{\int_k^{\ast} \left(f_j^\text{out}(k) a_j^\dagger(k)-f_j^{\text{out}\ast}(k) a_j(k)\right)}
	e^{-\int_k^{\ast} \left(f_j^\text{in}(k) a_j^\dagger(k)-f_j^{\text{in}\ast}(k) a_j(k)\right)}
	e^{i\operatorname{Re}2\delta (b)}
	|\text{\scriptsize IN}\rangle\,.
\end{equation}
In this new setup, the two dressings for initial and final states no longer commute, and  using the Baker--Campbell--Hausdorff formula $e^{A} e^{B}=e^{A + B} e^{+\frac12 [A,B]}$ one obtains
\begin{equation}\label{}
	|\text{\scriptsize OUT}\rangle 
	= 
	e^{\int_k^{\ast} \left(f_j(k) a_j^\dagger(k)-f_j^{\ast}(k) a_j(k)\right)}
	e^{\frac{1}{2}\int_k^{\ast} \left(f_j^{\text{out}\ast}(k) f_j^\text{in}(k)- f_j^\text{out}(k) f_j^{\text{in}\ast}(k)\right)+i\operatorname{Re}2\delta (b)}
	|\text{\scriptsize IN}\rangle\,,
\end{equation}
where
\begin{equation}\label{}
	f_j (k)= f_j^\text{out}(k) - f_j^\text{in}(k)
\end{equation}
and therefore, comparing with \eqref{eq:eiksrstatic}, we see that $|\text{\scriptsize OUT}\rangle = \mathcal{S}_{s.r} |\text{\scriptsize IN}\rangle$ provided the phase takes the value 
\begin{equation}\label{eq:tildedelta}
	2i\tilde\delta (b) = i\operatorname{Re}2\delta (b)-2i\delta^{\text{dr.}} (b)\,,\qquad
	2i\delta^{\text{dr.}} (b)= -\frac{1}{2}\int_k^{\ast}\!\!\!\! \left(f_j^{\text{out}\ast}(k) f_j^\text{in}(k)-f_j^\text{out}(k) f_j^{\text{in}\ast}(k) \right).
\end{equation}
The last equation, whose right-hand side is manifestly imaginary, identifies a contribution to the phase due to the dressing (hence the superscript ``dr.'').
The integral that we need to calculate in order to determine the resulting phase correction reads 
\begin{equation}\label{}
	2i\delta^\text{dr.}(b)
	=
	\frac12
	\sum_{\substack{n\in\text{out}\\m\in\text{in}}}
	\int_k^\ast
	\left[
	\frac{8\pi G m_n^2m_m^2(\sigma^2_{nm}-\tfrac12)}{(p_n\cdot k+i0)(p_m\cdot k-i0)}
	-
	\frac{8\pi G m_n^2m_m^2(\sigma^2_{nm}-\tfrac12)}{(p_n\cdot k-i0)(p_m\cdot k+i0)}
	\right].
\end{equation}
Sending $k^\mu\to -k^\mu$ in the second term, we see that this expression recombines as follows,
\begin{equation}\label{2iddr}
	2i\delta^\text{dr.}(b)
	=
	\frac12
	\sum_{\substack{n\in\text{out}\\m\in\text{in}}}
	8\pi G m_n m_m (\sigma^2_{nm}-\tfrac12)
	I_{nm}
\end{equation}
where, introducing the velocities $v_n^\mu$ according to \eqref{nvelocities} 
\begin{equation}\label{inm}
	I_{nm}
	=
	\int
	\frac{d^{4-2\epsilon}k}{(2\pi)^{4-2\epsilon}}
	\frac{2\pi\text{sgn}(k^0)\delta(k^2)\theta(\omega^\ast-|k^0|)}{(\eta_n v_n\cdot k+i0)(\eta_m v_m\cdot k-i0)}\,.
\end{equation}
Here $\operatorname{sgn}(k^0)=\theta(k^0)-\theta(-k^0)$ takes the value $+1$ (resp. $-1$) if $k^0>0$ ($k^0<0$).
The integrals $I_{nm}$ can be evaluated in a manner similar to \eqref{p_nkp_mk}, as we turn to illustrate.
Sending $\epsilon=\frac{4-D}{2}\to0$, which as we shall see leads to a finite leading-order contribution, and focusing on the rest frame of particle $n$ where $k^0=\omega_n=-v_n\cdot k$, we can decompose the integrated momentum according to \eqref{komeganomegam} to arrive at
\begin{equation}\label{}
	I_{nm}
	\!=\!
	\frac{1}{\sqrt{\sigma^2_{nm}-1}}
	\int_{-\omega^\ast}^{+\omega^\ast}\!\!\!
	\frac{\operatorname{sgn}(\omega_n)\,d\omega_n}{(-\eta_n\omega_n+i0)}
	\int
	\frac{ d\omega_m dk^2_\perp}{(-\eta_m\omega_m-i0)2(2\pi)^2}
	\delta\left(k_\perp^2-\frac{-\omega_n^2+2\omega_n\omega_m\sigma_{nm}-\omega_m^2}{\sigma_{nm}^2-1}\right)
\end{equation}
after taking the Jacobian determinant $1/\sqrt{\sigma_{nm}^2-1}$ into account.
Changing integration variables by letting $\omega_n=\omega$, $\omega_m=\omega x$ and $k_\perp^2= \omega^2 x_\perp^2$,
so that
\begin{equation}\label{}
	\operatorname{sgn}(\omega_n)\,	d\omega_n\,d\omega_m\,dk_\perp^2 = 
	\omega^3\,d\omega\,dx\,dx_\perp^2\,,
\end{equation}
and performing the integral over $x_\perp^2$ by means of the delta function,
\begin{equation}\label{}
	I_{nm}
	=
	\frac{1}{(2\pi)^2\sqrt{\sigma^2_{nm}-1}}
	\int_{-\omega^\ast}^{+\omega^\ast}\!
	\frac{d\omega}{(-\eta_n\omega+i0)}
	\int
	\frac{\omega\,dx}{2(-\eta_m\omega x-i0)}
	\theta\left(-x^2+2\sigma_{nm}x-1\right).
\end{equation}
The Heaviside $\theta$ function restricts the integration over $x$ to lie between the two \emph{positive} roots $\sigma_{nm}\pm\sqrt{\sigma_{nm}^2-1}$, so that the integrals over $\omega$ and $x$ factorize, leading to
\begin{equation}\label{}
	I_{nm} = \frac{\operatorname{arccosh}\sigma_{nm}}{(2\pi)^2\sqrt{\sigma_{nm}^2-1}} \int_{-\omega^\ast}^{\omega^\ast} \frac{\omega\,d\omega}{(-\eta_n\omega+i0)(-\eta_m\omega-i0)} \,.
\end{equation}
The new ingredient that we need to evaluate is thus the integral over ``positive and negative frequencies''
\begin{equation}\label{omeganomegam}
	\int_{-\omega^\ast}^{+\omega^\ast}\!\! \frac{\omega\,d\omega}{(-\eta_n\omega+i0)(-\eta_m\omega-i0)} =  -\frac{i\pi}{2}(\eta_n-\eta_m)\,.
\end{equation}
To see this, let us make the $i0$ prescription more manifest by introducing a small $\lambda>0$. Clearly the integral vanishes if $\eta_n=\eta_m$ because $\int_{-\omega^\ast}^{\omega^\ast}\omega d\omega/(\omega^2+\lambda^2)$ is zero by parity. If $\eta_n=-\eta_m$, the integral reduces instead to
\begin{equation}\label{}
	\begin{split}
		-\int_{-\omega^\ast}^{+\omega^\ast}\!\! \frac{\omega\,d\omega}{(\omega-i\lambda \eta_n)^2} 
		&=
		-\left[
		\log(\omega-i\lambda\eta_n)
		-\frac{i \lambda\eta_n}{\omega-i\lambda\eta_n}
		\right]_{-\omega^\ast}^{+\omega^\ast}
		\\
		&= -i\pi\eta_n + \frac{2i\eta_n\lambda\omega^\ast}{(\omega^\ast)+\lambda^2} \xrightarrow[\lambda\to0^+]{} -i\pi\eta_n\,,
	\end{split}
\end{equation} 
which shows \eqref{omeganomegam}, and this leads to the final expression for $I_{nm}$,
\begin{equation}\label{inmfinal}
	I_{nm} = - \frac{i\pi}{2}(\eta_n-\eta_m) \frac{\operatorname{arccosh}\sigma_{nm}}{(2\pi)^2\sqrt{\sigma_{nm}^2-1}}\,.
\end{equation}
Note that the result no longer depends on the cutoff $\omega^\ast$, so we may now send it to zero. What we obtain is thus a contribution localized at the zero-frequency end of the spectrum, i.e.~an effect intrinsically due to field configurations that are static in time domain.
Substituting \eqref{inmfinal} into \eqref{2iddr}, we obtain
\begin{equation}\label{}
	2i\delta^{\text{dr.}} (b)
	= -i G \sum_{\substack{n\in\text{out}\\m\in\text{in}}} 
	m_n m_m(\sigma^2_{nm}-\tfrac12)\frac{\operatorname{arccosh}\sigma_{nm}}{\sqrt{\sigma^2_{nm}-1}}
\end{equation}
so that, using \eqref{eq:sigmamn4p} and expanding for small deflections $Q = Q_{\text{1PM}}+\mathcal{O}(G^2)$,
\begin{equation}\label{2delta2}
	2i\delta^{\text{dr.}} (b)
	= \frac{iG Q_{\text{1PM}}^2}{2}\left[
	\frac{8-5\sigma^2}{3(\sigma^2-1)}
	+
	\frac{\sigma(2\sigma^2-3)\operatorname{arccosh}\sigma}{(\sigma^2-1)^{\frac{3}{2}}}
	\right]
	+\mathcal O(G^4)
	=i\operatorname{Re}2\delta_2^{\text{RR}} (b)
	+\mathcal O(G^4)\,,
\end{equation}
where we have recognized the 3PM radiation-reaction phase $\operatorname{Re}2\delta_2^\text{RR} (b)$.
We also neglected a $b$-independent $\mathcal O(G)$ contribution, which would be tantamount to a finite shift of the (IR-divergent) Shapiro time delay.
In this way, recalling that $2i\delta^\text{dr.} (b)$ defines the subtraction defining $2i\tilde\delta (b)$ in \eqref{eq:tildedelta}, we see that the inclusion of static modes in the dressing leads to an eikonal operator in which the overall phase $2i\tilde\delta (b)$ is the \emph{conservative} eikonal phase up to 3PM, i.e.~it does not include radiation-reaction effects. We shall see in later sections that the information about these effects is not lost, but is actually encoded in the ``soft factor'' $f_j(k)$ rather than in the explicit phase. 

Let us now turn to the prediction of this eikonal operator for two observables: the memory waveform and the angular momentum of the field.
Proceeding like for \eqref{hmunufmunu}, but taking the expectation on the $\mathcal{S}_{s.r.}|\Psi_{\text{in}}\rangle$ state including static effects, we find the following asymptotic limit for the gravitational field considering a large distance $r$ at fixed retarded time $u$ and angles given by $n^\mu$ (see also \ref{app:asymptoticlimit})
\begin{equation}\label{}
	h^{\mu\nu} \sim 
	\frac{1}{4\pi r}
	\int_{-\infty}^{+\infty} \frac{d\omega}{2i\pi}\, 
	f_\text{TT}^{\mu\nu}(\omega\, n)\, e^{-i\omega u}\,, \qquad
	f_\text{TT}^{\mu \nu} (k)
	=\Pi^{\mu \nu}_{\rho \sigma}(k)
	\sum_n \frac{\kappa\,  p_n^\rho p_n^\sigma}{p_n\cdot k-i0}\,,
\end{equation}
and performing the integral over $\omega$ by means of the Fourier representation of the Heaviside $\theta$ function,
\begin{equation}\label{thetatransform}
	\int_{-\infty}^{+\infty} \frac{d\omega}{i2\pi} \frac{e^{-i\omega u}}{-\eta_n \omega-i0}= \int_{-\infty}^{+\infty} \frac{d\omega}{i2\pi}\frac{e^{i\omega \eta_n u}}{ \omega-i0} = \theta(\eta_n u)\,,
\end{equation}
this leads to
\begin{equation}
	\label{eq:wavAg}
	W^{\mu\nu} = \kappa h^{\mu\nu} \sim  \frac{2G}{r} \Pi^{\mu\nu}_{\rho\sigma}(n) \sum_{a} \frac{p_a^\rho p_a^\sigma\, \theta(\eta_a u)}{(-\eta_a p_a\cdot n)}
	\,.
\end{equation}
Note that for $u>0$ ($u<0$) only the out (in) contributions $\eta_n>0$ ($\eta_n<0$) survive, so that the difference $\Delta W_{\mu\nu}$ yields the memory effect \eqref{memory}. 
Expanding \eqref{eq:wavAg} to leading order in the PM regime, where the out states are equal to the in states up to $\mathcal O(G)$ deflections, one instead finds 
\begin{equation}\label{Damourf1}
	W^{\mu\nu} \sim  \frac{2G}{r} \Pi^{\mu\nu}_{\rho\sigma}(n) \sum_{a\in\text{in}} \frac{p_a^\rho p_a^\sigma}{p_a\cdot n}
	+\mathcal O(G^2)\,,
\end{equation}
i.e.~the waveform has a $u$-independent $\mathcal O(G)$ contribution. This is the leading-order static field, whose inclusion was possible precisely thanks to the $-i0$ prescription in \eqref{softfactorF}. Before moving on to computing different contributions to the angular momentum of the process let us make a few remarks on the traditional GR perspective on this issue.

The issue of angular momentum and of its loss in GR is a subtle one and has a long history (see e.g. \cite{Ashtekar:2019rpv}). The approach based on the use of Bondi--Sachs coordinates \cite{Bondi:1960jsa}, \cite{Sachs:1962wk}, which is very convenient for describing the energy loss by the system via gravitational-wave emission, cannot be trivially extended to angular momentum. The problem is that the definition of  Bondi's angular momentum, $J_B$, suffers from a gauge ambiguity related to the possibility of performing supertranslations, a subgroup of the BMS transformations \cite{Bondi:1962px,Sachs:1962zza} that preserves the Bondi--Sachs coordinate conditions. 

As an example,
the $\mathcal O(G)$ static term in \eqref{Damourf1} can always be removed completely by performing a BMS supertranslation, 
\begin{equation}\label{}
	2\kappa \delta_T h_{AB} = 2\kappa T(n)\,\partial_u h_{AB} - r (2 D_A D_B -\gamma_{AB} D^C D_C)\,T(n)\,,
\end{equation}	
where $A,B,C,\ldots$ label angular directions and $\gamma_{AB} = (\partial_A n)\cdot (\partial_B n)$ is the metric on the sphere with covariant derivative $D_A$,
by choosing \cite{Veneziano:2022zwh}
\begin{equation}\label{}
	T(n) = 2 G \sum_{a\in\text{in}} (p_a\cdot n) \log(p_a\cdot n/m_a)\,.
\end{equation}
This raises the issue of whether such a static contribution has a physical meaning.
In \cite{Veneziano:2022zwh} it was argued that the true radiated angular momentum is the one computed by using a particular BMS gauge, called the canonical gauge in the GR literature, which amounts to setting the initial shear of the Bondi--Sachs metric to zero. In the post-Minkowskian expansion this contribution starts at ${\cal O}(G^3)$ and is insensitive to exactly zero frequency  gravitons. It was also proposed that the additional static contribution to angular-momentum loss corresponds to choosing a different Bondi gauge, dubbed ``intrinsic".

The issue has been further addressed in \cite{Chen:2021szm},  \cite{Javadinezhad:2022ldc} and, more recently, in \cite{Mao:2023evc} and \cite{Riva:2023xxm}. The outcome of these investigations is that both the radiative and the static contributions to the angular momentum loss by the two-body system can be defined in a BMS-invariant way. The former contribution is indeed the one computed in the canonical gauge, while the latter can also be given in gauge invariant form. 
In the approach of \cite{Veneziano:2022zwh}, \cite{Javadinezhad:2022ldc} it corresponds to a different gauge fixing of $J_B$, while in \cite{Chen:2021szm}, \cite{Mao:2023evc} and \cite{Riva:2023xxm}  the  mechanical angular momentum loss is identified with the difference between the final and initial Bondi angular momentum computed in {\em two different} Bondi gauges. In formulae:
\begin{equation}
	\label{eq:Bondi}
\Delta J_\text{mech} = J_B^{(+)}(u \to + \infty)  - 	J_B^{(-)}(u \to - \infty)
	\,,
\end{equation}
where the upper label $(-)$  refer to the usual canonical gauge choice at $u \to - \infty$,  while the label $(+)$ is defined by setting the final shear  to zero (because of the memory effect this actually means that at $u \to + \infty$ we are no longer in the canonical gauge). By adding and subtracting a term $J_B^{(-)}(u \to +\infty)$ one ends up with:
\begin{equation}
	\label{eq:Bondi1}
	\begin{aligned}
\Delta J_\text{mech} &= 	
\left(J_B^{(-)}(u \to +\infty) - J_B^{(-)}(u \to - \infty)\right) + \left(J_B^{(+)}(u \to + \infty)  - J_B^{(-)}(u \to +\infty)\right) \\
&
\equiv \Delta J_\text{mech}^\text{rad} + \Delta J_\text{mech}^\text{stat}
	\,.
	\end{aligned}
\end{equation}
Eq. \eqref{eq:Bondi1} makes it clear that the first term is insensitive to a $u$-independent (i.e. strictly zero-frequency) contribution to $J_B^{(-)}$ while the second term, being evaluated after radiation has stopped, is the difference of two constants and is proportional to the final shear once the initial shear is set to zero   \cite{Riva:2023xxm}. Such a difference is nothing but the well known memory and corresponds to the zero-frequency limit of the News tensor's Fourier transform. In \cite{Riva:2023xxm} the two prescriptions have been shown to agree at ${\cal O}(G^2)$.

If the decomposition \eqref{eq:Bondi1} can be confirmed at all orders in $G$, it would establish a clear and strict connection between the above BMS-based approach and the amplitude-based one discussed in this section where we distinguished strictly zero-frequency gravitons from those with any non-vanishing frequency.
 In \cite{Riva:2023xxm} it has been explicitly checked that the definition \eqref{eq:Bondi1} of the static contribution to the angular momentum loss is in full agreement with the one computed at ${\cal O}(G^3)$ in \cite{Manohar:2022dea, DiVecchia:2022owy} and discussed hereafter.
Perhaps another interesting remark before turning to the calculation is that the same technique to compute the static angular momentum loss applies also in the case of a massless scalar field (see \eqref{Jradscalar} below), for which no BMS ambiguities arise.

To discuss the angular momentum of the gravitational field, we take the expectation in the $\mathcal{S}_{s.r.}|\Psi_{\text{in}}\rangle$ state of the corresponding operator in De Donder gauge, which reads as follows
\begin{equation}\label{Jalphabetagrav}
	J_{\alpha\beta} = -i\int_k a_{\mu\nu}^\dagger(k) 
	\Big(
	D^{\mu\nu,\rho\sigma} k^{\phantom{[}}_{[\alpha} \frac{\overset{\leftrightarrow}{\partial}}{\partial k^{\beta]}}
	+
	2 \eta^{\mu\rho}\delta_{[\alpha}^\nu \delta^{\sigma}_{\beta]}
	\Big)
	a_{\rho\sigma}(k)\,,
\end{equation}
where 
\begin{equation}\label{}
	D^{\mu\nu,\rho\sigma} = \frac{1}{2}\left(
	\eta^{\mu\rho}\eta^{\nu\sigma} + \eta^{\mu\sigma}\eta^{\nu\rho} - \eta^{\mu\nu}\eta^{\rho\sigma} \right)
\end{equation}
is the tensor structure appearing in the gauge-fixed De Donder action (see \ref{appFeynRules}).
At this stage we face the problem of simplifying the intermediate state sum using the physical projector \eqref{TTprojector}. In particular, we would like to see how the dependence on the reference vector $\lambda^\alpha$ appearing in that projector eventually cancels out.
Relying on the transversality property $k_\mu F^{\mu\nu}=0$, we obtain the following result, expressed as a sum of orbital and spin contributions,
\begin{equation}\label{}
	\mathcal J_{\alpha\beta}
	=
	\mathcal L_{\alpha\beta} 
	+
	\mathcal S_{\alpha\beta}\,.
\end{equation}
Here the orbital angular momentum reads (with the soft $F^{\mu\nu}$ as in \eqref{softfactorF})
\begin{equation}\label{}
	i\mathcal L_{\alpha\beta} =\int_{k}^\ast
	\left(
	F^\ast_{\mu\nu} k^{\phantom{I}}_{[\alpha}\frac{\overset{\leftrightarrow}{\partial} F^{\mu\nu} }{\partial k^{\beta]}} 
	- \frac{F'^\ast}{D-2}
	k^{\phantom{I}}_{[\alpha}\frac{\overset{\leftrightarrow}{\partial}  F'}{\partial k^{\beta]}}
	+2 k^{\phantom{\ast}}_{[\alpha} F^\ast_{\beta]\mu}\lambda\cdot F^\mu 
	- 2 \lambda\cdot F^{\ast}_\mu k^{\phantom{\ast}}_{[\alpha} F_{\beta]}^\mu
	\right),
\end{equation}
with $\lambda\cdot F^\alpha=\lambda_\beta F^{\alpha\beta}$, $F' = \eta_{\rho\sigma} F^{\rho\sigma}$ and
\begin{equation}
	f  \frac{ \overset{\leftrightarrow}{\partial} }{\partial k^{\beta}}g \equiv \frac{1}{2}\left( f \frac{\partial g}{\partial k^\beta} - 
	\frac{\partial f}{\partial k^\beta}\,g\right),
	\label{doppiafreccia}
\end{equation}
while the spin angular momentum is given by
\begin{equation}\label{}
	i\mathcal S_{\alpha\beta} = \int_{k}^\ast
	\left(
	2F^{\ast\phantom{\mu}}_{\mu[\alpha}F^\mu_{\beta]}  
	+  2k^{\phantom{\mu}}_{[\beta}F^{\ast\phantom{\mu}}_{\alpha]\mu}\lambda\cdot F^\mu
	+
	2\lambda\cdot F^\ast_{\mu}k^{\phantom{\mu}}_{[\alpha} F^\mu_{\beta]}
	\right).
\end{equation}
Therefore, the dependence on the reference vector $\lambda^\alpha$ drops out in the sum and we obtain the simple expression
\cite{Manohar:2022dea,DiVecchia:2022owy}
\begin{equation}\label{JIJgraviton}
	\mathcal J_{\alpha\beta} = -i\int_k^\ast
	F^\ast_{\mu\nu} \left[
	\left(
	\eta^{\mu\rho}\eta^{\nu\sigma}-\tfrac1{D-2}\,\eta^{\mu\nu}\eta^{\rho\sigma}
	\right)
	k^{\phantom{I}}_{[\alpha}\frac{\overset{\leftrightarrow}{\partial}}{\partial k^{\beta]}}
	+2\eta^{\mu\rho}\delta^{\nu}_{[\alpha}\delta^{\sigma}_{\beta]}
	\right]
	F_{\rho\sigma}\,.
\end{equation}
The two terms within square brackets are reminiscent of the orbital and spin part of the gravitational angular momentum, in particular the factor of 2 is associated to dealing with a spin-2 field.
However, the two terms are not separately gauge invariant, as one can easily check, pointing to the fact that only their sum can be given a well defined physical meaning. Indeed, in order to obtain \eqref{JIJgraviton} where the intermediate transverse-traceless projections have dropped out, one needs to rely on nontrivial cancellations between the orbital and the spin term (we refer to \cite{DiVecchia:2022owy} for further details on this derivation).
The integrals appearing in \eqref{JIJgraviton} can be reduced to \eqref{inm} by first combining pieces with ``positive and negative frequencies'' like we did for \eqref{eq:tildedelta}.   
In this way we can write \eqref{JIJgraviton} 
\begin{equation}
	\mathcal J^{\alpha\beta}  = \frac{i\kappa^2}{2} \sum_{n,m} \left( \sigma_{nm}^2 - \tfrac{1}{2} \right) m_n^2 m_m^2 K^{[\alpha} p_m^{\beta]}- i \kappa^2 \sum_{n,m} (p_n p_m) p_n^{[\alpha} p_m^{\beta]} (p_m \cdot K) 
	\label{Jalphabeta}
\end{equation}
in terms of  the integral 
\begin{equation}
	K^\mu = \int \frac{d^4 k}{(2\pi)^4}
	\frac{ \operatorname{sgn}(k^0)  2\pi \delta (k^2) k^\mu \theta(\omega^\ast - |k^0|)}{(p_n k + i 0)(p_m k -i0)^2} =A p_n^\mu + B p_m^\mu\,.
	\label{Jalphabeta1}
\end{equation}
The integral  $(p_m K)$ can be expressed in terms of $I_{nm}$ in \eqref{inmfinal}
\begin{equation}
	(p_m K)=   \frac{  I_{nm}}{m_n m_m }=  \frac{i \pi}{2} 
	\frac{\eta_n -\eta_m}{ m_n m_m}   \frac{\eta_n \eta_m\operatorname{arccosh} \sigma_{nm}}{(2\pi)^2 \sqrt{\sigma_{nm}^2 -1}}\,,
	\label{Jalphabeta3}
\end{equation}
while $(p_nK)$ can be easily computed   
\begin{equation}
	(p_n K) =  \int \frac{d^4 k}{(2\pi)^4}
	\frac{ \operatorname{sgn}(k^0)  2\pi \delta (k^2)  \theta(\omega^\ast - |k^0|)}{(p_m k -i0)^2} = - \frac{i\pi \eta_m}{ (2\pi)^2 m_m^2}
	\label{Jalphabeta4}
\end{equation}
from the product of the two integrals
\begin{equation}
	\int_{-\omega^\ast}^{\omega^\ast} \frac{\omega d\omega}{(-\eta_m \omega-i 0)^2}= - i \pi \eta_m\,,\qquad
	\frac{1}{(2\pi)^2} \int_0^\pi \frac{\sin \theta d \theta}{(E_m - |{\vec{p}}_n | \cos \theta)^2} = \frac{1}{(2\pi)^2 m_m^2}\,.
	\label{Jalphabeta5}
\end{equation}
As a result the coefficient $A$ in \eqref{Jalphabeta1} is given by
\begin{equation}
	A= \frac{m_m^2 (p_n K) + (p_n p_m) p_m K}{(p_n p_m)^2 - m_n^2 m_m^2}=
	-\frac{i \pi }{2} \frac{  \eta_m+ \eta_n  + (\eta_n-\eta_m) \left(\sigma_{nm} \Delta_{nm} -1\right)}{(2\pi)^2 m_n^2 m_m^2 (\sigma_{nm}^2-1)} 
	\label{Jalphabeta2}
\end{equation}
with $\Delta_{nm}$ as in \eqref{cnmgraviton} below, $B$ is not needed because it will not contribute when inserted in the first term of \eqref{Jalphabeta}, by antisymmetry.  For the same reason, also the term with $\eta_n +\eta_m$ in \eqref{Jalphabeta2} does not contribute,
and we find the following expression for the angular momentum carried by the static gravitational field in $D=4$, after sending $\omega^\ast\to0$,
\begin{equation}\label{Jgraviton}
	\mathcal J^{\alpha\beta} = - \sum_{n\in\text{in}}\sum_{m\in\text{out}}
	c_{nm}\,p_n^{[\alpha}p_m^{\beta]}\,,
\end{equation}
with
\begin{equation}\label{cnmgraviton}
	c_{nm}
	=
	2G
	\left[
	\left(
	\sigma_{nm}^2-\frac12
	\right)
	\frac{\sigma_{nm}\Delta_{nm}-1}{\sigma_{nm}^2-1} 
	-2\sigma_{nm}\Delta_{nm}
	\right],\qquad
	\Delta_{nm}
	=
	\frac{\operatorname{arccosh}\sigma_{nm}}{\sqrt{\sigma^2_{nm}-1}}\,.
\end{equation}
The expression \eqref{Jgraviton} is Lorentz covariant, translation invariant and valid for arbitrary kinematics $p_n$ of the background hard process, i.e.~it holds regardless whether or not the outgoing momenta can be regarded as small deflections of the incoming ones. In fact, just like Weinberg's theorem, the formula holds independently of the number and of the specific details of the hard particles taking part in the background hard process, which may also carry spin  or be subject to tidal deformations: one need only assign their momenta. However, of course, it only captures the contribution to the angular momentum due to static/zero-frequency fields.

For a $2\to2$ process, $p_4^\mu= Q^\mu-p_1^\mu$, $p_3^\mu=-Q^\mu-p_2^\mu$ up to $\mathcal O(G^3)$ corrections, and thus
\begin{equation}\label{accurate}
	\mathcal{J}^{\alpha\beta} = 
	p_1^{[\alpha}Q^{\beta]} 
	\left(
	c_{13}-c_{14}
	\right)
	-
	p_2^{[\alpha}Q^{\beta]}
	\left(
	c_{24}-c_{23}
	\right)
	+
	\mathcal O(G^4)\,.
\end{equation}
where we have used that $\sigma_{13}= \sigma_{24}$.
In the PM expansion, using $c_{14}-c_{13} = \frac{G}2 \mathcal I(\sigma) + \mathcal O(G^3)$, $c_{23}-c_{24}=\frac{G}2 \mathcal I(\sigma) + \mathcal O(G^3)$ with
\begin{equation}\label{Isigma}
	\frac{1}{2} \mathcal{I}(\sigma)=\frac{8-5 \sigma^2}{3\left(\sigma^2-1\right)}+\frac{\sigma\left(2 \sigma^2-3\right) \operatorname{arccosh} \sigma}{\left(\sigma^2-1\right)^{3/2}}\,,
\end{equation}
we find
\begin{equation}\label{J2PMgen}
	\mathcal{J}^{\alpha\beta} = 
	-\frac{G}2
	(p_1-p_2)^{[\alpha}_{\phantom{P}}Q^{\beta]} 
	\mathcal I(\sigma)
	+
	\mathcal O(G^4)\,.
\end{equation}
In particular, substituting  \eqref{1PM} for the 1PM impulse and using
\begin{equation}
Q^\mu =- \frac{b^\mu}{b}Q_\text{1PM} + \mathcal O(G^3)\, ,
\label{Qmu}
\end{equation}
we get
\begin{equation}
\mathcal{J}^{\alpha\beta} =  \frac{G^2m_1m_2 (2\sigma^2-1) }{b^2 \sqrt{\sigma^2-1}^{\frac{1}{2}}} \mathcal I(\sigma) 
(p_1-p_2)^{[\alpha}_{\phantom{P}}b^{\beta]}
+
\mathcal O(G^3)\,,
\label{J2}
\end{equation}
which, in the center-of-mass frame where $-p_1 = (E_1, p)$ and $-p_2= (E_2, -p)$,
matches the $\mathcal O(G^2)$ results in Eq.~(4.6) of \cite{Damour:2020tta} and  in Eq.~(12) of \cite{Manohar:2022dea}.
However, as already emphasized, thanks to the universality of the soft theorem \eqref{J2PMgen} is also valid when the particles carry classical spin \cite{Alessio:2022kwv}, provided one substitutes the appropriate 1PM deflection.
Taking into account the 2PM correction to the deflection \eqref{delta1a} on top of that, eq.~\eqref{J2PMgen} also correctly accounts for the $\mathcal O(G^3)$ static angular momentum, as well tidal effects \cite{Heissenberg:2022tsn}. However, to that order, additional genuinely radiative terms also appear and we shall see in the next sections how they can be calculated in the eikonal framework.

\subsection{The ultrarelativistic and massless limits}
\label{sec:eikopsoftur}

It is very instructive to inspect the behavior of the spectrum \eqref{eq:d2gr} in several physically relevant limits. 
The standard PM regime requires that 
$Q^2 \sim (p \Theta)^2 \ll 2m_i^2$.
In this limit, one can extract the leading (3PM) contribution by Taylor-expanding the first line of~\eqref{eq:d2gr} in $Q^2$, while the second only give subleading contributions,
\begin{equation}
	\label{eq:standpm}
	\lim_{\omega \to0}\frac{dE}{d\omega} \simeq  \frac{2G }{\pi} Q^2
	\left[
	\frac{8-5\sigma^2}{3(\sigma^2-1)}
	+
	\frac{(2\sigma^2-3)\sigma\operatorname{arccosh}\sigma}{(\sigma^2-1)^{3/2}}
	\right].
\end{equation}
When rewritten in terms of the divergent part of $\operatorname{Im}2\delta$, using the simple link \eqref{eq:ZFLspectrum}, the result \eqref{eq:standpm} then reproduces the divergent part of $\operatorname{Im}2\delta_2$ in \eqref{3.2} (and agrees with Eq.~(5.14) of~\cite{DiVecchia:2021ndb}), once we use $Q\simeq Q_\text{1PM}$ (see \eqref{1PM} for the explicit expression).
Moreover, to the leading 3PM order, Eq.~\eqref{eq:standpm} has been explicitly shown to hold also if the colliding objects  carry spin, for generic spin alignments \cite{Alessio:2022kwv}. 
If we then consider in the formal ultra-relativistic limit $\sigma\to\infty$,  Eq.~\eqref{eq:d2gr} gives:
\begin{equation}
	\label{eq:standpmur}
	\lim_{\omega \to0}\frac{dE}{d\omega}  \simeq  \frac{4G }{\pi} Q^2  \left( \log\frac{s}{m_1 m_2} - \frac56 \right).
\end{equation}
However, the exact expression \eqref{eq:d2gr} shows that the PM approximation breaks down, even when $\Theta$ is small, when one of the two particles becomes ultra-relativistic, $p \gg m_i$. Indeed, while in this regime the first line in~\eqref{eq:d2gr}  can always be expanded to first order in $Q^2/s$, the second one presents instead a branch point at $Q^2 = -4m_i^2$, corresponding to the $t$-channel thresholds (outside the physical region). This implies that the PM expansion starts to diverge whenever for at least one index $i$, (see Eq.~\eqref{fzetaSERIES} and comments below it)
\begin{equation}
	\label{eq:KTbex}
	\frac{Q}{2 m_i} > 1\,.
\end{equation}
Since $Q\simeq p\,\Theta$ which  is of order $\sqrt{m_1m_2\sigma/2}\,\Theta$, we thus recover the bound first pointed out by D'Eath \cite{DEath:1976bbo} for the validity of the PM approximation,
\begin{equation}\label{}
	\operatorname{max}\left\{\sqrt{\frac{m_1}{m_2}},\sqrt{\frac{m_2}{m_1}}\right\}\,\frac{\sqrt{\sigma}\,\Theta}{2\sqrt 2} \lesssim 1\,.
\end{equation}
In view of this bound one could say that the PM approximation holds for weakly gravitating systems with ``generic'' but not entirely arbitrary speeds.

For 
the ZFL of the emitted energy spectrum, however, the ultrarelativistic regime, or equivalently the massless limit, can be obtained from \eqref{eq:d2gr} by considering $2p \to \sqrt{s}$ and $m_1, m_2\ll  Q = \sqrt{s} \sin\frac{\Theta}{2}$. The mass singularities appearing separately in each line of the formula neatly cancel against each other and then~\eqref{eq:d2gr} reduces to (see e.g.~\cite{Addazi:2019mjh} where the result is extended to an arbitrary number of external massless legs) 
\begin{equation}\label{eq:urd2}
	\lim_{\omega \to0}\frac{dE}{d\omega} \simeq  \frac{4 G }{\pi}	\left[
	s \log \frac{s}{s-Q^2}
	+
	Q^2\log\frac{s-Q^2}{Q^2}
	\right]
\end{equation}
or in terms of the deflection angle
\begin{equation}\label{yr}
	\lim_{\omega \to0}\frac{dE}{d\omega} \simeq - \frac{4 G }{\pi}s \left[\cos^2 \frac{\Theta}{2} \log \cos^2 \frac{\Theta}{2} + \sin^2 \frac{\Theta}{2} \log \sin^2 \frac{\Theta}{2} \right],
\end{equation}
which agrees with the leading soft limit of Eq.~(5.12) of \cite{Sahoo:2021ctw}. If we then consider the small $\Theta$ limit of~\eqref{yr}, after taking the ultrarelativistic limit, at leading order for $\Theta\ll 1$ we have
\begin{equation}
	\label{eq:lm0}
	\lim_{\omega \to0}\frac{dE}{d\omega}\simeq \frac{G s \Theta^2}{\pi} \left[1 + \log\frac{4}{\Theta^2}
	\right],
\end{equation}
which reproduces the result obtained in~\cite{Gruzinov:2014moa} (see also \cite{DEath:1976bbo}) within a classical GR approach and in~\cite{Ciafaloni:2018uwe} from a scattering amplitudes perspective. 
As anticipated, the true ultrarelativistic behavior of the spectrum exhibits a non-analytic dependence on $\Theta$ and thus on the Newton constant $G$, compatibly with the breakdown of the PM approximation.
These features are in fact shared by the leading-order angular momentum loss. For the component orthogonal to the scattering plane, one finds in the ultrarelativistic limit
\begin{equation}\label{}
	\mathcal J^{xy}
	\sim
	2Gs\,\sin\Theta\,\log\frac{\cos\frac{\Theta}{2}}{\sin\frac{\Theta}{2}}\,,
\end{equation}
which when further expanded for small $\Theta$ yields
\begin{equation}\label{Jhigh}
	\mathcal J^{xy}
	\sim
	Gs\,\Theta\,\log\frac{4}{\Theta^2}\,,
\end{equation}
again exhibiting the characteristic non-analyticity in the Newton constant.

\subsection{The \texorpdfstring{$\mathcal{N}=8$}{N=8} eikonal operator in the soft limit }
\label{sec:eikopsoftn8}

To include the presence of other massless fields (scalars and vectors), which are needed to discuss the case of $\mathcal{N}=8$ supergravity where the massive particles are described by KK modes, we can use the fact that $S$-matrix elements for soft emissions factorize in a way analogous to \eqref{eq:SMNSM}, with soft factors that instead of $w_j(k)$ are given by
\begin{equation}\label{wvecfsc}
	w_j^\text{vec}(k)=\sum_{n} \eta_n e_n \frac{\varepsilon^\ast_{\mu,j}(k) p_n^\mu}{p_n\cdot k},\qquad
	w^\text{sc}_j(k) = \sum_{n} \frac{g_n}{p_n\cdot k}
\end{equation}
for vectors and scalars respectively. As above, $\eta_n$ takes the value $+1$ for outgoing and $-1$ for incoming states, while $e_n$ and $g_n$ denote suitable couplings.
These new soft particles are easily accommodated in the eikonal operator by including in the exponent of~\eqref{eq:eiksr} the relevant operators $a_d$ for the dilaton (with coupling $g_n = -\kappa\,m_n^2/\sqrt{D-2}$)  and $a_{v_i,j}$ for two vectors ($e_n=\sqrt 2\,\kappa\, m_i$) and $a_{s_i}$ for two scalars ($g_n = \kappa\, m_i^2$). 
The corresponding ``memory waveforms'' are also obtained in a very similar way. 

In analogy with \eqref{Imdelta}, we find the following simple and general expressions for the energy emission spectra, 
\begin{align}\label{}
	\lim_{\omega \to0}\frac{dE^{\text{vec}}}{d\omega} &= \frac{1}{4 \pi^2} \sum_{n,m} e_n e_m (-\sigma_{nm}) F_{nm}\,,
	\label{Imdeltavec}
	\\
	\lim_{\omega \to0}\frac{dE^{\text{sc}}}{d\omega} &= \frac{1}{4 \pi^2}   \sum_{n,m} \frac{g_n g_m}{m_nm_m} F_{nm}\,.
	\label{Imdeltasc}
\end{align} 
Combining scalars, vectors and the graviton, we obtain a remarkably simple result for the $\mathcal{N}=8$ spectrum,
\begin{equation}\label{N8m}
	\lim_{\omega \to 0}\frac{dE^{\mathcal N=8}}{d\omega}
	=
	\frac{2G }{\pi}
	\sum_{n,m}
	m_n m_m (\sigma'_{nm})^2 F_{nm}\,,
\end{equation} 
where $\sigma_{nm}'=\sigma_{nm}-1$ if $n$ and $m$ have momenta compactified along the same KK direction (so that $m_n=m_m$) and  $\sigma_{nm}'=\sigma_{nm}$ otherwise.  
Specializing \eqref{N8m} to the $2\to2$ kinematics as in \eqref{eq:sigmamn4p}, we have
\begin{align}
	\label{eq:d2n8}
	\lim_{\omega \to0}\frac{dE^{{\cal N}=8}}{d\omega} &=  \frac{4 G }{\pi} 
	\Bigg[
	2m_1m_2 \sigma^2\, \frac{\operatorname{arccosh}\sigma}{\sqrt{\sigma^2-1}}
	-
	2m_1m_2 \sigma_Q^2\, \frac{\operatorname{arccosh}\sigma_Q}{\sqrt{\sigma_Q^2-1}}
	\\ \nonumber
	&-
	\frac{(Q^2)^2}{4 m_1^2}\,
	\frac{\operatorname{arccosh}\left(
		1+\tfrac{Q^2}{2m_1^2}
		\right)}{\sqrt{\left(
			1+\tfrac{Q^2}{2m_1^2}
			\right)^2-1}}
	-
	\frac{(Q^2)^2}{4 m_2^2}\,
	\frac{\operatorname{arccosh}\left(
		1+\tfrac{Q^2}{2m_2^2}
		\right)}{\sqrt{\left(
			1+\tfrac{Q^2}{2m_2^2}
			\right)^2-1}}
	\Bigg].
\end{align}
The standard relativistic PM regime, where $Q^2\ll 2m_i^2$ in the equation above, has an analytic PM expansion whose leading term reads
\begin{equation}
	\label{eq:standpmn8}
	\lim_{\omega \to0}\frac{dE^{{\cal N}=8}}{d\omega} \simeq \frac{4G\, Q^2}{\pi} 
	\left[
	\frac{\sigma^2}{\sigma^2-1}
	+
	\frac{(\sigma^2-2)\sigma\operatorname{arccosh}\sigma}{(\sigma^2-1)^{3/2}}
	\right].
\end{equation}
Approximating  $Q \simeq  p \Theta$ with $\Theta$ the leading deflection angle given in~\eqref{1PMN=8}, and exploiting the link \eqref{eq:ZFLspectrum}, this reproduces the divergent part of $\operatorname{Im}2\delta_2$ in \eqref{3.6}.

On the contrary, in the regime~\eqref{eq:KTbex}, the small-$\Theta$ expansion is non-analytic and interestingly, in extreme ultrarelativistic kinematics where the masses can be neglected, one obtains again~\eqref{eq:urd2}. Indeed, the contributions related to the dilaton, and the Kaluza--Klein scalars and vectors become negligible in this regime, as suggested by the fact that \eqref{Imdeltavec}, \eqref{Imdeltasc} scale with lower powers of $\sigma_{nm}$ compared to \eqref{Imdelta}, and the graviton provides the dominant behavior.
At ultra-high energies, in this way, only the contribution due to the emission of gravitons survives. This is a universal expression for two-derivatives theories in accordance with the expectation that gravity dominates the high-energy limit not just for the elastic scattering, as argued in~\cite{DiVecchia:2020ymx}, but also in the (soft) radiation sector. 

The inclusion of static modes for vector and scalar fields proceeds along similar lines as for the graviton case discussed above by means of the $-i0$ prescription, i.e.~by introducing the modified soft factors
\begin{equation}\label{fvecfsc}
	f_j^\text{vec}(k)=\sum_{n} \eta_n e_n \frac{\varepsilon^\ast_{\mu,j}(k)\, p_n^\mu}{p_n\cdot k-i0},\qquad
	f^\text{sc}(k) = \sum_{n} \frac{g_n}{p_n\cdot k-i0}
\end{equation}
using which one then finds the following vector and scalar waveforms,
\begin{equation}\label{}
	\langle A_\mu(x) \rangle \sim 
	\frac{1}{4\pi r} \Pi^{\mu}_{\nu}(\hat x)\,\sum_{n} \frac{e_n\,p_n^\nu\, \theta(\eta_n u)}{E_n-\vec k_n\cdot\hat x}\,,
	\quad
	\langle \Phi(x) \rangle \sim 
	\frac{1}{4\pi r} \sum_{n} \frac{g_n\, \theta(\eta_n u)}{E_n-\vec k_n\cdot\hat x}
\end{equation}
where $\Pi^{\mu\nu}$ is the transverse projector.
For the angular momenta one finds
\begin{equation}\label{}
	\mathcal J^\text{sc}_{\alpha\beta} 
	=
	-
	\frac{i}2
	\int_{\vec k} \left(f^\ast k_{[\alpha} \frac{\partial f}{\partial k^{\beta]}} - k_{[\alpha} \frac{\partial f^\ast}{\partial k^{\beta]}} f \right)
\end{equation}
and 
\begin{equation}\label{Jradscalar}
	(\mathcal J^{\text{sc}})^{\mu\nu} = \frac{1}{16\pi} \sum_{n,m}\frac{g_ng_m}{m_n^2m_m^2} \frac{\sigma_{nm}\Delta_{nm}-1}{\sigma_{nm}^2-1} (\eta_n-\eta_m)p_n^{[\mu}p_m^{\nu]}\,,
\end{equation}
for the scalar case, as well as
\begin{align}\label{JIJvecsc}
	\mathcal J^\text{vec}_{\alpha\beta} = - i \int_{\vec k}
	F^\ast_{\mu} \Big(
	\eta^{\mu\nu}
	k^{\phantom{I}}_{[\alpha}\frac{\overset{\leftrightarrow}{\partial}}{\partial k^{\beta]}}
	+\delta^{\mu}_{[\alpha}\delta^{\nu\phantom{\mu}}_{\beta]}
	\Big)
	F_{\nu}
\end{align} 
which evaluates to
\begin{align}\label{Jradvector}
	(\mathcal J^\text{vec})^{\alpha\beta}
	=
	\frac1{16\pi}
	\sum_{n,m}
	\frac{e_n e_m}{m_n m_m}\left[ -
	\sigma_{nm}\, \frac{\sigma_{nm}\Delta_{nm}-1}{\sigma_{nm}^2-1}
	+ \Delta_{nm}
	\right]
	(\eta_n-\eta_m)\,p_n^{[\alpha}p_m^{\beta]}
\end{align}
for the vector.
Combining this with the graviton result discussed above, one arrives at the following simple formula for the static angular momentum loss in $\mathcal{N}=8$,
\begin{equation}\label{JN8}
	\mathcal J^{\alpha\beta}_{\mathcal N=8}
	=
	\frac{G}{2}
	\sum_{n,m}
	\left[
	\sigma_{nm}'^{\,2} \frac{\sigma_{nm}\Delta_{nm}-1}{\sigma_{nm}^2-1}
	-2 \sigma'_{nm} \Delta_{nm}
	\right](\eta_n-\eta_m)p_n^{[\alpha}p_m^{\beta]}\,,
\end{equation}
where $\sigma'_{nm}=\sigma_{nm}-1$ if $n$ and $m$ corresponds to states compactified along the same direction, so that $m_n = m_m$, while it equals $\sigma'_{nm} = \sigma_{nm}$ otherwise.

\section{The eikonal operator beyond the soft limit}
\label{SemiclEik}

In this section, we discuss a framework that allows us to combine the dynamical information about graviton exchanges that is contained in the elastic eikonal, which, as we have seen, determines the  deflection of the colliding black holes, with the information about graviton emissions that is contained in the inelastic amplitudes. Building on the idea of introducing creation/annihilation operators for gravitons already introduced in the previous section, we construct here an eikonal operator that dictates the final state of the collision out of two ingredients: the eikonal phase and coherent graviton emissions with generic, i.e.~not necessarily soft, frequencies.
As we shall see, this provides a comprehensive strategy to calculate observable quantities associated to the scattering up to $\mathcal O(G^3)$: the waveforms themselves, the emitted linear and angular momentum, and the changes in the  linear and angular momentum of the colliding objects. In particular, this will allow us to explicitly check the corresponding balance laws.

\subsection{The elastic eikonal revisited}
\label{sec:eikwp}

So far the external states involved in the scattering have been described simply in terms of momentum eigenstates. Following~\cite{Kosower:2018adc}, we now introduce wavepackets $\Phi_i$ to describe classical particles, so the initial state for the $2\to 2$ scattering reads
\begin{equation}
	| \psi \rangle = \int_{-p_1}\int_{-p_2}
	\Phi (-p_1) \Phi (-p_2)  e^{i p_1 b_1 + ip_2 b_2} |- p_1 , - p_2 \rangle\;.
	\label{DO1}
\end{equation}
The wavepackets are peaked around the classical value (that with an abuse of notation we will still indicate with $-p_i$) and we use the following notation for the on-shell integrals
\begin{equation}
	\label{eq:onsint}
	\int_{p_{i}} = \int \frac{d^Dp_{i}}{(2\pi)^D}\,2\pi\theta(p_i^0)\delta(p_i^2+m_i^2)\;.
\end{equation}
By following~\cite{Cristofoli:2021jas}, we can formally write the final state $S_c |\psi\rangle =(1+i T) |\psi\rangle$ in the elastic case (the subscript $c$ stands for ``conservative'') as
\begin{equation}
	S_c |\psi \rangle = \int_{-p_1} \int_{-p_2} \int_{p_3}\int_{p_4}
	\Phi (-p_1) \Phi (-p_2)  e^{i p_1 b_1 + ip_2 b_2} |p_3, p_4 \rangle \langle p_3 , p_4 | S |- p_1 , - p_2 \rangle \;.
	\label{DO3}
\end{equation}
Of course we are interested only in the classical contribution to the $S$-matrix element which, as discussed in the previous sections, is more easily captured in an impact parameter representation, see~\eqref{fullampli}. Thus we write the momentum-space $S$-matrix element
\begin{equation}
	\langle p_3 , p_4 | i T | -p_1 , - p_2 \rangle  =(2\pi)^{D}\!\! \int d^D{Q}\,\delta^{(D)} (p_1+p_4 -{Q}) \delta^{(D)} (p_2+p_3 +{Q}) \, i {\cal A} (-(p_1+p_2)^2, {Q}^2)
	\label{DO4}
\end{equation}
as the inverse Fourier transform of the eikonal result
\begin{equation}
	2\pi \delta (2 {\bar{p}}_2 {Q}) \,2\pi\delta (2 {\bar{p}}_1 {Q}) \, i {\cal A} (-(p_1+p_2)^2, {Q}^2) = \int \!\! d^D {x}\, \left[(1+ 2i \Delta( {b})) \left( e^{2i \delta ( {b})}-1 \right)\right] e^{-i x {Q}} \;.
	\label{DO6}
\end{equation}
In this subsection, for ease of notation, we simply drop the imaginary part of $2\delta$.
Let us discuss the quantities ${b}$ and $\bar{p}_i$ in this relation. The momenta are simply defined as in \eqref{eq:barpi}, with $q^\mu$ replaced by $Q^\mu$,
\begin{equation}
	p_1 = -{\bar{p}}_1 + \frac{{Q}}{2}\,, \qquad p_2 =- {\bar{p}}_2 - \frac{{Q}}{2}\,,\qquad  
	p_4 = {\bar{p}}_1 + \frac{{Q}}{2}\,,\qquad p_3 ={\bar{p}}_2  - \frac{{Q}}{2}\,,
	\label{DO5}
\end{equation}
so the delta functions on the l.h.s. of~\eqref{DO6} follow from the fact that the $p_i$ are on-shell. Then in order to produce these delta functions from the r.h.s., the eikonal in the square parenthesis should not depend on the components of ${x}$ along $\bar{p}_{i}$, thus we introduced ${b}$ to indicate the components of ${x}$ orthogonal to $\bar{p}_i$ recovering the property ${b} {\bar{p}}_{1,2}=0$ of the eikonal impact parameter.
By using~\eqref{DO4} and~\eqref{DO6} in \eqref{DO3} we get
\begin{align}
	S_c |\psi \rangle 
	=& \int_{p_3} \int_{p_4} |p_3, p_4\rangle e^{-i b_1 p_4} e^{-ib_2 p_3}
	\int \frac{d^D {Q}}{(2\pi)^D} \int d^D {x} \nonumber \\
	\times &e^{ i {Q}(b_1-b_2)}e^{2i \delta ( {{b}})} e^{-i x {Q}} (1+2i \Delta({b})) 
	\Phi (p_4 -{Q}) \Phi ({Q}+p_3) \;.
	\label{DO10}
\end{align} 

\subsubsection{Saddle points}
The key idea of the eikonal approach is to approximate the integrals with 
the stationary phase contribution and this can be done in~\eqref{DO10} both for ${x}$ and ${Q}$~\cite{Cristofoli:2021jas}. From the condition on ${x}$ we obtain
\begin{equation}
	{Q}_\mu = \frac{\partial 2 \delta ({b})}{\partial x^\mu}=  \frac{\partial 2 \delta ({b})}{\partial b} \frac{{b}_\mu}{{b}}
	\label{DO11}
\end{equation}
where we used $\frac{\partial {b} }{\partial {{x}}^\mu}= \frac{{b}_\mu}{{b}}$. We thus recovered the second relation in Eq.~\eqref{bJbThetadelta}. In this approach, the relation between ${b}$ and the initial angular momentum follows from the other stationary phase condition
\begin{equation}
	(b_1 -b_2)^\mu - x^\mu 
	=- \frac{\partial 2  \delta ({b})}{\partial  {b}} \frac{b_{\nu}}{{b}}\frac{\partial {b}^\nu}{\partial {Q}^\mu}\;,
	\label{DO12}
\end{equation}
where the r.h.s. follows from the implicit dependence of ${b}$ on ${Q}$. In order to make this explicit, we need to decompose $x$ along $b$ and the space spanned by $\bar{p}_i$ 
\begin{equation}
	x^\mu =  {b}^\mu + ({\bar{p}}_1 +{\bar{p}}_2)^\mu A_1 +({\bar{p}}_1 -{\bar{p}}_2)^\mu  A_2 = {b}^\mu + (p_4 + p_3)^\mu A_1 +(p_4 -p_3-Q)^\mu  A_2\;,
	\label{DO13}
\end{equation}
which implies
\begin{equation}
	\frac{\partial {b}^\nu}{\partial Q^\mu} =-(\bar p_1+\bar p_2)^\nu\frac{\partial A_1}{\partial Q^\mu}-(\bar p_1-\bar p_2)^\nu\frac{\partial A_2}{\partial Q^\mu}+ \delta^\nu_\mu A_2  \;,
	\label{DO15}
\end{equation}
since the derivatives involved in the stationary point conditions are calculated by keeping $x$, $p_{3,4}$ fixed. By using \eqref{DO5}, \eqref{DO13} and~\eqref{DO15} in~\eqref{DO12}, and noting that only  the last term in \eqref{DO15} gives a nonvanishing contribution to it, we get 
\begin{equation}
	b^\mu_J \equiv (b_1-b_2)^\mu =  {b}^\mu  - (p_1 + p_2)^\mu A_1 - (p_1 -p_2)^\mu  A_2\,.
	\label{DO17}
\end{equation}
The classical values of $A_{1,2}$ can be determined by imposing that ${b} {\bar{p}}_{1,2}=0$ and $b_J {p}_{1,2}=0$ and by using~\eqref{DO5} in~\eqref{DO17}. We obtain 
\begin{equation}
	A_1 =\frac{ (m_1^2 -m_2^2) |Q| {b}}{4m_1^2 m_2^2 (\sigma^2-1)} ~,\quad A_2 = -\frac{ s |Q| {b}}{4m_1^2 m_2^2 (\sigma^2-1)} \;.
	\label{DO22}
\end{equation}
By contracting~\eqref{DO17} with $b_J$, we obtain $b_J^2=b_J b$. Contracting it instead with $b$, and recalling that $Q\cdot b=-|Q|b$, since the force is attractive for gravitational theories, we get
\begin{equation}
	b_J^2 = b^2 \left(1- \frac{s Q^2}{4m_1^2 m_2^2 (\sigma^2-1)} \right) = b^2  \cos^2 \frac{\Theta}{2} \;,
	\label{DO24}
\end{equation}
where in the final step we used the relation~\eqref{eq:thetap} between the value of $Q$ at the stationary point and the elastic scattering angle together with~\eqref{Ep}. We thus recover the first relation in Eq.~\eqref{bJbThetadelta} as well.

\subsubsection{Impulse and angular momentum}
\label{ssec:elasticimpulseandangularmomentum}

We now use the elastic eikonal operator \eqref{DO10} to calculate the impulse by taking expectations of the momentum operator for particle 1, 
\begin{equation}\label{}
	P_1^\alpha = \int_{p} p^\alpha a_1^\dagger(p)a_1(p)\,.
\end{equation}
Applying  this operator to the initial state \eqref{DO1}, we obtain
\begin{equation}\label{}
	P_1^\alpha|\psi\rangle
	=
	\int_{- p_1}\int_{- p_2}\Phi_1(-p_1)\Phi_2(-p_2)|-p_1,-p_2\rangle\, e^{ib_1\cdot p_1+ib_2\cdot p_2}(-p_1^\mu)
\end{equation}
after using that $|-p_1 \rangle = a_1^{\dagger} (-p_1) |0\rangle$
and therefore the expectation value,
\begin{equation}\label{}
	\langle \psi|P_1^\alpha|\psi\rangle= \int_{- p_1}|\Phi_1(-p_1)|^2\,(-p_1^\mu)\,.
\end{equation}
where we have used $\langle p| p' \rangle 2\pi \theta (p^0) \delta (p^2+m_1^2) =(2\pi)^D \delta^{(D)} (p -p')$ that follows from the canonical commutation relations.
For the final state, using \eqref{DO10}, we get
\begin{equation}\label{}
	\begin{split}
		P_1^\mu S_c|\psi\rangle
		&=
		\int_{ p_3}
		\int_{ p_4}
		e^{-ib_1\cdot p_4}\,\Phi_1(p_4-Q)
		e^{-ib_2\cdot p_3}\,\Phi_2(p_3+Q)\,p_4^\mu |p_3,p_4\rangle\\
		&\times
		\int\frac{d^DQ}{(2\pi)^D}\int d^Dx
		\,e^{i(b-x)\cdot Q+2i\delta(s,b)}
	\end{split}
\end{equation}
and for the expectation value calculated at the saddle point, 
\begin{equation}\label{}
	\langle \psi|S_c^\dagger P_1^\mu S_c|\psi\rangle
	=
	\int_{ p_3}
	\int_{ p_4}
	p_4^\mu
	\left|\Phi_1(p_4-Q)\right|^2
	\left|\Phi_2(p_3+Q)\right|^2.
\end{equation}
Formally performing the shifts $p_4^\mu=Q^\mu-p_1^\mu$, $p_3^\mu=-Q^\mu-p_2^\mu$,  we recover
\begin{equation}\label{}
	\langle \psi|S_c^\dagger P_1^\mu S_c|\psi\rangle
	-
	\langle \psi|P_1^\alpha|\psi\rangle
	=
	\int_{- p_1}
	\int_{- p_2}
	Q^\mu
	\left|\Phi_1(-p_1)\right|^2
	\left|\Phi_2(-p_2)\right|^2 \to Q^\mu.
\end{equation}
Of course for particle 2 we find
\begin{equation}\label{}
	\langle \psi|S_c^\dagger P_2^\mu S_c|\psi\rangle
	-
	\langle \psi|P_2^\alpha|\psi\rangle
	=
	- Q^\mu.
\end{equation}
The Jacobian for the change of variables considered in the last step, as well as the integral over the fluctuations around the saddle points, would require further study. We shall return to this point in the outlook section~\ref{sec:outloop}.

We will provide the analogous derivation for the angular momentum in the more complete eikonal framework discussed in the ensuing sections which will also include dissipative effects (see Eq.~\eqref{DeltaL1new} for the complete final result). However, let us isolate here the conservative portion of the answer corresponding in particular to~\eqref{DeltaL1new-cons} for particle 1, 
\begin{equation}\label{}
	\Delta L_{(1c)\alpha\beta} = b_{1[\alpha} Q_{\beta]} + p_{4[\alpha} \frac{\partial 2\delta(b)}{\partial p_4^{\beta]}}\,,\qquad
	\Delta L_{(2c)\alpha\beta} = - b_{2[\alpha} Q_{\beta]} + p_{3[\alpha} \frac{\partial 2\delta(b)}{\partial p_3^{\beta]}}\,.
\end{equation}
Focusing first on particle 1, we recall that $2\delta$ depends on $p_4^\mu$ both via $\sigma= -p_3\cdot p_4/(m_1m_2)$ and via the projection that relates $b_J^\mu$ to $b^\mu$ as in \eqref{DO13}. As a result, we find
\begin{equation}\label{L1cons}
	\Delta L_{(1c)\alpha\beta}
	=
	b_{1[\alpha}Q_{\beta]}\\
	+
	\left[
	p_{1[\alpha|}\,p_{2|\beta]}+
	\left(p_{1}+p_{2}\right)_{[\alpha} Q_{\beta]}
	\right]
	\frac{\partial 2\delta(s,b)}{\partial p_1\cdot p_2}+
	(A_1+A_2)
	p_{1[\alpha}\,Q_{\beta]}
\end{equation}
and similarly, for particle 2,
\begin{equation}\label{}
	\Delta L_{(2c)\alpha\beta}=
	-
	b_{2[\alpha}Q_{\beta]}
	-
	\left[p_{1[\alpha|}\,p_{2|\beta]}
	+
	\left(p_{1}+p_{2}\right)_{[\alpha} Q_{\beta]} \right]\frac{\partial 2\delta(s,b)}{\partial p_1\cdot p_2}
	+
	(A_1-A_2)
	p_{2[\alpha}\,Q_{\beta]}\,.
\end{equation}
As a result
\begin{equation}\label{L12cons}
	\Delta L_{(1c)\alpha\beta}
	+
	\Delta L_{(2c)\alpha\beta} = \left(b_J+A_1(p_1+p_2)+A_2(p_1-p_2)\right)_{[\alpha}Q_{\beta]}\,.
\end{equation}
Using \eqref{DO17}, we see that the combination between round brackets is precisely $b_\alpha$. Since $b^{[\alpha}Q^{\beta]}=0$ by \eqref{DO11}, we get
\begin{equation}\label{}
	\Delta L_{(1c)\alpha\beta}
	+
	\Delta L_{(2c)\alpha\beta} =0\,.
\end{equation}
Therefore, as expected, no mechanical angular momentum is lost by the two-body system in the conservative approximation.

To study \eqref{L1cons} on its own, let us choose a frame where the center-of-mass sits in the origin of the transverse plane
\begin{equation}\label{}
	E_1 b_1^\alpha + E_2 b_2^\alpha = 0\,,
\end{equation}
i.e.
\begin{equation}\label{}
	b_1^\alpha = \frac{E_2}{E}\,b_J^\alpha\,,\qquad
	b_2^\alpha = -\frac{E_1}{E}\,b_J^\alpha\,.
\end{equation}
Let us also decompose the transferred momentum using $p_1^\mu$, $p_2^\mu$ and $b_J^\mu$ as basis vectors,
\begin{equation}\label{Qdecompsimpl}
	Q^\mu = 
	\frac{2E_2}{E}\sin^2\frac{\Theta}{2}  p_1^\mu -\frac{2E_1}{E}\sin^2\frac{\Theta}{2}p_2^\mu -\frac{p}{b}\,\sin\Theta b_J^\mu\,.
\end{equation}
Using \eqref{Qdecompsimpl} we get
\begin{equation}\label{rewrb1A1A2}
	b_1^{[\alpha}Q^{\beta]} + (A_1+A_2)p_1^{[\alpha}Q^{\beta]}
	=
	-2\sin^2\frac{\Theta}{2}\,\frac{E_1E_2}{E^2}b^{[\alpha}(p_1+p_2)^{\beta]}+2\sin^2\frac{\Theta}{2}\,\tan\frac{\Theta}{2}\,\frac{E_1E_2}{E^2}\,\frac{b}{p}\,p_1^{[\alpha}p_2^{\beta]}\,.
\end{equation}
If we now go to the frame where the center-of-mass is also initially at rest,
\begin{equation}\label{}
	-p_1^{\alpha} = (E_1,p^I)\,,\qquad
	-p_2^{\alpha} = (E_2,-p^I)\,,
\end{equation}
we see that, if we restrict to spatial components, all terms in \eqref{L1cons} vanish after using \eqref{rewrb1A1A2},
\begin{equation}\label{DeltaL1=0}
	\Delta L_{(1c)}^{IJ}  = 0\,.
\end{equation}
This tells us that indeed angular momentum is conserved separately for each particle, in this special frame.
The mixed components $\Delta L^{0I}$ do not vanish, in general. However, they are actually not well defined, since they depend on the Shapiro time delay $\partial 2\delta_0/\partial E \propto \frac{1}{4-D}+\log(b/b_0)$, which is infinite in $D=4$ due to the long-range nature of the gravitational force. This requires a subtraction which leaves behind an arbitrary cutoff scale $b_0$ in the remaining logarithm.

\subsection{Coherent state approximation beyond the soft limit}
\label{sec:eikoppm}

As discussed in the previous section, in order to include radiation in the eikonal framework we need to introduce creation/annihilation operators describing the physical gravitons (or in general massless particles) that can be produced/absorbed in the scattering process. So far we focused on the soft limit which in our context means that the typical energy of the radiation quanta is much smaller than that of the potential gravitons exchanged between the two energetic particles: $\omega\ll \hbar v/b$. In the previous section we used the approach by Bloch-Nordsieck/Weinberg~\cite{Bloch:1937pw,Thirring:1951cz,Weinberg:1965nx} to describe soft radiation as an exponential dressing of the ``hard'' elastic scattering as done in several papers~\cite{Amati:1990xe,Ciafaloni:2015xsr,Ciafaloni:2018uwe,Laddha:2018vbn,Sahoo:2018lxl,Saha:2019tub,Addazi:2019mjh,Sahoo:2021ctw}. However, in the spirit of the eikonal exponentiation, we expect that classical radiation exponentiates at all frequencies at least in a PM expansion. The natural guess is that, to describe frequencies $\omega \gtrsim v/b$, the functions $w_j$ in~\eqref{eq:eiksr} should be derived from the classical limit of the whole 5-point function ${\cal A}^{(5)}$. We will use ${\mathcal W}_j$ to indicate these more general functions appearing in the operatorial part of the eikonal beyond the soft approximation.

We start by taking the most straightforward generalization of the approach followed in the elastic case and introduce the impact parameter representation of ${\cal A}^{(5)}$ by taking its Fourier transform 
\begin{equation}\label{A5til}
	\tilde{\mathcal{A}}^{(5) \mu\nu}(x_1,x_2,k) = \int \frac{d^Dq_1}{(2\pi)^{D-2}}\,\delta(2 p_1 q_1-q_1^2)\, \delta(2 p_2 q_2-q_2^2) e^{ix_1\cdot q_1+ix_2\cdot q_2}\mathcal A^{(5) \mu\nu}(q_1,q_2,k)\;,
\end{equation}
where $q_1+q_2+k=0$ and the delta functions enforce the on-shell condition for the energetic particles. As usual, in the classical limit we need to expand the full amplitude for small values of $q_i$ and $k$ with respect to hard momenta $p_i$ since the soft momenta are proportional to $\hbar$ in the classical regime. We expect also in the inelastic case the same pattern discussed in Section~\ref{sec:oneloop} for the elastic scattering: the leading tree-level contribution as $\hbar \to 0$ should exponentiate and account for the leading term in the classical limit at each loop level. Of course perturbative amplitudes at a fixed order in $G$ will also have terms that scale as the leading tree-level contribution: by comparing their form with the formal expansion in $G$ of the eikonal operator $\hat{\delta}$ one should be able to fix in principle the form of $\hat\delta$ order by order in the PM approximation. As a consistency check, contributions of order ${\cal O}(G^n)$ that are more divergent than the classical terms should be completely determined by lower order data. In our analysis we will focus on the leading PM contribution to the operator part of the eikonal and discuss how the approaches of Sections~\ref{sec:eikopwithout} and~\ref{sec:eikopwith} can be generalized beyond the soft approximation. In both cases, we take the following generalization of the elastic eikonal~\eqref{DO10}
\begin{align}
	S |\psi \rangle =& \int_{p_3} \int_{p_4} e^{-i b_1 p_4} e^{-ib_2 p_3}
	\int \frac{d^D {Q_1}}{(2\pi)^D} \int \frac{d^D {Q_2}}{(2\pi)^D}  \Phi (p_4 -{Q}_1) \Phi (p_3-Q_2)  \nonumber \\
	\times & \int\! d^D {x}_1  \int\! d^D {x}_2\, e^{ i (b_1-x_1)Q_1 + i (b_2-x_2)Q_2}e^{2i \hat\delta (x_1,x_2)}  |p_3, p_4,0\rangle \;,
	\label{DOopg}
\end{align}
where the operator $e^{2i\hat\delta}$ is determined by the $4$-point and $5$-point function as sketched above.

Let us stress that the final state~\eqref{DOopg} contains a large number of gravitons as it has the same coherent state structure of the soft eikonal operator~\eqref{eq:eiksr}. While this may not be surprising, since we are describing the classical radiation produced during the scattering, it is not obvious that the operator $e^{2i\hat\delta}$ can be determined just by the classical limit  of the $4$ and the $5$-point amplitudes order by order in $G$. It is argued in~\cite{Cristofoli:2021jas} that this is a consistency requirement because it ensures that the variance in the distribution of the emitted gravitons becomes negligible in the classical limit. A first check is of course to verify that the tree-level $6$-point function involving two gravitons in the final state does not yield classical contributions that should be included in the eikonal. This is discussed in~\cite{Cristofoli:2021jas} for the technically simpler case of QED and in~\cite{Britto:2021pud} for gravity: in both cases the assumption above holds. This supports the intuitive picture that the soft quanta are emitted one at the time when the energetic particles bend in the scattering. However, as mentioned in the outlook, this point deserves further investigation studying also the inelastic amplitudes at loop level. We expect that the eikonal operator involves also corrections that are non-linear in the creation/annihilation operators for the soft quanta. For instance, terms of this type are probably needed to capture contributions related to the Compton-like $4$-point amplitude with two gravitons and two energetic particles.

\subsubsection{Eikonal operator without static modes}
\label{sec:Ceikopwithout}

In order to define explicitly the eikonal operator~\eqref{DOopg}, we then need to combine the classical information extracted from the $4$ and the $5$-point functions. A first approach is to define~\cite{Cristofoli:2021jas}
\begin{equation}
	\label{eikopv1}
	e^{2i\hat{\delta}(x_1,x_2)}=  \int\!\frac{d^D Q}{(2\pi)^D} \int\! d^D x \, e^{-iQ(x-x_1+x_2)}
	e^{2i\delta_s(b)} e^{i\int_{k}\left[{\mathcal{W}}_j(x_1,x_2,k) a_j^\dagger(k)+{\mathcal{W}}_j^\ast(x_1,x_2,k) a_j(k)\right]}\,,
\end{equation}
where $\delta_s(b)$ is related to the elastic eikonal discussed in Section~\ref{sec:eikwp}, while $\mathcal{W}_j$ is derived from the classical limit of the $5$-point amplitude. Of course both objects will have a PM expansion and at 1PM and 2PM $\delta_s(b)$ is the same as the elastic eikonal discussed in Section~\ref{sec:treelevel} and~\ref{sec:oneloop}. The leading mixed term in this expansion, $2\delta_0 \mathcal W_{0,j}$ in the one-loop $2\to3$ amplitude has been recently checked to be consistent with the exponentiation \eqref{eikopv1} \cite{Brandhuber:2023hhy,Herderschee:2023fxh,Georgoudis:2023lgf}. We know that at 3PM $\delta_2$ develops an imaginary part, see Section~\ref{sec:radiation}, but here we do not need to include it as it will be automatically generated by the operator part we are adding: it is the same mechanism that in Section~\ref{sec:eikopsoft} yielded the divergent imaginary part of $\delta_2$ from the normal ordering of the soft eikonal operator, see~\eqref{eq:Imdeltadef}. Thus we identified $\delta_s(b)$ with the real part of the elastic eikonal and, following~\cite{DiVecchia:2022piu}, we take its dependence on the external momenta as follows,
\begin{equation}\label{deltas}
	2\delta_s(b) = \frac{1}{2}\left[
	\operatorname{Re}2\delta(\sigma_{12},b_e)+\operatorname{Re}2\delta(\sigma_{34},b_e)
	\right],
\end{equation}
where we symmetrize between incoming $\sigma_{12}=-\frac{p_1\cdot p_2}{m_1 m_2}$ and outgoing $\sigma_{34}=-\frac{p_3\cdot p_4}{m_1 m_2}$ momenta. This last step starts being relevant only at 4PM, so beyond the scope of this review, but we will briefly comment in Section~\ref{sec:outloop} on why the approach taken in~\eqref{deltas} is useful.

In the radiative sector we have a similar PM expansion where $\mathcal{W}_{j} = \mathcal{W}_{0j}+\mathcal{W}_{1j} + \ldots$ starts at order $G^{3/2}$. Since we will work at the first PM order for the operator part, we focus on $\mathcal{W}_{0j}$. This is obtained simply by using the classical tree-level 5-point function~\eqref{GGG2} in~\eqref{A5til} and then contracting it with physical polarization tensors such as the ${\varepsilon}_{\times,+}$ introduced in~\eqref{eq:6.154}. For our analysis we need just the leading term and so we can use the linearized version of the delta functions, $\delta(2 p_1 q_1)\, \delta(2 p_2 q_2)$, since $2 p_i q_i\gg q_i^2$ 
\begin{equation}\label{calW0}
	\mathcal{W}_0^{\mu\nu}(x_1,x_2,k)
	=
	\int \frac{d^Dq_1}{(2\pi)^D} \,e^{ix_1\cdot q_1+x_2\cdot q_2}
	2\pi\delta(2p_1\cdot q_1)	2\pi\delta(2p_2\cdot q_2)\\
	\,\mathcal{A}_0^{(5) \mu\nu}(q_1,q_2,k)\,,
\end{equation}
where, as before, $q_1+q_2+k=0$ and $\mathcal{A}_0^{(5) \mu\nu} (q_1, q_2, k)$ is given in~\eqref{GGG2}.
In the same spirit of Eq.~\eqref{deltas}, we take ${\mathcal{W}}^{\mu\nu}$ to depend on $\tilde p_1$ and $\tilde p_2$,
\begin{equation}\label{tildep}
	\tilde p_1 = \frac12(p_4-p_1)
	=p_4-\frac{Q_1}{2}\,,\qquad
	\tilde p_2 = \frac12(p_3-p_2)
	=p_3-\frac{Q_2}{2}\,,
\end{equation}
rather than on $p_1$ and $p_2$. In order to clarify the meaning of the integrations in~\eqref{DOopg}, it is convenient to change variables as follows $x_1=x_++\frac{x_-}{2}$, $x_2=x_+-\frac{x_-}{2}$, $Q_1=Q_e -\frac{P}{2}$ and $Q_2 = -Q_e-\frac{P}{2}$. 
By rewriting~\eqref{calW0} in terms of $x_\pm$ we see that $\mathcal{W}_0$ depends on $x_+$ only through an overall factor of $e^{-i x_+ k}$. When using this fact in~\eqref{DOopg}, one can see that the integration over $x_+$ simply implies that $P$  is equal to the sum of the momenta of the emitted gravitons, as it can be seen by expanding the last exponential with the creation modes. Therefore, at a generic order $(a^\dagger)^M$, we have $P = \sum_{m=1}^M k_m$. The factorized dependence  on $x_+$ implies also the following transformation property under translations
\begin{equation}\label{transtransf}
	x^\mu_{1,2}\to x^\mu_{1,2}+a^\mu,\qquad \mathcal{W}_0^{\mu\nu} \to  e^{-ia\cdot k} {\mathcal{W}}_0^{\mu\nu}\,.
\end{equation} 
The same property will hold also at the subleading PM orders as long as we start from a definition with the functional form of~\eqref{calW0}.

It is instructive to use the stationary phase approach to calculate the norm of the final state $\langle \psi|S^\dagger S|\psi\rangle$. Classical unitarity implies that it should be equal to the norm of the initial state $\langle \psi|\psi\rangle$ and here we would like to see that at least all large phases cancel at the stationary point. By using~\eqref{DOopg} and~\eqref{eikopv1} we have
\begin{align} \nonumber
	\langle \psi|S^\dagger S|\psi\rangle
	&=
	\int_{p_3, p_4}
	\int\!\frac{d^DQ'_i}{(2\pi)^D}\, d^Dx'_i \int\!\frac{d^DQ'}{(2\pi)^D} \,d^Dx'
	\,e^{-i \sum_r(b_r - x'_r)\cdot Q'_r +iQ'(x'-x'_1+x'_2)-i2\delta_s(b')}\\ \label{sds1}
	&\times
	\int\!\frac{d^DQ_i}{(2\pi)^D}\, d^Dx_i \int\!\frac{d^DQ}{(2\pi)^D}\, d^Dx
	\,e^{i \sum_r(b_r - x_r)\cdot Q_r  -iQ(x-x_1+x_2) +i2\delta_s(b)}\\ \nonumber
	&\times
	e^{-\frac{1}{2}\! \int_{k}{\mathcal{W}_j}^\ast(x_1,x_2,k){\mathcal{W}_j}(x_1,x_2,k)-\frac{1}{2}\! \int_{k}{\mathcal{W}_j}^\ast(x'_1,x'_2,k){\mathcal{W}_j}(x'_1,x'_2,k)+\int_{k}{\mathcal{W}_j}^\ast(x'_1,x'_2,k){\mathcal{W}_j}(x_1,x_2,k)}
	\\ \nonumber
	&\times
	\Phi_1^\ast(p_4-Q'_1)\,\Phi_2^\ast(p_3-Q'_2)
	\,\Phi_1(p_4-Q_1)\,\Phi_2(p_3-Q_2)\,.
\end{align}
The stationary conditions for $Q_i$ and $Q'_i$ yield 
\begin{align}\label{xibi}
	x_{i\mu} & =  b_{i\mu} + \frac{\partial 2\delta_s(b)}{\partial Q_i^\mu} + \int_k \left[ \frac{i}{2} \frac{\partial({\mathcal{W}_j}^\ast(x_1,x_2,k){\mathcal{W}_j}(x_1,x_2,k))}{\partial Q^\mu_i} - i\, {\mathcal{W}_j}^\ast(x'_1,x'_2,k)\frac{\partial{\mathcal{W}_j}(x_1,x_2,k)}{\partial Q^\mu_i}\right] \\\nonumber
	x'_{i\mu} & =  b_{i\mu} + \frac{\partial 2\delta_s(b')}{\partial {Q'}_i^{\mu}} - \int_k \left[ \frac{i}{2} \frac{\partial({\mathcal{W}_j}^\ast(x'_1,x'_2,k){\mathcal{W}_j}(x'_1,x'_2,k))}{\partial {Q'}^{\mu}_i} - i\, \frac{\partial{\mathcal{W}_j}^\ast(x'_1,x'_2,k)}{\partial {Q'}^{\mu}_i} {\mathcal{W}_j}(x_1,x_2,k)\right] \,.
\end{align}
Similarly from the variation of $x_i$ and $x'_i$, we have
\begin{align}\label{}
	Q_{i\mu} =  (-1)^{i+1} Q_\mu + \int_k \left[ \frac{i}{2} \frac{\partial({\mathcal{W}_j}^\ast(x_1,x_2,k){\mathcal{W}_j}(x_1,x_2,k))}{\partial x^\mu_i} - i\, {\mathcal{W}_j}^\ast(x'_1,x'_2,k)\frac{\partial{\mathcal{W}_j}(x_1,x_2,k)}{\partial x^\mu_i}\right] \\\nonumber
	Q'_{i\mu} =  (-1)^{i+1} Q'_\mu	- \int_k \left[ \frac{i}{2} \frac{\partial({\mathcal{W}_j}^\ast(x'_1,x'_2,k){\mathcal{W}_j}(x'_1,x'_2,k))}{\partial {x'}^{\mu}_i} - i\, \frac{\partial{\mathcal{W}_j}^\ast(x'_1,x'_2,k)}{\partial {x'}^{\mu}_i} {\mathcal{W}_j}(x_1,x_2,k)\right]
\end{align}
and from the variations of $x$, ${x'}$, $Q$ and $Q'$
\begin{align}
	x_\mu &= (x_1-x_2)_\mu  +\frac{\partial 2\delta_s(b)}{\partial {Q}^\mu}\;,~~
	{Q}_\mu = \frac{\partial 2\delta_s(b)}{\partial {x}^\mu}\;,~ \\
	x'_\mu &= (x'_1-x'_2)_\mu  +\frac{\partial 2\delta_s(b')}{\partial {Q'}^\mu}\;,~~
	{Q'}_\mu = \frac{\partial 2\delta_s(b')}{\partial {x'}^\mu}\;.   
\end{align}
We can satisfy these equations by requiring
\begin{equation}\label{Saddlensm}
	\begin{split}    
		x'_\mu & = x_\mu =  (x_1-x_2)_\mu + \frac{\partial 2\delta_s(b)}{\partial Q^\mu}\;,\quad {Q'}_\mu = {Q}_\mu = \frac{\partial 2\delta_s(b)}{\partial {x}^\mu}\;, \\
		Q'_{i\mu}  & =  Q_{i\mu} = (-1)^{i+1} Q_\mu
		-i\int_{k}{\mathcal{W}}_j^\ast(x_1,x_2,k)\frac{\overset{\leftrightarrow}{\partial}}{\partial x_i^\mu}{\mathcal{W}_j}(x_1,x_2,k)\,,
		\\   (x'_i-b_i)_\mu  & =  (x_i-b_i)_\mu  =
		\frac{\partial 2\delta(b)}{\partial Q_i^\mu}
		-i\int_{k}{\mathcal{W}_j}^\ast(x_1,x_2,k)\frac{\overset{\leftrightarrow}{\partial}}{\partial Q_i^\mu}{\mathcal{W}_j}(x_1,x_2,k) \,,
	\end{split}
\end{equation}
where $f\overset{\leftrightarrow}{\partial}g = (f{\partial}g-g{\partial}f)/2$. When~\eqref{Saddlensm} are satisfied, all large phases cancel in the evaluation of $\langle \psi|S^\dagger S|\psi\rangle$, so they define the classical values of various quantities. For instance, we can use~\eqref{Saddlensm} and the relations
\begin{equation}
	\label{eq:QiPi}
	p_1+p_4=Q_1\;,\quad p_2+p_3 = Q_2\;,
\end{equation}
which follow from the presence of the wavepackets $\Phi_i$, to express $Q_i$ in terms of the initial data $b_{1,2}$ and $p_{1,2}$.

As mentioned after~\eqref{tildep}, some of the integrals could have been performed exactly, but we preferred to treat all the variables on the same footing. By summing the results for the $Q_i$'s and using the factorized dependence on $x_+$ of ${\mathcal W}$, we recover 
\begin{equation}\label{}
	Q_1^\mu
	+
	Q_2^\mu
	=
	-\int_{k}{\mathcal{W}_j}^\ast(x_1,x_2,k)k^\mu{\mathcal{W}_j}(x_1,x_2,k)=-P^\mu
\end{equation}
ensuring momentum conservation.

\subsubsection{Eikonal operator including static modes}
\label{sec:Ceikopwith}

As we saw in Section~\ref{sec:eikopsoft}, for some applications it is convenient to include in the eikonal operator also the static contribution arising from the zero-frequency modes and this can be easily done also by following the formalism discussed in the previous section. To this end we introduce an auxiliary frequency scale $\omega^*$ such that the full result is well approximated by the soft limit for frequencies below $\omega^*$. Then we can write the eikonal operator simply by splitting the operator part in two terms, one for $\omega>\omega^*$, where it coincides with the formulation in~\eqref{eikopv1}, and the other for $\omega < \omega^*$ for which we follow the approach of Section~\ref{sec:eikopsoft}. This splitting is useful only in the intermediate steps so to implement the dressing with the zero-frequency modes discussed in Section~\ref{sec:eikopwith}, but on the final results we will take the limit $\omega^*\to 0$. Thus the differences between this approach and that of the previous section are limited to the static contributions related to zero-frequency contributions and the $-i0$ prescription, see the comments after Eq.~\eqref{eq:eiksrstatic}. In formulae we have
\begin{equation}
	\label{eikopv2a}
	\begin{split}
		&e^{2i\hat{\delta}(x_1,x_2)}   = \int\!\frac{d^D {Q}}{(2\pi)^D} \int\! d^D{x} \, e^{-i{Q}({x}-x_1+x_2)} e^{i 2 \delta_s (b)}   \\
		&\times e^{\int_k \theta(\omega^\ast-k^0) 
			\left[f_j^\text{out} a_j^\dagger 
			- f_j^{\text{out}\ast} a_j
			\right]} 
		e^{-\int_k \theta(\omega^\ast-k^0)\left[f_j^\text{in} a_j^\dagger 
			- f_j^{\text{in}\ast} a_j
			\right]}  \\
		&\times  e^{i\int_{k}\theta(k^0-\omega^\ast) \left[{\mathcal{W}}_j(x_1,x_2,k) a_j^\dagger(k)+{\mathcal{W}}_j^\ast(x_1,x_2,k) a_j(k)\right]}\,.
	\end{split}
\end{equation}
We can follow the same steps of Section~\ref{sec:eikopwith} and combine the oscillators in the second line of~\eqref{eikopv2a} into a single exponential. This is equivalent to simply redefining the eikonal phase as in~\eqref{eq:tildedelta}, that is 
\begin{equation}\label{}
	2i\tilde\delta (b)= 2i\delta_s (b) -2i\delta^{\text{dr.}} (b)
\end{equation}
with
\begin{equation}
	\label{eq::tildedelta2}
	2i\delta^{\text{dr.}}(b) = -\frac{1}{2}\int_k^{\omega^\ast}\!\!\!\! \left(f_j^{\text{out}\ast}(k) f_j^\text{in}(k)-f_j^{\text{in}\ast}(k) f_j^\text{out}(k)\right)=\frac{i}{4}G Q_{1PM}^2{\mathcal{I}} (\sigma)\,,
\end{equation}
where $2\delta^{\text{dr.}} (b)$ was derived in Eqs.~\eqref{eq:tildedelta} and \eqref{2delta2}, and has been reported again here just for convenience after using \eqref{Isigma}. Thus we can rewrite~\eqref{eikopv2a} as
\begin{equation} \label{eikopv2}
	\begin{split}
		& e^{2i\hat{\delta}(x_1,x_2)}   = \int\!\frac{d^D Q}{(2\pi)^D} \int\! d^Dx \, e^{-iQ(x-x_1+x_2)}  e^{i 2\tilde{\delta}(b)}  \\  & \times e^{\int_{k}\theta(\omega^\ast-k^0) \left[f_j a_j(k)^\dagger - f_j^\ast(k) a_j(k) \right]}  e^{i\int_{k}\theta(k^0-\omega^\ast) \left[{\mathcal{W}}_j(x_1,x_2,k) a_j^\dagger(k)+\mathcal{W}_j^\ast(x_1,x_2,k) a_j(k)\right]}\,,
	\end{split}
\end{equation}
where $f_j (k) = f_j^\text{out}(k) - f_j^\text{in}(k)$.
The operator \eqref{eikopv2} has to be used in~\eqref{DOopg} to obtain the full eikonal operator. The discussion of classical unitarity for this eikonal operator follows the same steps discussed from Eq.~\eqref{sds1}. One can check that the various integrals are dominated, in the classical limit, by the following values
\begin{subequations}
	\label{Saddlev2}
	\begin{align}
		\label{Saddle3}
		{x}_\mu &= (x_1-x_2)_\mu +\frac{\partial 2\tilde\delta(b)}{\partial {Q}^\mu}
		\,,
		\qquad
		{Q}_\mu = \frac{\partial 2\tilde\delta(b)}{\partial {x}^\mu}\,,
		\\ 
		\label{Saddle2}
		Q_{i\,\mu} &=
		(-1)^{i+1} {Q}_\mu -i\int_{k} {\mathcal{W}_j}^\ast(x_1,x_2,k)\frac{\overset{\leftrightarrow}{\partial}}{\partial x_i^\mu} {\mathcal{W}_j}(x_1,x_2,k)\;,~~
		\\
		\begin{split}
			\label{Saddle1}
			(x_i - b_i)_\mu &=
			\frac{\partial 2\tilde\delta(b)}{\partial Q_i}
			-i\int_{k} \theta(\omega^\ast-k^0) \;f_j^\ast(k)\frac{\overset{\leftrightarrow}{\partial}}{\partial Q_i^
				\mu} f_j(k) \\ & \quad 
			- i\int_{k}\theta(k^0-\omega^\ast)\; {\mathcal{W}_j}^\ast(x_1,x_2,k)\frac{\overset{\leftrightarrow}{
					\partial}}{\partial Q_i^\mu}{\mathcal{W}_j}(x_1,x_2,k)
			\,, 
		\end{split}
	\end{align}
\end{subequations}
where one should use \eqref{deltas} and \eqref{tildep}.

\subsubsection{The \texorpdfstring{${\cal N}=8$}{N=8} eikonal operator}
\label{sec:eikopn8}
It is straightforward to construct the eikonal operator \eqref{eikopv1} or \eqref{eikopv2} relevant to the classical limit of $\mathcal {N}=8$ amplitudes for $2\to2$ processes involving massive scalars plus additional emissions of gravitons, dilatons, and KK scalars and vectors. 
For the elastic eikonal phase, one should use the $\mathcal N=8$ result which we derived in the previous sections up to $\mathcal O(G^3)$.
For the inelastic portion of the operator, instead, one can introduce new ladder operators associated to the new emitted massless states, as discussed in Section~\ref{sec:eikopsoftn8}, paired with appropriate $2\to3$ amplitudes for the emission of such states. However, it is more convenient to adopt the approach of Section~\ref{sec:radiation} and simply promote the appropriate Lorentz indices $\mu,\nu,\ldots$ to 10-dimensional ones $M,N,\ldots$, employing of course the $\mathcal N=8$ coupling in Eq.~\eqref{betaGR}. This effectively captures all possible types of emissions and exchanges relevant to the classical limit.

\subsection{Computing observables from the eikonal operator}
\label{sec:eikopobs}

In this section we illustrate how the eikonal operator allows us to calculate in a systematic way  observable quantities 
that characterize gravitational collisions. The general strategy, as already discussed in the soft limit based on Eq.~\eqref{SOS}, is to take expectation values of the appropriate operators in the final state obtained by acting with the eikonal operator on the initial state.
Since in this way the final state now includes radiation, this will allow us to obtain not only the deflection angle, but also the leading-order (L.O.) gravitational waveform as well as the 3PM linear and angular momenta of the gravitational field generated by the collision.
Moreover, since the operator incorporates both field and the massive degrees of freedom in a dynamical fashion, it will also provide expressions for the corresponding changes of mechanical linear and angular momenta of each particle taking part in the collision that explicitly obey the corresponding balance laws.

\subsubsection{Waveforms}
\label{sec:wavforlo}

As we discussed in order to obtain \eqref{hmunufmunu} in the zero-frequency limit, starting from the canonically normalized  field operator $H_{\mu\nu}(x)$ and taking its expectation value in the final state dictated by the eikonal operator, we obtain the metric fluctuation $W_{\mu\nu}(x)$,
\begin{equation}\label{}
	g_{\mu\nu}(x)
	=
	\eta_{\mu\nu}
	+
	2
	W_{\mu\nu}(x)\,.
\end{equation}
Proceeding in the same way, but using this time the eikonal operator \eqref{eikopv1}, we find the following waveform  in $D=4$ that goes beyond the soft approximation
\begin{equation}\label{WcalW}
	W^{\mu\nu}(x) = \int_{-\infty}^{+\infty} \frac{d\omega}{2\pi}\,e^{-i\omega u} \widehat W^{\mu\nu}\left(\omega (1,\hat x)\right),\qquad
	\widehat W^{\mu\nu}(k)
	=
	\frac{2G}{r}
	\frac{\mathcal W^{\mu\nu}}{\sqrt{8\pi G}}\,,
\end{equation}
where $\omega$ is the frequency measured by an asymptotic detector, $u$ is the corresponding retarded time and $(1,\hat x)^\mu$ a null vector characterizing its direction with respect to the source. To obtain the L.O. waveform, we may employ $\mathcal{W}_0^{\mu\nu}$ given in \eqref{calW0}, which is the impact-parameter version of the classical $2\to3$ amplitude, with the further approximation $x_{1,2}^\alpha\simeq b_{1,2}^\alpha$ of the saddle-point conditions \eqref{Saddlensm}.
Using the kinematics conventions given in Subsection~\ref{sec:kinematics} (see in particular Eq.~\eqref{decomposition}, and Eqs.~\eqref{invariants} and following), we may cast it in the form
\begin{equation}\label{tointegrateW}
	\mathcal{W}^{\mu\nu}_0(k) = \frac{1}{4m_1m_2} \int \frac{d^4q_1}{(2\pi)^4}\,e^{ib_1\cdot q_1+i b_2\cdot q_2}2\pi\delta(v_1\cdot q_1)2\pi\delta(v_2\cdot q_2)\mathcal A_0^{(5)\mu\nu}(q_1,q_2,k) 
\end{equation}
where $q_1+q_2+k=0$. In this way we will recover the frequency-domain expression $\widehat{W}^{\mu\nu}(k)$ for the waveform \cite{Mougiakakos:2021ckm}. Performing the frequency Fourier transform \eqref{WcalW} leads to the expression ${W}^{\mu\nu}$ for the metric fluctuation in (retarded) time domain \cite{Kovacs:1977uw,Kovacs:1978eu,Jakobsen:2021smu}, which however we will not discuss explicitly here.

The starting point for computing the L.O. waveform is thus the five-point amplitude $\mathcal A_0^{(5)\mu\nu}$ in the classical limit given in Eq. \eqref{GGG2}. 
We find it convenient to rewrite the theory-dependent coefficient $\beta$ by factoring out the mass dependence according to 
\begin{equation}\label{}
	\beta= 2m_1^2 m_2^2 c_0\,,
\end{equation}
so that, in $D=4$,
\begin{equation}
	\beta_{GR}  = 4 m_1^2 m_2^2 \left( \sigma^2 - \tfrac{1}{2} \right),\qquad\beta_{\mathcal{N}=8} =4 m_1^2 m_2^2  \sigma^2 \,.
	\label{GGG3}
\end{equation}
and
\begin{equation}\label{}
	c_0^{\text{GR}} = 2\sigma^2-1\,,\qquad
	c_0^{\mathcal{N}= 8} = 2\sigma^2\,.
\end{equation}
We also recall that Eq.~\eqref{GGG2}  is symmetric $\mathcal A_0^{(5)\mu\nu}=\mathcal A_0^{(5)\nu\mu}$ and encodes gauge invariance via the important property
\begin{equation}
	k_\mu \mathcal A_0^{(5)\mu\nu} =0\,.
	\label{GGG4}
\end{equation}

In order to perform the Fourier transform, it can be useful to isolate the terms in $\mathcal A_0^{(5)\mu\nu}$ that have either $1/q_1^2$ or $1/q_2^2$ from those that have both factors, writing
\begin{equation}\label{12irr}
	\mathcal A_0^{(5)\mu\nu}
	=
	\frac{1}{q_1^2}
	\mathcal A_{0,1}^{(5)\mu\nu}
	+
	\frac{1}{q_2^2}
	\mathcal A_{0,2}^{(5)\mu\nu}
	+
	\frac{1}{q_1^2q_2^2}
	\mathcal A_{0,\text{irr}}^{(5)\mu\nu}
\end{equation}
where $\mathcal A_{0,1}^{(5)\mu\nu}$, $\mathcal A_{0,2}^{(5)\mu\nu}$, $\mathcal A_{0,\text{irr}}^{(5)\mu\nu}$ only contain $q_1$ and $q_2$ in the numerators.
For terms of the first kind, we may solve $q_2=-q_1-k$ in the integrand of \eqref{tointegrateW}, obtaining
\begin{equation}\label{}
	\mathcal{W}^{\mu\nu}_{0,1}(k) = \frac{e^{-ib_2\cdot k}}{4m_1m_2} \int \frac{d^4q_1}{(2\pi)^4}\,e^{ib\cdot q_1}2\pi\delta(v_1\cdot q_1)2\pi\delta(v_2\cdot q_1-\omega_2)\frac{1}{q_1^2}\mathcal A_{0,1}^{(5)\mu\nu}\Big|_{q_2=-q_1-k}
\end{equation}
with $b^\alpha=b_1^\alpha-b_2^\alpha$. Performing the decomposition \eqref{decomposition} of the integrated momentum $q_1^\mu$ and taking the Jacobian factor $1/\sqrt{\sigma^2-1}$ into account (see also \ref{usefulFT}), we see that the delta functions set $q_{1\parallel1}=0$ and $q_{1\parallel2}=-\omega_2$ (as already noted in \eqref{q1decompq2decomp}) so that
\begin{equation}\label{}
	\mathcal{W}^{\mu\nu}_{0,1}(k) = \frac{e^{-ib_2\cdot k}}{4m_1m_2\sqrt{\sigma^2-1}} \int \frac{d^{2}q_{1\perp}}{(2\pi)^{2}}\,e^{ib\cdot q_{1\perp}}\frac{1}{q_{1\perp}^2+\frac{\omega_2^2}{\sigma^2-1}}\mathcal A_{0,1}^{(5)\mu\nu}\Big|_{\scriptsize
		\!\!\!
		\begin{array}{rl}
			q_2	\!\!\!\!\!\!	   &=-q_1-k\\
			q_1	\!\!\!\!\!\!	   &=-\omega_2\check v_2+q_{1\perp}
		\end{array}
	}
\end{equation}
and this integral can be evaluated in terms of a modified Bessel function of the second kind $K_0(z)$,
\begin{equation}\label{}
	\int \frac{d^2q_{1\perp}}{(2\pi)^2}\frac{e^{ib\cdot q_{1\perp}}}{q_{1\perp}^2+\frac{\omega_2^2}{\sigma^2-1}} = \frac{1}{2\pi} K_0\left(
	\Omega_2
	\right),
\end{equation}
where we introduced the dimensionless combinations
\begin{equation}\label{}
	\Omega_1= \frac{\omega_1 b}{\sqrt{\sigma^2-1}}\,,
	\qquad
	\Omega_2= \frac{\omega_2 b}{\sqrt{\sigma^2-1}}\,,
\end{equation}
so that
\begin{equation}\label{W1pdb}
	\mathcal{W}^{\mu\nu}_{0,1}(k) = \frac{e^{-ib_2\cdot k}}{8\pi m_1m_2\sqrt{\sigma^2-1}} \mathcal A_{0,1}^{(5)\mu\nu}\Big|_{\scriptsize
		\!\!\!\!\!\!
		\begin{array}{rl}
			q_2	\!\!\!\!\!\!	   &=-q_1-k\\
			q_1	\!\!\!\!\!\!	   &=-\omega_2\check v_2+q_{1\perp}\\
			q_{1\perp}\!\!\!\!\!\!\! &=-i\partial_b
		\end{array}
	}\!\!\!
	K_0\left(
	\Omega_2
	\right).
\end{equation}
In the last step we have used the identity $q_{1\perp}^\mu e^{ib\cdot q_{1\perp}}= -i\partial_b^\mu e^{ib\cdot q_{1\perp}}$ to rewrite all instances of $q_{1\perp}^\mu$ appearing in $\mathcal A_{0,1}^{(5)\mu\nu}$ as derivatives with respect to the impact parameter $b^\mu$.
The discussion of the second kind of terms in \eqref{12irr} is entirely analogous, with the only difference that for those it is more convenient to solve the momentum conservation condition as $q_1^\mu=-q_2^\mu-k^\mu$, so that 
\begin{equation}\label{W2pdb}
	\mathcal{W}^{\mu\nu}_{0,2}(k) = \frac{e^{-ib_1\cdot k}}{8\pi m_1m_2\sqrt{\sigma^2-1}} \mathcal A_{0,2}^{(5)\mu\nu}\Big|_{\scriptsize
		\!\!\!\!\!\!
		\begin{array}{rl}
			q_1	\!\!\!\!\!\!	   &=-q_2-k\\
			q_2	\!\!\!\!\!\!	   &=-\omega_1\check v_1+q_{2\perp}\\
			q_{2\perp}\!\!\!\!\!\!\! &=-i\partial_b
		\end{array}
	}\!\!\!
	K_0\left(
	\Omega_1
	\right).
\end{equation}
Finally, we turn to the calculation of the third kind of terms in \eqref{12irr}, which, proceeding as above, can be cast in the form
\begin{equation}\label{}
	\mathcal{W}^{\mu\nu}_{0,\text{irr}}(k) = \frac{e^{-ib_2\cdot k}}{4m_1m_2\sqrt{\sigma^2-1}} \int \frac{d^{2}q_{1\perp}}{(2\pi)^{2}}\,e^{ib\cdot q_{1\perp}}\frac{	\mathcal A_{0,\text{irr}}^{(5)\mu\nu}\Big|_{\scriptsize
			\!\!\!
			\begin{array}{rl}
				q_2	\!\!\!\!\!\!&=-q_1-k\\
				q_1	\!\!\!\!\!\!&=-\omega_2\check v_2+q_{1\perp}
			\end{array}
	}}{\left(
		(q_{1\perp}+k_\perp)^2
		+\frac{\omega_1^2}{\sigma^2-1}
		\right)\left(q_{1\perp}^2+\frac{\omega_2^2}{\sigma^2-1}\right)}
\end{equation}
where $\mathcal P=-\omega_1^2+2\omega_1\omega_2\sigma-\omega_2^2$ as in \eqref{calP}.
The basic integral to be performed in this case is 
\begin{equation}\label{}
	\int \frac{d^{2}q_{1\perp}}{(2\pi)^{2}}
	\frac{e^{ib\cdot q_{1\perp}}}{\left(
		(q_{1\perp}+k_\perp)^2
		+\frac{\omega_1^2}{\sigma^2-1}
		\right)\left(q_{1\perp}^2+\frac{\omega_2^2}{\sigma^2-1}\right)}
	\equiv
	\frac{1}{2\pi} H\,.
\end{equation}
Introducing Schwinger parameters, one finds
\begin{equation}\label{}
	H
	=
	\int_{\mathbb R_+^2}\frac{dx_1 dx_2}{2 (x_1+x_2)}\,
	e^{-\frac{1}{x_1+x_2}\left(\frac{b^2}{4}+i b\cdot k x_1+\frac{x_1^2 \omega_1^2+2 \sigma  x_1 x_2 \omega_1 \omega_2+ x_2^2 \omega_2^2}{\sigma^2-1}\right)}.
\end{equation}
Letting $x_1= b^2x/(4\lambda)$, $x_2=b^2y/(4\lambda)$ with Feynman parameters $x,y$ obeying $x+y=1$, and using the integral representation
\begin{equation}\label{}
	\frac{K_1(z)}{z} = \int_0^\infty  e^{-\lambda-\frac{z^2}{4\lambda}}\, \frac{d\lambda}{4\lambda^2}\,,
\end{equation}
for the modified Bessel function of the second kind $K_1(z)$,
one can express $H$ as the following parametric integral,
\begin{equation}\label{}
	H = \frac{b^2}2 \int_0^1e^{-ix b\cdot k}
	\frac{K_1(\Omega(x))}{\Omega(x)}
	dx
\end{equation}
where, again letting $x+y=1$,
\begin{equation}\label{fx}
	\Omega(x) = \sqrt{\Omega_1^2x^2+2\Omega_1\Omega_2 \sigma x y+ \Omega_2^2 y^2}\,.
\end{equation} 
Therefore,
\begin{equation}\label{Wirrpdb}
	\mathcal{W}^{\mu\nu}_{0,\text{irr}}(k) = \frac{e^{-ib_2\cdot k}}{16\pi m_1m_2\sqrt{\sigma^2-1}} 
	\mathcal A_{0,\text{irr}}^{(5)\mu\nu}\Big|_{\scriptsize
		\!\!\!\!\!\!
		\begin{array}{rl}
			q_2	\!\!\!\!\!\!	   &=-q_1-k\\
			q_1	\!\!\!\!\!\!	   &=-\omega_2\check v_2+q_{1\perp}\\
			q_{1\perp}\!\!\!\!\!\!\! &=-i\partial_b
		\end{array}
	}\!\!\! 
	\Big[
	b^2\int_0^1e^{-ix b\cdot k}
	\frac{K_1(\Omega(x))}{\Omega(x)}
	dx
	\Big].
\end{equation}
To perform the remaining $b$-derivatives in \eqref{W1pdb}, \eqref{W2pdb} and \eqref{Wirrpdb}, it can be useful to recall that
\begin{equation}\label{}
	K_0'(z)=-K_1(z)\,,\qquad
	(z K_1(z))'=- z K_0 (z)\,.
\end{equation}

Proceeding in this way, and collecting the three pieces according to \eqref{12irr}, we obtain the following result for the covariant frequency-domain waveform \eqref{WcalW}, letting
\begin{equation}\label{w0leading}
	\widehat W_0^{\mu\nu}(k) = \widehat W^{\mu\nu}_{0,12}(k)
	+
	\widehat W^{\mu\nu}_{0,\text{irr}}(k)\,.
\end{equation}
Then 
\begin{equation}\label{}
	\widehat W^{\mu\nu}_{0,12}(k)
	=
	\frac{2G^2 m_1m_2}{b r \omega_1 \omega_2 (\sigma^2-1)}
	\sum_{j=1,2}
	\Big[
	A_j^{\mu\nu}
	e^{-ib_j\cdot k}
	\frac{b K_0(\Omega_j)}{\sqrt{\sigma^2-1}}
	-
	2i c_0
	B_j^{\mu\nu}
	e^{-ib_j\cdot k}
	K_1(\Omega_j)	
	\Big]
\end{equation}
with
\begin{align}\label{}
	\nonumber
	A_1^{\mu\nu}
	&=
	c_0 \left(\sigma ^2-1\right) v_1{}^{(\mu }k^{\nu )} \omega _2
	+
	v_1{}^{\mu }v_1{}^{\nu } \omega _2 \left(2 \sigma  \omega _2 \left(c_0-4(\sigma^2-1)\right)
	+2 c_0 \omega _1\right)
	\\
	\nonumber
	&+2 \sigma  v_1{}^{(\mu }v_2{}^{\nu )} \omega _1 \omega _2 \left(2(\sigma^2-1)-c_0\right)
	\,,
	\\
	B_1^{\mu\nu} 
	&= (b\!\cdot \!k) v_1{}^{\mu }v_1{}^{\nu } \omega _2+v_1{}^{(\mu }b^{\nu )} \omega _1 \omega _2
\end{align}
and $A_2^{\mu\nu}$, $B_2^{\mu\nu}$ are obtained by interchanging particle labels ($v_1^\mu\leftrightarrow v_2^\mu$, $b_1^\mu\leftrightarrow b_2^\mu$ and hence $b^\mu\leftrightarrow -b^\mu$). Here and in the following $\xi^{(\mu}\xi'^{\mu')}=\xi^{\mu}\xi'^{\mu'}+\xi^{\mu'}\xi'^{\mu}$ without additional factors. Moreover,
\begin{equation}\label{}
	\widehat W_\text{0,irr}^{\mu\nu}(k)
	=
	\frac{2G^2m_1 m_2}{r (\sigma^2-1)^{3/2}}
	\int_0^1e^{-ib(x)\cdot k}
	dx
	\Big[
	C^{\mu\nu}
	K_0(\Omega(x))
	+
	D^{\mu\nu}
	\frac{K_1(\Omega(x))}{\Omega(x)}
	\Big]
\end{equation}
where
\begin{equation}\label{}
	b^\mu(x)
	=
	x b^\mu_1+ y b^\mu_2
\end{equation}
(let us recall that $y=1-x$)
together with
\begin{align}\label{Cmunu}
	C^{\mu\nu}
	&=
	c_0(\sigma^2-1)	\left[
	2\eta ^{\mu \nu }
	-
	i  (x-y) b^{(\mu }k^{\nu )}
	\right]
	-2c_0 \left( v_1{}^{\mu }v_1{}^{\nu } -\sigma v_1{}^{(\mu }v_2{}^{\nu )} +  v_2{}^{\mu }v_2{}^{\nu }\right)
	\\
	\nonumber
	&
	\!-\! i b_{\phantom{1}}^{(\mu }v_1^{\nu )}\! \left(2 \sigma  \omega _1 \left(2 \left(\sigma^2\!-1\right)-c_0 y\right)-2 c_0 x \omega _2\right)
	\!-\! i v_2{}^{(\mu }b^{\nu )}\! \left(2 \sigma  \omega _2 \left(2\left(\sigma ^2\!-1\right)-c_0 x\right)-2 c_0 y \omega _1\right),
\end{align}
and
\begin{align}
	\label{Dmunu}
	D^{\mu\nu}
	&=
	c_0(\sigma^2-1)
	\Big(
	\frac{b^2}{2}(x-y)^2
	k^\mu k^\nu
	-\frac{2b^{\mu}b^{\nu}}{b^2} \Omega(x)^2
	\Big)
	\\
	\nonumber
	&
	+b^2 v_1{}^{(\mu }k^{\nu )} (x-y) \left(c_0 x \omega _1+\sigma  \omega _2 \left(c_0 y-2(\sigma ^2-1)\right)\right)
	\\
	\nonumber
	&
	+b^2 v_2{}^{(\mu }k^{\nu )} (y-x) \left(c_0 y \omega _2+\sigma  \omega _1 \left(c_0 x-2(\sigma ^2-1)\right)\right)
	\\
	\nonumber
	&
	+\frac{2 b^2 v_1{}^{\mu }v_1{}^{\nu }}{\sigma ^2-1}
	\left(c_0 \left(x \omega _1+\sigma y \omega _2\right){}^2+2 \left(\sigma ^2-1\right) \omega _2 \left(\omega _2 \left(\sigma ^2 (x-y)-1\right)-2 \sigma  x \omega _1\right)\right)
	\\
	\nonumber
	&
	+\frac{2 b^2 v_2{}^{\mu }v_2{}^{\nu }}{\sigma ^2-1}
	\left(c_0 \left(y \omega _2+\sigma x \omega _1\right){}^2+2 \left(\sigma ^2-1\right) \omega _1 \left(\omega _1 \left(\sigma ^2 (y-x)-1\right)-2 \sigma  y \omega _2\right)\right)
	\\
	\nonumber
	&
	+\frac{2 b^2 v_1{}^{(\mu }v_2{}^{\nu )} }{\sigma ^2-1}
	\Big(\sigma  x \omega _1^2 \left(2 \left(\sigma ^2-1\right)-c_0 x\right)
	\\
	\nonumber
	&+\omega_1 \omega_2 \left(2 \left(\sigma ^2-1\right)-c_0 \left(\sigma ^2+1\right) x y\right)+\sigma  y \omega _2^2 \left(2 \left(\sigma ^2-1\right)-c_0 y\right)\Big)
	.
\end{align}
As a cross check, one can verify that $k_\mu W_0^{\mu\nu}(k)=0$ by integrating by parts with respect to $x$, using $b\!\cdot\!k e^{ixb\cdot k} =-i \partial_{x} e^{ixb\cdot k}$. This causes the terms arising from integration by parts to cancel against the remaining integrated terms, while the boundary terms cancel against the non-integrated ones.
The expressions we obtained in this way are valid for both GR and $\mathcal N=8$ provided one takes into account the appropriate conventions discussed below Eq.~\eqref{GGG2}.

In order to extract two gravitational physical degrees of freedom encoded in the waveform, we construct a pair of orthogonal polarization tensors as follows. We keep $c_0$ generic so as to capture both gravitational-wave emissions in GR and in $\mathcal N=8$ supergravity.
We first introduce reference vectors $\tilde{e}_\theta^\mu$ and ${e}_\phi^\mu$ such that  (more explicit expressions for such reference vectors are given below in Eq.~\eqref{eeExplicit} and following)
\begin{equation}\label{properties1}
	\tilde{e}_\theta\cdot\tilde{e}_\theta= 1=e_\phi\cdot e_\phi\,,\qquad
	\tilde{e}_\theta\cdot e_\phi=0\,,\qquad
	\tilde e_\theta\cdot k=e_\phi\cdot k=0
\end{equation}
and
\begin{equation}\label{properties2}
	\tilde{e}_\theta\cdot b=0\,,\qquad
	e_\phi\cdot v_i=0\,,\qquad
	-\frac{v_1\cdot \tilde{e}_\theta}{\sigma\omega_1-\omega_2}
	=
	\frac{v_2\cdot \tilde{e}_\theta}{\sigma\omega_2-\omega_1}
	=
	\frac{1}{\sqrt{\mathcal{P}}} \,.
\end{equation}
Let us recall that $\mathcal{P}$ had been introduced in \eqref{calP} and is nonnegative.
We then construct the following transverse-traceless polarization tensors
\begin{equation}\label{eq:6.154}
	\varepsilon_\times^{\mu\nu} = \frac12(\tilde e_\theta^\mu e_\phi^\nu+\tilde e_\theta^\nu e_\phi^\mu)\,,\qquad
	\varepsilon_+^{\mu\nu} = \frac12(\tilde e_\theta^\mu \tilde e_\theta^\mu- e_\phi^\mu e_\phi^\nu)
\end{equation}
and define
\begin{equation}\label{}
	\widehat W_{\times}(k) = \varepsilon_{\times\mu\nu} \widehat W_0^{\mu\nu}(k)\,,\qquad
	\widehat W_+(k) = \varepsilon_{+\mu\nu} \widehat W_0^{\mu\nu}(k)\,.
\end{equation}
Isolating for convenience two contributions for each polarization according to
\begin{equation}\label{}
	\widehat W_{\times/+}(k) = \widehat W_{12,\times/+}(k) + \widehat W_{\text{irr},\times/+}(k)\,,
\end{equation}
we find the following expressions.
For the $\times$ polarization,
\begin{equation}\label{A12x}
	\widehat W_{12,\times}(k)
	=-\frac{4 i G^2 m_1 m_2 c_0}{br (\sigma ^2-1)}\, 
	b\!\cdot\! e_{\phi} 
	\left(e^{-i b_1\cdot k} K_1\left(\Omega _1\right) v_1\!\cdot \!\tilde{e}_{\theta }-e^{-i b_2\cdot k} K_1\left(\Omega _2\right) v_2\!\cdot\!\tilde{e}_{\theta }\right)
\end{equation}
and
\begin{equation}\label{Airrx}
	\widehat W_{\text{irr},\times}(k)
	=
	\frac{4iG^2m_1m_2}{r\sqrt{\sigma^2-1}}
	\left(\frac{c_0 \omega_1\omega_2}{\sqrt{\mathcal{P}}}-2\sigma \sqrt{\mathcal{P}}\right) \,b\!\cdot\! e_\phi
	\int_0^1\!\! e^{-ib(x)\cdot k} K_0(\Omega(x))dx
\end{equation}
with $b^\mu(x)$, $\Omega(x)$ as in \eqref{fx}.
For the $+$ polarization, we find 
\begin{align}\nonumber
	&\widehat W_{12,+}(k)\\
	\label{A12+}
	&=
	\frac{2 G^2 m_1 m_2}
	{r  \omega_1 \omega_2 \left(\sigma ^2-1\right)}
	\Big[
	i \frac{b\!\cdot\!k}{b} \,c_0 \left(e^{-i b_2\cdot k} K_1\left(\Omega _2\right) (v_2\!\cdot\!\tilde{e}_{\theta })^2 \omega _1-e^{-i b_1 \cdot k} K_1\left(\Omega _1\right) 
	(v_1\!\cdot\!\tilde{e}_{\theta })^2 \omega _2\right)
	\\
	&
	+
	\frac{e^{-i b_1\cdot k} K_0(\Omega_1)\,v_1\!\cdot\!\tilde{e}_{\theta } \omega _2  
		-
		e^{-i b_2\cdot k} 
		K_0(\Omega _2)
		v_2\!\cdot\!\tilde{e}_{\theta } \omega_1 
	}{\sqrt{\sigma^2-1}\sqrt{\mathcal P}} 
	\left(
	\left(\sigma ^2-1\right) \left(4 \mathcal P \sigma -c_0 \omega _1 \omega _2\right)-c_0 \mathcal P \sigma
	\right)
	\Big]
	\nonumber
\end{align}
and finally
\begin{equation}\label{Airr+}
	\begin{aligned}
		\widehat W_{\text{irr},+}(k)
		&=
		\frac{2 G^2 m_1 m_2}{r \sqrt{\sigma ^2-1}}
		\int_0^1
		\!\!
		dx\,
		e^{-ib(x)\cdot k}
		\Big[
		\frac{(b\!\cdot\!e_\phi)^2}{b^2}
		c_0
		K_1(\Omega(x)) \Omega(x)
		-
		c_0
		K_0(\Omega(x))
		\\
		&
		+
		\frac{b^2K_1(\Omega(x))}{\Omega(x) \mathcal P} 
		\left(
		c_0 \omega _1^2 \omega _2^2+2 \mathcal P^2-4  \sigma  \omega_1 \omega_2 \mathcal P \right)
		\Big].
	\end{aligned}
\end{equation}

Introducing an additional vector $\check{b}^\mu$ satisfying $\check b^2=b^2$ and $b\cdot \check{u}_{1,2}=b\cdot\check b=0$, the explicit decomposition of the reference vectors $\tilde e_\theta^\mu$, $e_\phi^\mu$ in terms of the basis vectors $v_1^\mu$, $v_2^\mu$, $b^\mu$, $\check b^\mu$ reads
\begin{equation}\label{eeExplicit}
	\tilde{e}_\theta^\mu= \frac{\omega_1 v_2^\mu - \omega_2 v_1^\mu}{\sqrt{\mathcal P}}\,,\qquad
	e_\phi^\mu= \frac{(\check b\cdot k) b^\mu-(b\cdot k)\check b^\mu}{b\sqrt{(b\cdot k)^2 + (\check b\cdot k)^2}}\,.
\end{equation}
In the center-of-mass (CM) frame where 
\begin{equation}\label{vvbCM}
	v_1^\mu = \frac{1}{m_1}(E_1,0,0,p)\,,
	\quad 
	v_2^\mu =\frac{1}{m_2}(E_2,0,0,-p)\,,
	\quad
	b^\mu= (0,0,b,0)\,,
	\quad
	\check b^\mu= (0,b,0,0)\,,
\end{equation}
we can introduce the explicit parametrization
\begin{equation}\label{kCM}
	k^\mu= \omega(1,\sin\theta\cos\phi,\sin\theta\sin\phi,\cos\theta)
\end{equation}
and the polarization vectors
\begin{equation}\label{eeCM}
	\tilde{e}_\theta^\mu= -\frac{1}{\sin\theta}(\cos\theta,0,0,1)\,,\qquad
	e_\phi^\mu= (0,-\sin\phi,\cos\phi,0),
\end{equation}
which satisfy \eqref{properties1}, \eqref{properties2}. In particular, $\mathcal P =\omega^2(\sigma^2-1)\sin^2\theta$.
We provide a more complete list of kinematic relations in \ref{KinRela1}, together with an expression for the $\times$ and $+$ projections in terms of reference vectors $e^\mu_\theta$, $e_\phi^\mu$ such that $e^\mu_\theta$ is not necessarily orthogonal to $b^\mu$.

Let us now explore the soft limit  $k^\mu \to 0$ of the full waveform $\widehat W^{\mu\nu}(k)$ \eqref{w0leading}.
To leading order in this limit, we find
\begin{equation}\label{lsoftw0}
	\begin{aligned}
		\widehat W_{L.}^{\mu\nu}(k)
		&=
		\frac{4 G^2 i m_1 m_2 c_0}{b^2 r \sqrt{\sigma ^2-1}\,\omega _1^2 \omega_2^2}\\
		&\times\left(
		v_2{}^{\mu }v_2{}^{\nu } \omega _1^2 (b\!\cdot\!k)
		-(b\!\cdot\!k) v_1{}^{\mu }v_1{}^{\nu } \omega _2^2-v_1{}^{(\mu }b^{\nu )} \omega _2^2 \omega _1+v_2{}^{(\mu }b^{\nu )} \omega _2 \omega _1^2
		\right)
	\end{aligned}
\end{equation}
which is in agreement with the PM limit of Weinberg's soft theorem (translated to $b$-space), using the leading-order deflection $Q_\text{1PM}=2G m_1m_2 c_0/(b\sqrt{\sigma^2-1})$.
Indicating by $\log(\omega b)$ a generic log-enhanced dependence on the graviton's overall scale,\footnote{For instance $\log(\Omega_1)=\log(\omega b)+\cdots$, $\log(\Omega_2)=\log(\omega b)+\cdots$ where the dots denote non-log-enhanced terms.} we find to subleading order in the soft limit
\begin{align}\nonumber
	\widehat W_{S.L.}^{\mu\nu}(k) 
	&= 
	\frac{4 G^2 m_1 m_2 \left(4(\sigma ^2-1)-c_0\right) \sigma  \log (\omega b)}{r \left(\sigma ^2-1\right)^{3/2} \omega _1 \omega_2}
	\left(
	v_1{}^{\mu }v_1{}^{\nu } \omega _2^2-v_1{}^{(\mu }v_2{}^{\nu )} \omega _1 \omega _2+\omega_1^2 v_2^{\mu } v_2^{\nu }
	\right)\\
	&
	-
	\frac{2G^2 m_1 m_2  c_0  \log(\omega b)}{r \sqrt{\sigma ^2-1} \omega _1 \omega _2}
	\left(
	k^{(\mu} \xi^{\nu)} - \eta^{\mu\nu}\xi\cdot k
	\right)
	\label{WSL}
\end{align}
with $\xi^\mu= \omega_2 v_1^\mu + \omega_1 v_2^{\nu}$.
Setting $c_0=c_0^{\text{GR}}$ to discuss the case of  particles interacting only gravitationally, we can compare with the subleading log-theorem \cite{Saha:2019tub,Sahoo:2021ctw}, 
which predicts the form of the trace-reversed metric perturbation $e^{\mu\nu}= \widehat W^{\mu\nu}-\frac12 \eta^{\mu\nu}\eta_{\alpha\beta} \widehat W^{\alpha\beta}$ according to
\begin{equation}\label{eSL}
	e^{\mu\nu}_{S.L.} (k)=
	\frac{4 G^2 m_1 m_2 \left(2\sigma^2-3\right) \sigma  \log (\omega b)}{r \left(\sigma ^2-1\right)^{3/2} \omega _1 \omega_2}
	\left(
	v_1{}^{\mu }v_1{}^{\nu } \omega _2^2-v_1{}^{(\mu }v_2{}^{\nu )} \omega _1 \omega _2+\omega_1^2 v_2^{\mu } v_2^{\nu }
	\right).
\end{equation}
We see that the first line of \eqref{WSL} agrees with \eqref{eSL}, while the second line of \eqref{WSL} vanishes when contracted with physical polarizations. 
Instead, we do not find agreement with the ``sub-subleading'' $\mathcal O(\omega\log\omega)$ soft contribution to waveforms emitted during collisions of massless objects \cite{Ciafaloni:2018uwe}, which lie above the bound \eqref{eq:KTbex}. For instance, for the $\times$ polarization, we find
\begin{equation}
\frac{\hat W_{\times L.}+\hat W_{\times, \mathcal O(\omega \log \omega)}}{\hat W_{\times L.}} \sim 
1+\frac{1}{4} (\omega b)^2 \log(\omega b)  (\sin\theta)^2
\end{equation}
in any center-of-mass frame, where $\vec p_1+\vec p_2=0$ (the above relation is instead independent of the translation frame), as $\sigma\to\infty$, while the result of \cite{Ciafaloni:2018uwe} gives rise to a similar relation but with the factor of $\frac14$ replaced by $\frac12$.

Finally, let us calculate the dilatonic waveform.
For this case, we set $c_0=c_0^{\mathcal{N}=8}$ in order to include not only dilaton emissions but also dilaton exchanges, and take the trace according to
\begin{equation}\label{}
	\widehat W = \frac{1}{\sqrt 2}\,\eta_{\mu\nu} \widehat W^{\mu\nu}\,.
\end{equation}
Contracting indices generates some terms proportional to $b\!\cdot\!k$ from $\widehat W_{\text{irr}}^{\mu\nu}$ that we can integrate by parts to obtain a more compact expression. The result is $\widehat W = \widehat W_{12} + \widehat W_\text{irr}$ with 
\begin{equation}\label{}
	\begin{split}
		&\widehat W_{12}(k) 
		=
		\frac{4 \sqrt{2}G^2m_1m_2\sigma^2}{r}
		\Bigg\{
		\frac{ i}{b \left(\sigma ^2-1\right)} 
		\left[ 
		\frac{b\!\cdot \! k}{\omega_1} e^{-i b_1\cdot k} K_1(\Omega_1)
		-
		\frac{b\!\cdot \! k}{\omega_2}
		e^{-i b_2\cdot k} K_1(\Omega_2)\right]\\
		&
		+
		\frac{ \left(\sigma^2-2\right) \omega _2+\sigma \omega _1}{\sigma\omega_1(\sigma^2-1)^{3/2}} e^{-i b_1\cdot k} K_0(\Omega_1)
		+
		\frac{\left(\sigma^2-2\right) \omega _1+\sigma\omega _2}{\sigma\omega_2(\sigma^2-1)^{3/2}} e^{-i b_2\cdot k} K_0(\Omega_2)
		\Bigg\}
	\end{split}
\end{equation}
and
\begin{equation}\label{}
	\widehat W_\text{irr}(k) = - \frac{4 \sqrt{2} b G^2 m_1 m_2 }{r}
	(\omega_1^2+\omega_2^2)  \int_0^1 e^{-ib(x)\cdot k}\frac{K_1(\Omega(x))}{\Omega(x)}\,.
\end{equation}

\subsubsection{Linear momentum}
\label{sec:linmomlo}

Once the waveforms are given one can easily write down the corresponding formulas for the spectrum of emitted energy and momentum for each polarization as a function of frequency or retarded time. In frequency domain,
which is more relevant for interferometers,  one has
\begin{equation}\label{boldPspectrum}
	d\boldsymbol P^\mu_{\times,+}
	=
	k^\mu
	\left|\omega {\mathcal{W}}_{\times,+}\right|^2 
	 \frac{d\omega\, d\Omega(n)}{2\omega(2\pi)^3}\,,
\end{equation}
where, as discussed in \ref{app:asymptoticlimit}, the measured frequency and the detector's angular positions are characterized by $k^\mu= \omega\, n^\mu$, while $d\Omega(n)$ is the angular measure.
In particular, using \eqref{kCM}, the gravitational energy emitted in a given direction $\theta, \phi$ takes the form (see e.g.~\cite{Kovacs:1977uw,Kovacs:1978eu})
\begin{equation}\label{dEdOmdom}
	d\boldsymbol P^{0}_{\times, +} =  \omega 
	\left|\omega {\mathcal{W}}_{\times,+}\right|^2 
	\sin\theta\,
	\frac{d\omega\,d\theta\, d\phi}{2\omega(2\pi)^3}\,. 
\end{equation}

These differential spectra, which characterize the angular and frequency distribution of the energy carried by gravitational waves, have been studied, for instance, in \cite{Kovacs:1977uw,Kovacs:1978eu} for massive objects at finite $\sigma$, and in \cite{DEath:1976bbo,Ciafaloni:2018uwe} for the ultra-relativistic case.
Leaving  to future work a more detailed study of the spectrum in different kinematical regimes, let us mention a few  of its broad features in the frequency domain after integration over the angular variables.

As we have already mentioned, in the limit $\omega \to 0$ the spectra tend to the well-known ZFL \cite{Smarr:1977fy}. 
	The characteristic frequency range for which a soft expansion  in powers of $\omega b$ is valid is of order $1/b$ and in $D=4$  contains logarithmically enhanced sub- and sub-sub-leading corrections. Such a mild $\omega$ dependence
	also  holds  above $1/b$ as long as  $ \omega b $ does not exceed $\sqrt{\sigma}$. Above this frequency the spectrum drops, first like a small power of $\omega$, and then exponentially when $\omega b  > \sigma^{\frac32}$. This qualitative feature of the spectrum is responsible for the  $\sigma$ dependence of the total radiated energy to be discussed below. It would be interesting to compare this spectrum with the one of the ultra-relativistic limit i.e.~above the bound \eqref{eq:KTbex}, $\Theta \sqrt{ \sigma } \gtrsim 1$, but, so far, this has only been done in the small-$\omega$ limit as already discussed in Section~\ref{sec:eikopsoft}.

Here we shall focus on the calculation of the total emitted energy and momentum, which is obtained by integrating \eqref{boldPspectrum} over the graviton phase space, to leading order in the coupling,
\begin{equation}\label{boldP}
	\boldsymbol P^\mu
	=
	\int_{ k} \tilde{\mathcal{A}}^{(5)} k^\mu \tilde{\mathcal{A}}^{(5)\ast}\,.
\end{equation}
Following the steps discussed in Section~\ref{sec:radiation}, this can be conveniently rewritten in terms of the Fourier transform of a three-particle cut, according to
\begin{equation}\label{}
	\boldsymbol P^\mu
	=
	\operatorname{FT}
	\int
	d(\text{LIPS})
	\,k^\mu
	\begin{gathered}
		\begin{tikzpicture}[scale=.5]
			\draw[<-] (-4.8,5.17)--(-4.2,5.17);
			\draw[<-] (-1,5.15)--(-1.6,5.15);
			\draw[<-] (-1,3.15)--(-1.6,3.15);
			\draw[<-] (-1,.85)--(-1.6,.85);
			\draw[<-] (-4.8,.83)--(-4.2,.83);
			\draw[<-] (-2.85,1.7)--(-2.85,2.4);
			\draw[<-] (-2.85,4.3)--(-2.85,3.6);
			\path [draw, thick, blue] (-5,5)--(-3,5)--(-1,5);
			\path [draw, thick, color=green!60!black] (-5,1)--(-3,1)--(-1,1);
			\path [draw] (-3,3)--(-1,3);
			\path [draw] (-3,1)--(-3,5);
			\draw[dashed] (-3,3) ellipse (1.3 and 2.3);
			\node at (-1,5)[right]{$k_1$};
			\node at (-1,3)[right]{$k$};
			\node at (-1,1)[right]{$k_2$};
			\node at (-5,5)[left]{$p_1$};
			\node at (-5,1)[left]{$p_2$};
			\node at (-2.8,4)[left]{$q_1$};
			\node at (-2.8,2)[left]{$q_2$};
			\draw[<-] (4.8,5.17)--(4.2,5.17);
			\draw[<-] (1,5.15)--(1.6,5.15);
			\draw[<-] (1,3.15)--(1.6,3.15);
			\draw[<-] (1,.85)--(1.6,.85);
			\draw[<-] (4.8,.83)--(4.2,.83);
			\draw[<-] (2.85,1.7)--(2.85,2.4);
			\draw[<-] (2.85,4.3)--(2.85,3.6);
			\path [draw, thick, red] (.5,0)--(.5,6);
			\path [draw, thick, blue] (5,5)--(3,5)--(1,5);
			\path [draw, thick, color=green!60!black] (5,1)--(3,1)--(1,1);
			\path [draw] (3,3)--(1,3);
			\path [draw] (3,1)--(3,5);
			\draw[dashed] (3,3) ellipse (1.3 and 2.3);
			\node at (5,5)[right]{$p_4$};
			\node at (5,1)[right]{$p_3$};
			\node at (2.8,4)[right]{$q_4$};
			\node at (2.8,2)[right]{$q_3$};
		\end{tikzpicture}
	\end{gathered},
\end{equation}
where FT stands for the Fourier transform \eqref{ft222diag}.
The integral on the right-hand side can be calculated using the reverse-unitarity method, i.e.~by reinterpreting the delta functions involved in the LIPS as cut propagators. 
The result in GR is \cite{Herrmann:2021lqe,Herrmann:2021tct,DiVecchia:2021bdo}
\begin{equation}\label{Prad}
	P_\text{rad}^\mu
	=
	\frac{G^3m_1^2m_2^2}{b^3} \left(\check v_1^\mu+\check v_2^\mu\right) \, \mathcal{E}(\sigma)\,,
\end{equation}
with 
\begin{equation}\label{calE}
	\begin{split}
		\frac{\mathcal E(\sigma)}{\pi} &= f_1(\sigma) + f_2(\sigma) \log\frac{\sigma+1}{2} + f_3(\sigma) \frac{\sigma\,\operatorname{arccosh}\sigma}{2\sqrt{\sigma^2-1}}\,, \\
		f_1(\sigma) &= \frac{210 \sigma ^6-552 \sigma ^5+339 \sigma ^4-912 \sigma ^3+3148 \sigma ^2-3336 \sigma +1151}{48(\sigma^2-1)^{3/2}}\,,\\
		f_2(\sigma) &= -\frac{35 \sigma^4+60 \sigma^3-150 \sigma^2+76 \sigma -5}{8\sqrt{\sigma^2-1}}\,,\\
		f_3(\sigma) &= \frac{\left(2 \sigma^2-3\right) \left(35 \sigma^4-30 \sigma^2+11\right)}{8 \left(\sigma^2-1\right)^{3/2}}\,.
	\end{split}
\end{equation}

Let us now check that this emission of energy and momentum is matched by corresponding radiative losses of energy-momentum of the colliding objects, i.e.~by the integrals in \eqref{Saddle2}, which we denote as follows to leading order in the PM expansion using the further saddle point conditions \eqref{Saddle3},
\begin{equation}\label{Q1RRstart}
	\boldsymbol Q_{1\mu} = \frac{1}{2} \int_{ k} \left[-i\frac{\partial \tilde{\mathcal{A}}^{(5)}}{\partial b_1^\mu} \tilde{\mathcal{A}}^{(5)\ast}
	+
	i\tilde{\mathcal{A}}^{(5)} \frac{\partial \tilde{\mathcal{A}}^{(5)\ast}}{\partial b_1^\mu} \right]\,.
\end{equation}
Proceeding in the by now familiar way (see Section~\ref{sec:radiation}), we can recast this as the following three-particle cut,
\begin{equation}\label{Q1RRintegral}
	\boldsymbol Q^{\mu}_{1}
	=
	\frac12
	\operatorname{FT}
	\int
	d(\text{LIPS})
	\,(q_1^\mu-q_4^\mu)\,
	\begin{gathered}
		\begin{tikzpicture}[scale=.5]
			\draw[<-] (-4.8,5.17)--(-4.2,5.17);
			\draw[<-] (-1,5.15)--(-1.6,5.15);
			\draw[<-] (-1,3.15)--(-1.6,3.15);
			\draw[<-] (-1,.85)--(-1.6,.85);
			\draw[<-] (-4.8,.83)--(-4.2,.83);
			\draw[<-] (-2.85,1.7)--(-2.85,2.4);
			\draw[<-] (-2.85,4.3)--(-2.85,3.6);
			\path [draw, thick, blue] (-5,5)--(-3,5)--(-1,5);
			\path [draw, thick, color=green!60!black] (-5,1)--(-3,1)--(-1,1);
			\path [draw] (-3,3)--(-1,3);
			\path [draw] (-3,1)--(-3,5);
			\draw[dashed] (-3,3) ellipse (1.3 and 2.3);
			\node at (-1,5)[right]{$k_1$};
			\node at (-1,3)[right]{$k$};
			\node at (-1,1)[right]{$k_2$};
			\node at (-5,5)[left]{$p_1$};
			\node at (-5,1)[left]{$p_2$};
			\node at (-2.8,4)[left]{$q_1$};
			\node at (-2.8,2)[left]{$q_2$};
			\draw[<-] (4.8,5.17)--(4.2,5.17);
			\draw[<-] (1,5.15)--(1.6,5.15);
			\draw[<-] (1,3.15)--(1.6,3.15);
			\draw[<-] (1,.85)--(1.6,.85);
			\draw[<-] (4.8,.83)--(4.2,.83);
			\draw[<-] (2.85,1.7)--(2.85,2.4);
			\draw[<-] (2.85,4.3)--(2.85,3.6);
			\path [draw, thick, red] (.5,0)--(.5,6);
			\path [draw, thick, blue] (5,5)--(3,5)--(1,5);
			\path [draw, thick, color=green!60!black] (5,1)--(3,1)--(1,1);
			\path [draw] (3,3)--(1,3);
			\path [draw] (3,1)--(3,5);
			\draw[dashed] (3,3) ellipse (1.3 and 2.3);
			\node at (5,5)[right]{$p_4$};
			\node at (5,1)[right]{$p_3$};
			\node at (2.8,4)[right]{$q_4$};
			\node at (2.8,2)[right]{$q_3$};
		\end{tikzpicture}
	\end{gathered}\,.
\end{equation}
We can then use the same techniques employed to calculate $P_\text{rad}^\mu$, based on reverse-unitarity, obtaining the following result
\begin{equation}\label{QRR1}
	\boldsymbol Q_{1}^{\mu} = - \frac{G^3m_1^2m_2^2}{b^3} \check v_2^\mu\,\mathcal{E}(\sigma)\,,
\end{equation}
with $\mathcal E(\sigma)$ as in \eqref{calE}, and similarly for particle 2
\begin{equation}\label{QRR2}
	\boldsymbol Q_{2}^{\mu} = - \frac{G^3m_1^2m_2^2}{b^3} \check v_1^\mu\,\mathcal{E}(\sigma)\,.
\end{equation}
We see that \eqref{Prad}, \eqref{QRR1}, \eqref{QRR2} obey the energy-momentum conservation condition
\begin{equation}\label{balanceB}
	\boldsymbol P^\mu + \boldsymbol Q_{1}^{\mu} + \boldsymbol Q_{2}^{\mu} = 0\,.
\end{equation}
This can be seen as a consequence at the level of integrals of the basic identity $q_1^\mu+q_2^\mu+k^\mu=0$ among the integrated momenta.
Taking into account that the additional $Q_\mu$ contribution enters the total impulse $Q_{i\mu}$ \eqref{Saddle2}  with opposite sign for the two particles, we thus recover the complete energy-momentum balance law,
\begin{equation}\label{balancetot}
	P^\mu +  Q_{1}^{\mu} +  Q_{2}^{\mu} = 0\,.
\end{equation}
Here we used that the only contribution to the energy-momentum of the gravitational field is the radiative one given in \eqref{boldP}, $P^\mu=\boldsymbol P^\mu$.

Let us comment on the difference between \eqref{Saddlensm} and \eqref{Saddlev2} when it comes to the non-radiative part of the impulse. In the former case,
$Q^\mu$ is determined by the derivative of the full eikonal phase $\operatorname{Re}2\delta$ up to 3PM, which therefore includes both conservative and radiation-reaction effects, as we discussed in Section~\ref{sec:twoloop},
\begin{equation}\label{QRRtobecomp}
	\mathcal Q_{1}^{\mu} =  - \frac{G}{2} Q^2_\text{1PM} {\mathcal{I}}(\sigma) \frac{b^\mu}{b^2}=-\mathcal Q_{2}^{\mu}\,.
\end{equation}
In the latter case, $Q^\mu$ is determined by the derivative of the ``reduced'' phase $2\tilde\delta$, which does \emph{not} include such effects, due to the subtraction of \eqref{eq::tildedelta2}. However, this difference is accompanied by the appearance of a new static contribution in the relation \eqref{Saddle1} linking $x^\mu$ to $b^\mu$, which we denote by
\begin{equation}\label{}
	\Delta x_\mu \equiv (x_1-x_2 - b_J)_{\mu}
	=
	- i\int_{k}\theta(\omega^\ast-k^0)\; F^\ast(k)\Big(\frac{\overset{\leftrightarrow}{\partial}}{\partial Q_1^\mu}
	-
	\frac{\overset{\leftrightarrow}{\partial}}{\partial Q_2^\mu}\Big) F(k)
	+
	\mathcal O(G^3)
	\,.
\end{equation}
To the order under consideration,
using the on-shell conditions and the same integrals involved in the evaluation of $\mathcal{J}^{\alpha\beta}$ in Section~\ref{sec:eikopsoft}, one obtains
\begin{equation}\label{QRRstaticFdF}
	\Delta x_\mu
	\simeq
	Q_\mu\,\mathcal G\,,\qquad 
	\mathcal G= \frac12 \left(c_{14}+c_{23}-c_{13}-c_{24}\right) 	\,.
\end{equation}
where $Q_\mu$ in this case  is the conservative part of the impulse.

This quantity contributes with an extra term to the relation  between $b$ and $b_J$ in \eqref{DO17}  and this implies that, in computing $Q^\mu$ in \eqref{Saddle3}, we can use the relation $b=x + \mathcal{G} Q (b=x)$ getting
\begin{equation}
Q^\mu= \frac{\partial 2\tilde{\delta} (x + \mathcal{G} Q(b=x) )}{\partial x^\mu}=\frac{\partial 2\tilde{\delta} (x )}{\partial x^\mu}+ \mathcal{G}Q \frac{\partial^2 2 \tilde{\delta} (x)}{\partial x \partial x^\mu}
+\mathcal O(G^4)\,.
\label{}
\end{equation}
The second term gives the  radiation reaction that can be written as
\begin{equation}
\mathcal Q_{1}^{\mu} = \frac{G \mathcal{I} (\sigma)}{4b} b^\mu \frac{\partial Q^2}{\partial b}= \frac{G \mathcal{I}(\sigma)}{4} \frac{\partial Q^2}{\partial b_\mu}= - \frac{G}{2} Q^2_\text{1PM} {\mathcal{I}}(\sigma) \frac{b^\mu}{b^2}
+\mathcal O(G^4)\,,
\label{}
\end{equation}
where, at leading order, we have put $x=b$, and  $\mathcal{G} = \frac{G}{2} \mathcal{I} (\sigma)$ by \eqref{QRRtobecomp}, as well as $Q=Q_\text{1PM}$.
In this way, the predictions of the two formalisms for the full impulse agree.

\subsubsection{Angular momentum }
\label{sec:angmomlo}

The angular momentum of the gravitational field sourced by the collision is given by the sum of two pieces,
\begin{equation}\label{}
	J^{\alpha\beta} = \boldsymbol{J}^{\alpha\beta} + \mathcal J^{\alpha\beta}.
\end{equation}
The second one, $\mathcal J^{\alpha\beta}$, is the contribution due to static modes, which starts at 2PM order and which we have already calculated in \eqref{Jgraviton}.
The first one, $\boldsymbol{J}^{\alpha\beta}$, is instead new. It starts at 3PM order and arises due to genuine gravitational wave emissions.
Its expression is the same as \eqref{JIJgraviton}, with the gravitational waveform $\tilde{\mathcal{A}}^{(5)\mu\nu}$ replacing the ``soft factor'' $F^{\mu\nu}$.
For convenience, let us break it down as follows 
\begin{equation}\label{JoJs}
	\boldsymbol J_{\alpha\beta} = \boldsymbol J^{(o)}_{\alpha\beta} + \boldsymbol J^{(s)}_{\alpha\beta}
	\,,\qquad
	\boldsymbol J^{(o)}_{\alpha\beta}
	=
	-i
	\int_{{k}}
	k_{[\alpha}
	\frac{\partial\tilde{\mathcal{A}}^{(5)}}{\partial k^{\beta]}}\tilde{\mathcal{A}}^{(5)\ast}
	\,,\qquad
	\boldsymbol J^{(s)}_{\alpha\beta}
	=
	i \int_{{k}}
	2\tilde{\mathcal{A}}^{(5)\mu}_{[\alpha} \tilde{\mathcal{A}}_{\beta]\mu}^{(5)\ast}\,.
\end{equation}
As already mentioned, the two terms in the previous equations are formally reminiscent of the orbital and spin terms of the gravitational angular momentum, but only their sum is physically meaningful and gauge invariant.
One of the main advantages of the expression \eqref{JoJs} for the angular momentum is that it is manifestly covariant under Lorentz transformation. To see how it transforms under translations, 
\begin{equation}\label{translation}
	b_{1,2}^\mu \mapsto b_{1,2}^\mu + a^\mu\,,
\end{equation}
let us first remark that, under \eqref{translation}, $\tilde{\mathcal{A}}^{(5)\mu\nu}(k)$ transforms according to
\begin{equation}\label{translA5}
	\tilde{\mathcal{A}}^{(5)\mu\nu}(k)\mapsto e^{-ia\cdot k}\tilde{\mathcal{A}}^{(5)\mu\nu}(k)\,,
\end{equation}
as is clear from the explicit expression \eqref{ft223DEF}.
Then, taking into account the differential operator in \eqref{JoJs}, we see that 
\begin{equation}\label{translJ}
	\boldsymbol J^{\alpha\beta} \mapsto \boldsymbol J^{\alpha\beta} + a^{[\alpha} \boldsymbol{P}^{\beta]}\,,
\end{equation} 
after comparing with the radiated energy-momentum \eqref{boldP}.  In this way, we see that \eqref{JoJs} is in fact Poincar\'e covariant.

Following the above strategy, we shall now recast these expression in a form which is amenable to the application of reverse-unitarity.
Let us start from $\boldsymbol J^{(s)}_{\alpha\beta}$, for which this manipulation is straightforward:
\begin{equation}\label{LL1Ls}
	\boldsymbol J^{(s)}_{\alpha\beta}
	=
	2i
	\operatorname{FT}
	\int
	d(\text{LIPS}) 	\left[ 
	\begin{gathered}
		\begin{tikzpicture}[scale=.5]
			\draw[<-] (-4.8,5.17)--(-4.2,5.17);
			\draw[<-] (-1,5.15)--(-1.6,5.15);
			\draw[<-] (-1,3.15)--(-1.6,3.15);
			\draw[<-] (-1,.85)--(-1.6,.85);
			\draw[<-] (-4.8,.83)--(-4.2,.83);
			\draw[<-] (-2.85,1.7)--(-2.85,2.4);
			\draw[<-] (-2.85,4.3)--(-2.85,3.6);
			\path [draw, thick, blue] (-5,5)--(-3,5)--(-1,5);
			\path [draw, thick, color=green!60!black] (-5,1)--(-3,1)--(-1,1);
			\path [draw] (-3,3)--(-1,3);
			\path [draw] (-3,1)--(-3,5);
			\draw[dashed] (-3,3) ellipse (1.3 and 2.3);
			\node at (-1,3)[below]{$k$};
			\node at (-5,5)[left]{$p_1$};
			\node at (-5,1)[left]{$p_2$};
			\node at (-2.8,4)[left]{$q_1$};
			\node at (-2.8,2)[left]{$-q_1-k$};
		\end{tikzpicture}
	\end{gathered}
	\right]_{[\alpha}^\mu
	\left[
	\begin{gathered}
		\begin{tikzpicture}[scale=.5]
			\draw[<-] (4.8,5.17)--(4.2,5.17);
			\draw[<-] (1,5.15)--(1.6,5.15);
			\draw[<-] (1,3.15)--(1.6,3.15);
			\draw[<-] (1,.85)--(1.6,.85);
			\draw[<-] (4.8,.83)--(4.2,.83);
			\draw[<-] (2.85,1.7)--(2.85,2.4);
			\draw[<-] (2.85,4.3)--(2.85,3.6);
			\path [draw, thick, blue] (5,5)--(3,5)--(1,5);
			\path [draw, thick, color=green!60!black] (5,1)--(3,1)--(1,1);
			\path [draw] (3,3)--(1,3);
			\path [draw] (3,1)--(3,5);
			\draw[dashed] (3,3) ellipse (1.3 and 2.3);
			\node at (5,5)[right]{$p_4$};
			\node at (5,1)[right]{$p_3$};
			\node at (2.8,4)[right]{$q-q_1$};
			\node at (2.8,2)[right]{$q_1+k-q$};
			\node at (1,3)[below]{$-k$};
		\end{tikzpicture}
	\end{gathered}
	\right]_{\beta]\mu}
\end{equation}
as follows by applying the same steps discussed in Section~\ref{sec:radiation}, carrying along the appropriate index contractions.
Let us now turn to $\boldsymbol{J}^{(o)}_{\alpha\beta}$, for which the presence of a derivative with respect to $k^\mu$ makes the connection between the $b$-space and the $q$-space representation less straightforward, unlike all quantities considered so far.
We first rewrite the defining expression \eqref{JoJs} in a frame where
$b_2^\alpha=0$ and thus $b^\alpha=b_1^\alpha$, which can always be attained by performing an appropriate translation according to Eqs.~\eqref{translation}, \eqref{translA5}, \eqref{translJ}. In this way we find
\begin{equation}\label{}
	\begin{split}
		i\boldsymbol{J}^{(o)}_{\alpha\beta}
		&=
		\int_{k} 
		k_{[\alpha}
		\frac{\partial}{\partial k^{\beta]}}
		\left[ 
		\int \frac{d^Dq_1}{(2\pi)^D}2\pi\delta(2p_1\cdot q_1)2\pi\delta(2p_2\cdot (q_1+k))
		e^{ib\cdot q_1}\!\!\!\!\!\!\!
		\begin{gathered}
			\begin{tikzpicture}[scale=.5]
				\draw[<-] (-4.8,5.17)--(-4.2,5.17);
				\draw[<-] (-1,5.15)--(-1.6,5.15);
				\draw[<-] (-1,3.15)--(-1.6,3.15);
				\draw[<-] (-1,.85)--(-1.6,.85);
				\draw[<-] (-4.8,.83)--(-4.2,.83);
				\draw[<-] (-2.85,1.7)--(-2.85,2.4);
				\draw[<-] (-2.85,4.3)--(-2.85,3.6);
				\path [draw, thick, blue] (-5,5)--(-3,5)--(-1,5);
				\path [draw, thick, color=green!60!black] (-5,1)--(-3,1)--(-1,1);
				\path [draw] (-3,3)--(-1,3);
				\path [draw] (-3,1)--(-3,5);
				\draw[dashed] (-3,3) ellipse (1.3 and 2.3);
				\node at (-1,3)[below]{$k$};
				\node at (-5,5)[left]{$p_1$};
				\node at (-5,1)[left]{$p_2$};
				\node at (-2.8,4)[left]{$q_1$};
				\node at (-2.8,2)[left]{$-q_1-k$};
			\end{tikzpicture}
		\end{gathered}
		\right] \\
		&\times\int \frac{d^Dq_4}{(2\pi)^D}2\pi\delta(2p_1\cdot q_4)2\pi\delta(2p_2\cdot (q_4-k))
		e^{ib\cdot q_4}
		\begin{gathered}
			\begin{tikzpicture}[scale=.5]
				\draw[<-] (4.8,5.17)--(4.2,5.17);
				\draw[<-] (1,5.15)--(1.6,5.15);
				\draw[<-] (1,3.15)--(1.6,3.15);
				\draw[<-] (1,.85)--(1.6,.85);
				\draw[<-] (4.8,.83)--(4.2,.83);
				\draw[<-] (2.85,1.7)--(2.85,2.4);
				\draw[<-] (2.85,4.3)--(2.85,3.6);
				\path [draw, thick, blue] (5,5)--(3,5)--(1,5);
				\path [draw, thick, color=green!60!black] (5,1)--(3,1)--(1,1);
				\path [draw] (3,3)--(1,3);
				\path [draw] (3,1)--(3,5);
				\draw[dashed] (3,3) ellipse (1.3 and 2.3);
				\node at (5,5)[right]{$p_4$};
				\node at (5,1)[right]{$p_3$};
				\node at (2.8,4)[right]{$q_4$};
				\node at (2.8,2)[right]{$k-q_4$};
				\node at (1,3)[below]{$-k$};
			\end{tikzpicture}
		\end{gathered}
	\end{split}
\end{equation}
where we have already appropriately ``flipped'' the $2\to3$ amplitude in the second line by rewriting it as a $3\to2$ one.
Shifting $q_4=q-q_1$, and using $\delta(2p_1\cdot q_1)\delta(2p_1\cdot q_4)=\delta(2p_1\cdot q_1)\delta(2p_1\cdot q)$, we have
\begin{align}\label{clearsub}
	\begin{split}
		&i\boldsymbol{J}^{(o)}_{\alpha\beta}
		=
		\int \frac{d^Dq}{(2\pi)^D}2\pi\delta(2p_1\cdot q)
		e^{ib\cdot q}
		\int_{k} 
		\int \frac{d^Dq_1}{(2\pi)^D}2\pi\delta(2p_1\cdot q_1)
		\\
		&\times
		k_{[\alpha}
		\frac{\partial}{\partial k^{\beta]}}
		\left[ 
		2\pi\delta(2p_2\cdot (q_1+k))
		\begin{gathered}
			\begin{tikzpicture}[scale=.5]
				\draw[<-] (-4.8,5.17)--(-4.2,5.17);
				\draw[<-] (-1,5.15)--(-1.6,5.15);
				\draw[<-] (-1,3.15)--(-1.6,3.15);
				\draw[<-] (-1,.85)--(-1.6,.85);
				\draw[<-] (-4.8,.83)--(-4.2,.83);
				\draw[<-] (-2.85,1.7)--(-2.85,2.4);
				\draw[<-] (-2.85,4.3)--(-2.85,3.6);
				\path [draw, thick, blue] (-5,5)--(-3,5)--(-1,5);
				\path [draw, thick, color=green!60!black] (-5,1)--(-3,1)--(-1,1);
				\path [draw] (-3,3)--(-1,3);
				\path [draw] (-3,1)--(-3,5);
				\draw[dashed] (-3,3) ellipse (1.3 and 2.3);
				\node at (-1,3)[below]{$k$};
				\node at (-5,5)[left]{$p_1$};
				\node at (-5,1)[left]{$p_2$};
				\node at (-2.8,4)[left]{$q_1$};
				\node at (-2.8,2)[left]{$-q_1-k$};
			\end{tikzpicture}
		\end{gathered}
		\right] 
		2\pi\delta(2p_2\cdot (q-q_1-k))
		\begin{gathered}
			\begin{tikzpicture}[scale=.5]
				\draw[<-] (4.8,5.17)--(4.2,5.17);
				\draw[<-] (1,5.15)--(1.6,5.15);
				\draw[<-] (1,3.15)--(1.6,3.15);
				\draw[<-] (1,.85)--(1.6,.85);
				\draw[<-] (4.8,.83)--(4.2,.83);
				\draw[<-] (2.85,1.7)--(2.85,2.4);
				\draw[<-] (2.85,4.3)--(2.85,3.6);
				\path [draw, thick, blue] (5,5)--(3,5)--(1,5);
				\path [draw, thick, color=green!60!black] (5,1)--(3,1)--(1,1);
				\path [draw] (3,3)--(1,3);
				\path [draw] (3,1)--(3,5);
				\draw[dashed] (3,3) ellipse (1.3 and 2.3);
				\node at (5,5)[right]{$p_4$};
				\node at (5,1)[right]{$p_3$};
				\node at (2.8,4)[right]{$q-q_1$};
				\node at (2.8,2)[right]{$q_1+k-q$};
				\node at (1,3)[below]{$-k$};
			\end{tikzpicture}
		\end{gathered}
	\end{split}
	\nonumber
\end{align}
When we let the differential operator act on the square bracket, we need to distinguish two types of terms.
For the terms where the derivative acts on the amplitude, we can use again
\begin{equation}\label{}
	\delta(2p_2\cdot(q_1+k))\delta(2p_2\cdot(q-q_1-k))=\delta(2p_2\cdot(q_1+k))\delta(2p_2\cdot q)\,.
\end{equation} 
Instead, for those where the derivative acts on the delta function we can use the following property,
\begin{equation}\label{}
	\begin{split}
		\frac{\partial\delta(2p_2\cdot (q_1+k))}{\partial k^\beta}\,\delta(2p_2\cdot (q-q_1-k))
		&=\frac{\partial\delta(2p_2\cdot (q_1+k))}{\partial k^\beta}\,\delta(2p_2\cdot q)
		\\
		&+
		\delta(2p_2\cdot (q_1+k))\,
		\frac{\partial\delta(2p_2\cdot q)}{\partial q^\beta}\,,
	\end{split}
\end{equation}
which follows from the distributional identity $\delta'(x)\delta(y-x)=\delta'(x)\delta(y)+\delta(x)\delta'(y)$.
In total we get three terms: the first one involves the standard Fourier transform in the usual transverse plane $\delta(2p_1\cdot q)\delta(2p_2\cdot q)$ and the derivative acts on the amplitude, the second one involves the standard Fourier transform and the derivative acts on the phase-space delta function $\delta(2p_2\cdot(q_1+k))$, while in the last one the derivative acts on one of the delta functions appearing in the Fourier transform, $\delta(2p_2\cdot q)$. Grouping together the first two terms, we thus find
\begin{equation}\label{Joalmost}
	\begin{split}
		i\boldsymbol{J}^{(o)}_{\alpha\beta}
		&=
		\operatorname{FT}
		\int
		\,k_{[\alpha}
		\frac{\partial }{\partial k^{\beta]}}
		\!\!	\left[ d(\text{LIPS})\!\!\!\!\!\!
		\begin{gathered}
			\begin{tikzpicture}[scale=.5]
				\draw[<-] (-4.8,5.17)--(-4.2,5.17);
				\draw[<-] (-1,5.15)--(-1.6,5.15);
				\draw[<-] (-1,3.15)--(-1.6,3.15);
				\draw[<-] (-1,.85)--(-1.6,.85);
				\draw[<-] (-4.8,.83)--(-4.2,.83);
				\draw[<-] (-2.85,1.7)--(-2.85,2.4);
				\draw[<-] (-2.85,4.3)--(-2.85,3.6);
				\path [draw, thick, blue] (-5,5)--(-3,5)--(-1,5);
				\path [draw, thick, color=green!60!black] (-5,1)--(-3,1)--(-1,1);
				\path [draw] (-3,3)--(-1,3);
				\path [draw] (-3,1)--(-3,5);
				\draw[dashed] (-3,3) ellipse (1.3 and 2.3);
				\node at (-1,3)[below]{$k$};
				\node at (-5,5)[left]{$p_1$};
				\node at (-5,1)[left]{$p_2$};
				\node at (-2.8,4)[left]{$q_1$};
				\node at (-2.8,2)[left]{$-q_1-k$};
			\end{tikzpicture}
		\end{gathered}
		\right]
		\begin{gathered}
			\begin{tikzpicture}[scale=.5]
				\draw[<-] (4.8,5.17)--(4.2,5.17);
				\draw[<-] (1,5.15)--(1.6,5.15);
				\draw[<-] (1,3.15)--(1.6,3.15);
				\draw[<-] (1,.85)--(1.6,.85);
				\draw[<-] (4.8,.83)--(4.2,.83);
				\draw[<-] (2.85,1.7)--(2.85,2.4);
				\draw[<-] (2.85,4.3)--(2.85,3.6);
				\path [draw, thick, blue] (5,5)--(3,5)--(1,5);
				\path [draw, thick, color=green!60!black] (5,1)--(3,1)--(1,1);
				\path [draw] (3,3)--(1,3);
				\path [draw] (3,1)--(3,5);
				\draw[dashed] (3,3) ellipse (1.3 and 2.3);
				\node at (5,5)[right]{$p_4$};
				\node at (5,1)[right]{$p_3$};
				\node at (2.8,4)[right]{$q-q_1$};
				\node at (2.8,2)[right]{$q_1+k-q$};
				\node at (1,3)[below]{$-k$};
			\end{tikzpicture}
		\end{gathered}
		\\
		&
		-
		\operatorname{FT}^{(2)}_{[\alpha}
		\int  d(\text{LIPS})
		k^{\phantom{(2)}}_{\beta]}
		\begin{gathered}
			\begin{tikzpicture}[scale=.5]
				\draw[<-] (-4.8,5.17)--(-4.2,5.17);
				\draw[<-] (-1,5.15)--(-1.6,5.15);
				\draw[<-] (-1,3.15)--(-1.6,3.15);
				\draw[<-] (-1,.85)--(-1.6,.85);
				\draw[<-] (-4.8,.83)--(-4.2,.83);
				\draw[<-] (-2.85,1.7)--(-2.85,2.4);
				\draw[<-] (-2.85,4.3)--(-2.85,3.6);
				\path [draw, thick, blue] (-5,5)--(-3,5)--(-1,5);
				\path [draw, thick, color=green!60!black] (-5,1)--(-3,1)--(-1,1);
				\path [draw] (-3,3)--(-1,3);
				\path [draw] (-3,1)--(-3,5);
				\draw[dashed] (-3,3) ellipse (1.3 and 2.3);
				\node at (-1,3)[right]{$k$};
				\node at (-5,5)[left]{$p_1$};
				\node at (-5,1)[left]{$p_2$};
				\node at (-2.8,4)[left]{$q_1$};
				\node at (-2.8,2)[left]{$-q_1-k$};
				\draw[<-] (4.8,5.17)--(4.2,5.17);
				\draw[<-] (1,5.15)--(1.6,5.15);
				\draw[<-] (1,3.15)--(1.6,3.15);
				\draw[<-] (1,.85)--(1.6,.85);
				\draw[<-] (4.8,.83)--(4.2,.83);
				\draw[<-] (2.85,1.7)--(2.85,2.4);
				\draw[<-] (2.85,4.3)--(2.85,3.6);
				\path [draw, thick, red] (.3,0)--(.3,6);
				\path [draw, thick, blue] (5,5)--(3,5)--(1,5);
				\path [draw, thick, color=green!60!black] (5,1)--(3,1)--(1,1);
				\path [draw] (3,3)--(1,3);
				\path [draw] (3,1)--(3,5);
				\draw[dashed] (3,3) ellipse (1.3 and 2.3);
				\node at (5,5)[above]{$q-p_1$};
				\node at (2.8,4)[right]{$q-q_1$};
				\node at (2.8,2)[right]{$k-q+q_1$};
			\end{tikzpicture}
		\end{gathered}
	\end{split}
\end{equation}
with $\operatorname{FT}_\alpha^{(2)}$ given by
\begin{equation}\label{FT2}
	\operatorname{FT}^{(2)}_{\alpha} \left[f_\mu(q)\right]
	\equiv
	\int \frac{d^Dq}{(2\pi)^D}\,2\pi\delta(2p_1\cdot q) \,2\pi\,\frac{\delta(2p_2\cdot q)}{\partial q^\alpha}\,e^{ib\cdot q}\,f_\mu(q)\,.
\end{equation}
This modified Fourier transform acts on a quantity that looks like the $q$-space expression of $\boldsymbol P^\alpha$. However, due to the derivative acting on $\delta(2p_2\cdot q)$, we can no longer calculate this quantity by assuming $v_2\cdot q=0$.  
To evaluate \eqref{FT2}, let us start by introducing velocities as in \eqref{eq:velocities}, $p_1^\mu=-m_1 v_1^\mu$ and $p_2^\mu=-m_2 v_2^\mu$ so that
\begin{equation}\label{}
	\operatorname{FT}^{(2)\alpha} \left[f_\mu(q)\right]
	=
	\frac{	u^\alpha_2}{4m_1m_2}
	\int \frac{d^Dq}{(2\pi)^D}\,2\pi\delta(v_1\cdot q) \,2\pi\delta'(v_2\cdot q)\,e^{ib\cdot q}\,f_\mu(q)\,.
\end{equation}
Proceeding as in \ref{usefulFT}, and in particular decomposing the integrated momentum $q^\mu$ as in \eqref{decomposition}, we find
\begin{equation}\label{}
	\operatorname{FT}^{(2)}_{\alpha} \left[f_\mu(q)\right]
	=
	-
	v_{2\alpha}
	\int \frac{d^{D-2}q_\perp}{(2\pi)^{D-2}}\,\frac{\delta(q_{\parallel1})dq_{\parallel1} \,2\pi\delta'(q_{\parallel2})dq_{\parallel2}}{4m_1m_2\sqrt{\sigma^2-1}}\,e^{ib\cdot q_\perp}\,f_{\mu}(q)\,.
\end{equation}
Integrating by parts (i.e.~using the definition of the $\delta'$ distribution) we can recast this as the ordinary Fourier transform of a derivative, obtaining
\begin{equation}\label{FT2toFT}
	\operatorname{FT}^{(2)}_{\alpha} \left[f_\mu(q)\right]
	=
	v_{2\alpha}
	\operatorname{FT}\left[
	\frac{\partial}{\partial q_{\parallel2}}f_{\mu}(q)
	\right].
\end{equation}
The second line of \eqref{Joalmost} can be simplified using \eqref{FT2toFT} so that
\begin{equation}\label{LL1Lp}
	\begin{split}
		i\boldsymbol{J}^{(o)}_{\alpha\beta}
		&=
		\operatorname{FT}
		\int_{ k}
		\,k_{[\alpha}
		\frac{\partial }{\partial k^{\beta]}}
		\!\!	\left[ d(\text{LIPS}')\!\!\!\!\!\!
		\begin{gathered}
			\begin{tikzpicture}[scale=.5]
				\draw[<-] (-4.8,5.17)--(-4.2,5.17);
				\draw[<-] (-1,5.15)--(-1.6,5.15);
				\draw[<-] (-1,3.15)--(-1.6,3.15);
				\draw[<-] (-1,.85)--(-1.6,.85);
				\draw[<-] (-4.8,.83)--(-4.2,.83);
				\draw[<-] (-2.85,1.7)--(-2.85,2.4);
				\draw[<-] (-2.85,4.3)--(-2.85,3.6);
				\path [draw, thick, blue] (-5,5)--(-3,5)--(-1,5);
				\path [draw, thick, color=green!60!black] (-5,1)--(-3,1)--(-1,1);
				\path [draw] (-3,3)--(-1,3);
				\path [draw] (-3,1)--(-3,5);
				\draw[dashed] (-3,3) ellipse (1.3 and 2.3);
				\node at (-1,3)[below]{$k$};
				\node at (-5,5)[left]{$p_1$};
				\node at (-5,1)[left]{$p_2$};
				\node at (-2.8,4)[left]{$q_1$};
				\node at (-2.8,2)[left]{$-q_1-k$};
			\end{tikzpicture}
		\end{gathered}
		\right]
		\begin{gathered}
			\begin{tikzpicture}[scale=.5]
				\draw[<-] (4.8,5.17)--(4.2,5.17);
				\draw[<-] (1,5.15)--(1.6,5.15);
				\draw[<-] (1,3.15)--(1.6,3.15);
				\draw[<-] (1,.85)--(1.6,.85);
				\draw[<-] (4.8,.83)--(4.2,.83);
				\draw[<-] (2.85,1.7)--(2.85,2.4);
				\draw[<-] (2.85,4.3)--(2.85,3.6);
				\path [draw, thick, blue] (5,5)--(3,5)--(1,5);
				\path [draw, thick, color=green!60!black] (5,1)--(3,1)--(1,1);
				\path [draw] (3,3)--(1,3);
				\path [draw] (3,1)--(3,5);
				\draw[dashed] (3,3) ellipse (1.3 and 2.3);
				\node at (5,5)[right]{$p_4$};
				\node at (5,1)[right]{$p_3$};
				\node at (2.8,4)[right]{$q-q_1$};
				\node at (2.8,2)[right]{$q_1+k-q$};
				\node at (1,3)[below]{$-k$};
			\end{tikzpicture}
		\end{gathered}
		\\
		&
		-
		v_{2[\alpha}
		\operatorname{FT} 
		\frac{\partial}{\partial q_{\parallel2}}
		\int  d(\text{LIPS})
		k^{\phantom{(2)}}_{\beta]}
		\begin{gathered}
			\begin{tikzpicture}[scale=.5]
				\draw[<-] (-4.8,5.17)--(-4.2,5.17);
				\draw[<-] (-1,5.15)--(-1.6,5.15);
				\draw[<-] (-1,3.15)--(-1.6,3.15);
				\draw[<-] (-1,.85)--(-1.6,.85);
				\draw[<-] (-4.8,.83)--(-4.2,.83);
				\draw[<-] (-2.85,1.7)--(-2.85,2.4);
				\draw[<-] (-2.85,4.3)--(-2.85,3.6);
				\path [draw, thick, blue] (-5,5)--(-3,5)--(-1,5);
				\path [draw, thick, color=green!60!black] (-5,1)--(-3,1)--(-1,1);
				\path [draw] (-3,3)--(-1,3);
				\path [draw] (-3,1)--(-3,5);
				\draw[dashed] (-3,3) ellipse (1.3 and 2.3);
				\node at (-1,3)[right]{$k$};
				\node at (-5,5)[left]{$p_1$};
				\node at (-5,1)[left]{$p_2$};
				\node at (-2.8,4)[left]{$q_1$};
				\node at (-2.8,2)[left]{$-q_1-k$};
				\draw[<-] (4.8,5.17)--(4.2,5.17);
				\draw[<-] (1,5.15)--(1.6,5.15);
				\draw[<-] (1,3.15)--(1.6,3.15);
				\draw[<-] (1,.85)--(1.6,.85);
				\draw[<-] (4.8,.83)--(4.2,.83);
				\draw[<-] (2.85,1.7)--(2.85,2.4);
				\draw[<-] (2.85,4.3)--(2.85,3.6);
				\path [draw, thick, red] (.3,0)--(.3,6);
				\path [draw, thick, blue] (5,5)--(3,5)--(1,5);
				\path [draw, thick, color=green!60!black] (5,1)--(3,1)--(1,1);
				\path [draw] (3,3)--(1,3);
				\path [draw] (3,1)--(3,5);
				\draw[dashed] (3,3) ellipse (1.3 and 2.3);
				\node at (5,5)[above]{$q-p_1$};
				\node at (2.8,4)[right]{$q-q_1$};
				\node at (2.8,2)[right]{$k-q+q_1$};
			\end{tikzpicture}
		\end{gathered}\,.
	\end{split}
\end{equation}
The dependence on $q_{\parallel2}$ in the integrand appearing in the second line comes from $v_2\cdot q$ and from the invariants that involve $q_{\parallel2}$ and a loop momentum, or in $q^2$ itself: $I(v_2\cdot q,q\cdot\ell_1,q\cdot \ell_2,q^2) $. Then, since
\begin{equation}\label{}
	\begin{split}
		v_2\cdot q &= -q_{\parallel2}\,,\\
		\ell_1\cdot q &= \ell_1\cdot q_\perp + \ell_1\cdot \check{v}_2\,q_{\parallel2}\,,\\
		\ell_2\cdot q &= \ell_2\cdot q_\perp + \ell_2\cdot \check{v}_2\,q_{\parallel2}\,,\\
		q^2&=q_\perp^2 + q_{\parallel2}^2
	\end{split}
\end{equation}
we find
\begin{equation}\label{}
	\frac{\partial I}{\partial q_{\parallel2}}
	=
	-\frac{\partial I}{\partial v_2\cdot q}
	+
	\ell_1\cdot\check{v}_2\,
	\frac{\partial I}{\partial (q\cdot\ell_1)}
	+
	\ell_2\cdot\check{v}_2\,
	\frac{\partial I}{\partial (q\cdot\ell_2)}\,,
\end{equation}
because $q^2$ is quadratic in $q_{\parallel2}$ so its derivative vanishes at $q_{\parallel2}=0$.

Let us comment as to why there are no ambiguities associated to derivatives of delta functions. As is clear from \eqref{ft223DEF}, one is allowed to choose different expressions for the five-point amplitude provided that they only differ by terms that vanish when 
\begin{equation}\label{}
	v_1\cdot\ell_1=0\quad \text{or}\quad
	v_2\cdot\ell_1+v_2\cdot\ell_2=0\,,
\end{equation}
thanks to the delta functions.
Different choices for these modifications will change each line in \eqref{LL1Lp} separately, but will not change the total sum, provided of course that the \emph{same} choice is made consistently in both lines.
The easiest option is to just treat $q_1$ and $k$ as completely independent integration variables, and let the derivative of the delta functions automatically take care of their interdependence on-shell due to the longitudinal components.

Although we discussed the above steps for $b^\alpha_2=0$, we present the final results in a frame where $b^\alpha_1+b^\alpha_2=0$, related to the previous one by a translation by $-b^\alpha/2$ (see Eqs.~\eqref{translation}, \eqref{translJ}), where particle-interchange symmetry is manifest.
Defining $\mathcal E_{\pm}$ and $\mathcal F$ in terms of the functions $\mathcal E$ in \eqref{calE} and  $\mathcal C$ given by
\begin{equation}
	\begin{split}
		\frac{\mathcal C}{\pi} & = g_1 + g_2 \log\frac{\sigma+1}{2} + g_3 \frac{\sigma\,\operatorname{arccosh}\sigma}{2\sqrt{\sigma^2-1}}\,, \\
		g_1 &= \frac{105\sigma^7-411 \sigma^6+240\sigma^5+537\sigma^4-683\sigma^3+111\sigma^2+386\sigma-237}{24(\sigma^2-1)^{2}}\,,\\
		g_2 &= \frac{35 \sigma ^5-90 \sigma ^4-70 \sigma ^3+16 \sigma ^2+155 \sigma -62}{4(\sigma^2-1)}\,,\\
		g_3 &= -\frac{(2 \sigma ^2-3) \left(35 \sigma ^5-60 \sigma ^4-70 \sigma ^3+72 \sigma ^2+19 \sigma -12\right)}{4 \left(\sigma^2-1\right)^{2}}
	\end{split}
\end{equation}
\cite{Herrmann:2021lqe,Herrmann:2021tct,Manohar:2022dea}
by letting 
\begin{equation}\label{}
	\mathcal C {\sqrt{\sigma^2-1}} = -\mathcal E_+ +\sigma \mathcal E_-\,,\qquad
	\mathcal F=\mathcal E_+-\tfrac12\,\mathcal E= - \mathcal E_- +\tfrac12 \mathcal E\,,
\end{equation}
we find
\begin{equation}\label{JabF}
	\boldsymbol{J}^{\alpha\beta}\simeq\frac{G^3m_1^2m_2^2}{b^3} \,\mathcal F  (\sigma) \left(
	b^{[\alpha}\check v_{1}^{\beta]}
	-
	b^{[\alpha}\check v_{2}^{\beta]}
	\right) , \qquad \mathcal{F} (\sigma) = \frac{(\sigma-1) \mathcal{E} - 2 \sqrt{\sigma^2-1} \mathcal{C}}{2(\sigma+1)} \,.
\end{equation}
After a translation  
\begin{equation}\label{}
	a^\mu=\frac{E_2 - E_1}{2(E_1+E_2)} b^\mu\,,
\end{equation}
which places the center of mass (or rather ``center of energy'') in the origin of the transverse plane,
Eq.~\eqref{JabF} becomes
	\begin{equation}
\begin{split}
\boldsymbol{J}^{\alpha\beta} &\simeq  \frac{G^3 m_1 m_2}{b^3} \Bigg[ b^{[\alpha} (m_1 p_2- m_2 p_1)^{\beta]} \frac{\mathcal{C} (\sigma)}{\sqrt{\sigma^2-1} }\\
 &+\frac{m_1 m_2}{E^2}  \mathcal{E} (\sigma) \frac{\sigma-1}{\sigma^2-1} 
\left(b^{[\alpha} p_1^{\beta]} (m_1+m_2 \sigma) - b^{[\alpha} p_2^{\beta]} (m_2+m_1 \sigma) \right) \Bigg]
\label{Jabcomple}
\end{split}
\end{equation}
that, in the center of mass system where $-p_1 = (E_1, p)$ and $-p_2 = (E_2, -p)$,  reproduces Eq.~(15) of Ref.~\cite{Manohar:2022dea} \emph{except} for the static (zero-frequency) modes. 
Adding \eqref{Jabcomple} with such contribution as given by \eqref{J2PMgen} expanded up to $\mathcal O(G^3)$, one reproduces the full result of Ref.~\cite{Manohar:2022dea},

Let us now provide the analogous formulas expressing the angular momentum that is lost by each particle, which we derive in full detail.
Acting with the angular momentum operator 
\begin{equation}\label{L1operator}
	L_{(1)\alpha\beta}  = -i \int_{ k_1}a^\dagger_1(k_1)k_{1[\alpha}\frac{\partial a_1(k_1)}{\partial k_1^{\beta]}}
\end{equation}
on the initial state \eqref{DO1} and using the commutation relations, one finds
\begin{equation}\label{}
	\begin{split}
		L_{(1)}^{\alpha\beta}|\psi\rangle  
		&= 
		\int_{- p_1}\int_{- p_2}
		\left(-ip_1^{[\alpha}\partial_{p_1}^{\beta]}\Phi_1(-p_1)\right)\Phi_2(-p_2)\,e^{ib_1\cdot p_1+ib_2\cdot p_2}|-p_1,-p_2\rangle
		\\
		&
		+\int_{- p_1}\int_{- p_2}b_1^{[\alpha}(-p_1^{\beta]})\Phi_1(-p_1)\Phi_2(-p_2)\,e^{ib_1\cdot p_1+ib_2\cdot p_2}|-p_1,-p_2\rangle\,.
	\end{split}
\end{equation}
The corresponding expectation value is therefore
\begin{equation}\label{Lin1}
	\langle \psi|L_{(1)}^{\alpha\beta}|\psi\rangle =
	\int_{- p_1}\Phi_1^\ast(-p_1)
	\left(-ip_1^{[\alpha}\partial_{p_1}^{\beta]}\Phi_1(-p_1)\right)+
	\int_{- p_1}b_1^{[\alpha}(-p_1^{\beta]})|\Phi_1(-p_1)|^2\,.
\end{equation}
Let us now act on the final state dictated by the eikonal operator. For notational simplicity, we discuss the action of the eikonal operator \eqref{eikopv1} which does not include the static-mode contributions involving $F^{\mu\nu}$. They can be easily reinstated by replacing $\mathcal W_j(x_1,x_2,k)$ with $\theta(k^0-\omega^\ast)\mathcal W(x_1,x_2,k)+ \theta(\omega^\ast-k^0)f_j(x_1,x_2,k)$. Doing this and replacing $\operatorname{Re}2\delta$ with $2\tilde\delta$, one obtains the result for \eqref{eikopv2}. 
For simplicity, we also suppress the additional integrals over $x$ and $Q$, which do not play a role in the derivation. 
With this proviso, we start by calculating
\begin{equation}\label{}
	\begin{split}
		L_{(1)\alpha\beta} S|\psi\rangle
		&=
		\int_{ p_3}
		\int_{ p_4}
		e^{-ib_1\cdot p_4-ib_2\cdot p_3}\, |p_3,p_4\rangle\\
		&\times
		\int\frac{d^DQ_1}{(2\pi)^D}\int\frac{d^DQ_2}{(2\pi)^D}\int d^Dx_1 \int d^Dx_2
		\,e^{i (b_1 - x_1)\cdot Q_1+ i(b_2-x_2)\cdot Q_2+2i\delta(b)}\\
		&\times\left(
		b_{1[\alpha|}p_{4|\beta]}
		+
		p_{4[\alpha}\frac{\partial 2\delta(b)}{\partial p_4^{\beta]}}
		+\int_{{k}}p_{4[\alpha}\frac{\partial}{\partial p_4^{\beta]}}\mathcal{W}(x_1,x_2,k)\,a^\dagger(k)
		\right)\\
		&\times
		e^{i\int_{{k}}\mathcal{W}(x_1,x_2,k)\,a^\dagger(k)}
		|0\rangle\,\Phi_1(p_4-Q_1)\,\Phi_2(p_3-Q_2)
		\\
		&+
		\int_{ p_3}
		\int_{ p_4}
		e^{-ib_1\cdot p_4-ib_2\cdot p_3}\, |p_3,p_4\rangle\\
		&\times
		\int\frac{d^DQ_1}{(2\pi)^D}\int\frac{d^DQ_2}{(2\pi)^D}\int d^Dx_1 \int d^Dx_2
		\,e^{i (b_1 - x_1)\cdot Q_1+ i(b_2-x_2)\cdot Q_2+2i\delta(b)}\\
		&\times\left(
		-i
		p_{4[\alpha}\frac{\partial}{\partial p_4^{\beta]}}\Phi_1(p_4-Q_1)
		\right)
		e^{i\int_{{k}}\mathcal{W}(x_1,x_2,k)\,a^\dagger(k)}
		|0\rangle\,\,\Phi_2(p_3-Q_2)
	\end{split}
\end{equation}
In order to make the expression slightly more compact, we used \eqref{3pcbspace} in the exponents to reabsorb the imaginary part of $2\delta$ arising from the reordering of the exponential factors.
In the last line we can use
\begin{equation}\label{stepQ1}
	p_{4[\alpha}\frac{\partial}{\partial p_4^{\beta]}}\Phi_1(p_4-Q_1)
	=
	(p_{4}-Q_1)_{[\alpha}\frac{\partial}{\partial p_4^{\beta]}}\Phi_1(p_4-Q_1) 
	- 
	Q_{1[\alpha}\frac{\partial}{\partial Q_1^{\beta]}}\Phi_1(p_4-Q_1) 
\end{equation}
and integrate by parts the second term.
Doing this and defining the operator
\begin{equation}\label{operatorO1}
	O_{(1)\alpha\beta} = 
	p_{4[\alpha}\frac{\partial}{\partial p_4^{\beta]}}+Q_{1[\alpha}\frac{\partial}{\partial Q_1^{\beta]}}
\end{equation}
we can write
\begin{equation}\label{}
	\begin{split}
		&L_{(1)\alpha\beta} S|\psi\rangle
		=
		\int_{ p_3}
		\int_{ p_4}
		e^{-ib_1\cdot p_4-ib_2\cdot p_3}\, |p_3,p_4\rangle\\
		&\times
		\int\frac{d^DQ_1}{(2\pi)^D}\int\frac{d^DQ_2}{(2\pi)^D}\int d^Dx_1 \int d^Dx_2
		\,e^{i (b_1 - x_1)\cdot Q_1+ i(b_2-x_2)\cdot Q_2+2i\delta(b)}\\
		&\times\left(
		x_{1[\alpha|} Q_{1|\beta]}
		+
		O_{(1)\alpha\beta} 2\delta(b)
		+
		\int_{{k}} O_{(1)\alpha\beta}\mathcal{W}(x_1,x_2,k)\,a^\dagger(k)
		\right)\\
		&\times
		e^{i\int_{{k}}\mathcal{W}(x_1,x_2,k)\,a^\dagger(k)}
		|0\rangle\,\Phi_1(p_4-Q_1)\,\Phi_2(p_3-Q_2)
		\\
		&+
		\int_{ p_3}
		\int_{ p_4}
		e^{-ib_1\cdot p_4-ib_2\cdot p_3}\, |p_3,p_4\rangle\\
		&\times
		\int\frac{d^DQ_1}{(2\pi)^D}\int\frac{d^DQ_2}{(2\pi)^D}\int d^Dx_1 \int d^Dx_2
		\,e^{i (b_1 - x_1)\cdot Q_1+ i(b_2-x_2)\cdot Q_2+2i\delta(b)}\\
		&\times\left(
		b_{1[\alpha|}(p_4-Q_1)_{|\beta]}
		-i
		(p_{4}-Q_1)_{[\alpha}\frac{\partial}{\partial p_4^{\beta]}}\Phi_1(p_4-Q_1)
		\right)
		e^{i\int_{{k}}\mathcal{W}(x_1,x_2,k)\,a^\dagger(k)}
		|0\rangle\,\,\Phi_2(p_3-Q_2)\,.
	\end{split}
\end{equation}
For the expectation value, we find
\begin{equation}\label{}
	\begin{split}
		&\langle \psi|S^\dagger L_{(1)\alpha\beta} S|\psi\rangle\\
		&=
		\int_{ p_3,  p_4}
		\int\frac{d^DQ'_1}{(2\pi)^D}\int\frac{d^DQ'_2}{(2\pi)^D}\int d^Dx'_1 \int d^Dx'_2
		\,e^{-i (b_1 - x'_1)\cdot Q'_1- i(b_2-x'_2)\cdot Q'_2-i\operatorname{Re}2\delta(b')}\\
		&\times
		\int\frac{d^DQ_1}{(2\pi)^D}\int\frac{d^DQ_2}{(2\pi)^D}\int d^Dx_1 \int d^Dx_2
		\,e^{i (b_1 - x_1)\cdot Q_1+ i(b_2-x_2)\cdot Q_2+i\operatorname{Re}2\delta(b)}
		\\
		&\times\left(
		x_{1[\alpha|} Q_{1|\beta]}
		+
		O_{(1)\alpha\beta} 2\delta(b)
		-i\int_{{k}}\mathcal{W}^\ast(x'_1,x'_2,k) O_{(1)\alpha\beta}
		\mathcal{W}(x_1,x_2,k)
		\right)|0\rangle
		\\
		&\times
		e^{-\operatorname{Im}2\delta(b')-\operatorname{Im}2\delta(b)+\int_{ k}\mathcal{W}^\ast(x'_1,x'_2,k)\mathcal{W}(x_1,x_2,k)}
		\\
		&\times
		\Phi_1^\ast(p_4-Q'_1)\,\Phi_2^\ast(p_3-Q'_2)
		\,\Phi_1(p_4-Q_1)\,\Phi_2(p_3-Q_2)
		\\
		&+
		\int_{ p_3,  p_4}
		\int\frac{d^DQ'_1}{(2\pi)^D}\int\frac{d^DQ'_2}{(2\pi)^D}\int d^Dx'_1 \int d^Dx'_2
		\,e^{-i (b_1 - x'_1)\cdot Q'_1- i(b_2-x'_2)\cdot Q'_2-i\operatorname{Re}2\delta(b')}\\
		&\times
		\int\frac{d^DQ_1}{(2\pi)^D}\int\frac{d^DQ_2}{(2\pi)^D}\int d^Dx_1 \int d^Dx_2
		\,e^{i (b_1 - x_1)\cdot Q_1+ i(b_2-x_2)\cdot Q_2+i\operatorname{Re}2\delta(b)}
		\\
		&\times\left(
		b_{1[\beta|}
		(p_{4}-Q_1)_{|\alpha]}\Phi_1(p_4-Q_1)
		-i
		(p_{4}-Q_1)_{[\alpha}\frac{\partial}{\partial p_4^{\beta]}}\Phi_1(p_4-Q_1)
		\right)
		\\
		&\times
		e^{-\operatorname{Im}2\delta(b')-\operatorname{Im}2\delta(b)+\int_{ k}\mathcal{W}(x'_1,x'_2,k)\mathcal{W}(x_1,x_2,k)}
		\\
		&\times
		\Phi_1^\ast(p_4-Q'_1)\,\Phi_2^\ast(p_3-Q'_2)\,\Phi_2(p_3-Q_2)\,.
	\end{split}
\end{equation}
At the saddle point,
\begin{equation}\label{}
	\begin{split}
		&\langle \psi|S^\dagger L_{(1)\alpha\beta} S|\psi\rangle
		\\
		&= 
		\int_{ p_3,  p_4} 
		|\Phi_1(p_4-Q_1)|^2\,|\Phi_2(p_3-Q_2)|^2\\
		&\times  \left(
		x_{1[\alpha|}Q_{1|\beta]}
		+
		O_{(1)\alpha\beta} \operatorname{Re}2\delta(b)
		-i\int_{{k}}\mathcal{W}^\ast(x_1,x_2,k) \overset{\leftrightarrow}{O}_{(1)\alpha\beta}\mathcal{W}(x_1,x_2,k)
		\right)\\
		&
		+\int_{ p_3,  p_4} 
		\Phi_1(p_4-Q_1)^\ast\,|\Phi_2(p_3-Q_2)|^2\\
		&\times  \left(
		b_{1[\beta|}
		(p_{4}-Q_1)_{|\alpha]}\Phi_1(p_4-Q_1)
		-i
		(p_{4}-Q_1)_{[\alpha}\frac{\partial}{\partial p_4^{\beta]}}\Phi_1(p_4-Q_1)
		\right)
	\end{split}
\end{equation}
and shifting the integration like in Section~\ref{ssec:elasticimpulseandangularmomentum}, $p_4=Q_1-p_1$, $p_3=Q_2-p_2$ and recognizing the last line as the angular momentum of the initial state \eqref{Lin1}, we finally obtain
\begin{equation}\label{langleLrangleQ1}
	\begin{split}
		&\langle \psi|S^\dagger L_{(1)\alpha\beta} S|\psi\rangle
		-
		\langle \psi| L_{(1)\alpha\beta} |\psi\rangle
		\\
		&= 
		\int_{ -p_1,  -p_2} 
		|\Phi_1(-p_1)|^2\,|\Phi_2(-p_2)|^2\\
		&\times  \left(
		x_{1[\alpha|}Q_{1|\beta]}
		+
		O_{(1)\alpha\beta} \operatorname{Re}2\delta(b)
		-i\int_{{k}}\mathcal{W}^\ast(x_1,x_2,k) \overset{\leftrightarrow}{O}_{(1)\alpha\beta}\mathcal{W}(x_1,x_2,k)
		\right)
	\end{split}
\end{equation}
with $O_{(1)\alpha\beta}$ the operator \eqref{operatorO1}.

Performing the appropriate replacements discussed below \eqref{Lin1}, if we are interested in the angular momentum up to $G^3$ including the static modes, we thus have
\begin{equation}\label{DeltaL1new}
	\Delta L_{(1)}^{\alpha\beta}
	=
	x_{1[\alpha|}Q_{1|\beta]}
	+
	p_{4[\alpha]}\frac{\partial 2\tilde\delta(b)}{\partial p_4^{\beta]}}
	-i\int_{{k}}\tilde{\mathcal{A}}^{(5)} p_{4[\alpha}\frac{\partial}{p_4^{\beta]}}\tilde{\mathcal{A}}^{(5)}
	-i\int_{{k}} F^\ast O_{(1)\alpha\beta} F + \mathcal O(G^4)\,,
\end{equation}
where we have neglected the second term in the operator \eqref{operatorO1} in $O_{(1)\alpha\beta}2\tilde\delta$ and similarly in $\tilde{\mathcal{A}}^{(5)\ast} O_{(1)\alpha\beta}\tilde{\mathcal{A}}^{(5)}$, because it would give $\mathcal O(G^4)$ effects since $\tilde{\mathcal{A}}^{(5)\ast}\tilde{\mathcal{A}}^{(5)}$ is already $\mathcal O(G^3)$. Instead, as we shall see, it is important to keep it in $F^\ast O_{(1)\alpha\beta} F$ because it grants the identity
\begin{equation}\label{FdtoGd}
	O_{(1)\alpha\beta} F^{\mu\nu} = p_{4[\alpha}\frac{\partial G^{\mu\nu}}{\partial p_4^{\beta]}}+p_{1[\alpha}\frac{\partial G^{\mu\nu}}{\partial p_1^{\beta]}}\,,
\end{equation}
where $G^{\mu\nu}(p_4,p_3,p_1,p_2)$ is the ``soft factor'' seen as a function of four independent hard momenta, i.e.
\begin{equation}\label{FtoGid}
	F^{\mu\nu}(p_4,p_3,Q_1,Q_2)
	=
	G^{\mu\nu}(p_4,p_3,Q_1-p_4,Q_2-p_3)\,.
\end{equation}
In the following, we will calculate \eqref{DeltaL1new}, breaking it down into the following terms
\begin{equation}\label{thisDeltaL}
	\Delta L_1^{\alpha\beta} = \Delta L_{(1c)}^{\alpha\beta} + \Delta \boldsymbol{L}^{\alpha\beta}_1 + \Delta \mathcal L^{\alpha\beta}_1
\end{equation}
and similarly for particle 2.
The conservative term reads as follows, using that the difference between  $x_1^\mu$ and $b_1^\mu$ is aligned with $Q^\mu$ as can be seen from the saddle point condition \eqref{Saddle3} and from Eqs.~\eqref{DO11}, \eqref{DO12} ,
\begin{equation}\label{DeltaL1new-cons}
	\Delta L_{(1c)}^{\alpha\beta}
	=
	x_{1[\alpha|}Q_{|\beta]}
	+
	p_{4[\alpha]}\frac{\partial\operatorname{Re}2\tilde\delta(b)}{\partial p_4^{\beta]}}
	=
	b_{1[\alpha|}Q_{|\beta]}
	+
	p_{4[\alpha]}\frac{\partial\operatorname{Re}2\tilde\delta(b)}{\partial p_4^{\beta]}}\,.
\end{equation}
We already calculated \eqref{DeltaL1new-cons} in Eq.~\eqref{L1cons} and following.
The radiative terms is given by
\begin{equation}\label{}
	\Delta \boldsymbol L_{i}^{\alpha\beta} = \operatorname{Im}\boldsymbol{J}_{i}^{\alpha\beta}+b_i^{[\alpha}  \boldsymbol{Q}^{\beta]}_{i}
\end{equation}
where $\boldsymbol Q^\alpha_i$ is the radiative contribution to the impulse as in \eqref{Q1RRstart} and we defined the shorthand
\begin{equation}\label{DeltaLExpl}
	\boldsymbol{J}_{i\alpha\beta}
	=
	\int_{{k}}
	p_{i[\alpha}
	\frac{\partial\tilde{\mathcal{A}}^{(5)}}{\partial p_i^{\beta]}}
	\,
	\tilde{\mathcal{A}}^{(5)\ast}\,.
\end{equation}
Focusing for definiteness on particle 2 in a frame where $b_2^\alpha=0$, and proceeding in a manner entirely analogous to what we did for $\boldsymbol J^{\alpha\beta}$, one arrives at the following expression in terms of three-particle cuts,
\begin{equation}\label{LJ2}
	\begin{split}
		\boldsymbol{J}_{2\alpha\beta}
		&=
		\operatorname{FT}
		\int
		v_{2[\alpha}
		\frac{\partial }{\partial v_2^{\beta]}}
		\left[ d(\text{LIPS})
		\begin{gathered}
			\begin{tikzpicture}[scale=.5]
				\draw[<-] (-4.8,5.17)--(-4.2,5.17);
				\draw[<-] (-1,5.15)--(-1.6,5.15);
				\draw[<-] (-1,3.15)--(-1.6,3.15);
				\draw[<-] (-1,.85)--(-1.6,.85);
				\draw[<-] (-4.8,.83)--(-4.2,.83);
				\draw[<-] (-2.85,1.7)--(-2.85,2.4);
				\draw[<-] (-2.85,4.3)--(-2.85,3.6);
				\path [draw, thick, blue] (-5,5)--(-3,5)--(-1,5);
				\path [draw, thick, color=green!60!black] (-5,1)--(-3,1)--(-1,1);
				\path [draw] (-3,3)--(-1,3);
				\path [draw] (-3,1)--(-3,5);
				\draw[dashed] (-3,3) ellipse (1.3 and 2.3);
				\node at (-1,3)[below]{$k$};
				\node at (-5,5)[left]{$p_1$};
				\node at (-5,1)[left]{$p_2$};
				\node at (-2.8,4)[left]{$q_1$};
			\end{tikzpicture}
		\end{gathered}
		\right]
		\begin{gathered}
			\begin{tikzpicture}[scale=.5]
				\draw[<-] (4.8,5.17)--(4.2,5.17);
				\draw[<-] (1,5.15)--(1.6,5.15);
				\draw[<-] (1,3.15)--(1.6,3.15);
				\draw[<-] (1,.85)--(1.6,.85);
				\draw[<-] (4.8,.83)--(4.2,.83);
				\draw[<-] (2.85,1.7)--(2.85,2.4);
				\draw[<-] (2.85,4.3)--(2.85,3.6);
				\path [draw, thick, blue] (5,5)--(3,5)--(1,5);
				\path [draw, thick, color=green!60!black] (5,1)--(3,1)--(1,1);
				\path [draw] (3,3)--(1,3);
				\path [draw] (3,1)--(3,5);
				\draw[dashed] (3,3) ellipse (1.3 and 2.3);
				\node at (2.8,4)[right]{$q-q_1$};
			\end{tikzpicture}
		\end{gathered}
		\\
		&	+
		v_{2[\alpha}
		\operatorname{FT} 
		\frac{\partial}{\partial q_{\parallel2}}
		\int  d(\text{LIPS})
		(q_1+k)^{\phantom{2}}_{\beta]}
		\begin{gathered}
			\begin{tikzpicture}[scale=.5]
				\draw[<-] (-4.8,5.17)--(-4.2,5.17);
				\draw[<-] (-1,5.15)--(-1.6,5.15);
				\draw[<-] (-1,3.15)--(-1.6,3.15);
				\draw[<-] (-1,.85)--(-1.6,.85);
				\draw[<-] (-4.8,.83)--(-4.2,.83);
				\draw[<-] (-2.85,1.7)--(-2.85,2.4);
				\draw[<-] (-2.85,4.3)--(-2.85,3.6);
				\path [draw, thick, blue] (-5,5)--(-3,5)--(-1,5);
				\path [draw, thick, color=green!60!black] (-5,1)--(-3,1)--(-1,1);
				\path [draw] (-3,3)--(-1,3);
				\path [draw] (-3,1)--(-3,5);
				\draw[dashed] (-3,3) ellipse (1.3 and 2.3);
				\node at (-1,3)[below]{$k$};
				\node at (-5,5)[left]{$p_1$};
				\node at (-5,1)[left]{$p_2$};
				\node at (-2.8,4)[left]{$q_1$};
				\draw[<-] (3.35,5.17)--(2.75,5.17);
				\draw[<-] (-.45,5.15)--(.15,5.15);
				\draw[<-] (-.45,3.15)--(.15,3.15);
				\draw[<-] (-.45,.85)--(.15,.85);
				\draw[<-] (3.35,.83)--(2.75,.83);
				\draw[<-] (1.4,1.7)--(1.4,2.4);
				\draw[<-] (1.4,4.3)--(1.4,3.6);
				\path [draw, thick, red] (-.7,0)--(-.7,6);
				\path [draw, thick, blue] (3.55,5)--(1.55,5)--(-.45,5);
				\path [draw, thick, color=green!60!black] (3.55,1)--(1.55,1)--(-.45,1);
				\path [draw] (1.55,3)--(-.45,3);
				\path [draw] (1.55,1)--(1.55,5);
				\draw[dashed] (1.55,3) ellipse (1.3 and 2.3);
				\node at (1.35,4)[right]{$q-q_1$};
			\end{tikzpicture}
		\end{gathered}
	\end{split}
\end{equation}
and employing reverse-unitarity we obtain the result
\begin{equation}\label{deltabL1}
	\begin{split}
		\Delta  \boldsymbol L_{1}^{\alpha \beta}\simeq \frac{G^{3} m_{1}^{2} m_{2}^{2}}{b^{3}}\left[+\frac{\mathcal{E}_{+} b^{[\alpha}v_{1}^{\beta]}}{\sigma-1} -\frac{1}{2}\,\mathcal{E}\, b^{[\alpha} \check{v}_{2}^{\beta]}\right],\\
		\Delta  \boldsymbol L_{2}^{\alpha \beta}\simeq \frac{G^{3} m_{1}^{2} m_{2}^{2}}{b^{3}}\left[-\frac{\mathcal{E}_{+} b^{[\alpha}v_{2}^{\beta]}}{\sigma-1} +\frac{1}{2}\,\mathcal{E}\, b^{[\alpha} \check{v}_{1}^{\beta]}\right].
	\end{split}
\end{equation}
Of course the expression for particle $1$ is obtained by interchanging particle labels everywhere.
The radiative quantities \eqref{JabF} and \eqref{deltabL1} obey the balance law
\begin{equation}\label{}
	\boldsymbol J^{\alpha\beta}
	+
	\Delta \boldsymbol L_{1}^{\alpha\beta}
	+
	\Delta \boldsymbol L_{2}^{\alpha\beta}
	=
	0\,.
\end{equation} 
Finally, the static-mode contributions to the changes in the angular momenta are given by the following formula valid up to $\mathcal O(G^3)$, 
\begin{equation}\label{DcalL}
	\begin{split}
		\Delta \mathcal L_{1}^{\alpha\beta} =
		\int_{{k}}
		F^\ast
		\left(
		p_{4[\alpha}
		\frac{\overset{\leftrightarrow}{\partial}}{\partial p_4^{\beta]}} 
		+
		Q_{1[\alpha}
		\frac{\overset{\leftrightarrow}{\partial}}{\partial Q_1^{\beta]}} 
		\right) F
		+
		b_1^{[\alpha}\mathcal{Q}^{\beta]}_{1}\,,
	\end{split}
\end{equation} 
where $\mathcal Q_1^\beta$ is the static contribution to the impulse, \eqref{QRRtobecomp}.
Under a translation \eqref{translation}, $\Delta \mathcal L_{i}^{\alpha\beta}\to \Delta \mathcal L_{i}^{\alpha\beta}+a_{\phantom{i}}^{[\alpha}\mathcal Q_{i}^{\beta]}$. Making use of the identity \eqref{FdtoGd}, we see that
\begin{equation}\label{}
-i\int_{{k}}
	F^\ast
	\left(
	p_{4[\alpha}
	\frac{\overset{\leftrightarrow}{\partial}}{\partial p_4^{\beta]}} 
	+
	Q_{1[\alpha}
	\frac{\partial}{\partial Q_1^{\beta]}} 
	\right) F
	=
	-i \int_{{k}}
	G^\ast
	p_{1[\alpha}\frac{\overset{\leftrightarrow}{\partial}}{\partial p_1^{\beta]}} G
	-i
	\int_{{k}}
	G^\ast
	p_{4[\alpha}\frac{\overset{\leftrightarrow}{\partial}}{\partial p_4^{\beta]}} G
	= 
	J_{(1)\alpha\beta} + J_{(4)\alpha\beta}\,,
\end{equation}
where we have defined
\begin{equation}\label{}
	J_{(n)\alpha\beta} = -i\int_{{k}}
	G^\ast
	p_{n[\alpha}\frac{\overset{\leftrightarrow}{\partial}}{\partial p_n^{\beta]}} G\,.
\end{equation}
Of course, a similar manipulation works for particle 2.
The integrals appearing in these combinations can be evaluated in the same way as those we discussed in the previous section and give the following result for $m=1,2,3,4$,
\begin{equation}\label{}
	2\eta_m J_{(m)}^{\alpha\beta}
	=
	\sum_{\eta_n=-\eta_m} 
	c_{nm}\,
	p_n^{[\alpha}p_m^{\beta]} 
	-
	\sum_{\substack{\eta_n=\eta_m\\n\neq m}}d_{nm} \, p_n^{[\alpha}p_m^{\beta]}\,,
\end{equation}
For instance,
\begin{equation}\label{}
	2 J^{\alpha\beta}_4 = c_{14} p_1^{[\alpha} p_4^{\beta]} + c_{24} p_2^{[\alpha} p_4^{\beta]} - d_{12} p_3^{[\alpha} p_4^{\beta]}\,.
\end{equation}
with $c_{nm}$ as defined in \eqref{cnmgraviton} and
\begin{equation}\label{}
	d_{nm} = 2G\,\frac{\sigma^2_{nm}-\tfrac12}{\sigma_{nm}^2-1}\,.
\end{equation}
In conclusion, we find the following result for \eqref{DcalL} and for the analogous expression for particle 2
\begin{equation}\label{}
	\Delta \mathcal L_{1}^{\alpha\beta} =
	J_{1}^{\alpha\beta}
	+
	J_{4}^{\alpha\beta}
	+
	b_{1}^{[\alpha} \mathcal Q^{\beta]}_{1}\,,\qquad
	\Delta \mathcal L_{2}^{\alpha\beta} =
	J_{2}^{\alpha\beta}
	+
	J_{3}^{\alpha\beta}
	+
	b_{2}^{[\alpha} \mathcal Q^{\beta]}_{2}\,.
\end{equation}
Here, $\mathcal Q_1^\alpha=-\mathcal Q_2^\alpha$ is the 3PM radiation-reaction contribution to the impulse,
\begin{equation}\label{}
	\mathcal Q_1^\alpha = - \frac{G Q_\text{1PM}^2 b^\alpha}{2b^2} \mathcal I(\sigma),\qquad
	Q_\text{1PM} = \frac{2Gm_1m_2(2\sigma^2-1)}{b\sqrt{\sigma^2-1}}
\end{equation}
with $\mathcal I (\sigma)$ as in \eqref{Isigma}. One can easily check that
\begin{equation}\label{key}
	J_{1}^{\alpha\beta}
	+
	J_{2}^{\alpha\beta}
	+
	J_{3}^{\alpha\beta}
	+
	J_{4}^{\alpha\beta}
	=
	-\mathcal J^{\alpha\beta}\,.
\end{equation}
Moreover, $	b_{1}^{[\alpha} \mathcal Q^{\beta]}_{1}+	b_{2}^{[\alpha} \mathcal Q^{\beta]}_{2}=(b_1-b_2)^{[\alpha}\mathcal Q^{\beta]}_{1}=b_J^{[\alpha}\mathcal Q^{\beta]}_{1}=0$ (up to $\mathcal O(G^4)$ corrections as follows from \eqref{DO17}) which vanishes by antisymmetry.
As a result, these static contributions obey the ``separate'' balance law
\begin{equation}\label{}
	\mathcal J^{\alpha\beta}
	+
	\Delta \mathcal L_{1}^{\alpha\beta}
	+
	\Delta \mathcal L_{2}^{\alpha\beta}
	=
	0
\end{equation}
so that, in conclusion,
\begin{equation}\label{}
	J^{\alpha\beta}
	+
	\Delta  L_{1}^{\alpha\beta}
	+
	\Delta  L_{2}^{\alpha\beta}
	=
	0\,.
\end{equation}
	
	\section{Summary and Outlook}
	\label{sec:outloop}
			
	Conventional perturbation theory provides, in the common practice, the main tool to approach scattering-amplitude calculations. 
	It is based on the assumption that, in the regime of interest, the relevant coupling constant can be regarded as sufficiently small. A textbook example is Quantum Electrodynamics, in which elementary charges couple via the fine-structure constant, $e^2/\hbar \approx 1/137$. The situation is dramatically different when it comes to gravitational interactions of classically sizable objects, such as black holes with masses (and, thus, energies) much larger than the Planck mass, whose effective coupling $GE^2/\hbar$ is extremely large. This seemingly renders our most effective techniques to calculate scattering amplitudes useless to study the classical limit of gravity.
	
	In this report, we illustrated how the eikonal exponentiation serves as a convenient, flexible and conceptually transparent tool to attack this problem in a variety of different theories and setups. This mechanism ``builds'' classical gravitational interactions by resumming the exchanges of many gravitons, in accordance with the intuition that their number, like any quantum number, should become large in the classical limit. Their contributions to the $2\to2$ amplitude ${\mathcal{A}}$ exponentiate, so that in the classical limit $1+i\tilde{\mathcal{A}} \simeq e^{2i\delta_0}$ with $2\delta_0(s,b)$ the leading eikonal phase, which is simply dictated by the single-graviton exchange and indeed proportional to $GE^2/\hbar$. In this way, perturbation theory comes back into the game since, by matching to the eikonal exponentiation, the calculation of higher-loop amplitudes allows one to obtain successively more refined approximations for this exponent, $2\delta_0+2\delta_1+\cdots$, weighted by the classical small PM parameter $GE/b$. In turn, the rapidly oscillating nature of the exponentiated amplitude fixes the values of the classical exchanged momentum, the impulse, via a saddle point condition. 
	
	After reviewing the combinatoric proof of the leading eikonal exponentiation, and illustrating its connection with the impulse and classical deflection angle, as well as the close relation with the phase shifts perhaps more familiar from elementary treatments of the angular momentum in quantum mechanics, we went on to discuss the single-exchange $\mathcal O(G)$ eikonal phase $2\delta_0$ and its manifold applications: minimally coupled massive particles in GR, both spinless and carrying a classical spin, massive particles in maximal supergravity, and string collisions.
	In many situations, proving the exponentiation to all orders is impractical, and one can resort to a more pragmatic approach of checking the constraints it imposes at each loop level. In this spirit, we presented the discussion of one-loop amplitudes in the classical limit both in order to retrieve the next-to-leading $\mathcal O(G^2)$ phase $2\delta_1$ and as a means of checking the $i(2\delta_0)^2/2!$ term dictated by the two-graviton exchanges. 
	A similar pattern holds for high-energy string scattering, and in the string setup new interesting phenomena arise. In particular, already the leading string eikonal is promoted to an operator: for instance, 
	excitations in the $s$-channel have to be included at short distances, while tidal deformations can become important also at large distance. We showed that the leading energy contribution of the one-loop elastic amplitude in string theory is consistent with the exponentiation of the leading eikonal operator, but currently a similar analysis for massive string states is lacking.

    After a detour into the interconnections between unitarity and the $b$-space exponentiation, we reviewed the two-loop, $\mathcal O(G^3)$, calculation focusing on maximal supergravity and GR. 
	The $\mathcal O(G^3)$ eikonal $2\delta_2$ is the lowest-order contribution to the classical exponent at which imprints of the dissipative nature of the scattering problem manifest themselves. It possesses both so-called radiation-reaction terms in its real part, which reflect into time-reversal-odd contributions to the deflection angle, and an infrared-divergent imaginary part. The latter is due to the fact that the $2\to2$ amplitude by its very nature neglects the emitted radiation, and, via unitarity, this corresponds to nontrivial three-particle cuts that the standard exponentiation does not account for.
	We therefore took a first step to amend this shortcoming by combining the eikonal exponentiation with the Weinberg exponentiation of soft quanta, which ought to be included in the description of any physical scattering event in order to obtain infrared-safe final quantities and  to restore manifest unitarity. The general nature of soft theorems also allowed us to take a peek beyond the conventional PM regime, by relaxing the assumption that the impulse be parametrically small compared to the masses. 
	The resulting exponential structure combining elastic collisions with inelastic emission processes takes the form of an operator, and in the last section we provided a self-contained discussion of this eikonal operator to $\mathcal O(G^3)$ beyond the soft approximation, where it neatly disentangles conservative and dissipative effects, and allows one to calculate a variety of different observables such as the emitted energy and angular momentum  during the collision.	
	
	Together with the emergence of these new theoretical structures, which we illustrated in detail in this report, the classical limit also presents simplifying features, mainly when it comes to evaluating loop integrals. In this limit, as long as the colliding objects stay far apart, one can focus on those contributions to loop integrals that are non-analytic in $q$-space and therefore relevant to the long-range behavior in $b$-space \cite{Bjerrum-Bohr:2002gqz,Holstein:2004dn}.  A key simplification in this respect is due to the method of region \cite{Beneke:1997zp,Smirnov:2001cm}, which permits to focus on such contributions directly, without the need to first evaluate the full integrals and then take their asymptotic expansions. We have briefly presented a few simple applications of this method, while dedicating more space to the physical lessons that can be extracted from the results of its application. Similarly, we have extensively applied, but only sketched, the powerful method of reverse-unitarity \cite{Anastasiou:2002yz,Anastasiou:2002qz,Anastasiou:2003yy,Anastasiou:2015yha,Herrmann:2021tct,Herrmann:2021lqe}, which allows one to calculate phase-space integrals from discontinuities of more conventional loop integrals.
	In summary, we hope to have been able to convey the main physical ideas behind the eikonal exponentiation of gravity amplitude up to $\mathcal O(G^3)$, and to have stimulated the reader's interest and curiosity towards this angle of approach to the problem of gravitational scattering, which not only led to the discovery of new patterns and to an improved theoretical understanding of gravity, but also to very concrete new predictions for gravitational observables.
	
\vspace{5pt}

	At this stage, several open challenges lie ahead, both from a conceptual and from a technical standpoint.	
	Remaining at ``low'' PM orders, all observables including the deflection angle and the emitted energy and angular momentum can be determined up to 3PM from the knowledge of the $2\to2$ amplitude up to two loops and of the tree-level $2\to3$ amplitude in the classical limit \cite{Herrmann:2021tct,DiVecchia:2021bdo,Herrmann:2021lqe,Manohar:2022dea,DiVecchia:2022piu}.
	However, as we have reviewed in the respective chapters, these observables behave very differently when one considers collisions with increasing center-of-mass energies.
	The deflection angle up to 3PM turns out to be perfectly smooth in the limit in which this energy $E$ (equivalently, the Lorentz factor $\sigma$) is taken to be large by keeping the leading deflection angle $G E/b$ small and fixed, i.e.~in the ultrarelativistic limit. This is actually crucial in order to ensure agreement with an early calculation of this quantity for collisions of massless objects \cite{Amati:1990xe}, and in order to clarify its universality. In this limit, the massless particle with the highest spin, the graviton, dominates, since it couples to the highest power of the energy, and the deflection angle becomes the same for any theory where this condition is satisfied. 
	In contrast, the 3PM emitted energy and angular momentum only make sense below the bound given by \eqref{eq:KTbex}, i.e.~provided $\sqrt{\sigma}(GE/b)$ is at most of order one
 \cite{DEath:1976bbo,Kovacs:1977uw,Kovacs:1978eu}. Trusting their 3PM expressions beyond this threshold leads to nonsensical conclusions, as the systems seems to be able to radiate much more energy and angular momentum than it possessed to begin with! A veritable ``energy crisis''.
	
	Guided by the example of the Zero-Frequency Limit of the energy emission spectrum, which is governed by soft theorems and can thus be calculated independently of the PM approximation \cite{DiVecchia:2022owy,DiVecchia:2022nna}, it is highly likely that this singularity arises due to the fact that we are attempting to calculate the coefficient of a power series, the ``expansion in powers of $G$'' (or PM expansion), of a quantity which is actually not analytic in $G$. This issue emerged very clearly in the ZFL of the spectrum \cite{DiVecchia:2022nna}, where the full expression starts exhibiting a branch singularity precisely at the bound \eqref{eq:KTbex}. In that case, the correct high-energy limit for the fraction of energy radiated at frequencies between 0 and $1/b$ does not behave, unphysically, like $(GE/b)^3 \log\sigma$, which is unbounded as $\sigma\to\infty$, but rather exhibits (bounded) non-analytic terms of the type $(GE/b)^3\log(GE/b)$.
	It is tempting to speculate that a similar mechanism may apply to the full emitted energy and angular momentum as well, in order to produce a possibly non-analytic but still physically sensible answer for their ultrarelativistic limits. This task is made challenging by the fact that soft theorems no longer apply to such integrated quantities, but it is also of great interest since it would provide a connection with the results obtained for the scattering of massless objects in Refs.~\cite{Ciafaloni:2015xsr,Ciafaloni:2018uwe,Addazi:2019mjh}.
	
	Scatterings of massless states have been the subject of intense studies up to two loops in string theory, already in the early days of the gravity eikonal \cite{Amati:1990xe} and, hopefully, the progress within a QFT approach discussed in this report will motivate new further studies in this string theory context. It would be interesting to derive the 2PM string eikonal operator ($2\hat{\delta}_1$ in our notation) and use the new results on higher-loop string amplitudes obtained over the past 20 years (see for instance~\cite{Berkovits:2022ivl} and references therein) to investigate systematically the eikonal exponentiation in string theory beyond the one-loop level, which is what has been included in this report.
	Moreover, little is known as far as the exponentiation of string amplitudes involving massive states is concerned, see however \cite{Bianchi:2011se}. Borrowing from the mileage and intuition gained in the field theory context, where the corresponding studies have reached three loops \cite{Bern:2021dqo,Bern:2021yeh}, in particular the focus on the non analytic terms that dominate the long range dynamics, will certainly provide a pivotal simplification, compared to the study of the full amplitude. Moreover, it could prove useful to introduce masses via Kaluza--Klein compactification rather than by considering excited string states.
	
	Moving on to a broader perspective, outstanding efforts have been recently devoted to the endeavor of bringing amplitude-based techniques for gravitational-wave physics closer to timely phenomenological applications, along two main directions: analytically continuing the results obtained from scattering amplitudes to the bound-system kinematics, and including all relevant effects beyond the point-particle approximation. Moving forward, it will of course be interesting to undertake similar analysis by means of the eikonal exponentiation as well.
	
	In the present report we have discussed a few basic steps towards making the information extracted from the amplitude more directly available for applications to bound systems, of which binary mergers represent the key relevant example for observational purposes, in the spirit of the so-called Boundary-to-Bound dictionary \cite{Kalin:2019rwq,Kalin:2019inp,Cho:2021arx,Saketh:2021sri}. This connection is essentially based on an analytic continuation, one could say, from ``positive'' to ``negative'' energy (after subtracting the rest mass in the CM frame). In addition, since most binary systems revolve along on quasi-circular trajectories, the resulting information also needs to be matched from a regime of large eccentricities, more directly accessible from the PM expansion, to small or vanishing eccentricities. We have illustrated how this can be done in practice, starting from 1PM and 2PM data in order to retrieve all PN data up to 2PN, while we leave the investigation of how 3PM and 4PM eikonal data may eventually combine to yield the 3PN and 4PN information for future work.
	
	String scattering represents a prototypical example where the colliding objects possess an internal structure, which manifests itself as a dependence of the eikonal operator on the string's excitation modes (see Section~\ref{ssec:seikop}). For phenomenological applications, one can similarly introduce such deformations, encoding in particular tidal Love numbers and higher-multiple modes in the amplitude approach \cite{Cheung:2020sdj,Bern:2020uwk,Cheung:2020gbf,Aoude:2020ygw,AccettulliHuber:2020dal} in an EFT spirit \cite{Goldberger:2004jt} by means of suitable additional parameters. This is particularly relevant for neutron stars, whose tidal deformability properties are expected to provide insights into their Quantum Chromodynamical origin, internal structure and on the resulting  equation of state.
	Including classical spin effects is also important and this poses a challenge as far as the amplitude approach is concerned. Indeed, while masses and tidal deformabilities can be simply encoded in a Lagrangian formulation via suitable continuous parameters, the intrinsically quantized nature of spin $s$ in the quantum world poses an obstacle to taking the classical limit, in which $\hbar s$ should become classically sizable and $s$ attain very large values. In this report we have presented a self-contained account of a simple strategy to include classical spin effects in the leading eikonal, i.e.~to first order in $GM/b$ but formally to all orders in the spin parameter. At present, the inclusion of spin from one-loop order onward relies on a strategy to break down the calculation into an expansion in spin multipoles. The missing ingredient to upgrade this to an all-order-in-spin calculation, the Compton-like amplitude with classical spin, remains partly elusive, despite encouraging recent developments \cite{Arkani-Hamed:2017jhn,Vines:2017hyw,Guevara:2018wpp,Chung:2018kqs,Maybee:2019jus,Guevara:2019fsj,Arkani-Hamed:2019ymq,Johansson:2019dnu,Chung:2019duq,Damgaard:2019lfh,Bautista:2019evw,Aoude:2020onz,Chung:2020rrz,Bern:2020buy,Guevara:2020xjx,Kosmopoulos:2021zoq,Aoude:2021oqj,Bautista:2021wfy,Chiodaroli:2021eug,Haddad:2021znf,Chen:2021kxt,Aoude:2022trd,Bern:2022kto,Alessio:2022kwv,FebresCordero:2022jts,Cangemi:2022bew,Bautista:2022wjf,Bjerrum-Bohr:2023jau,Kim:2023drc,Alessio:2023kgf,Bianchi:2023lrg} including possible ties with states built out of the string spectrum \cite{Cangemi:2022abk}. The only exception is  the calculation of the radiation reaction at two-loop order~\cite{Alessio:2022kwv} that, as in the spinless case discussed in~\cite{DiVecchia:2021ndb}, is only based on the soft limit of the five-point amplitude. The result obtained with this approach agrees in the small spin limit with the complete two-loop calculation for the spin one case fully described by an extended ${\cal{N}}=2$ world-line supersymmetry~\cite{Jakobsen:2022fcj}. 
	
	Coming to the frontier in the $G$-expansion \cite{Bern:2019nnu,Bern:2019crd,DiVecchia:2020ymx,Damour:2020tta,Dlapa:2021npj,Bern:2021yeh,Herrmann:2021tct,DiVecchia:2021bdo,Bern:2021dqo,Dlapa:2021vgp,Manohar:2022dea,Dlapa:2022lmu,Jakobsen:2023ndj}, a potentially loose end concerns the integrals over fluctuations around the saddle points characterizing the classical limit. Following the literature, we have presented the derivation of classical observables associated to the scattering by taking appropriate expectations of the relevant operators on the final state dictated by the action of the eikonal operator (minus their expectation on the initial state), and evaluated the integrals by means of the saddle point conditions. 
	In principle, one should be able to check that the Gaussian integrals for the fluctuations about such classical values eventually produce suppressed contributions, either weighted by powers of $\hbar$ or, perhaps more plausibly, of $GE/b$. In the latter case, it will be of course important to take them into account in the appropriate way when moving to higher orders in this parameter.
	Put another way, it should be possible to explicitly check the normalization condition for the final state, which, in a novel fashion, brings us back to the original issue of  unitarity restoration in the classical limit.
	
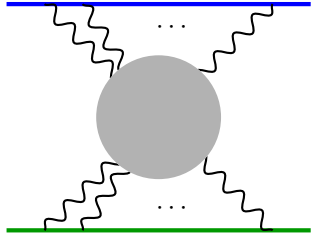
\begin{figure}[h!]
			\centering
			\begin{tikzpicture}
				\draw[color=green!60!black,ultra thick] (-2,0)--(2,0);
				\draw[color=blue,ultra thick] (-2,3)--(2,3);
				\draw[thick, style=decorate, decoration=snake] (-1.5,0) -- (0,1.5);
				\draw[thick, style=decorate, decoration=snake] (-1,0) -- (0,1.5);
				\draw[thick, style=decorate, decoration=snake] (1.5,0) -- (0,1.5);
				\draw[thick, style=decorate, decoration=snake] (-1.5,3) -- (0,1.5);
				\draw[thick, style=decorate, decoration=snake] (-1,3) -- (0,1.5);
				\draw[thick, style=decorate, decoration=snake] (1.5,3) -- (0,1.5);
				\filldraw[color=white, fill=white, very thick](0,1.5) circle (.8);
				\filldraw[color=gray!60!white, very thick](0,1.5) circle (.8);
				\node at (.2,.3){$\cdots$};
				\node at (.2,2.7){$\cdots$};
			\end{tikzpicture}
			\caption{Dominant diagrams contributing to the semiclassical eikonal phase. The shaded blob stands for a generic connected tree-level topology.}
			\label{fig:treeeintermediate}
		\end{figure}

 In this connection let us mention, for completeness, a partially successful attempt \cite{Amati:2007ak} at constructing a semiclassical unitary $S$-matrix in the case of ultrarelativistic (massless) collisions. In Subsections~\ref{ssec:string-brane-sup}, \ref{beyondtree} we have discussed what has been achieved in the two regimes labeled as I and II in Fig.~\ref{fig:StringRegimes}. Unfortunately, the most interesting regime, the one leading classically to gravitational collapse, is also the most difficult one to analyze.
		So far, it has only been approached in the point-particle limit and in $D=4$ (although going to $D>4$ one would avoid having to deal with some infrared divergences). The problem at hand is that the 
		PM expansion is an expansion in powers of $R/b$ and, precisely because of gravitational collapse, it is expected to break down at some critical value of ${\cal O}(1)$ for that ratio (see Fig.~\ref{fig:StringRegimes}).
		The only simplification that looks to be fully justified is that, order by order in the above expansion, the dominant diagrams contributing to the semiclassical eikonal phase consist of connected tree-diagrams involving the gravitons emitted by the two energetic particles as external lines (see Fig.~\ref{fig:treeeintermediate}). 
		Adding disconnected diagrams with the same external legs should exponentiate the eikonal phase in a standard way
		(much like the two-particle reducible contributions of Ref.~\cite{Brandhuber:2021eyq}). Summing tree-diagrams is supposed to be equivalent to solving a classical field theory. Using work by Lipatov \cite{Lipatov:1991nf} one can make a reasonable guess about the form of such field theory in the high-energy limit. After simplifying further Lipatov's action a two-dimensional field theory was proposed and studied in \cite{Amati:2007ak} with reasonable success in view of the rough approximations made.
		We refer to the literature concerning those developments and only mention here that  critical values for $R/b$ in good agreement with the classical expectations discussed in Section \ref{ssec:CTS} were indeed found \cite{Marchesini:2008yh,Veneziano:2008zb,Veneziano:2008xa}. On the other hand, in spite of considerable effort \cite{Ciafaloni:2011de,Ciafaloni:2017ort}, control of unitarity could not be achieved in any simple way above the critical value. At the same time,  no strong claim of unitarity loss below $b_c$ can be made in view of the many unjustified approximations made along the way.
	
        Coming back to the PM expansion, there still are aspects of the full 4PM result~\cite{Dlapa:2022lmu} that need to be investigated including a more detailed comparison with the PN approaches~\cite{Bini:2022enm,Almeida:2022jrv}. It will certainly be interesting to continue the analysis of the classical dynamics at three loops beyond the case of scalar particles, along the lines of~\cite{Jakobsen:2023ndj}, and also to calculate the angular momenta at 4PM. Recently the conservative classical dynamics at four loops was studied for the relativistic scattering in electrodynamics~\cite{Bern:2023ccb}, paving the way for the analysis of the GR case at 5PM. From the 4PM order,
               new exciting physical phenomena manifest themselves: tail and recoil effects. The tail effect is caused by an interaction between the two-body system and the gravitational energy that is first emitted and then reabsorbed by the system itself \cite{Blanchet:2013haa,Damour:2014jta,Galley:2015kus}. For this reason, it can be regarded as a ``globally conservative'' type of interaction which is however nonlocal in time, and thus very different from the typical ``potential'' interactions, whereby each objects feels the gravitational pull of the other one. 
	Recoil is caused by the net emission of linear momentum and represents a novel dissipative effect that combines with the more familiar back-reaction of radiation emissions on the relative motion to that order. Notably, recoil implies that, even if the system is observed from the initial center-of-mass frame, the two bodies will in general experience \emph{different} deflection angles compared to their original direction of motion.
	It will be important to investigate how such phenomena fit together in an improved version of the eikonal operator, yet to be formulated, in order to clarify the classical limit of the amplitude(s) up to 4PM order. A first step in this direction has been taken in~\cite{DiVecchia:2022piu} where the linear reaction~\cite{Bini:2012ji} at 4PM was derived from an expansion of the stationary phase conditions~\eqref{Saddlev2}, using in particular the prescription \eqref{deltas}, finding perfect agreement with the result of~\cite{Manohar:2022dea}. 
	In addition the high-energy behavior of the 4PM deflection angle(s) currently available in the literature points to a breakdown of the smoothness property enjoyed by it up to 3PM. Whether or not this is to be regarded as a manifestation of the bound \eqref{eq:KTbex} certainly deserves further investigation.
	
	This also relates to another open puzzle in massless $\mathcal N=8$ supergravity amplitudes, where it was observed that the exponentiation of the elastic $2\to2$ amplitude fails precisely starting to order $\mathcal O(G^4)$ due to a superclassical mismatch proportional to $\operatorname{Im}2\delta_2$ \cite{DiVecchia:2019myk,DiVecchia:2019kta}. Armed with the appropriate eikonal operator, whose phase is by construction manifestly real, it will hopefully be possible to move past this obstacle by appropriately including all relevant inelastic channels as well.
	
	As we have discussed in the last two sections of this report, the first step in this direction, which up to 3PM completely clarifies the structure of the exponentiation and allows for a systematic calculation of the observables, is to combine eikonalized (virtual) graviton exchanges with coherent (real) graviton emissions, effectively producing an operator that links elastic $2\to2$ with inelastic $2\to3$ amplitude.
	One cannot help but wonder whether the ingredients entering the next order will eventually turn out to be essentially the same, or if instead new ones are to be included. For instance it is possible that new amplitudes will need to be exponentiated, in particular the Compton-like graviton scattering in Fig.~\ref{fig:2g2ms}, as suggested by the fact that they appear as possible sub-diagrams or cuts of higher loop $2\to3$ and $2\to2$ amplitudes  \cite{Brandhuber:2023hhy,Herderschee:2023fxh,Elkhidir:2023dco,Georgoudis:2023lgf} (while of course analogous $1\to3$ processes with two final gravitons in the final states are kinematically forbidden).
	A closely related issue concerns the possible emergence of correlations among graviton emissions, in particular due to the constraint of total energy conservation. 
	Such effects, which ought to be encoded to leading order in the $2\to4$ amplitude with two graviton emissions, have been shown to be absent at tree level, and the picture of a coherent, uncorrelated emission has not received corrections so far \cite{Cristofoli:2021jas,Britto:2021pud}.
	Further research and careful scrutiny will hopefully help to clarify these points. Here we would like to mention some of the most recent developments which confirm the usefulness of the eikonal approach in the analysis of subleading PM corrections.

Extending the logic presented in this report, it is natural to generalize the eikonal operator in~\eqref{eikopv1} or~\eqref{eikopv2a} including in $\mathcal{W}_j$ also the subleading PM contributions that can be extracted from the classical limit of the $2\to3$ amplitude at 1-loop. It would then be possible to derive the NLO PM scattering waveform by following the same steps of Section~\ref{sec:wavforlo}. Starting from the result for the classical 1-loop 5-point amplitude~\cite{Brandhuber:2023hhy,Herderschee:2023fxh,Elkhidir:2023dco,Georgoudis:2023lgf}, this problem is being under intense investigation. First, the relation between the classical amplitude and the gravitational waveform becomes more subtle at NLO. As discussed in \cite{Caron-Huot:2023vxl}, the gravitational waveform is one of the asymptotic observables requiring a non-standard time-ordering from the S-matrix point of view, which, in the KMOC approach, is implemented by taking into account carefully some cut contributions. At NLO, these contributions implement a change of frame for the waveform \cite{Georgoudis:2023eke}, thus making the result very natural from the point of view of the eikonal exponentiation mentioned after~\eqref{eikopv1} and explaining the need of such a rotation to match the PM waveform against known PN~\cite{Bini:2023fiz} or soft data~\cite{Aoude:2023dui}. Beside this, it was realised that the choice of BMS frame affects the NLO waveform also beyond its delta-function contribution at zero-frequency \cite{Georgoudis:2023eke}. A detailed comparison between the NLO PM waveform and the PN results, including several non-linear effects, is discussed in~\cite{Georgoudis:2024pdz,Bini:2024rsy} finding perfect agreement. It is certainly interesting to extend this comparison to higher order to see whether the intuition based on the eikonal exponentiation can again provide useful guidance.

Besides the applications mentioned above, the key idea underlying the gravitational eikonal (resummation of leading contributions leading to an exponentiation) is likely to provide new insights also in other contexts. For instance, the AdS/CFT duality is the perfect setup to study some of the fundamental questions that motivated the early analysis of the gravitational eikonal as mentioned in the introduction. Advances in the analytic calculation of holographic correlators through a bootstrap approach~\cite{Bissi:2022mrs,Hartman:2022zik} can provide new tools to improve our understanding of the gravitational eikonal in this context and possibly use it to study the nature of black-hole microstates. The eikonal approach can be useful also to analyse the gravitational scattering in more general curved backgrounds, see for instance \cite{Gaddam:2020rxb,Gaddam:2021zka}, which again can encode interesting information about the nature of black holes as for instance pointed out in~\cite{Maldacena:2015waa}.
        
\subsection*{Acknowledgements} 
We would like to thank Francesco Alessio, Fabian Bautista, Brando Bellazzini, Marcus Berg, Zvi Bern, Massimo Bianchi, Donato Bini, Andreas Brandhuber, Emil Bjerrum-Bohr, Costas Bachas, Poul Henrik Damgaard, Thibault Damour, Stefano De Angelis, Maurizio Firrotta, Stefano Foffa, Claudio Gambino, Alessandro Georgoudis, Riccardo Gonzo, Enrico Herrmann, Henrik Johansson, Gregor K\"alin, David Kosower, Emil Martinec, Raffaele Marotta, Gustav Mogull, Alexander Ochirov, Donal O'Connell, Julio Parra-Martinez, Ludovic Plant\'e, Massimo Porrati, Rafael Porto, Fabio Riccioni, Massimiliano Riva, Michael Ruf, Chia-Hsien Shen, Riccardo Sturani, Gabriele Travaglini, Pierre Vanhove,  Filippo Vernizzi, Leong Khim Wong, Mao Zeng
for several enlightening discussions.

RR and GV are grateful to IHES for hospitality during the part of this work.
CH and RR are grateful to KITP, Santa Barbara, for hospitality during the workshop High-Precision Gravitational Waves. 

This research was partially supported in part by the National Science Foundation under Grant No.~NSF PHY-1748958. The research of RR is partially supported by the UK Science and Technology Facilities Council (STFC) Consolidated Grants ST/P000754/1 and ST/T000686/1. The research of CH (PDV) is fully (partially) supported by the Knut and Alice Wallenberg Foundation under grant KAW 2018.0116. Nordita is partially supported by Nordforsk.

\appendix

\section{Field Theory details and conventions }
\label{app:fieldtheory}

\subsection{Permutation Identities }
\label{identities}

Let us first show by induction that
\begin{equation}\label{idF}
	F(a_1,\ldots,a_n)\equiv
	\sum_{\sigma\in\mathcal S_{n}} \frac{1}{a_{\sigma_1}}\cdots \frac{1}{a_{\sigma_1}+\cdots+a_{\sigma_{n}}}
	=
	\frac{1}{a_1\cdots a_n}\,.
\end{equation}
For $n=2$, we find
\begin{equation}\label{}
	F(a_1,a_2) = \frac{1}{a_1(a_1+a_2)} + \frac{1}{a_2(a_2+a_1)} = \frac{1}{a_1a_2}\,.
\end{equation}
Moreover, denoting by $\mathcal S_n(j)$ the permutations such that $\sigma_n=j$, we can write
\begin{equation}\label{Finterm}
	F(a_1,\ldots,a_n) = \frac{1}{a_1+\cdots+a_n} \sum_{j=1}^n 
	\Big[ 
	\sum_{\sigma\in\mathcal S_{n}(j)}
	\frac{1}{a_{\sigma_1}}\cdots \frac{1}{a_{\sigma_1}+\cdots+a_{\sigma_{n-1}}}\Big].
\end{equation}
The quantity within brackets is equal to $F(a_1,\ldots,\widehat{a_j}, \ldots, a_n)$, where $\widehat{a_j}$ means that $a_j$ is omitted. Using the induction hypothesis we then find
\begin{equation}\label{}
	\sum_{j=1}^n F(a_1,\ldots,\widehat{a_j}, \ldots, a_n)
	=
	\sum_{j=1}^n	\frac{1}{a_1\cdots \widehat{a_j}\cdots a_n} = \frac{a_1+\cdots +a_n}{a_1\cdots a_n}\,,
\end{equation}
so that \eqref{Finterm} leads to \eqref{idF}.
The quantity $f$ in \eqref{myf} is related to $F$ by
\begin{equation}\label{}
	f(a_1,\ldots,a_n)
	=
	(a_1+\cdots +a_n)F(a_1,\ldots,a_n)\,,
\end{equation}
and therefore it is given by
\begin{equation}\label{idf}
	f(a_1,\ldots,a_n)=\sum_{j=1}^n	\frac{1}{a_1\cdots \widehat{a_j}\cdots a_n}\,.
\end{equation}

We shall now use \eqref{idf} together with
\begin{equation}\label{}
	2\pi \delta(\omega) = \int e^{i\omega t}dt\,,\qquad
	\frac{1}{\omega_k-i0} = \int  e^{-i(\omega_k-i0)t_k}\theta(t_k)\,i\,dt_k
\end{equation}
to simplify
\begin{equation}\label{}
	T(\omega_1,\ldots,\omega_n) \equiv 2\pi\delta(\omega_1+\cdots+\omega_n)
	f(\omega_1-i0,\ldots,\omega_n-i0),.
\end{equation}
Let us again start from the $n=2$ case, for simplicity.
We have
\begin{equation}\label{}
	T(\omega_1,\omega_2) = \int dt_1 dt_2 i\theta(t_1)e^{i\omega_1(t_2-t_1)}  e^{i\omega_2 t_2}+  \int dt_1 dt_2 i\theta(t_2)e^{i\omega_2(t_1-t_2)}  e^{i\omega_1 t_1}
\end{equation}
so that shifting $t_1\to t_2-t_1$ in the first term and $t_2\to t_1-t_2$ in the second leads to
\begin{equation}\label{}
	T(\omega_1,\omega_2) = \int dt_1 dt_2\,  i\left[\theta(t_2-t_1)+\theta(t_1-t_2)\right]e^{i\omega_1 t_1}  e^{i\omega_2 t_2}
\end{equation}
and thus
\begin{equation}
	\frac{1}{2\pi}\,T(\omega_1,\omega_2) =
	 2i\pi\, \delta(\omega_1) \delta(\omega_2)\,.
\end{equation}
A similar manipulation goes through for generic $n$, where 
\begin{equation}\label{}
	T(\omega_1,\ldots,\omega_n)=
	\int dt_1 \cdots dt_n
	\sum_{j=1}^n
	e^{i\omega_j t_j} \prod_{k\neq j} i\theta(t_k) e^{i\omega_k(t_j-t_k)}\,.
\end{equation}
In each term of the sum over $j$, we can send $t_k\to t_j-t_k$ for all $k\neq j$, to obtain
\begin{equation}\label{}
	T(\omega_1,\ldots,\omega_n)=
	i^{n-1}
	\int dt_1 \cdots dt_n\,
	e^{i(\omega_1 t_1+\cdots+\omega_n t_n)}
	\sum_{j=1}^n
	\prod_{k\neq j}
	\theta(t_j-t_k) \,.
\end{equation}
Noting that the factor involving the theta functions is 1 for any ordering of $t_1,\ldots,t_n$ finally yields
\begin{equation}\label{indentitydeltas}
	\frac{1}{2\pi}\,T(\omega_1,\ldots,\omega_n)
	= (2i\pi)^{n-1} \delta(\omega_1)\cdots \delta(\omega_n)\,.
\end{equation}

\subsection{Feynman rules } \label{appFeynRules}

In this appendix we collect useful Feynman rules.
We start from the classical action
describing a massless scalar $\phi$ minimally coupled to gravity, 
\begin{equation}\label{Sminimally}
	S
	=
	\int 
	\frac{R}{2\kappa^2} \sqrt{-g}\, d^Dx
	-
	\frac12
	\int
	\partial_\mu \phi\, g^{\mu\nu}\,\partial_\nu \phi\,
	\sqrt{-g}\,d^Dx\,,
\end{equation}
with $\kappa = \sqrt{8\pi G}$. Defining $g_{\mu\nu} = \eta_{\mu\nu} + 2\kappa\, h_{\mu\nu}$ and retaining quadratic terms only, one finds
\begin{align}
		S^{(2)}
		&=
		\frac12
		\int 
		h^{\mu\nu} 
		\left[
		\Box h_{\mu\nu}
		- 
		\partial_{\mu} \partial^\alpha h_{\alpha\nu}
		-
		\partial_{\nu} \partial^\alpha h_{\alpha\mu}
		+
		\partial_\mu \partial_\nu h^{\alpha}_\alpha
		- \eta_{\mu\nu} \left(
		\Box h^\alpha_\alpha-\partial^\alpha \partial^\beta h_{\alpha\beta}
		\right)
		\right]d^Dx
		\nonumber
		\\
		&
		+
		\frac12 
		\int 
		\phi\,
		\Box
		\phi\, d^Dx
		\,,
		\label{S2grav}
\end{align}
where indices are raised and lowered using the  Minkowski metric.
To quantize the theory we go to De Donder gauge,
\begin{equation}\label{}
	\partial^\alpha h_{\mu\alpha} = \frac12\, \partial_\mu h^\alpha_\alpha\,.
\end{equation}
This requires adding to the action a gauge-fixing term 
plus suitable ghost contributions, which however do not play any role in our classical analysis. The net effect is to replace $S^{(2)}$ with 
\begin{equation}\label{SDD2}
	S^{(2)}_{DD}
	=
	\frac12
	\int 
	h^{\mu\nu} 
	\left(
	\Box h_{\mu\nu}
	- \frac12 \eta_{\mu\nu} 
	\Box h^\alpha_\alpha
	\right)
	d^Dx
	+
	\frac12 \int \phi\,
	\Box
	\phi\,d^Dx\,.
\end{equation}
Rewriting the graviton part in terms of 
\begin{equation}\label{Gmunurhosigma}
	D_{\mu\nu,\rho\sigma} = \frac12\left(
	\eta_{\mu\rho}\eta_{\nu\sigma}+\eta_{\mu\sigma}\eta_{\nu\rho}-\eta_{\mu\nu}\eta_{\rho\sigma}
	\right),
\end{equation}
we find 
\begin{equation}\label{SDD}
	S^{(2)}_{DD}
	=
	\frac12
	\int 
	h^{\mu\nu} 
	D_{\mu\nu,\rho\sigma}\, \Box h^{\rho\sigma}
	d^Dx
	+
	\frac12 \int  \phi\,
	\Box
	\phi\,d^Dx\,.
\end{equation}
The inverse of $D_{\mu\nu,\rho\sigma}$ is 
\begin{equation}\label{Pmunurhosigma}
	P_{\mu\nu,\rho\sigma}
	=
	\frac12\left(
	\eta_{\mu\rho}\eta_{\nu\sigma}+\eta_{\mu\sigma}\eta_{\nu\rho}-\,\frac{2}{D-2}\eta_{\mu\nu}\eta_{\rho\sigma}
	\right),
\end{equation}
which satisfies
\begin{equation}\label{inverseDP}
	D^{\mu\nu,\rho\sigma}P_{\rho\sigma,\alpha \beta} = \frac{1}{2}\left(\delta^\mu_\alpha \delta^\nu_\beta + \delta^\nu_\alpha \delta^\mu_\beta\right),
\end{equation}
and this fixes the propagators to be
\begin{equation}\label{GGagrav}
	G_{\mu\nu,\rho\sigma}(k) = \frac{-i P_{\mu\nu,\rho\sigma}}{k^2-i0}\,,\qquad
	G(k) = \frac{-i}{k^2-i0}\,.
\end{equation}
The leading scalar-graviton interaction term
\begin{equation}\label{SIgrav}
	S_I = -\kappa \int
	h^{\mu\nu}
	\left[
	-\partial_\mu\phi\, \partial_\nu\phi +\frac12\,\eta_{\mu\nu} (\partial \phi)^2
	\right]d^Dx\,,
\end{equation}
which translates to the vertex
\begin{equation}\label{taupp}
	\tau^{\mu\nu}(p,p') =- i\kappa \Big[
	p^\mu p'^\nu + p^\nu p'^\mu
	- \eta^{\mu\nu} (p\cdot p')
	\Big],
\end{equation}
where the scalar lines are regarded as both outgoing.

In case the scalar has mass $m$,
\begin{equation}\label{}
	S
	=
	\int 
	\frac{R}{2\kappa^2} \sqrt{-g}\, d^Dx
	-
	\frac12
	\int
	\left(\partial_\mu \phi\, g^{\mu\nu}\,\partial_\nu \phi+m^2 \phi^2\right)
	\sqrt{-g}\,d^Dx\,,
\end{equation}
can be discussed along similar lines.
The massive scalar propagator is given by
\begin{equation}\label{}
	G_m(k) = \frac{-i}{k^2+m^2}
\end{equation}
while the leading scalar-graviton vertex in De Donder gauge reads
\begin{equation}\label{tauppmassive}
	\tau^{\mu\nu}(p,p') =- i\kappa \left[
	p^\mu p'^\nu + p^\nu p'^\mu
	- \eta^{\mu\nu} (p\cdot p' - m^2)
	\right].
\end{equation}
Note that this vertex is transverse with respect to the graviton momentum,
\begin{equation}\label{}
	\tau^{\mu\nu}(p,p')(p_\nu+p'_\nu)=0\,,
\end{equation}
whenever the scalar lines are on-shell $p^2=p'^2=-m^2$. Moreover, its trace reads
\begin{equation}\label{}
	\eta_{\mu\nu}\tau^{\mu\nu}(p,p')=-i\kappa \left[ \frac{2-D}{2}\,(p+p')^2+2m^2\right],
\end{equation}
which is zero in the massless case, up to terms that vanish on the graviton's mass shell.

\subsection{Useful Fourier transforms to impact parameter space}
\label{usefulFT}

In this appendix we collect a few useful properties of the Fourier transform into impact parameter space.
Taking the $D$-dimensional Fourier transform of the $S$-matrix element one obtains the invariant expression \eqref{FTexact} and this leads us to consider Fourier transforms of the following type,
\begin{equation}\label{invFT}
	\operatorname{FT}[f](b) 
	= 
	\int \frac{d^Dq}{(2\pi)^D}\,2\pi\delta(2p_1\cdot q-q^2)2\pi\delta(2p_2\cdot q+q^2)\,e^{ib\cdot q} f(q^2)\,.
\end{equation}
with $f(q^2)$ playing the role of the invariant amplitude $\mathcal A(s,-q^2)$.
In order to recast this in a more explicit way, let us start from the case in which $p_1$ and $p_2$ are massive momenta, rewriting $p_1^\mu=-m_1 v_1^\mu$ and $p_2^\mu=-m_2v_2^\mu$ as in \eqref{eq:velocities} so that
\begin{equation}\label{}
	\operatorname{FT}[f](b)
	=
	\frac{1}{4m_1m_2}
	\int \frac{d^Dq}{(2\pi)^D}\,2\pi\delta\left(v_1\cdot q+\frac{q^2}{2m_1}\right) \,2\pi\delta\left(v_2\cdot q-\frac{q^2}{2m_2}\right)\,e^{ib\cdot q}\,f(q^2)\,.
\end{equation}
It is convenient to change integration variables by decomposing the integrated momentum according to \eqref{decomposition} so that
\begin{equation}\label{}
	q^\mu = \check{v}^\mu_1 q_{\parallel 1} + \check{v}^\mu_2 q_{\parallel 2} + q_\perp^\mu
\end{equation}
where by definition $q_\perp\cdot p_{1,2}\equiv 0$.
Therefore
\begin{equation}\label{}
	(dq)^2 = \check{v}_1^2\,(dq_{\parallel1})^2
	+
	2\check{v}_1\cdot\check{v}_2\,dq_{\parallel1}dq_{\parallel2}
	+
	\check{v}_2^2\,(dq_{\parallel2})^2
	+
	(dq_\perp)^2
\end{equation}
and the determinant of this metric is given by
\begin{equation}\label{detgmassive}
	-\operatorname{det}g=
	(\check{v}_1\cdot\check{v}_2)^2
	- \check{v}_1^2\check{v}_1^2
	=\frac{1}{\sigma^2-1}\,.
\end{equation} 
Therefore,
\begin{equation}\label{}
	d^Dq \,= \frac{dq_{\parallel1}\,dq_{\parallel_1}}{\sqrt{\sigma^2-1}}\,
	d^{D-2}q_\perp
	\,,
\end{equation}
and we find
\begin{equation}\label{massiveFT}
	\operatorname{FT}[f](b)
	=
	\frac{1}{4 m_1 m_2}
	\int \frac{dq_{\parallel1}\,dq_{\parallel_1}}{\sqrt{\sigma^2-1}} \frac{d^{D-2}q_\perp}{(2\pi)^{D-2}}
	\delta\left(q_{\parallel1}-\frac{q^2}{2m_1}\right)
	\delta\left(q_{\parallel2}+\frac{q^2}{2m_2}\right) \,e^{ib\cdot  q}\,f(q^2)
	\,.
\end{equation}
We are only interested in evaluating this Fourier transform for very large $b$, orthogonal to $p_1$ and $p_2$ (up to $\mathcal O(G^2)$ corrections), and therefore we can expand the integrand for small $q\sim 1/b$. Using the Taylor expansion of the delta functions and noting that (cf.~\eqref{calP})
\begin{equation}\label{}
	q^2 = \frac{q_{\parallel1}^2-2\sigma q_{\parallel1}q_{\parallel2}+q_{\parallel2}^2}{\sigma^2-1}+q_\perp^2\,,
\end{equation}
we obtain
\begin{equation}\label{}
	\operatorname{FT}[f](b)
	=
	\frac{1}{4 m_1 m_2 \sqrt{\sigma^2-1}}
	\int \frac{d^{D-2}q_\perp}{(2\pi)^{D-2}}
	e^{ib\cdot  q_\perp}\,\left(f(q_\perp^2) + \frac{s}{4m_1^2m_2^2(\sigma^2-1)} \left[x^2 f(x)\right]_{x=q_\perp^2}'+\cdots
	\right)
\end{equation}
(the superscript $'$ stands for a total derivative with respect to the argument $x$)
or equivalently, using the identity \eqref{Ep},
\begin{equation}\label{4EpFT}
		\operatorname{FT}[f](b)
	=
	\frac{1}{4 Ep}
	\int \frac{d^{D-2}q_\perp}{(2\pi)^{D-2}}
	e^{ib\cdot  q_\perp}\,\left(f(q_\perp^2) + \frac{1}{4p^2} \left[x^2 f(x)\right]_{x=q_\perp^2}'+\cdots
	\right).
\end{equation}
In the massless case, $p_1^2=p_2^2=0$, we can instead adopt the decomposition
\begin{equation}\label{}
	q^\mu = \frac{1}{s} \left(p^\mu_1 x_1 + p_2^\mu x_2\right) + q_\perp^\mu\,,
\end{equation}
where again $q_\perp\cdot p_{1,2}=0$. Since $x_{1,2}=-2p_{2,1}\cdot q$, we then find
\begin{equation}\label{detgmassless}
	(dq)^2 = -\frac{dx_1 dx_2}{s}
	+
	(dq_\perp)^2\,,\qquad
	\sqrt{-\operatorname{det}g}
	=\frac{1}{2s}
\end{equation} 
and in this way \eqref{invFT} evaluates to
\begin{equation}\label{masslessFT}
	\operatorname{FT}[f](b)
	=
	\frac{1}{2 s}
	\int \frac{d^{D-2}q_\perp}{(2\pi)^{D-2}} \,e^{ib\cdot  q_\perp}\left(f(q_\perp^2) + \frac{1}{s} \left[x^2 f(x)\right]_{x=q_\perp^2}'+\cdots
	\right)\,.
\end{equation}
Recalling that in the massless case $E=2p=\sqrt{s}$, we see that the expression \eqref{4EpFT} is thus valid for both massive and massless setups.
It is easy to see that the first term on the right-hand side of \eqref{4EpFT} corresponds to simply dropping the $q^2$ in the arguments of the delta functions appearing in \eqref{invFT}, for which we adopt the notation
\begin{equation}\label{invFTLIN}
	\tilde{f}(b) = 
	\int \frac{d^Dq}{(2\pi)^D}\,2\pi\delta(2p_1\cdot q)2\pi\delta(2p_2\cdot q)\,e^{ib\cdot q} f(q^2)
	=
	\frac{1}{4 E p}
	\int \frac{d^{D-2}q_\perp}{(2\pi)^{D-2}}
	e^{ib\cdot  q_\perp}\,f(q_\perp^2)\,.
\end{equation}

Since in most applications $f(q^2)$ has a power-law dependence on $q^2$, the basic type of integrals that we need to calculate
is:  
\begin{equation}\label{B1}
 I_D(\nu) =  \int \frac{d^{D-2} q}{(2\pi)^{D-2}} \,{e}^{ib\cdot q} 
\left(q^2\right)^\nu  
= \frac{2^{2\nu}}{\pi^{1-\epsilon}} \frac{\Gamma (1+\nu -\epsilon)}{\Gamma (- \nu) (b^2)^{\nu +1 - \epsilon}} 
  \,,
  \qquad
  D=4-2\epsilon\,.
\end{equation}
A quick way to see this is to introduce Schwinger parameters, so that
\begin{equation}\label{}
	I_D(\nu)
	=
	\int_0^\infty dt\,\frac{t^{-1-\nu}}{\Gamma(-\nu)}
	\int  \frac{d^{D-2} q}{(2\pi)^{D-2}}\, e^{ib\cdot q-t q^2}
\end{equation}
and performing the Gaussian integral and letting $t=1/x$ leads to
\begin{equation}\label{}
	I_D(\nu)
	=
	\frac{1}{(4\pi)^{\frac{D-2}{2}}\Gamma(-\nu)}
	\int_0^\infty dx\,x^{-1+\nu+\frac{D-2}{2}}
	e^{-\frac{b^2}{4}\,x}\,.
\end{equation}
Recognizing the $\Gamma$-function in the last equation then yields Eq.~\eqref{B1}.

Expanding the identity \eqref{B1} around $\nu=0$ on both sides, one can also deduce the Fourier transform of any power of the logarithm $\left[\log(q^2)\right]^n$.
For instance expanding to linear order, we obtain
\begin{equation}\label{}
	\int\frac{d^{D-2}q}{(2\pi)^{D-2}}\,e^{ib\cdot q}\log\left(
	{q^2}
	\right)
	=
	-\frac{\Gamma(1-\epsilon)}{\pi(b^2)^{1-\epsilon}}\,.
\end{equation}

In the main text we also use the inverse Fourier transform (from $b$
to $q$-space) which can be easily obtained from \eqref{B1} by appropriately interchanging the roles of $q$ and $b$,
\begin{equation}
  \int\! { d^{D-2} b} \, {\rm e}^{-ib\cdot q} (b^2)^{-\nu} 
= \frac{\pi^{\frac{D-2}{2}}}{2^{2 \nu + 2 -D}} \frac{\Gamma\left(\frac{D}{2}-1-\nu\right)}{\Gamma(\nu)}
\left(q^2\right)^{1+\nu-\frac{D}{2}}\,.
  \label{B1bis}
\end{equation}

\section{The deflection angle in the probe limit }
\label{app:probelim}

The deflection angle is a key classical observable in the $2\to 2$ gravitational scattering which can be derived from scattering amplitudes thanks to the eikonal exponentiation. In the limit where the mass of one particle is much larger than any other energy scale in the problem, the result obtained from the eikonal approach should agree with a classical calculation describing the propagation of the other particle in the curved geometry produced by the heavy one. In order to carry out explicitly such a calculation, we need to know how the two particles couple to the massless fields in the theory under consideration. These couplings determine the classical solution for the massless fields produced by the heavy object, and this solution in turn determines the classical trajectory of the light probe in such background, \emph{neglecting} its backreaction.
In this appendix we will discuss some explicit examples of such probe-limit calculations for GR, brane scattering and $\mathcal{N}=8$ supergravity. 

\subsection{Geodesic motion in Schwarzschild}
\label{app:geoschw}

The simplest case is of course that of a scalar particle with a large mass. While it is possible to describe its gravitational field by following a diagrammatic approach~\cite{Duff:1974xx} (see also~\cite{KoemansCollado:2019ggb,Jakobsen:2020ksu,Mougiakakos:2020laz}), deriving the $D$-dimensional Schwarzschild metric order by order in the large distance expansions $R_s/r\ll 1$, here we will use directly the exact (Schwarzschild-Tangherlini) black-hole solution in $D$ spacetime dimensions,
\begin{equation}
	\label{eq:SchwarzD}
	ds^2 = - \left(1-\left(\frac{R_s}{r}\right)^{D-3} \right) dt^2 + \left(1-\left(\frac{R_s}{r}\right)^{D-3} \right)^{-1} dr^2 + r^2 d \Omega_{D-2}^2\,.
\end{equation}
The precise relation between the Schwarzschild radius $R_s$ and the mass of the heavy scalar $M$ is
\begin{equation}
	\label{eq:RsD}
	R_s^{D-3} = \frac{16\pi G M}{(D-2)\Omega_{D-2}}= \frac{8 \Gamma(\frac{D-1}{2}) G M}{\pi^{\frac{D-3}{2}}(D-2)}\;,
\end{equation}
where in the last step we used the area of the $n$-dimensional sphere of unit radius
$\Omega_{n}= \frac{2 \pi^{\frac{n+1}{2}}}{\Gamma ( \frac{n+1}{2})}$. One can write the action for a minimally coupled probe of mass $m_p$ as
\begin{equation}
	S = \frac{1}{2} \int d \tau \left( e(\tau)^{-1} \frac{d x^{\mu}}{d \tau} \frac{d x^{\nu}}{d \tau} g_{\mu \nu} -m_p^2 e(\tau) \right) \label{eq:gact01} \;,
\end{equation}
where $\tau$ parametrizes the trajectory and $e(\tau)$ is an auxiliary variable defining the world-line metric. The action is invariant under reparametrization $\tau\to \tau'(\tau)$ if $x^\mu$ is a scalar ($x'^{\mu}(\tau')=x^\mu(\tau)$) and $e'(\tau')\, d\tau' = e(\tau)\, d\tau$. Then we can choose a parametrization (a ```gauge'') in which $e(\tau)$ is constant and the equation of motions for $e(\tau)$ become a constraint. We shall also take into account that the motion takes place in the plane determined by the initial velocity and by the impact parameter, indicating with $\phi$ the angle in this plane. For later convenience, we write the constraint coming from the variation of $e(\tau)$ for a metric of the form~\eqref{eq:SchwarzD} but with generic functions of the radial coordinate $r$: $ds^2 = g_{tt} dt^2 + g_{rr} dr^2 + r^2 g_{\phi\phi} d\phi^2$, 
\begin{equation}
	\label{eq:etaueom}
	|g_{tt}| \left( \frac{dt}{d \tau} \right)^2 - g_{rr} \left( \frac{dr}{d \tau} \right)^2 - r^2 g_{\phi\phi} \left(\frac{d \phi}{d \tau} \right)^2 = e^2 m_p^2\;.
\end{equation}
Since we work with metrics that do not depend explicitly on the time and are spherically symmetric, we obtain the following conservation laws for the energy $E$ and the angular momentum $J$ of the probe
\begin{equation}
	\label{eq:clschw}
	e E_p = |g_{tt}| \frac{dt}{d\tau} ~,\quad \quad
	e J = r^2 g_{\phi\phi} \frac{d\phi}{d\tau} \;.
\end{equation}
By using these results in~\eqref{eq:etaueom} we obtain
\begin{equation}
	\label{eq:rdot}
	\frac{dr}{d\tau} = \pm \left[\frac{e^2 E_p^2}{|g_{tt}| g_{rr}} - \frac{e^2}{g_{rr}} \left( \frac{J^2}{r^2 g_{\phi\phi}} +  m_p^2\right)\right]^{\frac{1}{2}}\!,
\end{equation}
and 
\begin{equation}
	\label{eq:dphidr}
	\frac{d\phi}{dr} = \pm \frac{1}{r^2} \left[\frac{g_{\phi\phi}^2}{|g_{tt}| g_{rr}} \frac{ E_p^2}{J^2} - \frac{g_{\phi\phi}}{g_{rr}} \left( \frac{1}{r^2} + g_{\phi\phi} \frac{ m_p^2}{J^2 }\right)\right]^{-\frac{1}{2}}.
\end{equation}
Here $\pm$ refer to the incoming/outgoing portion of the trajectory, since a scattering process $r(\pm \tau)\to\infty$ and there is an inversion point $r_*$  corresponding to the largest root of $\frac{dr}{d\tau}$
\begin{equation}
	\label{eq:invpoint}
	\left[\frac{E_p^2}{|g_{tt}| g_{rr}} - \frac{1}{g_{rr}} \left( \frac{J^2}{r^2 g_{\phi\phi}} +  m_p^2\right)\right]_{r_*} = 0\,.
\end{equation}
Thus the scattering angle reads
\begin{align}
	\label{eq:cldangl}
	\Theta = & ~2 \int_{r_*}^\infty dr \left(\frac{d\phi}{d\tau}\right)\, \left(\frac{dr}{d\tau}\right)^{-1} - \pi \\ \nonumber = & ~2 J  \int_{r_*}^\infty \frac{dr}{r^2} \left[\frac{g_{\phi\phi}^2 E_p^2}{|g_{tt}| g_{rr}} - \frac{g_{\phi\phi}}{g_{rr}} \left(\frac{J^2}{r^2} + g_{\phi\phi} m_p^2\right)\right]^{-\frac{1}{2}}-\pi\;.
\end{align}
For the Schwarzschild metric we have $|g_{tt}|=g_{rr}^{-1}= 1-\left(\frac{R_s}{r}\right)^{D-3}$ and $g_{\phi\phi}=1$, so~\eqref{eq:cldangl} reduces to an incomplete elliptic integral.

For our purposes it is useful to write an explicit perturbative solution in the PM expansion. We first start from~\eqref{eq:invpoint} which for Schwarzschild's case reads
\begin{equation}
	\label{eq:invpointS}
	E_p^2 - \left(1-\left(\frac{R_s}{r_*}\right)^{D-3} \right) \left( \frac{J^2}{r_*^2} +  m_p^2\right) = 0\;.
\end{equation}
For general $D$ we can solve this constraint perturbatively: the leading contribution is obtained by ignoring the term proportional to $(R_s/r_*)^{D-3}$ and then it is straightforward to find the corrections in an expansion for large $J$
\begin{align}
	\begin{split}
	\label{eq:r*exp}
		r_* &= \frac{J}{\sqrt{E_p^2-m_p^2}} \left[1 + \sum_{n=1}^\infty \frac{E_p^2 \, c_n}{(E_p^2-m_p^2)} \,\left(\sqrt{E_p^2-m_p^2} \,\frac{R_s}{J} \right)^{n (D-3)} \right], \\
		c_1 &= -\frac{1}{2}\,,
		\quad
		c_2 = \frac{(5-2D) E_p^2+4m_p^2}{8(E_p^2-m_p^2)}\,,
		\\ 
		c_3 &= - \frac{(D-3)(3D-8) E_p^4-4(3D-8) E_p^2 m_p^2 + 8 m_p^4}{16(E_p^2-m_p^2)^2}\,,\quad \ldots
\end{split}
\end{align}
Then we can write~\eqref{eq:cldangl} in the Schwarzschild case after introducing the variable $u=r_*/r$
\begin{equation}
	\label{eq:cldanglS}
\Theta=
	~\frac{2 J}{r_*}  \int_{0}^1 d u  \left[E_p^2 - \left(1-\left(\frac{R_s u }{r_*}\right)^{D-3}\right) \left(\frac{J^2 u^2}{r_*^2} +  m_p^2\right)\right]^{-\frac{1}{2}}-\pi\;.
\end{equation}
We can then use~\eqref{eq:r*exp} in the equation above to rewrite the integrand in term of $J$ instead of $r_*$ and then expand it for small values $\sqrt{E_p^2-m_p^2} \, R_s/J$. The integral in $u$ can be performed in terms of Euler's Beta and we obtain
\begin{equation}
	\label{eq:chiperd}
	\Theta = \sum_{n=1}^\infty \Theta_n \left(\frac{\sqrt{E_p^2-m_p^2}\, R_s}{J} \right)^{n(D-3)}
\end{equation}
whose explicit terms up to 3PM are
\begin{align} \label{eq:chi1g}
	\Theta_1 & =  \frac{\sqrt{\pi } \Gamma \left(\frac{D}{2}-1\right) \left[(D-2) E_p^2-m_p^2\right]}{2 \Gamma \left(\frac{D-1}{2}\right) (E_p^2-m_p^2)} \,,\\
	\label{eq:chi2g} \Theta_2 & =  \frac{\sqrt{\pi } \Gamma \left(D-\frac{5}{2}\right)}{8 \Gamma (D-2)} \frac{(2 D-5)(2 D-3) E_p^4-6 (2 D-5) E_p^2 m_p^2+3 m_p^4}{(E_p^2-m_p^2)^2}\,, \\
	\begin{split}
	\Theta_3 & = \frac{\sqrt{\pi } \Gamma \left(\frac{3 D}{2}-4\right)}{48 \Gamma \left(\frac{3 D}{2}-\frac{7}{2}\right)} (E_p^2-m_p^2)^{-3}  \\
	& \times \bigl[(3 D-8) (3D-6) (3 D-4) E_p^6- 15 (3 D-8) (3D-6) E_p^4 m_p^2  \label{eq:chi3g}\\
	& + 45 (3 D-8) E_p^2 m_p^4- 15 m_p^6\bigr] \;.
	\end{split}
\end{align}
For convenience, let us write out the $D=4$ expressions as well: 
\begin{align} \label{eq:chi1g4D}
	\Theta_1 & =  \frac{2 E_p^2-m_p^2}{E_p^2-m_p^2} \,,\\
	\label{eq:chi2g4D} 
	\Theta_2 & = \frac{3 \pi  \left(5 E_p^2-m_p^2\right)}{16 \left(E_p^2-m_p^2\right)}\,, \\
	\begin{split}\label{eq:chi3g4D}
	\Theta_3 & = \frac{-120 E_p^4 m_p^2+60 E_p^2 m_p^4+64 E_p^6-5 m_p^6}{12 \left(E_p^2-m_p^2\right){}^3}\,.
	\end{split}
\end{align}

It is also instructive to study the result for the deflection angle in the PN (as opposed to PM) expansion, restricting for simplicity to $D=4$.
For this purpose, it is convenient to introduce the variables $v_\infty$, $j_\text{PN}$ (as in \eqref{eq:PNdef}) and $\alpha$ according to,
\begin{equation}\label{}
	E_p =  m_p \sqrt{1+v_\infty^2}\,,\qquad v_\infty= \frac{1}{j_\text{PN}\alpha}\,,\qquad
	J = G M m_p j_\text{PN}\,.
\end{equation}
In these variables, Eq.~\eqref{eq:invpointS} for the inversion point takes the following form
\begin{equation}\label{eq:invpointPN}
	1+\frac{1}{(j_\text{PN}\alpha)^2} - \left(1+\frac{G^2 M^2 j_\text{PN}^2}{r_\ast^2}\right)\left(1-\frac{2G M}{r_\ast}\right)=0\,.
\end{equation}
The PN limit can be then introduced by considering the scaling limit
\begin{equation}\label{PNlimitprobe}
	j_\text{PN} \sim \mathcal O(c)\,,\qquad
	G \sim \mathcal O(c^{-2})\,,\qquad
	\alpha \sim \mathcal O(1)\qquad
	\text{as }c\to\infty\,,
\end{equation}
which reflects the fact that the angular momentum becomes large and the velocity small, while keeping $\alpha = G M m_p/(J v_\infty)$ of order one. 
It is straightforward to  solve \eqref{eq:invpointPN} perturbatively in the limit \eqref{PNlimitprobe}, obtaining
\begin{equation}\label{}
	r_\ast = G M \left[ j_\text{PN}^2 \alpha(\sqrt{1+\alpha^2}-\alpha) - \left(1+\frac{\alpha}{\sqrt{1+\alpha^2}}\right) \right] + \mathcal O(c^{-4})
\end{equation}
for the 0PN and 1PN contributions. Substituting into \eqref{eq:cldanglS}, expanding in the same limit and performing the resulting elementary integrals, one then obtains the following PN-expanded deflection angle
\begin{equation}\label{}
	\Theta = 2\arctan\alpha + \frac{2}{j_\text{PN}^2} \left[
	3\left(\arctan\alpha+\frac{\pi}{2}\right)+\frac{3\alpha^2+2}{\alpha(1+\alpha^2)} 
	\right]
	+ \mathcal O(c^{-4})\,.
\end{equation}
This calculation actually reproduces the deflection angle up to 1PN independently of the probe-limit approximation, i.e.~the first two lines of Eq.~(45) of \cite{Bini:2017wfr}.

\subsection{Geodesic motion in D-brane metric}
\label{app:geoDp}

For the analysis of the string-brane scattering it is useful to discuss the geodetic motion in the gravitational backreaction of a stack of D$p$-branes. It is straightforward to adapt the analysis of the previous section to the case where the metric is given by~\eqref{eq:pbrsol}. By writing the transverse space in polar coordinate this metric takes the following form
\begin{equation}
	\label{eq:Dpmetpol}
	ds^2 =  \left[{H(r)}\right]^{-\frac{1}{2}}\, \eta_{\alpha\beta}dx^\alpha dx^\beta +  \left[{H(r)}\right]^{\frac{1}{2}} \left(dr^2 + r^2 d\phi^2 d\Omega^2_{d-p-3} \right)\;,
\end{equation}
where $H(r)$ is the harmonic function given in~\eqref{eq:Hhf}. We can then read the components of the metric involved in the geodesic equations~\eqref{eq:etaueom} and~\eqref{eq:clschw}
\begin{equation}
	\label{eq:metcomDp}
	|g_{tt}|^{-1} = g_{rr} = g_{\phi\phi} = \sqrt{H}\;, 
\end{equation}
Since we are focusing on the motion of light string state with a large kinetic energy, we can neglect the terms involving the probe mass and from~\eqref{eq:dphidr} we obtain
\begin{equation}
	\label{eq:4.2}
	\frac{d\phi}{dr} = \frac{ \pm b_J}{r^2\sqrt{1-\left(\frac{b_j}{r}\right)^2+\left(\frac{R_p}{r}\right)^{d-p-3}}} =\pm \frac{b_J}{r^2 F}\;,\quad \Leftrightarrow\quad
	\frac{d\phi}{d\rho} = \frac{\pm\hat{b}_J}{\sqrt{1-\hat{b}_J^2 \rho^2+\rho^{d-p-3}}}\;,
\end{equation}
where $F=\pm\sqrt{H(r) -\frac{b_J}{r^2}}$, $\rho=R_p/r$ and $\hat{b}_J=b_J/R_p$. One can follow the same steps as in the previous section to find the scattering angle by integrating~\eqref{eq:cldangl}, where the turning point $r_*$ (or $\rho_*$) is defined as in~\eqref{eq:invpoint}. At the leading PM order we get
\begin{equation}
	\label{eq:3.10}
	\Theta_1 = \sqrt{\pi}\, \frac{\Gamma\left(\frac{d-p-2}{2}\right)}{\Gamma\left(\frac{d-p-3}{2}\right)} \left(\frac{R_p}{r}\right)^{d-p-3}\,.
\end{equation}
For completeness let us quote also some exact result when the integral in~\eqref{eq:cldangl} can be expressed in terms of elementary functions
\begin{equation}
	\label{eq:4.5}
	p=d-4~\Rightarrow \Theta = 2\arctan\left(\frac{1}{2\hat{b}_J}\right)\;,\quad
	p=d-5~\Rightarrow \Theta = \frac{\pi \hat{b}_J}{\sqrt{\hat{b}_J^2-1}} -\pi\;
\end{equation}
and when~\eqref{eq:cldangl} reduces to a complete elliptic integral such as the case of a stack of D$3$-branes in type II theories
\begin{equation}
	\label{eq:4.6}
	p=d-7 \Rightarrow \Theta = 2 \sqrt{1+k_3^2}\;, K(k_3) -\pi \;, \quad k_3 = \frac{\hat{b}_J}{2}\left( \hat{b}_J - \sqrt{\hat{b}_J^2-4}\right) - 1\;.
\end{equation}
Here $K$ is the complete elliptic integral of the first kind
\begin{equation}
	\label{eq:4.6bis}
	K(k_3)  = \int_0^1 \frac{dx}{\sqrt{(1-x^2)(1-k_3^2 x^2)}}\;.
\end{equation}
Notice that when we have $3$ transverse directions (i.e.~$p=d-4$) the full deflection angle in the probe limit is determined by the leading eikonal as a consequence of supersymmetry as pointed out in~\cite{Caron-Huot:2018ape}.

We conclude this appendix by providing some details on the integrals in~\eqref{eq:cxy0} relevant for the semiclassical dynamics of a string probe (rather than just a point-like object) in the D-brane metric. Starting from~\eqref{eq:cxy0b}, it is convenient to proceed as done for~\eqref{eq:cxy0a}: we can neglect the second term in the square parenthesis since it scales as $(R_p/b)^{2(d-p-3)}$ and rewrite the first term as a total derivative with respect to the radial coordinate. Then we have
\begin{equation}
	\label{eq:intmuy}
	\frac{1}{2} \int_{-\infty}^\infty \!\!du \; {\cal G}_{\hat{y}} \simeq \int_0^\infty \! d\bar{r}\,\partial_{\bar{r}} \left[\frac{F}{\sqrt{H}} \partial_{\bar{r}} \ln\left(\bar{r} \sin\bar\phi H^{\frac{1}{4}} \right)\right]=-\left.\frac{b_J \cos\bar\phi}{\bar{r}^2 \sqrt{H} \sin\bar\phi}\right|_{\bar{r}=r_*}\!\!\!\simeq -\frac{\Theta_1}{b}\;.
\end{equation}
The only non-trivial contribution arises when the derivative inside the square parenthesis acts on $\ln\sin\bar\phi$, since this produces a factor of $\frac{d\bar\phi}{dr}$ which, thanks to~\eqref{eq:4.2}, cancel the overall factor of $F$ and yields a finite contribution as $r\to r_*$. In the final step we used $b_J\simeq b$, $\bar\phi(r_*)= \frac{\Theta+\pi}{2}$ and~\eqref{eq:3.10}. Then one can check that the result in~\eqref{eq:intmuy} agrees with the one in~\eqref{eq:cxy0b} with $\mu_{\hat{y}}^2$ defined in~\eqref{eq:mu0muy}. When evaluating~\eqref{eq:cxy0c}, it is more convenient to evaluate explicitly the derivatives for both terms in the square parenthesis obtaining
\begin{equation}
	\label{eq:c0erA0}
	\frac{1}{2} \int_{-\infty}^\infty \!\!du\;  {\cal G}_{0}=  \int_0^\infty \! du\,\left[\frac{\partial_u^2 \left(\bar{r} F H^{-\frac{1}{4}}\right)}{\bar{r} F H^{-\frac{1}{4}}}\right] =  \int_{r_*}^\infty \frac{d\bar{r}}{\bar{r} F H^{-\frac{1}{4}}}\;\partial_{\bar{r}}\left[\frac{F}{H^{\frac{1}{2}}} \partial_{\bar{r}} \left(\bar{r} F H^{-\frac{1}{4}}\right)\right]\;.
\end{equation}
Then it is straightforward to evaluate the expression above to 1PM order obtaining
\begin{equation}
	\label{eq:c0erA}
	\frac{1}{2} \int_{-\infty}^\infty \!\!du\;  {\cal G}_{0} \simeq \int_{b}^\infty d\bar{r} \frac{d-p-3}{4 \bar{r} \sqrt{\bar{r}^2-b^2}}\left[(d-p-4) - (d-p-1) \frac{b^2}{\bar{r}^2}\right] \left(\frac{R_p}{\bar{r}}\right)^{d-p-3},
\end{equation}
where we used again $r_*\simeq b$ at 1PM. Thanks to the relation
\begin{equation}
	\label{eq:c.9}
	\int_b^\infty \!\!dr \frac{r^{1-2x}}{r^2-b^2} = \frac{\sqrt{\pi} b^{1-2x} \Gamma\left(x-\frac{1}{2}\right)}{2 \Gamma(x)}\;,
\end{equation}
one can check that, after performing the integral,~\eqref{eq:c0erA} reproduces~\eqref{eq:cxy0c} with $\mu_0^2$ defined in~\eqref{eq:mu0muy}.

\subsection{Probe limit of the \texorpdfstring{$\mathcal N=8$}{N=8} case}
\label{app:geon8}

In order to perform the classical calculation for the massive $\mathcal N=8$ case it is easier to start from the setup discussed in Section~\ref{sec:n8tree} and dualize it to a frame where all the massive objects are 
D$p$-branes. This can be done by the following chain of dualities:
\begin{itemize}
	\item lift the setup to M-theory and then go back to type IIA by compactifying along the $9^{\rm th}$ direction; in this way we the first particle is transformed into a bound state of $n_1$ D$0$-branes, while the second particle is unchanged;
	\item perform T-dualities along the directions $7,\;8,\;9$ so that the external states become a bound state of $n_1$ D$3$-branes and a F$1$-string wrapped $n_2$ times along the direction $8$;
	\item since we are now in type IIB string theory we can perform a $S$-duality and obtain bound states of $n_1$ D$3$-branes and $n_2$ D$1$-branes;
	\item for convenience we can T-dualize back to type IIA along the $8^{\rm th}$ direction to get $n_1$ D$2$-branes and $n_2$ D$0$-branes.
\end{itemize}
All objects in each step are point-like in the noncompact directions and the string frame metric produced by the bound state of D$2$-branes is
\begin{equation}
	\label{eq:D2bsm}
	ds^2 = \left(1+\frac{4GM}{r} \right)^{-\frac{1}{2}} (-dt^2 + dx_7^2+dx_8^2) + \left(1+\frac{4GM}{r} \right)^{\frac{1}{2}} dx_\perp^2\;.
\end{equation}
The classical solution includes also a non-trivial RR 2-form, which does not play any role in this problem, and a dilaton\footnote{If the number of noncompact direction is $D$ we have
	\begin{equation}
		\label{eq:genDN8}
		\frac{\Gamma\left(\frac{D-3}{2}\right)}{\pi^{\frac{D-1}{2}}} \frac{\kappa_D T_2 n_1}{2} = \frac{\Gamma\left(\frac{D-3}{2}\right)}{\pi^{\frac{D-3}{2}}} 4G_D M\;,
	\end{equation}
	and the harmonic function is $1+\frac{\Gamma\left(\frac{D-3}{2}\right)\,4G_D M}{\pi^{\frac{D-3}{2}}\, r^{D-3}}$.
}
\begin{equation}
	\label{eq:D2bsd}
	e^{\phi} = \left(1 + \frac{4GM}{r} \right)^{\frac{1}{4}}\;,
\end{equation}
where $M$ is the mass of the D$2$-brane bound state. The action for the D$0$-brane probes involves both the fields above and read
\begin{equation}
	\label{eq:d0pb}
	S = -m_{p=0} \int d\tau e^{-\phi} \sqrt{\left|g_{\mu \nu} \frac{d x^{\mu}}{d \tau} \frac{d x^{\nu}}{d \tau}\right|}\;,
\end{equation}
where of course we restrict the motion in the noncompact space. This action is equivalent to~\eqref{eq:gact01} with an effective metric $g^{\rm eff}_{\mu\nu} = e^{-2\phi} g_{\mu\nu}$. We can then calculate the deflection angle by using~\eqref{eq:cldangl} with
\begin{equation}
	\label{eq:gn8}
	g^{\rm eff}_{00} = \left(1 + \frac{4GM}{r} \right)^{-1}\;,\quad  g^{\rm eff}_{rr} =  g^{\rm eff}_{\phi\phi} = 1\;.
\end{equation}
Then in this case the deflection angle is given by a circular integral
\begin{align}
	\label{eq:dangln8}
	\Theta =&~  ~2 J  \int_{r_*}^\infty \frac{dr}{r^2} \left[\left(1 + \frac{4GM}{r} \right) E_p^2 - \left(\frac{J^2}{r^2} + m_p^2\right)\right]^{-\frac{1}{2}}-\pi\nonumber\\
	= &~ 2 \arctan\left( \frac{2GM}{b_J} \frac{\sigma^2}{\sigma^2-1}\right) ,
\end{align}
where used $E_p=m_p \sigma$ and $J = m_p b\sqrt{\sigma^2-1}$.

\section{Some relations satisfied by the Levi-Civita tensor in \texorpdfstring{$D=4$}{D=4} }
\label{LeviCivita}

In this Appendix we provide some relations satisfied by the product of two the four-dimensional  Levi-Civita tensors.
We list them here
\begin{equation}
\epsilon^{\mu \nu \rho \sigma} \epsilon_{\mu \nu \rho \sigma} = - 4!
\label{LC1}
\end{equation}
\begin{equation}
\epsilon^{\mu \nu \rho \sigma} \epsilon_{\mu \nu \rho \delta} = -6 \delta^{\sigma}_{\delta} 
\label{LC2}
\end{equation}
\begin{equation}
\epsilon^{\mu \nu \rho \sigma} \epsilon_{\mu \nu \gamma \delta} = - 2 \delta^{\rho \sigma}_{\gamma \delta}
\label{LC3}
\end{equation}
 \begin{equation}
\epsilon^{\mu \nu \rho \sigma} \epsilon_{\mu \beta \gamma \delta}=  - \delta^{\nu \rho \sigma}_{\beta \gamma \delta} 
\label{LC4}
\end{equation}
and
\begin{equation}
\epsilon^{\mu \nu \rho \sigma} \epsilon_{\alpha \beta \gamma \delta} = - \delta^{\mu}_{\alpha} \delta^{\nu \rho \sigma}_{\beta \gamma \delta} + \delta^{\mu}_{\beta} \delta^{\nu \rho \sigma}_{\alpha \gamma \delta} -  \delta^{\mu}_{\gamma} \delta^{\nu \rho \sigma}_{\alpha \beta \delta} +  \delta^{\mu}_{\delta} \delta^{\nu \rho \sigma}_{\alpha \beta \gamma}  
\label{LC5}
\end{equation}
where
\begin{equation}
\delta^{\rho \sigma}_{\gamma \delta} = \delta^{\rho}_{\gamma} \delta^{\sigma}_{\delta} - \delta^{\rho}_{\delta} \delta^{\sigma}_{\gamma} 
\label{CL6}
\end{equation}
and
\begin{equation}
 \delta^{\nu \rho \sigma}_{\beta \gamma \delta} = \delta^{\nu}_{\beta} 
 \left(\delta^{\rho}_{\gamma} \delta^{\sigma}_{\delta} -\delta^{\rho}_{\delta} \delta^{\sigma}_{\gamma}\right) -
 \delta^{\nu}_{\gamma} 
 \left(\delta^{\rho}_{\beta} \delta^{\sigma}_{\delta} -\delta^{\rho}_{\delta} \delta^{\sigma}_{\beta}\right) +
  \delta^{\nu}_{\delta} 
 \left(\delta^{\rho}_{\beta} \delta^{\sigma}_{\gamma} -\delta^{\rho}_{\gamma} \delta^{\sigma}_{\beta}\right)\,.
\label{CL7}
\end{equation}
We use the convention where $\epsilon_{0123} =1$.

\section{String theory background}
\label{app:string}

In this appendix we provide some details on how to derive the relevant amplitudes in the context of the bosonic theory which is technically simpler than the superstring case while capturing all the main features.

\subsection{String theory conventions}
\label{app:conventions}

Here we collect our string conventions and a short discussion about the boundary state that is needed to describe the stack of D$p$-branes used in the main text.

\subsubsection{Scales and coupling constants}

Free string theory in flat spacetime is described by the Nambu-Goto action:
	\begin{align}
		S_\text{string} &= - T \int d \tau \int_0^\pi d\sigma \sqrt{({\dot{x}}\cdot x')^2 - {\dot{x}}^2 {x'}^2}\,, \nonumber \\ \dot{x}& \equiv \partial_{\tau} x(\sigma,\tau)\,, \qquad x' \equiv \partial_{\sigma} x(\sigma,\tau)\,, \qquad  \dot{x}\cdot x' \equiv  \eta_{\mu\nu} \dot{x}^{\mu} x'^{\nu} \,,
		\label{NaGo}
	\end{align}
	where the double integral is nothing but  the area swept by the string and $T$ is the classical string tension with dimensions of an energy per unit length (in units in which $c=1$). At the classical level (e.g. in the context of cosmic strings)  $T$ is the only free parameter and the classical equations of motion are obviously independent of  it.\footnote{When moving  in a non trivial geometry  the size of the string relative to the characteristic scale of the geometry does instead matter at the classical level.}
The inverse of $T$ has the same dimensions as an angular momentum per squared mass (recall that we have set $c=1$) and has been denoted traditionally by $2 \pi\alpha'$ since it first made its appearance in hadronic physics as the slope of the linear Regge trajectories: $\alpha(t) = \alpha(0) + \alpha' t$.

In a quantum context what matters is the action in Planck units, a dimensionless quantity, and this necessarily introduces a fundamental area $\ell_s^2$ as the natural replacement of Planck's constant in string theory \cite{Veneziano:1986zf}. In this review we adopt the following relation between these various quantities:
\begin{equation}
  \frac{S_\text{string}}{\hbar} = \frac{T} {\hbar}\, ({\rm Area~Swept}) = \frac{{\rm Area~Swept}}{2 \pi \ell_s^2 }\,,\qquad 
  \ell_s = \sqrt{\alpha' \hbar} = \sqrt{\frac{\hbar} {2 \pi T} }\;, 
	\label{basicrel}
\end{equation} 
where we should note the analogy with the definition of the Planck length in four dimensions, $\ell_P = \sqrt{G \hbar}$, with $G$ replacing the role of $\alpha'$. The similarity goes even further: classical GR has no intrinsic length (or mass) scale while quantum gravity does.
Physically, $\ell_s$ plays the role of a minimal length scale, a minimal size for fundamental quantum strings, and a short-distance cutoff regularizing quantum corrections. Its inverse, $M_s \equiv \frac{\hbar}{\ell_s}$ is the energy/mass scale associated with string excitations and with the cutoff in momentum space. In string theory it is thus natural to take $\ell_s$ as the basic unit of length and to express any other quantity, according to its dimensionality, in terms of $\ell_s$,  $c$, $\hbar$, and of dimensionless numbers. The latter, when not directly fixed, are associated with the value of dimensionless scalar fields, called moduli.

 In the rest of this Appendix, and elsewhere in this review, we will follow the common practice of using interchangeably $\alpha' $ and $\ell_s^2$ (i.e.~set $\hbar =1$) except when the distinction is physically relevant. As an example, $\alpha(t)$ is dimensionless (being an exponent characterizing power-like Regge behavior) and therefore $\alpha'$  cannot be just the inverse of $T$. What appears in the above expression for $\alpha(t)$  is actually $\frac{\alpha' t}{\hbar} = \frac{t}{M_s^2}$, but we shall omit the $\hbar^{-1}$ factor throughout.

In perturbative string theory the strength of the gravitational interaction depends on  $\alpha'$ and on the moduli of the theory as follows:
\begin{equation}
	\label{eq:kappad}
	2\kappa^2_d \hbar=16 \pi G_d \hbar = 16 \pi \ell_d^{d-2} = 2^{-\frac{d-10}{2}} (2 \pi)^{d-3} g^2_s\, (\alpha ' \hbar)^{\frac{d-2}{2}} =  2^{-\frac{d-10}{2}} (2 \pi)^{d-3} g^2_s\, \ell_s^{d-2} \,,
\end{equation}
where $\ell_d$ is the $d$-dimensional Planck length, $g_s$, the string coupling, is related to the expectation value of the dilaton and of course we have $d=10$ for critical superstring theory and $d=26$ for the bosonic theory. Note that at very weak string coupling $\ell_d/\ell_s \ll 1$. This physically means that string-size corrections intervene well before quantum gravity loops get out of control. That does not mean, however, that straight perturbation theory is always reliable at $g_s^2 \ll1$: although there is a formal loop expansion in powers of $g_s^2$, these can be enhanced by large kinematical factors such as powers of the energy or infrared singularities as discussed in the main body of this report.

When some of the directions are compactified on a manifold of volume $V_c$ and only $D$ directions are non-compact, it is useful to introduce the Newton constant $\kappa_D^2 = 8 \pi G_D$  for the $D$-dimensional theory appropriate for describing physics at distances much larger than the size of the compact dimensions:
\begin{equation}
	\label{eq:kappaD}
	\kappa^2_D=\frac{\kappa^2_d}{V_c} = 8 \pi \ell_D^{D-2} \,.
\end{equation}
It follows that the ratio $\ell_D/\ell_s$ is further reduced if $V_c  \gg \ell_s^{d-D}$, the case of ``large extra dimensions". 
In this report only the case of string-size extra dimensions will be considered.

The spectrum of type II and bosonic string theories also contains  D$p$-branes, non-perturbative objects which can support the end-points of open strings~\cite{Polchinski:1995mt}. In its simplest configuration a D$p$-brane enforces Neumann boundary conditions on the string fields  along $p$ spatial directions and time, and Dirichlet boundary conditions along the remaining $d-(p+1)$ directions. In other words, the end-points of an open string move on a $(p+1)$-dimensional Minkowskian submanifold. The tension (i.e. energy per unit $p$-dimensional surface) of these objects is 
\begin{equation}
	\label{eq:TpTen}
	\tau_p = \frac{T_p}{\kappa_d}\;, \quad \mbox{with} \quad T_p = 2^{-\frac{d-10}{4}} \sqrt{\pi} (2\pi \sqrt{\alpha'})^{\frac{d-2p-4}{2}}\;,
\end{equation}
which of course fixes its coupling to gravity. In the superstring case the D$p$-branes are minimally coupled to the $(p+1)$ Ramond--Ramond potential with a RR charge density $\mu_p$ given by $\mu_p = \sqrt{2}T_p$ (i.e.~charge per unit $p$-volume). 

In Section~\ref{sec:string-disk}, we use a stack of $N$ D$p$-branes as a target in a (thought) scattering experiment with fundamental strings. It is then convenient to introduce the scale $R_p$ of the geometry sourced by the D$p$-branes
\begin{equation}
	\label{eq:Rpscale}
	R_p^{d-p-3} =  \frac{ \Gamma\left( \frac{d-p-3}{2}\right) }{ \pi^{\frac{d-p-1}{2}}  }
	\frac{g_s N}{4}  \frac{ (2 \pi \sqrt{\alpha'} )^{d-p -3}}{2^{\frac{d-10}{2}}} \,\, .
\end{equation}
In terms of the tension $T_p$ we have
\begin{equation}
	\label{eq:RpTp}
	\frac{\kappa_d T_p N}{2} = \frac{\pi^{\frac{d-p-1}{2}}  R_{p}^{d-p-3}}{ \Gamma\left( \frac{d-p-3}{2}\right)} \ .
\end{equation}

\subsubsection{String mode expansion}
\label{app:modeexp}

Our conventions on the string coordinates and their mode expansion in the closed string case are
\begin{equation}
	\label{eq:XLRc}
	\begin{aligned}
		X^M(z,\bar{z}) & =  X^M_L(z) + X^M_R(\bar{z})\;, \\
		X^M_L(z) & = q_L - i \frac{\alpha'}{2} p_L^M \ln z + i \sqrt{\frac{\alpha'}{2}} \sum_{n\not=0} \frac{\alpha_n^M}{n} z^{-n} \;,\\
		X^M_R(\bar{z}) & = q_R - i \frac{\alpha'}{2} p_R^M \ln \bar{z} + i \sqrt{\frac{\alpha'}{2}} \sum_{n\not=0} \frac{\bar\alpha_n^M}{n} \bar{z}^{-n} \;.
	\end{aligned}
\end{equation}
In the closed string sector the modes $\alpha_n$ and $\bar\alpha_n$ are independent and, after canonical quantization, satisfy the commutation relations
\begin{equation}
	\label{eq:comalpha}
	{} [\alpha^M_n, \alpha_{m}^N]= \eta^{MN} n\, \delta_{n+m}\;,\quad [\bar\alpha^M_n, \bar\alpha_{m}^N] = \eta^{MN} n \,\delta_{n+m}\;,\quad [\alpha^M_n, \bar\alpha_{m}^N] = 0\;.
\end{equation}
It is  convenient to take also the left/right center of mass and momentum operators to be independent
\begin{equation}
	\label{eq:qpcom}
	[q^M_L,p^N_L] = \frac{i}{2} \eta^{MN} \;,\quad [q^M_R,p^N_R] = \frac{i}{2} \eta^{MN}\,; \quad [q^M_L,p^N_R] = [q^M_R,p^N_L] = 0
\end{equation}
and impose, for the non-compact directions, an identification on their eigenvalues of the physical states $p_L^M = p_R^M = p^M$, where $p^M$ is the total momentum. It is  also convenient to introduce $\sqrt{\frac{\alpha'}{2}}p_L^M= \alpha_0^M$ and $\sqrt{\frac{\alpha'}{2}}p_R^M= \bar\alpha_0^M$, so we have
\begin{equation}
	\label{eq:dXLRc}
	\partial X^M = - i \sqrt{\frac{\alpha'}{2}} \sum_{n} {\alpha_n^M} z^{-n-1}\;, \quad
	\bar\partial X^M = - i \sqrt{\frac{\alpha'}{2}} \sum_{n} {\bar\alpha_n^M} \bar{z}^{-n-1}\;.
\end{equation}
Then we have the following Operator Product Expansions (OPE)
\begin{equation}
	\label{eq:OPEbos}
	\begin{gathered}
		\partial X^M(z_1) \partial X^N(z_2) \sim -\frac{\alpha'}{2} \frac{\eta^{MN}}{(z_1-z_2)^2}+\ldots \\ e^{i k_1 X_L(z_1)} e^{i k_2 X_L(z_2)} \sim (z_1-z_2)^{\frac{\alpha'}{2} k_1 k_2} e^{i (k_1+k_2) X_L(z_2)}+\ldots 
	\end{gathered}
\end{equation}
and similarly for the anti-holomorphic part. 

\subsubsection{Normalizations for string amplitudes}
\label{app:stringnormaliszations}

We associate a factor of $\frac{\kappa_d}{2\pi}$ to each string vertex operator and indicate with $C_{S_2}$ the normalization of the closed string tree-level amplitudes where the worldsheet has the topology of a sphere. Imposing perturbative unitarity on the factorization of a 4-point function into two 3-point amplitudes one obtains
\begin{equation}
	\label{eq:Cs2kap}
	C_{S_2}\left(\frac{\kappa_d}{2\pi}\right)^2 \frac{\alpha'}{8\pi} = 1\;.
\end{equation}
The factor of $\alpha'/(8 \pi)$ is related to the closed string propagator
\begin{equation}
	\label{eq:clospr}
	P(a) = \frac{\alpha'}{2} (L_0 + \tilde{L}_0 -2a)^{-1} = \frac{\alpha'}{8\pi} \int\limits_{|z|\leq 1} 
	\frac{d^2z}{|z|^{2}}  z^{L_0-{a}} \bar{z}^{\tilde{L}_0-{a}}~,
\end{equation}
where $a=1$ in the bosonic theory and $a=\frac{1}{2}$ in the NS-NS sector of type II theories.
The overall normalization of the left-hand side is chosen so to have residue one at various mass poles (for instance, close to the tachyon pole, behaves as $1/(p^2+m_T^2)$, with $m_T^2=-4/\alpha'$), while on the right-hand side we use the definition of the measure $d^2z = i\, d z \wedge d \bar{z}= 2\, d {\rm Re}(z)  d{\rm Im}(z)$ as in~\cite{Polchinski:1998rq}. 

\subsubsection{The boundary state}
\label{app:boundarystate}

In this subsection we give a short introduction to the boundary state that describes the D$p$-branes. For more details see for instance~\cite{DiVecchia:1999mal}. 
D$p$-branes  are extended $p$ dimensional objects  characterized by
the fact that open strings  have their endpoints attached to them.
The open string with the endpoints at $\sigma=0, \pi$ attached to  two parallel D$p$-branes
satisfies the usual Neumann boundary conditions along the directions
longitudinal to the world volume of the branes
\begin{equation}
\partial_{\sigma} X^{\alpha}|_{\sigma=0, \pi} =0 \hspace{2cm} \alpha=0, 1, \ldots,
p
\label{neu1}
\end{equation}
and  Dirichlet boundary conditions along the directions transverse to the
brane
\begin{equation}
X^{i} |_{\sigma=0} = y^i~~ ;~~X^{i} |_{\sigma=\pi} = w^i  \hspace{2cm} i= p+1, \ldots, d-1
\label{dir1}
\end{equation}
where $y^i$ and $w_i$  are the coordinates of the two D$p$-branes and we take $\sigma$ and $\tau$ in the two intervals $0\leq \sigma \leq \pi$ and $0\leq \tau \leq T$.

The previous are the conditions in the so-called open string channel, but those characterizing the boundary state are instead those in the so-called closed string channel. This nomenclature follows from the fact that the annulus diagram can be constructed in two ways: as one-loop of open strings or as a tree diagram
with a closed string propagators connecting two boundary states.  These two descriptions are connected by a conformal transformation in terms of the variable $\zeta\equiv\sigma+i\tau$:
\begin{equation}
\label {ctz}
\zeta = \sigma + i \tau \rightarrow -i\zeta=\tau - i\sigma~.
\end{equation}
After the inversion $\sigma\rightarrow -\sigma$ the previous conformal
transformation
simply amounts to exchange $\sigma$ with $\tau$ and viceversa
\begin{equation}
\label{ctst}
(\sigma, \tau)\rightarrow (\tau, \sigma)~.
\end{equation}
Finally in order to have the closed string variables $\sigma$ and $\tau$ to
vary in the  intervals $\sigma\in[0, \pi]$ and  $\tau \in [0, \hat{T}]$
corresponding to a closed string propagating between the two D branes one
must
perform the following conformal rescaling
\begin{equation}
\label{res}
\sigma \rightarrow \frac{\pi}{T}\sigma ~~~~~~~~~~
\tau\rightarrow \frac{\pi}{T}\tau~,
\end{equation}
with  $\hat{T}=\pi^2/T.$ 

The equations characterizing the boundary state are obtained by applying the conformal transformation previously constructed to the
boundary conditions for the open string given in (\ref{neu1}) and
(\ref{dir1}). At $\tau =0$ we get the following conditions:
\begin{equation}
\label {bc1c}
\partial_{\tau}X^\alpha|_{\tau=0}|B_X \rangle =0 ~~~~~~~~~~\alpha =0,...,p~,
\end{equation}
\begin{equation}
\label {bc2c}
X^i|_{\tau=0}|B_X \rangle =y^i ~~~~~~~~~~i = p+1,..., d-1~.
\end{equation}
Analogous conditions can be obtained for the D$p$-branes at
$\tau = {\hat{T}}$.

The previous equations can be easily written in terms of the closed string
oscillators by making use of the expansion in eq.(\ref{eq:XLRc}), obtaining
\begin{gather}
\label{over1}
(\alpha_n^\alpha+\widetilde\alpha_{-n}^\alpha)|B_X \rangle  =0\;,\quad
(\alpha_n^i-\widetilde\alpha_{-n}^i)|B_X \rangle  =0\quad \forall n\neq 0\,,
\nonumber\\
{\hat{p}}^\alpha|B_X \rangle  = 0 \qquad({\hat{q}}^i-y^i)|B_X \rangle =0\,,
\end{gather}
where ${\hat{q}}= q_L+q_R$ and ${\hat{p}}= p_L=p_R$.
Introducing the reflection matrix
\begin{equation}
\label{matS}
\mathcal{R}_p^{\mu\nu}=(\eta^{\alpha\beta},-\delta^{ij})~,
\end{equation}
the equations for the non-zero modes can be rewritten as
\begin{equation}
\label {over1s}
(\alpha_n^\mu+\mathcal{R}\indices{_p^\mu_\nu}\widetilde\alpha_{-n}^\nu)|B_X \rangle
=0~~~~~\forall \,\, n \neq 0
~.
\end{equation}
The state satisfying the previous equations can easily be determined to be
\begin{equation}
\label{b1}
|B_X \rangle  =\frac{T_p}{2} \delta^{d-p-1}({\hat q}^i-y^i) \left(\prod_{n=1}^\infty
e^{-\frac{1}{n}
\alpha_{-n}
\mathcal{R}_p \widetilde\alpha_{-n}}\right)|0\rangle _{\alpha}|0\rangle
_{\widetilde\alpha}
|p=0\rangle ~,
\end{equation}
where the normalization $\frac{T_p}{2}$ is fixed  by imposing that the computation of  the annulus diagram in the open and in the closed string channel gives the same result.

\subsection{String-brane scattering: the bosonic theory at tree-level}
\label{ssec:string-brane-bos-1}

The simplest possible setup where we can study the string theory leading eikonal is the scattering of two closed string tachyons of momenta $p_1$ and $p_2$ off a stack of $N$ D$p$-branes in bosonic string theory. The tree-level string amplitude capturing the scattering process mentioned above is 
\begin{equation}
	\label{eq:TTtree}
	{\cal A}_0^{T} = \frac{\kappa_{d} N T_p}{2} \frac{\Gamma (-\alpha' E_s^2-1) \Gamma\left(-\frac{\alpha'}{4} t -1\right)}{\Gamma\left( -\alpha' E_s^2 - \frac{\alpha'}{4} t -2\right)}~,
\end{equation}
where the kinematics is identical to the one discussed in the superstring case after~\eqref{eq:sstree}. It is useful to provide some detail on the derivation of Eq.~\eqref{eq:TTtree} as this will help in clarifying the key physics novelty of the string eikonal with respect to the field theory setup.

As standard we start from the vertex operators describing the emission of the external states. For the closed string tachyon we have
\begin{equation}
	\label{eq:tachvo}
	{\cal V}_{T}(z_i,\bar{z}_i) = \frac{\kappa_{d}}{2\pi} \hat{\cal V}_{T}(z_i,\bar{z}_i) = \frac{\kappa_{d}}{2\pi}  e^{i p_i X(z_i,\bar{z}_i)}\;,
\end{equation}
where $X^M(z_i,\bar{z}_i)$ are the string embedding coordinates given in \eqref{eq:XLRc}, the momenta are on-shell $p_i^2 = \frac{4}{\alpha'}$ (see~\ref{app:conventions} for our string theory conventions) and the exponentials are understood to be normal ordered. The tree-level amplitude corresponds to world-sheet with the topology of the disk with the two closed string insertions at two points ($z_1$, $z_2$) in its interior
\begin{equation}
	\label{eq:TTDp}
	{\cal A}_0^{T} = C_{S_2} \frac{\alpha' \kappa_d}{8\pi} N \int \frac{d^2z_1 d^2z_2}{dV_{SL(2,R)}} \langle 0| {\cal V}_{T}(z_1,\bar{z}_1) {\cal V}_{T}(z_2,\bar{z}_2) | B \rangle\;.
\end{equation}
Here $|B\rangle$ is the boundary state describing the stack of D$p$-branes, see Eq.~\eqref{b1}, $|0\rangle$ is the $SL(2,C)$ invariant vacuum and our conventions on the string normalizations are summarized in~\ref{app:stringnormaliszations}. The effect of $|B\rangle$ on the world-sheet fields is to identify the holomorphic and the anti-holomorphic parts through the reflection matrix $\mathcal{R}_p$ which is the identity along the D$p$-branes and minus the identity in the transverse directions. After the identification, the anti-holomorphic fields are placed at $1/\bar{z}_2$ and $1/\bar{z}_1$, while the holomorphic ones are inside the disk of unit radius at $z_1$ and $z_2$ and $|z|=1$ represents the world-sheet boundary. Finally the measure can be easily obtained by fixing three punctures, say $1/\bar{z}_2$, $1/\bar{z}_1$ and $z_2$ and by integrating over the remaining one inserting the contribution of the $c$-ghost
\begin{align}
	\frac{d^2z_1 d^2z_2}{dV_{SL(2,R)}} & = dz_1 \langle c\left(\frac{1}{\bar{z}_2}\right) c\left(\frac{1}{\bar{z}_1}\right)  c\left({z}_2\right)\rangle \\ \nonumber & = dz_1 \left(\frac{1}{\bar{z}_2}-\frac{1}{{\bar{z}}_1}\right) \left(\frac{1}{\bar{z}_1}-z_2\right) \left(\frac{1}{\bar{z}_2}-z_2\right) = \left(\frac{1}{\bar{z}_2}-z_1\right)^2 \left(\frac{1}{\bar{z}_1}-z_2\right)^2 dx 
\end{align}
where in the last step we introduced the $SL(2,R)$ invariant cross-ratio
\begin{equation}
	\label{eq:x-cr}
	x = \frac{(z_1-z_2)\left(\frac{1}{\bar{z}_2}-\frac{1}{\bar{z}_1}\right)}{\left(\frac{1}{\bar{z}_1}-z_2\right) \left(\frac{1}{\bar{z}_2}-z_1\right)}\;.
\end{equation}
The cross-ratio $x$ is manifestly real and, since the $z_i$'s are inside the disk of unit radius, it lies between zero and one. In particular, the limit $x\to 0$ corresponds to $|z_1-z_2|\to 0$ describing a world-sheet where the two external states interact first via a 3-point closed string vertex with one leg glued to the disk. Then by using the on-shell conditions, momentum conservation $(1 + \mathcal{R}_p) p_1 + (1 + \mathcal{R}_p) p_2=0$ and the ordering $ \frac{1}{|z_2|}> \frac{1}{|z_1|}>|z_1| > |z_2|$ we have
\begin{align}
	\label{eq:TTder}
	{\cal A}^{T}_0 & =  \frac{\kappa_{d} T_p}{2} N \int_0^1 dx \left(\frac{1}{\bar{z}_1} - z_2 \right)^2 \left(\frac{1}{\bar{z}_2} -  z_1\right)^2\left[
	\left( \frac{1}{\bar{z}_1} - z_1 \right)^{ \frac{\alpha'}{2} p_1 \mathcal{R}_p p_1  } 
	\left( \frac{1}{\bar{z}_1} - z_2 \right)^{\frac{\alpha'}{2} p_1 \mathcal{R}_p p_2  }\right. \nonumber \\  & \left.
	\left(\frac{1}{\bar{z}_2} - \frac{1}{\bar{z}_1}\right)^{\frac{\alpha'}{2} p_1  p_2  }  \left(z_1 - z_2\right)^{\frac{\alpha'}{2} p_1  p_2  } 
	\left(\frac{1}{\bar{z}_2} - z_1 \right)^{\frac{\alpha'}{2} p_1 \mathcal{R}_p p_2  }
	\left(\frac{1}{\bar{z}_2} - z_2 \right)^{\frac{\alpha'}{2} p_2 \mathcal{R}_p p_2  } \right] \\ \nonumber
	& =  \frac{\kappa_{d} T_p}{2} N \int_0^1 dx \,\,(1-x)^{-\alpha' E_s^2-2} x^{-\frac{\alpha'}{4} t -2} ~,
\end{align}
where we used~\eqref{eq:Cs2kap} the normalization factor $\frac{T_p}{2}$ of the boundary state (see Eq. \eqref{b1}) to simplify the overall normalization and
\begin{equation}
	\label{eq:kinbran}
	p_1 p_2 = \frac{- t + m_1^2 + m_2^2}{2} \, , \quad p_r \mathcal{R}_p p_r = - 2 E_s^2 + m_r^2 \, , \quad
	p_1 R p_2 = 2 E_s^2 +  \frac{t - m_1^2 - m_2^2}{2}  \; , 
\end{equation}
where $m_i^2=-\frac{4}{\alpha'}$. Thanks to the integral representation of the Euler Beta function one obtains~\eqref{eq:TTtree}.

Exactly as in the field theory setup the eikonal is captured by the limit $E_s^2\gg |t|$ and by using~\eqref{eq:gstirl}
\begin{equation}
	\label{eq:TTeikL}
	{\cal A}_0^{T} \simeq \frac{\kappa_{d} N T_p}{2} (-\alpha' E_s^2)^{1+\frac{\alpha' t}{4}} \Gamma\left(-1-\frac{\alpha' t}{4}\right) = \frac{\kappa_{d} N T_p}{2} (\alpha' E_s^2)^{1+\frac{\alpha' t}{4}} \frac{{\rm e}^{-i \pi \frac{\alpha' t}{4}}}{-\frac{\alpha' t}{4}} \frac{\Gamma\left(1-\frac{\alpha' t}{4}\right)}{1+\frac{\alpha' t}{4}} \;.
\end{equation}
Notice that this result can be directly derived directly from the last line~\eqref{eq:TTder}. When $\alpha' E_s^2 \gg 1$ then the $x$-integral can be performed by focusing on the region $x\to 0$ with $\alpha' E_s^2 x$ finite: by expanding the integrand in this regime and finally extending the region of integration to infinity, we have
\begin{equation}
	\label{eq:zto1ap}
	{\cal A}_0^{T} \simeq \frac{\kappa_{d} T_p}{2} N \int_0^\infty dx \, e^{\alpha' E_s^2 x } x^{-\frac{\alpha' t}{4}-2} = \frac{\kappa_{d} T_p}{2} N (-\alpha' E_s^2)^{1+\frac{\alpha' t}{4}}\, \Gamma\left(-\frac{\alpha' t}{4}-1\right)\;,
\end{equation}
which agrees with~\eqref{eq:TTeikL}. This shows explicitly that the result is entirely captured by the limit discussed before~\eqref{eq:TTder}. We can now follow the same approach used in Sect.~\ref{ssec:string-brane} for the superstring and perform the Fourier Transform~\eqref{eq:DbraneFT} of the bosonic result. Then we obtain
\begin{equation}
	\label{eq:incg1}
	2\delta_0 = \frac{\kappa_d N T_p E_s}{4\pi}\, \frac{\Gamma\left(1+\frac{\alpha' \nabla^2_b}{4}\right)}{1-\frac{\alpha'  \nabla^2_b}{4}} \left[(\pi b^2)^{-\frac{d-4-p}{2}} \gamma\left(\frac{d-4-p}{2},\frac{b^2}{\bY}\right)\right]\;,
\end{equation}
which is very similar to~\eqref{eq:incg2} except for the tachyonic pole and the fact that the critical dimension is $d=26$.
We recall that $\gamma(z,a)$ denotes the incomplete $\Gamma$-function.

\subsubsection{The Reggeon vertex formalism for bosonic strings}
\label{app:reggeonb}

In this section we follow~\cite{Ademollo:1989ag,Ademollo:1990sd,Brower:2006ea} and show that the $t$-channel pole in~\eqref{eq:TTeikL} is related to a particular class of closed string states in the leading Regge trajectory. Actually it is not difficult to present the argument for the scattering of general closed string states instead of focusing on tachyons~\eqref{eq:tachvo}, so we consider directly the case where we have generic physical vertex operators ${\cal V}_1$ and ${\cal V}_2$ in~\eqref{eq:TTDp} at the place of two ${\cal V}_{T}$. Notice that our discussion does not require the two vertex operators to be equal and this will be exploited later in Section~\ref{ssec:seikop}.

As suggested by the discussion after Eq.~\eqref{eq:TTeikL}, the eikonal result can be obtained by focusing on the degeneration channel where the world-sheet looks like a vertex involving the three closed string states with one of them being off-shell and propagating until it interacts with the stack of D$p$-branes. By inserting a complete set of states between the 3-point interaction and the boundary state we have
\begin{equation}
	\label{eq:factre1}
	{\cal A}_0^{(12)} =  C_{S_2} \frac{\alpha' \kappa_d}{8\pi} N \left[\sum_{{\cal Q}_\ell} \int \frac{d^2 z}{2\pi} (z \bar{z})^{\ell-2-\frac{\alpha' t}{4}} \langle {\cal V}_{1} {\cal V}_{2} {\cal Q}_\ell \rangle  \langle {\cal Q}_\ell| B \rangle\right] \;,
\end{equation}
where ${\cal Q}_\ell$ is a generic state of momentum $q=(p_1+p_2)$ at level $\ell$ and the sum, of course, covers all possible levels $\ell=0,1,\ldots$; the integral over the phase of $z$ is trivial, as the level matching condition has already been implemented, and the factor involving $z \bar{z}\equiv x$ follows from the closed string propagator~\eqref{eq:clospr}. We now need to see how the contributions of different states ${\cal Q}_\ell$ in~\eqref{eq:factre1} scale with the energy of the incident string and, in the eikonal limit, we would like to focus on the leading terms. In order to do so it is convenient to decompose the polarizations of the states ${\cal Q}_\ell$ by using the vectors $(e^\pm)$ defined in~\eqref{eq:epmlc} and the directions orthogonal to them. Factors of $E_s$ can arise from the contractions between $\partial^r X^+ = e^+_M\, \partial^r X^M$ in ${\cal Q}_\ell$ and the universal exponential factors $e^{i p_{1,2} X}$ in ${\cal V}_{1,2}$. Thus, in the high-energy regime, the decomposition~\eqref{eq:factre1} is dominated by the states
\begin{equation}
	\label{eq:calQQQ}
	{\cal Q}_\ell = \frac{1}{\ell!}  \left(i\sqrt{\frac{2}{\alpha'}} \partial X^+ \right)^\ell  \left(i\sqrt{\frac{2}{\alpha'}} \bar\partial X^+ \right)^\ell e^{i q X}\;.
\end{equation}
Below we summarize the contractions between the world-sheet coordinates in ${\cal Q}_\ell$ and ${\cal V}_{1,2}$ that are relevant for evaluating~\eqref{eq:factre1}
\begin{align}
	\label{eq:dXpeix}
	i \sqrt{\frac{2}{\alpha'}} \partial^r X^+(z) e^{ip_1 X(w)} & \sim (\sqrt{\alpha'} E_s)\, \partial^{r-1} \left(\frac{1}{z-w}\right) e^{i p_1 X(w)} \\ \nonumber i \sqrt{\frac{2}{\alpha'}} \partial^r X^+(z) e^{i p_2 X(w)} & \sim - (\sqrt{\alpha'} E_s)\, \partial^{r-1} \left(\frac{1}{z-w}\right) e^{i p_2 X(w)}\;.
\end{align}
Another source of factors of $E_s$ is the contraction of $\partial^r X^+(z)$ with the tensor part of the external vertex operator when this describe a massive state. In this case we have
\begin{equation}
	\label{eq:dXdX}
	i \sqrt{\frac{2}{\alpha'}} \partial^r X^+(z)\; i \partial^s X^M (w) \sim (\sqrt{\alpha'} E_s)\, \frac{v_i^M}{m} \partial^{r-1}_z \partial^{s-1}_w \frac{1}{(z-w)^2}\;,
\end{equation}
where $v_i^M$ indicates the longitudinal polarization for a massive state of spatial momentum $\vec{p}_i$. It is then clear that the leading contributions to~\eqref{eq:factre1} come from the states ${\cal Q}_\ell$ that have the highest number of factors of $X^+$, so, at level $\ell$, it is convenient to have $(\partial X^+)^\ell$ rather then structures with higher derivatives such as $\partial^r X^+$. Thus in summary, in the high-energy regime, the relevant states exchanged between the incident string and the stack of D$p$-branes are\footnote{The overall factor of $1/\ell!$ ensures that the two point function is normalized to one.}
\begin{equation}
	\label{eq:lreVbos}
	{\cal Q}^R_\ell = \frac{1}{\ell !} \left(i \sqrt{\frac{2}{\alpha'}} \partial X^{+}\right)^\ell \left(i \sqrt{\frac{2}{\alpha'}} \bar\partial X^+ \right)^\ell e^{i q X(z,\bar{z})}\;.
\end{equation}
Beside being off-shell these states do not satisfy exactly the BRST constraint also because $e^+ q \sim 1/E$, but in the high-energy regime this violation becomes irrelevant so it does not affect our calculation. Notice that at level $\ell$ such state yields a contribution proportional to $(\sqrt{\alpha'} E_s)^{2\ell}$, so only the graviton ($\ell=1$) yields a result compatible with a classical eikonal contribution, while all contributions of higher spin states ($\ell\geq 2$) grow too quickly with the energy. It is only after resumming the contributions of the whole tower of states in the leading Regge trajectory that one finds the eikonal result~\eqref{eq:zto1ap}.

The basic idea of the Reggeon vertex approach is to perform the sum over the states~\eqref{eq:lreVbos} formally at the operator level. We first notice that the scalar product $\langle {\cal Q}^R_\ell| B \rangle$ is independent of $\ell$ thanks to the normalization in~\eqref{eq:lreVbos} and one is left with only the normalization of the boundary state in \eqref{b1}. Thus, when focusing on the contributions of the states~\eqref{eq:lreVbos}, Eq.~\eqref{eq:factre1} reads
\begin{align}
	{\cal A} & \simeq \frac{N T_p \kappa_d}{2} \langle \hat{\cal V}_{1} \hat{\cal V}_{2} \int_0^\infty\!\!\! d x\, \sum_{\ell=0}^\infty \frac{1}{\ell!}\left[x \left(i \sqrt{\frac{2}{\alpha'}} \partial X^{+} i \sqrt{\frac{2}{\alpha'}} \bar\partial X^{+}\right)\right]^{\ell} x^{-2-\frac{\alpha' t}{4}}\,e^{i q X} \rangle \nonumber \\ \label{eq:sumell} & = \frac{N T_p \kappa_d}{2}  \langle \hat{\cal V}_{1} \hat{\cal V}_{2} \int_0^\infty\!\!\! d x\, e^{x \frac{2}{\alpha'} i \partial X^{+} i \bar\partial X^{+}} x^{-2-\frac{\alpha' t}{4}}\, e^{i q X} \rangle \\ \nonumber 
	&=  \frac{N T_p \kappa_d}{2} {\langle \hat{\cal V}_{1} \hat{\cal V}_{2} \left[i \sqrt{\frac{2}{\alpha'}} \partial X^{+}\, i \sqrt{\frac{2}{\alpha'}} \bar\partial X^{+} \right]^{1+\frac{\alpha' t}{4}} e^{i q X} \rangle \Gamma\!\left(-1-\frac{\alpha' t}{4}\right)} \nonumber \\ 
     \nonumber  & = \frac{N T_p \kappa_d}{2} e^{-i \frac{\alpha' t}{4}}\, \Gamma\!\left(-1-\frac{\alpha' t}{4}\right) \langle \hat{\cal V}_{1} \hat{\cal V}_{2} {\cal V}_R\rangle = \Pi^{D_p}_R \langle \hat{\cal V}_{1} \hat{\cal V}_{2} {\cal V}_R\rangle  \ .
\end{align}
where the hat on the vertices means that they are stripped of their normalization as in~\eqref{eq:tachvo}. As already mentioned, $x= z \bar{z}$ and, at high energy, we can extend the integral over $x$ from the interval $(0,1)$ to $(0,\infty)$. In the manipulations above we treated the combination appearing the square parenthesis as a positive quantity as its leading contribution when inserted in a correlator is $\alpha E_s^2$ as follows from~\eqref{eq:dXpeix}. Then the integral over $x$  has to be evaluated via an analytic continuation on $E_s^2$ and this is the origin of the phase in the final step. The symbols $\Pi^{D_p}_R$ and $ {\cal V}_{R}$ indicate the Reggeon propagator glued to the boundary state and the Reggeon vertex
\begin{gather}
	\Pi^{D_p}_R = \frac{N T_p}{2} \, e^{-i \frac{\alpha' t}{4}}\, \Gamma\!\left(-1-\frac{\alpha' t}{4}\right)\, \;, \label{eq:VrPra}  \\ \label{eq:VrPrb}
	{\cal V}_R =  \kappa_d \, \left[\sqrt{\frac{2}{\alpha'}} {i\partial X^{+}}\, \sqrt{\frac{2}{\alpha'}} {i\bar\partial X^{+}} \right]^{1+\frac{\alpha' t}{4}}\!\!\! e^{i q X}\,.    
\end{gather} 
It is convenient to separate a Reggeon propagator and its coupling to the D$p$-branes $\Pi^{D_p}_R = \Pi_R \langle {\cal V}_R|B\rangle$ as follows
\begin{equation} 
	\label{eq:PiBspl}
	\Pi_R =\frac{1}{2\pi} e^{-i \frac{\alpha' t}{4}} \frac{\Gamma\!\left(-1-\frac{\alpha' t}{4}\right)}{\Gamma\!\left(2+\frac{\alpha' t}{4}\right)} \;, \qquad
	\langle {\cal V}_R|B\rangle = 2\pi\,\frac{N T_p}{2} \Gamma\!\left(2+\frac{\alpha' t}{4}\right)\;.
\end{equation}
The motivation for doing so is that the Reggeon vertex $\hat{\cal V}_R$ and propagator $\Pi_R$ can be used to derive the high-energy limit of pure closed string amplitudes, as done in Section~\ref{ssec:string-brane-sup}. At this stage the split in~\eqref{eq:PiBspl} is somewhat arbitrary, but it can be justified by sketching how to adapt the steps in~\eqref{eq:sumell} to the tree-level amplitudes with four external closed strings. In this case, the integrand in~\eqref{eq:factre1} takes the schematic form $(z \bar{z})^{\ell-2-\frac{\alpha' t}{4}} \langle {\cal V}_{1} {\cal V}_{2} {\cal Q}_\ell \rangle  \langle {\cal Q}_\ell  {\cal V}_{3} {\cal V}_{4}\rangle$ and the holomorphic and the anti-holomorphic sectors are independent. It is then convenient to factorize the calculation by summing independently over the number of the holomorphic and the anti-holomorphic insertions since the level matching condition is imposed at the end by the integral over the phase of $z$. By focusing on the operator part, we need to consider the integral
\begin{align}
	\int d^2 z \sum_{\ell,\bar\ell=0}^\infty & \left[\frac{1}{\sqrt{\ell!}}\left(z\, i \sqrt{\frac{2}{\alpha'}} \partial X^{+}\right)^{\ell} \frac{1}{\sqrt{\bar{\ell}!}}\left(\bar{z} \,i \sqrt{\frac{2}{\alpha'}} \bar\partial X^{+}\right)^{\bar{\ell}}\right] (z \bar{z})^{-2-\frac{\alpha' t}{4}} \nonumber\\ &\left[\frac{1}{\sqrt{\ell!}}\left(i \sqrt{\frac{2}{\alpha'}} \partial X^{+}\right)^{\ell} \frac{1}{\sqrt{\bar{\ell}!}}\left(i \sqrt{\frac{2}{\alpha'}} \bar\partial X^{+}\right)^{\bar{\ell}}\right]\;.
	\label{eq:clsPi}
\end{align}
A standard approach is to rewrite the factor of $(z \bar{z})^{-2-\frac{\alpha' t}{4}} $ as an exponential by introducing a Schwinger parameter
\begin{equation}
	\label{eq:zsp}
	(z \bar{z})^{-2-\frac{\alpha' t}{4}} =\int_0^\infty \!d\tau\, e^{-|z|^2 \tau} \frac{ \tau^{1+\frac{\alpha' t}{4}}}{\Gamma\left(2+\frac{\alpha' t}{4}\right)}\;.
\end{equation}
Then the 2D integral over $z$ in~\eqref{eq:clsPi} is Gaussian and the final integral over $\tau'=1/\tau$ yields a $\Gamma$-function. Then one can see that the tree-level amplitude with four closed strings is captured, at high energy, by the correlator $ \langle \hat{\cal V}_{1} \hat{\cal V}_{2} \hat{\cal V}_R\rangle \Pi_R \langle \hat{\cal V}_R \hat{\cal V}_{3} \hat{\cal V}_{4} \rangle$ involving the same Reggeon vertex~\eqref{eq:VrPrb} as before and the propagator $\Pi_R$ in~\eqref{eq:PiBspl}.

It is straightforward to follow the steps discussed in Section~\ref{ssec:seikop} and obtain the bosonic eikonal operator from the Reggeon vertex~\eqref{eq:VrPrb}. The final result has the same structure as in the superstring case~\eqref{eq:Ahat0}, just with the overall factor which follows from~\eqref{eq:PiBspl}
		\begin{equation}
			\label{eq:Ahat0bos}
			\hat{\cal A}_0 \simeq \frac{N T_p \kappa_d}{2} e^{-i \frac{\alpha' t}{4}}\, \Gamma\!\left(-1-\frac{\alpha' t}{4}\right) \, (\alpha' E^2_s )^{1+\frac{\alpha' t}{4}} \int_0^{2\pi}\! \frac{d\sigma}{2\pi} :e^{i q \hat{X}}:\;.
                \end{equation}

\subsection{String-brane scattering: the bosonic theory at one-loop}
\label{ssec:string-brane-bos-2}

In the closed string channel the annulus amplitudes can be evaluated
using two boundary states. For instance the one-loop correction to the amplitude in~\eqref{eq:TTDp} is
\begin{equation}
	\label{eq:diam1l}
	{\cal A}^{T}_1 = C_{S^2} \left(\frac{\alpha' \kappa_d}{8\pi} N\right)^2 \left(\frac{\kappa_{d}}{2\pi}\right)^2 \frac{1}{4\pi} \int \frac{d^2q}{|q|^2} d^2z_1 d^2z_2 \langle B |\hat{\cal V}_T(z_1) \hat{\cal V}_T(z_2)  q^{L_0-1} \bar{q}^{\bar{L}_0-1} |B \rangle~,
\end{equation}
where, as in~\eqref{eq:TTDp}, we inserted a factor of $\frac{\alpha' \kappa_d}{8\pi} N$ for each closed string propagator $P$ \eqref{eq:clospr}, a factor of $\frac{\kappa_{d}}{2\pi}$ for each vertex and a symmetry factor $\frac{1}{4\pi}$ related to the residual symmetries of the annulus. 

The contribution of the zero modes $q^\mu$ and $p^\mu$ to~\eqref{eq:diam1l} is
\begin{equation}
	\label{Zerom}
	(2 \pi)^{p+1} \delta^{(p+1)} (p_1 + p_2) (2 \pi^2 \alpha' \lambda)^{-\frac{d-p-1}{2}} {\rm e}^{\frac{\alpha'}{2 \pi\lambda} \left[
		\left(E^2_s+ \frac{4}{\alpha'}\right)   \left(\log \frac{|z_1|}{|z_2|}\right)^2 - t \log |z_1| \log |z_2|  \right]} ~,
\end{equation}
where we introduced $\lambda$ via $\log|q|=-\pi \lambda$ and used the on-shell conditions for the external tachyons $m_1^2=m_2^2=-4/\alpha'$. The contribution of the non-zero modes has two effects. First it yields the usual annulus measure
\begin{equation}
	\label{eq:meas1l}
	d\mu_1 = 2\pi d\lambda \frac{1}{ |q|^2} \prod_{n=1}^{\infty} \frac{1}{\left(1 - |q|^{2n}\right)^{d-2}}~.
\end{equation}
Then it transforms the disk Green function $\log(z_i-z_j)$ into the annulus one, which can be
expressed in terms of the prime form $\log E(z_i,z_j)$, where
\begin{equation}
	E(z_i, z_j ) =   (z_i - z_j)  \prod_{n=1}^{\infty} 
	\frac{  \left( 1 - |q|^{2n} \frac{z_{i}}{z_{j}} \right) 
		\left(1 - |q|^{2n} \frac{z_j}{z_i}  \right) }{( 1 - |q|^{2n})^2 }    \ , 
	\label{Ez1z2}
\end{equation}
or in terms of the Jacobi $\theta$-function~\eqref{eq:th1P}
\begin{equation}
	E(z_i , z_j) = 2 \pi i  \,\,{\rm e}^{i \pi (\nu_i + \nu_j)} 
	\frac{\theta_{1} ( \nu_i - \nu_j | i\lambda )  }{\theta_{1} ' (0 | i\lambda) }
	\label{Eab}~.
\end{equation}
Then we can generalize the disk integrand derived in~\ref{ssec:string-brane-bos-1} to the annulus topology. As done in Section~\ref{ssec:1lstrng} we introduce the variables
\begin{equation}
	z_{i } = {\rm e}^{2 \pi i \nu_{i}}\equiv {\rm e}^{2 \pi i (i \lambda \rho_1 - \omega_1)}  ~,~~~z_{j} = {\rm e}^{2 \pi i \nu_j}\equiv {\rm e}^{2 \pi i (i \lambda \rho_2 - \omega_2)} ~
	\label{eq:zrn}
\end{equation}
and obtain the following result for the one-loop amplitude
\begin{align}
	\label{eq:diam1l2}
	{\cal A}^{T}_1 & =  \left(\frac{\kappa_{d} T_p N}{2}\right)^2 \frac{\alpha'}{16\pi}  (2 \pi^2 \alpha')^{-\frac{d-p-1}{2}} (2\pi)^4 
	\int_{0}^{\infty}   \frac{d \lambda}{\lambda^{\frac{d-p-5}{2}}} \frac{1}{ |q|^2} \prod_{n=1}^{\infty} \frac{1}{\left(1 - |q|^{2n}\right)^{d-2}}
	\\ \nonumber  \times &
	4 \int_0^{\frac{1}{2}} \!\! d \rho_1 \! \int_0^{\frac{1}{2}} \!\! d \rho_2 \! \int_0^1 \!\! d\omega_1 \! \int_0^1 \!\! d\omega_2  \,\, {\rm e}^{- (\alpha' E^2_s +2) V_s - \frac{\alpha' t}{4} V_t } \frac{ (\theta_1 ' (0 | i \lambda))^4\,\, {\rm e}^{ 4 \pi \lambda \rho^2}  }{(2\pi)^4 \theta_1^2 ( i \lambda \rho - \omega |  i \lambda )  \theta_1^2 (- i \lambda \rho - \omega | i \lambda)}~,
\end{align}
where the factor $4$ in the second line comes from the normalization of $d^2 z_1$ and $d^2 z_2$ discussed after \eqref{eq:clospr}, we switched to the variables introduced in~\eqref{eq:rhoomega}
and the functions $V_s$, $V_t$ are exactly those appearing in the superstring amplitude defined in~\eqref{eq:Vsti}.
As in the superstring case, the kinematic configuration we are interested in (large $E_s$ and small $R_p/b$) implies that the integral is dominated by the region of small $\rho$ and large $\lambda$. In this limit we have
\begin{align}
	{\cal A}^{T}_1 \sim & \left(\frac{\kappa_{d} T_p N}{2}\right)^2 \frac{\alpha'}{16\pi}  (2 \pi^2 \alpha')^{-\frac{d-p-1}{2}} (2\pi)^4
	\int_{0}^{\infty}   \frac{d \lambda}{\lambda^{\frac{d-p-5}{2}}} \,\,\, {\rm e}^{2   \pi \lambda }  
	\int_0^1 d \zeta \int_0^1 d \omega  \nonumber \\ \label{rho=0}
	\times & 2 \int_{\mathcal R(\zeta)}\!\!\! d \rho \,\, {\rm e}^{2 \pi \alpha' E^2_s \lambda \rho^2}
	{\rm e}^{ 2 \pi \lambda \zeta (1 - \zeta) \frac{\alpha' t}{4} } ( 4 \sin^2 \pi \omega)^{- \frac{\alpha' t}{4}}
	\\ \nonumber
	\times &
	\exp \left[4 \alpha' E^2_s \sin^2 (\pi \omega) \left( {\rm e}^{- 2 \pi \lambda \zeta} + {\rm e}^{- 2 \pi \lambda(1-  \zeta)} \right) \right]  \left(4 \sin^2 (\pi \omega)\right)^{-2}~,
\end{align}
where the region of integration ${\mathcal R}(\zeta)$ is defined after~\eqref{eq:BVVB2bis}. The final factor in the last line comes from the last fraction in Eq.~(\ref{eq:diam1l2}) and is not present in the superstring case as one can see by comparing~\eqref{eq:diam1l2}
and~\eqref{eq:BVVB2bis}. As we will soon see, this difference is related to the presence of a tachyonic state in the bosonic theory. The integral over $\rho\sim 0$ is Gaussian (after a Wick rotation $E_s\to i E_e$). By writing the exponential in the last line as a double series of terms proportional to $e^{-2\pi n \lambda \zeta}$ and $e^{-2\pi m \lambda (1-\zeta)}$ we obtain an expression very similar to the integrand $I_1$ in Appendix~A of~\cite{DAppollonio:2010ae}. Then the integral over $\omega$ can also be performed and one obtains
\begin{eqnarray}
	{\cal A}^{T}_1 & \sim & \left(\frac{\kappa_{d} T_p N}{2}\right)^2 \frac{\alpha'}{8\pi}  (2 \pi^2 \alpha')^{-\frac{d-p-1}{2}} (2\pi)^4
	\int_0^1 d \zeta     \frac{i}{\sqrt{2 \alpha' E_s^2}}  (\alpha' E^2_s)^{2} 4^{- \frac{\alpha' t}{4}}    \nonumber \\
	&\times &    \sum_{n,m=0}^{\infty} \frac{1}{n! m!}  \left( 4 \alpha' E^2_s \right)^{(n-1) + (m-1)}     
	\frac{1}{\pi} B\left(\frac{1}{2} , n+m-2  - \frac{\alpha't}{4} + \frac{1}{2} \right) \label{gty4}
	\\
	&\times &  \Gamma\left( -\frac{d-p-6}{2} \right) \pi^{\frac{d-p-6}{2}} \left[ -2  \zeta (1- \zeta) \frac{\alpha' t}{4} +  2(n-1) \zeta
	+ 2 (m-1)  (1- \zeta) \right]^{\frac{d-p-6}{2}} \! . \nonumber
\end{eqnarray}
where the last line comes from the integral over $\lambda$.
This expression is very similar to the superstring case, 
except that $n,~m$ are shifted to $n-1,~m-1$, due 
to the presence of the 
tachyon pole ($e^{2\pi\lambda}$) in the first line of~\eqref{rho=0} and of the last factor ($sin^{-4} (\pi \omega)$) in the final line of the same equation. We can trade the integral over $\zeta$ for a momentum integral in $D=d-p-2$ dimensional space by using the identity
\begin{eqnarray}
	\int_0^1 d \zeta \,\, \Gamma\left( -\frac{d-p-6}{2} \right)   \left[ -2   \zeta (1- \zeta) \frac{\alpha' t}{4} +  2 (n-1) \zeta + 2  (m-1)  (1- \zeta) \right]^{\frac{d-p-6}{2}} \nonumber \\
	= (2 \pi \alpha' )^{\frac{d- p-2}{2}}  \int  \frac{ d^{d-p-2} \mathbf{k}}{ (2\pi)^{d-p-2}} 
	\left[ 2(n-1) + \frac{\alpha'}{2} \mathbf{k}^2\right]^{-1} \left[ 2(m-1) + \frac{\alpha'}{2} (\mathbf{k}-\mathbf{q})^2\right]^{-1},
	\label{polesr}
\end{eqnarray}
where we used bold symbols to indicate the transverse $(d-2)$ vectors following the convention introduced in~\eqref{Breitframe}. We can rewrite the sums as integrals by using
\begin{equation}
	\label{eq:sommer}
	\sum_{m=0}^\infty \frac{1}{m!} \frac{f(m) s^m}{m+t} = 
	- \int_{\cal C}\frac{dm}{2\pi i} {\rm e}^{-i \pi m} \Gamma(-m)  \frac{f(m) s^m}{m+t}~,
\end{equation}
where the contour includes all the poles in the $\Gamma(-m)$ and not the other ones. We can then
deform the contour and focus on the poles of the propagators in~\eqref{polesr}, which are
the only ones that contribute to the leading term in the energy. We find
\begin{eqnarray}
	{\cal A}^{T}_1 & \sim &
	\left(\frac{\kappa_{d} T_p N}{2}\right)^2 \frac{\alpha'}{8\pi}  (2 \pi^2 \alpha')^{-\frac{d-p-1}{2}} (2\pi)^4 
	\frac{i \pi^{\frac{d-p-6}{2}}}{\sqrt{2 \alpha' E^2_s}} \frac{1}{4} (\alpha' E^2_s)^{2} 4^{- \frac{\alpha' t}{4}}    \nonumber \\
	&\times & (2 \pi \alpha' )^{\frac{d-p-2}{2}}  \int  \frac{ d^{d-p-2} \mathbf{k}}{ (2\pi)^{d-p-2}}   
	\Gamma \left( -1 + \frac{\alpha'}{4} \mathbf{k}^2\right ) \Gamma \left( -1 + \frac{\alpha'}{4} (\mathbf{q}-\mathbf{k})^2 \right) {\rm e}^{i \pi \frac{\alpha'}{4} \mathbf{k}^2 + i \pi \frac{\alpha'}{4} (\mathbf{q}-\mathbf{k})^2}  \nonumber \\
	&\times & \left( 4 \alpha' E^2_s \right)^{- \frac{\alpha'}{4} \mathbf{k}^2 -   \frac{\alpha'}{4}{(\mathbf{q}-\mathbf{k})^2}}   \frac{1}{\pi} B\left(\frac{1}{2} , - \frac{\alpha'}{4} \mathbf{k}^2  -  \frac{\alpha'}{4} (\mathbf{q}-\mathbf{k})^2   - \frac{\alpha't}{4} + \frac{1}{2}\right) + \ldots \;,
	\label{fibuy}
\end{eqnarray}
where the dots stand for the contributions of the other poles that involve also the Euler Beta-function in the second line of~\eqref{gty4}. These contributions are needed to cancel spurious singularities in~\eqref{fibuy}. For instance the last line has a pole when $\alpha'(\mathbf{k}^2-\mathbf{kq})\simeq 1$ which does not correspond to the propagation of a physical state. However, this region is suppressed by a factor of $(\alpha' E_s^2)^{-\frac{1}{2}}$ with respect to the $\alpha' \mathbf{k}^2,\,\alpha' \mathbf{q}^2\ll 1$ since we have $\left( 4 \alpha' E_s^2 \right)^{- \frac{\alpha'}{4} \mathbf{k}^2 -   \frac{\alpha'}{4}{(\mathbf{q}-\mathbf{k})^2}} \sim \left( 4 \alpha' E_s^2 \right)^{- \frac{\alpha'}{4} \mathbf{q}^2-\frac{1}{2} }$, thus it is not reliably captured by~\eqref{gty4}. In particular at this subleading order one should add the contribution neglected in~\eqref{gty4}. Selecting a pole from the Euler Beta-function and the other from~\eqref{polesr}, one obtains  a structure that exactly cancels the spurious pole in~\eqref{gty4} mentioned above. We thus focus on the leading contribution written in~\eqref{gty4} and recast the Euler Beta-function as the correlator introduced~\eqref{eq:evBX}. By using Eq.~\eqref{eq:TTeikL} we can rewrite~\eqref{fibuy} in the following factorized form
\begin{align}
	{\cal A}^{T}_1 \sim &  \left[\frac{\alpha'}{8\pi}  (2 \pi^2 \alpha')^{-\frac{d-p-1}{2}} (2\pi)^4 \frac{i \pi^{\frac{d- p-6}{2}}}{\sqrt{2 \alpha' E^2_s}} \frac{1}{4} (2 \pi \alpha' )^{\frac{d-p-2}{2}}\right]
	\int  \frac{ d^{d-p-2} \mathbf{k} }{ (2\pi)^{d-p-2}} 
	\\ \nonumber \times &\;
	{\cal A}^{T}_0(E_s,\mathbf{k}) {\cal A}^{T}_0(E_s,\mathbf{q}-\mathbf{k}) \langle 0| \prod_{i=1}^{2} \int\limits_0^{2 \pi} \frac{ d \sigma_i}{2 \pi} :{\rm e}^{i \mathbf{k} {\hat{X}}(\sigma_1)} :\,:{\rm e}^{i (\mathbf{q}-\mathbf{k}) {\hat{X}}(\sigma_2)} : | 0 \rangle~.
\end{align}
Since the square parenthesis on the first line is just $i/(4 E_s)$, we indeed obtain~\eqref{eq:Ahstr} for $h=2$ for the case of external tachyon states in bosonic string theory.

\section{More on the \texorpdfstring{$2\to3$}{2->3} kinematics and on the \texorpdfstring{$\times$, $+$}{x, +} waveforms}
\label{KinRela1}

In this appendix we provide a list of relations that apply to the $2\to3$ kinematics discussed at the end of Subsection~\ref{sec:kinematics} and which can be useful, in particular, to manipulate the waveforms presented in Subsection~\ref{sec:wavforlo}.
We also present alternative expressions for the $\times$, $+$ projections of the $b$-space amplitude $\tilde{\mathcal A}^{(5)\mu\nu}_{0}$, i.e.~the leading-order gravitational waveform, that are equivalent to those presented in Subsection~\ref{sec:wavforlo}.

Let us begin by recalling that the incoming momentum vectors $p_1$ and $p_2$ define to the incoming particle's velocities  $v_1^\mu$ and $v_2^\mu$ by \eqref{eq:velocities}, which satisfy $v_1^2=-1$, $v_2^2=-1$, and are related to the Lorenz factor $\sigma$ by $\sigma=-v_1\cdot v_2$ as in \eqref{eq:sigma}.
The projector onto the plane spanned by $v_1^\mu$ and $v_2^\mu$ can be read off from \eqref{decomposition},
\begin{equation}
	P\indices{^\mu_\nu} = - \check v_{1}^{\mu} v_{1\nu} - \check v_2^\mu v_{2\nu} = \frac{1}{\sigma^2-1} \left[ v_{1}^{\mu} v_{1\nu} - \sigma ( v_{1}^{\mu} v_{2\nu} + v_{2}^{\mu} v_{1\nu} ) + v_{2}^{\mu}v_{2\nu}\right],
	\label{Pmunu}
\end{equation}
where $\check v_1^\mu$, $\check v_2^\mu$ are the dual velocities defined by \eqref{check12}.
This projector of course satisfies
\begin{equation}\label{}
	P\indices{^\mu_\nu}v_{1}^\nu= v_1^\mu\,,\qquad
	P\indices{^\mu_\nu}v_{2}^\nu= v_2^\mu\,,\qquad
	P\indices{^\mu_\rho}P\indices{^\rho_\nu} = P\indices{^\mu_\nu}\,.
\end{equation}
For any vector $\xi^\mu$, we may then rewrite its decomposition \eqref{decomposition} into longitudinal and transverse projections as follows,
\begin{equation}\label{}
	\xi^\mu = \xi_\parallel^\mu + \xi_\perp^\mu\,,\qquad  \xi_\parallel^\mu= P\indices{^\mu_\nu} \xi^\nu\,,\qquad
	\xi_\parallel\cdot \xi_\perp =0\,.
\end{equation}
In terms of the invariant products $\sigma=-v_1\cdot v_2\ge1$, $\omega_1=-v_1\cdot k\ge0$ and $\omega_2=-v_2\cdot k$ introduced in \eqref{invariants},
one can explicitly compute the longitudinal projections of various vectors. 
Since we always work in a regime in which the momentum transfers $q_1$, $q_2$ are small, and thus \eqref{p1q1p2q2} hold to leading order, we recover \eqref{q1decompq2decomp},
\begin{equation}
	q_{1\parallel}^\mu \approx -\omega_2 \check v_2^\mu = \frac{\omega_2 (v_{2}^\mu - \sigma v_{1}^\mu)}{\sigma^2-1}
	\,,\qquad
	q_{2\parallel}^{\mu} \approx - \omega_1 \check v_1^\mu = \frac{\omega_1 (v_{1}^{\mu} - \sigma v_{2}^{\mu})}{\sigma^2-1}
	\label{q1paq2pa}
\end{equation}
and define
\begin{equation}
	\Delta^\mu_\parallel \equiv \frac{1}{2 (\sigma^2-1)} \left[ -v_1^\mu (\omega_1+ \sigma \omega_2) + v_2^\mu (\omega_2+\sigma \omega_1) \right] \approx \frac{1}{2} (q_1-q_2)_{\parallel}^\mu
	\label{q1-q2pa}
\end{equation}
for later convenience.
From now on we shall not distinguish between $\approx$ and $=$ signs, for simplicity, since we always work to leading order in the approximation \eqref{p1q1p2q2}.
Then from the following relations we find
\begin{equation}
	b^2 q_{1\parallel}^2= \frac{b^2\omega_2^2}{\sigma^2-1} \equiv \Omega_2^2\,,\qquad
	b^2 q_{2\parallel}^2= \frac{b^2\omega_1^2}{\sigma^2-1} \equiv \Omega_1^2
	\label{q1q2pa}
\end{equation}
and
\begin{equation}
	k\cdot q_{1\parallel} = \frac{\omega_2 (-\omega_2+ \sigma \omega_1)}{\sigma^2-1}\,,\qquad
	k\cdot q_{2\parallel} = \frac{\omega_1 (-\omega_1+ \sigma \omega_2)}{\sigma^2-1}\,.
	\label{kq1q2pa}
\end{equation}
When saturating \eqref{q1-q2pa} with the polarization vector ${\tilde{e}}^\mu_\theta$ defined in \eqref{eeExplicit}, that is
\begin{equation}
	{\tilde{e}}_\theta^\mu = \frac{\omega_1 v_2^\mu - \omega_2 v_1^\mu}{\sqrt{{\cal{P}}}} \,,
\end{equation}
with $\mathcal{P} = - \omega_1^2 + 2\omega_1\omega_2\sigma-\omega_2^2$  as in \eqref{calP}, we find
\begin{equation}\label{}
	\frac{1}{2}(q_1-q_2)_{\parallel} \cdot {\tilde{e}}_{\theta} = \frac{\omega_1 \omega_2}{\sqrt{{\cal{P}}}}\,.
	\label{GGG26k}
\end{equation}

Let us now give the explicit expressions of the previously introduced quantities in terms of $\sigma$, in the center-of-mass frame, choosing the following explicit parametrization for the graviton momentum $k^\mu$ defined in \eqref{kCM},
\begin{equation}
	k^\mu = \omega n^\mu\,,\qquad n^\mu = (1, \sin \theta \cos \phi, \sin \theta \sin \phi, \cos \theta) \,.
	\label{komega}
\end{equation}
We get
\begin{equation}
	\omega_1 = \frac{\omega \left( {{m}}_1 + {{m}}_2 \sigma -\cos \theta {{m}}_2 \sqrt{\sigma^2-1}\right) }{ \sqrt{s}}\,,\qquad 
	\omega_2 = \frac{\omega \left( {{m}}_2 + {{m}}_1 \sigma +\cos \theta {{m}}_1 \sqrt{\sigma^2-1}\right) }{ \sqrt{s}}
	\label{ome1ome2}
\end{equation}
and\footnote{In order to compare with the formalism used in Section~6 of Ref.~\cite{DiVecchia:2021bdo} we need to use the following identities $ c_{1,2}^2 {\mathbf{k}}^2= q^2_{1,2 \parallel}, d_{1,2}{\mathbf{k}}^2 = (k q_{1.2 \parallel})$ and
	$b^2 {\mathbf{k}}^2 f (x) = \Omega^2 (1-x)$ where $\Omega(x)$ is given in \eqref{GGG45p}.}
\begin{equation}
	\begin{split}
		\sqrt{q_{1\parallel}^2} &= \frac\omega{\sqrt s} \frac{m_2+m_1 \sigma + m_1 \sqrt{\sigma^2-1} \cos \theta}{\sqrt{\sigma^2-1}}= \frac{\omega_2}{\sqrt{\sigma^2-1}}\,,  \\
		\sqrt{q_{2\parallel}^2} &=  \frac\omega{\sqrt s} \frac{m_1+m_2 \sigma - m_2 \sqrt{\sigma^2-1} \cos \theta}{\sqrt{\sigma^2-1}}= \frac{\omega_1}{\sqrt{\sigma^2-1}}  
		\label{M1M2a}
	\end{split}
\end{equation}
together with 
\begin{align}
	\begin{split}
		k \cdot q_{1\parallel} &= \omega^2 \frac{ \left( m_2+m_1 \sigma+m_1 \cos \theta  \sqrt{\sigma^2-1} \right)\left( m_2 \sqrt{\sigma^2-1} - (m_1+m_2 \sigma) \cos \theta \right)}{s\sqrt{\sigma^2-1}} \,,  \\
		k \cdot q_{2\parallel} &= \omega^2  \frac{ ( m_1+m_2 \sigma- m_2 \cos \theta \sqrt{\sigma^2-1})(m_1 \sqrt{\sigma^2-1}+ (m_2+m_1 \sigma)\cos \theta)}{  s \sqrt{\sigma^2-1}} \,.
		\label{N1N2a}
	\end{split}
\end{align}
Adopting the notation 
\begin{equation}\label{}
	\mathbf k^\mu = k_\perp^\mu\,,
\end{equation}
we also find
\begin{equation}
	{\cal{P}}= -\omega_1^2 + 2\omega_1 \omega_2 \sigma -\omega_2^2   =(\sigma^2-1) {\mathbf{k}}^2\,,\qquad
	{\mathbf{k}}^2=\omega^2 \sin^2 \theta 
	\label{SLT17}
\end{equation}
In particular, the last two relations make it obvious that $\mathcal P \ge 0$, consistently with \eqref{calP}.

We conclude this appendix by presenting expressions for the $\times$, $+$ projections of the impact-parameter space $2\to3$ amplitude in terms of polarization vectors $e_\theta^\mu$, $e_\phi^\mu$ such that 
\begin{equation}\label{orthogonalitiesE}
	e_\theta\cdot k = 0\,,\qquad e_\phi\cdot k=0\,, \qquad
	e_\phi\cdot v_{i}=0\, ,
\end{equation}
but such that $e_\theta^\mu$ is not necessarily orthogonal to $b^\mu$ (in contrast with the vector $\tilde e_\theta^\mu$ employed in Subsection~\ref{sec:wavforlo}, see Eq.~\eqref{properties2}). 
We thus start from the five-point amplitude
given in Eq.~\eqref{GGG2} (using $M=\mu$, $N=\nu$ and $\beta=\beta^\text{GR}$ as in \eqref{eq:betan8gr} as appropriate for GR),
perform the Fourier transform \eqref{tointegrateW}, thus obtaining $\mathcal W^{\mu\nu}_0 = \tilde{\mathcal{A}}_0^{\mu\nu}$ by following the steps detailed in Subsection~\ref{sec:wavforlo}, and project it along the two polarizations according to
\begin{eqnarray}
	\tilde{\mathcal A}^{(5)}_{\times} = \tilde{\mathcal A}^{(5)\mu\nu}_0 e_{\phi\mu} e_{\theta\nu}\,,\qquad
	\tilde{\mathcal A}^{(5)}_{+} = \tilde{\mathcal A}^{(5)\mu\nu}_0\frac{1}{2} \left(e_{\theta\mu} e_{\theta\nu} - e_{\phi\mu} e_{\phi\nu}\right)\,.
	\label{GGG10}
\end{eqnarray}
Explicitly, in the center-of-mass defined by \eqref{vvbCM} and \eqref{kCM},
where the longitudinal directions are $0$ and $3$, the velocities have vanishing components along the $1$, $2$ axis, while
\begin{equation}
	\begin{aligned}
		v_1^0=& \frac{m_1+m_2 \sigma}{\sqrt{s}}\,,\qquad v_1^3= \frac{m_2\sqrt{\sigma^2-1}}{\sqrt{s}}\,, \\
		v_2^0=& \frac{m_2+m_1 \sigma}{\sqrt{s}}\,,\qquad v_2^3 =-  \frac{m_1\sqrt{\sigma^2-1}}{\sqrt{s}}\,.
		\label{u1u2}
	\end{aligned}
\end{equation}
Then, we choose
\begin{equation}\label{GGG9-1}
	\begin{aligned}
e_{\theta}^\mu =  (0, \cos \theta \cos \phi, \cos \theta \sin \phi, - \sin \theta)\,,\qquad
e_\phi^\mu= (0, -\sin \phi, \cos \phi, 0)\,,
	\end{aligned}
\end{equation}
so that \eqref{orthogonalitiesE} hold and
\begin{equation}\label{GGG9-2}
	e_\theta^2 = 1\,,\qquad
	e_\theta\cdot e_\phi=0\,,\qquad
	e_\phi^2 = 1\,.
\end{equation}
For the  $\times$ polarization, we get  
\begin{eqnarray}
  \tilde{\mathcal A}^{(5)}_{\times} & = & \frac{(8\pi G)^{\frac{3}{2}}}{4Ep (2\pi)}i ({\hat{b}} e_\phi) 
                 \Bigg\{ \beta \Bigg[  \left( \frac{(p_1 e_{\theta}) \Omega_1}{(p_1k) b} e^{ -i b k/2} K_1 ( \Omega_1)  - \frac{(p_2 e_{\theta}) \Omega_2}{(p_2k) b } e^{ i b k/2} K_1 ( \Omega_2)  \right) \nonumber \\
                                    &+ &  i  \int_0^1 dx \,e^{ -i \frac{kb}{2} (2x-1)}
              \Bigg(  ({\hat{b}} e_\theta) \Omega (x)  K_1 (\Omega (x)) + i  (x-\frac{1}{2}) ({\bf k} e_\theta) b K_0 ( \Omega (x))  \Bigg)
	\nonumber \\
	&+&  (e_\theta  \Delta_\parallel)  b \int_0^1 dx \,e^{-i \frac{kb}{2} (2x-1)} K_0 (\Omega (x))  \Bigg] \nonumber \\
	&-& 4  p_1p_2   ((p_1k) (p_2 e_\theta) - (p_2k) (p_1 e_\theta)) 
	\int_0^1 dx \,e^{-i \frac{kb}{2} (2x-1)} b K_0 ( \Omega (x))  \Bigg\} ,
	\label{GGG20}
\end{eqnarray}
where for GR we have (see Eq.~\eqref{eq:betan8gr})
\begin{equation}
	\beta= \beta^\text{GR} =  2m_1^2 m_2^2 (2\sigma^2-1)\,,
	\label{betaGR}
\end{equation}
while, for the polarization $+$ we get
\begin{eqnarray}
  \tilde{\mathcal A}^{(5)}_{+} & = & \frac{(8\pi G)^{\frac{3}{2}}}{4Ep (4\pi)}
	\Bigg\{ \beta \Bigg[ \frac{(p_1 e_\theta)^2  }{(p_1k)^2}
	e^{- i k b/2}  \left( (k q_{2\parallel})  K_0 ( \Omega_1) - i (k\hat{b})\, \frac{\Omega_1}{b} K_1 ( \Omega_1)  \right)
	\nonumber \\
	& + &  \frac{(p_2 e_\theta)^2 }{(p_2k)^2 } e^{ i k b/2}   \left( (kq_{1\parallel}) K_0 ( \Omega_2) + i (k\hat{b})\, \frac{\Omega_2}{b}  K_1 ( \Omega_2)  \right) \nonumber \\
	&+ & \frac{p_1 e_\theta}{p_1k}e^{- i b k/2}  \left((- ( e_\theta {\bf k}) +2 (e_\theta \Delta_\parallel)   )  K_0 (\Omega_1)   + 2i  (e_\theta {\hat{b}})  \frac{\Omega_1}{b}  K_1 (\Omega_1)\right) \nonumber \\
	&- & \frac{p_2 e_\theta}{p_2k} e^{ i b k/2}  \left( (( e_\theta {\bf k}) + 2(e_\theta \Delta_\parallel)   )  K_0 ( \Omega_2)  +2i  (e_\theta {\hat{b}})  \frac{\Omega_2}{b} K_1 (\Omega_2)\right) \nonumber \\
	&+ & 2 (e_\theta \Delta_\parallel)^2    \int_0^1 dx \,e^{-i \frac{kb}{2} (2x-1)}  \left(\frac{b^2}{2}\frac{K_1 (\Omega (x))}{ \Omega (x)}\right)  + 2 (e_\theta \Delta_\parallel)   \nonumber \\
	& \times & \int_0^1 dx  e^{-i \frac{kb}{2} (2x-1)}\Bigg(   (e_\theta {\bf k})(1-2 x)  \left( 
	\frac{b^2}{2}  \frac{K_1 (\Omega (x))}{\Omega (x)} \right) + i (e_\theta {\hat{b}})
	b K_0 (\Omega (x))\Bigg) \nonumber \\
	&+ & \frac{1}{2} \int_0^1 dx  e^{-i \frac{kb}{2} (2x-1)} \Bigg[ 2 (e_{\theta \perp} e_{\theta\perp})  K_0 (\Omega (x)) - 2 (e_\theta {\hat{b}})^2 \Omega (x) K_1 (\Omega (x) )
	\nonumber \\
	&+ &  (e_\theta {\bf k})^2(1-2x)^2 \left( \frac{b^2}{2}  \frac{K_1 (\Omega(x))}{\Omega (x)} \right)  +2i (e_\theta {\bf k}) (e_\theta {\hat{b}}) (1-2x)  b K_0 (\Omega (x))\Bigg] \nonumber \\
	&+ &\int_0^1 dx \,e^{-i \frac{kb}{2} (2x-1)} \left[ - (e_\phi e_\phi) K_0 ( \Omega (x)) +  (e_\phi {\hat{b}})^2  \Omega (x) K_1 (\Omega (x) )\right] \Bigg]\nonumber \\ 
	&+&8 \left( (p_1k)   (p_2 e_\theta)  - (p_2 k) (p_1 e_\theta) \right)^2 \int_0^1 dx \,e^{-i \frac{kb}{2} (2x-1)}  \left(\frac{b^2}{2}\frac{K_1 (\Omega (x))}{\Omega (x)}\right)  \nonumber \\
	&+& (8p_1p_2) \left[\left( (p_1 e_\theta)^2  \frac{kp_2}{kp_1}- (p_1 e_\theta)  (p_2 e_\theta) \right) e^{- i b k/2} K_0 ( \Omega_1)  \right. \nonumber \\
	&+& \left( (p_2 e_\theta)^2   \frac{kp_1}{kp_2}-  (p_1 e_\theta) (p_2 e_\theta)\right) e^{ i b k/2} K_0 (\Omega_2) 
	\nonumber \\
	&+ &  \left((p_1 k) (p_2 e_\theta)  -(p_2 k) (p_1 e_\theta)  \right) \int_0^1 dx \,e^{-i \frac{kb}{2} (2x-1)} 
	\left( - i (e_\theta {\hat{b}}) b K_0 ( \Omega (x))  \right.
	\nonumber \\
	&- & \left.  (e_\theta \mathbf{k}) (\frac{1}{2}-x) \frac{ b^2 K_1 ( \Omega (x))}{\Omega (x)} 
	\left. - 2(e_\theta \Delta_\parallel)  \left(\frac{b^2}{2} \frac{K_1 ( \Omega (x))}{\Omega (x)}\right)  \right) \right] 
	\Bigg\},
	\label{GGG45}
\end{eqnarray}
where
\begin{equation}
	\Omega (x) = \sqrt{\Omega_1^2 x^2 + \Omega_2^2 (1-x)^2 + 2 \Omega_1 \Omega_2 \sigma x(1-x)}\,.
	\label{GGG45p}
\end{equation}
Note that the vector $\tilde e_{\theta}^\mu$ employed in Subsection~\ref{sec:wavforlo} and the vector $e_\theta^\mu$ employed here differ by a vector proportional to $k^\mu = \omega \, n^\mu$ 
\begin{equation}
	{\tilde{e}}^\mu_{\theta}=  
	(- \cot \theta, 0, 0, -1/\sin \theta)
	=
	e_{\theta}^\mu - \cot \theta \, n^\mu 
	\,.
	\label{GGG9-3}
\end{equation}
Therefore, by gauge invariance, either choice must actually yield the same projected waveforms. We have verified explicitly, as a check of the correctness of our the expressions, that the sum of Eqs.~\eqref{A12x}, \eqref{Airrx} for the $\times$ polarization agrees with Eq.~\eqref{GGG20} and the sum of Eqs.~\eqref{A12+}, \eqref{Airr+} for the $+$ polarization agrees with Eq.~\eqref{GGG45}, as a consequence of $k^\mu \tilde{\mathcal{A}}_0^{\mu\nu}=0$ (taking into account the overall factor in \eqref{WcalW}).

\section{From the field to the asymptotic waveform }
\label{app:asymptoticlimit}
From the eikonal operator \eqref{eikopv1}, or from its soft versions \eqref{eq:eiksr} or \eqref{eq:eiksrstatic}, we obtain expressions for the expectation of the canonically normalized field \eqref{eq:theGravitonField} in the final state of the form
\begin{equation}\label{hmunu}
	h^{\mu\nu}(x)
	=
	\int_k \left[
	e^{ik\cdot x}
	i
	\tilde{\mathcal{T}}^{\mu\nu}(k)
	-
	e^{-ik\cdot x}
	i
	\tilde{\mathcal{T}}^{\mu\nu}(k)^\ast
	\right]
\end{equation}
with $\tilde{\mathcal T}^{\mu\nu}(k)=\mathcal W_\text{TT}^{\mu\nu}(k)$ or $\tilde{\mathcal T}^{\mu\nu}(k)=-i w_\text{TT}^{\mu\nu}(k)$ or $\tilde{\mathcal T}^{\mu\nu}(k)=-i F_\text{TT}^{\mu\nu}(k)$ respectively in those three cases. An important property that can be seen to hold in all three applications considered in the text is
\begin{equation}\label{Hproperty}
	\tilde{\mathcal{T}}^{\mu\nu}(-k)=\tilde{\mathcal{T}}^{\ast\mu\nu}(k)\,.
\end{equation}
The prediction \eqref{hmunu} in only reliable for the gravitational field sourced by the collision far away from the sources, and one thus faces the issue of taking the asymptotic limit of \eqref{hmunu} in a null direction. This amounts to considering a detector placed, spatially, very far away for retarded times comparable to the one at which the scattering event takes place. To this end, we introduce the decomposition 
\begin{equation}\label{xdecomp}
	x^\mu = u \, t^{\mu} + r \, n^\mu\,,
\end{equation}
where $t^\mu$ is the detector's four-velocity, $t^2 = -1$, and ${n}^\mu$ is a future-directed null vector, such $ n^2=0$ and $-t\cdot n=1$, characterizing its direction. 
In this way, $u$ describes the retarded time of the asymptotic detector and $r>0$ represents its distance from the source, so we want to take the limit
\begin{equation}\label{}
	r\to\infty\,,\qquad
	u,\ n^\mu \text{ fixed}.
\end{equation}

It is convenient to change integration variable in \eqref{hmunu} in a similar way, letting
$k^\mu = \rho\, t^\mu + \omega\, m^\mu$
with $\omega>0$, and such that $m^2=0$ and $-t\cdot m=1$.
Taking into account that the corresponding metric reads
\begin{equation}\label{}
	(dk)^2 = -d \rho^2 +2 d\rho\,d\omega+\omega^2(dm)^2\,,
\end{equation}
we find that Eq.~\eqref{hmunu} takes the following form
\begin{equation}\label{}
	h^{\mu\nu}(x) = i\int_0^\infty \frac{d\omega}{2\omega}\,\omega^{D-2}\int\frac{d^{D-2}m}{(2\pi)^{D-1}}\, \tilde{\mathcal{T}}^{\mu\nu}(\omega  m)\, e^{-i \omega  u + i \omega r \,n\cdot m}+ \text{(c.c.)}\,,
\end{equation}
(where c.c. stands for ``complex conjugate'').
The fact that $r\to\infty$ in the last term in the  exponent can be compensated by letting $m = n+z$, where the new integration variable $z$ must obey $t\cdot z=0$ and $z\cdot n=-z^2/2$, and focusing on the region  $z\sim \mathcal O(1/\sqrt{r})$. Then, to leading order we can approximate $m\sim n$ everywhere else, obtaining
\begin{equation}\label{}
	h^{\mu\nu}(x) \sim i\int_0^\infty \frac{d\omega}{2\omega}\,\omega^{D-2}\tilde{\mathcal{T}}^{\mu\nu}(\omega  n)\,e^{-i \omega  u} \int\frac{d^{D-2}z}{(2\pi)^{D-1}}\, e^{-\frac{i}{2}\omega r z^2}+ \text{(c.c.)}\,.
\end{equation}
Performing the Gaussian integral over $z$, we thus obtain the desired asymptotic limit,
\begin{equation}\label{}
	h^{\mu\nu}(x) \sim \frac{i}{2(2i \pi r)^{\frac{D-2}{2}}}\int_{0}^\infty\frac{d\omega}{2\pi}\,\omega^{\frac{D-4}{2}}\tilde{\mathcal{T}}^{\mu\nu}(\omega n) e^{-i \omega  u}+ \text{(c.c.)}\,.
\end{equation}
Although the previous expression holds in arbitrary dimensions, we can focus on $D=4$, 
\begin{equation}\label{}
	h^{\mu\nu}(x) \sim \frac{1}{4\pi r}\int_{0}^\infty\frac{d\omega}{2\pi}\,\tilde{\mathcal{T}}^{\mu\nu}(\omega\,n) e^{-i \omega  u}+ \text{(c.c.)}
\end{equation}
and finally, using \eqref{Hproperty}, we note that the two terms combine to yield a single integral over ``positive and negative frequencies'',
\begin{equation}\label{}
	h^{\mu\nu}(x) \sim \frac1{4 \pi r}\int_{-\infty}^\infty\frac{d\omega}{2\pi}\,\tilde{\mathcal{T}}^{\mu\nu}(\omega  n) e^{-i {u}\omega}\,.
\end{equation}

\biboptions{compress}

\end{document}